\documentclass[fleqn,usenatbib]{mnras}

\usepackage[T1]{fontenc}

\DeclareRobustCommand{\VAN}[3]{#2}
\let\VANthebibliography\thebibliography
\def\thebibliography{\DeclareRobustCommand{\VAN}[3]{##3}\VANthebibliography}

\usepackage{graphicx}	% Including figure files
\usepackage{amsmath}	% Advanced maths commands
\usepackage{amssymb}	% Extra maths symbols
\usepackage{xspace}     % adds a space after a macro
\usepackage{booktabs}   % extra commands for tables
\usepackage{pdflscape}  % display pages in landscape mode
\makeatletter
\let\LS@vtryfc\@vtryfc
\g@addto@macro\landscape{%
  \def\@vtryfc#1{%
    \LS@vtryfc{#1}%
    \LS@rot
  }%
}
\makeatother
\usepackage{makecell}   % line breaks in table cells
\usepackage{subdepth}  % equalise the height of subscripts
\usepackage{longtable}  % additional format for long tables
\usepackage{newfloat}   % to force figures in appendices to appear below section name
\DeclareFloatingEnvironment[placement={!ht},name=Figure, within=section]{myfloat}
\usepackage[font=small, singlelinecheck=off]{caption} % to use with long figures and tables in Appendices B, C and D. Suggested command, \contcaption, produced misalignment of figures.
\usepackage{newtxtext}
\usepackage[varvw]{newtxmath}

\newcommand{\po}{\phantom{1}}
\newcommand{\pp}{\phantom{$-$}}
\newcommand{\pd}{\phantom{.}}

\newcommand{\kms}{\,km\,s$^{-1}$\xspace} % km/s
\newcommand{\Msun}{$\,M_\odot$\xspace}  % solar mass
\newcommand{\Rsun}{$\,R_\odot$\xspace}  % solar radius
\newcommand{\lc}{LCT\xspace}    % lines combination test
\newcommand{\rvf}{\texttt{rvfit}\xspace} % rvfit software

\defcitealias{sana+12}{S12}
\defcitealias{kobulnicky+14}{K14}
\defcitealias{almeida+17}{TMBM}
\newcommand{\sana}{\citetalias{sana+12}\xspace}
\newcommand{\kobu}{\citetalias{kobulnicky+14}\xspace}
\newcommand{\tmbm}{\citetalias{almeida+17}\xspace}

\newcommand{\orcid}[1]{\href{https://orcid.org/#1}{\includegraphics[width=8pt]{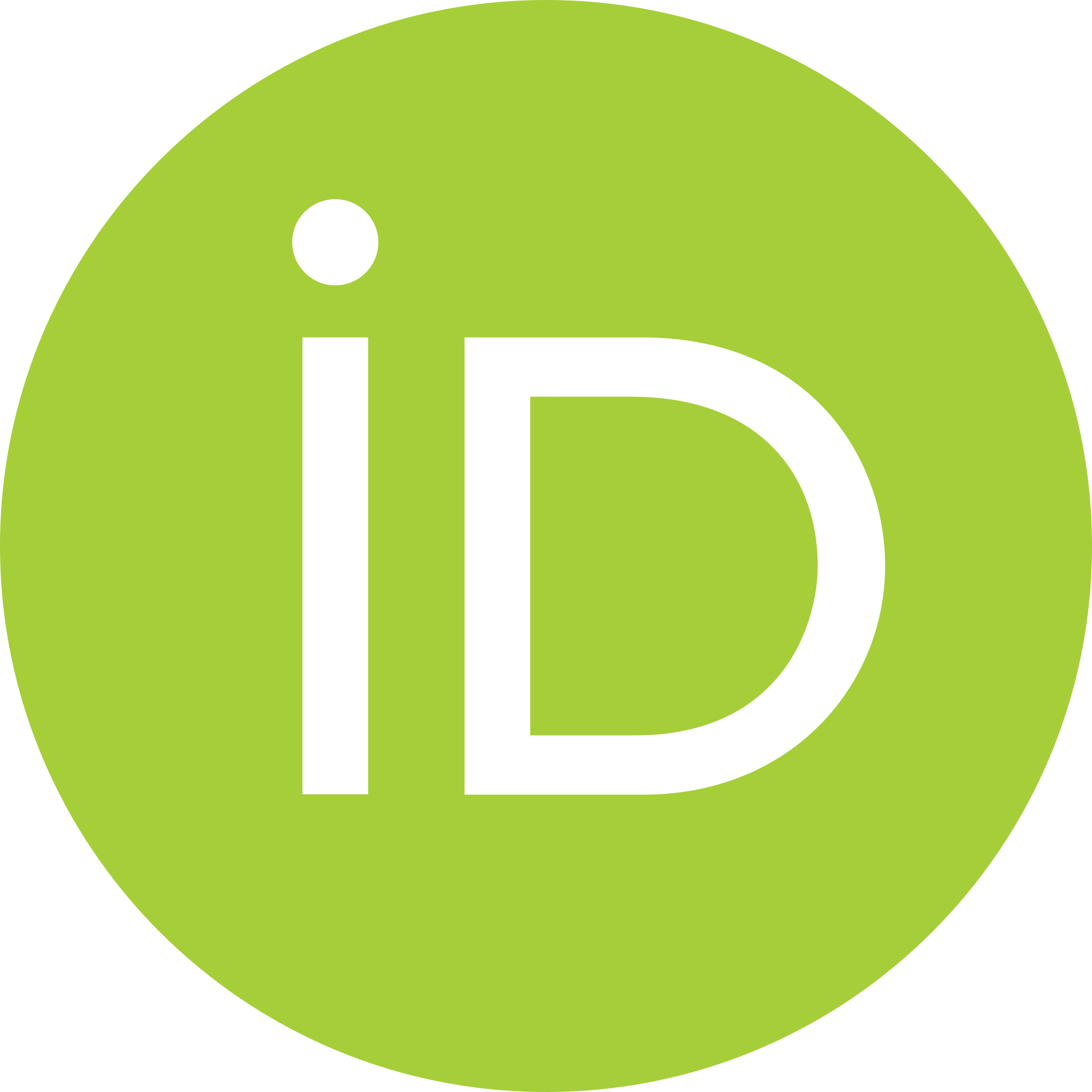}}}

\newlength{\VSpaceBeforeTabBib}
\newlength{\VSpaceBeforeTabFoot}
\newcommand\tablefoot[1]{\VSpaceBeforeTabBib=1ex%
  \par\vspace{\VSpaceBeforeTabFoot}
  \noindent
  \begin{minipage}{\linewidth}
    {\small\bfseries Notes.}~%
    \small
    \ignorespaces
    #1%
  \end{minipage}%
}
\newcolumntype{H}{>{\setbox0=\hbox\bgroup}c<{\egroup}@{}} % To hide a column 

\title[The BBC Programme]{The B-type Binaries Characterisation Programme \\ I. Orbital solutions for the 30 Doradus population}

\author[J.~I. Villaseñor et al.]{
J.~I. Villaseñor\textsuperscript{\orcid{0000-0002-7984-1675}},$^{1}$\thanks{E-mail: jvi@roe.ac.uk}
        W.~D. Taylor,$^{2}$
        C.~J. Evans,$^{2,1}$
        O.~H. Ram\'{i}rez-Agudelo\textsuperscript{\orcid{0000-0002-9379-5409}},$^{3}$
        H. Sana,$^{4}$
        \newauthor
        L.~A. Almeida\textsuperscript{\orcid{0000-0002-3817-6402}},$^{5,6}$
        S.~E. de Mink\textsuperscript{\orcid{0000-0001-9336-2825}},$^{7,8,9}$
        P.~L. Dufton$^{10}$
        and N. Langer$^{11,12}$\vspace{2mm}
\\
    $^{1}$Institute for Astronomy, University of Edinburgh, Royal Observatory, Blackford Hill, Edinburgh, EH9 3HJ, UK\\
    $^{2}$UK Astronomy Technology Centre, Royal Observatory, Blackford Hill, Edinburgh, EH9 3HJ, UK\\
    $^{3}$German Aerospace Center (DLR), Institute for the Protection of Terrestrial Infrastructures, Rathausallee 12, Sankt Augustin, D-53757, Germany\\
    $^{4}$Institute of Astrophysics, KU Leuven, Celestijnenlaan 200D, 3001 Leuven, Belgium\\
    $^5$ Escola de Ci\^encias e Tecnologia, Universidade Federal do Rio Grande do Norte, Natal, RN 59072-970, Brazil\\
    $^6$ Departamento de F\'isica, Universidade do Estado do Rio Grande do Norte, Mossor\'o, RN 59610-210, Brazil\\
    $^{7}$Max-Planck-Institut für Astrophysik, Karl-Schwarzschild-Stra{\ss}e 1, 85740 Garching bei München, Germany\\
    $^{8}$Anton Pannekoek Institute for Astronomy, University of Amsterdam, Science Park 904, 1098XH Amsterdam, The Netherlands\\
    $^{9}$Harvard-Smithsonian Center for Astrophysics, Harvard University, 60 Garden St, Cambridge, MA 02138, USA\\
    $^{10}$Astrophysics Research Centre, School of Mathematics \& Physics,  Queen’s University, Belfast, BT7 1NN, UK\\
    $^{11}$Argelander-Institut f\"ur Astronomie, Universit\"at Bonn, Auf dem H\"ugel 71, 53121 Bonn, Germany\\
    $^{12}$Max-Planck-Institut f\"ur Radioastronomie, Auf dem H\"ugel 69, 53121 Bonn, Germany
}

\date{Accepted 2021 July 20. Received 2021 July 20; in original form 2021 January 18}

\pubyear{2021}

\begin{document}
\label{firstpage}
\pagerange{\pageref{firstpage}--\pageref{lastpage}}
\maketitle

% Abstract of the paper
\begin{abstract}
We present results from the B-type Binaries Characterisation (BBC) programme, a multi-epoch spectroscopic study of 88 early B-type binary candidates in the 30 Doradus region of the Large Magellanic Cloud (LMC). From radial-velocity analysis of 29 observational epochs we confirm the binary status of 64 of our targets, comprising 50 SB1 and 14 SB2 B-type binaries. A further 20 systems (classified as SB1*) show clear signs of periodicity but with more tentative periods. Orbital solutions are presented for these 84 systems, providing the largest homogeneous sample to date of the binary properties of early B-type stars. Our derived orbital-period distribution is generally similar to those for samples of more massive (O-type) binaries in both the LMC and the Galaxy. This similarity with the properties of the more massive O-type binaries is important as early B-type stars are expected to account for the majority of core-collapse supernovae. Differences in the period distributions of the different samples start to increase above 4\,d, and are also present between the earliest (B0-0.7) and later-type (B1-2.5) systems within the BBC sample, although further study is required to understand if this is an observational bias or a real physical effect. We have examined the semi-amplitude velocities and orbital periods of our sample to identify potential candidates that could hide compact companions. Comparing with probability distributions of finding black hole companions to OB-type stars from a recent theoretical study, we have found 16 binaries in the higher probability region that warrant further study.
\end{abstract}


%%%%%%%%%%%%%%%%% BODY OF PAPER %%%%%%%%%%%%%%%%%%

%%%%%%%%%%%%%%%%%%%%%%%%%%
\section{Introduction}\label{sec:intro}
%%%%%%%%%%%%%%%%%%%%%%%%%%

Studies in the Galaxy and the LMC over the past decade have confirmed that the majority of massive stars are members of binary or multiple systems \citep[e.g.][]{kobulnicky+fryer07, mason+09, kiminki+kobulnicky12, chini+12, sana+12, sana+13, sana+14, kobulnicky+14, dunstall+15}. 
Much of the work done on the properties of these massive binaries has targeted O-type stars as these have the most extreme physical parameters. With masses ranging from $\sim$16\Msun up to well over 200\Msun \citep{martins+05, crowther+10, crowther+16}, these stars drive the chemical evolution of their host galaxies by enriching the interstellar medium (ISM) with chemically-processed material from their strong winds and by exploding as core-collapse supernovae (CCSN). They also have strong ultraviolet fluxes which are capable of ionising large volumes of gas in the ISM. 

In contrast, B-type stars have received less attention. Although less exotic than O-type stars, they are substantially more numerous. The spectral range between B0 and B3 (with initial masses of $\sim$6--15\Msun ) deserves special attention as they account for the majority of CCSN progenitors ($\sim$70\% if assuming single-star evolution with a \citet{kroupa01} initial mass function), ending their lives as neutron stars (NSs). B-type binaries, after one of the stars exploded as a SN, might become high-mass X-ray binaries \citep[HMXB,][]{reig11}. Since the discovery of the first pulsar in a binary \citep{hulse+taylor75}, HMXBs have been proposed as a channel to produce binary NSs \citep[BNSs, see for example][]{flannery+vandenheuvel75, tauris+17}. Furthermore, BNS are thought to be progenitors of interesting phenomena when merging, such as short gamma ray bursts \citep[SGRBs, ][]{fong+berger13, berger14, abbott+17d}, magnetars \citep{duncan+thompson92, dai+06, rowlinson+13}, and the recently confirmed gravitational waves \citep{abbott+17b} and their optical counterparts, kilonovae \citep{coulter+17, smartt+17}.

Notably, the most numerous class of HMXBs are Be/X-ray binaries \citep{reig11, tauris+17}, where Be stars are the optical companions of the NS. We can also anticipate similar systems, where the B-type (or Be) star was originally the secondary but now appears as the primary, accompanied by a X-ray quiet compact object.
\citet{langer+20} recently computed a large grid of evolutionary models to investigate the role of mass transfer and mergers of massive binaries ($M_1=$ 10--40\Msun) in forming OB-type stars with a BH companion. They found that the masses of OB-type stars at the moment of formation of the BH are mostly concentrated between 8 and 25\Msun with a peak near 14\Msun. They also predicted 120 OB+BH binaries in the Large Magellanic Cloud (LMC), from which about half would be B+BH binaries (mostly Be stars). However, only one O+BH system is known in the LMC to date \citep[LMC~{X-1},][]{orosz+09}, arguing for a large number of OB+BH/NS systems that are X-ray dim and potentially undetected in existing studies of OB-type stars.
\smallskip

Past efforts to study the multiplicity characteristics of large samples of Galactic B-type stars \citep{abt+levy78,wolf78,levato+87,abt+90,raboud96} were mostly focused on late B-type stars and therefore are not representative of the population of CCSN progenitors. The intrinsic distributions of orbital periods for early B-type eclipsing binaries (EBs) in the Galaxy, LMC and SMC were investigated photometrically by \citet{moe+distefano13,moe+distefano15}, with a wider review
given by \citet{moe+distefano17}. However, there remains a lack of spectroscopic studies with sufficient cadence to determine orbital solutions for large samples of early-type binaries \citep[e.g.][]{duchene+kraus13}.

Here we present results from the B-type Binaries Characterisation (BBC) programme (P.I. Taylor, 096.D-0825), which was conceived to better characterise the properties of binaries in the important early B-type domain. This builds on work from the VLT-FLAMES Tarantula Survey \citep[VFTS,][]{evans+11}, which observed more than 800 OB-type massive stars (with $V$\,$<$\,17\,mag) in the 30 Doradus region of the LMC, one of the brightest and most active star-forming regions in the Local Group.

With a focus on multiplicity, six epochs of VFTS observations were used to estimate the intrinsic binary fraction of the population of O- \citep{sana+13} and B-type stars \citep{dunstall+15}. Both studies found a high multiplicity fraction after bias correction; 51$\pm$4\% for the O-type stars and 58$\pm$11\% for the B-type stars (in good agreement within the uncertainties) for periods of up to about 8 yr ($10^{3.5}$ d). However, more comprehensive monitoring was required to characterise the orbital properties of the candidate binaries. With that intention, the Tarantula Massive Binary Monitoring project (TMBM) obtained 32 epochs of spectroscopy for an unbiased subset of the O-type binary sample of the VFTS. From analysis of these data, \citet{almeida+17} presented orbital solutions for 82 systems together with the distributions of their orbital parameters. The BBC observations were designed to enable similar follow-up of 88 of the candidate B-type binaries in the VFTS from \citet{dunstall+15}.

This article is structured as follows: Sect.~\ref{sec:observations} describes the observational sample, Sect.~\ref{sec:meth} outlines our methods to estimate the stellar radial velocities (RVs), orbital periods and determine full orbital solutions, Sect.~\ref{sec:results} presents our results and compares them with published distributions for O-type stars, Sect.~\ref{sec:disc} discussed further aspects of our results, and Sect.~\ref{sec:summary} gives a brief summary of our findings.

%%%%%%%%%%%%%%%%%%%%%%%%%%%%%%%%%%%
\section{Observations}\label{sec:observations} % Sect. 2
%%%%%%%%%%%%%%%%%%%%%%%%%%%%%%%%%%%

\subsection{Background}

The VFTS observed 438 B-type stars, with spectral classifications
and estimates of stellar RVs presented by \citet{evans+15}. In parallel, a detailed RV variability analysis was undertaken by \citet{dunstall+15}, who employed a cross-correlation technique to look for RV shifts between the (typically six) FLAMES-Giraffe spectra of each target obtained with the LR02 setting (spanning 3960--4564\,\AA). The spectra of all but 30 targets were sufficiently good to
enable estimates of RVs, yielding results for 361 dwarfs and giants, and 47 supergiants.

The statistical criteria used to assess the RV variability of the O-type stars in the VFTS \citep{sana+13} were adapted by \citet{dunstall+15} to investigate the RVs of the B-type spectra. For a target to be considered as a candidate binary system, the difference between at least one pair of RVs had to be larger than four times the combined uncertainty of the two measurements and simultaneously exceed their adopted threshold of $\Delta$RV$_{\rm min}$\,$=$\,16\kms. Objects with statistically-significant variations in the range 5\,$<$\,$\Delta$RV\,$<$\,16\kms that fulfilled the first condition were considered as `RV variables' (see \citeauthor{dunstall+15} for further discussion on selection criteria).

They found 96 candidate single-lined binaries (SB1s) and five double-lined (SB2) systems. These comprised 90 unevolved objects (i.e. the dwarfs and giants) and 11 supergiants, giving observed spectroscopic binary fractions of 25 and 23\%, respectively. In addition, 23 unevolved stars and 17 supergiants were classified as RV variables.

\begin{figure*}
\centering
  \includegraphics[width=0.725\hsize]{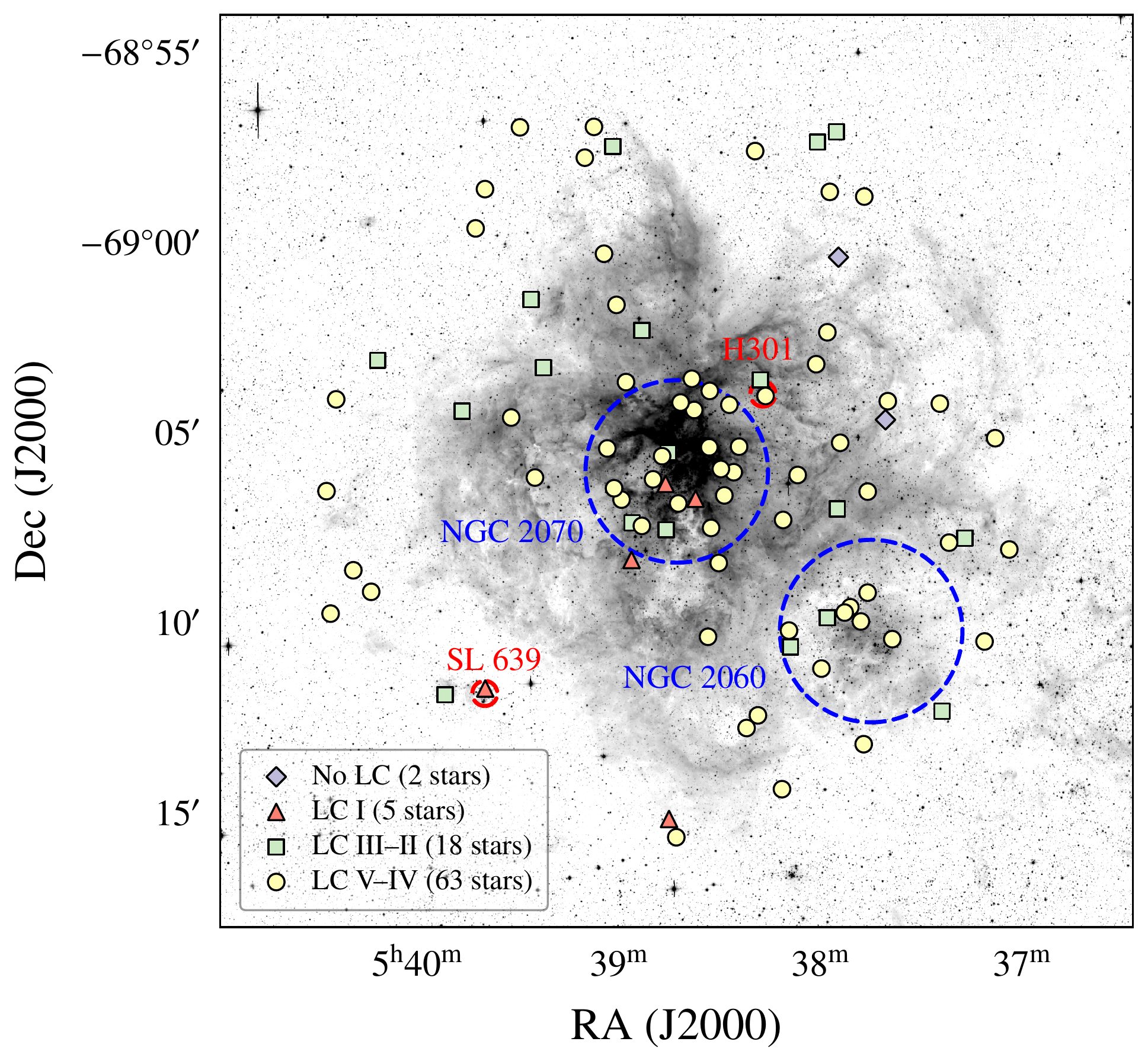}
  \caption{Spatial distribution of the 88 B-type binary candidates with new spectroscopy. The sample has been plotted by luminosity class (LC, see legend). The blue-dashed circles (with radii of 2\farcm4) highlight the approximate extent of the NGC\,2070 and NGC\,2060 clusters, whereas SL 639 and Hodge 301 (with radii of 0\farcm33) are represented in red.}
     \label{fig:FoV}
\end{figure*}

\subsection{Monitoring campaign}

The approach for the BBC campaign was to observe as many as possible of the 101 candidate binaries from \citeauthor{dunstall+15} in one FLAMES fibre configuration. The input list also included seven further targets that were flagged by \citeauthor{dunstall+15} as candidate binaries but where absolute RVs could not be estimated (see their footnote~4). The FLAMES Fibre Positioning Observation Support Software ({\sc fposs}) was used to maximise the number of targets assigned to fibres from the 108 candidate systems, resulting in observations of the 88 systems listed in Table~\ref{tab:tabB1}\footnote{Comprised of 78 of the 96 candidate SB1s from \citeauthor{dunstall+15}, four of the previously classified SB2 systems, and six of the seven additional candidates without absolute RVs.}.

To obtain coverage of a selection of hydrogen, helium and metallic lines for RV estimates of each target, we adopted the same strategy as the TMBM campaign in using the Medusa mode of FLAMES with the LR02 set-up. This provides a spectral resolving power $R=7000$, which is sufficient to obtain individual RVs precise to better than 5\,\kms.

To securely identify cosmic rays, each one-hour observing block (OB) was comprised of three back-to-back science exposures of 894\,s. These were observed over the period 2015 October to 2016 December, where execution of the 29 OBs from the service queue was (loosely) constrained to ensure a varying cadence (from daily-intervals up to the long baseline of observations more than one year after the start of the programme). The Heliocentric Julian Dates (HJDs) of the start of each triplet of exposures are listed in Table~\ref{tab:hjd}.

With the Medusa fibres assigned to the candidate binaries, we used the remaining fibres to obtain further observations of other notable VFTS targets. These included candidate O-type runaways \citep{walborn+14}, candidate B-type runaways \citep{evans+15}, the extreme rotator VFTS\,102 \citep{dufton11}, the X-ray bright VFTS\,399 \citep{clark15}, the peculiar B[e]-like supergiant VFTS\,698 \citep{dunstall+12}, and several stars of interest in the context of the nitrogen abundances from \citet{grin17}. Although not the focus of this article, details of these additional targets are given in Table~\ref{tab:extras}.

\begin{table}
\caption{Heliocentric Julian Dates (HJD) at the start of each Observing
Block (OB) for the BBC campaign.}
\label{tab:hjd}
\center
\begin{tabular}{cc|cc}
%\begin{tabular*}{0.3\textwidth}{@{\extracolsep{\fill}}lcc}
\toprule
\toprule
OB & HJD & OB & HJD \\
\midrule
01 &	2457299.745 & 16 &	2457417.563 \\
02 &	2457332.733 & 17 &	2457418.611 \\
03 &	2457335.782 & 18 &	2457420.622 \\
04 &	2457339.698 & 19 &	2457421.695 \\
05 &	2457339.780 & 20 &	2457423.598 \\
06 &	2457366.724 & 21 &	2457427.607 \\
07 &	2457379.551 & 22 &	2457432.517 \\
08 &	2457394.652 & 23 &	2457622.866 \\
09 &	2457398.723 & 24 &	2457651.760 \\
10 &	2457400.577 & 25 &	2457681.710 \\
11 &	2457402.602 & 26 &	2457692.743 \\
12 &	2457410.692 & 27 &	2457698.748 \\
13 &	2457411.579 & 28 &	2457724.707 \\
14 &	2457415.547 & 29 &	2457726.710 \\
15 &    2457416.559 & & \\
\bottomrule
\end{tabular}
\end{table}

\subsection{Data reduction}\label{ssec:datared}

The data were reduced using the ESO CPL FLAMES/GIRAFFE pipeline v.2.1.5 for bias and dark subtraction, correction for the flat-field response, wavelength calibration, and (summed) extraction of the spectra. Each spectrum was then corrected to the heliocentric frame. An error spectrum was also produced by the pipeline for each fibre, which records the statistical error arising from 
different stages in the reduction for each wavelength bin.

Mean spectra were obtained for each epoch by combining the three exposures
weighted by the associated error spectra. A 5$\sigma$-clip was also included when taking the average to reject outlying cosmic rays. The combined data for each epoch were then normalised by division of a low-order polynomial fit to the continuum. 

All but six of our OBs were obtained in grey or dark time (fractional lunar illumination of $<$\,70\,\%). With the exception of two OBs (12, 13), the background continuum level was generally low in the dozen sky fibres allocated across the FLAMES field. We experimented with subtracting the median sky spectrum for each epoch from each science spectrum. Although it led to a marginal improvement in S/N, it then led to complications with over-/under-subtraction of the nebular emission lines because some of the sky fibres had (weak) nebular emission. Given our approach to model the nebular emission as part of our RV analysis (see Sect.~\ref{ssec:nebem}) and for a consistent approach to all of the spectra we decided not to subtract the sky spectra from our data. From tests with and without sky subtraction for objects without significant nebular emission, this did not have a significant impact on the final RV estimates. The exceptions were OBs 12 and 13, where they were only useful for estimates for a couple of the brightest stars. Subtracting the sky spectra for these two OBs alone does improve their S/N but then caused problems with the nebular components, so we did not pursue this further. We add that two OBs (10, 18) had generally low S/N for many of the targets due to other observational factors; these were also only used for some of the brighter targets.

\subsection{The B-type binary candidates} \label{ssec:Bcandidates}

\begin{figure}
    \centering
    \includegraphics[width=0.95\hsize]{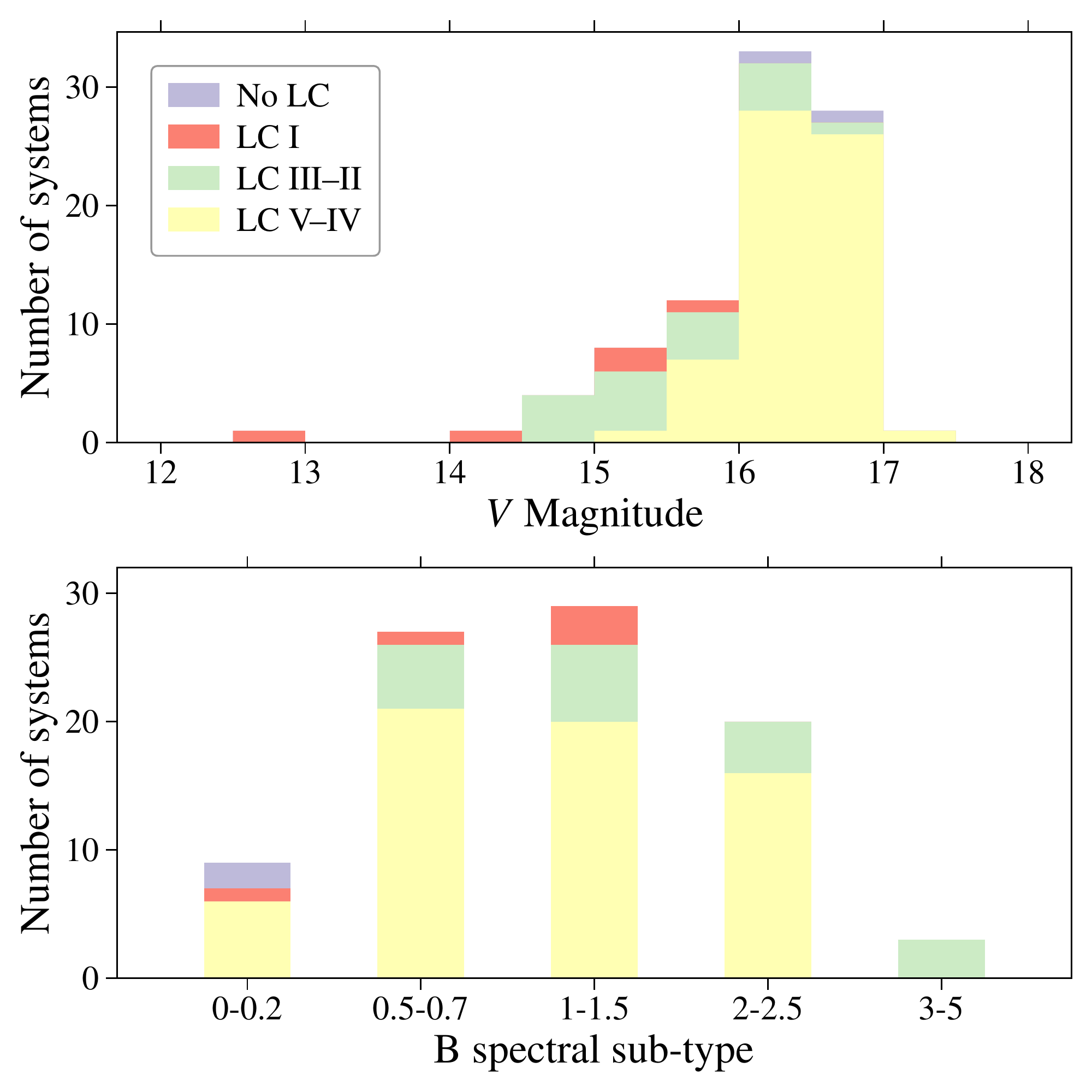}
    \caption{Histograms of $V-$band magnitudes (upper panel) and spectral types (lower panel) of the BBC sample (spectral classifications from \citealt{evans+15}).} 
    \label{fig:mag+SpT}
\end{figure}

The spatial distribution of the B-type sample observed by the BBC campaign in the 30~Doradus region is shown in Fig.~\ref{fig:FoV}; the luminosity class (LC) of each target is also shown. The central cluster, NGC\,2070, contains a quarter of the observed sample, including 27\% of the dwarfs (LC V--IV) and two of the five supergiants (LC I). In contrast, only seven dwarfs and two giants (LC III--II) are within NGC\,2060, the older association to the southwest. The two older, smaller stellar clusters, Hodge~301 and SL\,639, contain only two and one BBC targets, respectively. Most of the stars in our sample are field stars (62\%), comprising 60\% of our dwarf targets and 67\% of the giants in the field, plus an additional supergiant and two stars without LCs.

The distribution of $V$-band magnitudes and spectral types for the sample are shown in Fig.~\ref{fig:mag+SpT}, with their LC colour-coded as in Fig.~\ref{fig:FoV}. Our B-type systems are dominated by faint ($V$\,$\sim$\,16--17\,mag) main-sequence (MS) stars (72\% of the sample), with typical S/N ratios of below 40 for most of our spectra (see Sect. \ref{ssec:meth-sn}). Systems over the spectral range B0.5 to B2.5 are well distributed, but there is a drop at the earliest subtypes, with only nine members in the B0--0.2 bin (of which two do not have a LC). For the latest types (B3--5), dwarfs were too faint for the VFTS magnitude limit \citep[$V=17$, see][]{evans+11}, so only three giants/bright giants were observed.

From visual inspection of the spectra and the subsequent RV analysis we noted 14 targets as SB2 (or candidate SB2) systems, of which half were previously identified. The SB2s thus comprise 16\% of the sample, lower than the 33\% fraction of binaries that were SB2s in the TMBM sample. This is probably (partly) a consequence of the fainter magnitudes of the B-type stars and hence the lower S/N.

%%%%%%%%%%%%%%%%%%%%%%%%%%%%%%%%%%%%%%%%%%%%%%%
\section{Methodology} \label{sec:meth} % Sect. 3
%%%%%%%%%%%%%%%%%%%%%%%%%%%%%%%%%%%%%%%%%%%%%%%

\subsection{Stellar radial velocities} \label{ssec:RVs} % Sect. 3.1

Following the approach taken by \citet{sana+13} and \citet{almeida+17}, we determined RVs from shifts of the spectral lines due to the Doppler effect. We have employed a line-by-line Gaussian fitting method to each spectrum that uses \texttt{lmfit} \citep{lmfit}, a non-linear least-squares minimisation package available in {\sc python}.

\begin{figure}
    \centering
    \includegraphics[width=0.95\hsize]{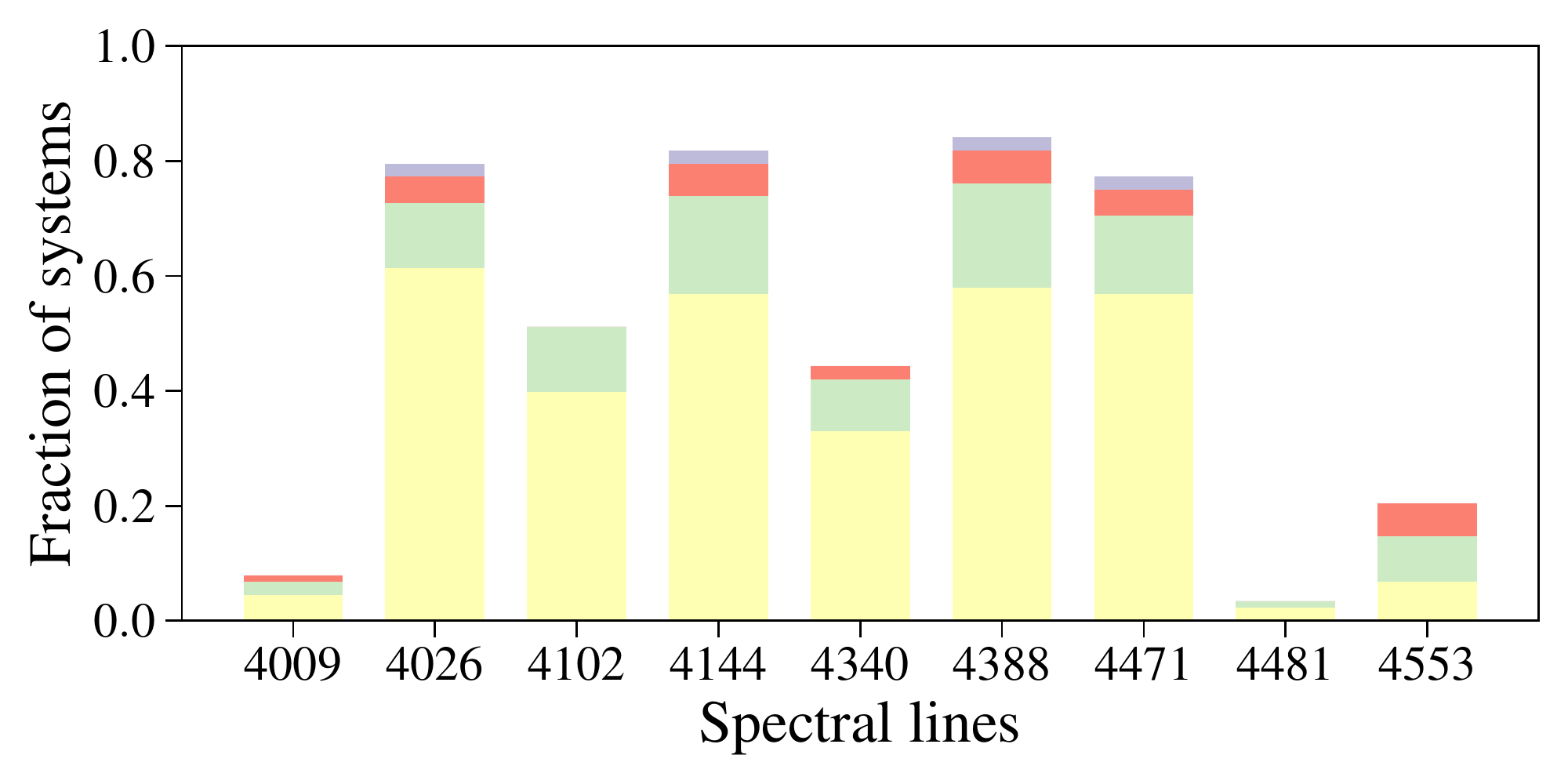}
    \caption{Fraction of systems for which each spectral line was used to estimate radial velocities. The colours are as in Figures~\ref{fig:FoV} and~\ref{fig:mag+SpT} }
    \label{Fig:SLhisto}
\end{figure}

The dominant absorption lines in the LR02 region of early B-type spectra are the Balmer series and a set of \ion{He}{I} lines. These are complemented by a range of weak metallic features (e.g. \ion{Si}{III}, \ion{Si}{IV}, and \ion{Mg}{II}) and, at the earliest types, weak \ion{He}{II} absorption.
Of the metallic lines, given the S/N of our spectra and the distribution of spectral types, only \ion{Si}{III} $\lambda$4553 offers a useful RV diagnostic for a reasonable number of our targets. \ion{Si}{IV} is only present at the earliest types (and blended with an \ion{O}{II} line in the case of \ion{Si}{IV} $\lambda$4089), while \ion{Mg}{II} is only present in the latest types.

For the majority of our sample we therefore used seven lines to estimate the RVs: H$\delta$, H$\gamma$, \ion{He}{I} $\lambda\lambda$4026, 4144, 4388, 4471, and \ion{Si}{III} $\lambda$4553. For the small number of stars classified earlier than B0.5, we also attempted fits to the  \ion{He}{II} $\lambda\lambda$4200,4542 lines, but these did not yield useful information in most cases given the weakness of the features combined with the S/N of the spectra from each epoch. In general, the weaker \ion{He}{I} ($\lambda\lambda$4009, 4121) lines were too weak to provide useful fits. However, for six systems where we otherwise struggled to find a period, the \ion{He}{I} $\lambda$4009 and \ion{Mg}{II} $\lambda$4481 lines were strong enough to provide useful information and led to improved results.

To ensure a sufficient number of lines for each target it was also necessary to separately take into account the nebular contamination (see Sect.~\ref{ssec:nebem}). The fraction of the sample for which each diagnostic line was used in estimating the final RVs of each target is shown in Fig.~\ref{Fig:SLhisto}. 

The line fits for the RV estimates were based on a set of given initial parameters. Each line was fitted from an initial estimate of the centre ($\mu$), amplitude ($A$) and full width at half maximum (FWHM), employing Gaussian profiles for the helium and silicon lines, and a Lorentzian for the Balmer lines. In the case of slow rotators (mainly evolved stars) the profiles of the \ion{He}{I} lines will be dominated by the instrument profile which is Gaussian, whereas for rapid rotators the spectral line profile is dominated by rotational broadening which is not Gaussian. However, the S/N of our sample tends to be lower for MS stars (i.e. generally the faster rotators), so the line profile can be fitted sufficiently well by Gaussians. We tested both Gaussian and Lorentzian fits to the spectra of stars with different LCs finding that the Gaussian fits did a better job of fitting the wings of the lines in all cases.

\begin{figure}
    \centering
    \includegraphics[width=0.95\hsize]{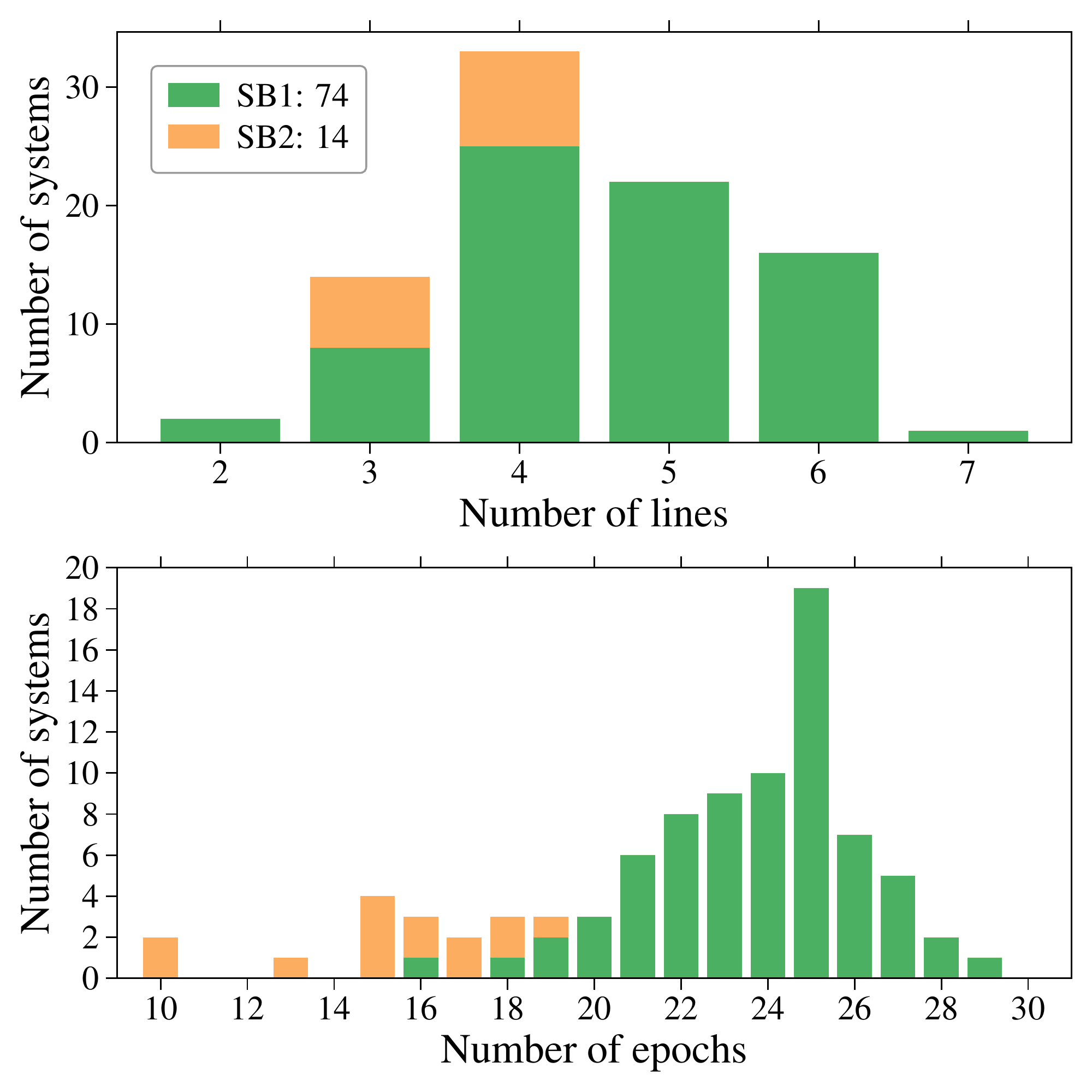}
    \caption{{\it Top panel:} Number of spectral lines used to determine RVs for each system. {\it Bottom panel:} Final number of epochs used to determine the orbital periods after removing epochs with low S/N or with large RV errors. In both panels the SB2 systems are shown in orange.}
    \label{fig:nlin+neps}
\end{figure}

All the lines were fitted independently for each epoch of each star, obtaining values and uncertainties for $A$, $\mu$ and FWHM. A median-absolute-deviation \citep[MAD, see e.g.][]{hampel74} test was then applied to these uncertainties to identify the diagnostic lines with better fits (smaller errors) for each system. Given that the errors on $\mu$ and $A$ were typically smaller, the FWHM errors were used in most cases for the MAD test, as these proved to be more sensitive to erroneous fits (e.g. misidentification of the line due to low S/N, solar features, or intrinsic weakness of the spectral line). The MAD test is a robust measure of dispersion in a data set and is commonly used to find and reject outliers. It is defined as the median of the absolute differences of the data from their median:
\begin{equation}\label{eq:mad2}
    \mathrm{MAD}(X) = k\,\,\mathrm{med}(|X-\mathrm{med}(X)|)\,,
\end{equation}
where med stands for median, $X$ is a variable denoting the data values and $k$ is a `consistency' constant dependant on the distribution. Here we have assumed a normal underlying distribution for our errors for which $k$ takes a value of 1.4826 \citep{rousseeuw+croux93}. A rejection criterion (RC) must be set such that:
\begin{equation}
    \frac{X-\mathrm{med}(X)}{\mathrm{MAD}(X)} > \mathrm{RC}
\end{equation}
is considered an outlier. In our case the medians of the uncertainties for the fits in the 29 epochs are computed for each spectral line and the MAD test is applied to them; spectral lines with a value given by Eq.~\ref{eq:mad2} larger than the adopted RC are considered outliers and therefore rejected. A RC value between 2 and 3 is usually used, with a higher value being a more relaxed outlier cutoff. The remaining set of lines was then used to compute the RV of each observation.

With the lines selected, the individual spectra were also checked to reject spectra with low S/N (see Sect.~\ref{ssec:meth-sn}) to avoid undue influence of misidentifications of line centres on the final RV estimate. Erroneous fits can be identified by their characteristically large errors, so a MAD test similar to that used to select the diagnostic lines was applied to detect epochs with poor fits. For each epoch, a mean uncertainty was computed for each fitted parameter considering the lines selected in the previous step. The MAD test again searches for outliers in the mean uncertainties, i.e. epochs with large errors, and their RVs were not included in the computation of the orbital period.
Figure~\ref{fig:nlin+neps} shows the number of spectral lines (top) and number of epochs (bottom) used in the RV analysis (with the SB1 and SB2 systems highlighted). For $\sim$80\% of the systems at least four lines (up to a maximum of seven) were used, while the number of epochs fluctuates around a mean of 24, with a strong peak at 25. Fewer epochs were used for the SB2 systems, with a mean of only 15 (see Sect.~\ref{ssec:sb2}).

\begin{table}
\caption[Absorption lines used in the determination of RVs with their respective rest wavelength from the NIST Atomic Spectra Database.]{Absorption lines used in the determination of RVs with their respective rest wavelengths from the NIST Atomic Spectra Database\footnotemark.}
\label{tab:tab1}
\center
\begin{tabular}{lcc}
%\begin{tabular*}{0.3\textwidth}{@{\extracolsep{\fill}}lcc}
\toprule
\toprule
Ion & 
$\lambda$ (\AA) &    
$\lambda_{\mathrm{rest}}$ (\AA) \\
\midrule
\ion{He}{I}     & 4009 & 4009.256 \\
\ion{He}{I}     & 4026 & 4026.191 \\
H$\delta$       & 4102 & 4101.734 \\
\ion{He}{I}     & 4144 & 4143.761 \\
H$\gamma$       & 4340 & 4340.472 \\
\ion{He}{I}     & 4388 & 4387.930 \\
\ion{He}{I}     & 4471 & 4471.480 \\
\ion{Mg}{II}    & 4481 & 4481.130 \\
\ion{Si}{III}   & 4553 & 4552.620 \\
\bottomrule
\end{tabular}
\end{table}
\footnotetext{https://physics.nist.gov/PhysRefData/ASD/lines\_form.html}

RVs ($v_{\rm rad}$) were calculated for the selected epochs and lines from:
    \begin{equation}
        %\centering
        v_{\mathrm{rad}} = \frac{\mu-\lambda_{\mathrm{rest}}}{\lambda_{\mathrm{rest}}}c \,,
    \end{equation}
where the adopted values of $\lambda_{\mathrm{rest}}$ are given in Table \ref{tab:tab1} and $c$ is the speed of light. For each epoch of each system a weighted mean was calculated from the RVs of the selected spectral lines using 
    \begin{equation}\label{eq:wgtmean}
        %\centering
        \bar{v}_\mathrm{r} = \frac{\sum\limits_i \omega_i v_i}{\sum\limits_i \omega_i}\,\,\,\, \mathrm{with}\,\,\,\,
        \omega_i = \frac{1}{(\sigma_i)^2} \,,
        % \omega_i = \frac{1}{(\Delta v_i)^2} \,,
    \end{equation}
where $v_i$ are the individual RVs calculated from each of the spectral lines available, while $\omega_i$ are the weights chosen as the inverse of the uncertainty squared that \texttt{lmfit} computes from the covariance matrix. The uncertainty of the RV weighted mean was calculated from
    \begin{equation}
        %\centering
        \sigma_{\bar{v}_\mathrm{r}} = \frac{1}{\sum\limits_i \omega_i} \sqrt{\sum\limits_i (\omega_i \sigma_i)^2} \,,
    \end{equation}
which follows from the definition of variance and Eq. \ref{eq:wgtmean}. The choice of a weighted mean was to include as many epochs as possible, to exploit any useful information we can from the lower S/N observations.

\subsubsection{Double-lined systems} \label{ssec:sb2} % Sect. 3.1.1

\begin{figure}
    \centering
    \includegraphics[width=0.95\hsize]{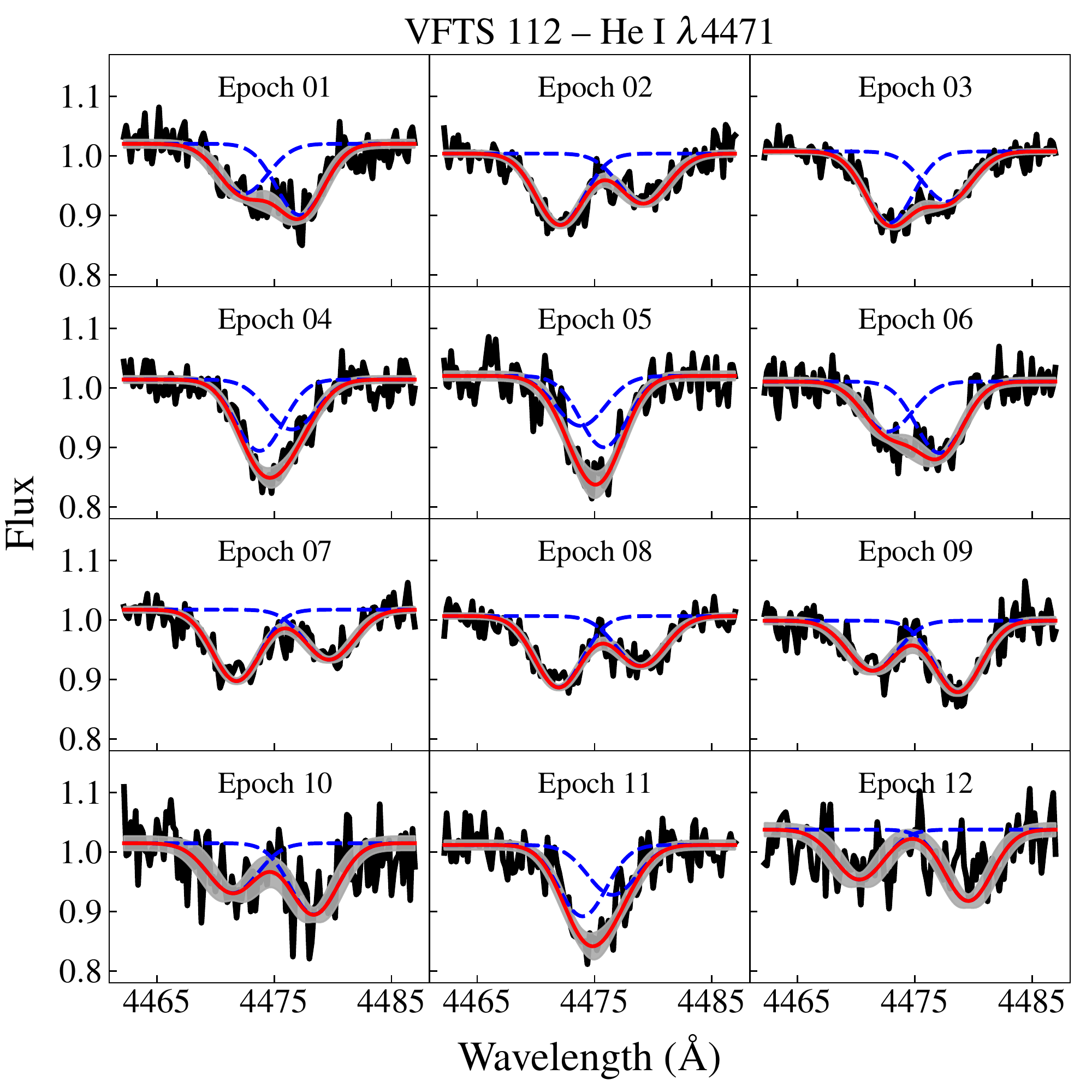}
    \caption{Example fits for the SB2 system VFTS\,112. Blue-dashed lines indicate individual spectral lines for primary and secondary, red-solid lines are the resulting fits, with the grey-shaded areas representing the 3$\sigma$ uncertainties.}
    \label{fig:sb2fit}
\end{figure}

Double Gaussian fits were implemented in the code to measure the RVs of both components of the identified SB2 systems. A key difference with the (single-line) fits for the SB1 systems is when the RV difference of the two components is close to zero. This impacts on the quality of the fit as it becomes difficult to identify the contribution of each star to the blended spectral line, as shown by epochs 4, 5 and 11 in Fig.~\ref{fig:sb2fit}. To avoid erroneous RVs from such observations, the code uses a minimum separation for the components (of 2 to 4 \AA, depending on the line width) below which epochs are rejected. The rejected epochs will therefore be those with RVs close to the systemic velocity, i.e., where the velocities of the two components intersect in the final orbital solution (see Fig.~C3, available as supplementary material).

In some spectra the two components can be separated sufficiently well, but the code occasionally misidentifies the two components, e.g. if the primary is redshifted, the code can associate the blueshifted line to the primary and vice versa. Effects such as low S/N spectra and systems with comparable intensities of the two components are the main cause of such misidentifications. Thus, the code checks for discrepancies in the shifts of the primary component in an epoch. For example, for a set of four spectral lines, if the primary (stronger) component is redshifted in three and blueshifted in the remainder, the code inverts the primary and secondary component in the discrepant line, so that the primary is now redshifted in all four cases. However, if the fits to the primary component were redshifted and blueshifted the same number of occasions (two each in the example), it is not possible to distinguish both components and the code rejects that epoch. 
One last check acts on noisy observations, where the code applies a MAD test to the errors of the calculated centre of the lines, rejecting spectra with large errors, as with the SB1 systems. 

These three filters reduced the number of observations used to estimate the orbital period for the SB2 systems to an average of 15 epochs (in contrast to an average of
$\sim$24 for the SB1 systems, see Fig. \ref{fig:nlin+neps}). 
The majority of the SB2 systems are short-period binaries, thus we expect the orbits to be mostly circular or to have small eccentricities. This means that a robust estimate of the period is possible with fewer epochs in the case of SB2 systems, and that little information is lost regarding the eccentricity as long as the rejected epochs are close to the systemic velocity and the separation between RV measurements is not extreme. 

\subsubsection{Nebular contamination} \label{ssec:nebem} % Sect. 3.1.2

\begin{figure}
    \centering
    \includegraphics[width=0.95\hsize]{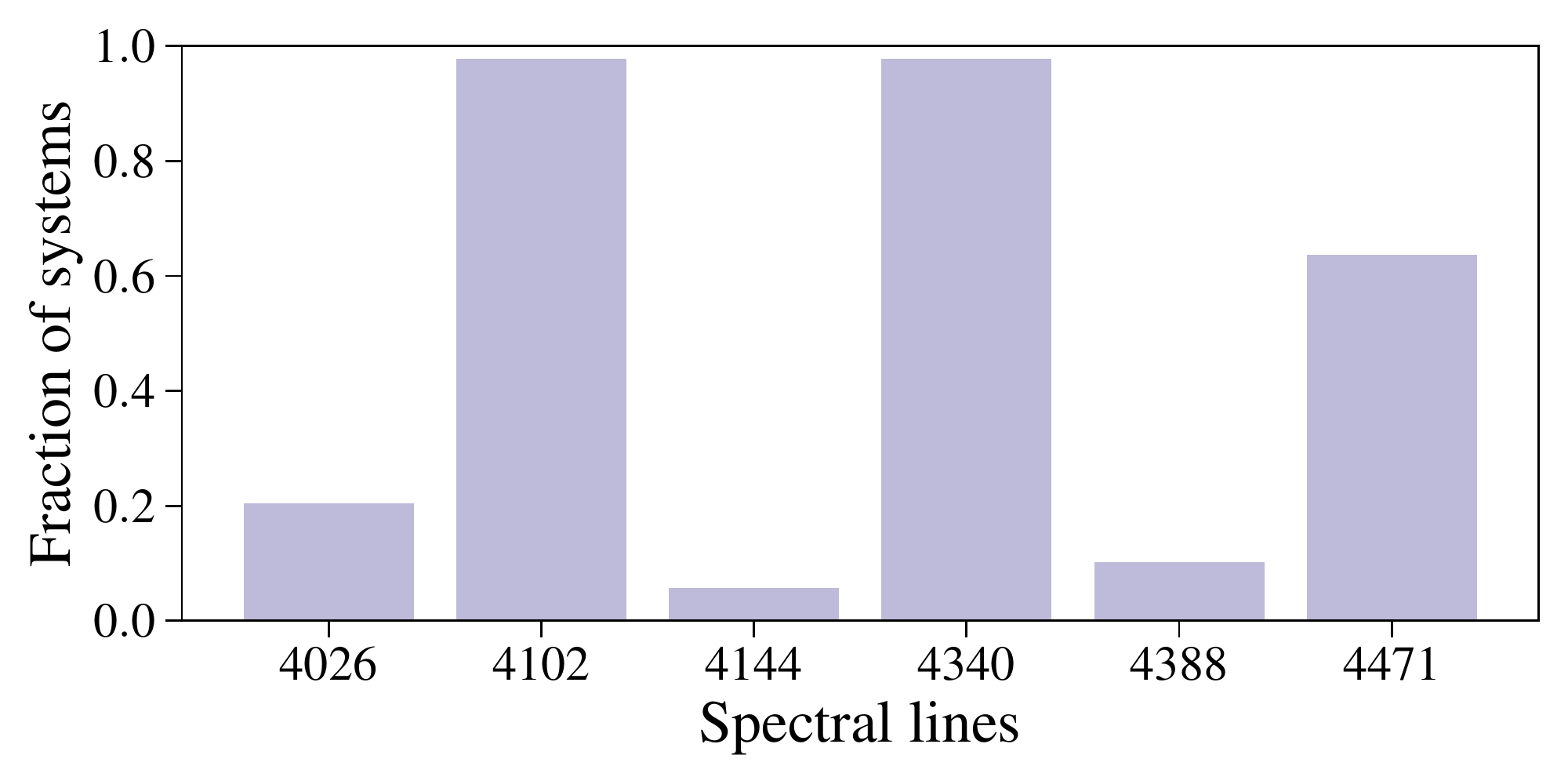}
    \caption{Fraction of systems for key diagnostic lines where the 
    nebular contamination was sufficiently bad that the radial-velocity
    analysis included a model emission component (see Sect.~\ref{ssec:nebem} for details).}
    \label{fig:histnebem}
\end{figure}

Nebular contamination of the spectra varies significantly, from being completely absent in some cases to very strong in a large fraction of the sample. This is most notable in the Balmer lines, where the nebular emission can be several times stronger than the stellar absorption. Figure~\ref{fig:histnebem} shows the frequency of nebular contamination for our diagnostic lines. The Balmer lines of nearly all of our targets displayed some degree of contamination and $\sim$60\% of the \ion{He}{I} $\lambda$4471 profiles are also significantly affected. It is less of a problem for the other helium lines, with only 20, 10, and 6\% significantly affected for \ion{He}{I} $\lambda$4026, $\lambda$4388 and $\lambda$4144, respectively.

\begin{figure*}
    \centering
    \includegraphics[width=0.99\hsize]{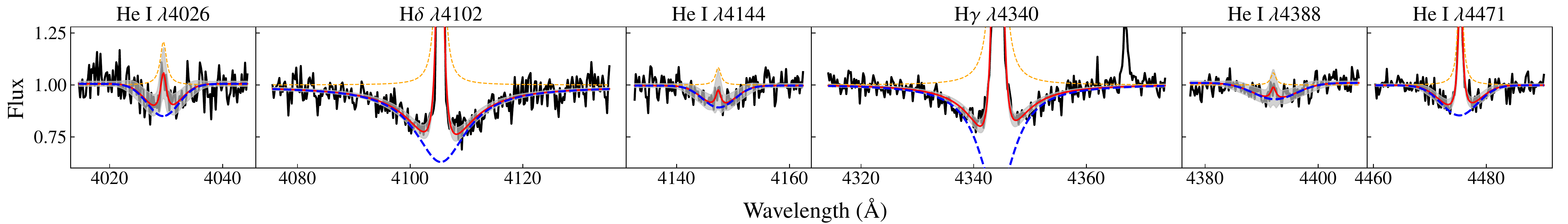}
    \caption{Example fits for VFTS\,337 that take into account contamination by nebular emission. Although classified as a Be star \citep[B2: V-IIIe+ from][]{evans+15}, from inspection of the [O~{\scriptsize III}] and [N~{\scriptsize II}] lines in the LR03 and HR15N data from the VFTS, the nebular emission dominates the Be contribution. The two components, stellar absorption (blue) and nebular emission (orange), are shown together with the resulting fit (red) to illustrate our procedure, and the varying degree of contamination in the different lines.}
    \label{fig:nebem_fit}%
\end{figure*}

To have enough lines for robust RV estimates it was necessary to include the Balmer lines and to account for nebular emission. 
We fitted the spectra with a superposition of a Gaussian/Lorentzian profile (for the contaminated \ion{He}{I}/Balmer absorption lines) and of a Lorentzian profile (nebular and/or Be-type stellar disk emission). In other words, we fit both the spectral absorption line and the nebular/disk emission, as shown by the example in Fig.~\ref{fig:nebem_fit}. Due to the variation in the relative strength of the nebular emission between epochs
(arising from variations in the astronomical seeing impacting differently on point and extended sources in terms of the flux at the fibre aperture), it was not possible to develop an automated method of detection, so the lines affected by nebulosity were identified by eye.

\subsubsection{Signal to Noise} \label{ssec:meth-sn} % Sect. 3.1.3

A second difficulty with our adopted methods was the low S/N of many of the spectra as mentioned in Sect.~\ref{ssec:Bcandidates}, including variations in the S/N of repeated observations of each target given changes in the observing conditions, airmass, lunar illumination etc. The S/N per pixel of our spectra are generally in the range of 15 to 60, with the exception of the supergiant VFTS\,591 for which the spectra have S/N\,$\sim$\,120. The mean S/N of the sample is 33 (with a median of 30), with 83\% of the spectra having S/N\,$<$\,40 and 53\% with S/N\,$<$\,30, down to a minimum of S/N\,$\sim$\,15. This is even more pronounced in the dwarfs, with all of the spectra having S/N\,$<$\,50 and 65\% with S/N\,$<$\,30. 
Although low S/N increases the challenge of successful fits to the spectra (and to identify SB2 systems), it has less impact on the period search (see Fig.~\ref{fig:SN-LSP} and discussion in Sect.~\ref{ssec:detectability}).

%_______________________________________
\subsection{Orbital periods}\label{ssec:meth-Porb} % Sect. 3.2

The Lomb-Scargle (LS) periodogram \citep{lomb76, scargle82} is a widely used tool in the astronomical community to search for signals in unevenly sampled time series. We implemented the LS routines from {\sc astropy} \citep{astropy:2013, astropy:2018} to search for orbital periods within the weighted-mean RVs for each epoch of each system. The BBC programme was designed to find periods between 1 day and a year, but a frequency grid covering a range between 0.4 and 1000 days was used for the LS analysis because three systems showed broad peaks in the periodogram close to 500 d. We tested our code on three B-type systems that were also observed as part of the TMBM programme (VFTS\,225, 779, 827) and found differences of less than 0.4\% between our estimated periods and those from \citet{almeida+17}.

\begin{table}
\caption{Velocity results for the four targets classified as RV variables, i.e. with significant RV
shifts but uncertain periods.}
\label{tab:RVvar}
\center
\begin{tabular}{cccc}
%\begin{tabular*}{0.3\textwidth}{@{\extracolsep{\fill}}lcc}
\toprule
\toprule
VFTS & 
$\Delta$RV$_{\rm max}$\,(\kms)  & $\sigma_{\rm RV}$\,(\kms)    &
Possible $P_{\rm orb}$\,(d) \\
\midrule
144     & 58.77$\pm$8.05  & 9.78  & 171.37\po \\
391     & \po60.25$\pm$10.24 & 10.21\po & \po1.12 \\
591     & 25.44$\pm$2.02  & 0.89  & 478.68\po \\
890     & 35.23$\pm$3.74  & 5.23  & 18.92 \\
\bottomrule
\end{tabular}
\end{table}

To determine the significance of the periods found, we computed the false-alarm probability (FAP) to estimate the probability of having a peak in the periodogram of a certain height at a given frequency if the data show no periodic signal. While the FAP does not tell us if we are correctly identifying the true period of the system, it is a measurement of the significance of the signal \citep{vanderplas18}. We considered a peak as corresponding to the true period if its FAP exceeded a false-alarm level (FAL) of 0.1\%. Systems that satisfy this condition and that also have an orbital solution were classified as SB1.

Beyond these cases, there were further systems that show clear signals in their periodograms. These include cases with a FAP between 0.1 and 1\% (or just below 0.1\%), and some examples with double peaks of similar strength. Even if they present convincing orbital solutions with a low $\chi^2$, to differentiate them from those with more robust periods, we classified them as SB1*. Periodograms for each of the systems classified as SB1, SB1* and SB2 are available as supplementary material in Appendix C.

A last group of systems from our targets are those with FAPs larger than 1\% and low LS powers ($\lesssim$2). We could not find a reliable orbital period for these targets and they were classified as RV variables (RV var). Nevertheless, the possible orbital periods obtained for the RV var systems are given in Table~\ref{tab:RVvar}, together with maximum RV variation ($\Delta$RV$_{\rm max}$) and standard deviation of the mean RV ($\sigma_{\rm RV}$). These periods should only be considered as tentative at best but, considering the significant RV variations in some of them, it is possible that these are the true period and other factors are interfering with the signal in the periodogram (see Sect.~\ref{ssec:detectability}).

\begin{figure}
    \centering
    \includegraphics[width=0.99\hsize]{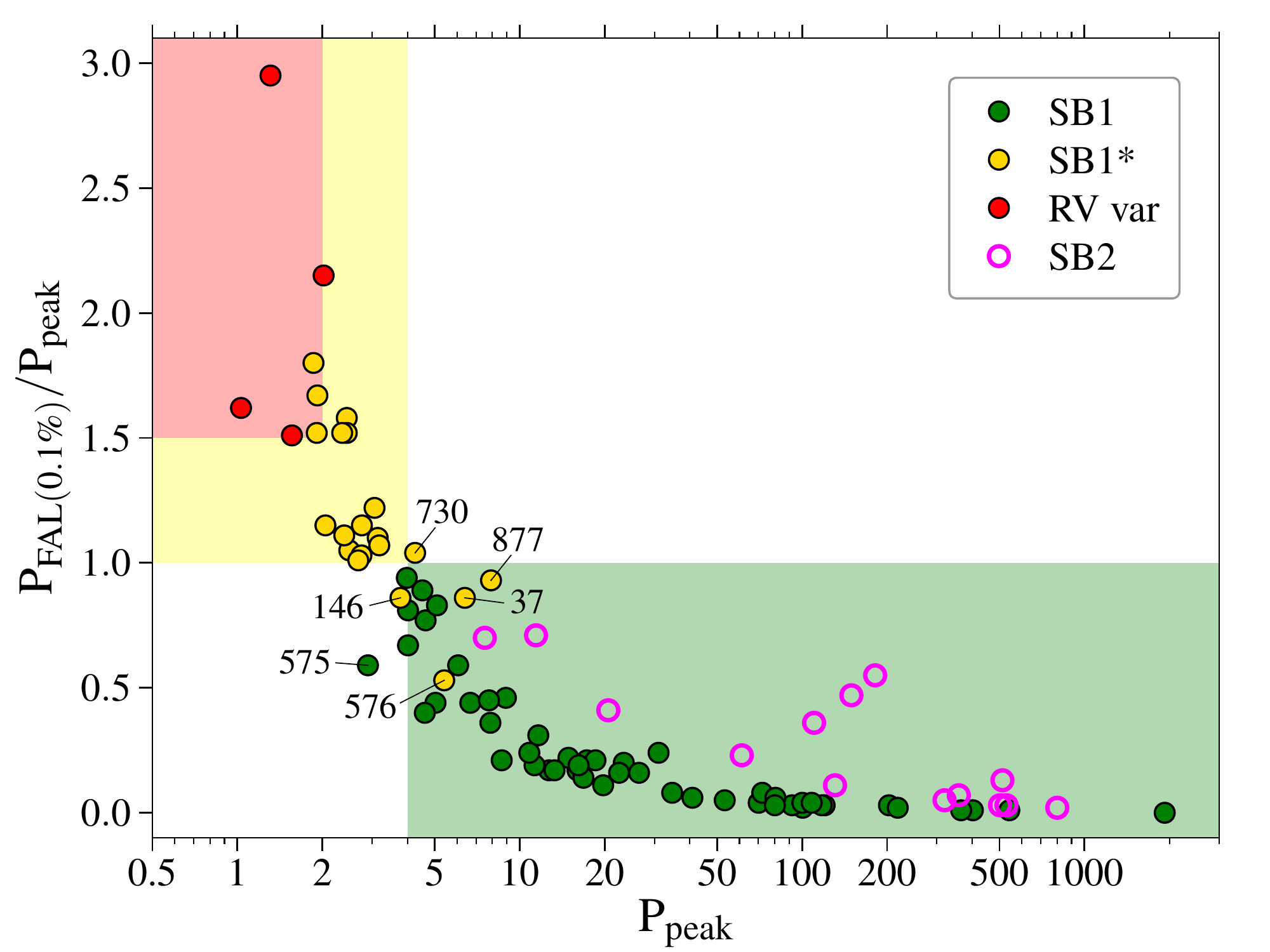}
    \caption{Classification of the SB1 systems. The ratio of the Lomb-Scargle (LS) power at the 0.1\% false-alarm level (FAL) to the peak value (P$_{\rm peak}$, used to identify the potential period) for each system as a function of P$_{\rm peak}$. Systems with P$_{\rm peak}$ above the 0.1\% FAL therefore have a ratio $<$\,1 (i.e. a strong signal), which defines the green area with robust periods (73\% of the sample). The yellow area includes 15 systems with less certain periods that are classified as SB1* (with some exceptions outside this area from considering other factors). Systems that did not exceed the 1\% FAL were classified as RV variables and are indicated in red.}
    \label{fig:classcrit}%
\end{figure}

Our classification criteria are illustrated in Fig.~\ref{fig:classcrit}. The $x$-axis corresponds to the LS power of the peak in the periodogram and the $y$-axis is the ratio of the power at 0.1\% FAL and the power of the peak (i.e. the value in the $x$-axis). The green-shaded area indicates a region with LS power $\geq4$ and above 0.1\% FAL (or P$_{\mathrm{FAL}}/$P$_{\mathrm{peak}}<1$). A total
of 64 binaries, including all the SB2 systems, are in this region (i.e. 73\% of the BBC sample). The labelled objects are the exceptions to our criteria, but we emphasise these are not strict limits and their binary classification took into account orbital solutions, morphology and other aspects. For notes on the individual systems see Appendix~\ref{apx:notes}.

\subsubsection{Non-sinusoidal signals}\label{ssec:nss} % Sect. 3.3.1

The LS periodogram searches for a period by fitting a sinusoidal model to each frequency of the grid. Of course, the best model will not always be a sinusoid, e.g. due to eccentricity of the orbit. Nevertheless, 
given our time-sampling, we expect most of our systems to be relatively short-period binaries with near circular orbits. For example, 60\% of the O-type systems observed by the TMBM programme had periods of less than 20\,d, and 40\% had eccentricities of less than 0.1 \citep{almeida+17}.

For eccentric systems, where the signal is not sinusoidal, higher harmonics are expected to be present in the periodogram. For a system with a true period ($P_{\mathrm{true}}$) at a frequency $f_0$, higher harmonics could be present at $mf_0$, with $m$ a positive integer \citep{vanderplas18}. However, this will not prevent the sinusoidal model finding the correct period if it can closely fit the data.

\subsubsection{Other aliasing effects} % Sect. 3.3.2

\begin{figure}
\centering
  \includegraphics[width=0.95\hsize]{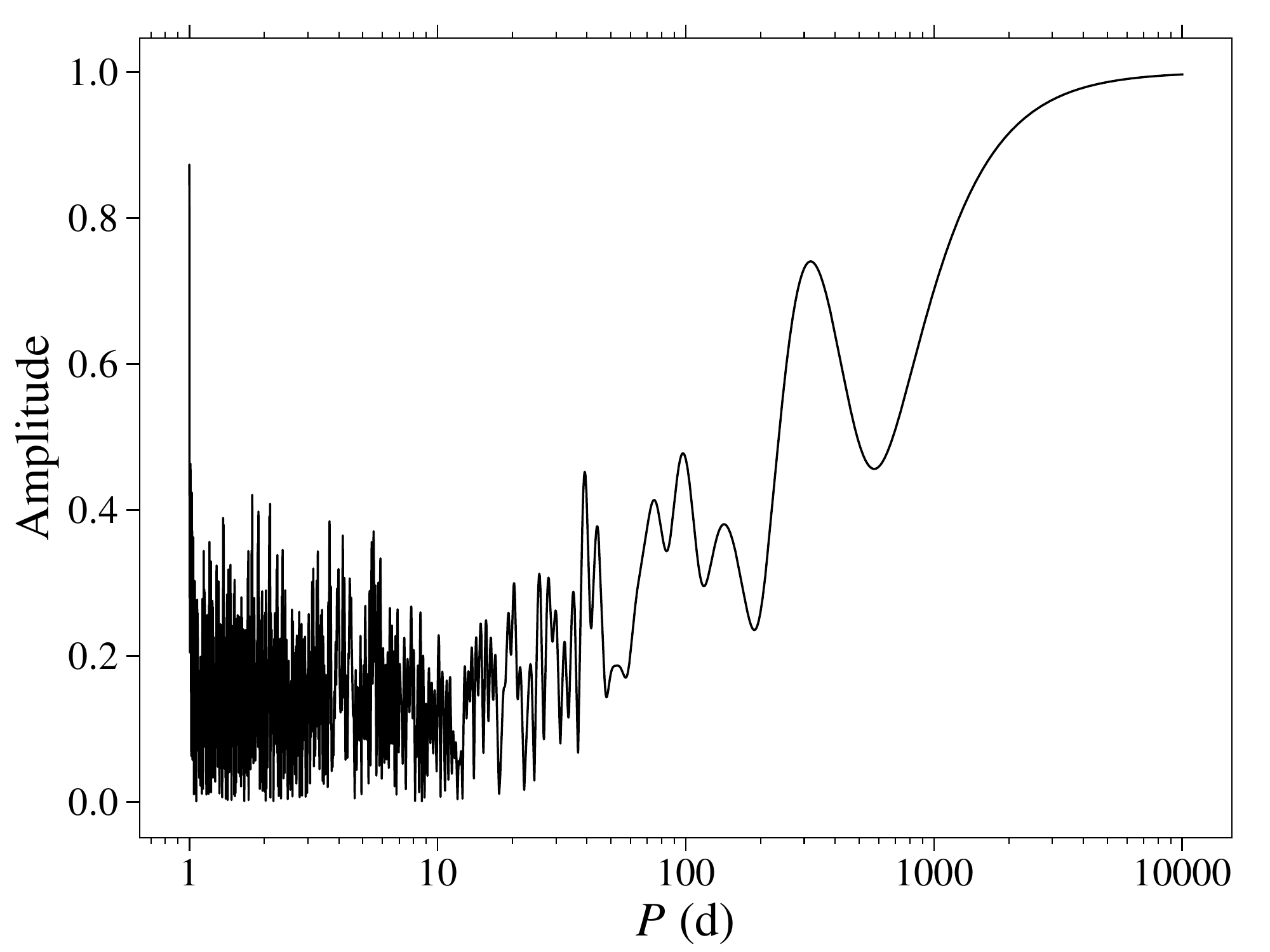}
  \caption{Power spectrum of the BBC observing window. Our observational strategy enabled us to determine orbital periods ranging from one day up to a year.}
     \label{Fig:PSwin}
\end{figure}

The power spectrum of our observing window is shown in Fig.~\ref{Fig:PSwin} where two clear peaks are visible. The larger one above 300\,d corresponds to the 1 yr alias and the peak at 1\,d can be explained as a consequence of the day and night cycle. Such features can create aliases at $f_0\,\pm\,n\,\delta f$, where $\delta f$ in the case of nocturnal observations (typical of ground-based observatories) would be 1 cycle d$^{-1}$ and $ n \in\mathbb{Z}^+$ \citep[see][and references therein]{vanderplas18}. We have identified such aliases in most of our sample and they have been taken into account when determining the periods. 
The power spectrum demonstrates that we have very good detectability for periods of up to 50\,d. Beyond this our detectability declines, which significantly limits the accuracy of periods above 100\,d, while becoming increasingly difficult or infeasible to detect periods up to the baseline of the BBC programme (427\,d). Periods longer than this can not be trusted since not even a full cycle is covered.

\subsubsection{Checks on the orbital periods}\label{ssec:lcomb}

Most of our adopted lines for RV estimates are affected by Stark broadening and three of our helium lines are diffuse. 
To test the robustness of the orbital periods determined by our LS analysis, we implemented a brute-force method that takes all the available lines in each spectrum and combines them to form subsets of a minimum of three lines (two in a few difficult cases and the SB2s). If we take all nine lines in Table~\ref{tab:tab1}, this gives 466 possible sets of lines. We then tested each of the combinations with the LS routine, to estimate the fraction of the different combinations that returned the same (or a different) period from our initial analysis. This method proved useful to
identify weak (but robust) signals in some of the systems where we had otherwise struggled to estimate a period, as well as confirming the majority of the periods of the SB1 systems.

%______________________________________________
\subsection{Orbital solutions}\label{ssec:orbsol} % subsection 3.3

Full orbital solutions were obtained for the complete sample with the IDL code \rvf \citep{iglesias-marzoa+15}. \rvf uses an adaptive simulated annealing algorithm \citep{ingber00} to fit Keplerian orbits to RV data using seven parameters: orbital period $P_{\mathrm{orb}}$, time of periastron passage $T_p$, eccentricity $e$, argument of the periastron $\omega$, systemic velocity $\gamma$, semi-amplitude of the primary $K_1$ and, in the case of SB2 systems, the semi-amplitude of the secondary star $K_2$. The code minimises the $\chi^2$ function:
\begin{equation}
\chi^2=\sum_{i=1}^{N}\left[\left(\frac{\tilde{v}_1(t_i)-v_1(t_i)}{\sigma_1(t_i)}\right)^2 + \left(\frac{\tilde{v}_2(t_i)-v_2(t_i)}{\sigma_2(t_i)}\right)^2 \right]\,,
\end{equation}
where $N$ is the number of RV measurements, $v_1$ and $v_2$ are the RVs of the primary and secondary stars (when available) at time $t_i$ and $\sigma_1$ and $\sigma_2$ are their associated uncertainties. Finally, $\tilde{v}_{1,2}$ is the RV the code computes by solving the following set of equations:
\begin{align}
& \tilde{v}_1(t_i)=\gamma+K_1 \left[cos(\theta(t_i)+\omega) + e\cos\omega \right] \\[1.ex] 
& \tilde{v}_2(t_i)=\gamma+K_2 \left[cos(\theta(t_i)+\omega') + e\cos\omega' \right] \\[1.ex] 
& \omega' = \omega + \pi \\
& \theta(t_i)=2\arctan\left[ \sqrt{\frac{1+e}{1-e}}\tan \left( \frac{E(t_i)}{2} \right) \right] \\[1.ex] 
& E(t_i) - e\sin E(t_i)=M(t_i) \label{eq:kepler} \\[1.ex] 
& M(t_i) = \frac{2\pi}{P}(t_i-T_p) \,,
\end{align}
where $\theta$ is the true anomaly, $E$ is the eccentric anomaly which must be obtained by solving Kepler's equation (Eq.~\ref{eq:kepler}), $M$ is the mean anomaly and $T_p$ is the time of periastron passage.
Once \rvf selects the best model from the minimisation of the $\chi^2$, it returns the orbital parameters with their respective uncertainty, which are computed from the covariance matrix (see \citeauthor{iglesias-marzoa+15} for details).

We folded the RVs to the estimated periods from the LS tests ($P_{\mathrm{orb}}$) and fitted sinusoids to the RV curves. From the sinusoidal fits we obtained initial values for $\gamma$, $K_1$ and $K_2$ (for SB2s), in which we used the median value of the HJDs of our observations as the initial parameter for $T_p$. We then used these as input to the \rvf analyses, adopting a search range of $\pm$20\% on the initial values for $P_{\mathrm{orb}}$, $\gamma$, $K_1$ and $K_2$, while $T_p$ was constrained by the initial and end dates of the observing campaign. For $e$ and $\omega$ we used initial values of 0.1 and 0 respectively, letting them vary in (almost) the full parameter space (0--0.999 for $e$ and 0--360$^{\circ}$ for $\omega$).

The estimated periods from the \rvf analyses are in excellent agreement with the LS results, with a median difference of only 0.04\% (when comparing to the initial LS periods) for the whole sample. We also found small differences for the other parameters, with a median of 0.2\% for $\gamma$ and 5.6\% for $K_1$, where the latter was expected to be larger given the inclusion of eccentricity. Results for all the orbital parameters are given in Table~\ref{tab:tabB3} for the SB1s, in Table~\ref{tab:tabB4} for the SB1* systems, and in Table~\ref{tab:tabB5} for SB2s. The minimum masses, mass ratios and minimum orbital separations of the SB2 systems are listed in Table~\ref{tab:tabB6}. RV curves obtained from solutions given by \rvf are available as supplementary material for all systems in Appendix D.

%%%%%%%%%%%%%%%%%%%%%%%%%%%%%%%%%%%%%%%%%%%%%%%
\section{Results}\label{sec:results}
%%%%%%%%%%%%%%%%%%%%%%%%%%%%%%%%%%%%%%%%%%%%%%%

In summary, from the analysis of the 88 candidate B-type binaries observed by the BBC programme, we have found reliable periods for 64 binaries and found clear signs of periodicity for 20 further systems (i.e. estimated periods for 95\% of the sample). The remaining four systems display RV variations but without significant periodic signals. We now describe our results and compare them with published results for other samples of massive binaries.

\begin{figure}
\centering
  \includegraphics[width=0.94\hsize]{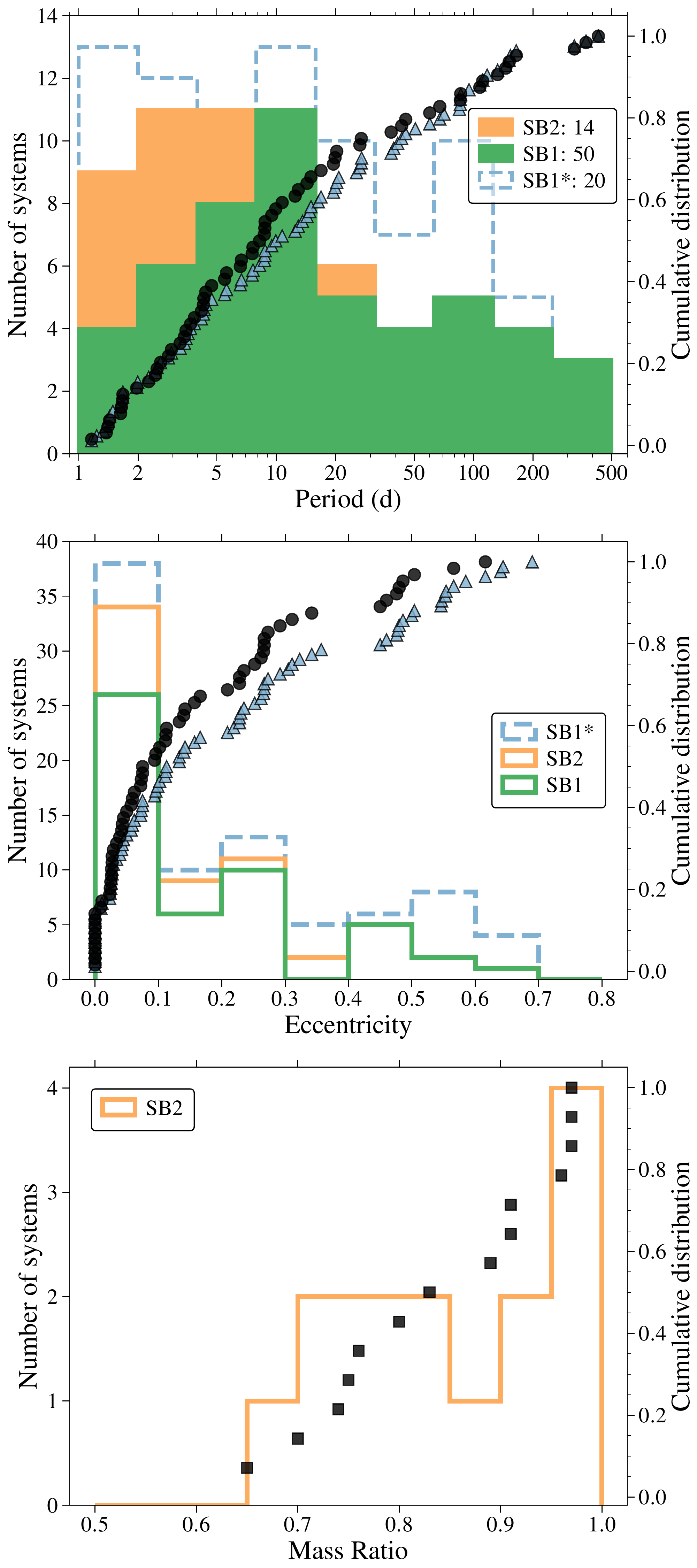}
  \caption{Distribution of orbital parameters ($P_{\rm orb}$, $e$, $q$) for the SB1 and SB2 systems; the dashed-blue histograms include systems with possible but unconfirmed periods (SB1*). The cumulative distributions of the combined SB1 and SB2 systems are indicated by black circles in the upper two panels, while the blue triangles show the distributions with the addition of the SB1* systems. Black squares in the lower panel show the cumulative distribution of mass ratios for the SB2 systems.}
     \label{fig:P_histo.pdf}
\end{figure}

\subsection{Distribution of orbital parameters}

\subsubsection{Orbital periods}

The observed distribution of orbital periods is shown in Fig.~\ref{fig:P_histo.pdf} (top panel). The number of short-period SB1 systems increases from four systems with $P_{\rm orb}$\,$<$\,2\,d, up to 11 with 8\,$<$\,$P_{\rm orb}$\,$<$\,16\,d. The SB2 sample substantially increases the number of binaries with $P_{\rm orb}$\,$<$\,8\,d, making the distribution almost constant in log space up to $P_{\rm orb}$\,$\sim$\,16\,d. This range presents a large accumulation of systems, in fact the majority of the BBC sample are short-period binaries; from the cumulative distribution function (cdf, black circles in Fig.~\ref{fig:P_histo.pdf}),
40\% of the BBC sample have $P_{\rm orb}$\,$<$\,5\,d, 75\% have $P_{\rm orb}$\,$<$\,30\,d, and $\sim$85\% have $P_{\rm orb}$\,$<$\,100\,d. The SB1* group increases the number of systems with short periods ($P$\,$<$\,2\,d) but includes more systems in the range 10--100\,d (to be expected given that it becomes harder to secure robust detections of longer periods with the cadence of our observations). 

To determine if the contribution of the SB1* systems is statistically significant we have compared the distribution of SB1+SB2 systems with respect to the full sample (SB1+SB2+SB1*) with a k-sample Anderson-Darling (AD) test and a 2-sample Kuiper (K) test. The AD test statistic returned a negative value which was lower than the critical value for the 25\% significance level (at which the value is capped) meaning that the results are not significant at a significance level of 25\%, so we cannot reject the null hypothesis that the two samples are drawn from the same distribution. This is supported by the K test which returned a false-positive probability (fpp) value of 0.997, i.e. there is a K-test probability of 99.7\% of obtaining two samples this different from the same distribution. These results are summarised in Table~\ref{tab:tab2}. Given the similarity of the two distributions, we use the SB1+SB2 distribution (i.e. black circles in Fig.~\ref{fig:P_histo.pdf}) in the following discussion.

\subsubsection{Eccentricities}

\begin{figure}
\centering
  \includegraphics[width=0.95\hsize]{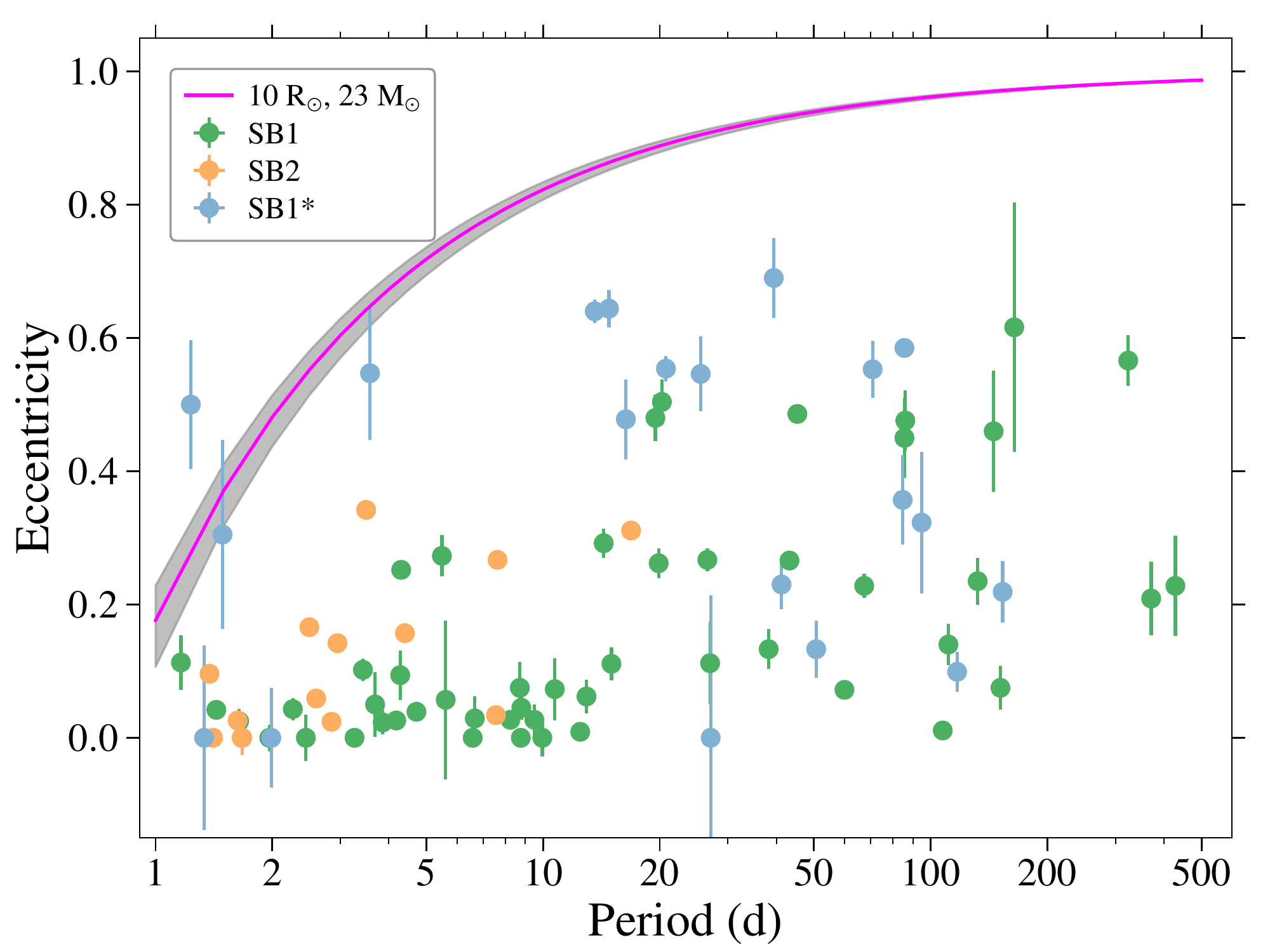}
  \caption{Eccentricities versus orbital periods for the BBC sample. The solid-magenta line represents the curve of contact for typical B-type binaries with a combined radius ($R=R_1+R_2$) of 10\Rsun and total mass of $23\pm5$\Msun with uncertainties represented by the grey-shaded area.}
  \label{fig:evP.pdf}
\end{figure}

Results for eccentricities ($e$) are also shown in the middle panel of Fig.~\ref{fig:P_histo.pdf}. There is a concentration of systems with $e$\,$<$\,0.1 (55\% of the sample), confirming that many of the binaries have circular or close-to-circular orbits. Figure~\ref{fig:evP.pdf} shows that low eccentricities ($e<0.2$) are preferred by the shorter-period systems ($P_{\rm orb}<20$\,d). From the cumulative distribution in Fig.~\ref{fig:P_histo.pdf}, close to 90\% of the detected binaries have $e$\,$<$\,0.4, with only two systems with $e$\,$>$\,0.55 and none with $e$\,$>$\,0.65. The systems classified as SB1* contain ten high-eccentricity systems ($e>0.4$) but, as shown by Fig.~\ref{fig:evP.pdf}, the uncertainties on $e$ for several of these are typically larger than the rest of the sample, including three possible short-period systems close to the curve of contact for a typical B-type-star radius (which gave us further cause to classify them as SB1*). As for the orbital periods, we conducted AD and K tests (K-test probability of 98.9\%), which argue that the samples are not significantly different (i.e. SB1+SB2 vs. SB1+SB2+SB1*).

\subsubsection{Mass ratios}\label{sec:q}

The distribution of mass ratios ($q$) for the 14 SB2 systems is shown in the lower panel of Fig.~\ref{fig:P_histo.pdf}. There are no systems with $q$\,$<$\,0.65 and half the detected systems have near equal-mass components (with $q$\,$>$\,0.90). These results are not unexpected given the observational limitations where the quality of the data, and the rapid drop in luminosity for lower-mass companions, makes it easier to detect near equal-mass systems. For the O-type binaries from \citet{almeida+17}, companions were detected down to $q$ as low as 0.35, but a companion with such a small mass ratio would be undetectable with our observations of the lower-mass, B-type systems.

\subsection{Comparison of spectral properties}

\begin{figure}
\centering
  \includegraphics[width=0.95\hsize]{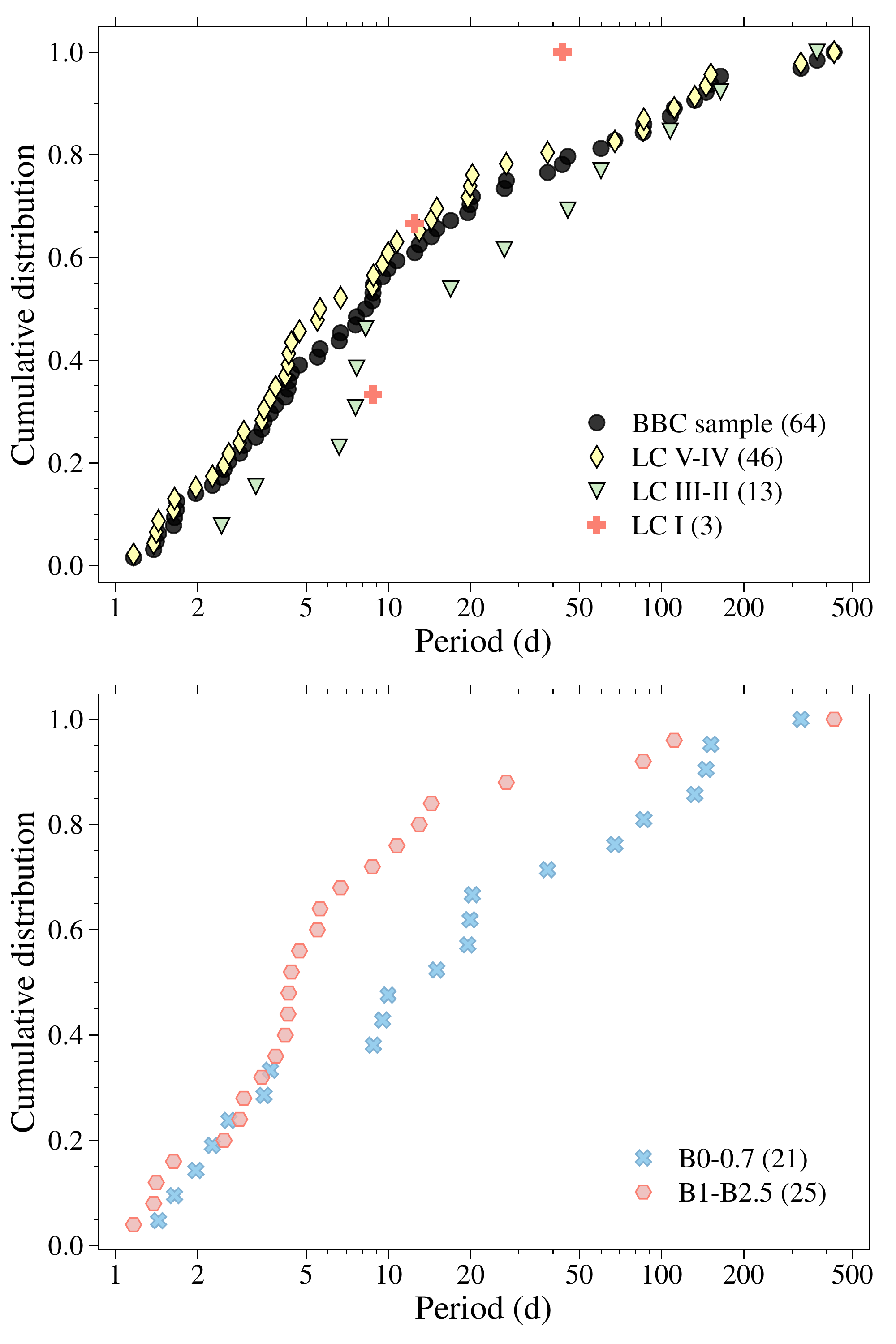}
  \caption{{\it Top panel:} Cumulative distribution of orbital periods of the BBC sample and those for subsets by luminosity class (LC). (Note that there were two targets without luminosity classifications.) {\it Bottom panel:} Cumulative distribution of the main-sequence stars split into the earliest and later B-type systems in our sample.}
     \label{fig:cd_ST.pdf}
\end{figure}

We investigated the period distribution of the SB1+SB2 sample with respect to spectral type and luminosity class (serving as proxies for the mass and evolutionary status, respectively, of the primaries). The upper panel of Fig.~\ref{fig:cd_ST.pdf} shows the cumulative period distribution for the whole sample compared with distributions grouped by LC. Our sample is dominated by dwarfs so it is difficult to say much about the more evolved systems. The giants (classes III and II) are present across most of the range of periods, with a small peak in the 5-10\,d bin. There are only three (out of the initial five) supergiants, distributed between 8 and 50\,d and, as one might expect given their evolutionary status, absent at the shortest periods (in which a merger or a common-envelope configuration would have already occurred given their large radii). While the number of evolved stars is too low to draw any conclusions, we have compared the full sample with that of the dwarfs using the AD and K tests and found no statistically significant difference. 

To investigate the properties of the dwarfs further, the lower
panel of Fig~\ref{fig:cd_ST.pdf} shows the period distributions for the earliest (B0-0.7) and later (B1-2.5) types in the sample.
There is a notable rise of later-type systems around 4\,d, with the fraction of later-type binaries about 30\% higher than the earlier-type systems for $P_{\rm orb}$\,$<$\,8\,d. Interestingly, this contrasts with the results from \citet{almeida+17} who found a tendency towards short periods for early O-type binaries in comparison to later O-types. 

Our results could be explained by a bias in being less sensitive to detection of lower-mass secondaries in the later-type systems. Assuming a flat distribution of mass ratios, late-type primaries have less massive companions in comparison to early-type primaries. At longer periods, binaries with less massive secondaries will be more difficult to detect as they present smaller semi-amplitude velocities. For example, for two systems with primary masses of 14 and 8\Msun, for $q=0.3$, $P_{\mathrm{orb}}=100$\,d, $e=0$, and $i=45^\circ$, their semi-amplitude velocities are 19.7 and 16.3\kms respectively, just above our detection limit. This small difference is unlikely to explain the relative dearth of late B-type stars with longer periods on its own, but probably combines with the fact that we will be more sensitive to smaller mass ratios for the more massive primaries.

Our statistical tests indicate that these two sub-samples are different at a significance level of 9\% from the AD test and a probability of 13\% from the K test, so not formally significant.
Another possible explanation for the later types favouring shorter periods could be that it was harder to detect longer periods for such stars from the original VFTS data, presuming they were generally fainter (i.e. lower S/N), such that they were not included in the BBC sample. However, the range of spectral types is relatively small, and the magnitudes of the targets in the two bins overlap significantly (e.g. due to differences in line-of-sight extinction), suggesting this might not be a strong factor in our results. Moreover, there did not appear to be a significant difference in the estimated binary fractions as a function of magnitude from the VFTS analysis \citep[see Fig.~4 from][although again extinction is probably a limiting factor]{dunstall+15}. We note that the observed binary fraction of B0-0.7 stars from \citeauthor{dunstall+15} was 32\%, while it was only 21\% for those classified B1-2.5. Further investigation beyond the scope of this study is required to ascertain if the binary fraction is truly lower in the later-type bin, or if observational factors have influenced the results.

\subsection{Comparison with published samples}

\begin{figure}
\centering
  \includegraphics[width=0.95\hsize]{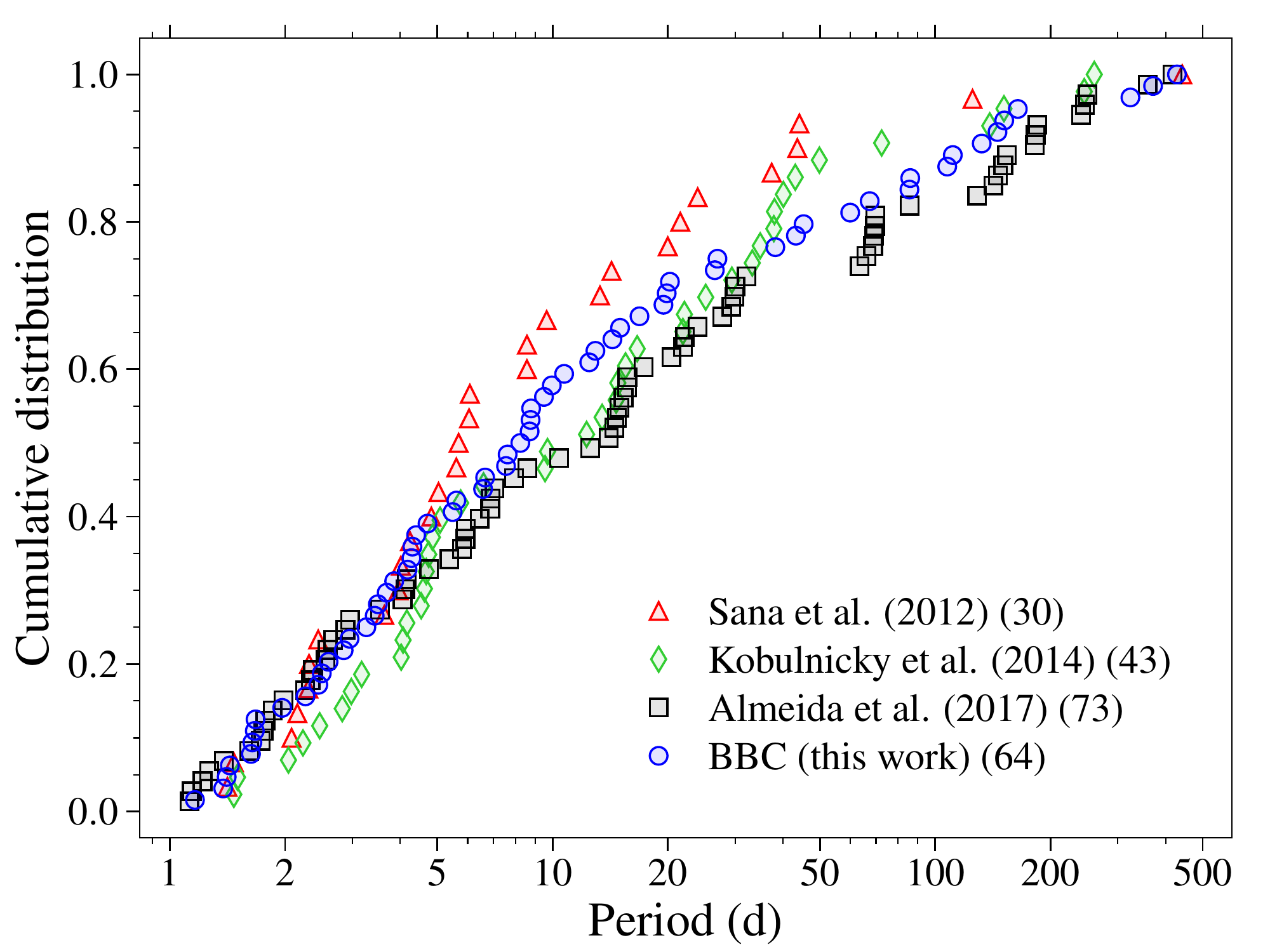}
  \caption{Cumulative distribution of orbital periods from the BBC systems (blue circles) compared with multiplicity studies of OB-type binaries in the Galaxy \citep[][red triangles and green diamonds, respectively]{sana+12, kobulnicky+14} and the LMC \citep[][black squares]{almeida+17}. The three published samples have been truncated at periods of 500\,d to enable better comparison with the BBC results.}\label{fig:cumdis}
\end{figure}

In Fig.~\ref{fig:cumdis} we compare the orbital period distribution of our B-type binaries with those  from three published studies:
\begin{itemize}
\item{{\it S12:} O-type stars from \citet{sana+12}, comprised of 40 O-type (O3-9.7) binaries in six young (1-4\,Myr) Galactic clusters, with the analysis based on (typically) 20 epochs of observations from previous studies.}
\item{{\it K14:} Results for 48 (23 O- and 25 B-type) stars in the Cygnus OB2 association (3-4 Myr) from \citet{kobulnicky+14}. Their analysis was based on 14 epochs obtained at $R$\,$=$\,2500 to 4500 and good S/N (60 to 200), with RV estimates from \ion{He}{I}~$\lambda5876$, and extended with observations from previous campaigns with different telescopes.}
\item{{\it TMBM:} Results for VFTS follow-up of 93 O- and 7 B-type binaries in 30~Dor in the LMC by \citet{almeida+17}, using the same FLAMES LR02 set-up as here.} 
\end{itemize}

Given the absence of systems with periods larger than 500\,d in our sample, we truncated the published samples at this period for a better comparison. Each of these campaigns followed similar observational strategies but we note that the distributions do not account for the different observational biases in the final observed samples. To quantify the statistical significance of the differences seen in Fig.~\ref{fig:cumdis} we performed AD and K tests on the different samples. The results from these tests are included in Table~\ref{tab:tab2}.

\begin{table}
\caption{Results from Anderson-Darling (AD) and Kuiper (K) tests on the distributions of orbital parameters from the BBC, and other Galactic and LMC samples of O- and B-type binaries.}
\label{tab:tab2}
\centering
\resizebox{\columnwidth}{!}{%
\begin{tabular}{llcc|cc}
%\begin{tabular*}{0.3\textwidth}{@{\extracolsep{\fill}}lcc}
\toprule
\toprule
Sample 1 & Sample 2 & \multicolumn{2}{c|}{{AD test}} & \multicolumn{2}{c}{{K test}}\\
& & Stat & SL & D & fpp \\
\midrule
\multicolumn{6}{c}{{BBC subsamples}}\\
\midrule
\multicolumn{6}{l}{{Periods:}}\\
BBC & BBC+SB1* & $-$0.947 & 25\% & 0.110 & 99.7\% \\
\multicolumn{6}{l}{{Eccentricities:}}\\
BBC & BBC+SB1* & \pp0.195 & 25\% & 0.120 & 98.9\% \\
%\midrule
\multicolumn{6}{l}{{Luminosity class and spectral type:}}\\
BBC & Dwarfs only & $-$1.149 & 25\% & 0.086 & \pd100\% \\
Early-B dwarfs & Late-B dwarfs & \pp1.375 & 8.8\% & 0.427 & 13.4\% \\
\midrule
\multicolumn{6}{c}{{Comparison with other OB-type samples}}\\
\midrule
\multicolumn{6}{l}{{30~Dor O-type period distribution:}}\\
BBC & TMBM & $-$0.481 & 25\% & 0.173 & 73.8\% \\
Early-B dwarfs & TMBM dwarfs & $-$1.072 & 25\% & 0.208 & 92.0\% \\
Late-B dwarfs & TMBM dwarfs & \pp1.921 & 5.2\% & 0.373 & \po6.4\% \\
\midrule
\multicolumn{6}{l}{{Galactic OB-type period distribution:}}\\
BBC & S12 & $-$0.490 & 25\% & 0.226 & 69.9\% \\
BBC & K14 & $-$0.353 & 25\% & 0.221 & 55.3\% \\
S12 & TMBM & \pp0.872 & 14.3\%\ & 0.311 & 16.7\% \\
% \multicolumn{6}{l}{{Unevolved S12 sample:}}\\
Early-B dwarfs & S12 dwarfs & \pp0.449 & 21.7\% & 0.376 & 24.2\% \\
Late-B dwarfs & S12 dwarfs & $-$0.497 & 25\% & 0.267 & 73.5\% \\
\midrule
\multicolumn{6}{l}{{Galactic B-type period distributions:}}\\
BBC dwarfs & K14 dwarfs & $-$0.024 & 25\% & 0.333 & 34.4\% \\
Early-BBC & Late-K14 & $-$0.464 & 25\% & 0.419 & 31.3\% \\
Late-BBC & Late-K14 & \pp3.028 & 1.90\% & 0.507 & 6.74\% \\
\midrule
\multicolumn{6}{l}{{Eccentricity distributions:}}\\
BBC & S12 & $-$0.673 & \pd25\% & 0.239 & 70.3\% \\
BBC & K14 & \pp5.650 & 0.20\% & 0.519 & 0.34\% \\
BBC & TMBM & \pp3.908 & 0.87\% & 0.359 & 0.30\% \\
%\multicolumn{6}{l}{{Mass ratio distributions:}}\\
%BBC & S12 & \pp5.377 & 0.26\% & 0.640 & 0.48\% \\
%BBC & K14 & \pp2.389 & 3.39\% & 0.429 & 61.8\% \\
%BBC & TMBM & \pp0.112 & 25\% & 0.329 & 66.6\% \\
\bottomrule
% \multicolumn{6}{p\columnwidth}{{\bf Notes. }\footnotesize{The `BBC sample' here refers to the SB1+SB2 results. The four sub-column entries are: (1) AD test statistic (Stat); (2) significance level (SL); (3) K test statistic (D); (4) false-positive probability (fpp).}} \\
\end{tabular}}
\tablefoot{The `BBC sample' here refers to the SB1+SB2 results. The four sub-column entries are: (1) AD test statistic (Stat); (2) significance level (SL); (3) K test statistic (D); (4) false-positive probability (fpp).}
\end{table}

\subsubsection{O-type binaries in 30~Dor}
Our first test was to compare our results for the B-type systems with those for the O-type binaries in 30~Dor from the \tmbm study. Both campaigns had similar observational strategies, targeting the full population of CCSN progenitors in 30~Dor between them.

For $P_{\rm orb}$\,$<$\,4\,d the distributions are near identical, although with a slightly larger fraction of the O-type binaries at the shortest periods ($P_{\textrm{orb}}<1.4$ d). 
There is a small difference at $P_{\rm orb}$\,$\sim$\,5\,d, with larger differences evident for 8\,$<$\,$P_{\rm orb}$\,$<$\,30\,d. For instance, at $P_{\rm orb}$\,$=$\,11\,d, the BBC fraction is at 60\% while it is 48\% for the \tmbm results. Aside from these differences, the two distributions are generally similar, and both of our statistical tests found no evidence to reject the null hypothesis, with an AD-test significance level of 25\% and a K-test probability of 73.8\%.

\subsubsection{Galactic samples}
The period distributions of the published Galactic samples display more differences with the BBC results. At 2\,d, the O-type binaries (\sana) and Cyg-OB2 (\kobu) samples have a lower fraction of systems, with a comparable fraction for all four samples at $P_{\rm orb}$\,$\sim$\,5\,d. It is remarkable that 40\% of the systems in three of the samples have $P_{\rm orb}$\,$<$\,6\,d (and with 55\% of systems from \sana).

Differences between the distributions start to more clearly appear after $P_{\rm orb}$\,$\sim$\,8\,d. For $P_{\rm orb}$\,$=$\,13\,d, the binary faction is 70\% for \sana, 60\% for BBC, and more like 50\% for \kobu (comparable to the \tmbm fraction). These differences are reduced by 20\,d, but increase again at $\sim$40\,d, with a larger fraction of the \kobu sample having $P_{\rm orb}$ of 30-50\,d. All four samples have fractions above 80\% for $P_{\rm orb}$\,$<$\,70\,d, with both Galactic samples reaching 90\% by a $P_{\rm orb}$\,$\sim$\,50\,d, beyond which their slopes flatten off, with relatively few systems with periods in the range of 200 to 500\,d.

As summarised in Table~\ref{tab:tab2}, our formal comparisons of the BBC results with those from \sana and \kobu reveal no statistically-significant differences.  For completeness, we also compared the two O-type distributions (i.e. S12 and \tmbm), finding a K-test probability of 16.7\%, which is lower than the value of 27\% from \citet{almeida+17} which only included the O-type stars (omitting the small number of early B-type objects) and
also considered the longer-period systems omitted here. Similarly, the AD test returns a value of 14\%, which is still not statistically significant.

\subsubsection{Dwarf samples}

\begin{figure}
\centering
  \includegraphics[width=0.95\hsize]{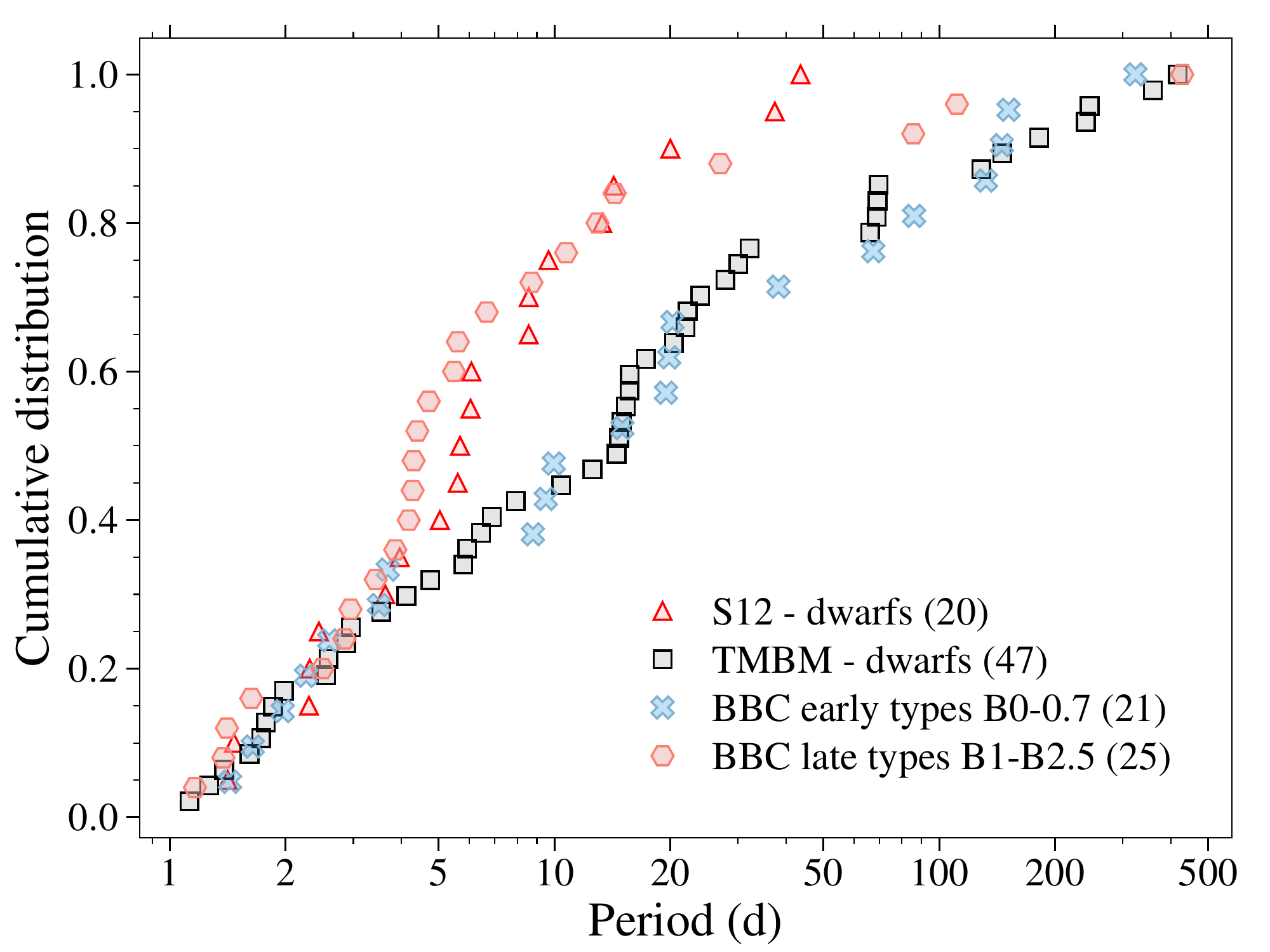}
  \caption{As in the bottom panel of Fig.~\ref{fig:cd_ST.pdf} with the inclusion of dwarfs from the \sana and \tmbm samples.}\label{fig:cdf_dwarfs}
\end{figure}

The Galactic and LMC period distributions of O-type dwarfs are compared with the two BBC sub-samples from Fig.~\ref{fig:cd_ST.pdf} (i.e. B0-0.7 and B1-2.5) in Fig.~\ref{fig:cdf_dwarfs}. The earliest BBC dwarfs have a 92\% probability of being drawn from the same distribution as the \tmbm sample of O-type dwarf systems. 
This suggests that the more massive B-type stars in the LMC follow the same period distribution as the more massive, main-sequence (and sub-giant) O-type binaries. 

In contrast, the period distribution of the later B-type binaries  more closely follows the distribution of the galactic sample of O-type binaries from \sana (K-test probability of 73.5\%). As discussed above, observational factors might influence the results for the later-type BBC systems, and the \sana study was not sensitive to longer-period systems. More thorough investigation of the biases in the BBC sample to quantify the intrinsic distributions for the B-type stars will be explored in a future paper.

\begin{figure}
\centering
  \includegraphics[width=0.95\hsize]{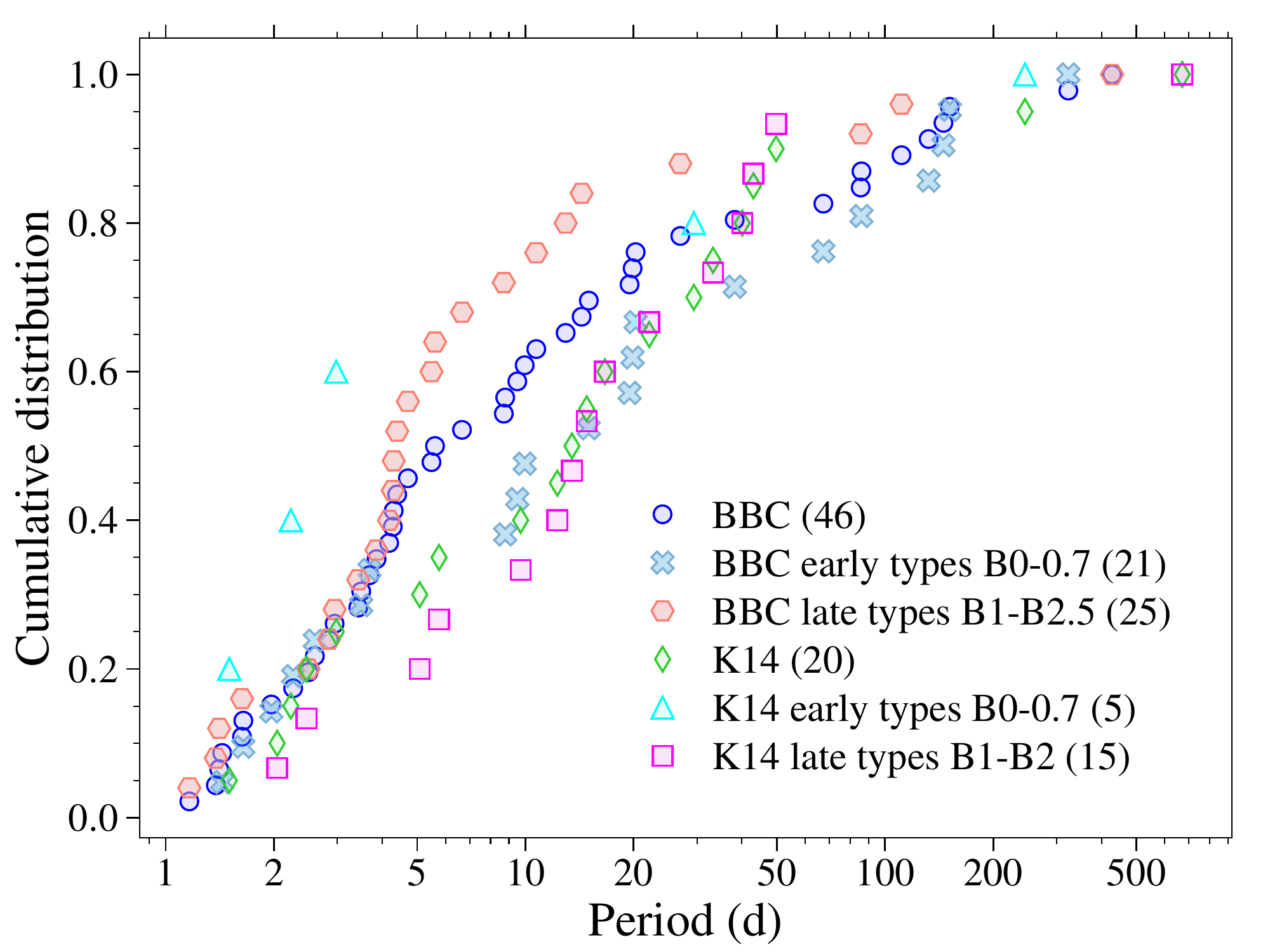}
  \caption{Comparison of cumulative distributions of periods between our sample and the B-type binaries from \citet{kobulnicky+14}. Blue circles and green diamonds represent the distributions of dwarfs for each sample. In both cases we have separated the samples in early and later-B-type stars with almost identical spectral ranges.}
     \label{fig:cdf_P_K14}
\end{figure}

\subsubsection{B-type samples}
We took a closer look at the results for the B-type binaries from \kobu compared to our period distribution (Fig.~\ref{fig:cdf_P_K14}). The distributions are qualitatively quite different, with the largest difference ($\sim$20\%) at $P_{\rm orb}$\,$=$\,10\,d. Nonetheless, these are not statistically significant (in part given the sample sizes), with a K-test probability of 34.4\% of drawing these samples from the same distribution. 

We also split the \kobu sample into early/late types in Fig.~\ref{fig:cdf_P_K14} as we did earlier for the BBC results. This results in only five members in the earlier spectral bin for the \kobu results, so we only consider the later-type objects further (albeit still with a limited sample size of only 15). The later-type, B-type Galactic binaries appear to more closely follow the distribution of the earlier-type systems from the BBC results (at least for periods of up to a month). The K-test returned a probability of 31.3\% between the early BBC and late \kobu samples, which drops further to 6.7\% when comparing the two groups of later-type systems. However, both selection effects and small sample sizes limit meaningful conclusions.

A study of the B-type binary population of NGC~6231 (Banyard et al. in prep.) will considerably increase the number of B-type binaries characterised in young Galactic clusters, enabling a better comparison between the properties of Galactic and LMC B-type binaries in the near future.

\subsubsection{Eccentricities and mass ratios}

\begin{figure}
\centering
  \includegraphics[width=0.95\hsize]{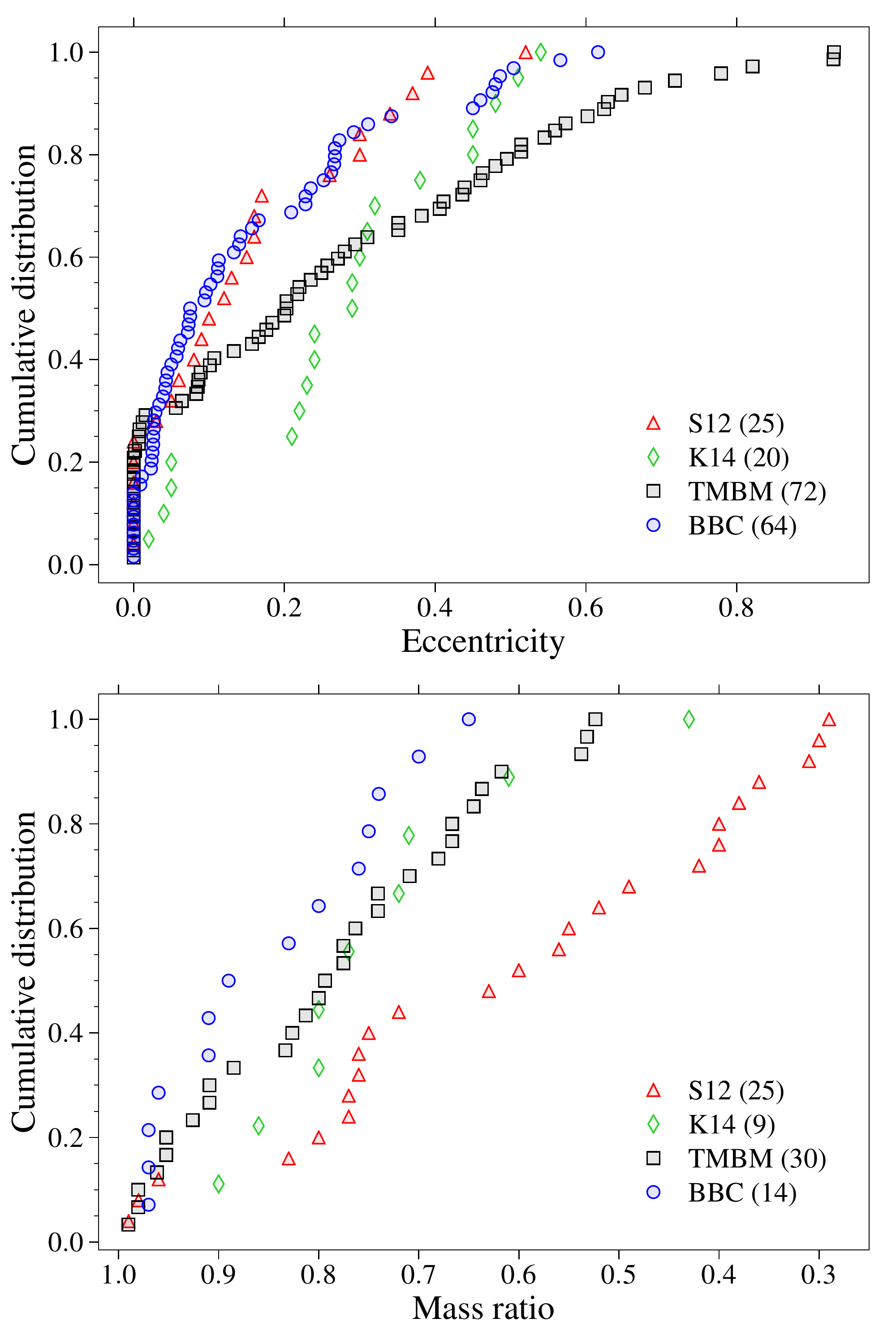}
  \caption{Cumulative distributions of eccentricities (top) and mass ratios for the BBC SB2 systems (bottom) compared with the three published samples.}
     \label{fig:cdf_eq}
\end{figure}

The distributions of eccentricities and mass ratios are shown in the upper and lower panels of Fig.~\ref{fig:cdf_eq}, respectively, for the BBC results compared with the other three samples. In both figures the \kobu results are limited to only the B-type systems, and we note that \sana were not able to estimate either parameter for five of their systems.

Close to 30\% of all the systems have $e$\,$<$\,0.03, with the exception of the B-type results from \kobu where 80\% of the systems have 0.2\,$<$\,$e$\,$<$\,0.6. Once above 30\%, the \tmbm distribution also differs from the BBC results due to the much larger eccentricities for some of the O-type systems (even some with $e$\,$>$\,0.9). For both the \tmbm and \kobu results, the probability of the BBC results being drawn from the same distribution is only 0.3\%. The BBC distribution is similar to that from \sana, although our sample has a larger number with $e$\,$>$\,0.4
(and the \sana sample is relatively small in comparison). 

Meaningful comparison of the mass ratios is not possible due to the small samples and the different observational factors influencing the B-type SB2s compared to the O-type systems (see Sect.~\ref{sec:q}) but we have included the distributions of all samples in the bottom panel of Fig.~\ref{fig:cdf_eq} for completeness.

\section{Discussion}\label{sec:disc}

\subsection{Period detectability}\label{ssec:detectability}

We were unable to confirm the orbital period for the 20 systems that we labelled as SB1* in Sect.~\ref{ssec:meth-Porb}, which include five giants (three of which have spectral types of B2 or later), five (from six) of the Be-type stars, and one supergiant (see Table~\ref{tab:tabB4}). These 20 systems constitute 23\% of the BBC sample, and there is a further 5\% (four targets) for which we did not find signs of periodicity
(Table~\ref{tab:RVvar}). Possible explanations for these include low-amplitude RV shifts from low-mass companions, fast rotators, low S/N of our targets, and that our spectral range exactly spans the temperature domain of pulsating B-type stars, e.g. $\beta$~Cep \citep{stankov+handler05} and even slowly-pulsating stars \citep{waelkens+98}. These points are now discussed in turn.

\begin{figure}
\centering
  \includegraphics[width=0.95\hsize]{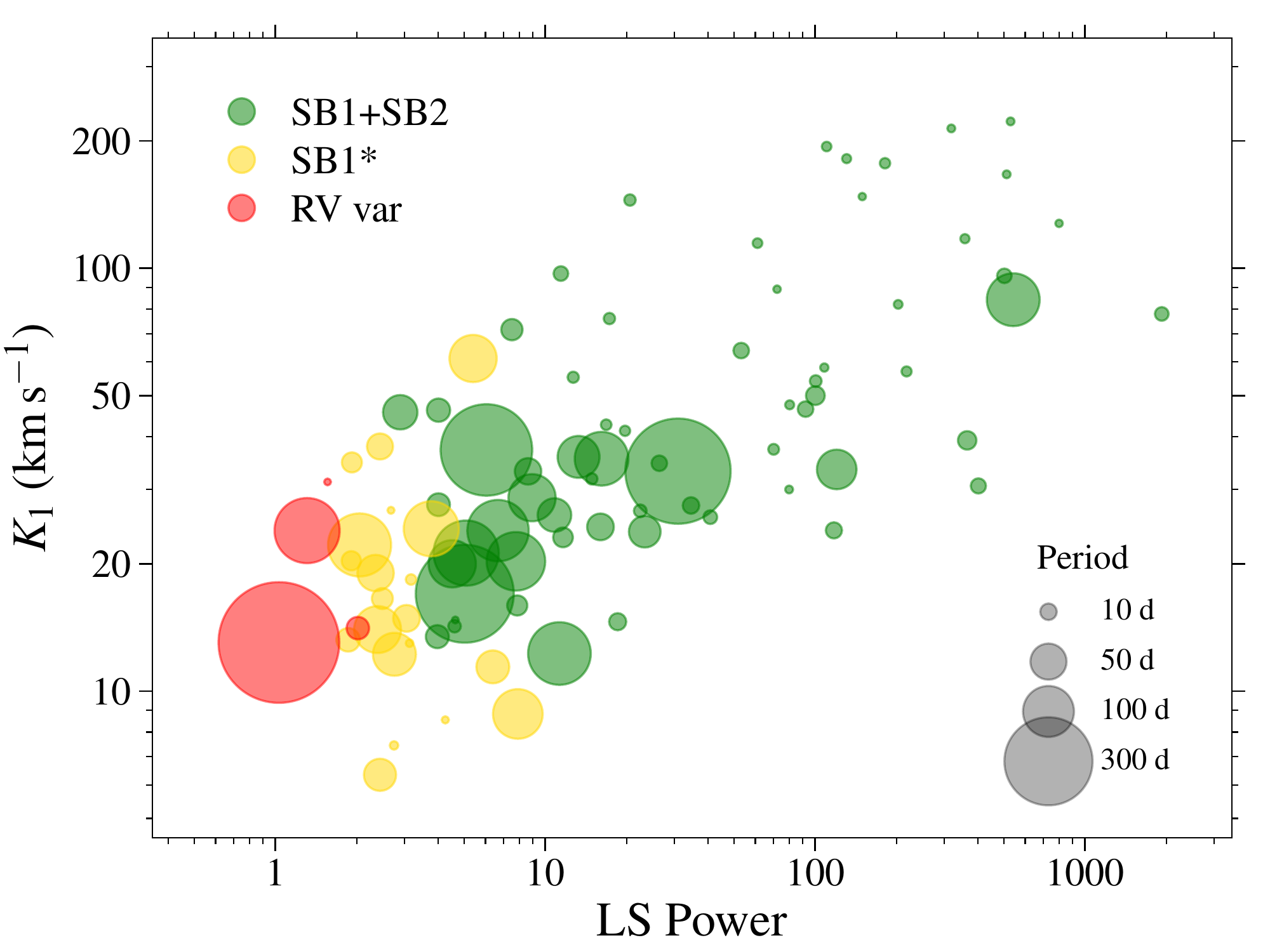}
  \caption{Comparison of semi-amplitude velocities ($K_1$) and Lomb-Scargle (LS) power. The three groups classified by the reliability of their estimated periods show that low LS powers are correlated with low $K_1$ values.  Circle sizes are scaled by the orbital periods as shown.}
     \label{fig:dRV}
\end{figure}

\subsubsection{Effects of semi-amplitude velocities and rotation}

Figure~\ref{fig:dRV} offers clues for the non-negligible fraction of systems without confirmed orbital periods. There is a clear trend of semi-amplitude velocities with LS power: systems with larger $K_1$ present stronger signals in the periodograms and vice versa, albeit with some notable exceptions, as discussed in Appendix~\ref{apx:notes}. In fact, 85\% of the SB1* systems have $K_1$\,$\lesssim$\,30\kms. It is also interesting to note that the systems with larger periods ($P_{\mathrm{orb}}\gtrsim100$) have LS powers of less than 20 in most cases, an indication of the difficulties in finding long-period binaries from our observational campaign. 

\begin{figure}
\centering
  \includegraphics[width=0.95\hsize]{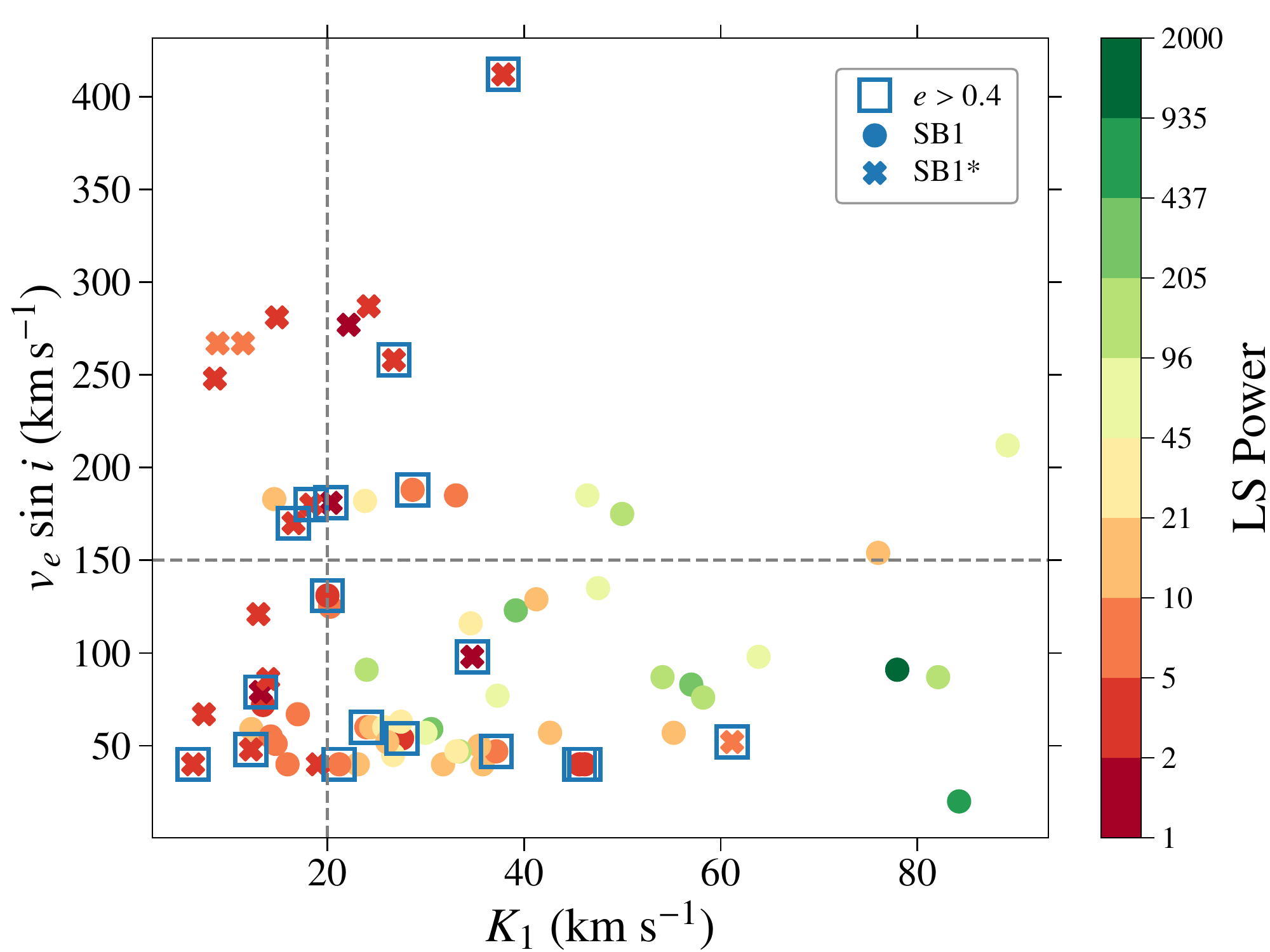}
  \caption{Rotational velocities as a function of the semi-amplitude velocities for the SB1 and SB1* samples. Systems with high eccentricities are marked by blue squares. The diagram suggests that all SB1* systems have less significant signals in the LS periodogram due to their high $v_e\sin\,i$, low $K_1$, and/or high eccentricities. LS power is indicated by the log-scale colour bar, with the tick labels rounded to the closest integer.}
     \label{fig:vsini}
\end{figure}

Moreover, an analysis of the projected rotational velocities ($v_e\sin\,i$) of our sample \citep[taken from][see Table~\ref{tab:tabB1}]{dufton+13,garland+17} shows that 40\% of the SB1* systems have $v_e\sin\,i > 230$\kms, from which three were classified as Be stars by \citet{evans+15}. Figure~\ref{fig:vsini} shows the $v_e\sin\,i$ values vs. semi-amplitude velocities; all SB1* systems are either relatively fast rotators ($v_e\sin\,i > 150$\kms), have low semi-amplitude velocities ($K_1<20$\kms), have high eccentricities ($e>0.4$), or a mixture of these properties. 

This evidence might suggest a few challenges: ({\it i}) for systems with low $K_1$, we do not have enough S/N to accurately determine the RV shifts; ({\it ii}) small RV shifts might be more affected by stellar pulsations; ({\it iii}) an important fraction of systems with small RV shifts are high-eccentricity (and/or long-period) binaries; ({\it iv}) systems with large $v_e\sin\,i$ values might need higher S/N to determine robust RV shifts and periods; ({\it v}) a combination of all these points.

\subsubsection{S/N of the BBC campaign}

\begin{figure}
\centering
  \includegraphics[width=0.95\hsize]{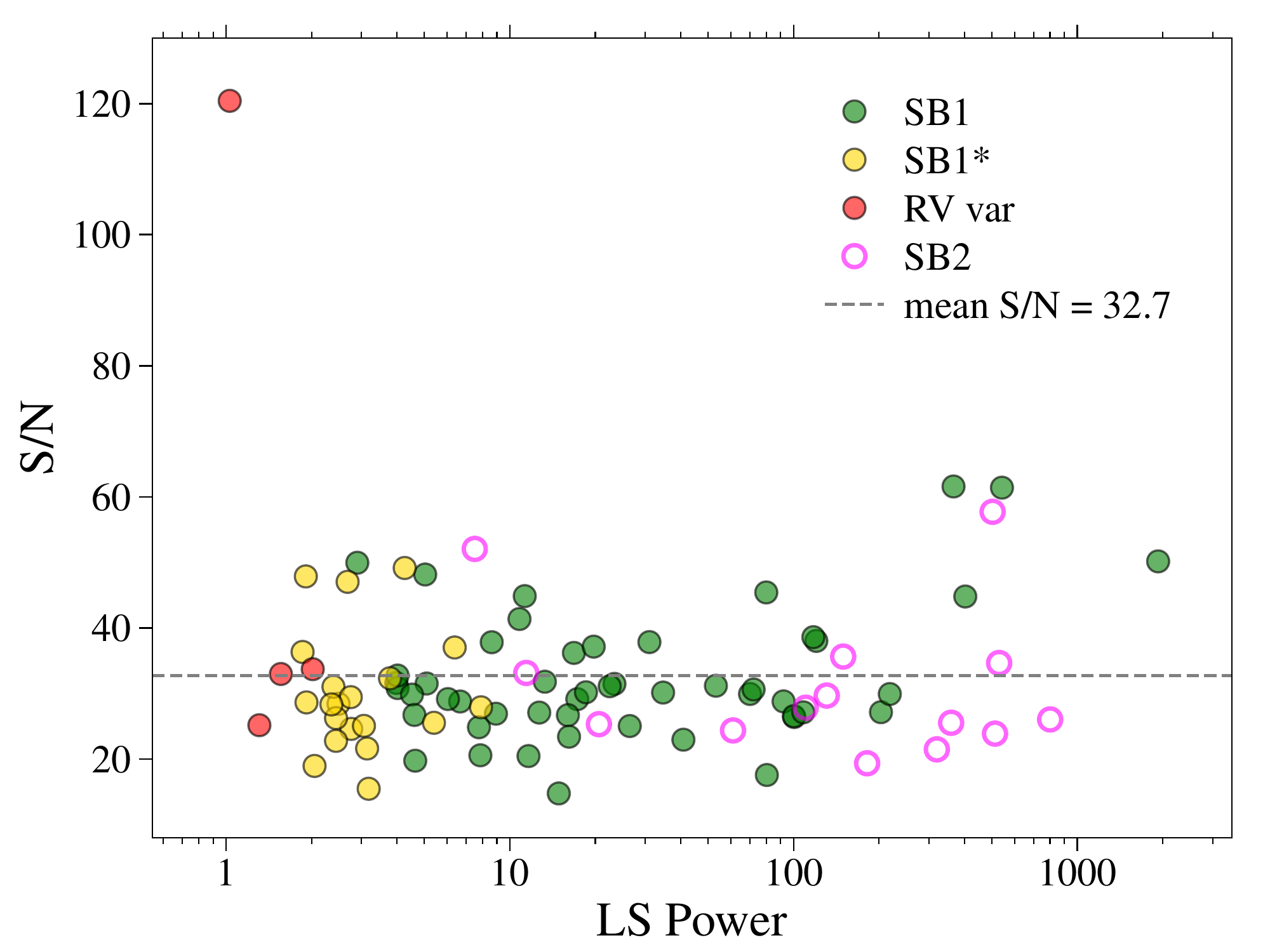}
  \caption{Correlation between S/N of our targets and LS power. }
     \label{fig:SN-LSP}
\end{figure}

The lower S/N of the BBC sample in comparison to the \tmbm and Galactic samples might have limited the period estimates of some of our targets. However, this does not appear to be the case given our results. Figure~\ref{fig:SN-LSP} shows the S/N values for the BBC targets against the LS power of the the maximum peak in the periodograms. There is no strong correlation between these values, although SB1 systems with the highest S/N have also the highest LS power.
It appears that S/N did not play a significant role in the determination of orbital periods for most of our sample, but it remains as a possible source of uncertainty in the periods of systems with high $v_e\sin\,i$ and low semi-amplitude velocities, and it is still crucial in identifying SB2 systems with fainter secondaries.

\subsubsection{Be stars and non-radial pulsations}

Most Be stars are thought to present non-radial pulsations (NRPs), with periods of 0.5 to 2\,d and amplitudes of $\lesssim 20$\kms \citep{rivinius+13}. If some of our period determinations are affected by NRP, this would be expected to happen for the short-period and low-$K$ systems.

In this context, we examined the semi-amplitude velocities and periods for the BBC sample. As expected, the short-period systems generally have higher $K$-values, especially the SB2 systems (see Fig.~\ref{fig:K1-P}). When we consider the SB1 and SB1* systems there are a handful of systems with $P_{\rm orb}<2$\,d and $K_1<30$\kms (lower left in  Fig.~\ref{fig:K1-P}). These are two SB1 (VFTS 179 and 324) and four SB1* systems (VFTS 662, 697, 730 and 847). VFTS 697 and 847 are both Be stars, and VFTS 730 has a large rotational velocity ($v_e\sin\,i$\,$=$\,248\,\kms). Of the two SB1 systems, VFTS 324 has the most robust period and a well constrained orbital solution, whereas VFTS 179 shows evidence of a possible alternative period of $\sim$7\,d (see individual notes in Appendix~\ref{apx:notes}). Except for VFTS 324, each of these systems could be displaying periodicity from NRP. These
results are particularly interesting for the Be stars as there is 
lack of known Be stars with MS companions. If not due NRP, the low $K$-velocities of the Be stars would require a MS secondary or a compact companion. We note that a more detailed study of the Be stars from the VFTS (including those discussed here) is underway (Dufton et al. in prep), including revisiting the estimates of $v_e\sin\,i$.

\subsection{Eclipsing binaries}

Eleven of our systems (including seven of the SB2s) were reported as EBs from analysis of the OGLE imaging survey \citep{pawlak+16}. For 10 of the 11 systems, the OGLE and BBC periods show excellent agreement, with absolute differences ranging from 0.0042 to 0.000014 days (6 min to 1.2 s). Our periods for VFTS 112 and 189 are also in excellent agreement with results using photometry from the EROS survey \citep{muraveva+14}.

The only exception compared to the OGLE results is the SB2 system VFTS 883, where the published period is 2.00014 times our estimate. This can happen due to a mis-identification of the components. In fact, VFTS 883 has the lowest S/N ($\sim$20) of the SB2 systems, affecting the number of epochs used in our LS analysis and, apparently, the correct identification of the components of the binary.

To search for such effects in our other SB2 systems, we conducted the LS analysis on the absolute difference between RVs of the two components in the SB2 systems. This test should produce signals that are half the orbital period \citep{almeida+17}. Beyond the seven EBs, we confirmed the period estimates of four of our SB2 systems, whereas we did not find significant signals for the three remaining systems (VFTS 520, 589, 686). The latter two of these are the more eccentric systems, which probably limited this additional test, although given the good agreement between the overlapping systems in BBC and OGLE, our periods for these appear robust.

\subsection{Searching for compact companions}

\begin{figure}
\centering
  \includegraphics[width=0.95\hsize]{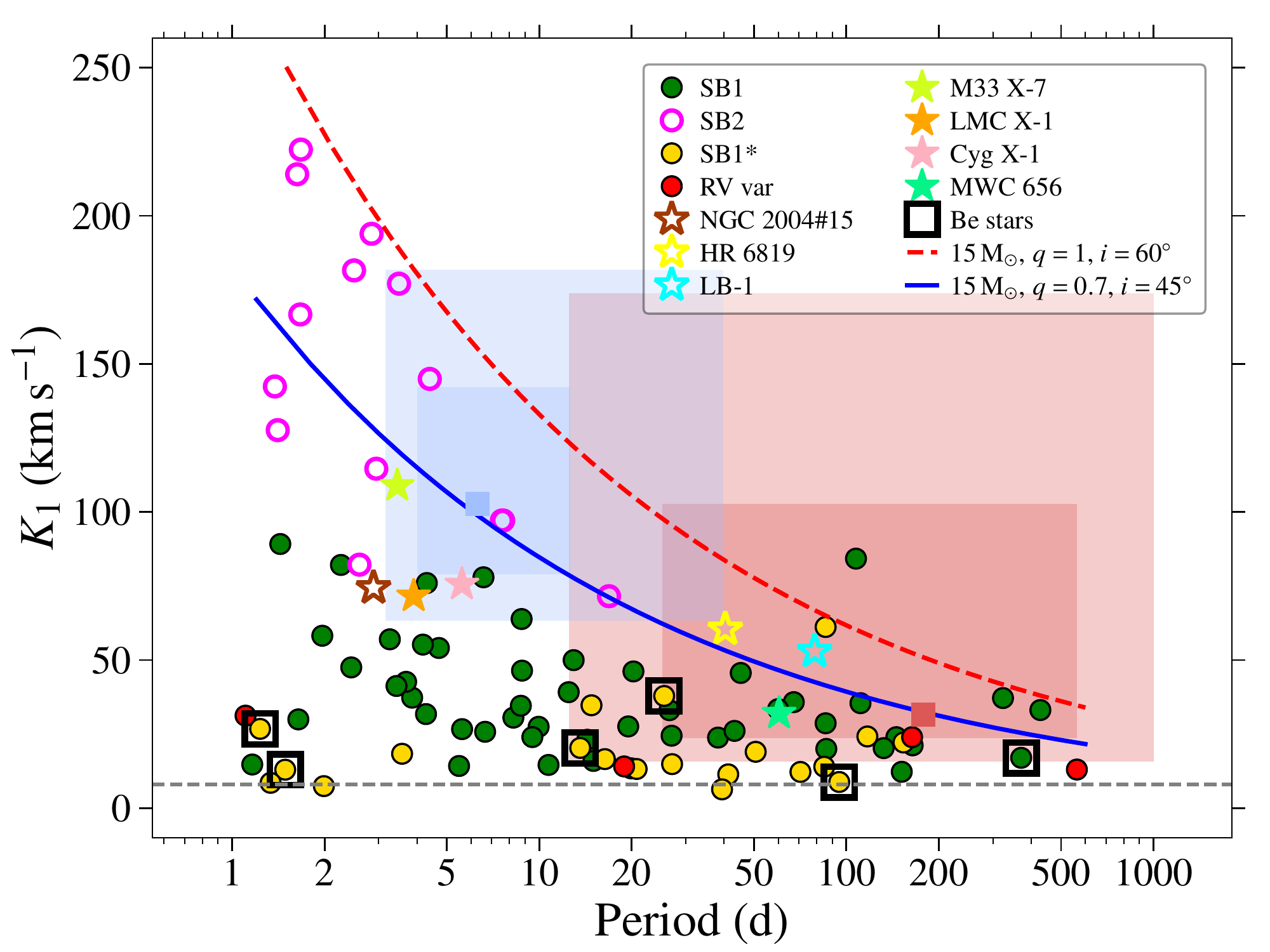}
  \caption{Semi-amplitude velocity ($K_1$) vs. period ($P$) for the BBC sample, compared with results for known and candidate BH systems from the literature. The red-solid line represents systems with $M_1=15$\Msun, $q=1$ and $i=\pi/6$, whereas the blue-solid line corresponds to $M_1=15$\Msun, $q=0.7$ and $i=\pi/4$. Shaded blue and red areas indicate the most probable regions for a BH companion from \citet{langer+20}, see text for details. The dashed line indicates the 8~km\,s$^{-1}$ threshold ($\Delta\mathrm{RV}_{\mathrm{min}}=16$~km\,s$^{-1}$) chosen by \citet{dunstall+15}.}
     \label{fig:K1-P}
\end{figure}

 The $K$-$P$ diagram in Fig.~\ref{fig:K1-P} is an interesting way of examining the sample because SB1 systems with a high $K_1$ are potential candidates to have compact companions. It is worth mentioning that these are semi-amplitudes of the projected orbital velocities meaning that we only see a fraction of the true velocity.

Two recent studies claimed to have found BHs in binary/multiple systems \citep{liu+19,rivinius+20}, but subsequent studies have proposed a different configuration involving a primary helium star and a Be secondary in both cases \citep{shenar+20, bodensteiner+20b}. We have included these two objects, LB-1 and HR~6819, in Fig.~\ref{fig:K1-P} with values from \citet{shenar+20} and \citet{bodensteiner+20b}, respectively. We also include a third, shorter-period system of interest in this context, NGC 2004\#115 \citep[Lennon et al. in prep., from the FLAMES data of][]{evans+06}. These three systems each have $K_1$\,$>$\,50\kms. We have also plotted in Fig.~\ref{fig:K1-P} four of the known OB-type stars with BH companions \citep{orosz+07, orosz+09, orosz+11, casares+14}\footnote{M33-X7 is a 70\Msun O6 III with a 15.6\Msun BH companion \citep{pietsch+06,orosz+07}. LMC X-1 contains a 31.8\Msun O8(f)p primary with a 10.91\Msun BH \citep{orosz+09, walborn+10a}. Cyg X-1 is a 19.2\Msun O9.7 Iabpvar star with a 14.8\Msun BH secondary \citep{orosz+11, sota+11}. Finally, MWC 656 is the only known Be star (B1.5 IIIe, $M$\,=\,10\,--16\Msun) with a BH companion of 3.8--6.9\Msun \citep{casares+14}.}. The known BH companions and recent candidates tend to have large $K_1$ velocities relative to the bulk of the BBC results for a given period.
Several of the BBC systems are found in similar locations in the figure, with five SB1 systems with $K_1$\,$>$\,70\kms. Additional data for these systems are required to see if they are simply outliers to the distribution or if they might also harbour BH companions.

\citet{langer+20} recently suggested that about 1.5\% of the B-type stars in the LMC with masses above 10\Msun could have BH companions, which would translate to about 60 B+BH systems. Given the current population of HMXBs in the LMC \citep{vanjaarsveld+18}, binaries with compact companions are probably X-ray quiet, as previously discussed by \citet{casares+14}. 
\citeauthor{langer+20} also predicted the distributions of 
periods and semi-amplitudes (their Fig.~12) for systems resulting from 
Case~A and B mass-transfer scenarios. For the Case~A systems they estimated the maximum probability of finding a BH companion for a given system at just above 2\% at a peak of 6.3\,d in the orbital period distribution, and a probability of $\sim$1.5\% at the peak of the semi-amplitude velocities (130\kms).  The probabilities for the
Case~B systems are higher, with almost 8\% at a peak of 178\,d, and $>$4.5\% for peak velocities of 40\kms. This implies that most of the OB binaries with BH companions are expected to come from Case~B evolution and, in fact, the period distribution of their Case~B systems agrees well with that of Galactic Be/X-ray binaries.

The maximum probabilities, the peak in the distributions from Fig.~12 in \citet{langer+20}, are included in our $K$-$P$ diagram (Fig.~\ref{fig:K1-P}) as small blue (Case~A, $P_{\mathrm{orb}}=6.3\,\mathrm{d}, K_1=102.7\,\mathrm{km~s}^{-1}$) and small red (Case~B, $P_{\mathrm{orb}}=177.8\,\mathrm{d}, K_1=31.6\,\mathrm{km~s}^{-1}$) squares, reduced by 21\% for $K$ as suggested by \citeauthor{langer+20} due to projection effects. The larger blue and red rectangles in the figure indicate probabilities between 0--1\% (largest blue rectangle) and $>$1\% (middle blue rectangle) for both $P$ and $K_1$ in the Case~A systems, and probabilities between 1--3\% and 0--3\% (largest red rectangle) for $P$ and $K_1$ respectively, and $>$3\% (middle red rectangle) for both $P$ and $K_1$ for the Case~B systems. These predicted regions match the location in the diagram of the known BH companions to OB-type stars and the other candidates; 33 BBC systems are also within those regions, with 16 located in the higher-probability region for the case B systems.

Curves for two different sets of parameters are plotted in Fig.~\ref{fig:K1-P} to help guide the discussion. The red curve indicates the position of systems with $M_1$\,$=$\,15\Msun, $q$\,$=$\,1 and $i$\,$=$\,$\pi$/3, effectively giving an upper boundary for the SB2 systems. The blue line was plotted for $M_1$\,$=$\,15\Msun, $q$\,$=$\,0.7 and $i$\,$=$\,$\pi$/4; this is a more typical value for $i$, but $q=0.7$ is close to the limit at which we can detect SB2 systems (albeit the curve here is for circular orbits, and the limit depends on $e$ as well as $P_{\rm orb}$). Nonetheless, for a flat $q$-distribution we expect the area below the curve for a sample such as ours to be mostly populated by SB1 systems, as observed in Fig.~\ref{fig:K1-P}. Another reason to choose $q$\,$=$\,0.7 is that it was the average value at $\sim$15\Msun predicted by \citeauthor{langer+20} for B+BH systems, with a rather narrow $q$-distribution. We therefore expect B+BH binaries to be close to this line, which also passes through the regions of highest probability from the simulations. Furthermore, the locations of known and candidate BHs are well traced by the curve, with several BBC binaries close to the curve and four systems above it. The analysis of massive binaries through the $K$-$P$ diagram and the predictions from \citet{langer+20} open interesting possibilities for future studies of this and other samples of massive stars. All BBC candidates to B+BH systems will be investigated in more detail by Villase\~nor et al. (in prep.).

\section{Summary}\label{sec:summary}

We have presented comprehensive multi-epoch spectroscopy of 88 candidate B-type binaries in the 30~Dor region of the LMC, analysed in the framework of the B-type Binaries Characterisation (BBC) programme. Using profile fits of (up to) nine absorption lines in the FLAMES spectra, we have estimated stellar RVs for each target for each epoch, and used LS tests to search these for periodicities from binary motion. For those with robust periods we estimated orbital properties (periods, eccentricity, mass ratios for the SB2 systems) using the \rvf code, and then compare these with published distributions for more massive O-type stars in the Galaxy and LMC. Our findings include:
\begin{itemize}
    \item{Robust period estimates for 50 SB1 and 14 SB2 systems ranging between 1.16 and 428\,d, which comprises the largest homogeneous study of binary B-type stars to date. We have less secure periods for a further 20 systems (classified here as `SB1*'), with only four targets where our period search did not find a robust value.}\smallskip
    \item{Determination of full orbital orbital solutions, including eccentricities and mass ratios for SB2 systems. The observed distributions show a preference for low eccentricities ($e<0.4$) and mass ratios close to one, although observational biases could have an important effect, particularly for the latter.}\smallskip
    \item{The period distributions of the earliest B-type dwarfs (B0-0.7 types) compared to the dwarfs with later types (B1-2.5) are qualitatively quite different. Our statistical tests show there is only a $\sim$10\% chance of these being drawn from the same parent distribution. Although of weak significance, the later-type objects have relatively few systems with longer-periods ($>$10-20\,d). Whether this is an observational bias or is physically meaningful is not clear at present.}\smallskip
    \item{The period distribution from the BBC results shows no significant differences to those for O-type stars in the Galaxy \citep{sana+12}, in 30~Dor \citep{almeida+17}, nor the combined Galactic OB sample in Cyg OB2 \citep{kobulnicky+14}. Of particular note, the period distribution of the early-type (B0-0.7) BBC results is in excellent agreement with that for O-type stars in 30~Dor \citep{almeida+17}.}\smallskip
    \item{We have investigated the semi-amplitude velocities ($K_1$) of our confirmed binaries as a function of their orbital periods ($P_{\rm orb}$) in the $K$-$P$ diagram (see Fig.~\ref{fig:K1-P}), compared to those of known/candidate BH-binaries and theoretical predictions. We have identified several targets that have moderately large $K_1$ velocities that merit further study to investigate the nature of their companions (Villaseñor et al. in prep.).}
\end{itemize}
In summary, our study suggests that the general properties of binaries in the early B-type domain are largely similar to those for their more massive O-type cousins, at both Galactic and LMC metallicities. This apparent invariance with mass and metallicity for massive stars is an important result for population-synthesis models that include binary evolution in the context of the rates of CCSNe and compact remnants. However, detailed modelling of the different observational biases affecting the BBC sample will need to be computed to determine the universality of our findings. 

\section*{Acknowledgements}

Based on observations at the European Southern Observatory Very Large Telescope in programme 096.D-0825. We thank Liz Bartlett for helpful discussions regarding the period analysis of these data, and Danny Lennon for important suggestions in the determination of orbital solutions. We are grateful to the referee for their thoughtful and constructive comments. JIV acknowledges support from CONICYT-Becas Chile, ``Doctorado en el extranjero'' programme, Grant No. 72170619. O.H.R.A. acknowledges funding from the European Union’s Horizon 2020 research and innovation programme under the Marie Skłodowska-Curie grant agreement No 665593 awarded to the Science and Technology Facilities Council. SdM was funded in part by the European Union’s Horizon 2020 research and innovation program from the European Research Council (ERC, Grant agreement No. 715063), and by the Netherlands Organization for Scientific Research (NWO) as part of the Vidi research program BinWaves with project number 639.042.728 and the National Science Foundation under Grant No. (NSF grant number 2009131).

\section*{Data Availability}
 
Reduced and normalised data available on request.

%%%%%%%%%%%%%%%%%%%% REFERENCES %%%%%%%%%%%%%%%%%%

\bibliographystyle{mnras}
\bibliography{bibfile} 

%%%%%%%%%%%%%%%%% APPENDICES %%%%%%%%%%%%%%%%%%%%%

\appendix

\section{Notes on individual objects}\label{apx:notes}

The following notes describe difficulties we had finding periods for specific systems (e.g. particularities found in the spectra and/or in the periodograms of some of our targets) and further details on the tests conducted, specifically our lines combination test (LCT) described in Sect.~\ref{ssec:lcomb}.

\begin{itemize}
%\subsubsection*{BBC 018}
\item {\it VFTS\,018:} Possible period at 70.8\,d but the peak does not surpass the 0.1\% FAP. The \lc analysis found 46\% of the line combinations with the peak at 70.8\,d, but also found peaks at 3.9\,d (16\% of line sets) and 61\,d (6\%).
\smallskip

%\subsubsection*{BBC 037}
\item {\it VFTS\,037:} Weak nebular emission is present in all the helium lines and we carried out the analysis with and without fitting the nebular emission. Two peaks were detected at 41 and 71\,d using different sets of lines, with the latter period slightly favoured by fits without nebular emission. We adopted $P_{\rm orb}$\,$=$\,41\,d as its peak is above the 0.1\% FAP, but flag it as only a `possible' period given the other peak. Our \lc for fits including nebular emission recovered $P_{\rm orb}$\,$=$\,41\,d for 20\% of the possible line sets (in which all included a larger peak at 1\,d, most likely an alias), and where the peak at 41\,d was above the 0.1\% FAP for two combinations. The possible period of 71\,d was only recovered for 12\% of the line sets, where it neither surpassed the 0.1\% FAP nor the strength of the alias close to 1\,d. A third peak at 46\,d was also present in the periodogram and presented a feasible solution with \rvf, but its peak in the periodogram was always weaker than that at 41\,d.
\smallskip

%\subsubsection*{BBC 097}
\item {\it VFTS\,097:} Although the RV curve looks noisy and its peak in the periodogram is slightly above the 0.1\% FAL, we confirmed the estimated 19.8\,d period for 55\% of line combinations from the LCT.
\smallskip

%\subsubsection*{BBC 106}
\item {\it VFTS\,106:} Possible period of 16.4 days found for 24\% of the different sets of lines, all of which included \ion{He}{I} $\lambda$4471 (which presents weak nebular emission). Its RV data is well fit by \rvf with one of the lowest $\chi^2$ values among the SB1* systems.
\smallskip

%\subsubsection*{BBC 144}
\item {\it VFTS\,144:} Two peaks are present at $P_{\rm orb}$\,$\sim$\,171 and $\sim$351\,d but neither surpasses the 1\% FAP threshold. 
\smallskip

%\subsubsection*{BBC 146}
\item {\it VFTS\,146:} The LCT found $P_{\rm orb}$\,$=$\,117\,d (that we missed in our initial analysis) for 52\% of the line sets. Given the weak signal in periodogram and resulting RV curve, we have therefore classified this system as SB1*.
\smallskip

%\subsubsection*{BBC 155}
\item {\it VFTS\,155:} Two peaks at $\sim$154 and $\sim$335\,d, but $P_{\rm orb}$\,$\sim$\,154\,d is adopted as it exceeds the 1\% FAP in 12\% of the cases when running the \lc. We have included this system in the SB1* group.
\smallskip

%\subsubsection*{BBC 162}
\item {\it VFTS\,162:} The spectra indicate line-profile variability or a very weak secondary. Periods of 98 and 144\,d were estimated from different sets of lines, with the longer period the most common and prominent. The longer period presents the clearer periodogram with the stronger peak and was adopted as the orbital period of this system.
\smallskip

%\subsubsection*{BBC 179}
\item {\it VFTS\,179:} The shortest-period binary in our sample with $P_{\rm orb}$\,$=$\,1.16\,d. A second, weaker peak is present in the periodogram at 7\,d, but the shorter period had the maximum peak in the periodogram for 62\% of the line combinations and is adopted as the period. However, we note that we found a plausible orbital solution for the 7\,d period, with $e=0.28$, $K_1=18.9$\kms and $\chi^2_{\rm red}=1.15$, but it relies on RVs from the weak \ion{Mg}{ii}. 
\smallskip

%\subsubsection*{BBC 204}
\item {\it VFTS\,204:} We found $P_{\rm orb}$\,$=$\,157\,d for this system only after checking with our \lc. Although we found some scatter, with periods between 148 and 161 days, the 157\,d period was recovered for 36\% of the possible line combinations. The \rvf analysis computed a final period of 164\,d, with the highest eccentricity value among the SB1 systems. 
\smallskip

\item {\it VFTS\,206:} Given the observed profile in the stronger \ion{He}{i} $\lambda4026$ and $\lambda4471$ lines, it is possible that the primary is a fast rotator with a B8 secondary. In addition to the solution reported in Table~\ref{tab:tabB6}, we also investigated the scenario of a potentially smaller mass ratio. The signal remains very strong for the same 
period (7.56\,d), and we found a orbital solution with $q=0.19$. For a typical B8 mass of 3.5\Msun, this implies $M_1\sim18$\Msun. However, knowing the limitations of our observational campaign to detect such mass ratios, we favoured the solution with more similar masses.
\smallskip

%\subsubsection*{BBC 213}
\item {\it VFTS\,213:} A possible period of 13.5 d was determined for this Be star from a peak above 1\% FAP present for 57\% of the possible sets of lines. However, two stronger peaks at 0.5 and 1\,d are present in the periodograms of all the cases found with the \lc, that might be aliases of the longer period. Its RV curve presents significant scatter and appears to have a high eccentricity.
\smallskip

%\subsubsection*{BBC 218}
\item {\it VFTS\,218:} Presents line-profile variability in its helium lines, most noticeably in \ion{He}{I} $\lambda\lambda$4388, 4471. The variability shifts between the red and blue wing which might be caused by a weak secondary. A possible period of 20.8 d was found from a weak signal above the 1\% FAP, which was also found in 19\% of the cases with our \lc, and for an additional 55\% as a secondary peak. A weaker peak is also present in the periodogram at 80\,d that does not improve the orbital solution. As in the case of VFTS\,213, its RV curve presents significant scatter, high eccentricity and a high $\chi^2$ value.
\smallskip

%\subsubsection*{BBC 240}
\item {\it VFTS\,240:} A SB2 system with a weak secondary component only visible in a few epochs that makes it difficult to fit. We have confirmed its 1.38\,d period by fitting it also as a SB1 and including the hydrogen lines.
\smallskip

%\subsubsection*{BBC 255}
\item {\it VFTS\,255:} This SB2 system presents nebular emission in \ion{He}{I} $\lambda$4026 and $\lambda$4471. We confirmed its period by both using and excluding these lines from the RV measurements to determine the period. 
\smallskip 

%\subsubsection*{BBC 257}
\item {\it VFTS\,257:} Although its RV curve presents some scatter, 71\% of line combinations indicated $P_{\rm orb}$\,$=$\,132\,d. 
\smallskip

% \subsubsection*{BBC 291}
\item {\it VFTS\,291:} Presents some line-profile variability in He I $\lambda$4026 which looks to be mild nebular emission. This does not appear to be the case from inspection of He I $\lambda$4471, which could suggest signs of a weak secondary. It has a 107.6\,d period with a circular orbit.
\smallskip

% \subsubsection*{BBC 299}
\item {\it VFTS\,299:} Another system displaying line-profile variability. We are unable to determine if its variability is due to weak nebular emission or a secondary. Its period of 20.3\,d was recovered for 100\% of the different sets of lines with our \lc, although the RV curve presents some scatter and high eccentricity.
\smallskip

% \subsubsection*{BBC 337}
\item {\it VFTS\,337:} One of our six Be stars. It presents emission in all the spectral lines available for RV measurements and the weak helium lines are probably also broadened by rotation. We were initially unable to find a period for this system but further checks with our \lc showed a peak at 27\,d for 12\% of the possible lines sets, which we have included as a possible period.
\smallskip

% \subsubsection*{BBC 359}
\item {\it VFTS\,359:} Line-profile variability is noticeable in the wings of He I $\lambda$4388, especially for epochs 6, 21 and 23. It is unclear if these features are from lunar contamination (i.e. linked to the background subtraction), have a nebular origin or hint at a faint secondary. A peak at 19.5\,d is present for 67\% of the possible lines combinations confirming our initial period.
\smallskip

% \subsubsection*{BBC 364}
\item {\it VFTS\,364:} As with VFTS 255, we have confirmed the period of this SB2 system with and without the use of He I $\lambda$4026 and $\lambda$4471 which present clear nebular contamination.
\smallskip

% \subsubsection*{BBC 383}
\item {\it VFTS\,383:} We have confirmed the SB2 status of this binary due to the presence of \ion{Si}{III} $\lambda$4553 from the secondary spectrum, which was otherwise difficult to determine given the nebular contamination in the He I $\lambda$4026 and $\lambda$4471 lines.
\smallskip

% \subsubsection*{BBC 388}
\item {\it VFTS\,388:} Three different peaks were recovered using the \lc at 14.8, 20.4 and 61.7\,d for line sets of 24, 2 and 31\%, respectively. In most of these cases it was a secondary peak, smaller than the 0.5 and 1\,d aliases. We adopt $P_{\rm orb}$\,$=$\,14.8\,d as the tentative period for the system since it has the strongest peak and is the only case with a primary peak.
\smallskip

% \subsubsection*{BBC 391}
\item {\it VFTS\,391:} Nebular emission was present in the available Balmer and helium lines, whereas the metallic lines were too weak to be fitted. We did not find signs of periodicity. %Two periods at 1.12 (3 sets) and 10.17 (1 set) d but too weak.
\smallskip

% \subsubsection*{BBC 396}
\item {\it VFTS\,396:} Subtle nebular emission in He I $\lambda$4026 and $\lambda$4388. The 85\,d period was found for 48\% of the different sets of lines and seems robust from the periodogram with a peak above the 0.1\% FAL. We have found some dispersion in the period with our \lc between 83.7 and 92.1\,d, similar to the case of VFTS\,204, and a final period of 86\,d with \rvf. 
\smallskip

% \subsubsection*{BBC 442}
\item {\it VFTS\,442:} Possible period at 27.1\,d revealed by our \lc for 21\% of the possible line sets. While the peak in the periodogram is clear, the RVs present larger errors and scatter in comparison to the semi-amplitude velocity in the blue-shifted (lower) portion of the RV curve.
\smallskip

% \subsubsection*{BBC 501}
\item {\it VFTS\,501:} The spectra show evidence that this could be a SB2 system. However, strong nebular emission in He I $\lambda$4026 and $\lambda$4471 suggest that the apparent secondary component seen in the weaker He lines ($\lambda$4144 and $\lambda$4388) might be an artefact of weaker nebular emission. We measured the centre of the emission and determined that it is moving with a maximum difference of 0.8\,\AA\ in He I $\lambda$4026. We analysed the system as a SB2 finding a period of 99\,d with unrealistic minimum masses of 127 and 59\Msun, so we decided to treat it as a SB1, for which we determined a period of 151\.d.
\smallskip

% \subsubsection*{BBC 575}
\item {\it VFTS\,575:} The 46\,d period was recovered for 99\% of the possible line combinations indicating a robust period. Regardless of the low LS power in the periodogram (P$_{\mathrm{peak}}<3$) the RV data is remarkably well fitted by the solution found with \rvf.
\smallskip

% \subsubsection*{BBC 576}
\item {\it VFTS\,576:} Two similar peaks above 0.1\% FAL are present in the periodogram at 115.8 and 85.2\,d. The Balmer lines are affected by broad nebular emission, interfering with the fit. Our \lc was inconclusive, so we are not able to securely determine its period nor explain the presence of a second peak, so this system was classified as SB1*. Nonetheless, we adopt the 85\,d period as it presents a far better orbital solution from comparison of RV curves. VFTS 576 presents a high semi-amplitude velocity compared with other systems with similar periods and a large mass function.
\smallskip

% \subsubsection*{BBC 591}
\item {\it VFTS\,591:} This is the brightest target (B0.2 Ia) and its spectra have the highest S/N in our sample. A period of 468\,d is generally present in the periodogram for different sets of lines, but the more significant peaks (above 1\% FAL) arise from including the Balmer lines, which suffer slight nebular contamination (which is challenging to fit). Given the high S/N ratio of the data we can estimate accurate RVs without including the Balmer lines, but we were unable to obtain a robust period estimate. Moreover, inclusion of the original VFTS spectra did not improve the signal in the periodogram (at which point the peak close to 500\,d reported in Table \ref{tab:tabB4} disappeared). Its RV variability might come from a non-binary origin.
\smallskip

% \subsubsection*{BBC 606}
\item {\it VFTS\,606:} Besides the peak at 86.6\,d there was a slightly stronger alias at 1\,d and a third peak at 73\,d, which for 7\% of lines sets in our \lc was the primary peak. For these reasons we have classified this system as SB1*.
\smallskip

% \subsubsection*{BBC 662}
\item {\it VFTS\,662:} Presents very shallow helium lines and double-peaked emission in its Balmer lines. Classified as B3-5 III \citep{evans+15}, it is one of the latest-type objects in our sample. We were unable to find a reliable period as its helium lines are broad and very weak. It is evident that \ion{He}{I} $\lambda4471$ is filled-in by double-peaked emission, which could also be the case for the other helium lines. Using \ion{Mg}{II} $\lambda4481$ we determined a possible period of 1.99\,d. We relaxed our condition of a minimum of three lines to two in our \lc for this system, and from the 120 possible sets, the 1.99\,d period was only retrieved in 4\% of the cases. However, the LS peak is close to the 0.1\% FAL which gives some confidence in this period, although its RV curve presents some scatter and a low amplitude.
\smallskip

% \subsubsection*{BBC 665}
\item {\it VFTS\,665:} Clear peak around 330\,d but with considerable uncertainty between 302 and 366. However, 86\% of combinations from our \lc yielded a period in this range. We have determined a period of 324\,d with \rvf.  
\smallskip

% \subsubsection*{BBC 686}
\item {\it VFTS\,686:} Arguably the least obvious SB2 system in our sample. The second component is not resolved in the data and its presence is inferred from the shape of the wings of the helium lines of the giant primary. A peak at 16.9\,d is present for every combination of lines arguing that this is a reliable period. \citet{garland+17} also identified this system as a SB2 from the combined LR03 VFTS spectra.
\smallskip

% \subsubsection*{BBC 697}
\item {\it VFTS\,697:} Presents strong nebular emission in its Balmer lines and \ion{He}{I} $\lambda4471$. Using only the three helium lines free (or less affected) from nebular contamination, it was possible to retrieve a short period of 1.49\,d. From our \lc, this period was found for only 7\% of the cases, but this might by due to the strong nebular contamination. Its RV curve displays considerable scatter and it was classified only as SB1*.
\smallskip

% \subsubsection*{BBC 730}
\item {\it VFTS\,730:} We found a short possible period of 1.33\,d for only one combination of lines. We considered it as `possible' given that the corresponding peak in the periodogram is close to the 0.1\% FAL. We note that this system has a considerably higher $\gamma$-velocity than the rest of the sample.
\smallskip

% \subsubsection*{BBC 784}
\item {\it VFTS\,784:} We could not find signs of periodicity for this system. However, lowering our requirement for the minimum number of lines for RV determination to two, the \lc showed a peak above the 1\% FAP level at 3.6\,d, with a stronger alias at 0.8\,d. This period was only recovered for 5\% of the possible line sets and the periodogram presented was obtained using two lines (H$\delta$ and \ion{He}{I} $\lambda$4388), so we consider this only as a possible period and classified it as SB1*.
\smallskip

% \subsubsection*{BBC 792}
\item {\it VFTS\,792:} Only 14 epochs of observations were obtained for VFTS\,792, from which two were unusable due to low S/N. Our \lc suggested an orbital period larger than 400 d but the spread in possible periods was large (from 460 to 940\,d) as a consequence of our limited sensitivity at such large periods and due to the small number of epochs for this particular system. Given its possible long period, we included the original observations taken by the VFTS to increase the number of epochs to 17. We have also used \ion{He}{I} $\lambda4009$ and \ion{Mg}{II} $\lambda4481$ which were strong enough to be fitted in this case. We retrieved $P_{\rm orb}$\,$=$\,431.7\.d with a clear and strong signal with our LS test and a final period of 428\,d with \rvf.
\smallskip

% \subsubsection*{BBC 799}
\item {\it VFTS\,799:} After first tests, we expanded our set of spectral lines to include \ion{He}{I} $\lambda4009$ and \ion{Mg}{II} $\lambda4481$ and ran our \lc with a set of nine lines. Our test retrieved a peak at 39.4\,d but only for 2\% of the possible combinations (eight sets) which are all related to the metallic \ion{Si}{III} and \ion{Mg}{II} lines. This is one of the most highly-eccentric binaries in the SB1* group ($e=0.69$) and with the lowest semi-amplitude velocity ($K_1=6.34$\,\kms).
\smallskip

% \subsubsection*{BBC 847}
\item {\it VFTS\,847:} This was classified as B0.7-1 IIIne by \citet{evans+15} but Si III $\lambda$4553 is absent from the BBC data  or indistinguishable from noise. Considering the rotational broadening, this seems to be an earlier-B main sequence star. A peak at 1.2\,d was recovered for only 4\% of line combinations, with a secondary peak at 5.2\,d that might be an alias of the former period or the true period. Given the small number of lines from which it was possible to determine a period, the presence of a strong alias and the significant scatter in the RV curve, we classified this system as SB1*.
\smallskip

% \subsubsection*{BBC 850}
\item {\it VFTS\,850:} The \lc, including \ion{He}{I} $\lambda4009$ and \ion{Mg}{II} $\lambda4481$, for this initially `aperiodic' system found a weak but persistent signal at 50.5\,d, usually accompanied by a secondary peak around 111\,d. The former peak was present in 43\% of possible line sets and was the primary peak in 5\% of the cases (23 different sets). However, given the low power of the LS peak and the large scatter in the RV curve, we have classified this system as SB1*.
\smallskip

% \subsubsection*{BBC 874}
\item {\it VFTS\,874:} A long period of 371\,d was found for this system with \rvf, however a second peak around 190\,d was also present in the periodogram. The shorter peak is probably an alias of the true period and presents a more noisy RV curve. The \lc showed that 80\% of combinations favoured the longer period, with a dispersion in the range of 360 and 407\,d.
\smallskip

% \subsubsection*{BBC 877}
\item {\it VFTS\,877:} Spectra of this Be star present deep narrow cores in the Balmer lines and \ion{He}{I} $\lambda4471$ from epoch 1 to 22, after which it disappears. This variability might be associated with the dynamics of the decretion disk \citep[e.g.][]{doazan+86,clark+03}, but requires further investigation. Two periods around 83 and 93\,d were recovered, with the latter being the most significant. Its LS periodogram shows a clear peak just above the 0.1\% FAL. However, given the issues regarding the Balmer and \ion{He}{I} $\lambda$4471 lines (and use of other lines did not yield any orbital period), we have only classified the system as SB1*.
\smallskip

%\subsubsection*{BBC 883}
\item {\it VFTS\,883:} For SB2 systems we have applied the LS test to the difference between RV measurements of primary and secondary components as for the TMBM sample. This test should retrieve a period that is half the orbital period. VFTS 883 was the only case where we found the same period in both cases which is an indication that the true period could be twice our determined period. If this were the case, going from 3.5 to 7\,d will not have any major impact in the orbital period distribution.
\smallskip

% \subsubsection*{BBC 890}
\item {\it VFTS\,890:} Displays a strong, good S/N, helium spectrum, however, we were not able to determine a robust orbital period. A signal of 18.9\,d was found with the \lc in 56\% of the possible line sets, but only as the primary peak in 16\% of all possible cases, and in none of these cases is the peak is above the 1\% FAP level. We report this system only as a RV variable. 
\end{itemize}
\onecolumn

% Appendix B

\begin{landscape}
\section{Tables}\label{apx:tables}
%\begin{table*}
\begin{longtable}{rlllcrrrrrrrrcc}
\caption{Properties of the BBC sample. Spectral classifications, magnitudes and colours are from \citet{evans+15}. Physical parameters, including projected rotational velocities ($v_e \sin i$), are from \citet{dufton+13} and \citet{garland+17} for dwarfs (LC V-III) and from \citet{mcevoy+15} for the supergiants. SB classifications are from this work.}\label{tab:tabB1}\\
%\centering
\toprule
\toprule
% \hline\hline
 VFTS &          SpT &           LC & SB &      $V$ &     $B-V$ & \multicolumn{1}{c}{$v_e \sin i$} & log L/L$_{\sun}$ &  $T_\text{eff}$ &  \multicolumn{1}{c}{$\log g$} & \multicolumn{1}{c}{M$_\text{s}$} & \multicolumn{1}{c}{M$_\text{e}$} & \multicolumn{1}{c}{\po\po$t$} &   $\alpha$ (2000) &      $\delta$ (2000) \\
 & & & & & & [\kms] & & [K] & [cm\,s$^{-2}$] & [\Msun] & [\Msun] & [Myr] \\
\midrule
% \hline
\endfirsthead
\caption{continued.}\\
\toprule
\toprule
% \hline\hline
 VFTS &          SpT &           LC & SB &      $V$ &     $B-V$ & \multicolumn{1}{c}{$v_e \sin i$} & log L/L$_{\sun}$ &  $T_\text{eff}$ &  \multicolumn{1}{c}{$\log g$} & \multicolumn{1}{c}{M$_\text{s}$} & \multicolumn{1}{c}{M$_\text{e}$} & \multicolumn{1}{c}{\po\po$t$} &   $\alpha$ (2000) &      $\delta$ (2000) \\
 & & & & & & [\kms] & & [K] & [cm\,s$^{-2}$] & [\Msun] & [\Msun] & [Myr] \\
\midrule
% \hline
 \endhead
% \hline
\midrule
 \endfoot
%  \hline
 \endlastfoot
    9 &    B1-1.5 &            V &      SB1 &  16.18 &     0.11 &        87 &         - &      - &     - &   - &   - &      - &  05 37 04.26 &  $-$69 08 05.65 \\
   15 &      B0.5 &            V &      SB1 &  16.20 &     0.04 &       185 &         - &      - &     - &   - &   - &      - &  05 37 08.56 &  $-$69 05 10.03 \\
   18 &      B1.5 &            V &     SB1* &  16.62 &     0.17 &        48 &      4.20 &      - &     - &   - &   - &      - &  05 37 11.46 &  $-$69 10 30.97 \\
   27 &        B1 &       III-II &      SB1 &  14.71 &     0.09 &        91 &      4.63 &  20500 &  3.10 &  12 &  14 &  13.18 &  05 37 17.37 &  $-$69 07 48.05 \\
   33 &    B1-1.5 &            V &      SB1 &  16.20 &     0.07 &        77 &      4.26 &  24000 &  3.90 &   - &   - &      - &  05 37 22.05 &  $-$69 07 55.13 \\
   37 &        B2 &         III: &     SB1* &  15.79 &     0.06 &       267 &         - &      - &     - &   - &   - &      - &  05 37 23.95 &  $-$69 12 21.27 \\
   41 &       B2: &            V &      SB1 &  16.94 &     0.14 &  $\leq$40 &      3.95 &      - &     - &   - &   - &      - &  05 37 24.89 &  $-$69 04 15.73 \\
   97 &        B0 &           IV &      SB1 &  16.21 &     0.16 &        72 &         - &      - &     - &   - &   - &      - &  05 37 38.75 &  $-$69 10 27.70 \\
  106 &      B0.2 &            V &     SB1* &  16.43 &     0.05 &       170 &         - &      - &     - &   - &   - &      - &  05 37 40.22 &  $-$69 04 12.11 \\
  112 &   EarlyBs &      - &      SB2 &  16.19 &     0.08 &       408 &         - &      - &     - &   - &   - &      - &  05 37 40.89 &  $-$69 04 41.52 \\
  144 &     B0.7: &            V &   RV var &  16.81 &     0.00 &       154 &         - &      - &     - &   - &   - &      - &  05 37 46.08 &  $-$69 09 14.37 \\
  146 &       B2: &            V &     SB1* &  16.24 &     0.25 &       287 &         - &      - &     - &   - &   - &      - &  05 37 46.10 &  $-$69 06 34.93 \\
  155 &     B0.7: &            V &     SB1* &  16.95 &     0.08 &       277 &         - &      - &     - &   - &   - &      - &  05 37 47.18 &  $-$69 13 13.60 \\
  157 &        B2 &            V &      SB1 &  16.69 &  $-$0.08 &       175 &         - &      - &     - &   - &   - &      - &  05 37 47.23 &  $-$68 58 49.25 \\
  162 &      B0.7 &            V &      SB1 &  16.49 &     0.07 &        60 &      4.22 &      - &     - &   - &   - &      - &  05 37 48.06 &  $-$69 09 59.88 \\
  179 &        B1 &            V &      SB1 &  16.93 &     0.09 &        51 &      4.03 &  27000 &  4.40 &   - &   - &      - &  05 37 51.18 &  $-$69 09 37.44 \\
  189 &     B0.7: &            V &      SB1 &  16.31 &     0.10 &       212 &         - &      - &     - &   - &   - &      - &  05 37 52.86 &  $-$69 09 45.87 \\
  195 &      B0.5 &            V &      SB1 &  16.86 &     0.06 &  $\leq$40 &      4.06 &  28000 &  3.90 &   - &   - &      - &  05 37 54.22 &  $-$69 05 17.89 \\
  199 &   EarlyBs &          - &      SB2 &  16.91 &  $-$0.01 &         - &         - &      - &     - &   - &   - &      - &  05 37 54.78 &  $-$69 00 24.99 \\
  204 &        B2 &          III &      SB1 &  16.13 &     0.31 &  $\leq$40 &      4.41 &  22500 &  3.50 &   - &   - &      - &  05 37 55.04 &  $-$69 07 02.20 \\
  206 &        B3 &         III: &      SB2 &  14.99 &     0.15 &       <40 &         - &      - &     - &   - &   - &      - &  05 37 55.43 &  $-$68 57 06.99 \\
  211 &        B1 &            V &      SB1 &  16.56 &  $-$0.12 &       188 &         - &      - &     - &   - &   - &      - &  05 37 57.34 &  $-$68 58 41.89 \\
  213 &        B2 &        III:e &     SB1* &  15.50 &     0.04 &       181 &         - &      - &     - &   - &   - &      - &  05 37 57.96 &  $-$69 09 54.11 \\
  215 &      B1.5 &            V &      SB1 &  16.58 &  $-$0.03 &       154 &         - &      - &     - &   - &   - &      - &  05 37 58.03 &  $-$69 02 23.54 \\
  218 &      B1.5 &            V &     SB1* &  15.63 &     0.38 &        79 &      4.89 &      - &     - &   - &   - &      - &  05 37 59.71 &  $-$69 11 14.29 \\
  225 &    B0.7-1 &       III-II &      SB1 &  15.07 &  $-$0.01 &  $\leq$40 &      4.53 &  24500 &  3.25 &   - &   - &      - &  05 38 00.96 &  $-$68 57 23.28 \\
  227 &        B2 &            V &      SB1 &  16.51 &  $-$0.09 &       183 &         - &      - &     - &   - &   - &      - &  05 38 01.31 &  $-$69 03 13.91 \\
  240 &      B1-2 &            V &      SB2 &  15.85 &     0.01 &        77 &         - &      - &     - &   - &   - &      - &  05 38 06.75 &  $-$69 06 08.76 \\
  246 &        B1 &          III &      SB1 &  16.83 &     0.45 &       135 &         - &      - &     - &   - &   - &      - &  05 38 08.86 &  $-$69 10 40.04 \\
  248 &       B2: &            V &      SB2 &  16.49 &     0.00 &       268 &         - &      - &     - &   - &   - &      - &  05 38 09.32 &  $-$69 10 14.33 \\
  255 &       B2: &            V &      SB2 &  16.68 &     0.02 &       274 &         - &      - &     - &   - &   - &      - &  05 38 11.05 &  $-$69 07 19.70 \\
  257 &  B0.7-1.5 &            V &      SB1 &  16.70 &  $-$0.04 &       125 &         - &      - &     - &   - &   - &      - &  05 38 11.40 &  $-$69 14 24.65 \\
  278 &      B2.5 &            V &      SB1 &  16.82 &  $-$0.07 &        60 &         - &      - &     - &   - &   - &      - &  05 38 16.21 &  $-$69 04 04.01 \\
  291 &        B5 &        II-Ib &      SB1 &  14.85 &     0.12 &        20 &      4.31 &  13500 &  2.35 &   6 &  14 &  12.88 &  05 38 17.73 &  $-$69 03 38.66 \\
  299 &      B0.5 &            V &      SB1 &  16.36 &  $-$0.05 &  $\leq$40 &      4.14 &  28000 &  4.25 &   - &   - &      - &  05 38 18.52 &  $-$69 12 28.11 \\
  305 &       B2: &            V &      SB1 &  16.59 &  $-$0.10 &        57 &         - &      - &     - &   - &   - &      - &  05 38 19.22 &  $-$68 57 37.48 \\
  324 &      B0.2 &            V &      SB1 &  15.53 &  $-$0.13 &        57 &      4.42 &  28500 &  3.90 &   - &   - &      - &  05 38 21.89 &  $-$69 12 48.05 \\
  334 &      B0.7 &            V &      SB1 &  16.26 &  $-$0.06 &       182 &         - &      - &     - &   - &   - &      - &  05 38 24.02 &  $-$69 05 23.99 \\
  337 &       B2: &      V-IIIe+ &     SB1* &  16.72 &     0.14 &       412 &         - &      - &     - &   - &   - &      - &  05 38 25.60 &  $-$69 06 04.31 \\
  342 &        B1 &            V &      SB1 &  16.94 &  $-$0.04 &  $\leq$40 &      3.84 &      - &     - &   - &   - &      - &  05 38 26.86 &  $-$69 04 17.82 \\
  351 &      B0.5 &            V &      SB1 &  15.98 &     0.08 &  $\leq$40 &      4.47 &  28500 &  4.00 &   - &   - &      - &  05 38 28.39 &  $-$69 06 40.89 \\
  359 &      B0.5 &            V &      SB1 &  16.30 &     0.03 &        54 &      4.27 &  28000 &  4.00 &   - &   - &      - &  05 38 29.39 &  $-$69 05 59.19 \\
  364 &     B2.5: &            V &      SB2 &  16.82 &     0.03 &       140 &         - &      - &     - &   - &   - &      - &  05 38 30.07 &  $-$69 08 27.60 \\
  383 &     B0.5: &            V &      SB2 &  16.10 &     0.20 &       164 &         - &      - &     - &   - &   - &      - &  05 38 32.31 &  $-$69 07 32.22 \\
  388 &      B0.5 &            V &     SB1* &  16.42 &     0.06 &        98 &         - &      - &     - &   - &   - &      - &  05 38 32.63 &  $-$69 03 55.93 \\
  391 &     B0.5: &            V &   RV var &  16.32 &  $-$0.02 &       260 &         - &      - &     - &   - &   - &      - &  05 38 32.80 &  $-$69 05 24.92 \\
  396 &      B0.5 &            V &      SB1 &  16.32 &     0.11 &       131 &         - &      - &     - &   - &   - &      - &  05 38 33.27 &  $-$69 10 24.00 \\
  430 &      B0.5 &  Ia+((n))Nwk &      SB1 &  15.11 &     0.64 &        98 &      5.56 &  24500 &  2.65 &  18 &   - &      - &  05 38 36.87 &  $-$69 06 46.08 \\
  434 &     B1.5: &            V &      SB1 &  16.13 &     0.16 &        45 &      4.38 &      - &     - &   - &   - &      - &  05 38 37.22 &  $-$69 04 25.98 \\
  442 &      B1-2 &            V &     SB1* &  16.66 &     0.08 &       281 &         - &      - &     - &   - &   - &      - &  05 38 37.89 &  $-$69 03 36.76 \\
  501 &      B0.5 &            V &      SB1 &  15.74 &     0.08 &        59 &      4.57 &      - &     - &   - &   - &      - &  05 38 41.23 &  $-$69 04 14.22 \\
  520 &       B1: &            V &      SB2 &  16.69 &     0.08 &        53 &      4.11 &      - &     - &   - &   - &      - &  05 38 41.99 &  $-$69 06 53.46 \\
  534 &        B0 &           IV &      SB1 &  15.66 &     0.21 &        57 &      4.82 &  29000 &  3.75 &   - &   - &      - &  05 38 42.80 &  $-$69 15 40.29 \\
  575 &      B0.7 &          III &      SB1 &  15.11 &     0.02 &  $\leq$40 &      4.57 &  26000 &  3.75 &   - &   - &      - &  05 38 44.91 &  $-$69 05 33.10 \\
  576 &        B1 &        IaNwk &     SB1* &  15.67 &     0.78 &        52 &      5.31 &  20000 &  2.50 &  16 &  32 &   4.90 &  05 38 44.94 &  $-$69 15 11.45 \\
  589 &      B0.5 &            V &      SB2 &  15.83 &     0.15 &  $\leq$40 &      4.63 &  27500 &  4.00 &   - &   - &      - &  05 38 45.59 &  $-$69 07 34.86 \\
  591 &      B0.2 &           Ia &   RV var &  12.55 &     0.23 &        48 &      5.91 &  25000 &  2.80 &  53 &  48 &   3.31 &  05 38 45.69 &  $-$69 06 22.45 \\
  606 &    B0-0.5 &         V(n) &     SB1* &  16.60 &     0.06 &        86 &         - &      - &     - &   - &   - &      - &  05 38 46.71 &  $-$69 05 38.78 \\
  637 &      B1-2 &     V+EarlyB &      SB2 &  16.61 &     0.04 &         - &         - &      - &     - &   - &   - &      - &  05 38 49.40 &  $-$69 06 15.28 \\
  662 &      B3-5 &         III: &     SB1* &  16.12 &     0.08 &        67 &      3.94 &  17500 &  3.60 &   - &   - &      - &  05 38 52.55 &  $-$69 02 20.30 \\
  665 &      B0.5 &            V &      SB1 &  16.43 &     0.11 &        47 &      4.33 &  28000 &  4.15 &   - &   - &      - &  05 38 52.74 &  $-$69 07 28.71 \\
  686 &      B0.7 &          III &      SB2 &  14.99 &     0.17 &  $\leq$40 &      4.83 &  24000 &  3.60 &   - &   - &      - &  05 38 55.63 &  $-$69 07 23.77 \\
  687 &      B1.5 &   Ib((n))Nwk &      SB1 &  14.29 &     0.28 &       123 &      4.95 &  20000 &  2.65 &  10 &  25 &   6.31 &  05 38 55.83 &  $-$69 08 22.38 \\
  697 &      B1-2 &           Ve &     SB1* &  16.55 &     0.18 &       121 &         - &      - &     - &   - &   - &      - &  05 38 57.28 &  $-$69 03 41.68 \\
  705 &      B0.7 &            V &      SB1 &  16.43 &     0.07 &        87 &         - &      - &     - &   - &   - &      - &  05 38 58.71 &  $-$69 06 47.10 \\
  715 &        B1 &            V &      SB1 &  16.64 &  $-$0.10 &       116 &         - &      - &     - &   - &   - &      - &  05 39 00.01 &  $-$69 01 39.76 \\
  718 &      B2.5 &          III &      SB1 &  15.99 &  $-$0.06 &       185 &         - &      - &     - &   - &   - &      - &  05 39 00.89 &  $-$68 57 29.65 \\
  719 &        B1 &            V &      SB1 &  17.00 &     0.08 &        50 &      3.99 &      - &     - &   - &   - &      - &  05 39 00.91 &  $-$69 06 29.16 \\
  723 &      B0.5 &            V &      SB1 &  16.19 &     0.16 &        63 &      4.50 &  27500 &  3.90 &   - &   - &      - &  05 39 02.85 &  $-$69 05 26.35 \\
  730 &        B1 &        IV(n) &     SB1* &  15.41 &  $-$0.10 &       248 &         - &      - &     - &   - &   - &      - &  05 39 03.61 &  $-$69 00 18.98 \\
  742 &        B2 &            V &      SB1 &  16.93 &  $-$0.02 &        60 &      3.73 &      - &     - &   - &   - &      - &  05 39 06.41 &  $-$68 56 58.70 \\
  752 &        B2 &            V &      SB2 &  16.48 &  $-$0.12 &       194 &         - &      - &     - &   - &   - &      - &  05 39 09.12 &  $-$68 57 47.26 \\
  779 &        B1 &        II-Ib &      SB1 &  15.46 &     0.19 &        47 &      4.73 &  23500 &  3.20 &  11 &  17 &  10.00 &  05 39 21.53 &  $-$69 03 18.36 \\
  784 &       B1: &            V &     SB1* &  16.83 &     0.19 &       180 &         - &      - &     - &   - &   - &      - &  05 39 24.22 &  $-$69 06 11.72 \\
  788 &        B1 &          III &      SB1 &  16.15 &     0.09 &        83 &         - &      - &     - &   - &   - &      - &  05 39 25.10 &  $-$69 01 30.72 \\
  792 &        B2 &            V &      SB1 &  15.96 &  $-$0.06 &        47 &      4.07 &      - &     - &   - &   - &      - &  05 39 28.08 &  $-$68 56 58.89 \\
  799 &  B0.5-0.7 &            V &     SB1* &  16.86 &     0.10 &  $\leq$40 &      4.13 &  26500 &  4.00 &   - &   - &      - &  05 39 31.13 &  $-$69 04 36.80 \\
  821 &        B0 &         V-IV &      SB1 &  16.03 &  $-$0.14 &        91 &         - &      - &     - &   - &   - &      - &  05 39 38.43 &  $-$68 58 36.04 \\
  827 &      B1.5 &           Ib &      SB1 &  15.34 &     0.31 &        52 &      5.03 &  21000 &  3.10 &  28 &  15 &  12.02 &  05 39 39.27 &  $-$69 11 44.20 \\
  837 &        B1 &            V &      SB1 &  16.07 &  $-$0.09 &       129 &         - &      - &     - &   - &   - &      - &  05 39 41.25 &  $-$68 59 37.94 \\
  847 &    B0.7-1 &        IIIne &     SB1* &  15.48 &  $-$0.06 &       258 &         - &      - &     - &   - &   - &      - &  05 39 45.58 &  $-$69 04 26.23 \\
  850 &        B1 &          III &     SB1* &  16.15 &     0.18 &  $\leq$40 &      4.34 &  24000 &  3.75 &   - &   - &      - &  05 39 51.16 &  $-$69 11 53.59 \\
  874 &      B1.5 &        IIIe+ &      SB1 &  15.37 &     0.02 &        62 &      4.37 &      - &     - &   - &   - &      - &  05 40 10.33 &  $-$69 03 04.95 \\
  877 &      B1-3 &      V-IIIe+ &     SB1* &  16.36 &     0.18 &       267 &         - &      - &     - &   - &   - &      - &  05 40 12.81 &  $-$69 09 10.30 \\
  883 &      B0.5 &            V &      SB2 &  16.49 &     0.13 &       300 &         - &      - &     - &   - &   - &      - &  05 40 18.03 &  $-$69 08 36.06 \\
  888 &      B0.5 &            V &      SB1 &  16.18 &  $-$0.07 &        76 &      4.18 &  27000 &  4.15 &   - &   - &      - &  05 40 22.62 &  $-$69 04 06.07 \\
  890 &        B2 &            V &   RV var &  16.13 &     0.02 &       160 &         - &      - &     - &   - &   - &      - &  05 40 24.86 &  $-$69 09 44.14 \\
  891 &        B2 &            V &      SB1 &  16.48 &     0.07 &        55 &      4.04 &      - &     - &   - &   - &      - &  05 40 25.73 &  $-$69 06 30.78 \\
\bottomrule
\end{longtable}
\tablefoot{Definition of column headers: (1) VFTS identifier; (2) spectral types; (3) luminosity classes; (4) spectroscopic binary classification; (5) $V$-band magnitude; (6) $B-V$ colour; (7) projected rotational velocity; (8) luminosity; (9) effective temperature; (10) logarithmic surface gravity; (11) spectroscopic mass; (12) evolutionary mass; (13) age; (14) right ascension; (15) declination. \citet{evans+15} identified two SB2 systems, VFTS 112 and 199, that were classified as early B$+$early B (noted as `EarlyBs' above); \citeauthor{evans+15} also identified VFTS 637 has having an early B-type companion. Projected rotational velocities, in the case of SB2 systems, should only be considered as upper limits since the majority of the systems were not classified as SB2 at the time of analysis from \citet{dufton+13} and \citet{garland+17}.}
%\end{tabular}
%\end{table*}

\end{landscape}

\begin{table*}
\caption{\label{tab:tabB2} Spectral lines used for RV measurements of each target. The systems where fewer epochs were used are generally the SB2 systems (see discussion in Sect.~\ref{ssec:sb2}). He{\,\scriptsize I} $\lambda4009$ and Mg{\,\scriptsize II} $\lambda4481$ were used only in systems where we could not find a convincing period with our main set of lines and if we were able to perform an appropriate fit to them given the S/N of the spectra.} 
\centering
\begin{tabular}{rcccccccccccc}
\toprule
\toprule
%\hline\hline
%  VFTS &          SpT &           LC &         SB2 & He I+II $\lambda$4026 & H$\delta$ $\lambda$4102 & He I $\lambda$4144 & H$\gamma$ $\lambda$4340 & He I $\lambda$4388 & He I $\lambda$4471 & Si III $\lambda$4553 &  n of lines &  n of epochs \\
  \thead{VFTS} &           
  \thead{He{\,\scriptsize I} \\ $\lambda$4009} & 
  \thead{He{\,\scriptsize I}+{\,\scriptsize II} \\ $\lambda$4026} & 
  \thead{H$\delta$ \\ $\lambda$4102} & 
  \thead{He{\,\scriptsize I} \\ $\lambda$4144} & 
  \thead{H$\gamma$ \\ $\lambda$4340} & 
  \thead{He{\,\scriptsize I} \\ $\lambda$4388} & 
  \thead{He{\,\scriptsize I} \\ $\lambda$4471} & 
  \thead{Mg{\,\scriptsize II} \\ $\lambda$4481} & 
  \thead{Si{\,\scriptsize III} \\ $\lambda$4553} &  
  \thead{$n$ lines} &  \thead{$n$ epochs} \\
\midrule
    9 &    - &    x &    - &    x &    x &    x &    x &    - &    - &           5 &           25 \\
   15 &    - &    x &    x &    - &    x &    x &    x &    - &    - &           5 &           25 \\
   18 &    - &    x &    x &    x &    - &    x &    - &    - &    - &           4 &           24 \\
   27 &    - &    x &    x &    x &    - &    x &    x &    - &    x &           6 &           25 \\
   33 &    - &    x &    - &    x &    - &    x &    x &    - &    - &           4 &           27 \\
   37 &    - &    - &    x &    - &    x &    x &    x &    - &    - &           4 &           20 \\
   41 &    - &    x &    x &    x &    - &    x &    x &    - &    - &           5 &           25 \\
   97 &    - &    - &    - &    x &    x &    x &    x &    - &    - &           4 &           24 \\
  106 &    - &    x &    - &    - &    x &    - &    x &    - &    - &           3 &           25 \\
  112 &    - &    x &    - &    x &    - &    x &    x &    - &    - &           4 &           15 \\
  144 &    - &    x &    x &    x &    - &    - &    x &    - &    - &           4 &           25 \\
  146 &    - &    x &    x &    x &    x &    x &    x &    - &    - &           6 &           20 \\
  155 &    - &    x &    - &    x &    x &    x &    x &    - &    - &           5 &           23 \\
  157 &    - &    x &    - &    x &    - &    x &    x &    - &    - &           4 &           24 \\
  162 &    - &    x &    x &    - &    x &    x &    - &    - &    - &           4 &           22 \\
  179 &    - &    x &    x &    - &    - &    x &    x &    - &    - &           4 &           22 \\
  189 &    - &    x &    x &    x &    x &    x &    x &    - &    - &           6 &           27 \\
  195 &    - &    x &    x &    x &    - &    x &    x &    - &    x &           6 &           23 \\
  199 &    - &    x &    - &    x &    - &    x &    x &    - &    - &           4 &           13 \\
  204 &    - &    x &    - &    x &    - &    x &    - &    - &    x &           4 &           21 \\
  206 &    - &    x &    - &    x &    - &    - &    x &    - &    - &           3 &           14 \\
  211 &    - &    x &    - &    x &    - &    x &    x &    - &    - &           4 &           26 \\
  213 &    - &    - &    x &    - &    - &    x &    x &    - &    - &           3 &           25 \\
  215 &    - &    - &    x &    x &    x &    x &    x &    - &    - &           5 &           26 \\
  218 &    - &    x &    - &    - &    - &    x &    x &    - &    - &           3 &           25 \\
  225 &    - &    x &    x &    x &    x &    x &    x &    - &    x &           7 &           25 \\
  227 &    - &    x &    x &    x &    x &    x &    x &    - &    - &           6 &           20 \\
  240 &    - &    x &    - &    x &    - &    x &    x &    - &    - &           4 &           10 \\
  246 &    - &    - &    x &    x &    x &    x &    x &    - &    - &           5 &           23 \\
  248 &    - &    x &    - &    x &    - &    x &    x &    - &    - &           4 &           17 \\
  255 &    - &    x &    - &    x &    - &    x &    - &    - &    - &           3 &           16 \\
  257 &    - &    x &    x &    x &    - &    - &    x &    - &    - &           4 &           22 \\
  278 &    - &    x &    - &    x &    x &    x &    x &    - &    - &           5 &           25 \\
  291 &    - &    - &    x &    x &    x &    x &    x &    - &    - &           5 &           26 \\
  299 &    - &    x &    x &    - &    x &    - &    - &    - &    - &           3 &           25 \\
  305 &    - &    x &    x &    x &    - &    x &    x &    - &    - &           5 &           26 \\
  324 &    - &    x &    x &    x &    - &    x &    x &    - &    x &           6 &           28 \\
  334 &    - &    x &    x &    - &    x &    x &    x &    - &    - &           5 &           22 \\
  337 &    - &    - &    x &    x &    x &    x &    x &    - &    - &           5 &           19 \\
  342 &    - &    x &    x &    x &    x &    x &    x &    - &    - &           6 &           24 \\
  351 &    - &    x &    x &    x &    x &    x &    x &    - &    - &           6 &           25 \\
  359 &    - &    x &    x &    - &    x &    x &    - &    - &    - &           4 &           23 \\
  364 &    - &    x &    - &    x &    - &    x &    x &    - &    - &           4 &           18 \\
  383 &    - &    - &    - &    x &    - &    x &    x &    - &    - &           3 &           15 \\
  388 &    - &    - &    x &    x &    - &    - &    x &    - &    - &           3 &           23 \\
  391 &    - &    - &    - &    x &    x &    x &    x &    - &    - &           4 &           22 \\
  396 &    - &    - &    x &    x &    - &    x &    x &    - &    - &           4 &           22 \\
  430 &    - &    x &    - &    x &    - &    x &    x &    - &    x &           5 &           27 \\
  434 &    - &    x &    - &    x &    - &    x &    x &    - &    - &           4 &           24 \\
  442 &    - &    x &    x &    x &    - &    x &    - &    - &    - &           4 &           24 \\
  501 &    - &    x &    - &    x &    x &    - &    - &    - &    x &           4 &           25 \\
  520 &    - &    x &    - &    x &    - &    x &    - &    - &    - &           3 &           17 \\
  534 &    - &    x &    x &    x &    - &    - &    x &    - &    x &           5 &           26 \\
  575 &    - &    x &    - &    x &    - &    x &    x &    - &    x &           5 &           28 \\
  576 &    - &    x &    - &    x &    - &    x &    x &    - &    x &           5 &           23 \\
  589 &    - &    - &    - &    x &    - &    x &    - &    - &    x &           3 &           18 \\
\bottomrule
\end{tabular}
\end{table*}

\begin{table*}
\contcaption{} 
\centering
\begin{tabular}{rcccccccccccc}
\toprule
\toprule
  \thead{VFTS} &           
  \thead{He{\,\scriptsize I} \\ $\lambda$4009} & 
  \thead{He{\,\scriptsize I}+{\,\scriptsize II} \\ $\lambda$4026} & 
  \thead{H$\delta$ \\ $\lambda$4102} & 
  \thead{He{\,\scriptsize I} \\ $\lambda$4144} & 
  \thead{H$\gamma$ \\ $\lambda$4340} & 
  \thead{He{\,\scriptsize I} \\ $\lambda$4388} & 
  \thead{He{\,\scriptsize I} \\ $\lambda$4471} & 
  \thead{Mg{\,\scriptsize II} \\ $\lambda$4481} & 
  \thead{Si{\,\scriptsize III} \\ $\lambda$4553} &  
  \thead{$n$ lines} &  \thead{$n$ epochs} \\
\midrule
  591 &    x &    - &    - &    x &    - &    x &    - &    - &    x &           4 &           29 \\
  606 &    - &    x &    x &    x &    - &    - &    x &    - &    - &           4 &           24 \\
  637 &    - &    x &    - &    x &    - &    x &    x &    - &    - &           4 &           15 \\
  662 &    - &    - &    x &    x &    - &    - &    - &    x &    - &           3 &           23 \\
  665 &    - &    - &    x &    x &    x &    - &    x &    - &    - &           4 &           21 \\
  686 &    x &    - &    - &    x &    - &    x &    - &    - &    - &           3 &           19 \\
  687 &    - &    x &    - &    x &    x &    x &    x &    - &    x &           6 &           27 \\
  697 &    - &    x &    - &    x &    - &    x &    - &    - &    - &           3 &           25 \\
  705 &    - &    x &    x &    - &    x &    x &    x &    - &    - &           5 &           25 \\
  715 &    - &    x &    x &    x &    - &    x &    x &    - &    - &           5 &           23 \\
  718 &    - &    x &    x &    x &    - &    x &    x &    - &    - &           5 &           27 \\
  719 &    - &    x &    x &    x &    x &    x &    x &    - &    - &           6 &           25 \\
  723 &    - &    x &    x &    x &    - &    x &    x &    - &    x &           6 &           21 \\
  730 &    x &    x &    - &    x &    - &    - &    - &    - &    - &           3 &           18 \\
  742 &    - &    x &    - &    - &    x &    x &    x &    - &    - &           4 &           23 \\
  752 &    - &    x &    - &    x &    - &    x &    x &    - &    - &           4 &           16 \\
  779 &    - &    x &    - &    x &    x &    x &    x &    - &    x &           6 &           25 \\
  784 &    - &    - &    x &    - &    - &    x &    - &    - &    - &           2 &           21 \\
  788 &    - &    x &    x &    x &    x &    x &    x &    - &    - &           6 &           24 \\
  792 &    x &    x &    - &    x &    - &    x &    - &    x &    - &           5 &           16 \\
  799 &    - &    x &    x &    - &    x &    - &    - &    x &    x &           5 &           22 \\
  821 &    - &    x &    - &    x &    - &    x &    x &    - &    - &           4 &           24 \\
  827 &    - &    x &    - &    x &    x &    x &    x &    - &    x &           6 &           26 \\
  837 &    - &    x &    - &    x &    x &    x &    x &    - &    - &           5 &           24 \\
  847 &    x &    - &    - &    x &    - &    - &    - &    - &    - &           2 &           26 \\
  850 &    - &    - &    - &    x &    x &    x &    x &    - &    - &           4 &           21 \\
  874 &    - &    x &    x &    - &    x &    x &    - &    - &    x &           5 &           25 \\
  877 &    - &    x &    - &    x &    x &    - &    x &    - &    - &           4 &           19 \\
  883 &    - &    x &    - &    x &    - &    x &    x &    - &    - &           4 &           10 \\
  888 &    - &    x &    x &    x &    x &    x &    x &    - &    - &           6 &           22 \\
  890 &    x &    x &    x &    - &    x &    x &    x &    - &    - &           6 &           21 \\
  891 &    - &    x &    - &    x &    - &    x &    x &    - &    - &           4 &           25 \\
\bottomrule
\end{tabular}
\end{table*}

\clearpage

\begin{landscape}
% \begin{table*}
{\small
\begin{longtable}{clHrcccccccr}
\caption{Orbital parameters determined with \texttt{rvfit} for the SB1 sample. Eccentricities without errors have been fixed to zero due to their low values ($e<0.001$). In those cases, the argument of the periastron ($\omega$) is set to 90$^{\circ}$. The same applies to Tables \ref{tab:tabB4} and \ref{tab:tabB5}.} 
\label{tab:tabB3} \\
% \centering
% \begin{tabular}{cHHccccccHHr}
\toprule
\toprule
VFTS &   SpT & SB & \multicolumn{1}{c}{$P_{\mathrm{orb}}$} & $T_p$ & $e$ & $\omega$ & $\gamma$ & $K_1$ & $a_1\sin i$ & $f(m_1,m_2)$ & $\chi^2_{\rm red}$ \\
 & &  & \multicolumn{1}{c}{[d]} & [HJD] & & [deg] & [\kms] & [\kms] & [\Rsun] & [10$^{-2}$\Msun] & \\
\midrule
\endfirsthead
\caption{continued.}\\
\toprule
\toprule
VFTS &   SpT & SB & \multicolumn{1}{c}{$P_{\mathrm{orb}}$} & $T_p$ & $e$ & $\omega$ & $\gamma$ & $K_1$ & $a_1\sin i$ & $f(m_1,m_2)$ & $\chi^2_{\rm red}$ \\
 & &  & \multicolumn{1}{c}{[d]} & [HJD] & & [deg] & [\kms] & [\kms] & [\Rsun] & [10$^{-2}$\Msun] & \\
\midrule
 \endhead
\midrule
 \endfoot
 \endlastfoot
 009 &          B1-1.5 V &  SB1 &     4.711676 $\pm$ 0.000298 &   2457302.43 $\pm$ 1.33 &  0.039 $\pm$ 0.010 &  229.90 $\pm$ 100.79 &      279.36 $\pm$ 0.73 &    54.08 $\pm$ 1.72 &        5.03 $\pm$ 0.16 &                 7.705 $\pm$ 0.769 &                1.3 \\
 015 &            B0.5 V &  SB1 &     8.787956 $\pm$ 0.002010 &   2457307.15 $\pm$ 0.54 &  0.045 $\pm$ 0.018 &    55.16 $\pm$ 22.47 &      274.33 $\pm$ 0.69 &    46.43 $\pm$ 0.75 &        8.06 $\pm$ 0.13 &                 9.083 $\pm$ 0.653 &                1.2 \\
 027 &         B1 III-II &  SB1 &     6.582728 $\pm$ 0.000202 &   2457303.48 $\pm$ 0.01 &              0.000 &                90.00 &      284.55 $\pm$ 0.26 &    77.95 $\pm$ 0.44 &       10.14 $\pm$ 0.06 &                32.298 $\pm$ 0.543 &                0.6 \\
 033 &          B1-1.5 V &  SB1 &     3.855850 $\pm$ 0.000381 &   2457300.65 $\pm$ 0.31 &  0.023 $\pm$ 0.018 &   125.43 $\pm$ 28.24 &      281.98 $\pm$ 0.58 &    37.30 $\pm$ 0.49 &        2.84 $\pm$ 0.04 &                 2.072 $\pm$ 0.140 &                0.8 \\
 041 &             B2: V &  SB1 &    14.333256 $\pm$ 0.010703 &   2457303.15 $\pm$ 0.20 &  0.292 $\pm$ 0.022 &     59.83 $\pm$ 5.94 &      268.76 $\pm$ 0.51 &    23.09 $\pm$ 0.43 &        6.26 $\pm$ 0.12 &                 1.600 $\pm$ 0.147 &                0.6 \\
 097 &             B0 IV &  SB1 &    19.870379 $\pm$ 0.024970 &   2457317.55 $\pm$ 0.34 &  0.262 $\pm$ 0.022 &      7.21 $\pm$ 6.91 &      277.69 $\pm$ 0.56 &    13.45 $\pm$ 0.13 &        5.10 $\pm$ 0.06 &                 0.451 $\pm$ 0.035 &                2.0 \\
 157 &              B2 V &  SB1 &    12.940303 $\pm$ 0.007045 &   2457307.44 $\pm$ 0.65 &  0.062 $\pm$ 0.025 &    62.49 $\pm$ 17.81 &      275.98 $\pm$ 0.90 &    49.97 $\pm$ 0.82 &       12.76 $\pm$ 0.21 &                16.634 $\pm$ 1.494 &                0.6 \\
 162 &            B0.7 V &  SB1 &   145.430440 $\pm$ 1.497402 &   2457390.26 $\pm$ 4.03 &  0.460 $\pm$ 0.091 &    295.95 $\pm$ 8.04 &      277.37 $\pm$ 0.73 &    23.96 $\pm$ 0.39 &       61.17 $\pm$ 3.45 &                14.521 $\pm$ 5.078 &                1.0 \\
 179 &              B1 V &  SB1 &     1.162719 $\pm$ 0.000119 &   2457299.75 $\pm$ 0.04 &  0.113 $\pm$ 0.041 &   312.57 $\pm$ 12.52 &      272.35 $\pm$ 0.73 &    14.73 $\pm$ 0.20 &        0.34 $\pm$ 0.00 &                 0.038 $\pm$ 0.005 &                1.4 \\
 189 &           B0.7: V &  SB1 &     1.434800 $\pm$ 0.000026 &   2457300.79 $\pm$ 0.03 &  0.042 $\pm$ 0.013 &    112.47 $\pm$ 8.79 &      268.83 $\pm$ 0.80 &    89.18 $\pm$ 1.18 &        2.53 $\pm$ 0.03 &                10.516 $\pm$ 0.581 &                2.2 \\
 195 &            B0.5 V &  SB1 &    15.009641 $\pm$ 0.010377 &   2457308.69 $\pm$ 2.75 &  0.111 $\pm$ 0.025 &   181.40 $\pm$ 65.30 &      260.15 $\pm$ 0.70 &    15.94 $\pm$ 0.83 &        4.70 $\pm$ 0.25 &                 0.618 $\pm$ 0.107 &                1.2 \\
 204 &            B2 III &  SB1 &   164.340390 $\pm$ 5.546343 &  2457316.68 $\pm$ 18.70 &  0.616 $\pm$ 0.187 &   320.84 $\pm$ 25.43 &      248.96 $\pm$ 4.62 &    21.20 $\pm$ 0.36 &      54.24 $\pm$ 10.31 &                 7.927 $\pm$ 7.201 &                1.5 \\
 211 &              B1 V &  SB1 &    85.596408 $\pm$ 0.474383 &   2457333.99 $\pm$ 1.07 &  0.450 $\pm$ 0.060 &    153.53 $\pm$ 9.02 &      246.82 $\pm$ 1.24 &    28.66 $\pm$ 1.01 &       43.31 $\pm$ 2.13 &                14.875 $\pm$ 3.718 &                0.4 \\
 215 &            B1.5 V &  SB1 &     4.303465 $\pm$ 0.000371 &   2457301.35 $\pm$ 0.04 &  0.252 $\pm$ 0.014 &     10.66 $\pm$ 2.85 &      288.74 $\pm$ 0.70 &    76.01 $\pm$ 0.92 &        6.26 $\pm$ 0.08 &                17.755 $\pm$ 1.004 &                2.2 \\
 225 &     B0.7-1 III-II &  SB1 &     8.235572 $\pm$ 0.000823 &   2457299.83 $\pm$ 0.21 &  0.027 $\pm$ 0.012 &    140.71 $\pm$ 9.28 &      264.52 $\pm$ 0.23 &    30.56 $\pm$ 0.57 &        4.97 $\pm$ 0.09 &                 2.433 $\pm$ 0.161 &                0.7 \\
 227 &              B2 V &  SB1 &    10.720421 $\pm$ 0.008120 &   2457306.55 $\pm$ 0.63 &  0.073 $\pm$ 0.047 &   279.64 $\pm$ 19.59 &      257.69 $\pm$ 0.72 &    14.59 $\pm$ 0.45 &        3.08 $\pm$ 0.10 &                 0.342 $\pm$ 0.058 &                0.6 \\
 246 &            B1 III &  SB1 &     2.442677 $\pm$ 0.000157 &   2457300.87 $\pm$ 0.01 &              0.000 &                90.00 &      270.09 $\pm$ 0.93 &    47.52 $\pm$ 0.78 &        2.29 $\pm$ 0.04 &                 2.716 $\pm$ 0.133 &                0.6 \\
 257 &        B0.7-1.5 V &  SB1 &   132.251400 $\pm$ 1.446657 &   2457377.11 $\pm$ 5.90 &  0.235 $\pm$ 0.035 &    66.46 $\pm$ 14.37 &      243.45 $\pm$ 1.06 &    20.27 $\pm$ 0.43 &       51.50 $\pm$ 1.31 &                10.479 $\pm$ 1.358 &                0.6 \\
 278 &            B2.5 V &  SB1 &    26.968116 $\pm$ 0.040315 &   2457326.62 $\pm$ 1.61 &  0.112 $\pm$ 0.062 &    54.43 $\pm$ 21.61 &      259.69 $\pm$ 0.96 &    24.45 $\pm$ 0.81 &       12.95 $\pm$ 0.44 &                 4.009 $\pm$ 0.859 &                1.0 \\
 291 &          B5 II-Ib &  SB1 &   107.464580 $\pm$ 0.093519 &   2457390.07 $\pm$ 2.42 &  0.011 $\pm$ 0.006 &    303.48 $\pm$ 8.07 &      262.00 $\pm$ 0.41 &    84.24 $\pm$ 0.46 &      178.93 $\pm$ 0.99 &              665.593 $\pm$ 15.638 &                1.7 \\
 299 &            B0.5 V &  SB1 &    20.258164 $\pm$ 0.030188 &   2457311.44 $\pm$ 0.23 &  0.504 $\pm$ 0.034 &     91.04 $\pm$ 3.85 &      282.31 $\pm$ 1.36 &    46.16 $\pm$ 0.52 &       15.96 $\pm$ 0.41 &                13.284 $\pm$ 1.872 &                1.9 \\
 305 &             B2: V &  SB1 &     4.178925 $\pm$ 0.000327 &   2457300.10 $\pm$ 0.08 &  0.026 $\pm$ 0.012 &    290.02 $\pm$ 7.01 &      287.75 $\pm$ 0.64 &    55.22 $\pm$ 0.97 &        4.56 $\pm$ 0.08 &                 7.283 $\pm$ 0.465 &                9.2 \\
 324 &            B0.2 V &  SB1 &     1.642798 $\pm$ 0.000045 &   2457300.45 $\pm$ 0.03 &  0.025 $\pm$ 0.018 &    181.95 $\pm$ 7.10 &      265.68 $\pm$ 0.30 &    29.97 $\pm$ 0.81 &        0.97 $\pm$ 0.03 &                 0.458 $\pm$ 0.045 &                2.0 \\
 334 &            B0.7 V &  SB1 &    38.202653 $\pm$ 0.058381 &   2457302.05 $\pm$ 1.01 &  0.133 $\pm$ 0.030 &    292.96 $\pm$ 9.83 &      262.86 $\pm$ 0.74 &    23.82 $\pm$ 0.37 &       17.83 $\pm$ 0.28 &                 5.210 $\pm$ 0.532 &                0.5 \\
 342 &              B1 V &  SB1 &     4.277640 $\pm$ 0.001352 &   2457301.03 $\pm$ 0.38 &  0.094 $\pm$ 0.037 &   360.00 $\pm$ 29.93 &      250.52 $\pm$ 0.53 &    31.76 $\pm$ 3.70 &        2.67 $\pm$ 0.31 &                 1.402 $\pm$ 0.515 &                2.0 \\
 351 &            B0.5 V &  SB1 &    67.399134 $\pm$ 0.090763 &   2457316.50 $\pm$ 1.23 &  0.228 $\pm$ 0.018 &     18.58 $\pm$ 5.62 &      253.01 $\pm$ 0.39 &    35.79 $\pm$ 0.64 &       46.41 $\pm$ 0.85 &                29.534 $\pm$ 2.278 &                0.8 \\
 359 &            B0.5 V &  SB1 &    19.485809 $\pm$ 0.018241 &   2457310.17 $\pm$ 0.25 &  0.480 $\pm$ 0.035 &     18.81 $\pm$ 5.25 &      275.21 $\pm$ 0.89 &    27.59 $\pm$ 0.45 &        9.32 $\pm$ 0.26 &                 2.862 $\pm$ 0.417 &                1.1 \\
 396 &            B0.5 V &  SB1 &    86.029718 $\pm$ 0.864490 &   2457367.00 $\pm$ 1.36 &  0.476 $\pm$ 0.045 &    260.54 $\pm$ 9.75 &      260.58 $\pm$ 1.16 &    20.00 $\pm$ 1.39 &       29.91 $\pm$ 2.26 &                 4.852 $\pm$ 1.324 &                0.7 \\
 430 &  B0.5 Ia+((n))Nwk &  SB1 &     8.760079 $\pm$ 0.000943 &   2457300.62 $\pm$ 0.02 &              0.000 &                90.00 &      246.95 $\pm$ 0.56 &    63.85 $\pm$ 0.71 &       11.06 $\pm$ 0.12 &                23.626 $\pm$ 0.793 &                4.1 \\
 434 &           B1.5: V &  SB1 &     5.605693 $\pm$ 0.002017 &   2457302.75 $\pm$ 0.63 &  0.057 $\pm$ 0.119 &    80.37 $\pm$ 42.66 &      263.86 $\pm$ 0.84 &    26.68 $\pm$ 1.01 &        2.95 $\pm$ 0.11 &                 1.097 $\pm$ 0.411 &                1.1 \\
 501 &            B0.5 V &  SB1 &   151.455600 $\pm$ 1.435372 &  2457330.37 $\pm$ 17.79 &  0.075 $\pm$ 0.033 &    65.57 $\pm$ 42.85 &      252.45 $\pm$ 0.61 &    12.27 $\pm$ 0.25 &       36.64 $\pm$ 0.82 &                 2.877 $\pm$ 0.334 &                0.6 \\
 534 &             B0 IV &  SB1 &     3.685536 $\pm$ 0.000273 &   2457299.75 $\pm$ 0.10 &  0.050 $\pm$ 0.049 &   105.13 $\pm$ 10.16 &      262.92 $\pm$ 0.46 &    42.64 $\pm$ 0.39 &        3.10 $\pm$ 0.03 &                 2.949 $\pm$ 0.445 &               11.5 \\
 575 &          B0.7 III &  SB1 &    45.296391 $\pm$ 0.014320 &   2457327.55 $\pm$ 0.11 &  0.486 $\pm$ 0.007 &    207.99 $\pm$ 1.10 &      252.87 $\pm$ 0.28 &    45.64 $\pm$ 0.46 &       35.71 $\pm$ 0.39 &                29.790 $\pm$ 1.209 &                0.8 \\
 665 &            B0.5 V &  SB1 &  323.539960 $\pm$ 10.747288 &   2457423.21 $\pm$ 2.85 &  0.566 $\pm$ 0.038 &    289.36 $\pm$ 9.48 &      276.46 $\pm$ 2.29 &    37.17 $\pm$ 0.66 &      196.01 $\pm$ 9.64 &               96.533 $\pm$ 17.298 &                3.0 \\
 687 &   B1.5 Ib((n))Nwk &  SB1 &    12.468230 $\pm$ 0.001659 &   2457307.27 $\pm$ 0.61 &  0.009 $\pm$ 0.011 &    68.95 $\pm$ 17.73 &      274.85 $\pm$ 0.26 &    39.16 $\pm$ 0.45 &        9.65 $\pm$ 0.11 &                 7.754 $\pm$ 0.377 &                1.2 \\
 705 &            B0.7 V &  SB1 &     2.260548 $\pm$ 0.000127 &   2457300.09 $\pm$ 0.09 &  0.043 $\pm$ 0.017 &    55.74 $\pm$ 13.69 &      245.99 $\pm$ 0.78 &    82.11 $\pm$ 1.07 &        3.67 $\pm$ 0.05 &                12.931 $\pm$ 0.834 &                1.5 \\
 715 &              B1 V &  SB1 &     8.709452 $\pm$ 0.002346 &   2457307.19 $\pm$ 0.36 &  0.075 $\pm$ 0.039 &   301.27 $\pm$ 14.95 &      262.56 $\pm$ 0.74 &    34.57 $\pm$ 0.55 &        5.93 $\pm$ 0.10 &                 3.697 $\pm$ 0.465 &                1.4 \\
 718 &          B2.5 III &  SB1 &    26.539014 $\pm$ 0.023709 &   2457319.26 $\pm$ 0.33 &  0.267 $\pm$ 0.017 &    347.86 $\pm$ 4.60 &      275.88 $\pm$ 0.57 &    33.08 $\pm$ 0.80 &       16.72 $\pm$ 0.41 &                 8.910 $\pm$ 0.815 &                1.2 \\
 719 &              B1 V &  SB1 &   111.149350 $\pm$ 0.218268 &   2457367.57 $\pm$ 4.97 &  0.140 $\pm$ 0.031 &   160.10 $\pm$ 14.81 &      268.79 $\pm$ 0.70 &    35.44 $\pm$ 0.84 &       77.09 $\pm$ 1.86 &                49.765 $\pm$ 5.887 &                1.5 \\
 723 &            B0.5 V &  SB1 &     9.954140 $\pm$ 0.003285 &   2457336.20 $\pm$ 0.06 &              0.000 &                90.00 &      245.83 $\pm$ 0.52 &    27.47 $\pm$ 0.77 &        5.41 $\pm$ 0.15 &                 2.139 $\pm$ 0.181 &                2.1 \\
 742 &              B2 V &  SB1 &     6.663969 $\pm$ 0.001457 &   2457306.11 $\pm$ 0.36 &  0.029 $\pm$ 0.033 &   266.86 $\pm$ 18.42 &      277.54 $\pm$ 0.62 &    25.78 $\pm$ 1.08 &        3.39 $\pm$ 0.14 &                 1.181 $\pm$ 0.188 &                0.7 \\
 779 &          B1 II-Ib &  SB1 &    59.934898 $\pm$ 0.041291 &   2457299.75 $\pm$ 1.62 &  0.072 $\pm$ 0.012 &     76.83 $\pm$ 9.91 &      258.24 $\pm$ 0.31 &    33.43 $\pm$ 0.60 &       39.50 $\pm$ 0.70 &                23.020 $\pm$ 1.476 &                0.2 \\
 788 &            B1 III &  SB1 &     3.261387 $\pm$ 0.000145 &   2457301.31 $\pm$ 0.01 &              0.000 &                90.00 &      264.99 $\pm$ 0.58 &    57.00 $\pm$ 1.17 &        3.67 $\pm$ 0.08 &                 6.258 $\pm$ 0.385 &                1.2 \\
 792 &              B2 V &  SB1 &   428.182100 $\pm$ 1.503608 &  2454939.09 $\pm$ 26.43 &  0.228 $\pm$ 0.075 &   290.38 $\pm$ 19.23 &      295.30 $\pm$ 1.35 &    33.12 $\pm$ 2.72 &     272.94 $\pm$ 22.96 &              148.809 $\pm$ 50.818 &                0.4 \\
 821 &           B0 V-IV &  SB1 &     9.495538 $\pm$ 0.002847 &   2457304.46 $\pm$ 0.44 &  0.027 $\pm$ 0.023 &    18.56 $\pm$ 16.63 &      257.32 $\pm$ 0.44 &    23.99 $\pm$ 0.73 &        4.50 $\pm$ 0.14 &                 1.357 $\pm$ 0.155 &                0.6 \\
 827 &           B1.5 Ib &  SB1 &    43.228625 $\pm$ 0.020966 &   2457334.17 $\pm$ 0.33 &  0.266 $\pm$ 0.012 &     98.51 $\pm$ 3.29 &      248.42 $\pm$ 0.28 &    26.08 $\pm$ 0.26 &       21.48 $\pm$ 0.22 &                 7.116 $\pm$ 0.354 &                0.7 \\
 837 &              B1 V &  SB1 &     3.428811 $\pm$ 0.000265 &   2457302.96 $\pm$ 0.05 &  0.102 $\pm$ 0.017 &    193.22 $\pm$ 5.30 &      270.31 $\pm$ 0.43 &    41.25 $\pm$ 0.34 &        2.78 $\pm$ 0.02 &                 2.455 $\pm$ 0.141 &                7.1 \\
 874 &        B1.5 IIIe+ &  SB1 &   370.823220 $\pm$ 5.679991 &  2457308.23 $\pm$ 11.75 &  0.209 $\pm$ 0.055 &   318.25 $\pm$ 15.02 &      265.34 $\pm$ 0.41 &    16.99 $\pm$ 0.12 &      121.76 $\pm$ 2.53 &                17.614 $\pm$ 3.083 &                2.9 \\
 888 &            B0.5 V &  SB1 &     1.965846 $\pm$ 0.000045 &   2457301.38 $\pm$ 0.01 &              0.000 &                90.00 &      255.08 $\pm$ 0.59 &    58.21 $\pm$ 1.14 &        2.26 $\pm$ 0.04 &                 4.017 $\pm$ 0.236 &                2.2 \\
 891 &              B2 V &  SB1 &     5.480890 $\pm$ 0.002054 &   2457304.80 $\pm$ 0.16 &  0.273 $\pm$ 0.031 &    145.97 $\pm$ 9.16 &      261.20 $\pm$ 0.58 &    14.24 $\pm$ 0.22 &        1.48 $\pm$ 0.03 &                 0.146 $\pm$ 0.016 &                1.7 \\
\bottomrule
% \end{tabular}
\end{longtable}}
% \end{table*}

\begin{table}
\caption{\label{tab:tabB4} Orbital parameters determined with \texttt{rvfit} for the SB1* stars.}
\centering
% \small
\begin{tabular}{clHrcccccccr}
\toprule
\toprule
VFTS &   SpT & SB & \multicolumn{1}{c}{$P_{\mathrm{orb}}$} & $T_p$ & $e$ & $\omega$ & $\gamma$ & $K_1$ & $a_1\sin i$ & $f(m_1,m_2)$ & $\chi^2_{\rm red}$ \\
 & &  & \multicolumn{1}{c}{[d]} & [HJD] & & [deg] & [\kms] & [\kms] & [\Rsun] & [10$^{-2}$\Msun] & \\
\midrule
 018 &        B1.5 V &  SB1* &   70.862401 $\pm$ 0.192531 &  2457301.86 $\pm$ 1.53 &  0.553 $\pm$ 0.043 &   264.95 $\pm$ 11.03 &      279.39 $\pm$ 0.59 &    12.22 $\pm$ 0.15 &       14.25 $\pm$ 0.53 &                 0.774 $\pm$ 0.148 &                1.6 \\
 037 &       B2 III: &  SB1* &   41.242466 $\pm$ 0.230584 &  2457317.96 $\pm$ 0.83 &  0.230 $\pm$ 0.037 &   233.36 $\pm$ 11.66 &      287.96 $\pm$ 0.59 &    11.41 $\pm$ 0.35 &        9.05 $\pm$ 0.29 &                 0.585 $\pm$ 0.087 &                1.3 \\
 106 &        B0.2 V &  SB1* &   16.352137 $\pm$ 0.016438 &  2457305.73 $\pm$ 0.57 &  0.478 $\pm$ 0.060 &   212.85 $\pm$ 17.56 &      269.02 $\pm$ 0.96 &    16.55 $\pm$ 0.49 &        4.70 $\pm$ 0.22 &                 0.520 $\pm$ 0.130 &                0.3 \\
 146 &         B2: V &  SB1* &  117.035250 $\pm$ 0.645774 &  2457352.79 $\pm$ 4.45 &  0.099 $\pm$ 0.030 &   320.50 $\pm$ 14.54 &      280.79 $\pm$ 0.81 &    24.20 $\pm$ 0.36 &       55.71 $\pm$ 0.90 &                16.936 $\pm$ 1.710 &                3.8 \\
 155 &       B0.7: V &  SB1* &  153.434020 $\pm$ 1.464121 &  2457432.74 $\pm$ 5.77 &  0.219 $\pm$ 0.046 &   113.12 $\pm$ 14.84 &      279.41 $\pm$ 1.40 &    22.16 $\pm$ 0.66 &       65.57 $\pm$ 2.16 &                16.070 $\pm$ 2.752 &                1.3 \\
 213 &      B2 III:e &  SB1* &   13.592696 $\pm$ 0.006953 &  2457310.50 $\pm$ 0.20 &  0.640 $\pm$ 0.018 &    128.31 $\pm$ 8.04 &      275.06 $\pm$ 0.66 &    20.33 $\pm$ 0.27 &        4.20 $\pm$ 0.10 &                 0.536 $\pm$ 0.054 &                5.3 \\
 218 &        B1.5 V &  SB1* &   20.751590 $\pm$ 0.021051 &  2457309.78 $\pm$ 0.37 &  0.554 $\pm$ 0.019 &    323.96 $\pm$ 4.92 &      246.56 $\pm$ 0.58 &    13.22 $\pm$ 0.14 &        4.51 $\pm$ 0.09 &                 0.286 $\pm$ 0.026 &                4.8 \\
 337 &   B2: V-IIIe+ &  SB1* &   25.505882 $\pm$ 0.064516 &  2457301.93 $\pm$ 0.62 &  0.546 $\pm$ 0.056 &   175.56 $\pm$ 11.58 &      234.49 $\pm$ 2.41 &    37.88 $\pm$ 1.94 &       15.99 $\pm$ 1.08 &                 8.435 $\pm$ 2.401 &                0.5 \\
 388 &        B0.5 V &  SB1* &   14.792479 $\pm$ 0.010906 &  2457308.70 $\pm$ 0.23 &  0.644 $\pm$ 0.028 &    213.27 $\pm$ 6.30 &      260.38 $\pm$ 1.11 &    34.73 $\pm$ 0.68 &        7.77 $\pm$ 0.28 &                 2.874 $\pm$ 0.444 &                1.2 \\
 442 &        B1-2 V &  SB1* &   27.069231 $\pm$ 0.114201 &  2457317.19 $\pm$ 0.86 &              0.000 &                90.00 &      252.53 $\pm$ 1.13 &    14.86 $\pm$ 0.64 &        7.95 $\pm$ 0.34 &                 0.920 $\pm$ 0.118 &                0.7 \\
 576 &      B1 IaNwk &  SB1* &   85.596184 $\pm$ 0.105916 &  2457351.52 $\pm$ 0.48 &  0.585 $\pm$ 0.010 &    269.61 $\pm$ 1.75 &      265.75 $\pm$ 0.60 &    61.16 $\pm$ 0.75 &       83.93 $\pm$ 1.26 &               108.276 $\pm$ 6.245 &                2.3 \\
 606 &   B0-0.5 V(n) &  SB1* &   84.706776 $\pm$ 0.638226 &  2457382.03 $\pm$ 3.93 &  0.357 $\pm$ 0.067 &   210.02 $\pm$ 18.64 &      245.45 $\pm$ 1.12 &    13.98 $\pm$ 1.48 &       21.87 $\pm$ 2.40 &                 1.956 $\pm$ 0.770 &                0.5 \\
 662 &     B3-5 III: &  SB1* &    1.989458 $\pm$ 0.000671 &  2457300.11 $\pm$ 0.06 &              0.000 &                90.00 &      273.99 $\pm$ 0.75 &     7.44 $\pm$ 0.29 &        0.29 $\pm$ 0.01 &                 0.008 $\pm$ 0.001 &                0.4 \\
 697 &       B1-2 Ve &  SB1* &    1.488723 $\pm$ 0.000522 &  2457300.65 $\pm$ 0.77 &  0.305 $\pm$ 0.142 &  360.00 $\pm$ 206.17 &      262.34 $\pm$ 4.45 &    12.98 $\pm$ 4.14 &        0.36 $\pm$ 0.12 &                 0.029 $\pm$ 0.031 &                0.5 \\
 730 &      B1 IV(n) &  SB1* &    1.334263 $\pm$ 0.000492 &  2457333.21 $\pm$ 0.06 &              0.000 &                90.00 &      318.57 $\pm$ 1.29 &     8.55 $\pm$ 0.38 &        0.23 $\pm$ 0.01 &                 0.009 $\pm$ 0.001 &                0.3 \\
 784 &         B1: V &  SB1* &    3.575900 $\pm$ 0.004914 &  2457301.30 $\pm$ 0.34 &  0.547 $\pm$ 0.100 &    108.45 $\pm$ 7.86 &      265.83 $\pm$ 2.68 &    18.36 $\pm$ 0.39 &        1.09 $\pm$ 0.09 &                 0.135 $\pm$ 0.058 &                0.4 \\
 799 &    B0.5-0.7 V &  SB1* &   39.374872 $\pm$ 0.133568 &  2457305.08 $\pm$ 0.96 &  0.690 $\pm$ 0.060 &    47.07 $\pm$ 18.90 &      263.98 $\pm$ 0.87 &     6.34 $\pm$ 0.13 &        3.57 $\pm$ 0.29 &                 0.039 $\pm$ 0.014 &                0.4 \\
 847 &  B0.7-1 IIIne &  SB1* &    1.233195 $\pm$ 0.000133 &  2457300.01 $\pm$ 0.06 &  0.500 $\pm$ 0.097 &    96.35 $\pm$ 36.84 &      291.97 $\pm$ 3.38 &    26.76 $\pm$ 1.06 &        0.56 $\pm$ 0.04 &                 0.159 $\pm$ 0.065 &                0.4 \\
 850 &        B1 III &  SB1* &   50.662273 $\pm$ 0.072595 &  2457311.41 $\pm$ 1.38 &  0.133 $\pm$ 0.043 &    205.40 $\pm$ 8.07 &      244.00 $\pm$ 0.51 &    19.02 $\pm$ 0.24 &       18.88 $\pm$ 0.26 &                 3.516 $\pm$ 0.477 &               10.0 \\
 877 &  B1-3 V-IIIe+ &  SB1* &   94.783970 $\pm$ 2.332124 &  2457375.52 $\pm$ 3.30 &  0.323 $\pm$ 0.106 &   211.82 $\pm$ 16.26 &      248.53 $\pm$ 0.55 &     8.83 $\pm$ 0.11 &       15.65 $\pm$ 0.74 &                 0.572 $\pm$ 0.205 &                0.4 \\
\bottomrule
\end{tabular}
\end{table}
\end{landscape}
\clearpage
\begin{table}
\caption{\label{tab:tabB5} Orbital parameters determined with \texttt{rvfit} for the SB2 sample.}
\centering
\begin{tabular}{cHcccccccr}
\toprule
\toprule
VFTS &  SB &  
$P_{\mathrm{orb}}$ &    
$T_p$  &    $e$ &   
$\omega$ & 
$\gamma$  & 
$K_1$  & 
$K_2$  & $\chi^2_{\rm red}$ \\
 &  & [d] & [HJD] & & [deg] & [\kms] & [\kms] & [\kms] & \\
\midrule
 112 &  SB2 &   1.674083 $\pm$ 0.000030 &  2457300.43 $\pm$ 0.00 &              0.000 &               90.00 &      264.02 $\pm$ 1.24 &   222.35 $\pm$ 1.66 &   266.80 $\pm$ 2.58 &                4.8 \\
 199 &  SB2 &   1.666830 $\pm$ 0.000041 &  2457300.84 $\pm$ 0.00 &              0.000 &               90.00 &      281.41 $\pm$ 1.02 &   166.69 $\pm$ 1.30 &   174.47 $\pm$ 2.68 &                3.9 \\
 206 &  SB2 &   7.556457 $\pm$ 0.000906 &  2457305.59 $\pm$ 0.27 &  0.029 $\pm$ 0.012 &  315.73 $\pm$ 13.02 &      262.54 $\pm$ 0.91 &    97.10 $\pm$ 1.94 &   108.62 $\pm$ 1.16 &                5.4 \\
 240 &  SB2 &   1.377795 $\pm$ 0.000060 &  2457300.43 $\pm$ 0.04 &  0.096 $\pm$ 0.014 & \po0.00 $\pm$ 8.59 &      255.51 $\pm$ 1.27 &   142.38 $\pm$ 3.07 &   147.09 $\pm$ 1.76 &                7.5 \\
 248 &  SB2 &   2.494240 $\pm$ 0.000058 &  2457299.86 $\pm$ 0.02 &  0.166 $\pm$ 0.007 &   183.46 $\pm$ 2.40\po &      278.48 $\pm$ 0.90 &   181.55 $\pm$ 1.22 &   238.41 $\pm$ 1.89 &               11.8 \\
 255 &  SB2 &   2.844891 $\pm$ 0.000100 &  2457335.35 $\pm$ 0.06 &  0.024 $\pm$ 0.009 &   111.52 $\pm$ 7.67\po &      251.27 $\pm$ 1.26 &   193.92 $\pm$ 2.16 &   200.14 $\pm$ 2.28 &                8.2 \\
 364 &  SB2 &   4.399906 $\pm$ 0.000202 &  2457302.31 $\pm$ 0.04 &  0.157 $\pm$ 0.007 &    58.08 $\pm$ 3.50 &      250.16 $\pm$ 0.96 &   144.92 $\pm$ 1.84 &   149.39 $\pm$ 1.61 &               20.9 \\
 383 &  SB2 &   2.597446 $\pm$ 0.000159 &  2457300.76 $\pm$ 0.08 &  0.059 $\pm$ 0.011 &   201.74 $\pm$ 9.83\po &      249.04 $\pm$ 0.81 &    82.21 $\pm$ 1.69 &   117.36 $\pm$ 0.87 &                2.5 \\
 520 &  SB2 &   2.947548 $\pm$ 0.000123 &  2457300.83 $\pm$ 0.05 &  0.142 $\pm$ 0.013 &   309.17 $\pm$ 6.25\po &      256.74 $\pm$ 1.12 &   114.60 $\pm$ 1.31 &   176.82 $\pm$ 3.09 &                8.0 \\
 589 &  SB2 &   7.629761 $\pm$ 0.000505 &  2457332.83 $\pm$ 0.03 &  0.267 $\pm$ 0.006 &    65.56 $\pm$ 1.61 &      276.33 $\pm$ 0.50 &    97.11 $\pm$ 0.62 &   121.52 $\pm$ 1.15 &                7.0 \\
 637 &  SB2 &   1.628692 $\pm$ 0.000033 &  2457300.91 $\pm$ 0.03 &  0.026 $\pm$ 0.008 &    97.80 $\pm$ 7.68 &      240.20 $\pm$ 1.12 &   214.02 $\pm$ 1.69 &   241.65 $\pm$ 2.65 &                6.1 \\
 686 &  SB2 &  16.869613 $\pm$ 0.002899\po &  2457304.83 $\pm$ 0.13 &  0.311 $\pm$ 0.007 &   117.54 $\pm$ 2.18\po &      253.69 $\pm$ 0.50 &    71.56 $\pm$ 0.68 &   101.93 $\pm$ 1.89 &               14.9 \\
 752 &  SB2 &   1.407603 $\pm$ 0.000028 &  2457300.07 $\pm$ 0.00 &              0.000 &               90.00 &      248.93 $\pm$ 0.70 &   127.63 $\pm$ 1.03 &   140.60 $\pm$ 1.31 &                2.4 \\
 883 &  SB2 &   3.495192 $\pm$ 0.000330 &  2457302.86 $\pm$ 0.02 &  0.359 $\pm$ 0.013 &   161.48 $\pm$ 3.66\po &      248.28 $\pm$ 1.80 &   170.74 $\pm$ 3.21 &   226.81 $\pm$ 3.87 &                7.1 \\
\bottomrule
\end{tabular}
\end{table}
\begin{table}
\caption{\label{tab:tabB6} Derived orbital parameters for the SB2 sample.}
\centering
\begin{tabular}{clcccccc}
\toprule
\toprule
VFTS & SpT &
\multicolumn{1}{c}{$M_1\sin ^3i$} & 
\multicolumn{1}{c}{$M_2\sin ^3i$} &    
\multicolumn{1}{c}{$q$} & 
\multicolumn{1}{c}{$a_1\sin i$} & 
\multicolumn{1}{c}{$a_2\sin i$} & 
\multicolumn{1}{c}{$a\sin i$} \\
  &  & [\Msun] & [\Msun] & [$M_2/M_1$] & [\Rsun] & [\Rsun] & [\Rsun]\\
\midrule
 112 &  EarlyBs &   11.07 $\pm$ 0.24\po &   9.23 $\pm$ 0.16 &  0.83 $\pm$ 0.01 &        7.36 $\pm$ 0.05 &   8.83 $\pm$ 0.09 &     16.19 $\pm$ 0.10 \\
 199 &  EarlyBs &   3.51 $\pm$ 0.11 &       3.35 $\pm$ 0.07 &  0.96 $\pm$ 0.02 &        5.49 $\pm$ 0.04 &   5.75 $\pm$ 0.09 &     11.24 $\pm$ 0.10 \\
 206 &  B3 III: &   3.59 $\pm$ 0.10 &       3.21 $\pm$ 0.13 &  0.89 $\pm$ 0.02 &       14.50 $\pm$ 0.29\po &   16.22 $\pm$ 0.17\po &     30.71 $\pm$ 0.34 \\
 240 &  B1-2 V &    1.74 $\pm$ 0.06 &       1.68 $\pm$ 0.07 &  0.97 $\pm$ 0.02 &        3.86 $\pm$ 0.08 &   3.99 $\pm$ 0.05 &      \po7.85 $\pm$ 0.10 \\
 248 &  B2: V &     10.42 $\pm$ 0.19\po &   7.93 $\pm$ 0.13 &  0.76 $\pm$ 0.01 &        8.83 $\pm$ 0.06 &   11.59 $\pm$ 0.09\po &     20.42 $\pm$ 0.11 \\
 255 &  B2: V &     9.15 $\pm$ 0.23 &       8.87 $\pm$ 0.22 &  0.97 $\pm$ 0.02 &       10.90 $\pm$ 0.12\po &   11.25 $\pm$ 0.13\po &     22.15 $\pm$ 0.18 \\
 364 &  B2.5: V &   5.68 $\pm$ 0.14 &       5.51 $\pm$ 0.15 &  0.97 $\pm$ 0.02 &       12.45 $\pm$ 0.16\po &   12.83 $\pm$ 0.14\po &     25.28 $\pm$ 0.21 \\
 383 &  B0.5: V &   1.25 $\pm$ 0.03 &       0.88 $\pm$ 0.03 &  0.70 $\pm$ 0.02 &        4.21 $\pm$ 0.09 &   6.01 $\pm$ 0.04 &     10.23 $\pm$ 0.10 \\
 520 &  B1: V &     4.45 $\pm$ 0.18 &       2.88 $\pm$ 0.09 &  0.65 $\pm$ 0.01 &        6.61 $\pm$ 0.08 &   10.20 $\pm$ 0.18\po &     16.81 $\pm$ 0.19 \\
 589 &  B0.5 V &    4.11 $\pm$ 0.09 &       3.28 $\pm$ 0.06 &  0.80 $\pm$ 0.01 &       14.11 $\pm$ 0.09\po &   17.66 $\pm$ 0.17\po &     31.78 $\pm$ 0.19 \\
 637 &  B1-2 V+EarlyB &   8.46 $\pm$ 0.20 & 7.49 $\pm$ 0.14 &  0.89 $\pm$ 0.01 &        6.89 $\pm$ 0.05 &   7.78 $\pm$ 0.09 &     14.66 $\pm$ 0.10 \\
 686 &  B0.7 III &  4.61 $\pm$ 0.19 &       3.23 $\pm$ 0.09 &  0.70 $\pm$ 0.01 &       22.68 $\pm$ 0.22\po &   32.31 $\pm$ 0.60\po &     54.99 $\pm$ 0.64 \\
 752 &  B2 V &      1.48 $\pm$ 0.03 &       1.34 $\pm$ 0.02 &  0.91 $\pm$ 0.01 &        3.55 $\pm$ 0.03 &   3.91 $\pm$ 0.04 &      \po7.46 $\pm$ 0.05 \\
 883 &  B0.5 V &    10.56 $\pm$ 0.45\po &   7.95 $\pm$ 0.34 &  0.75 $\pm$ 0.02 &       11.01 $\pm$ 0.22\po &   14.63 $\pm$ 0.26\po &     25.64 $\pm$ 0.34 \\
\bottomrule
\end{tabular}
\end{table}\clearpage
\begin{table*}
\caption{Additional stars observed as part of the BBC programme.}\label{tab:extras}
\centering
\begin{tabular}{llccl}
\toprule
\toprule
VFTS & Classification & $V$ & $B-V$ & Comments \\
\midrule
\multicolumn{5}{l}{{\it Candidate O-type runaways:}}\\
072 & O2 V-III(n)((f*)) & 13.70 & $-$0.14 & \\
138 & O9 Vn             & 15.63 & $-$0.09 & \\
249 & O8 Vn             & 15.52 & $-$0.03 & \\
285 & O7.5 Vnnn         & 15.63 & $-$0.06 & \\
356 & O6: V(n)z         & 15.87 & \pp0.16 & \\
660 & O9.5 Vnn          & 15.92 & \pp0.03 & \\
706 & O6-7 Vnnz         & 15.77 & \pp0.14 & \\
722 & O7 Vnnz           & 15.04 & $-$0.13 & \\
746 & O6 Vnn            & 15.38 & \pp0.12 & \\
751 & O7-8 Vnnz         & 16.32 & \pp0.15 & \\
755 & O3 Vn((f*))       & 15.04 & \pp0.14 & \\
768 & O8 Vn             & 16.10 & \pp0.20 & \\
770 & O7 Vnn            & 15.79 & \pp0.08 & \\
%    &                   &       &         & \\
    \midrule
\multicolumn{5}{l}{{\it Other VFTS targets:}}\\
%\hline
091 & O9.5 IIIn         & 15.98 & \pp0.20 & N-enriched \citep{grin17}\\
102 & O9: Vnnne$+$      & 15.70 & \pp0.35 & Extreme rotator \citep{dufton11} \\
267 & O3 III-I(n)f*     & 13.49 & $-$0.05 & N-enriched \citep{grin17} \\
298 & B1-2 V-IIIe$+$    & 16.68 & \pp0.24 & Candidate runaway \citep{evans+15} \\
328 & O9.5 III(n)       & 15.90 & $-$0.11 & N-enriched \citep{grin17} \\
358 & B0.5: V           & 16.87 & \pp0.00 & Candidate runaway \citep{evans+15}\\
368 & B1-3 V            & 16.68 & \pp0.01 & Candidate runaway \citep{evans+15}\\
399 & O9 IIIn           & 15.83 & \pp0.08 & X-ray bright \citep{clark15} \\
456 & Onn               & 15.46 & \pp0.13 & Rapid rotator \citep{ora13} \\
467 & B1-2 Ve$+$        & 16.91 & \pp0.38 & Candidate runaway \citep{evans+15}\\
574 & O9.5 IIIn         & 15.89 & $-$0.12 & N-enriched \citep{grin17}\\
698 & mid-B $+$ early-B & 13.68 & \pp0.44 & Peculiar B[e]-like system \citep{dunstall+12} \\
703 & O7: V: $+$ O8: V: & 16.91 & \pp0.30 & O-type SB2 \citep[not observed by][]{almeida+17} \\
704 & O9.2 V(n)         & 16.76 & $-$0.07 & Moderately rapid rotator (240-300\,\kms) \\
\bottomrule
\end{tabular}
\tablefoot{Classifications are from \citet{walborn+14} and \citet{evans+15}, for O- and B-type spectra, respectively. Photometry is from \citet{evans+11}.}
\end{table*}

\bsp	% typesetting comment
\clearpage

\renewcommand\themyfloat{\thesection\arabic{myfloat}}   
%\setcounter{postfigure}{2}
%\setcounter{myfloat}{2}

% Appendix C
\section{PERIODOGRAMS}\label{apx:Pgrams}
\setcounter{figure}{1}
\begin{myfloat}
    \centering
    \includegraphics[width=0.31\textwidth]{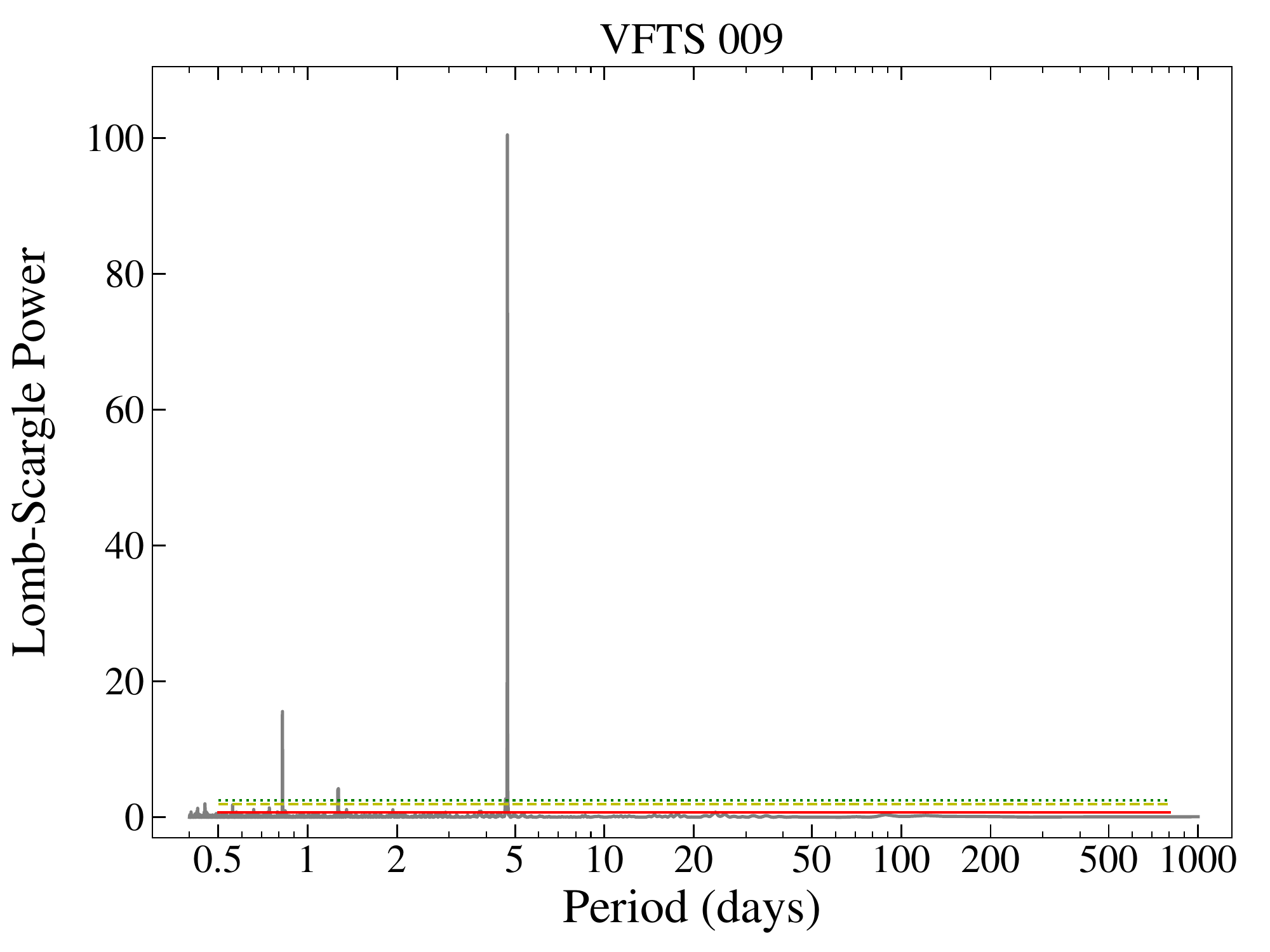}\hfill
    \includegraphics[width=0.31\textwidth]{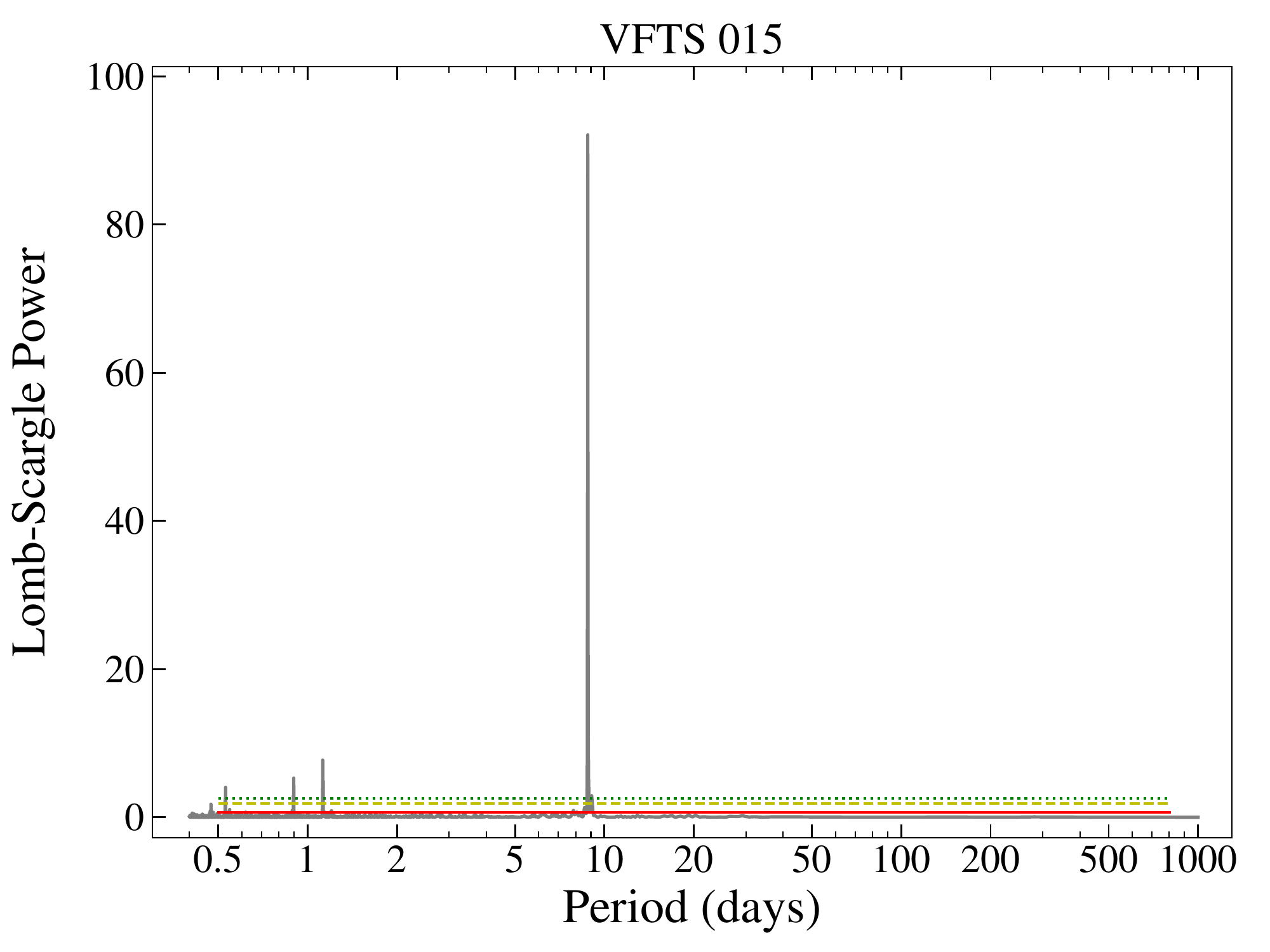}\hfill
    \includegraphics[width=0.31\textwidth]{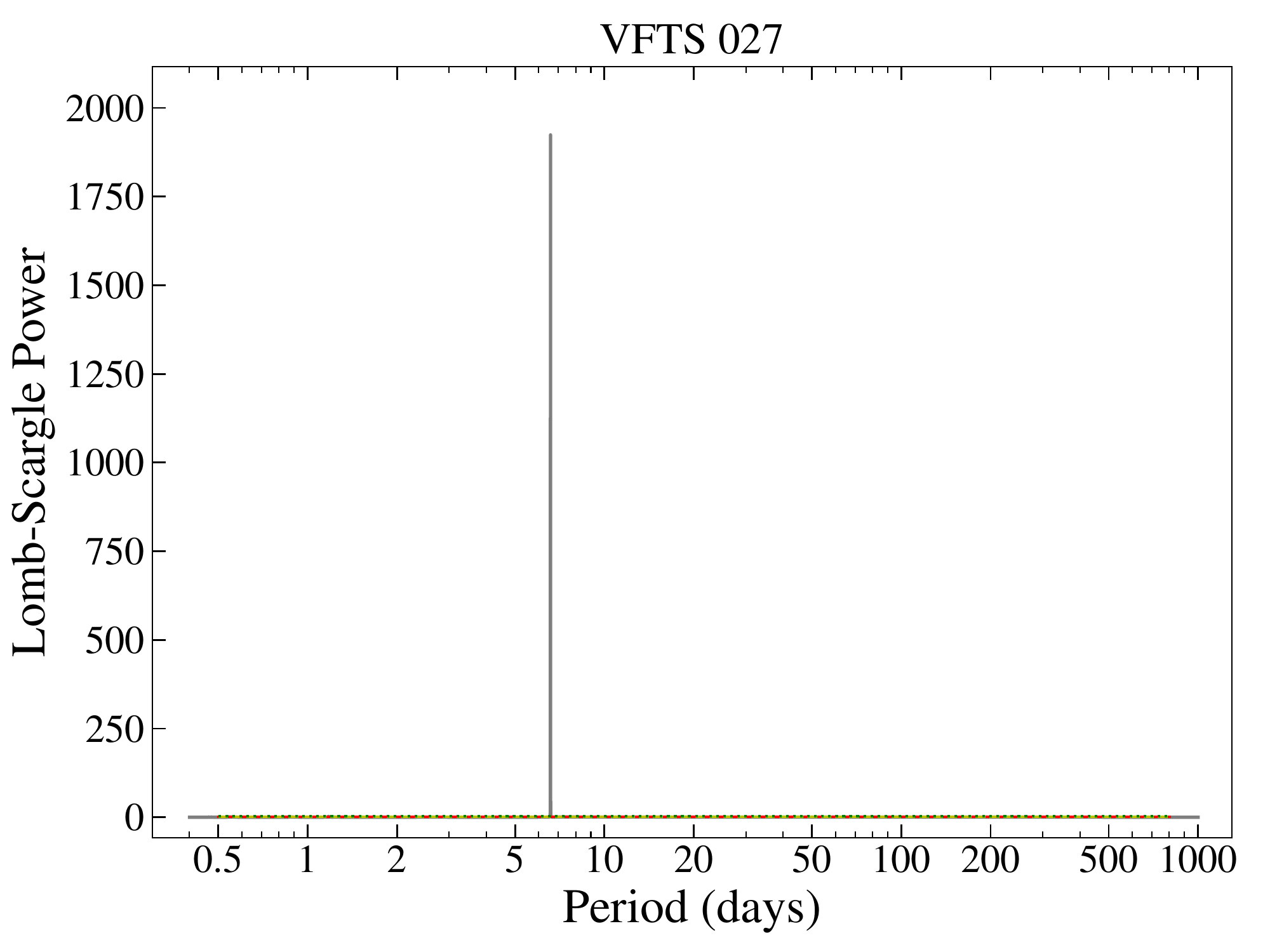}\hfill
    \includegraphics[width=0.31\textwidth]{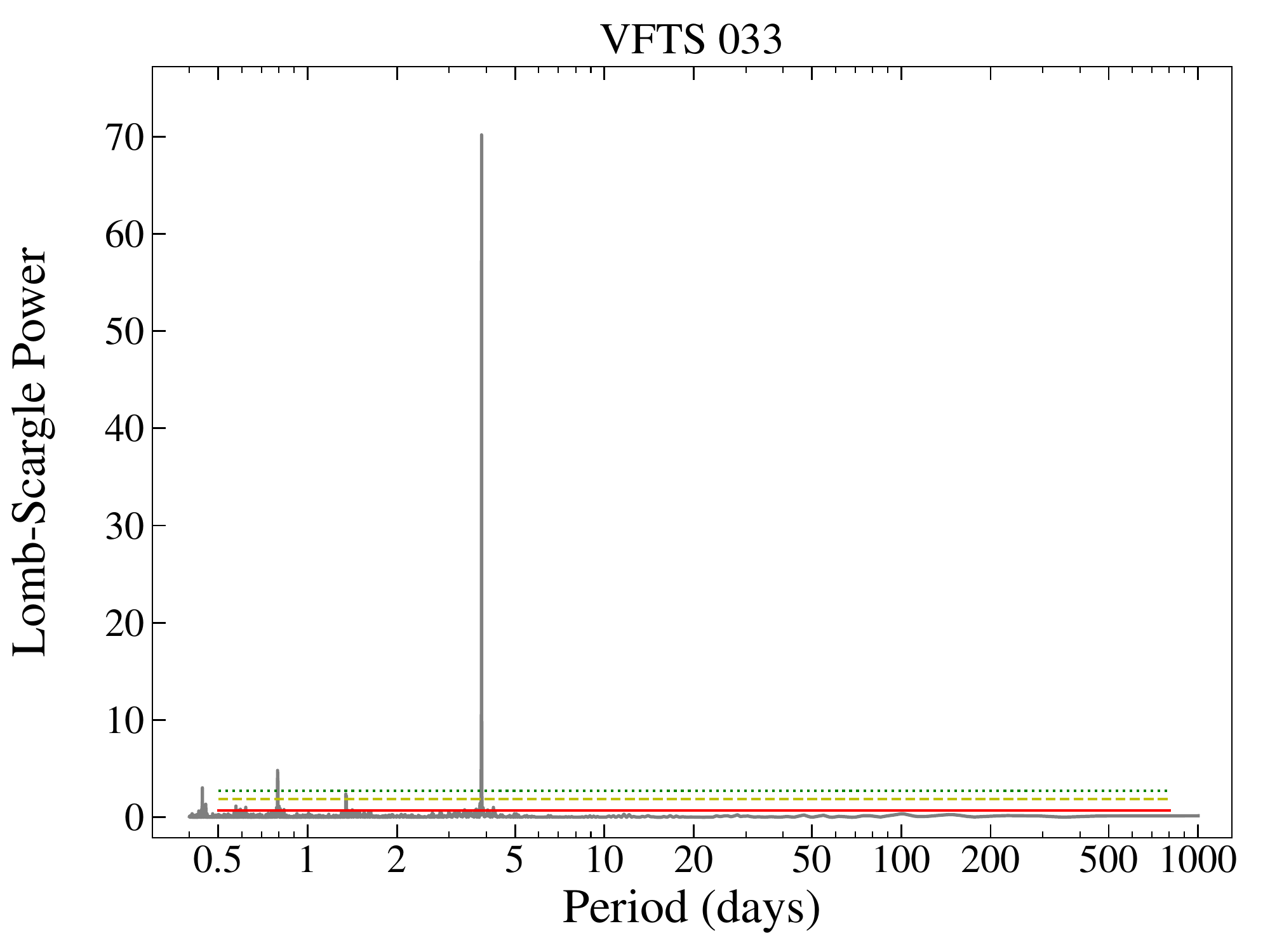}\hfill
    \includegraphics[width=0.31\textwidth]{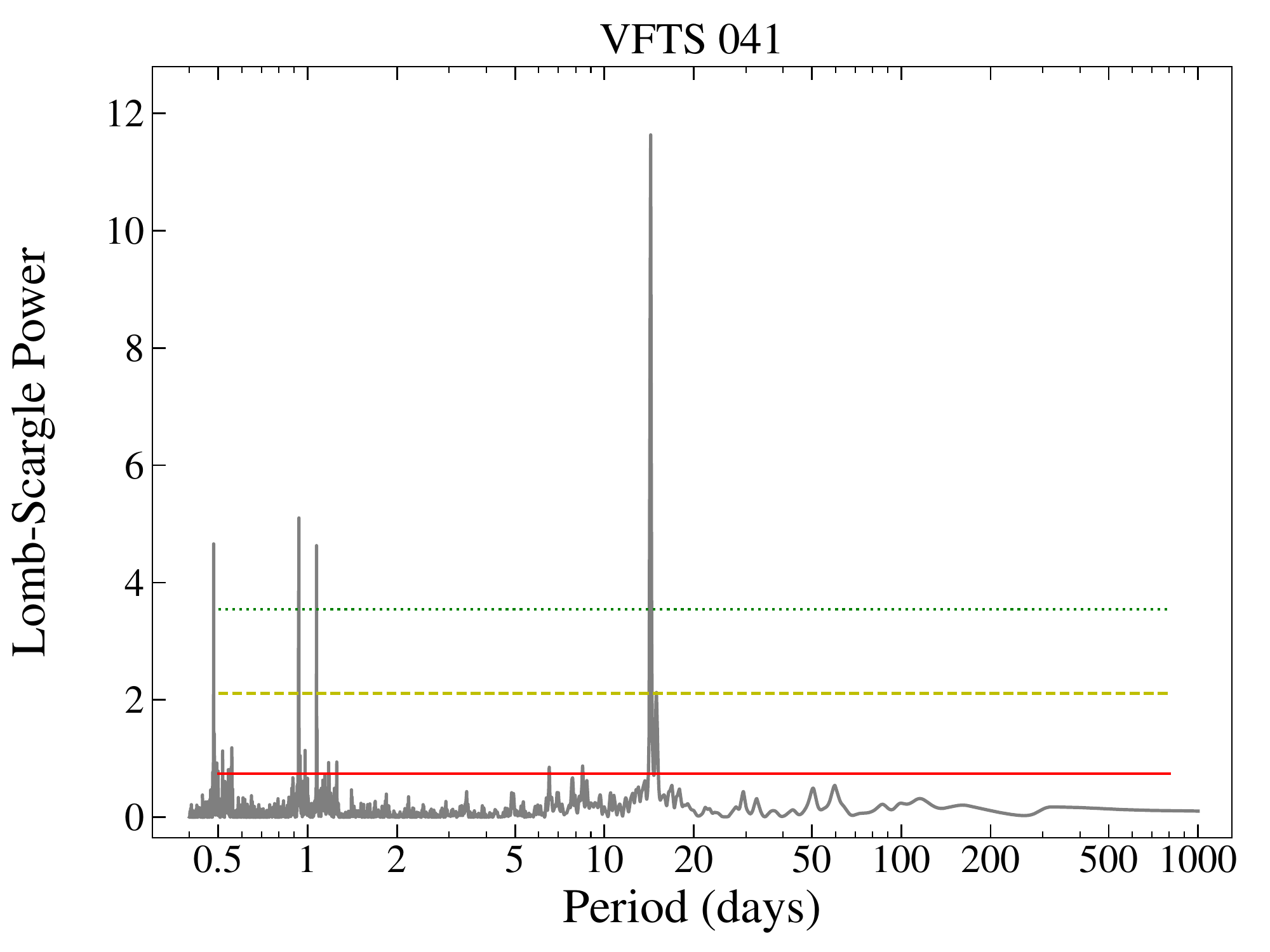}\hfill
    \includegraphics[width=0.31\textwidth]{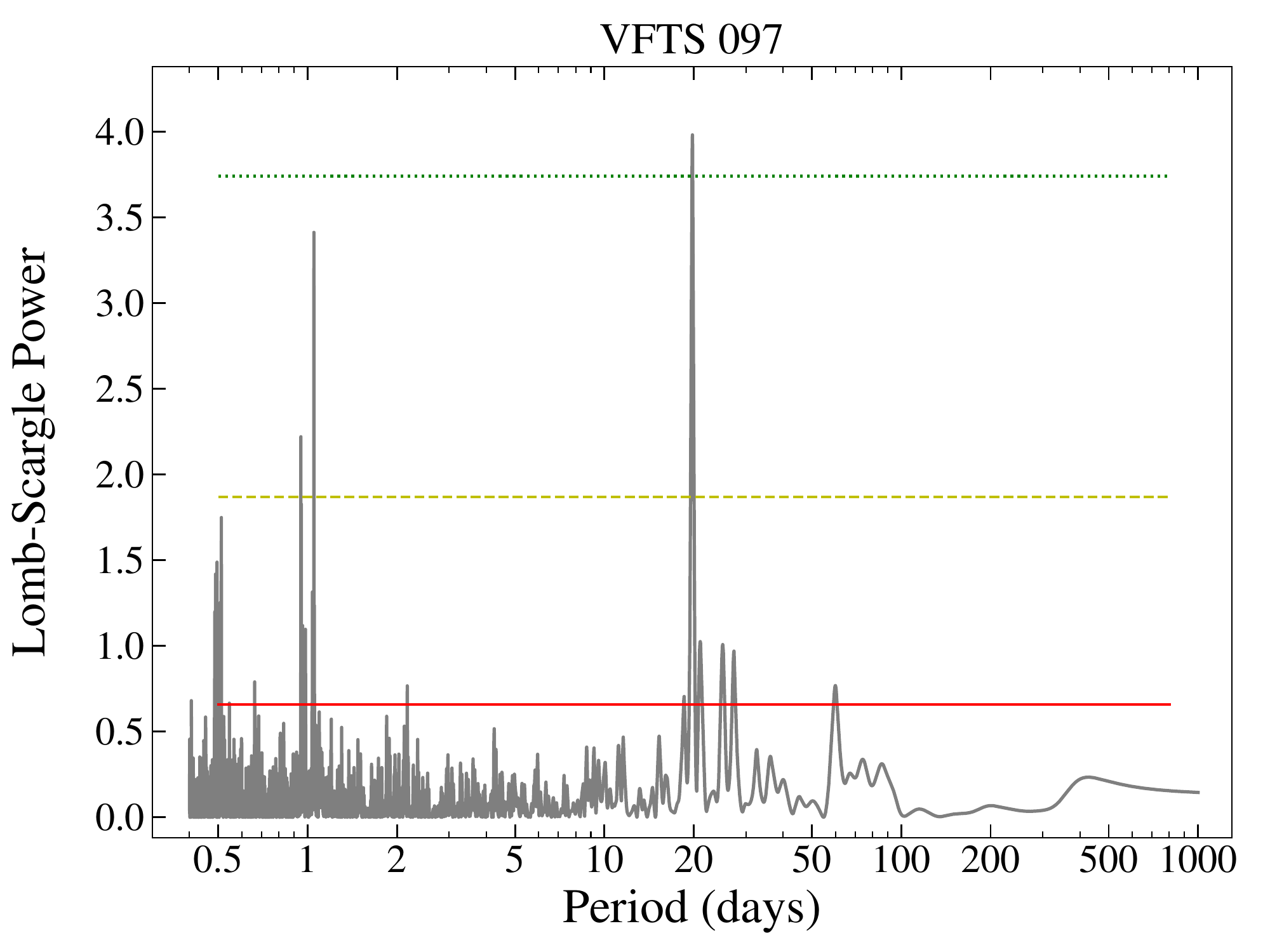}\hfill
    \includegraphics[width=0.31\textwidth]{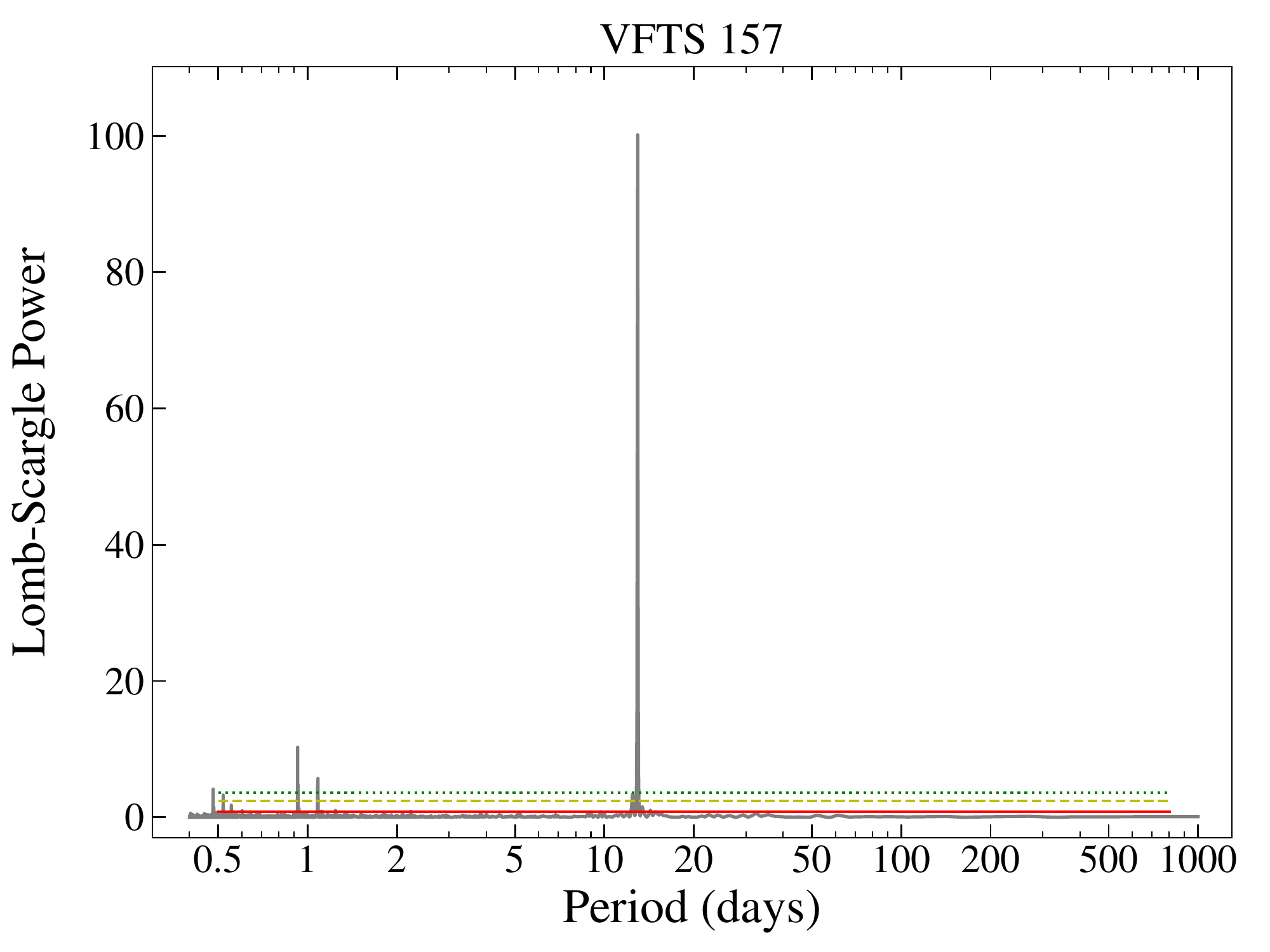}\hfill
    \includegraphics[width=0.31\textwidth]{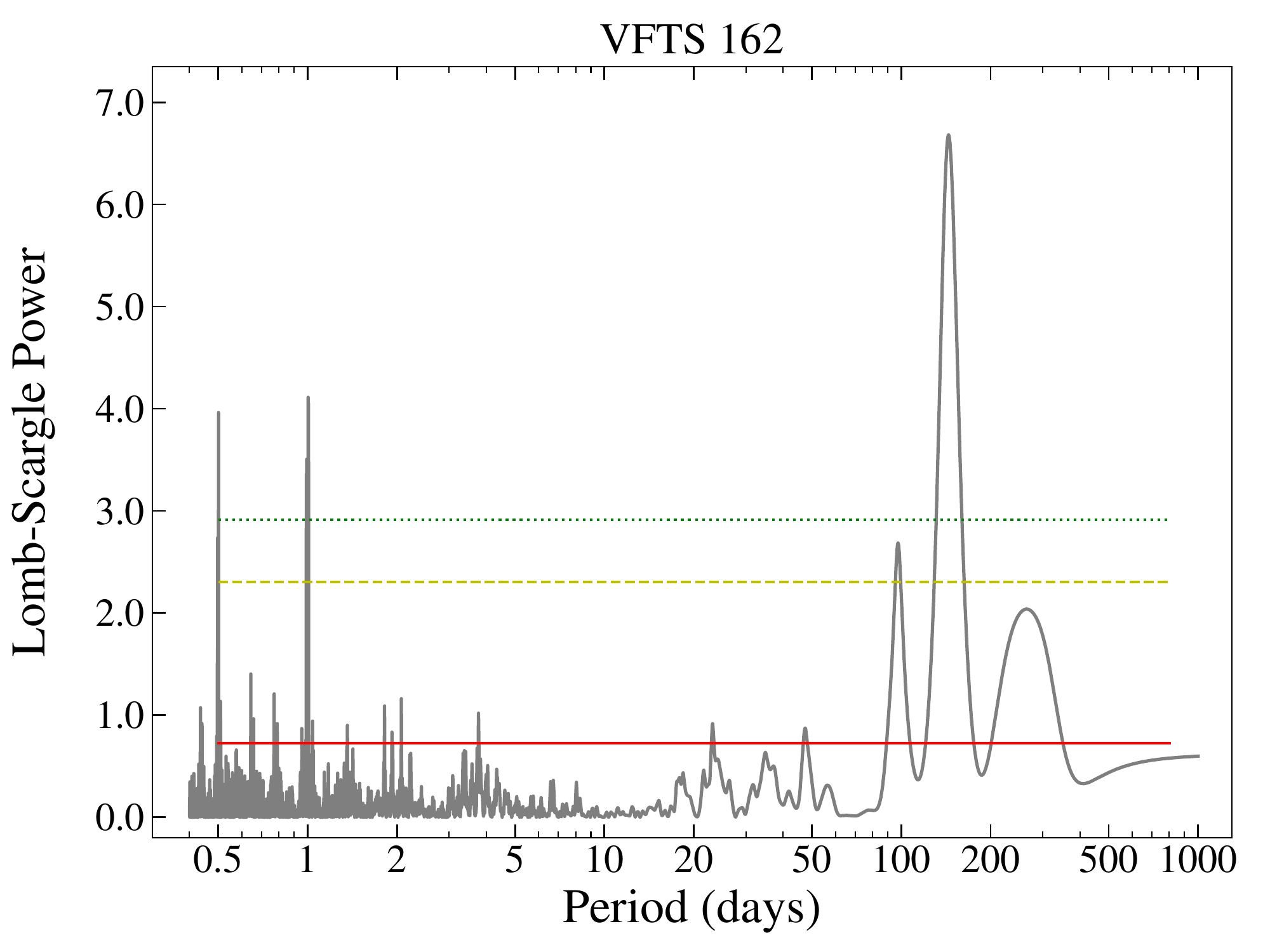}\hfill
    \includegraphics[width=0.31\textwidth]{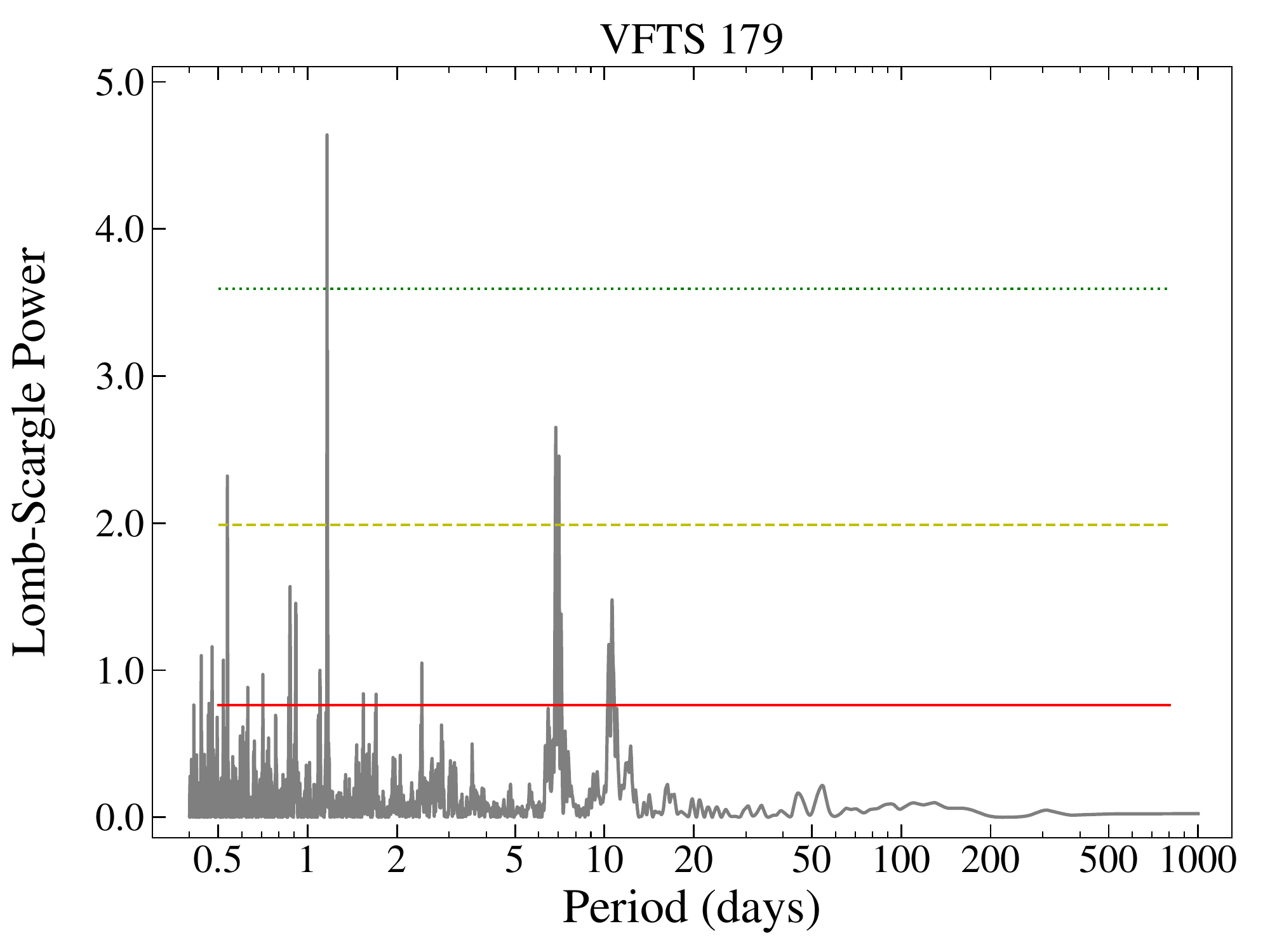}\hfill
    \includegraphics[width=0.31\textwidth]{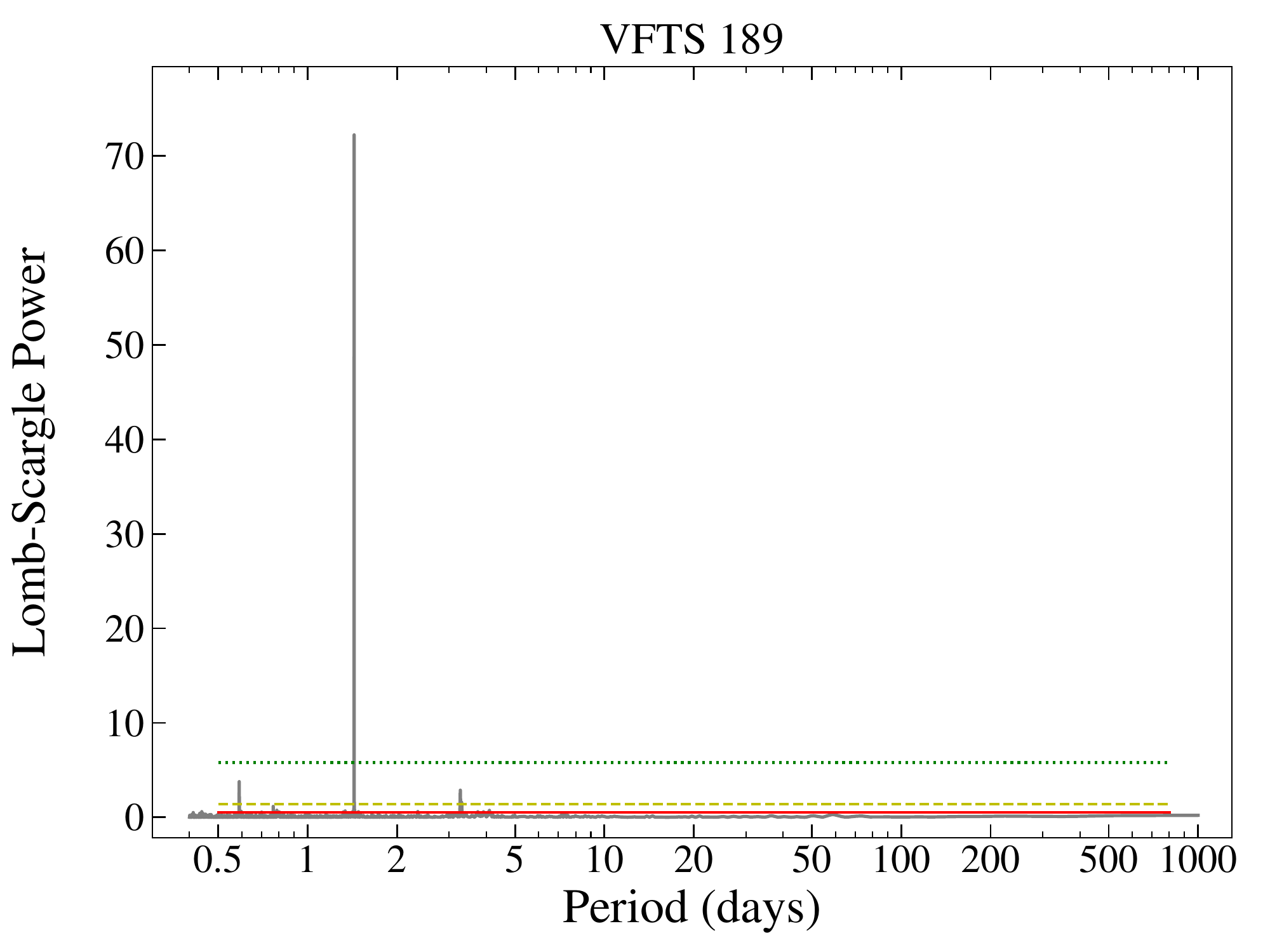}\hfill
    \includegraphics[width=0.31\textwidth]{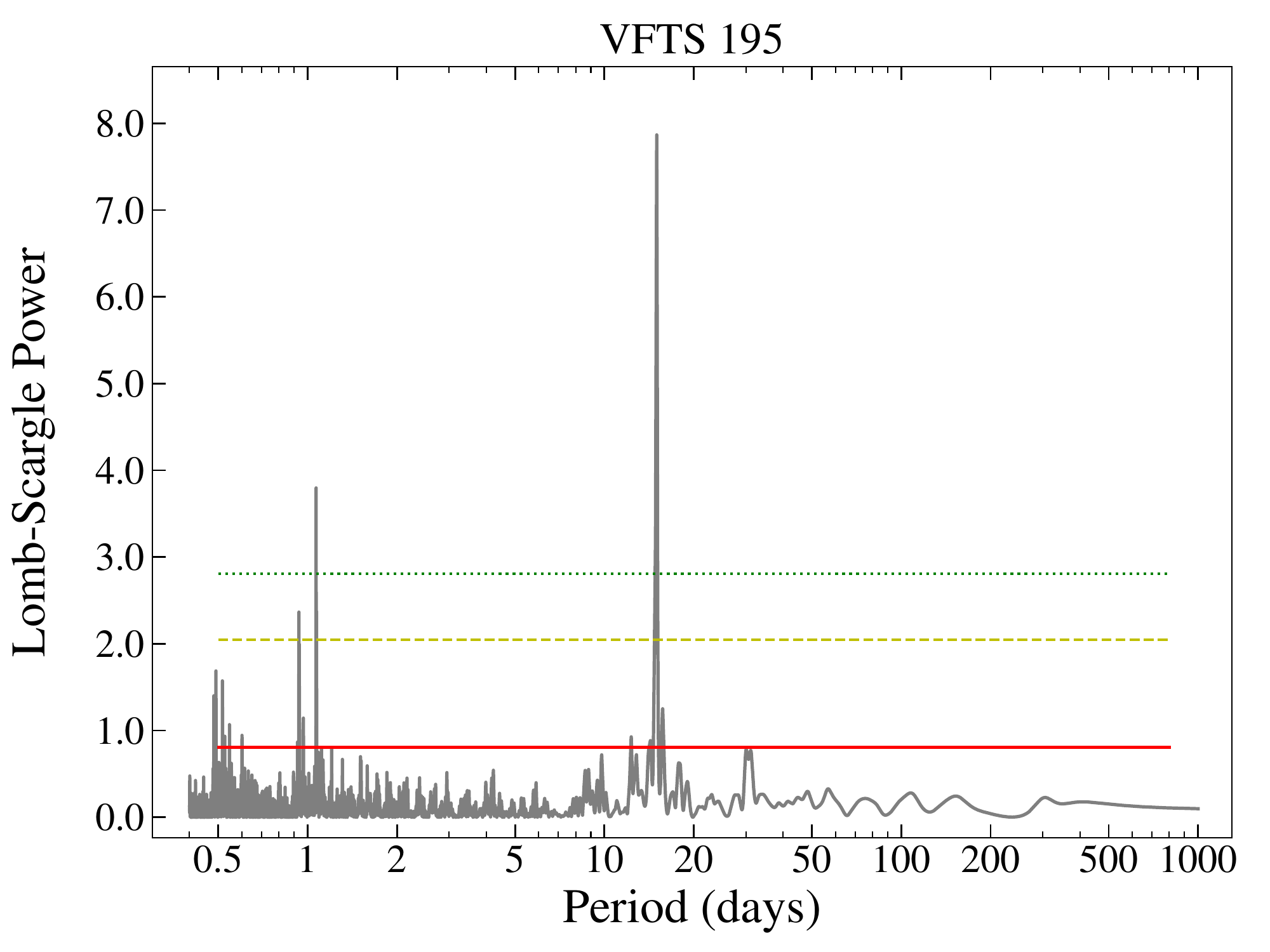}\hfill
    \includegraphics[width=0.31\textwidth]{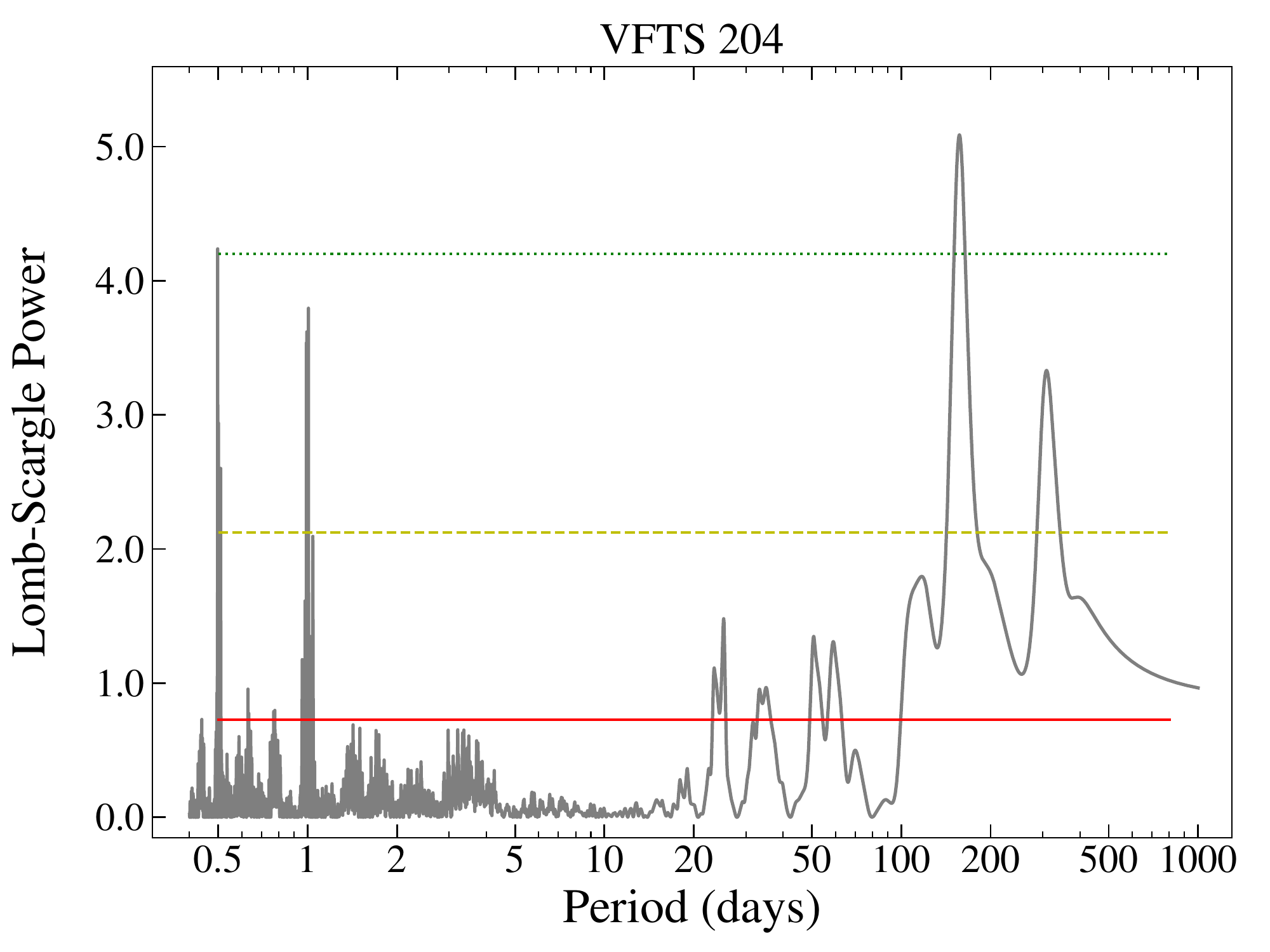}\hfill
    \includegraphics[width=0.31\textwidth]{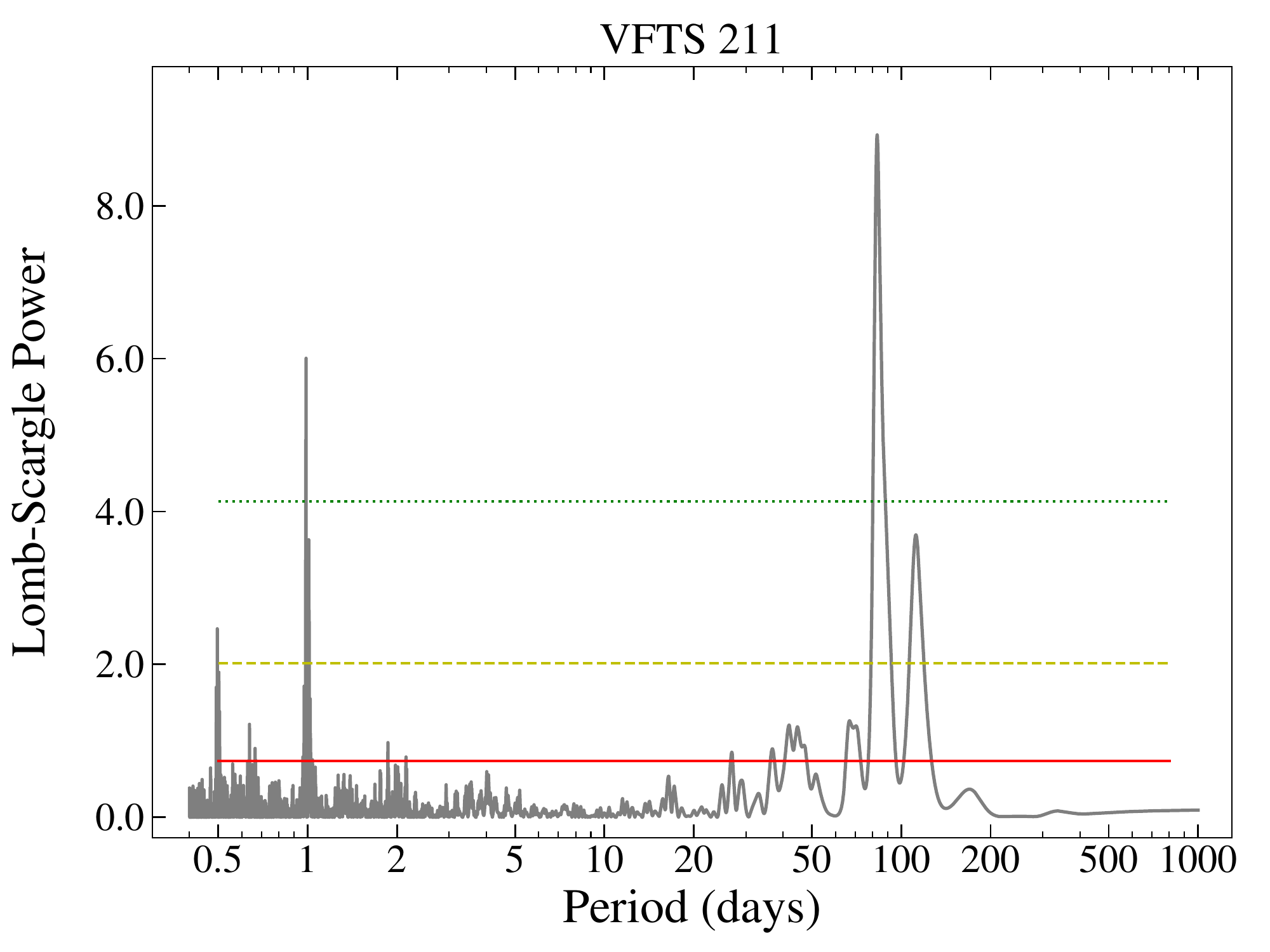}\hfill
    \includegraphics[width=0.31\textwidth]{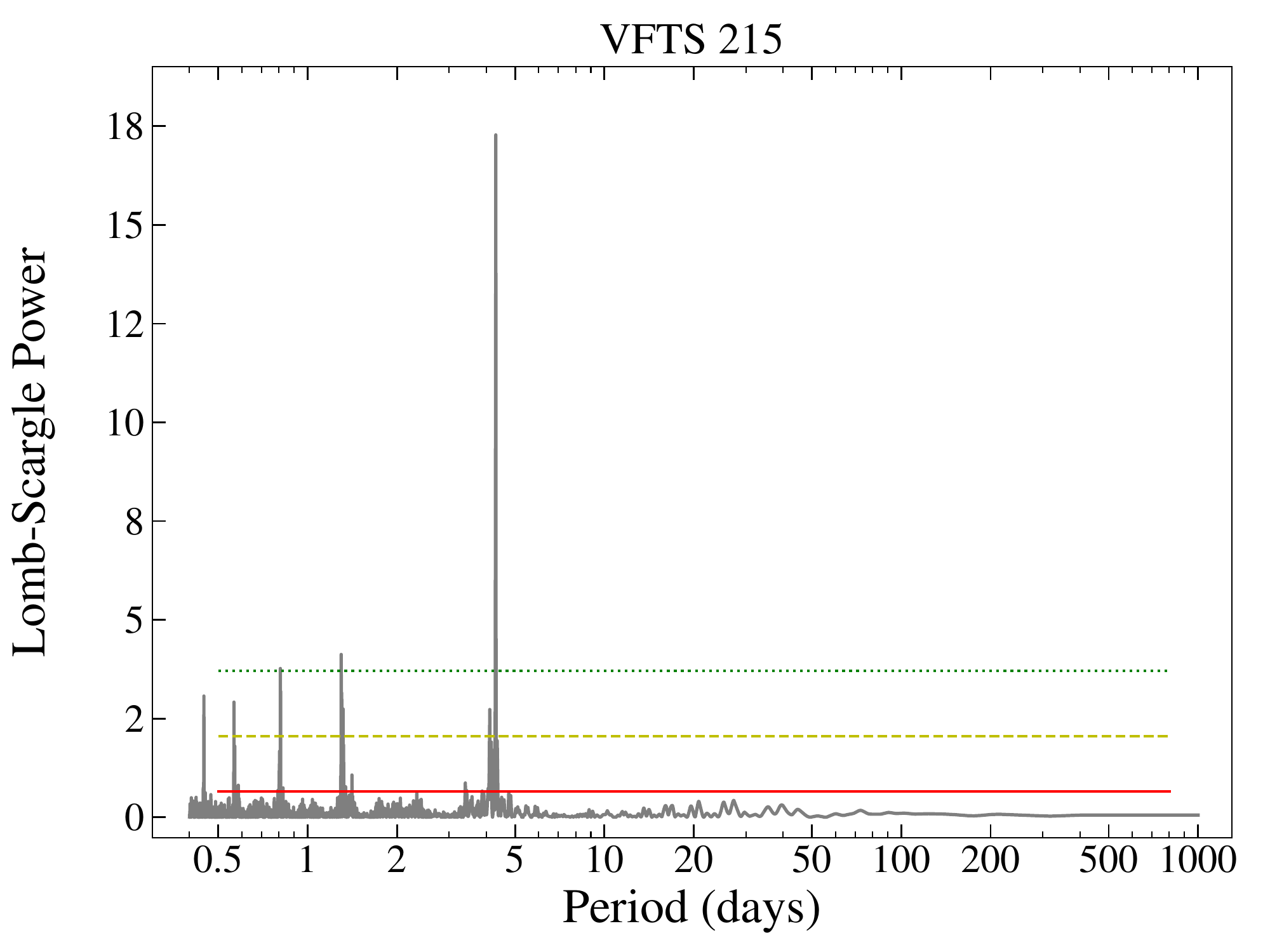}\hfill
    \includegraphics[width=0.31\textwidth]{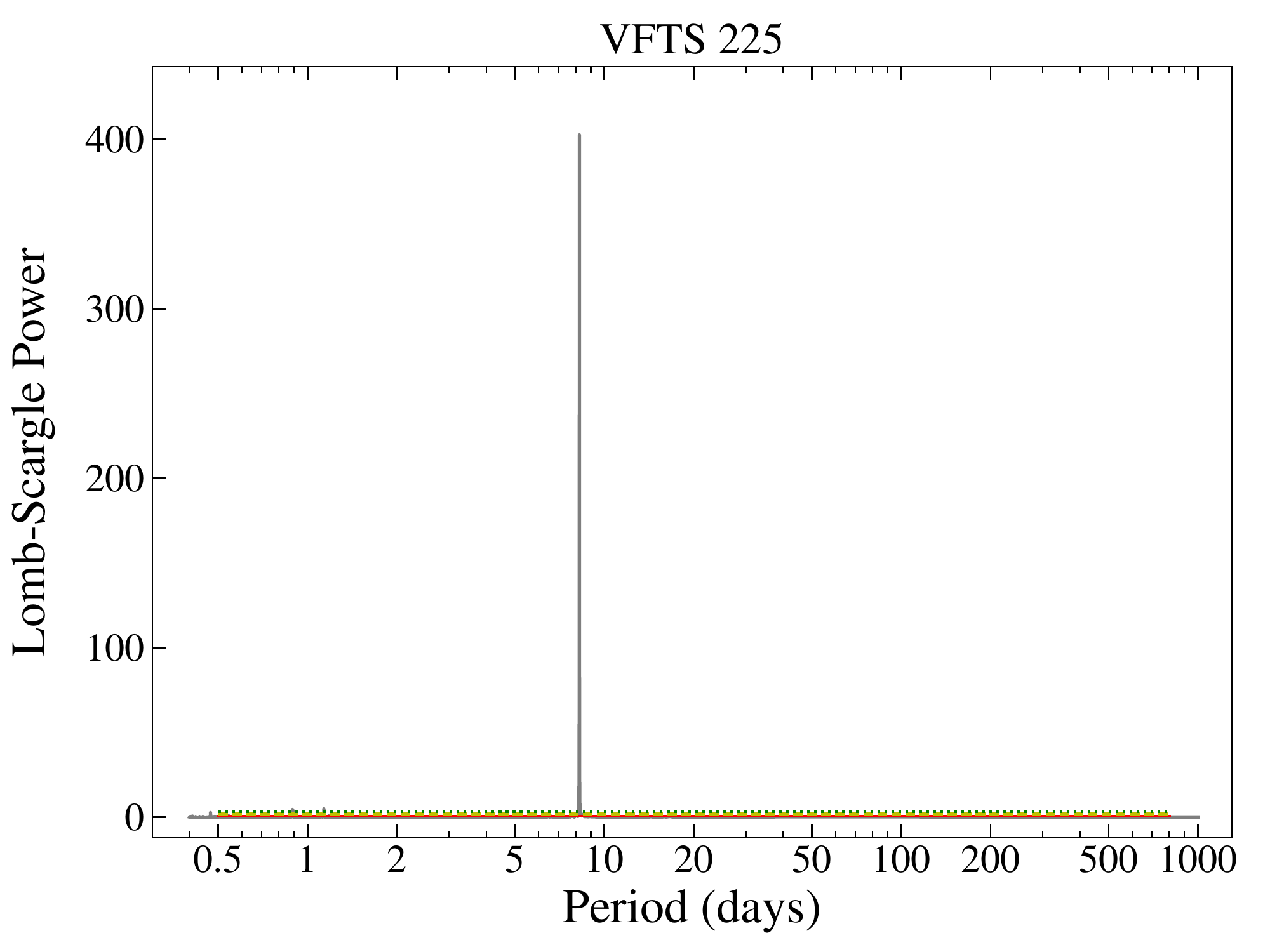}\hfill
    \caption{Periodograms of the SB1 systems.}\label{figAp:LSsb1}
\end{myfloat}

\begin{myfloat}
\ContinuedFloat
    \centering
    \includegraphics[width=0.31\textwidth]{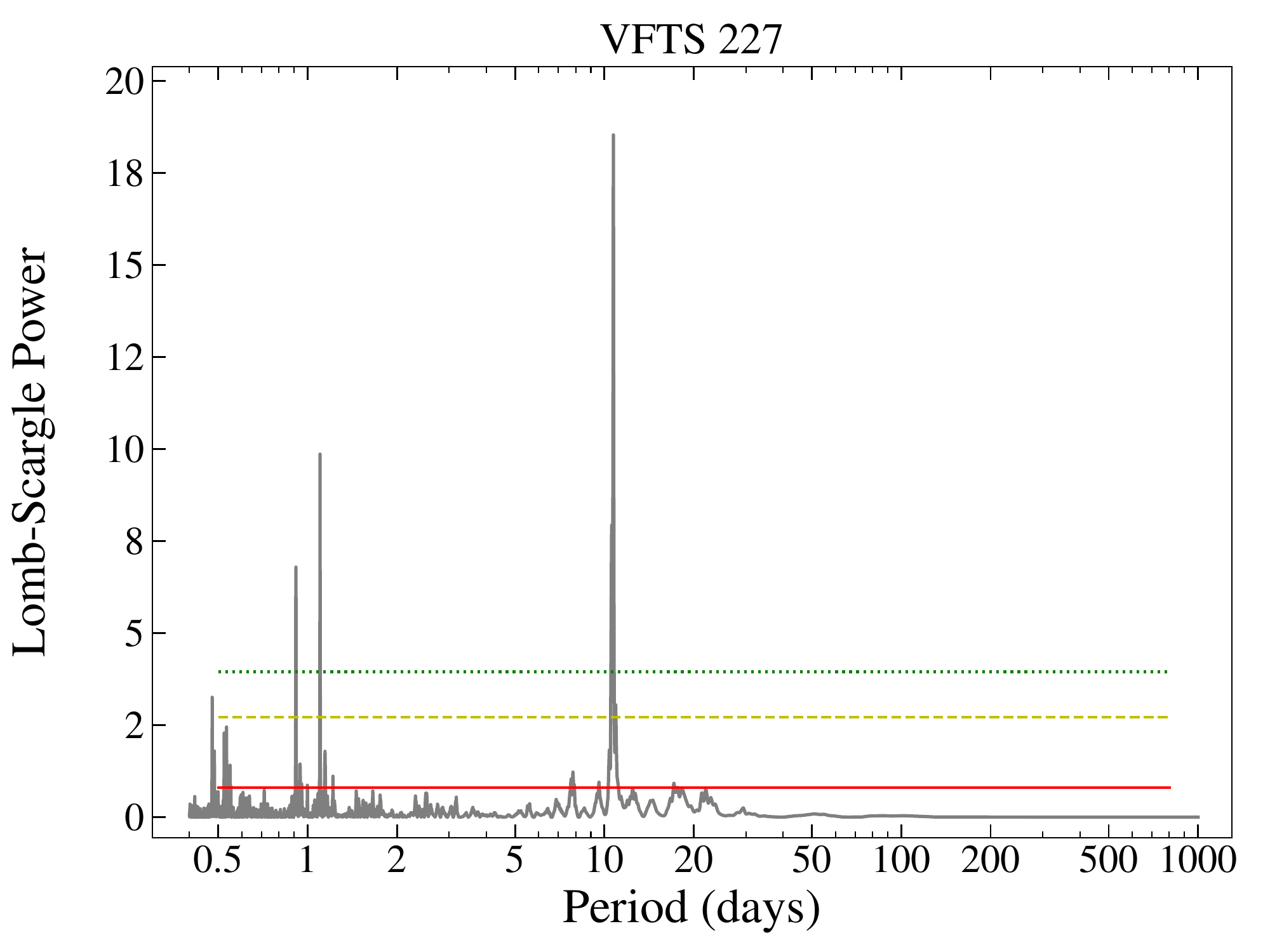}\hfill
    \includegraphics[width=0.31\textwidth]{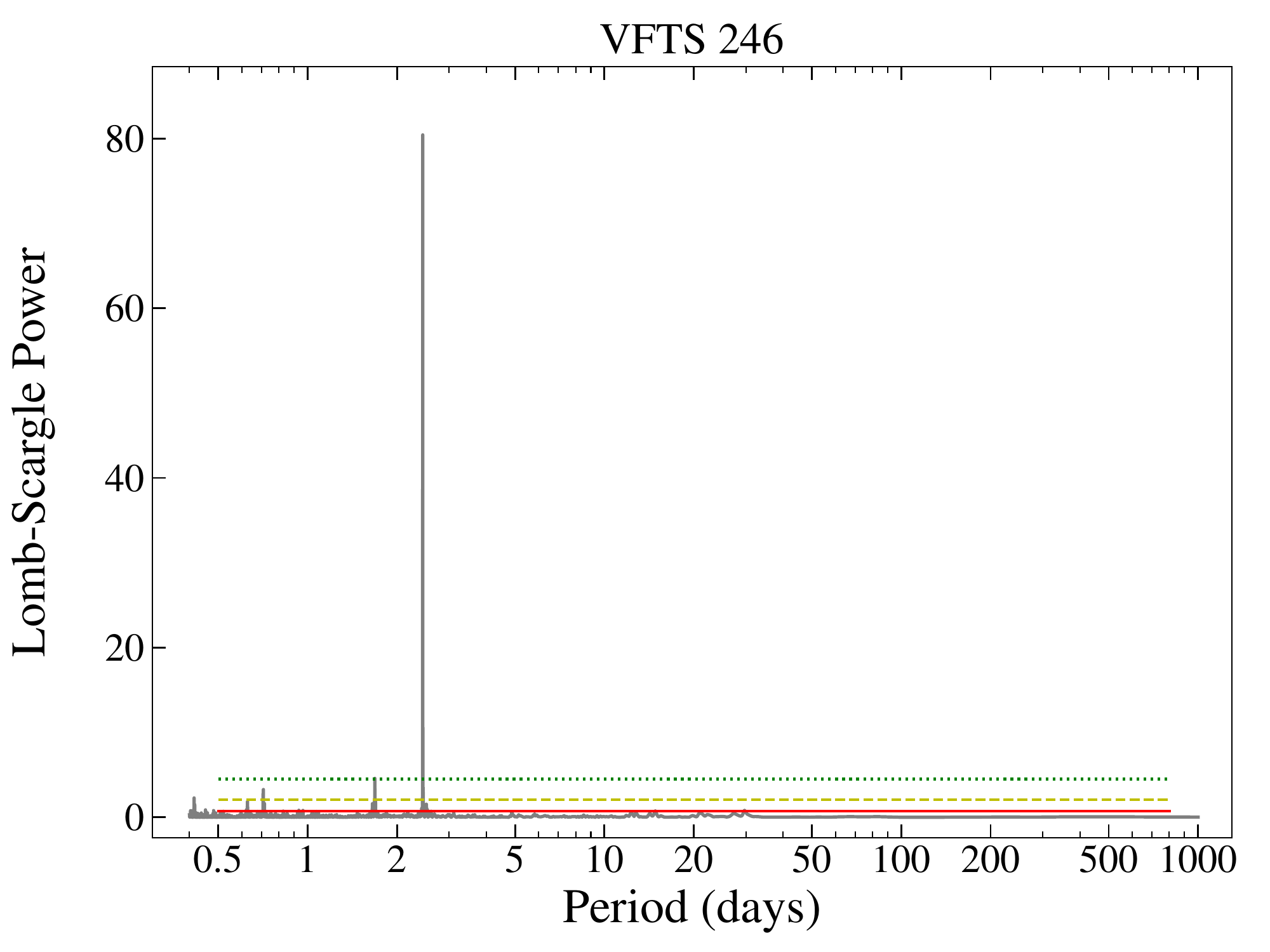}\hfill
    \includegraphics[width=0.31\textwidth]{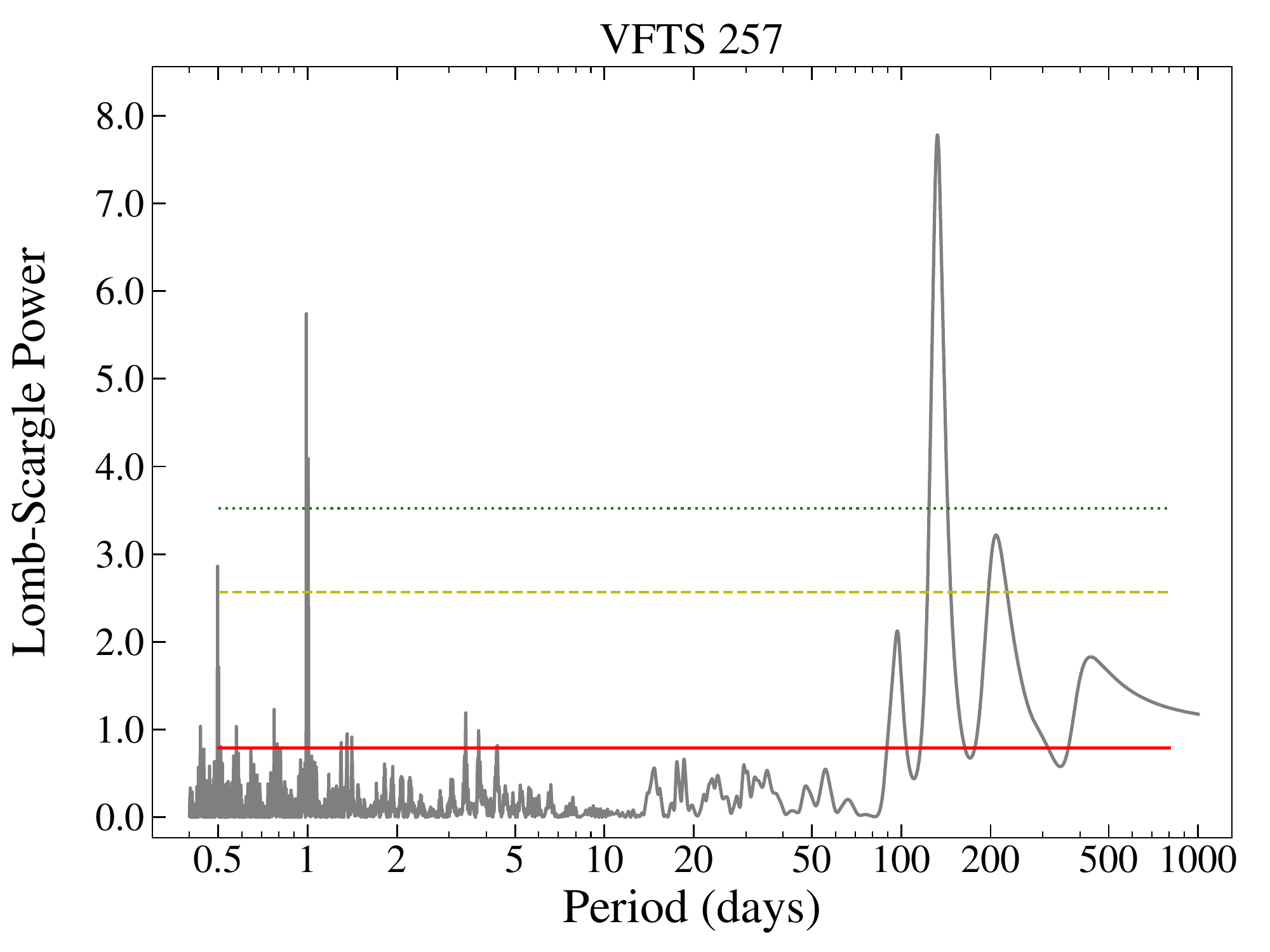}\hfill
    \includegraphics[width=0.31\textwidth]{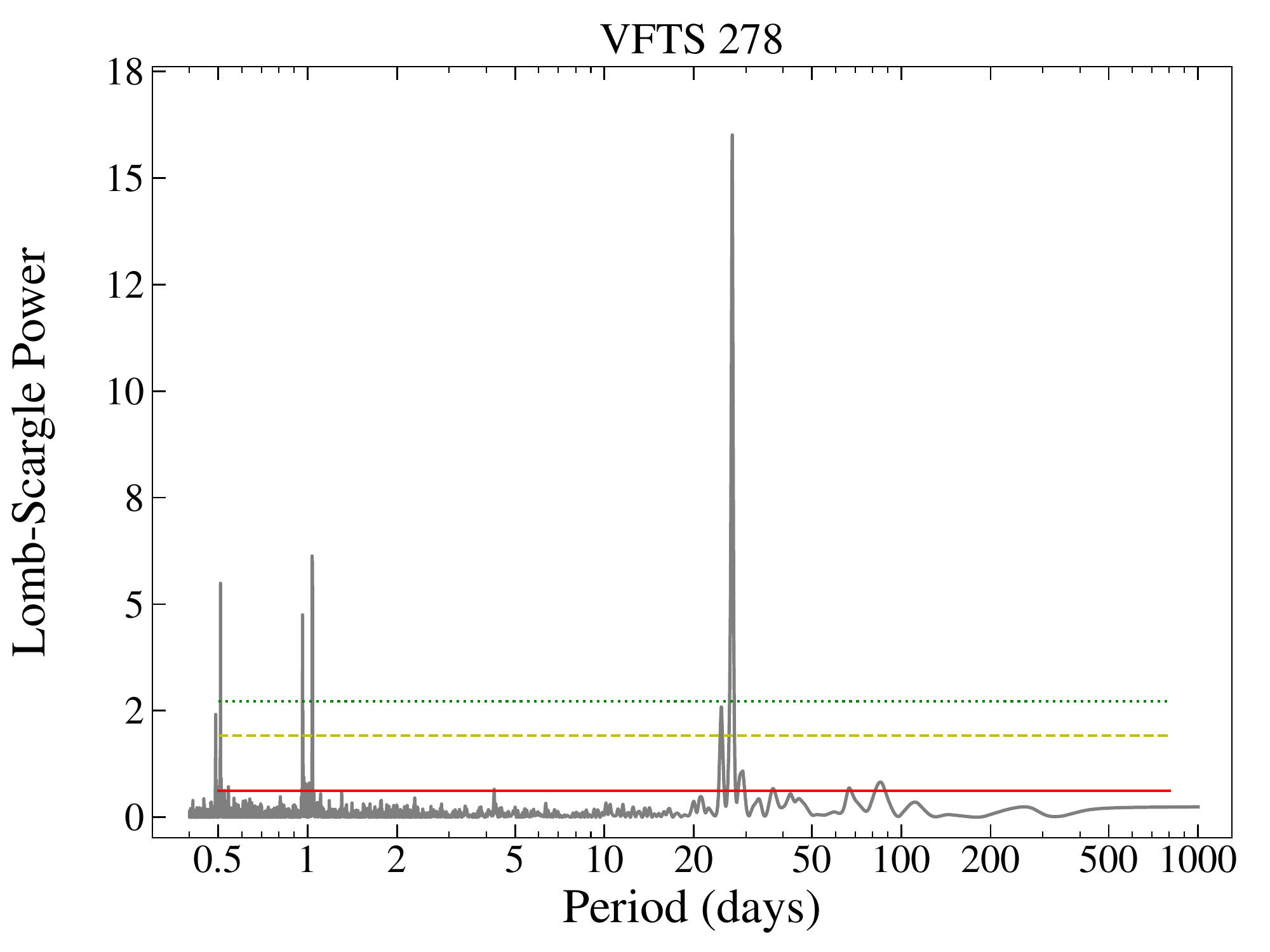}\hfill
    \includegraphics[width=0.31\textwidth]{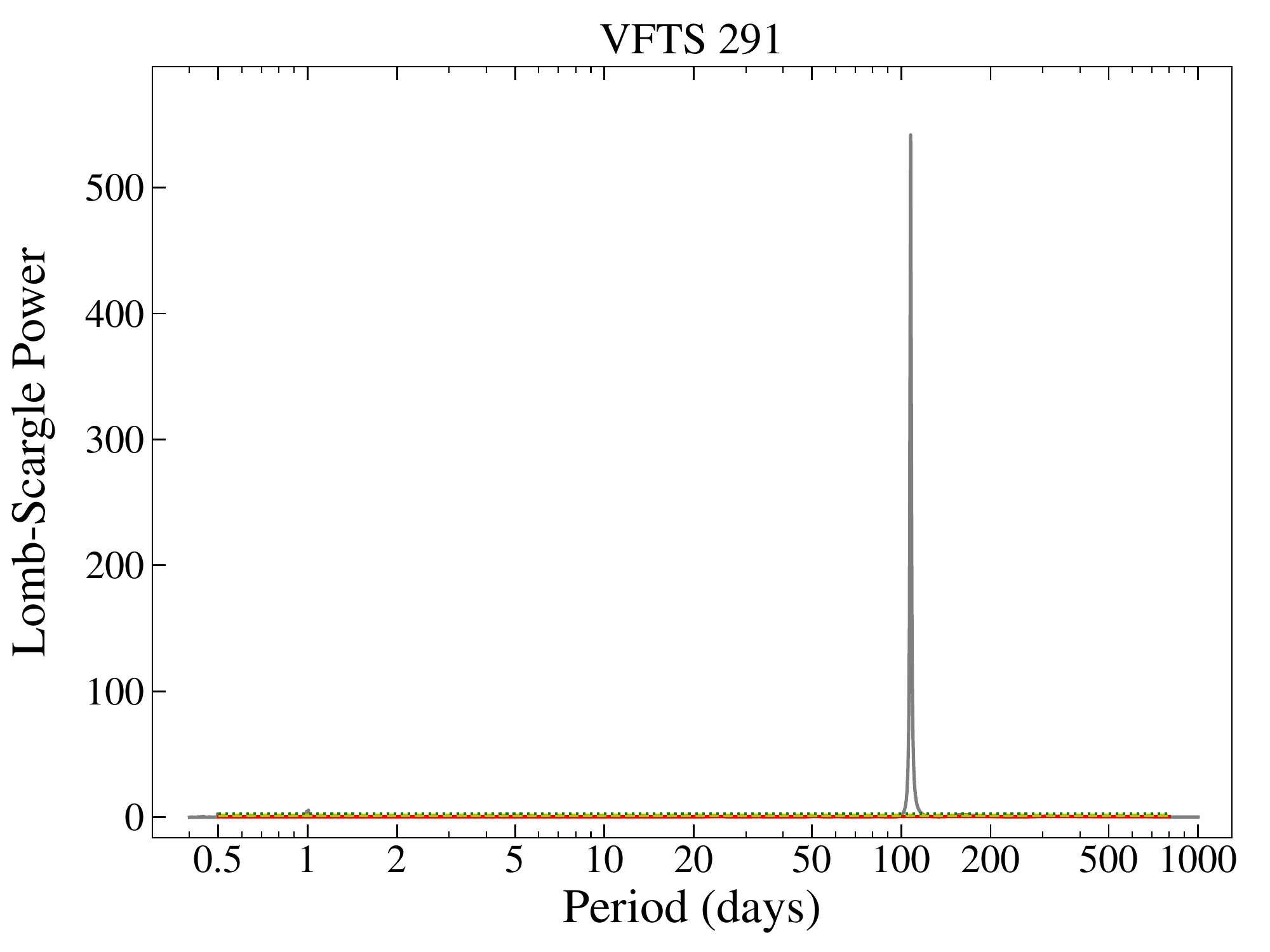}\hfill
    \includegraphics[width=0.31\textwidth]{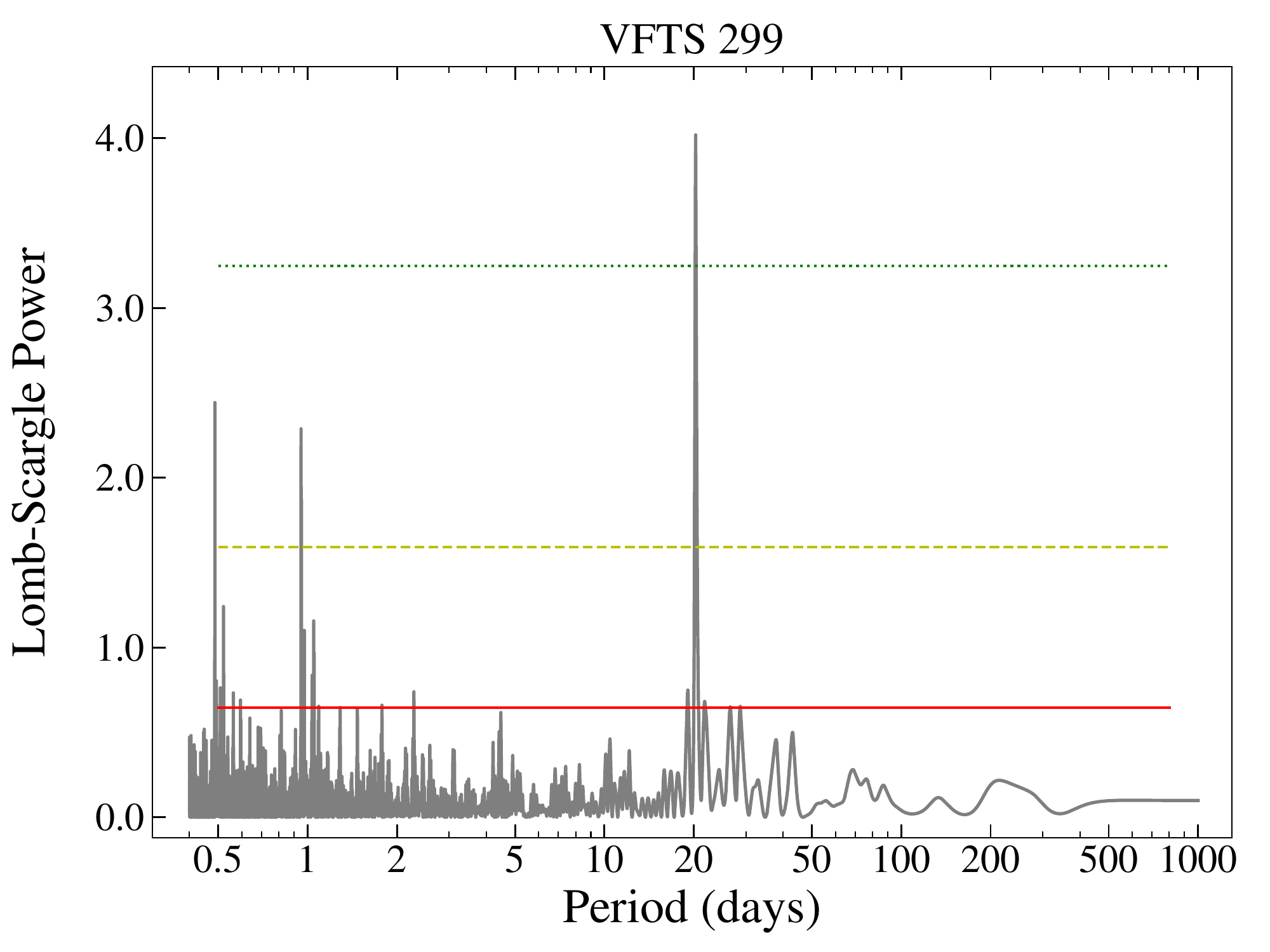}\hfill
    \includegraphics[width=0.31\textwidth]{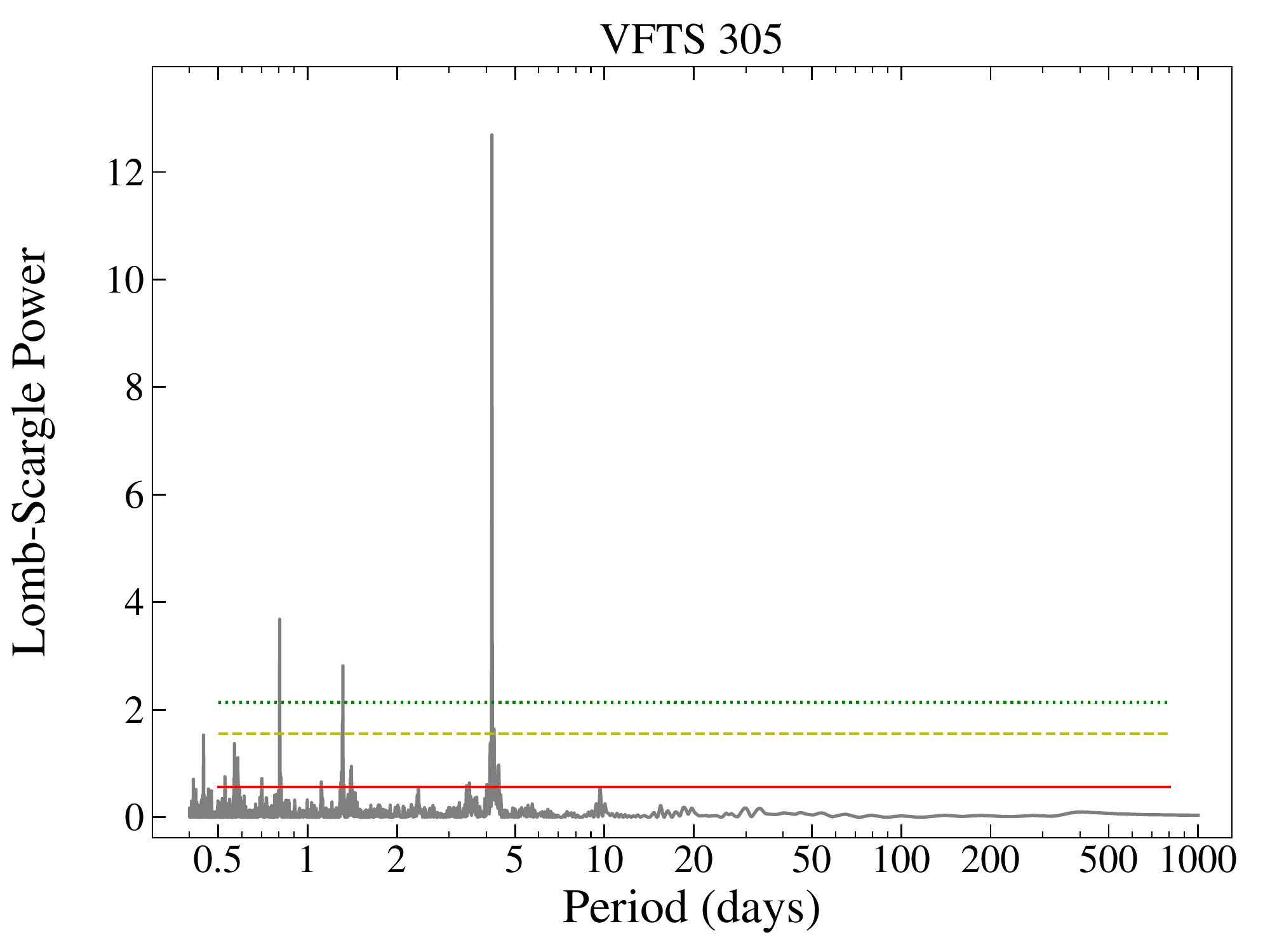}\hfill
    \includegraphics[width=0.31\textwidth]{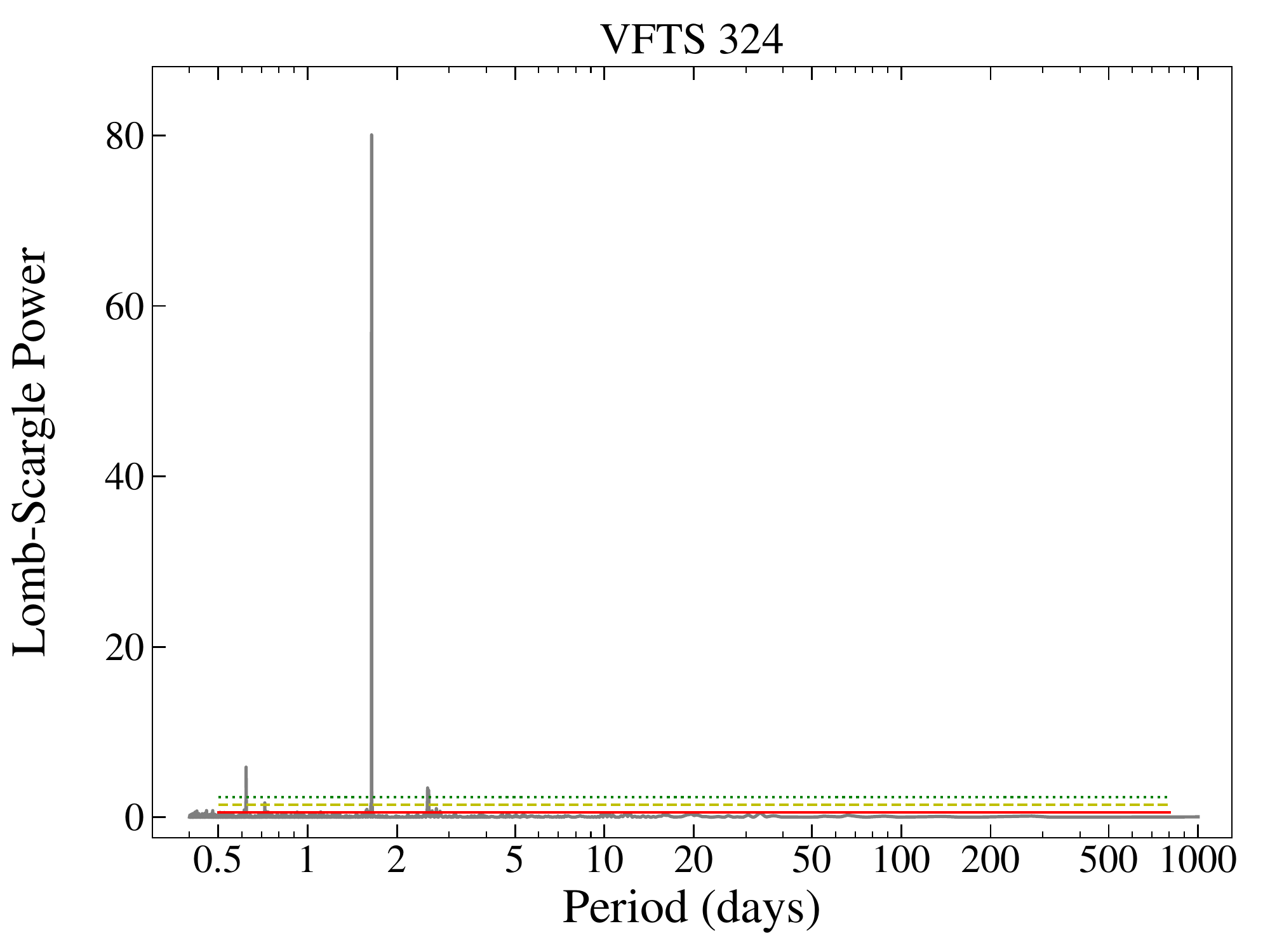}\hfill
    \includegraphics[width=0.31\textwidth]{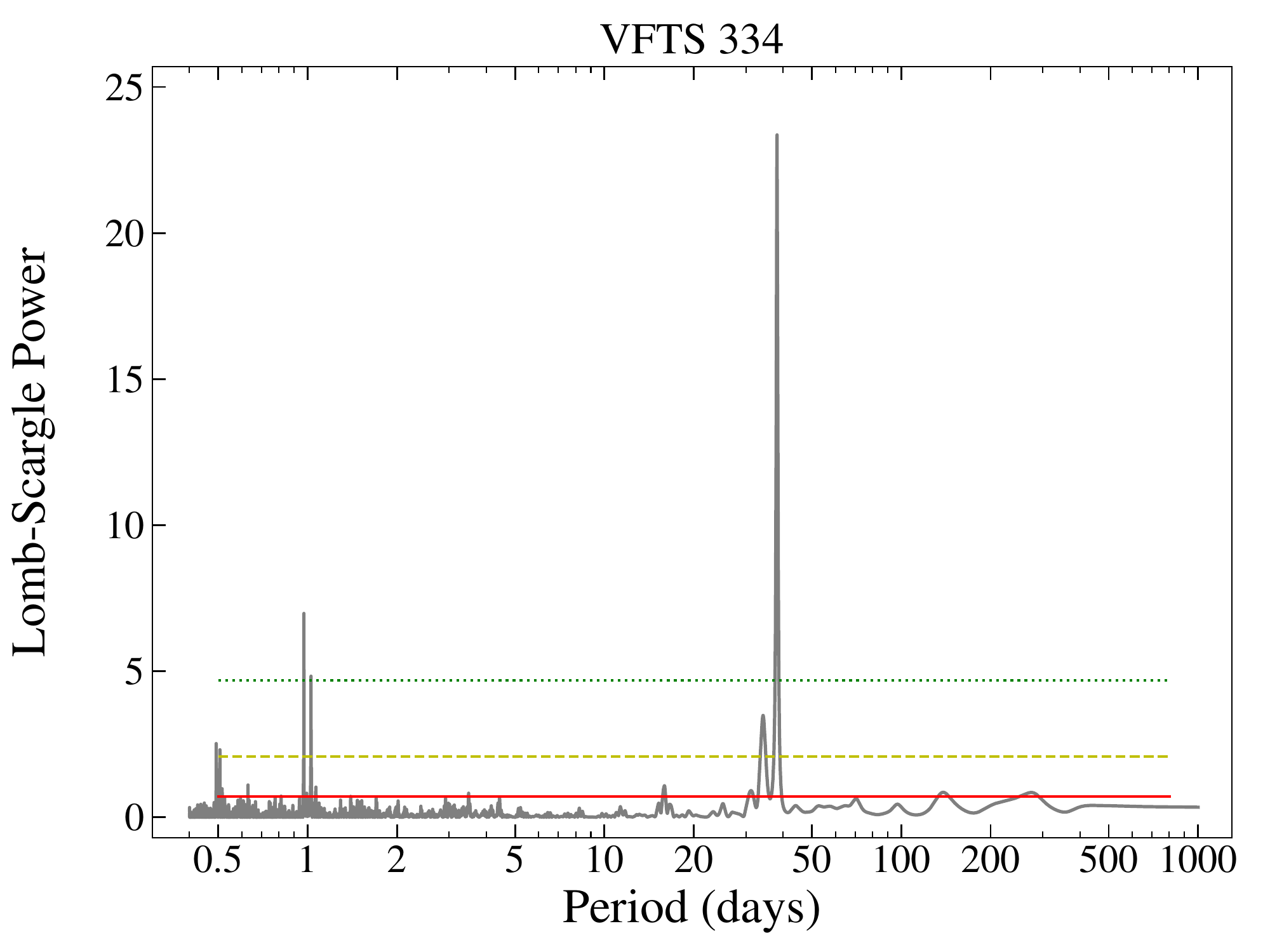}\hfill
    \includegraphics[width=0.31\textwidth]{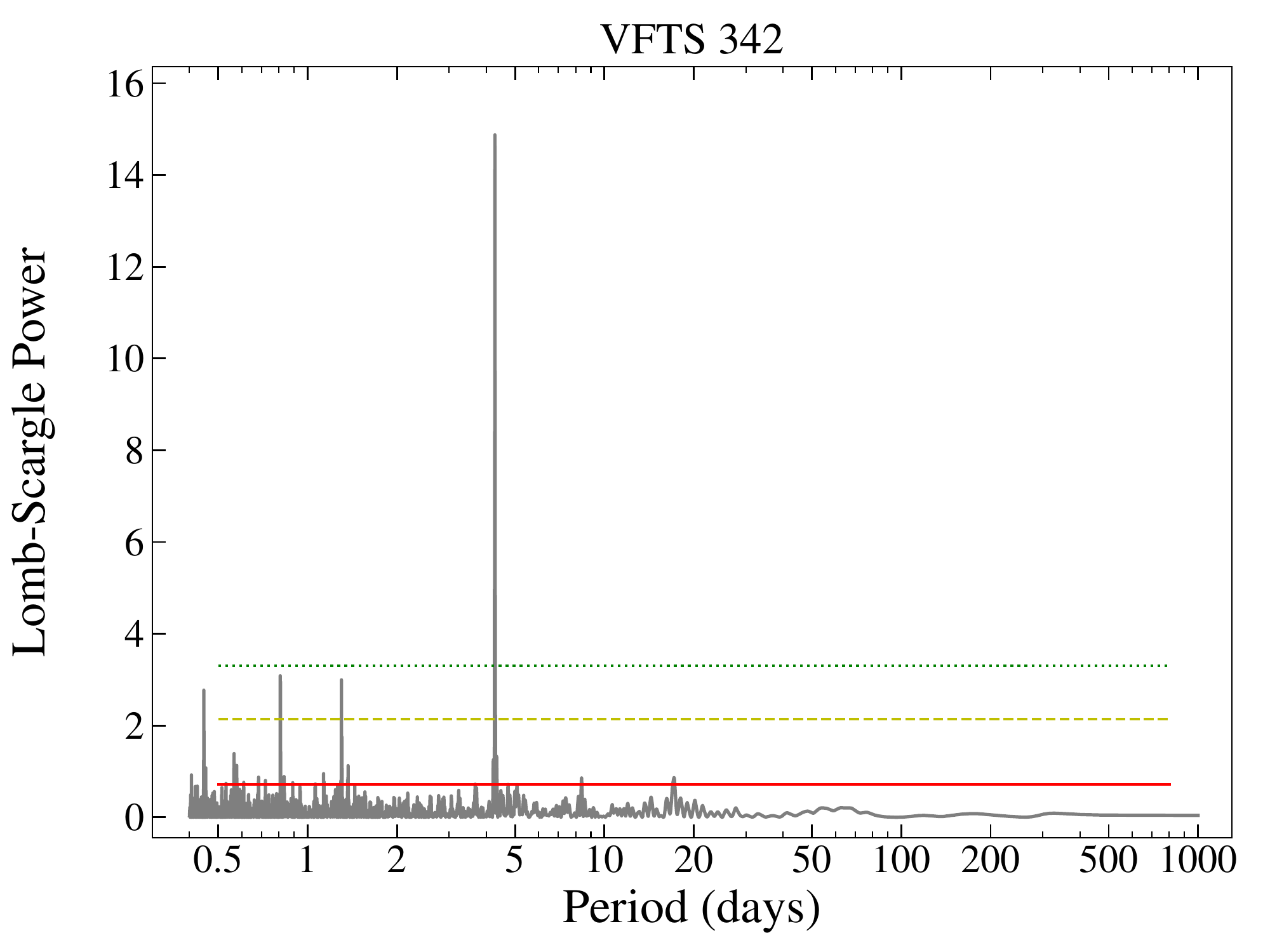}\hfill
    \includegraphics[width=0.31\textwidth]{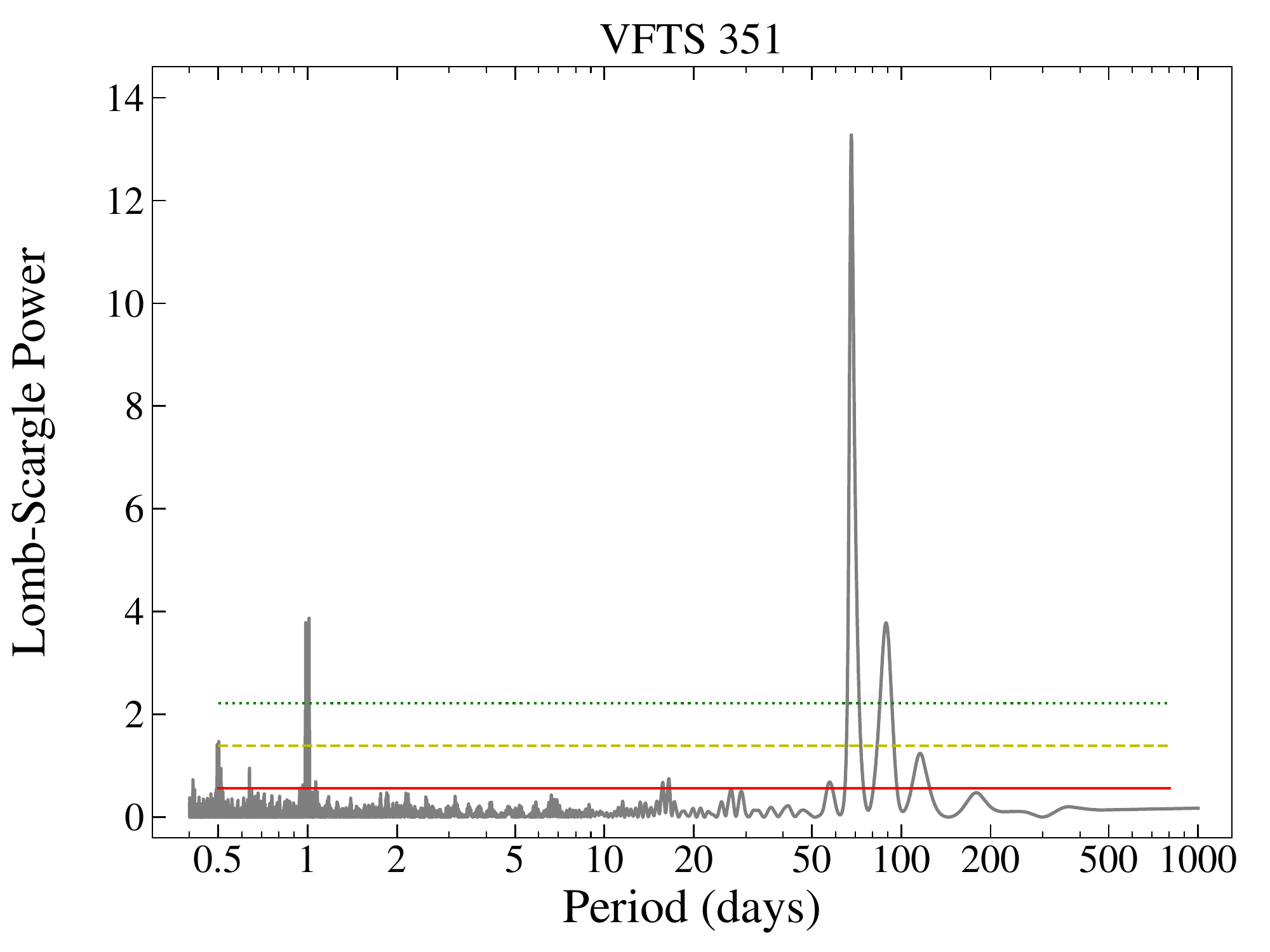}\hfill
    \includegraphics[width=0.31\textwidth]{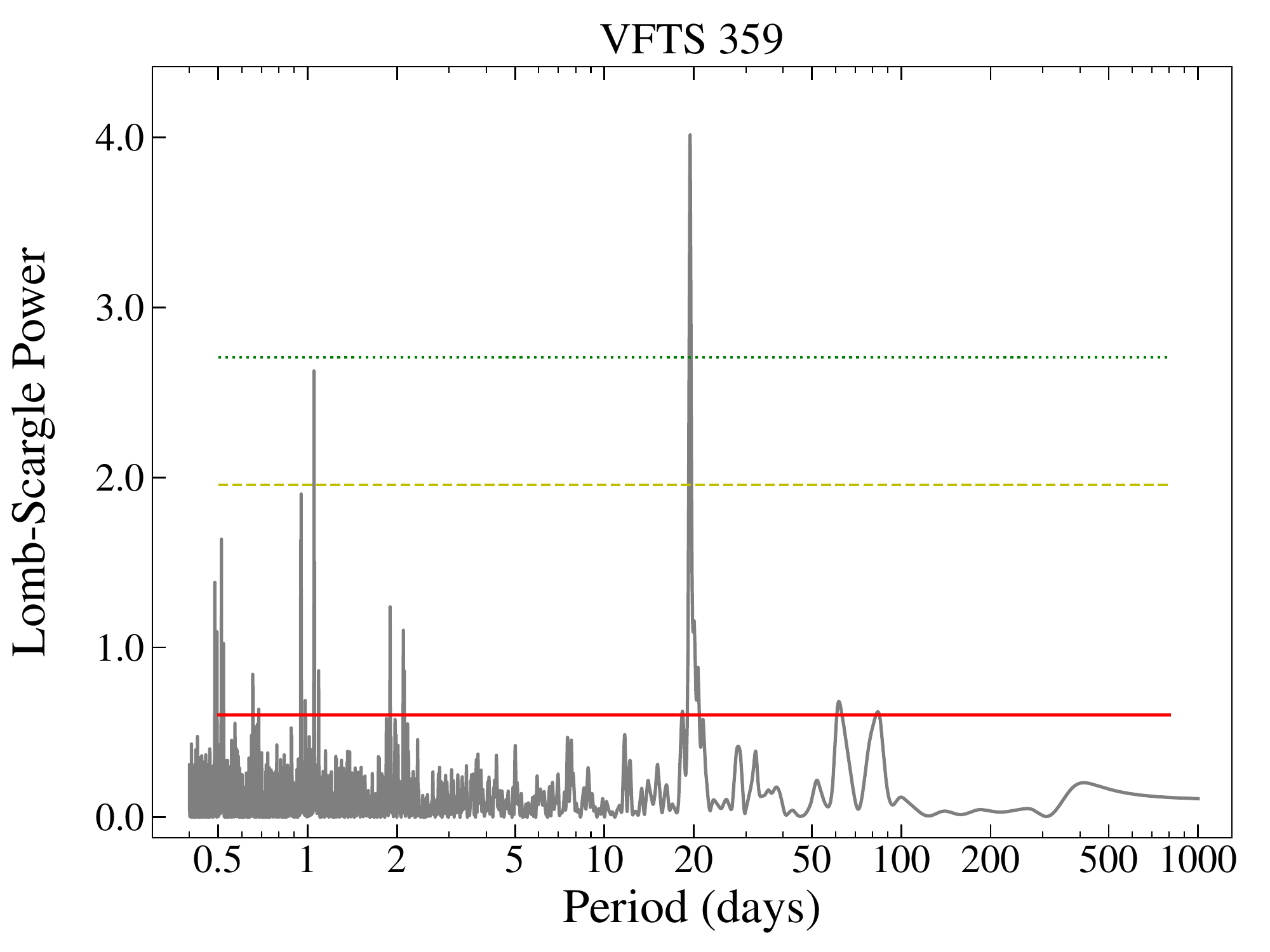}\hfill
    \includegraphics[width=0.31\textwidth]{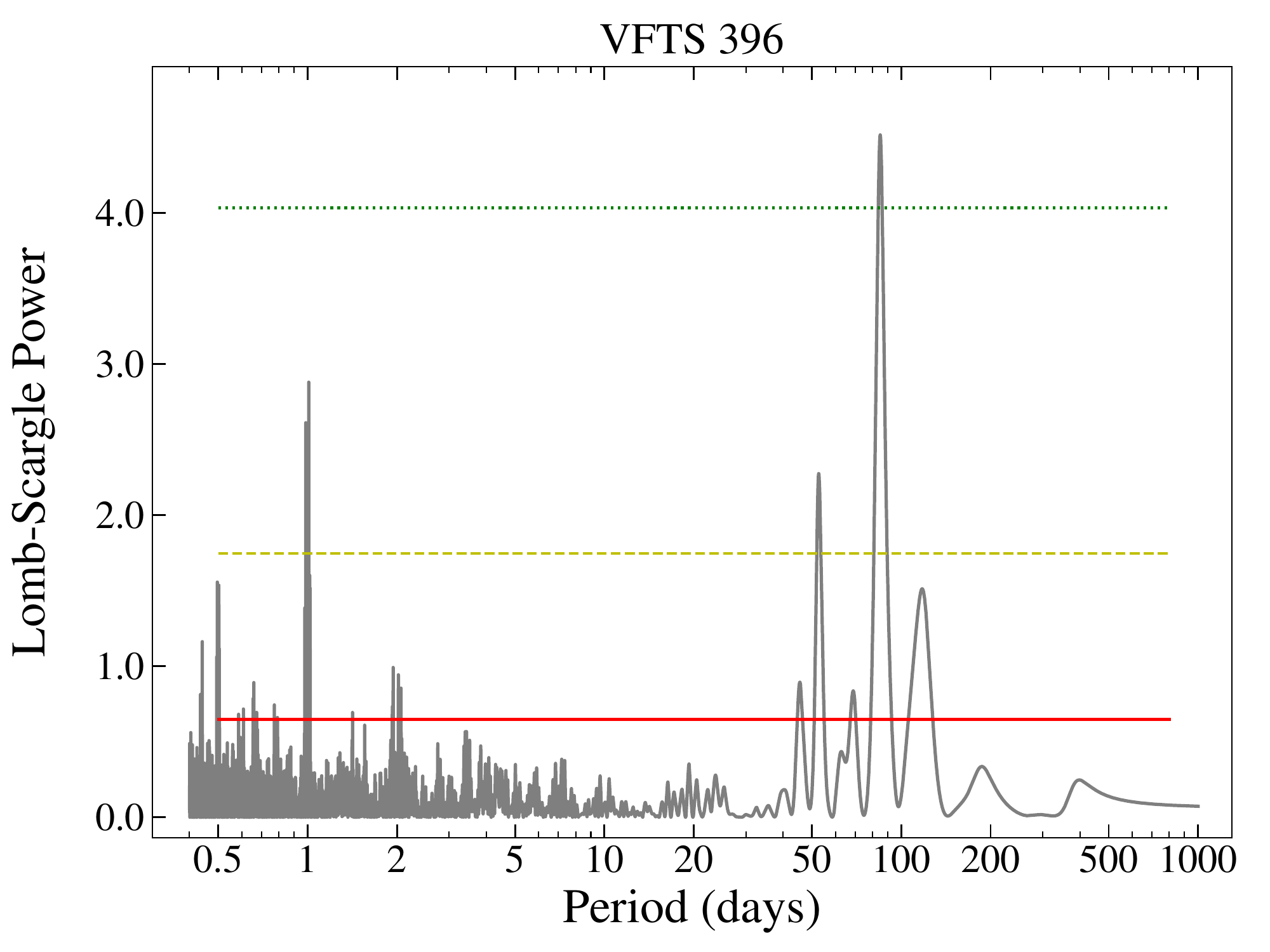}\hfill
    \includegraphics[width=0.31\textwidth]{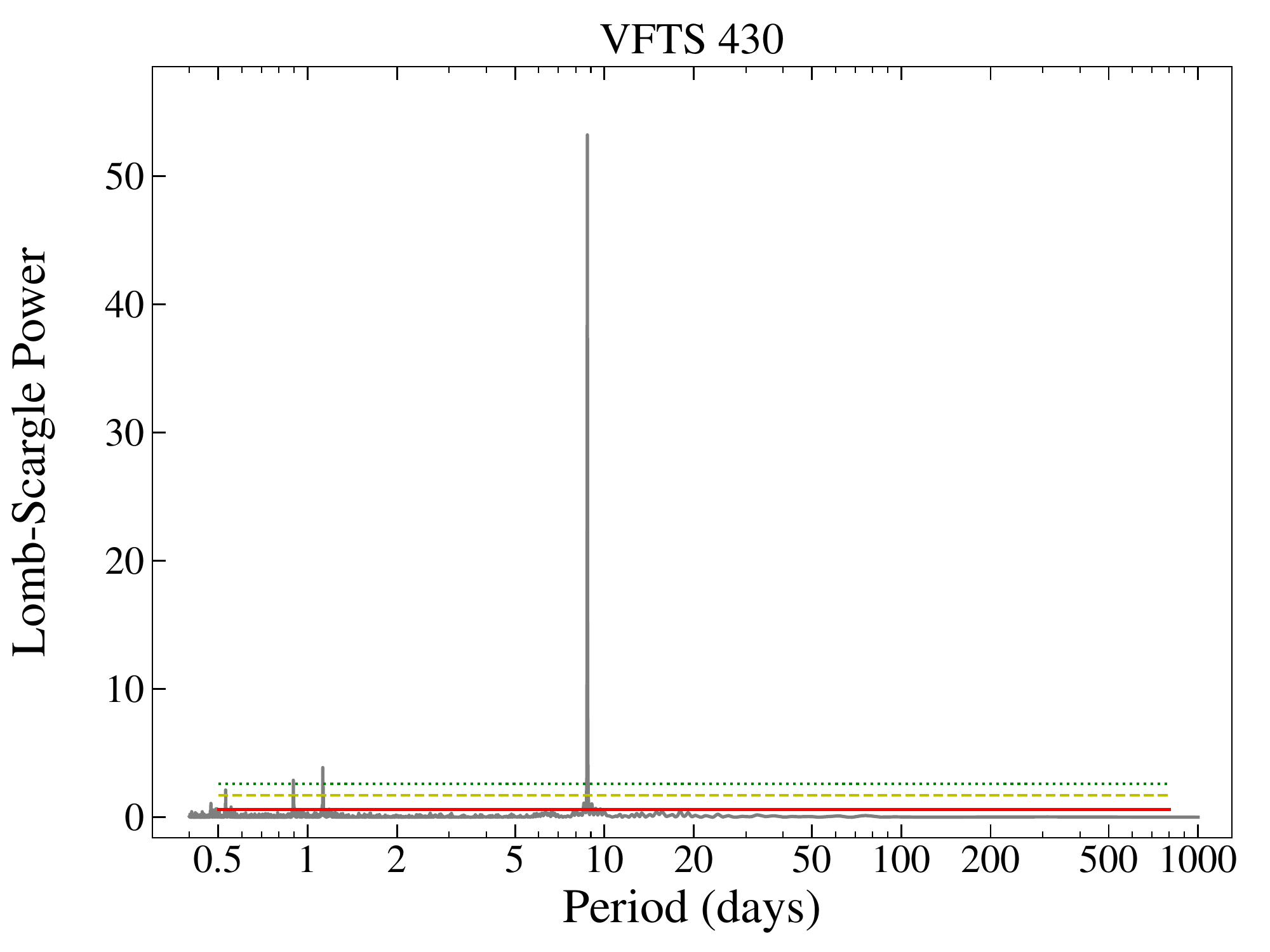}\hfill
    \includegraphics[width=0.31\textwidth]{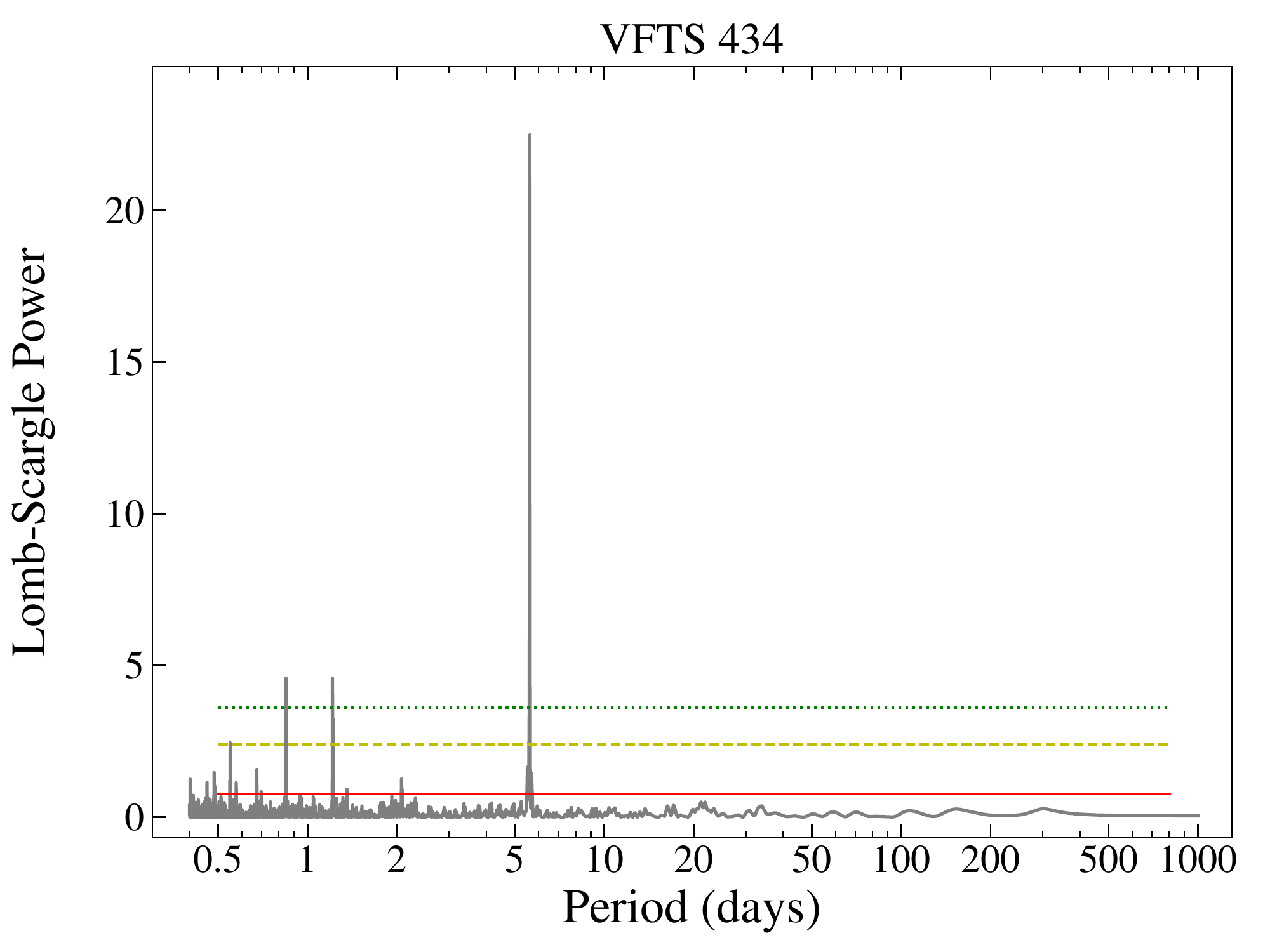}\hfill
    \caption{$-$ \it continued}
\end{myfloat}

\begin{myfloat}
\ContinuedFloat
    \centering
    \includegraphics[width=0.31\textwidth]{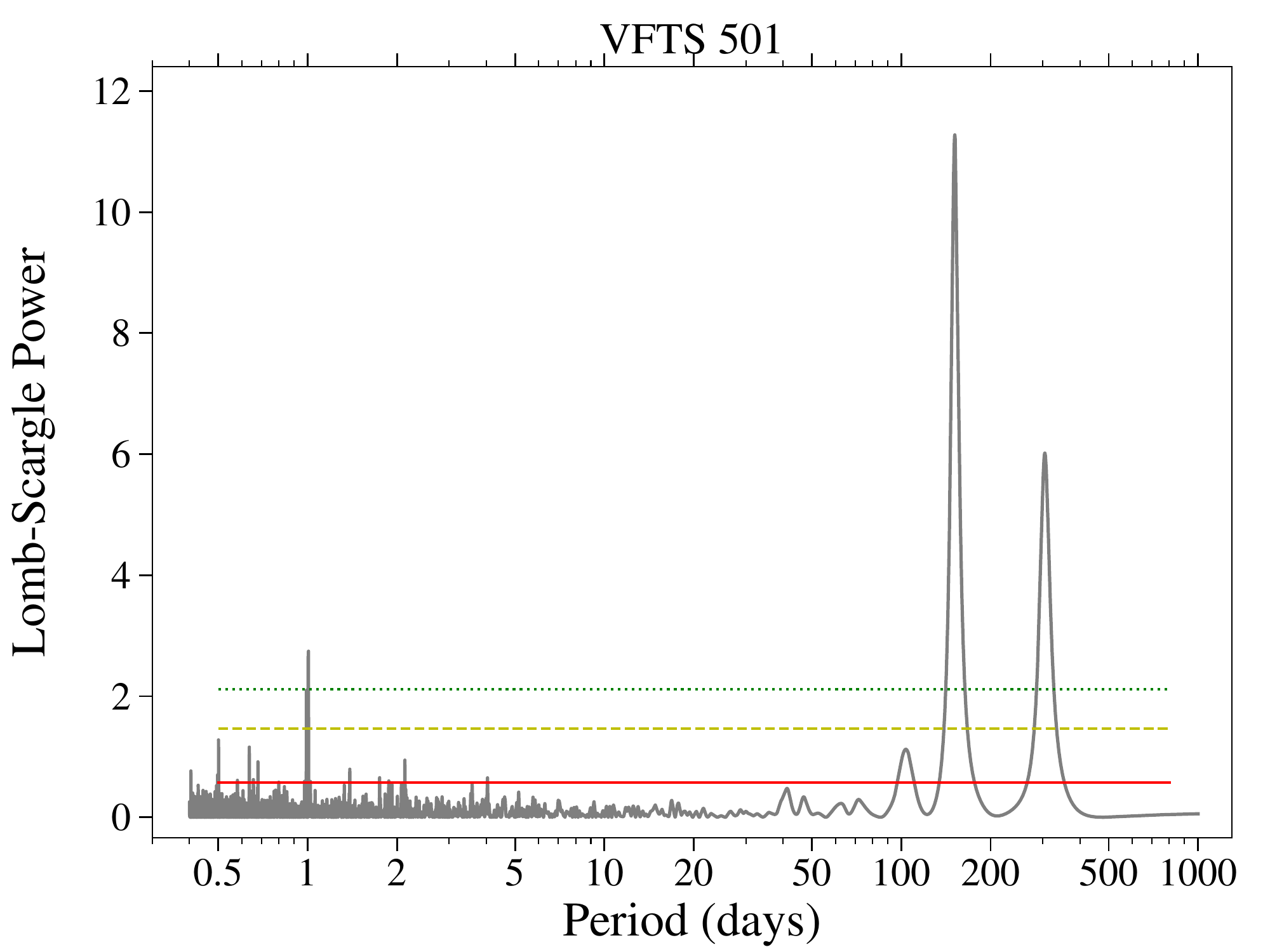}\hfill
    \includegraphics[width=0.31\textwidth]{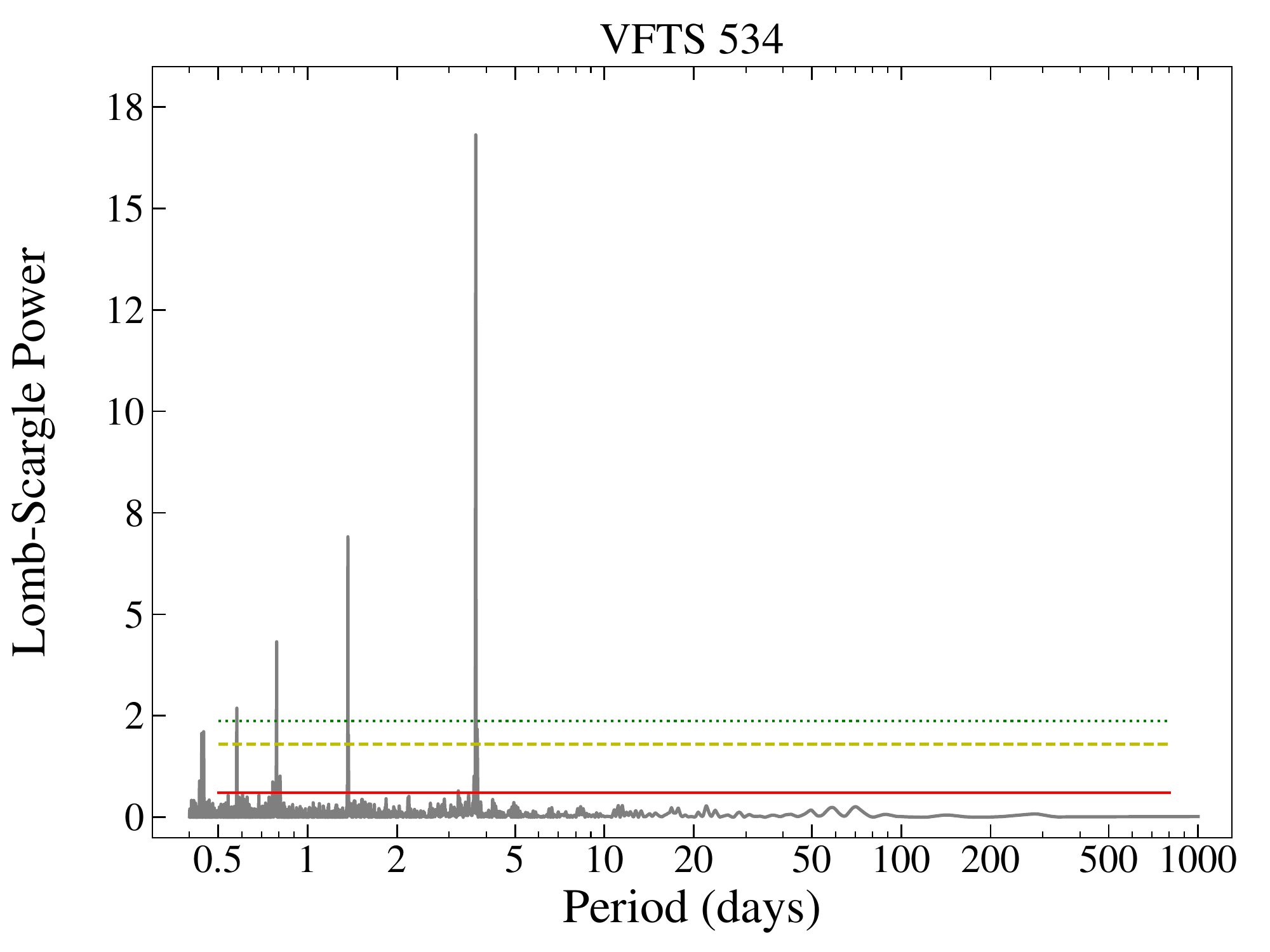}\hfill
    \includegraphics[width=0.31\textwidth]{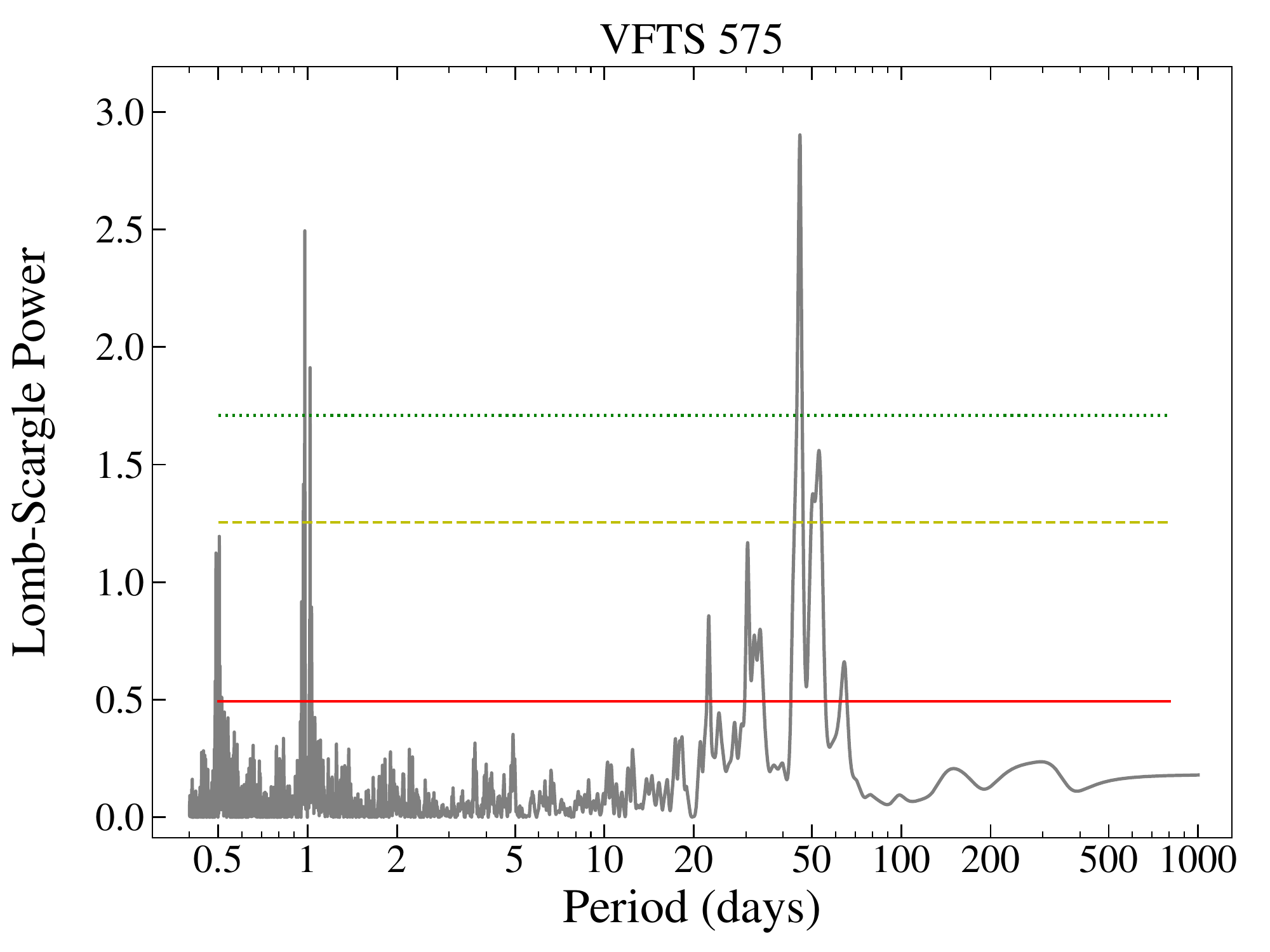}\hfill
    \includegraphics[width=0.31\textwidth]{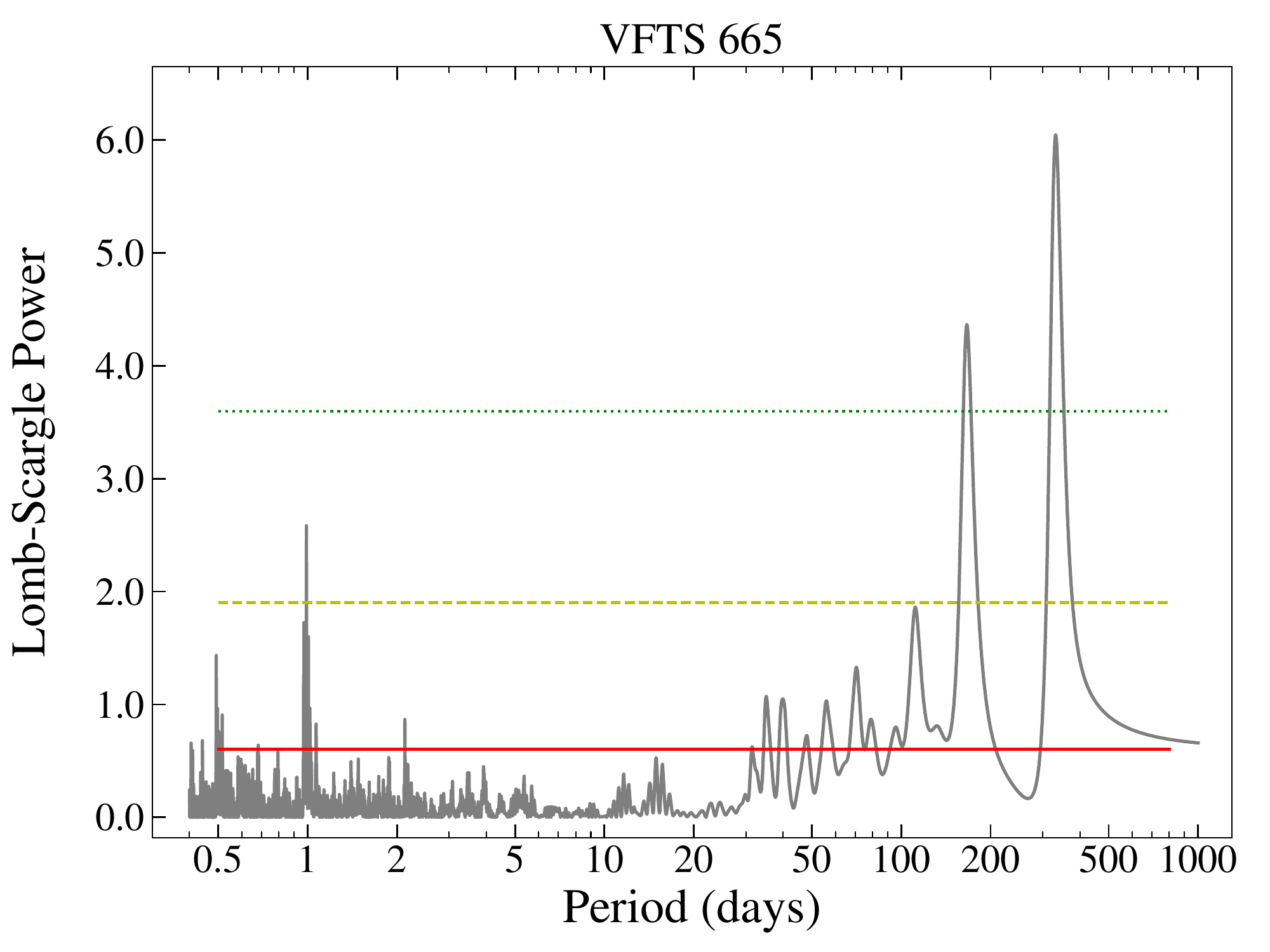}\hfill
    \includegraphics[width=0.31\textwidth]{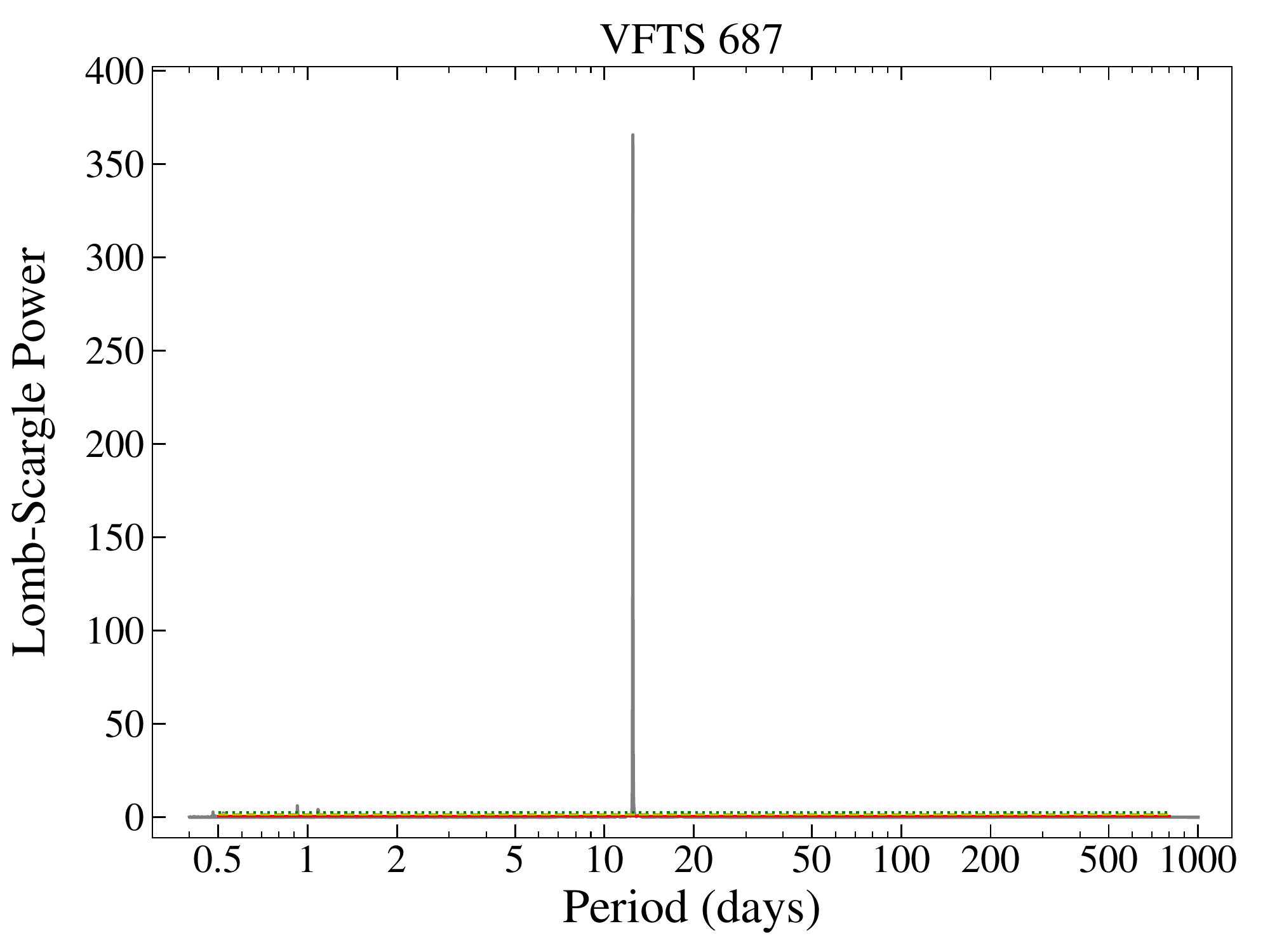}\hfill
    \includegraphics[width=0.31\textwidth]{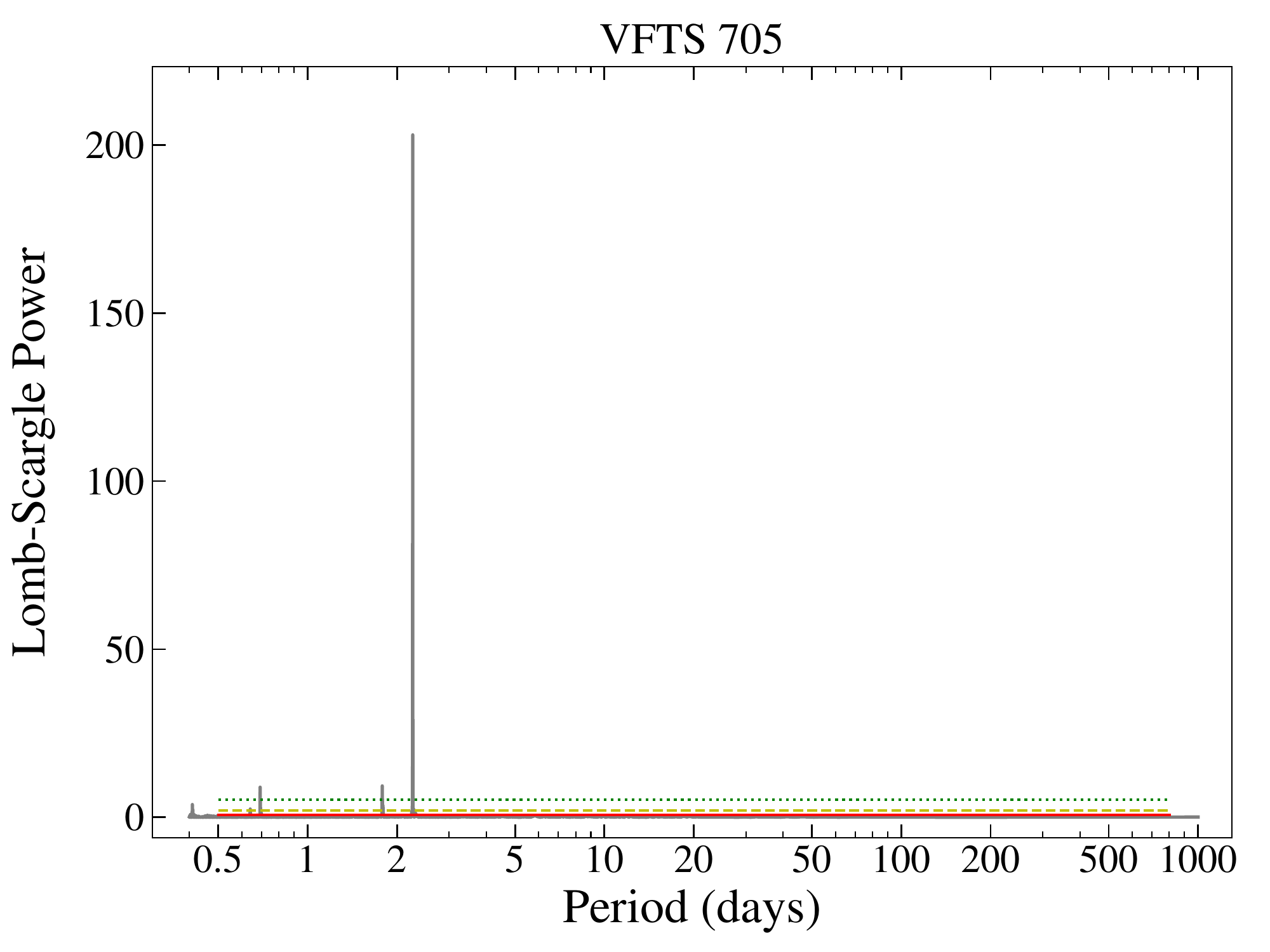}\hfill
    \includegraphics[width=0.31\textwidth]{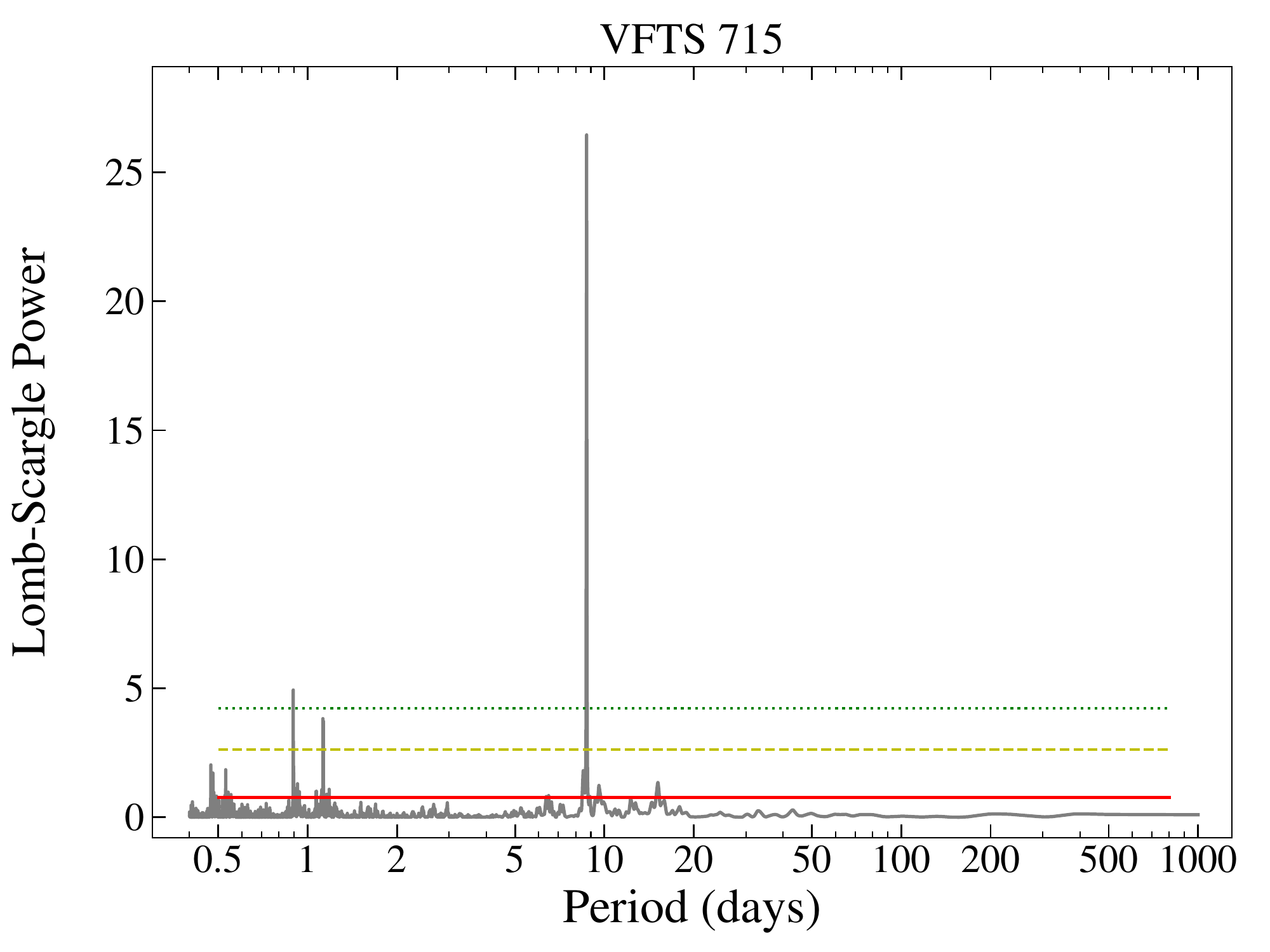}\hfill
    \includegraphics[width=0.31\textwidth]{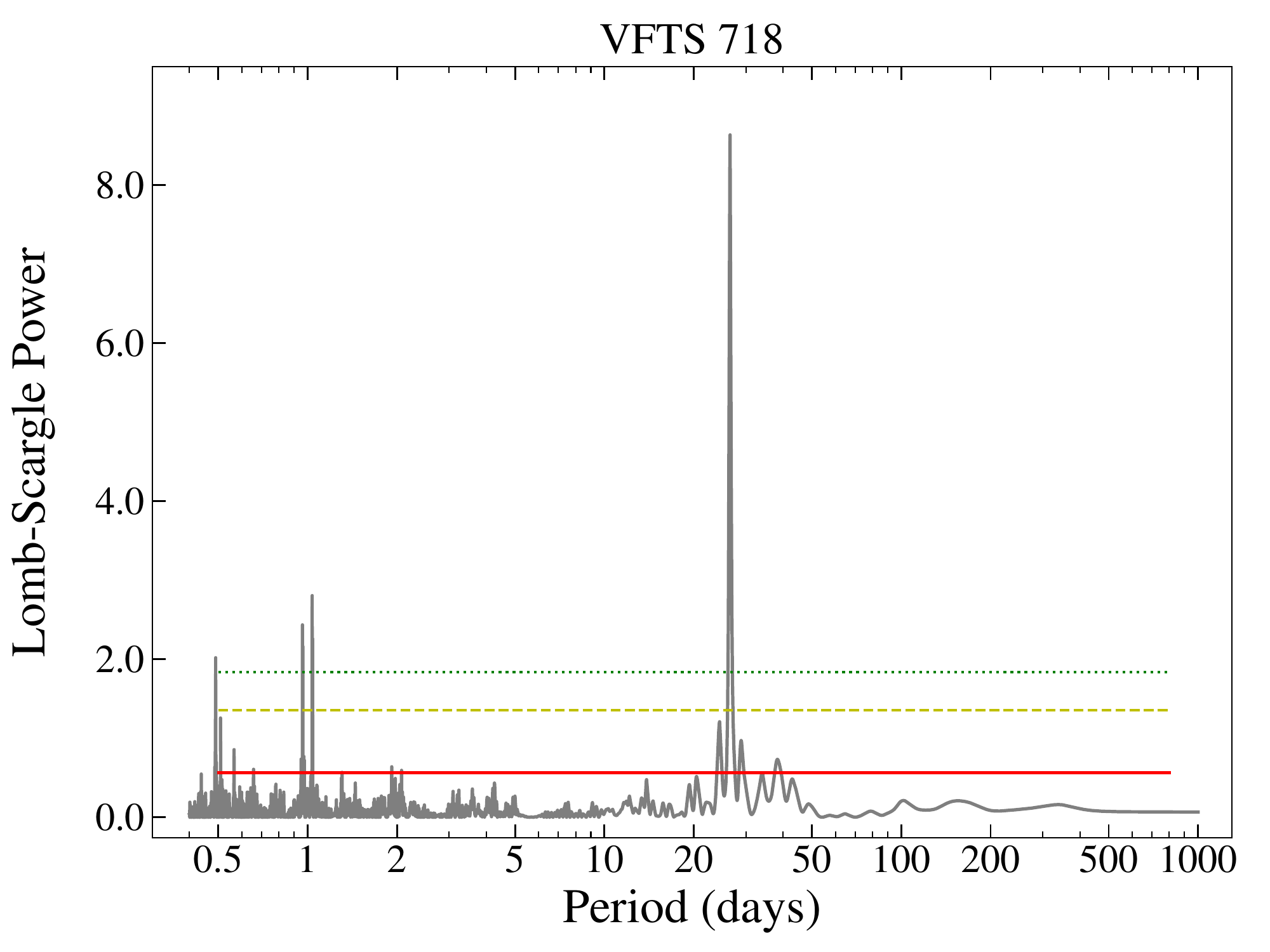}\hfill
    \includegraphics[width=0.31\textwidth]{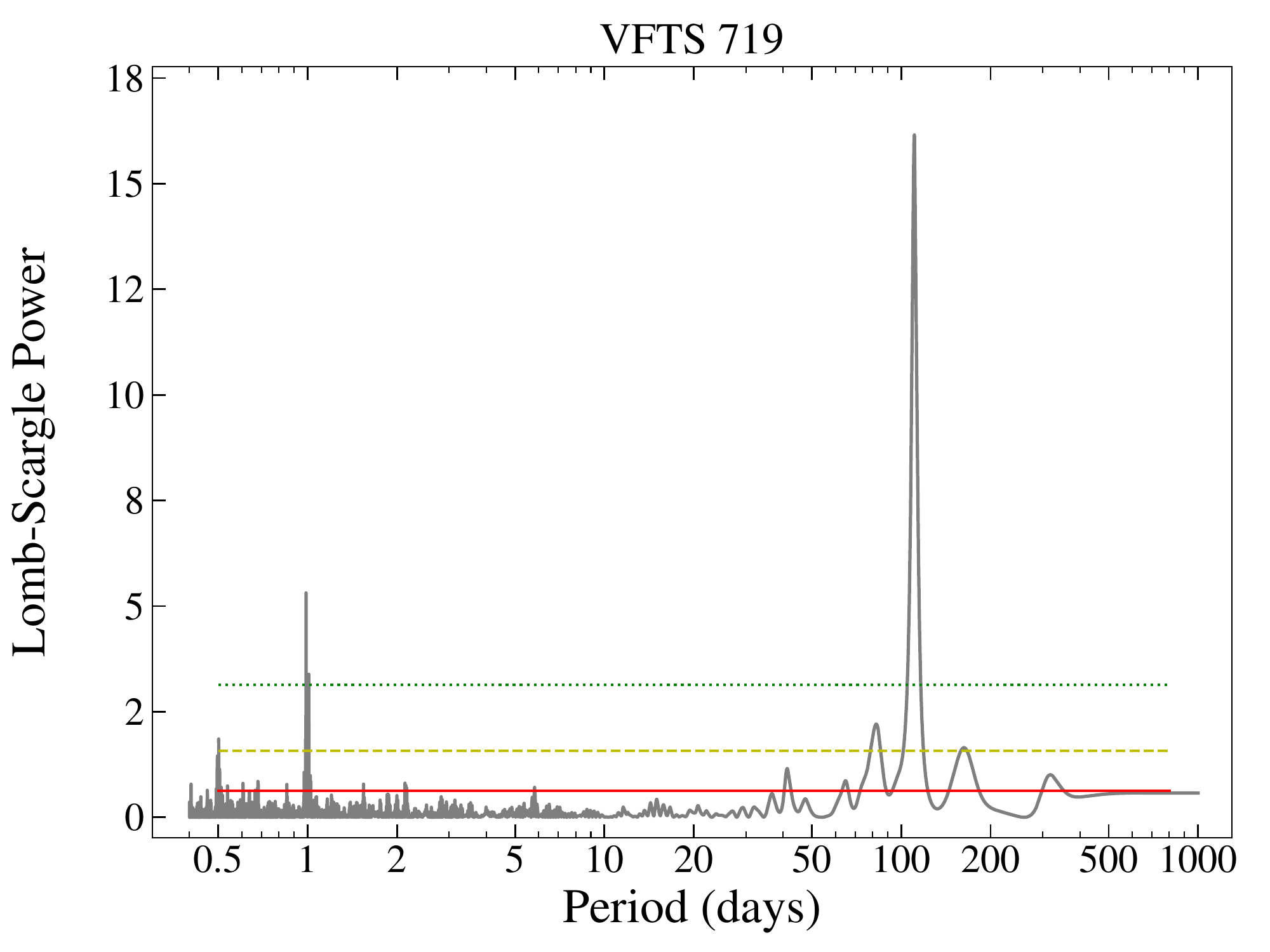}\hfill
    \includegraphics[width=0.31\textwidth]{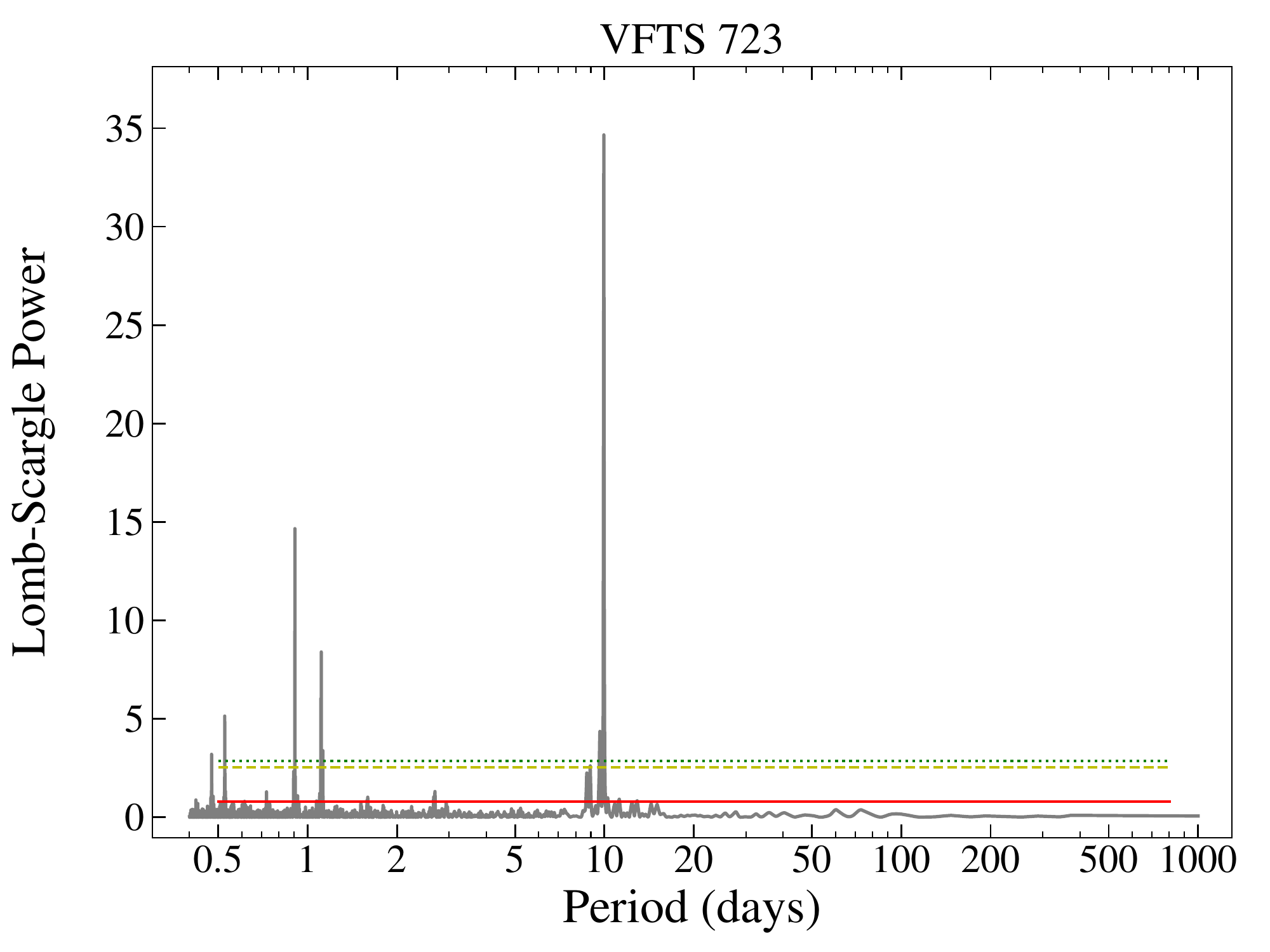}\hfill
    \includegraphics[width=0.31\textwidth]{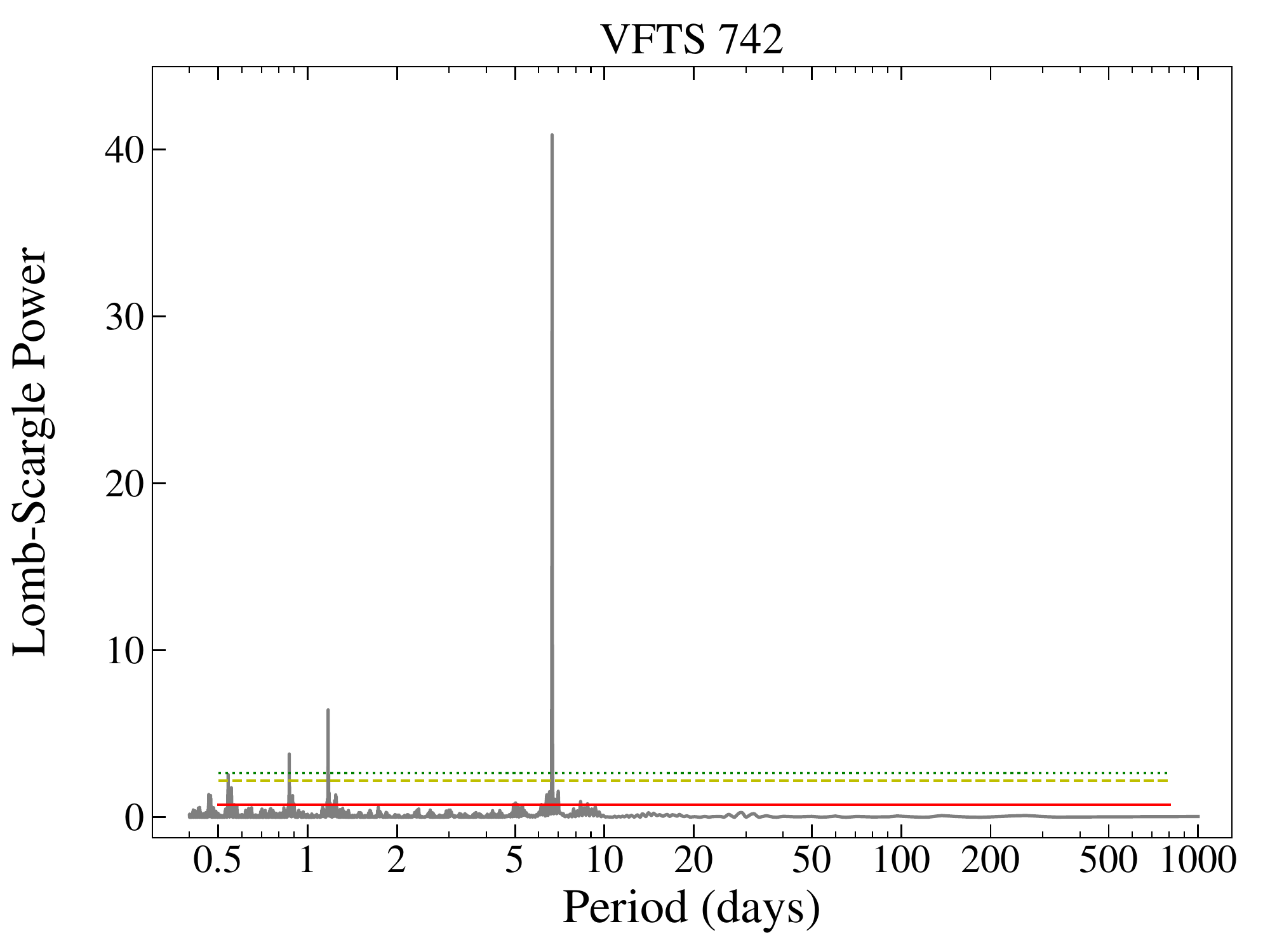}\hfill
    \includegraphics[width=0.31\textwidth]{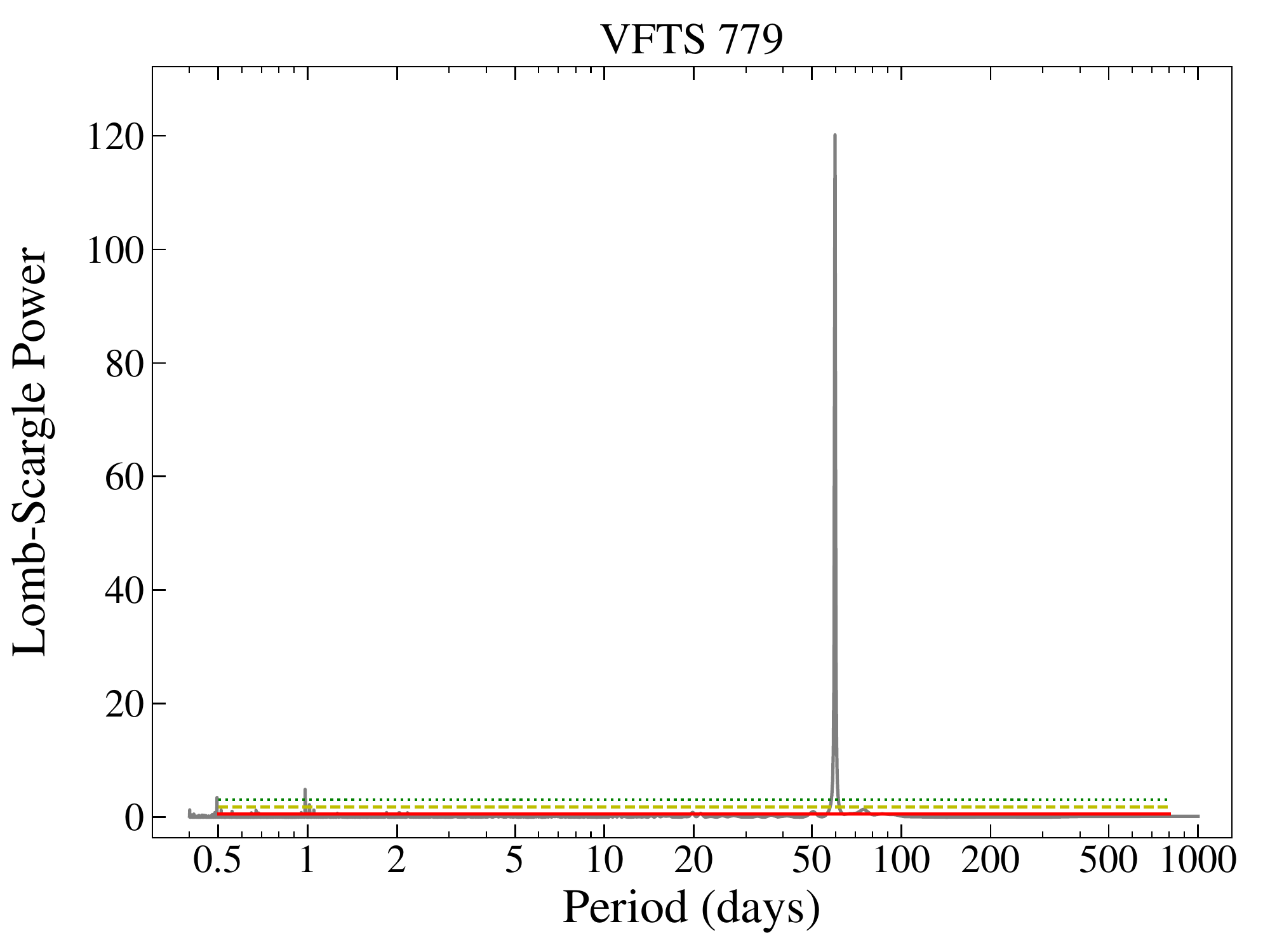}\hfill
    \includegraphics[width=0.31\textwidth]{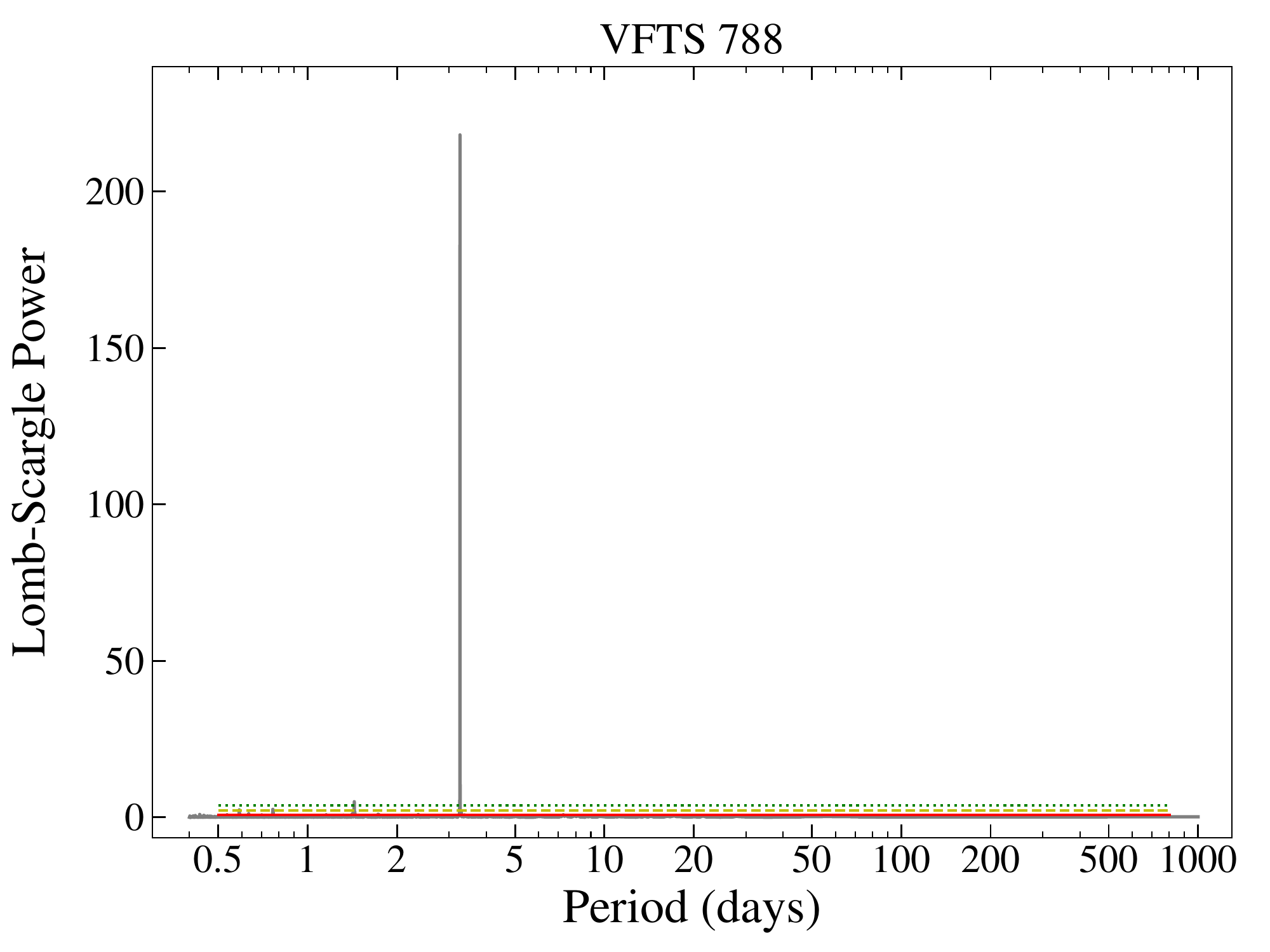}\hfill
    \includegraphics[width=0.31\textwidth]{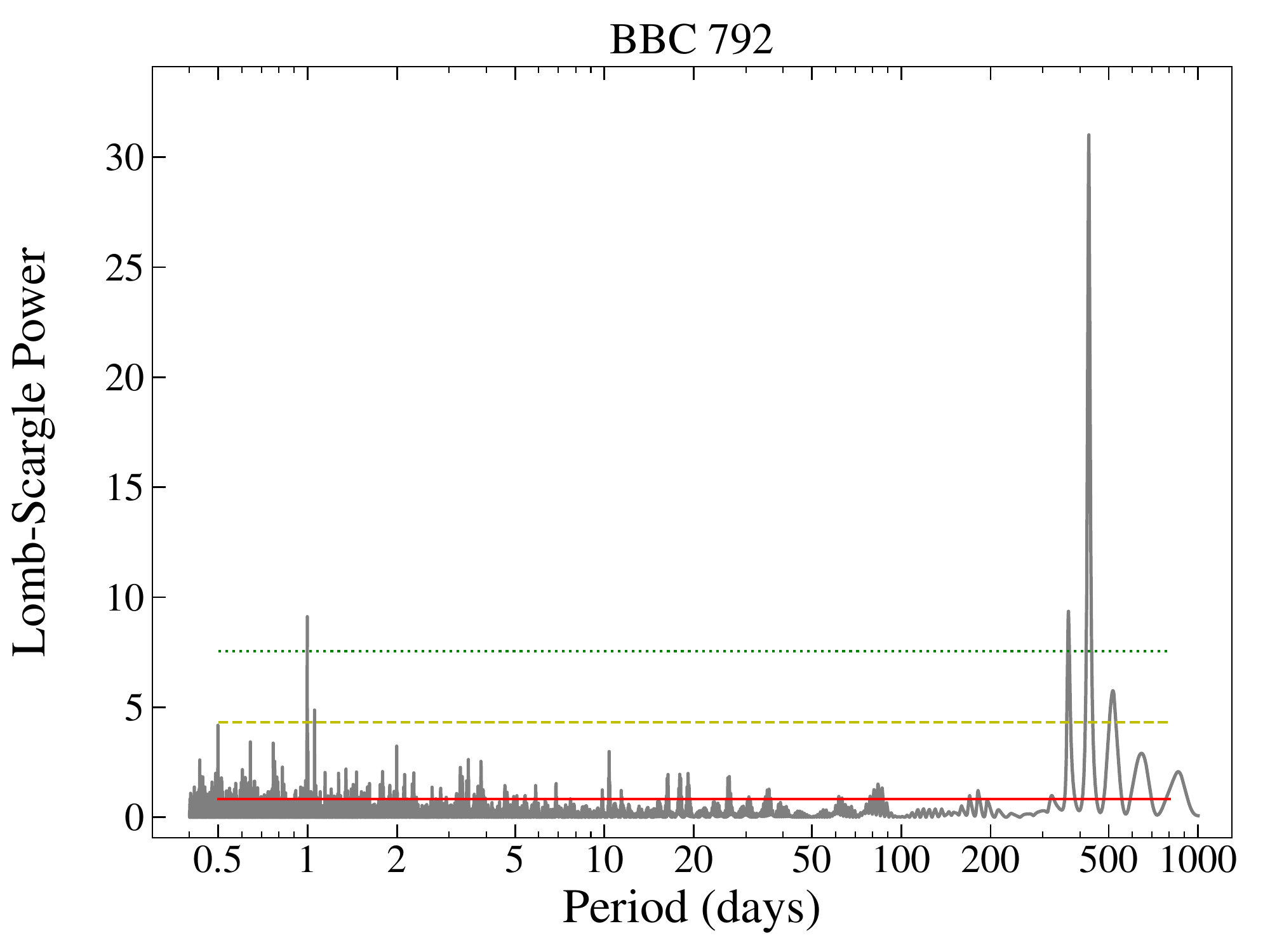}\hfill
    \includegraphics[width=0.31\textwidth]{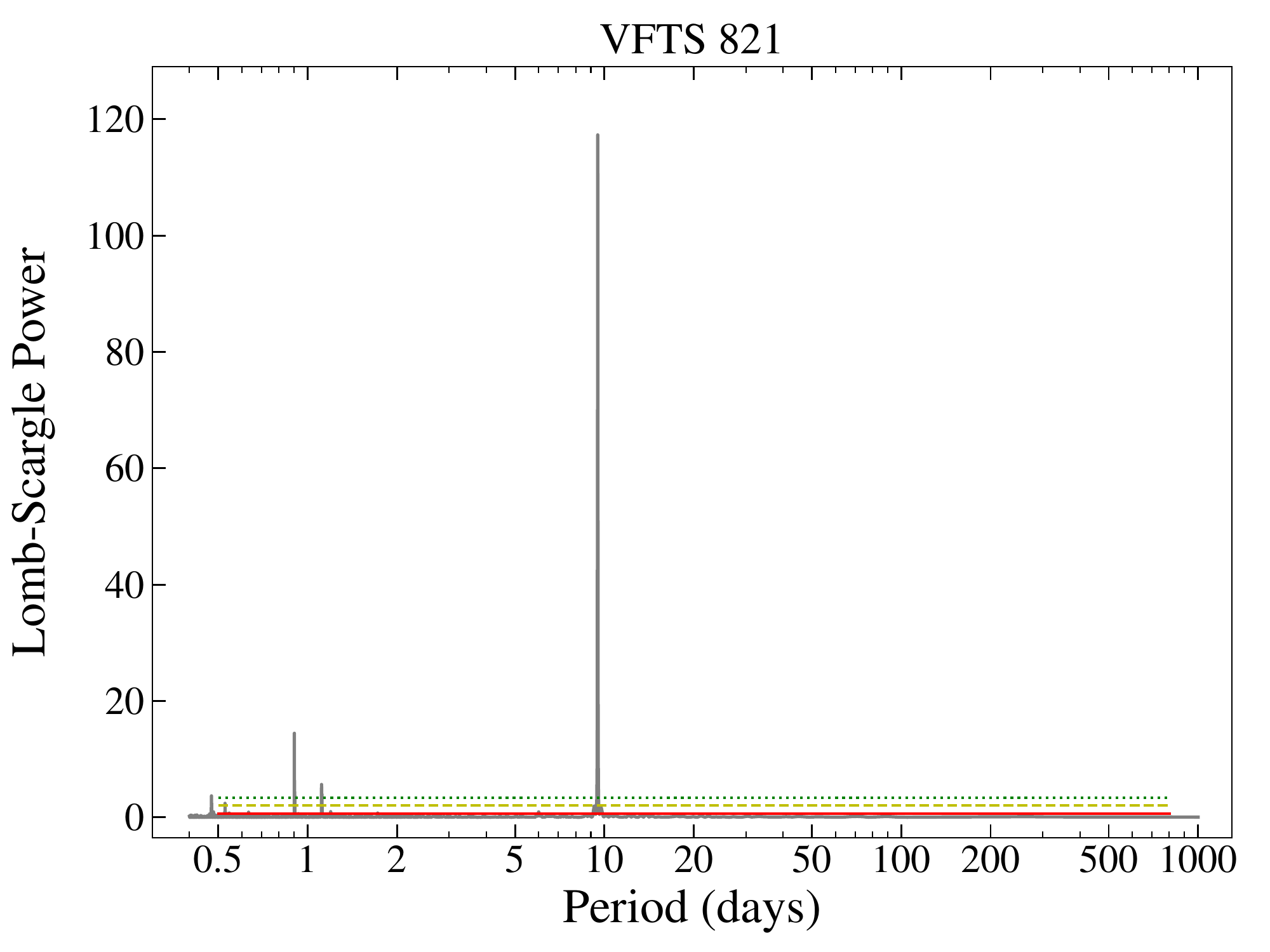}\hfill
    \caption{$-$ \it continued}
\end{myfloat}

\begin{myfloat}
\ContinuedFloat
    \centering
    \includegraphics[width=0.31\textwidth]{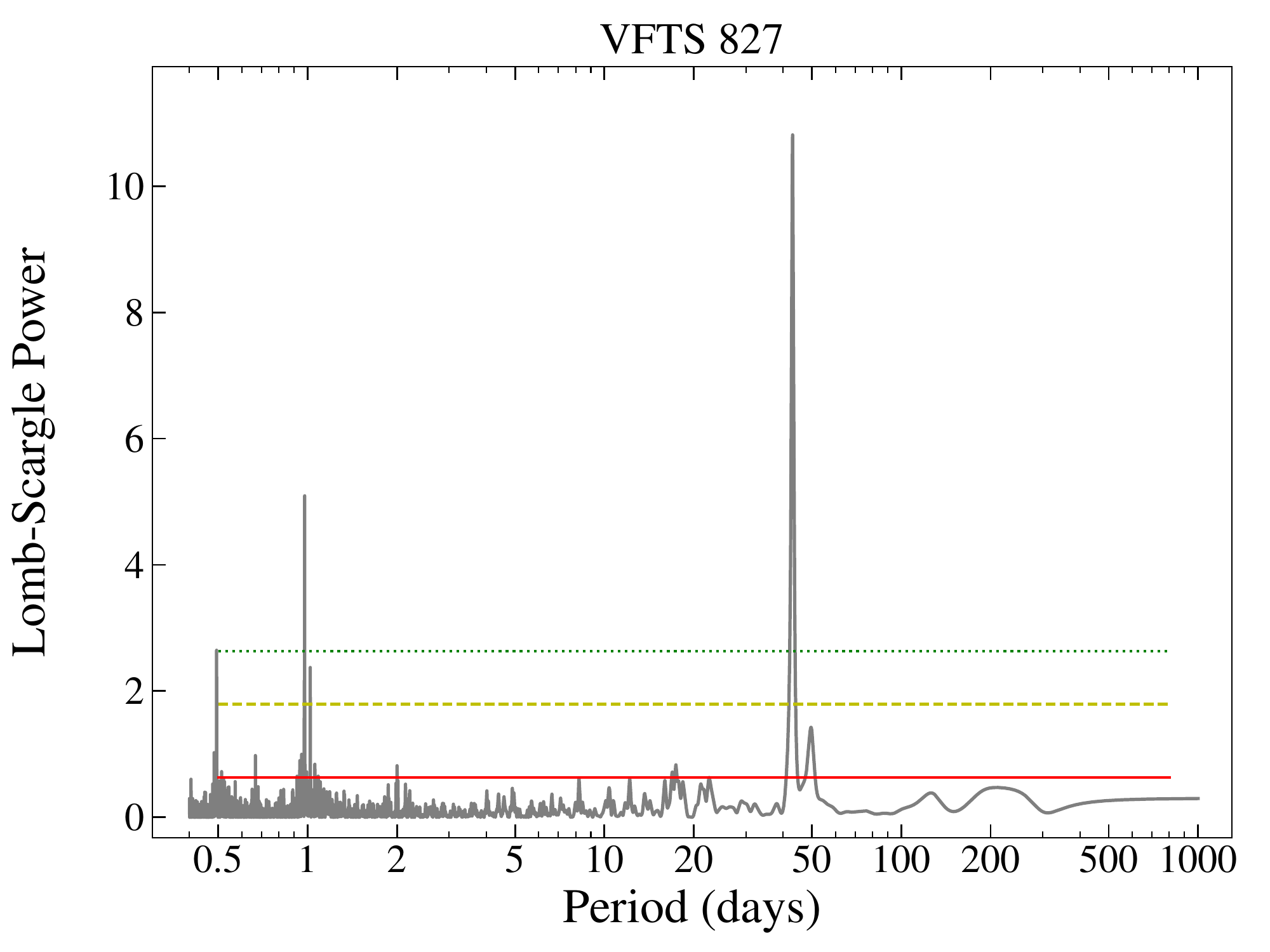}\hfill
    \includegraphics[width=0.31\textwidth]{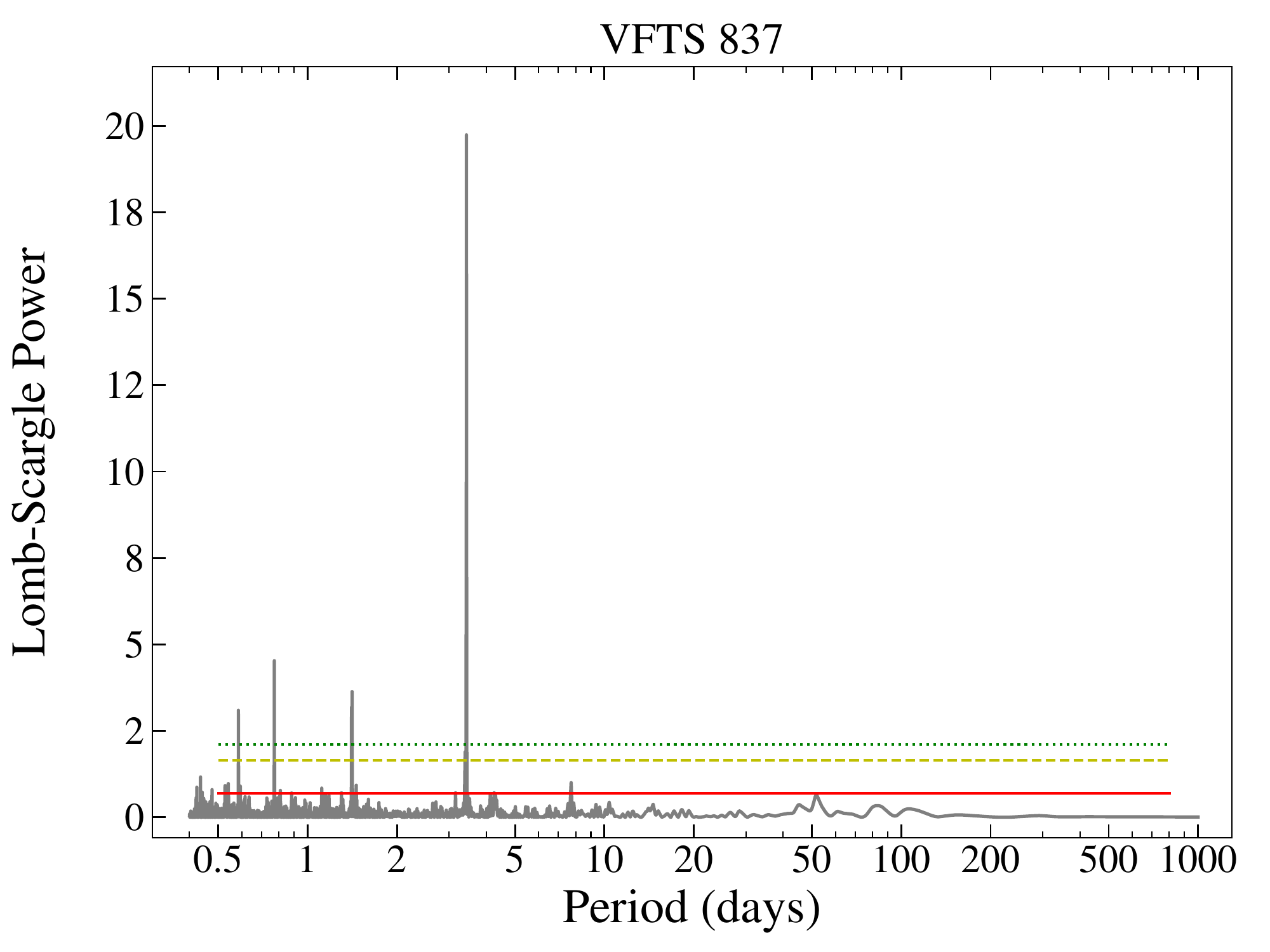}\hfill
    \includegraphics[width=0.31\textwidth]{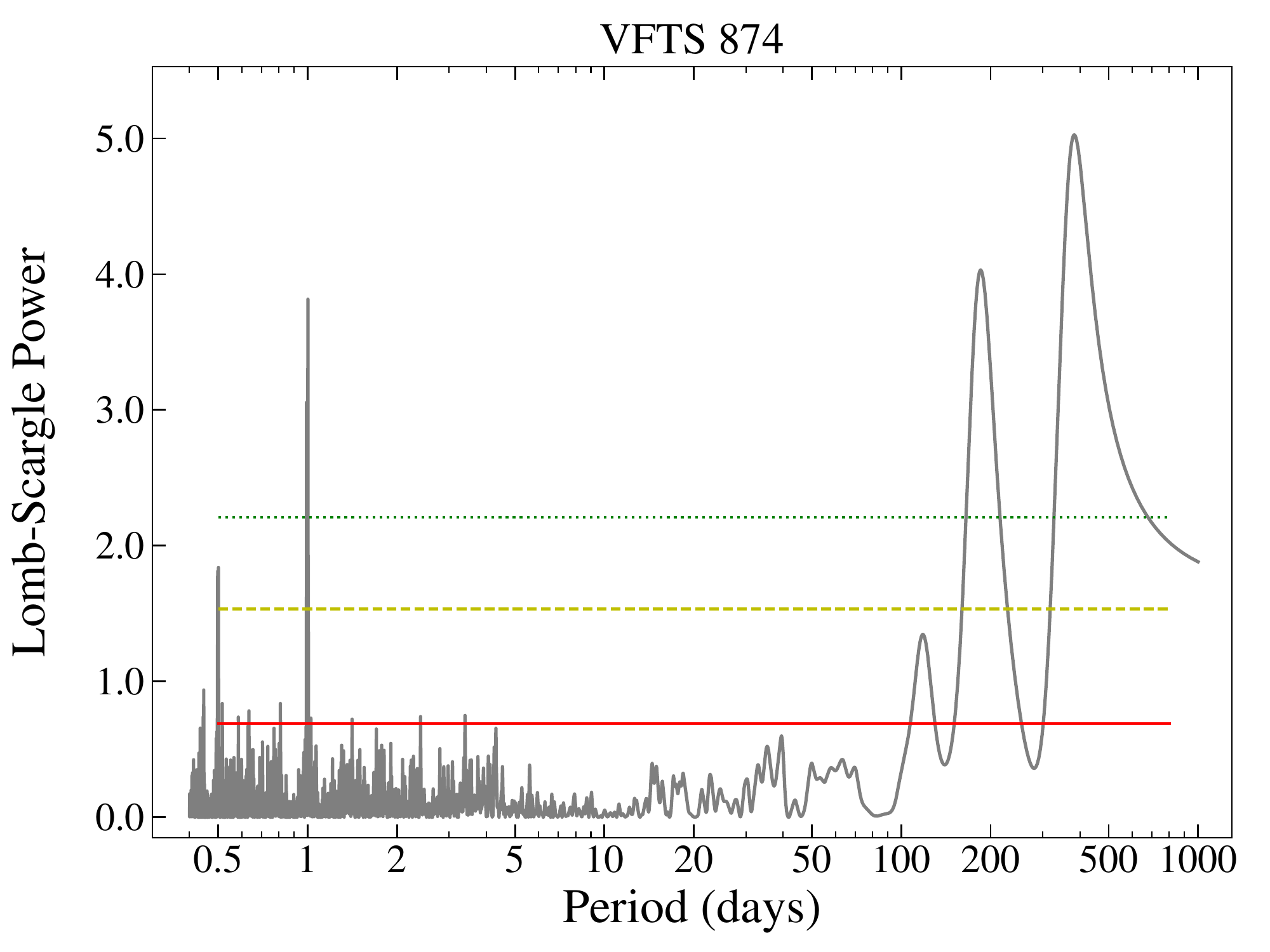}\hfill
    \includegraphics[width=0.31\textwidth]{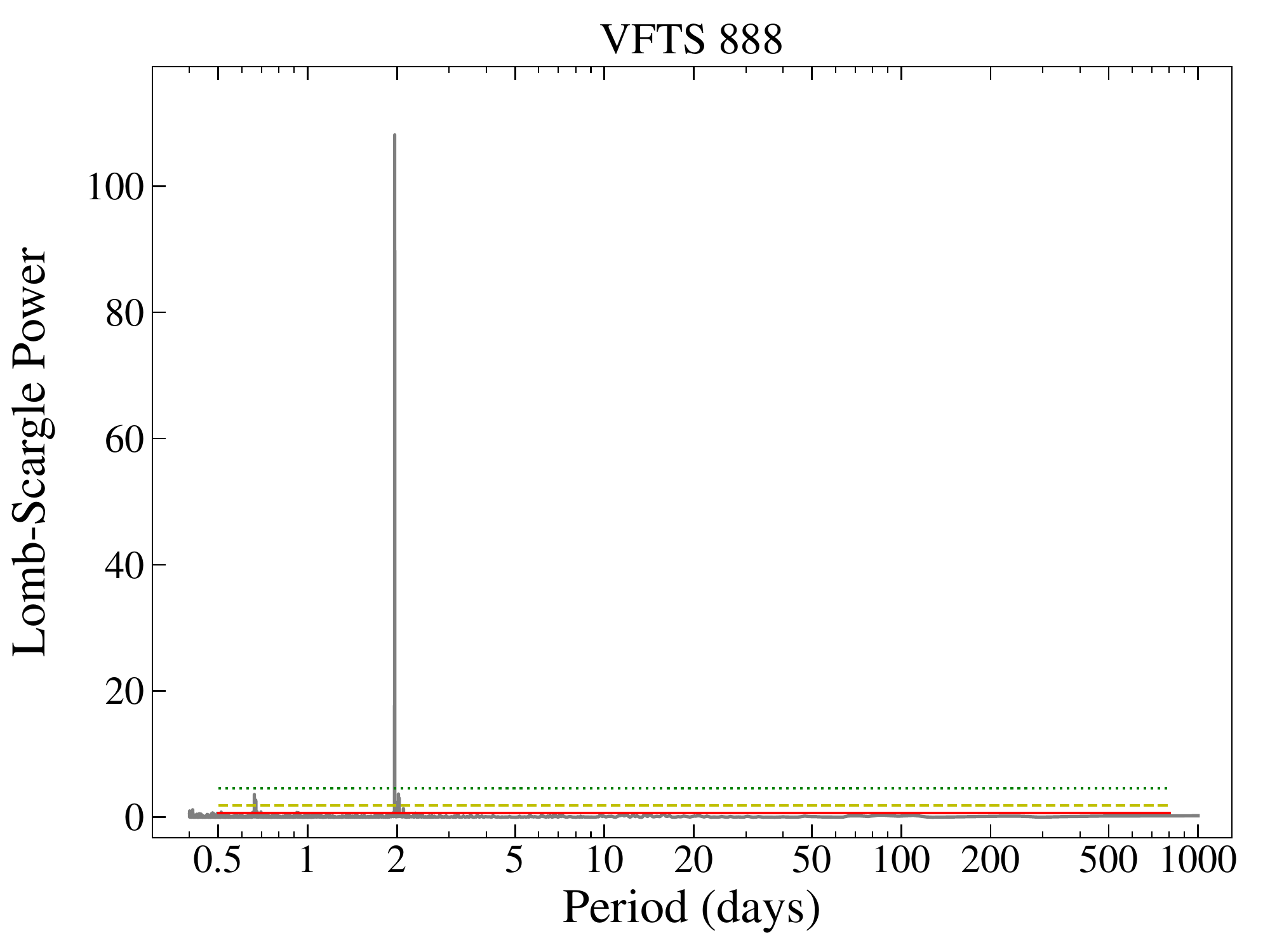}\hspace{0.03\textwidth}
    \includegraphics[width=0.31\textwidth]{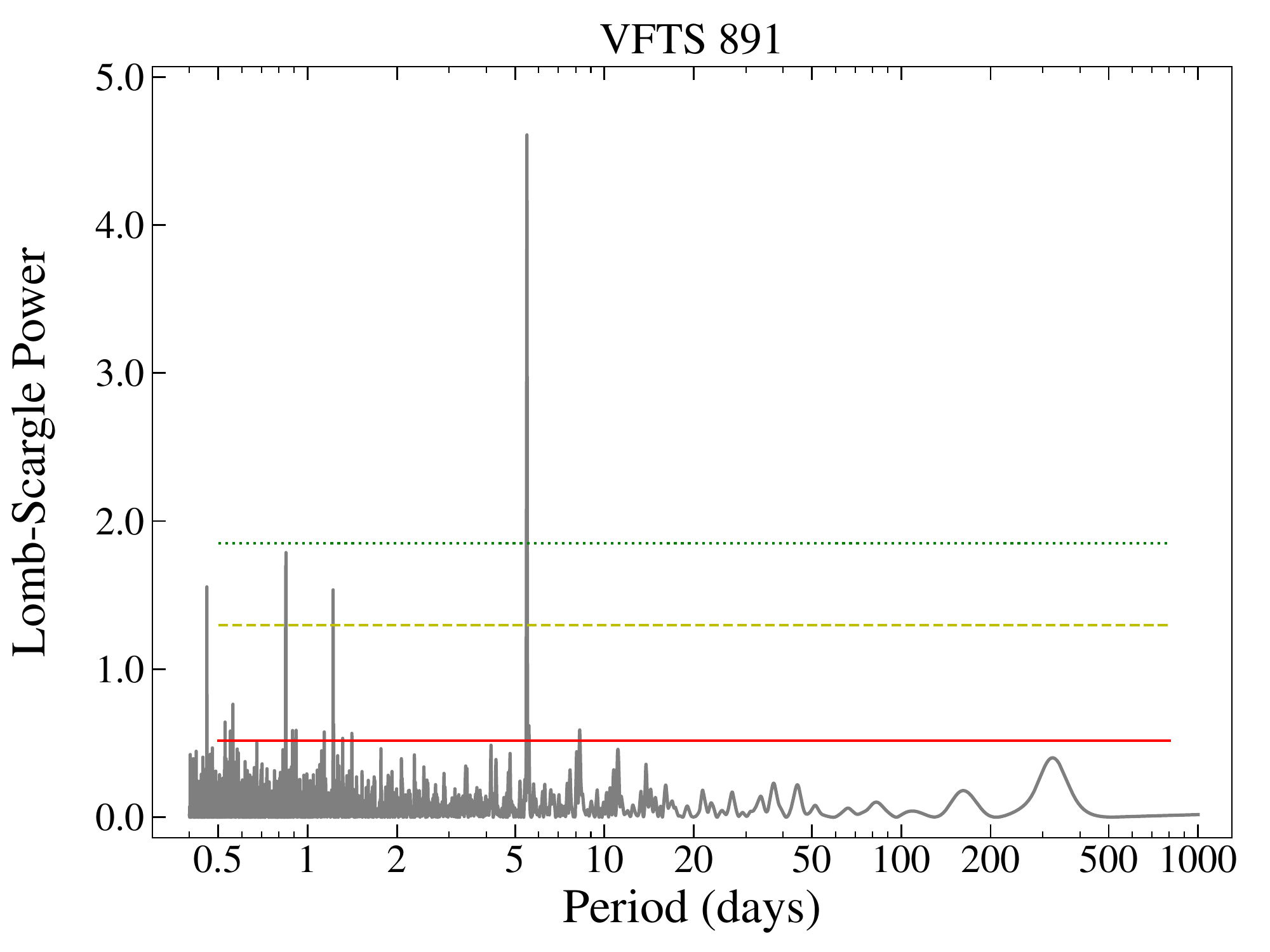}
    \caption{$-$ \it continued}
\end{myfloat}
\clearpage
\begin{figure*}
    \centering
    \includegraphics[width=0.31\textwidth]{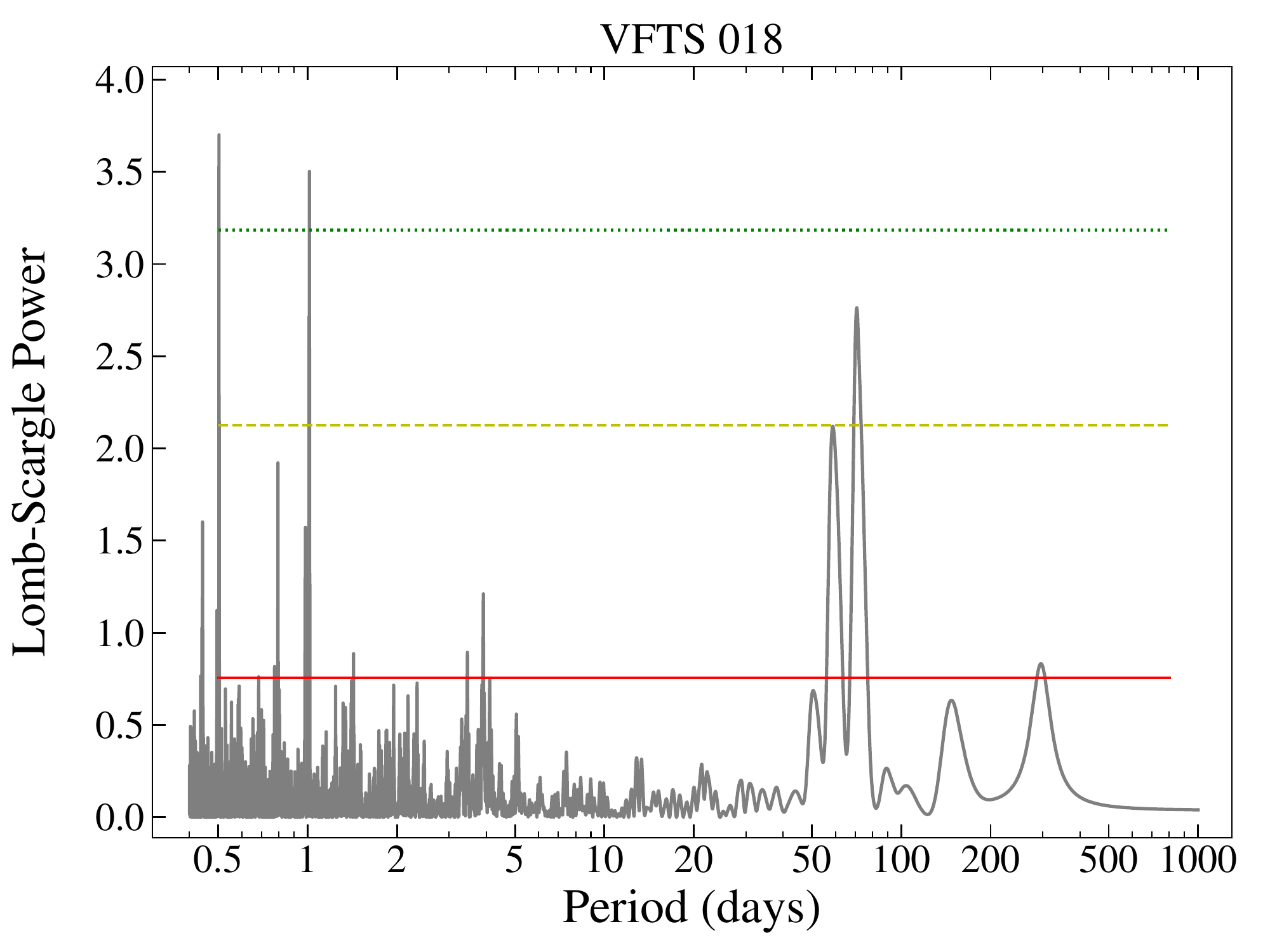}\hfill
    \includegraphics[width=0.31\textwidth]{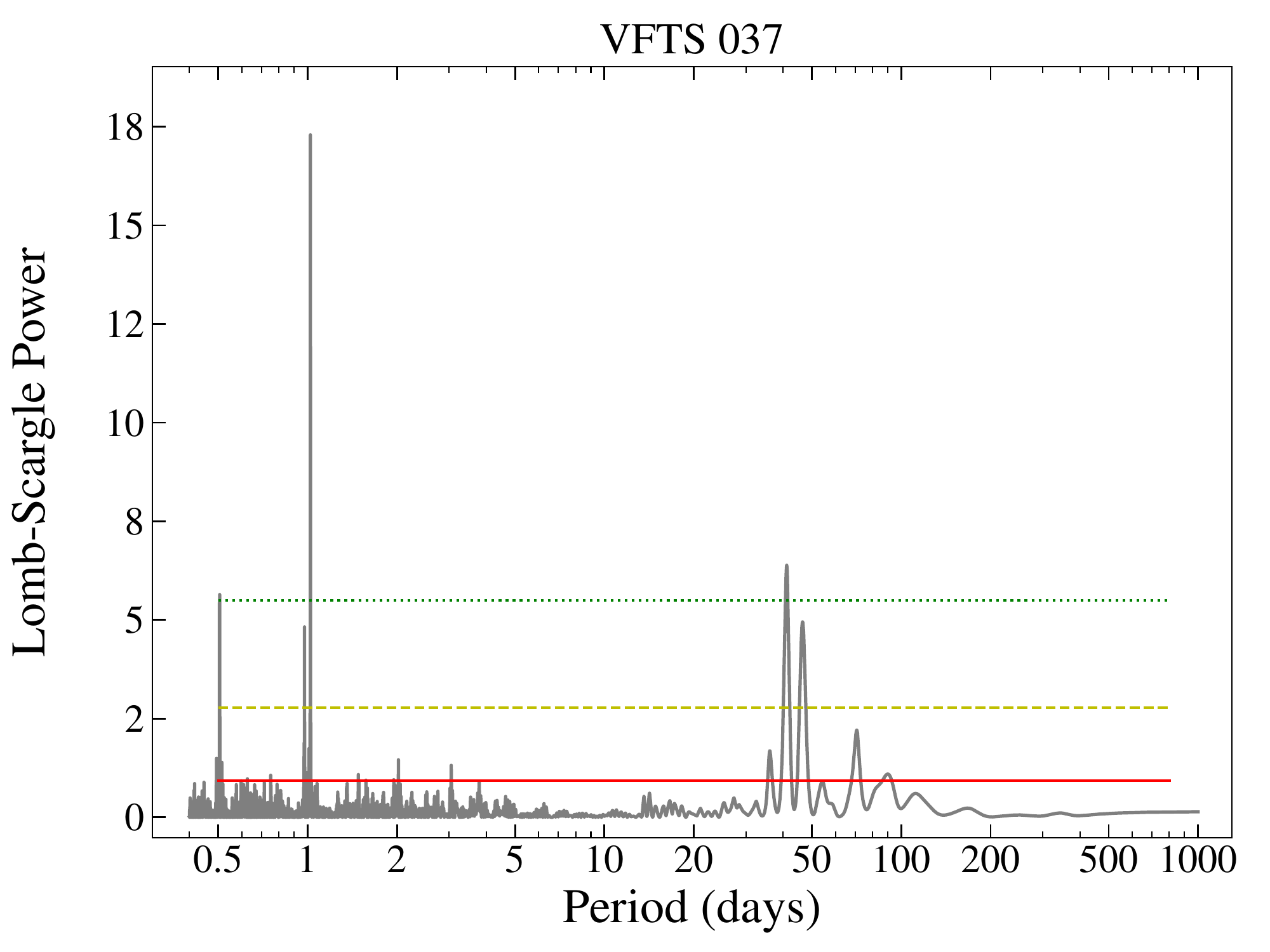}\hfill
    \includegraphics[width=0.31\textwidth]{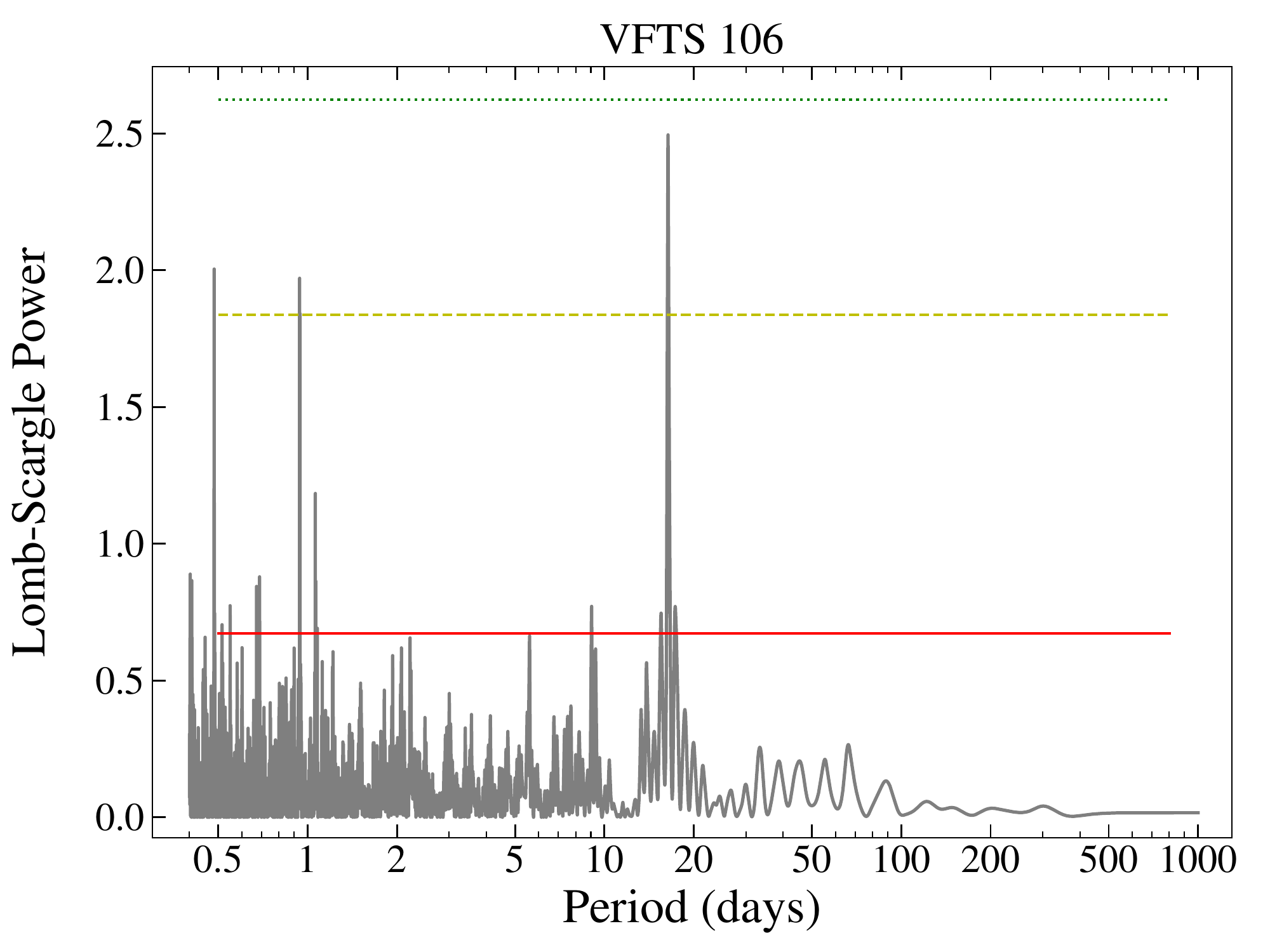}\hfill
    \includegraphics[width=0.31\textwidth]{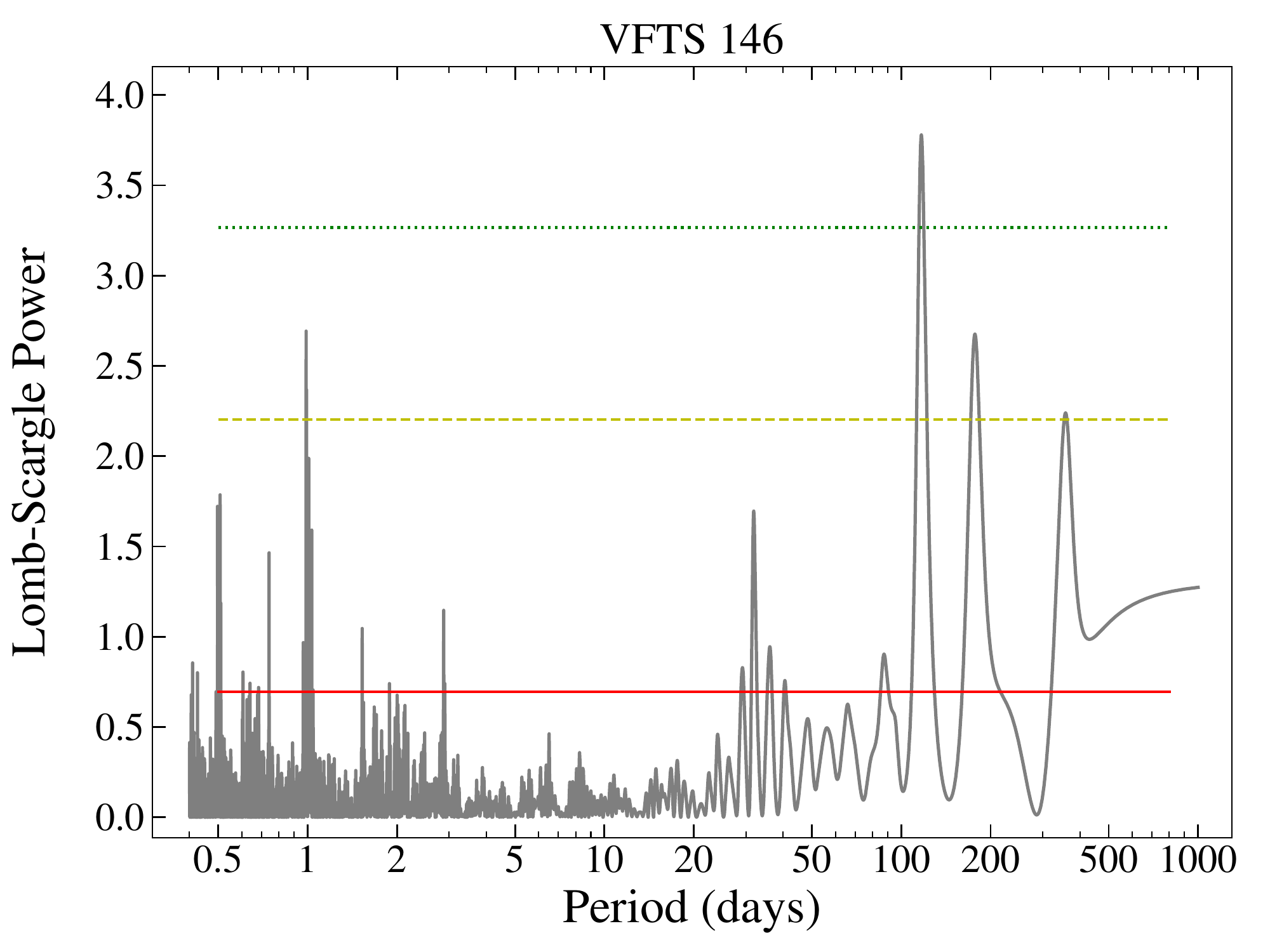}\hfill
    \includegraphics[width=0.31\textwidth]{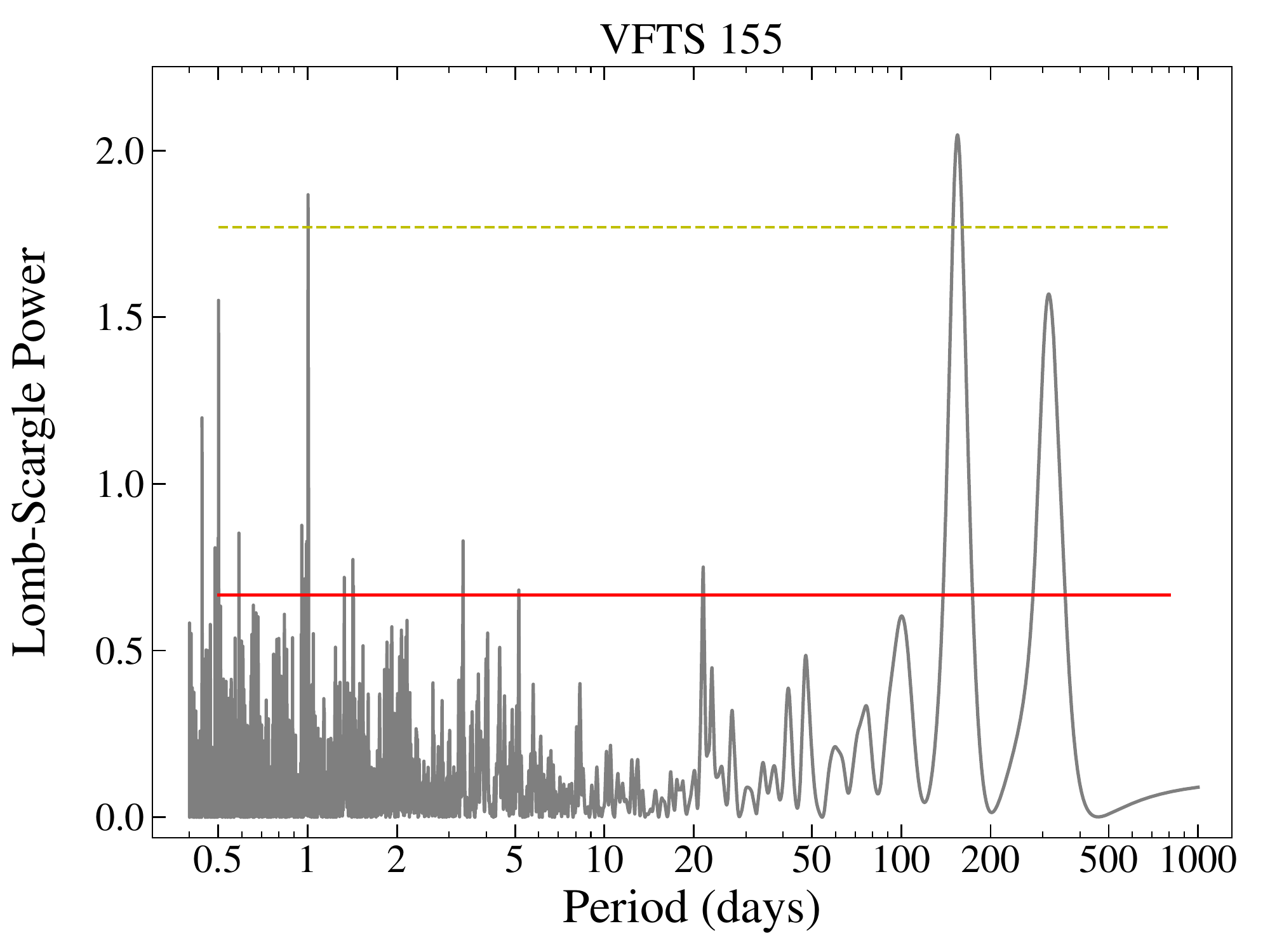}\hfill
    \includegraphics[width=0.31\textwidth]{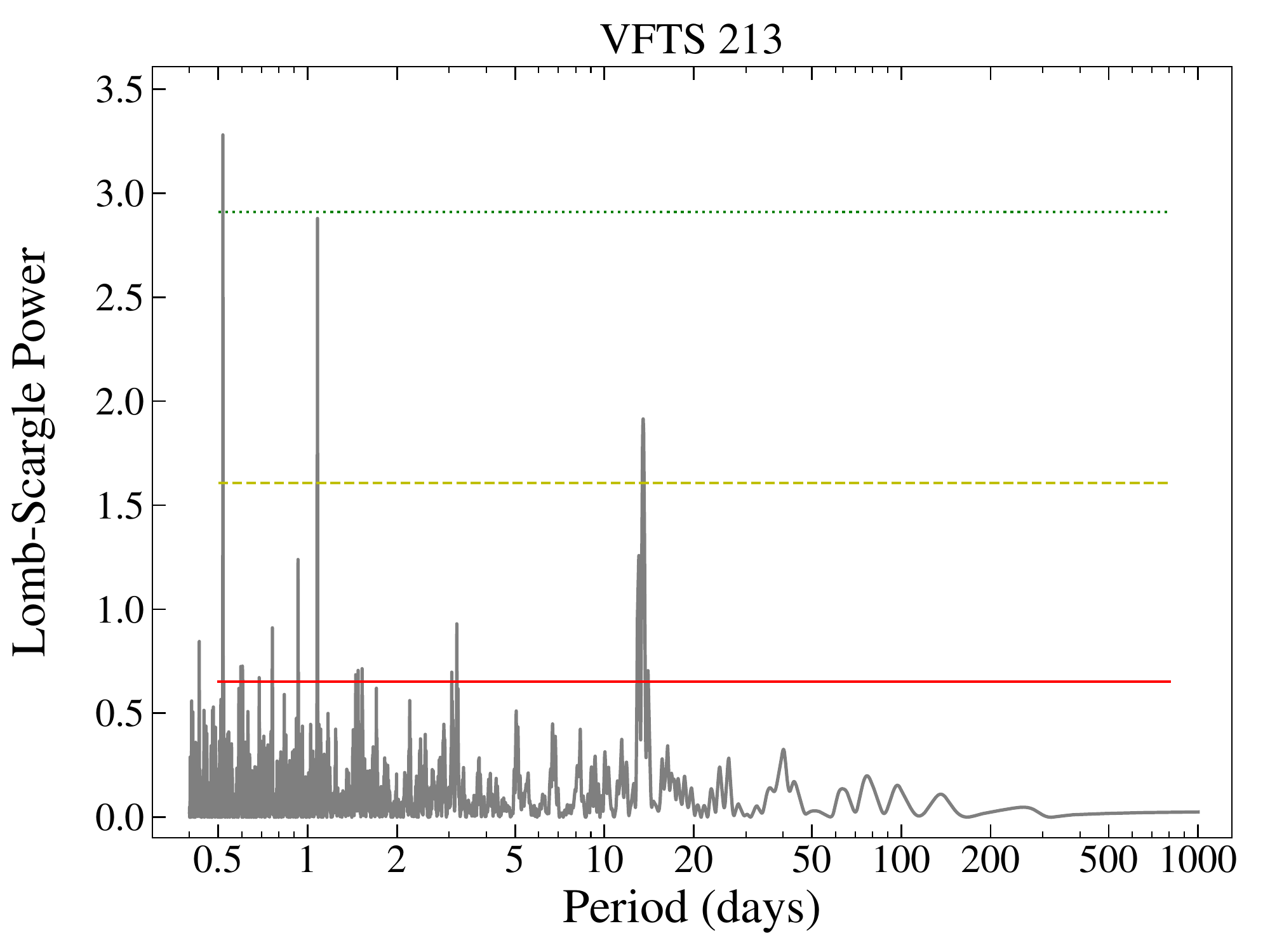}\hfill
    \includegraphics[width=0.31\textwidth]{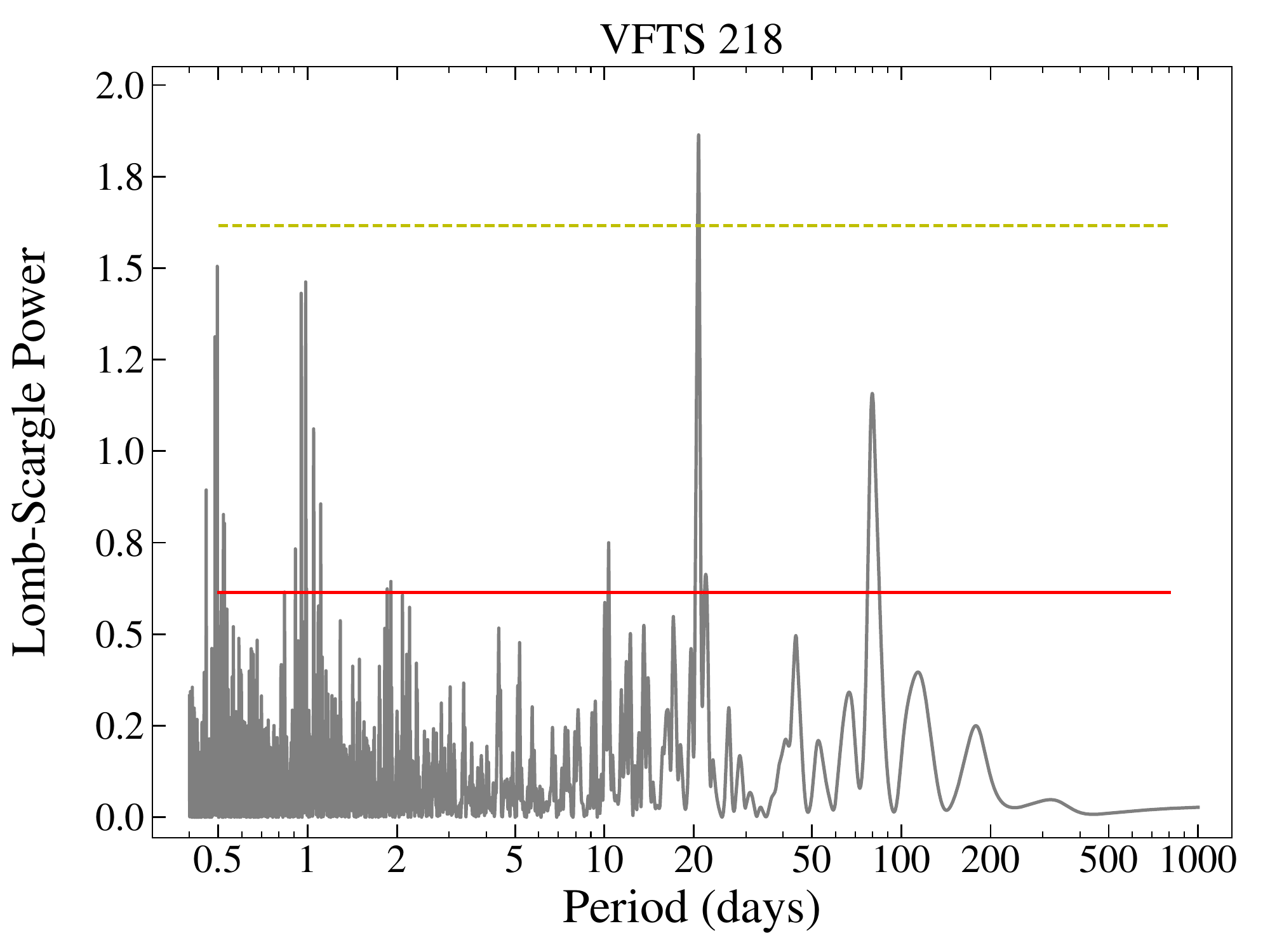}\hfill
    \includegraphics[width=0.31\textwidth]{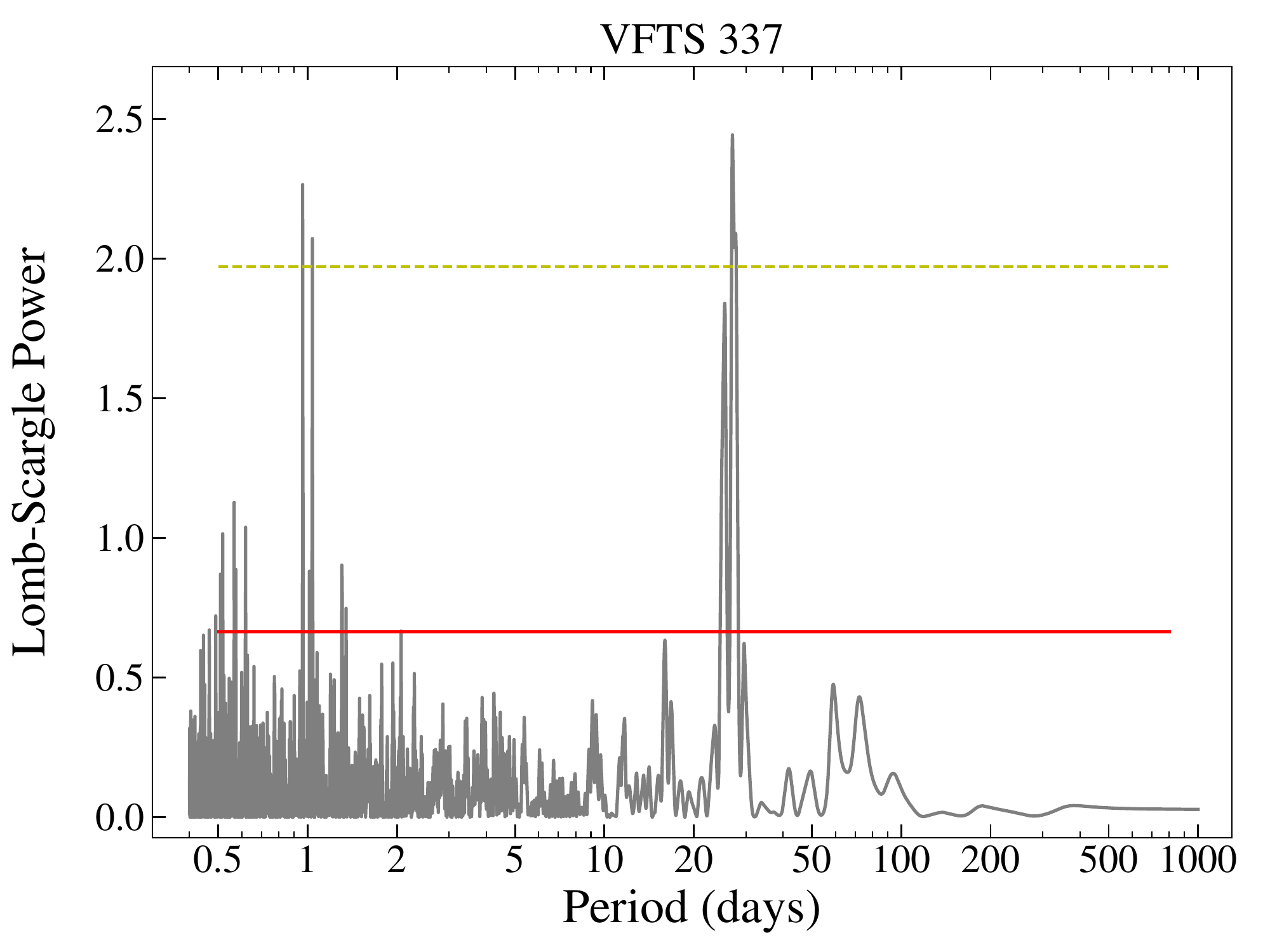}\hfill
    \includegraphics[width=0.31\textwidth]{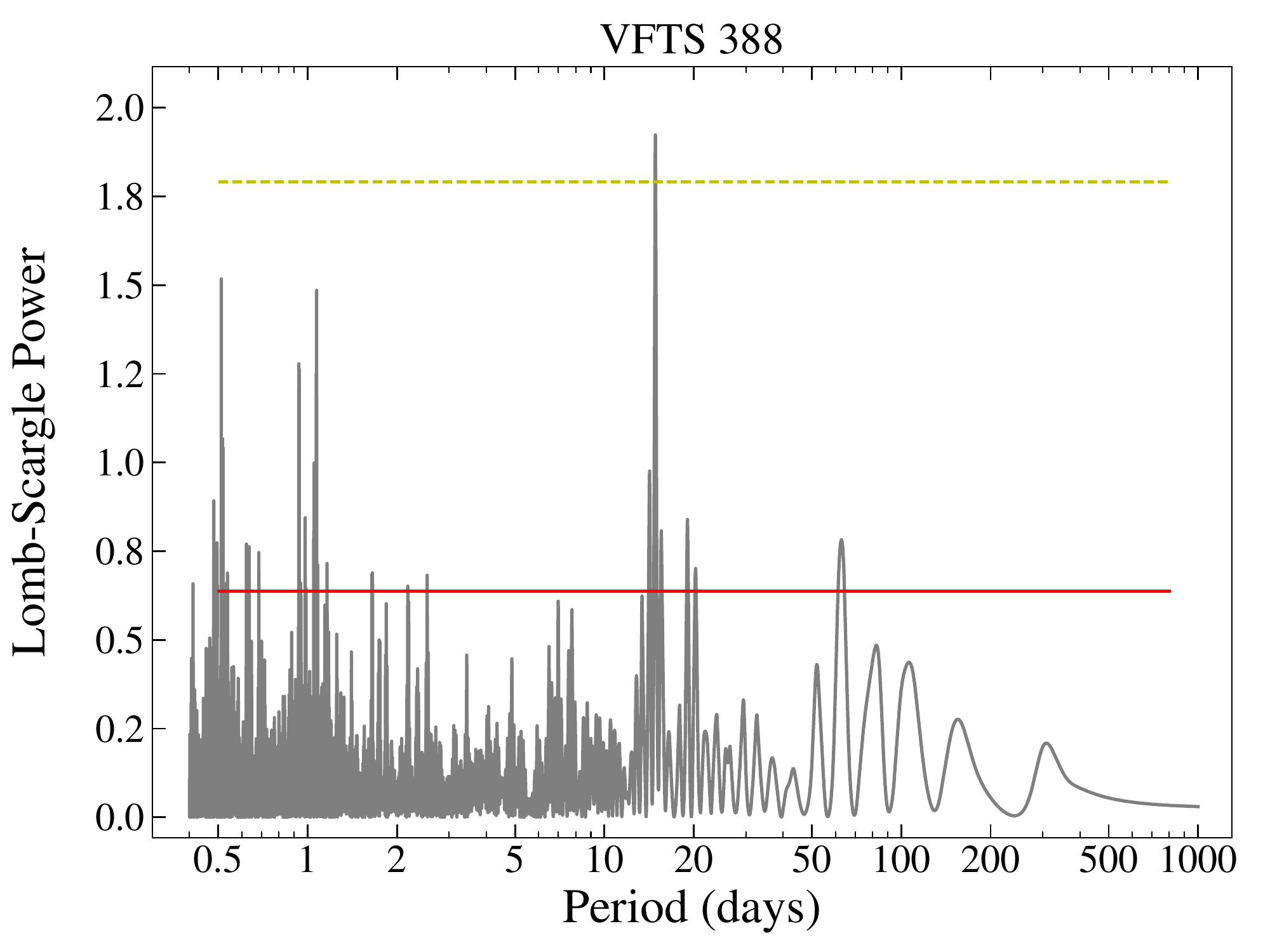}\hfill
    \includegraphics[width=0.31\textwidth]{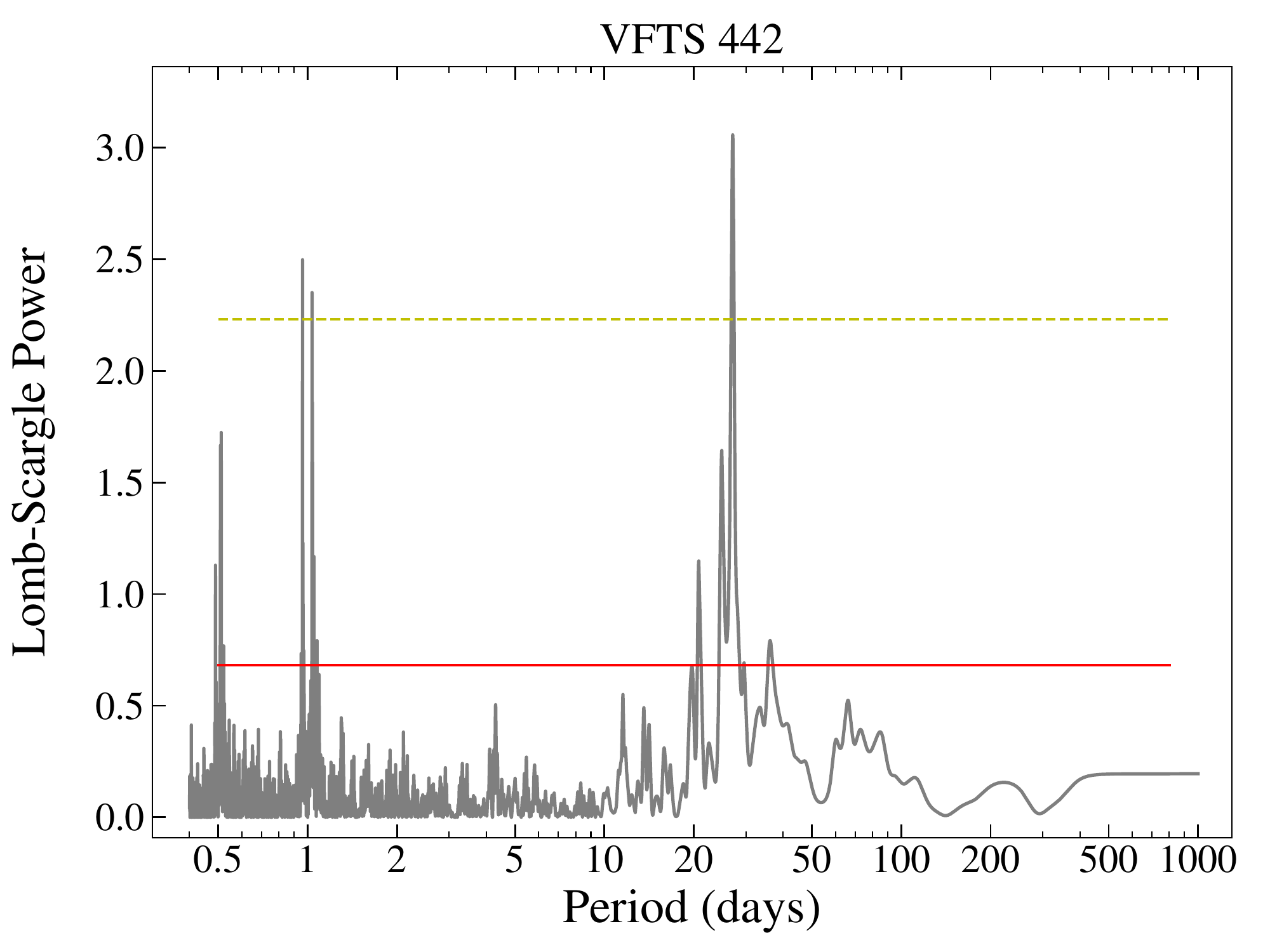}\hfill
    \includegraphics[width=0.31\textwidth]{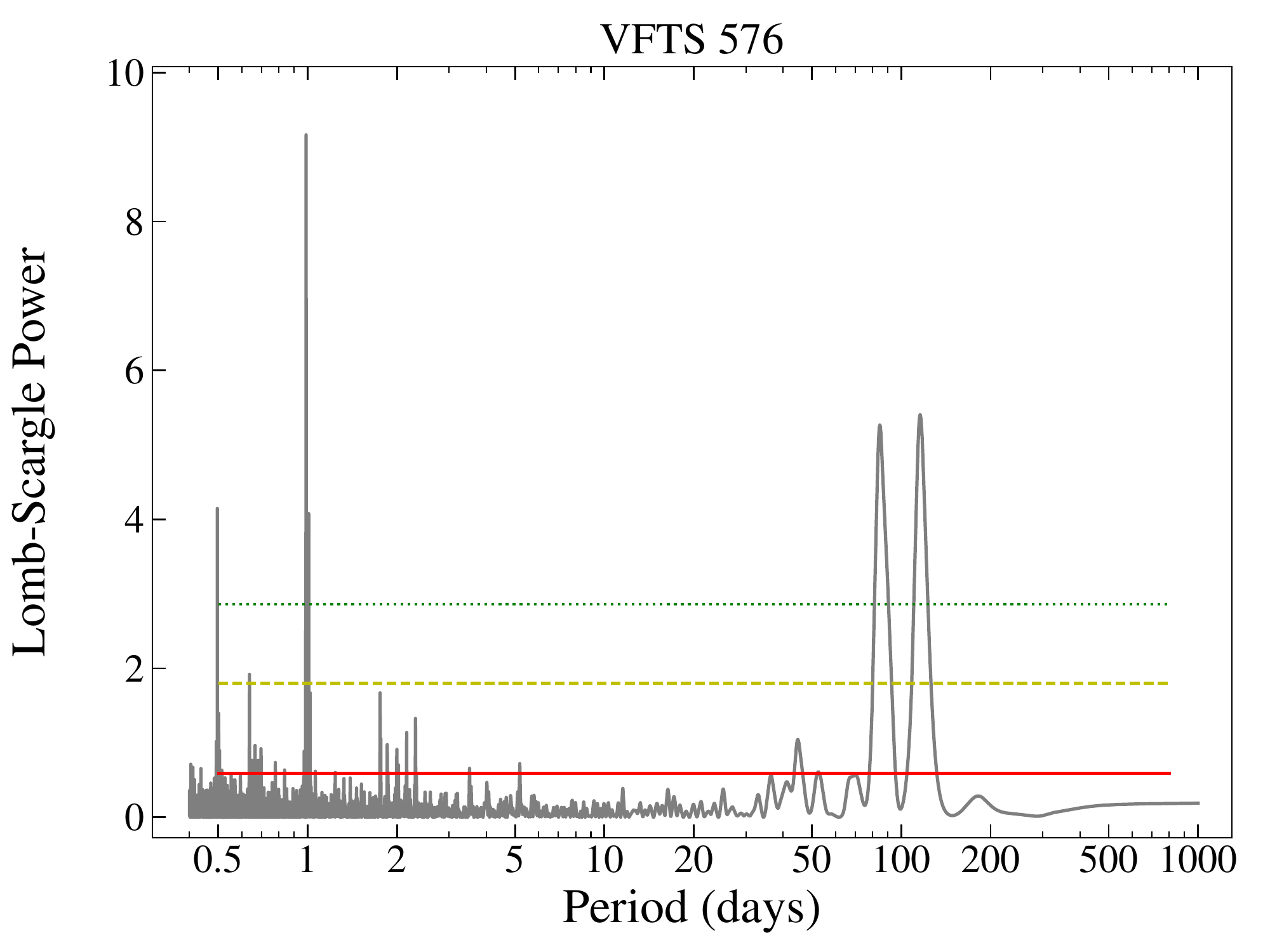}\hfill
    \includegraphics[width=0.31\textwidth]{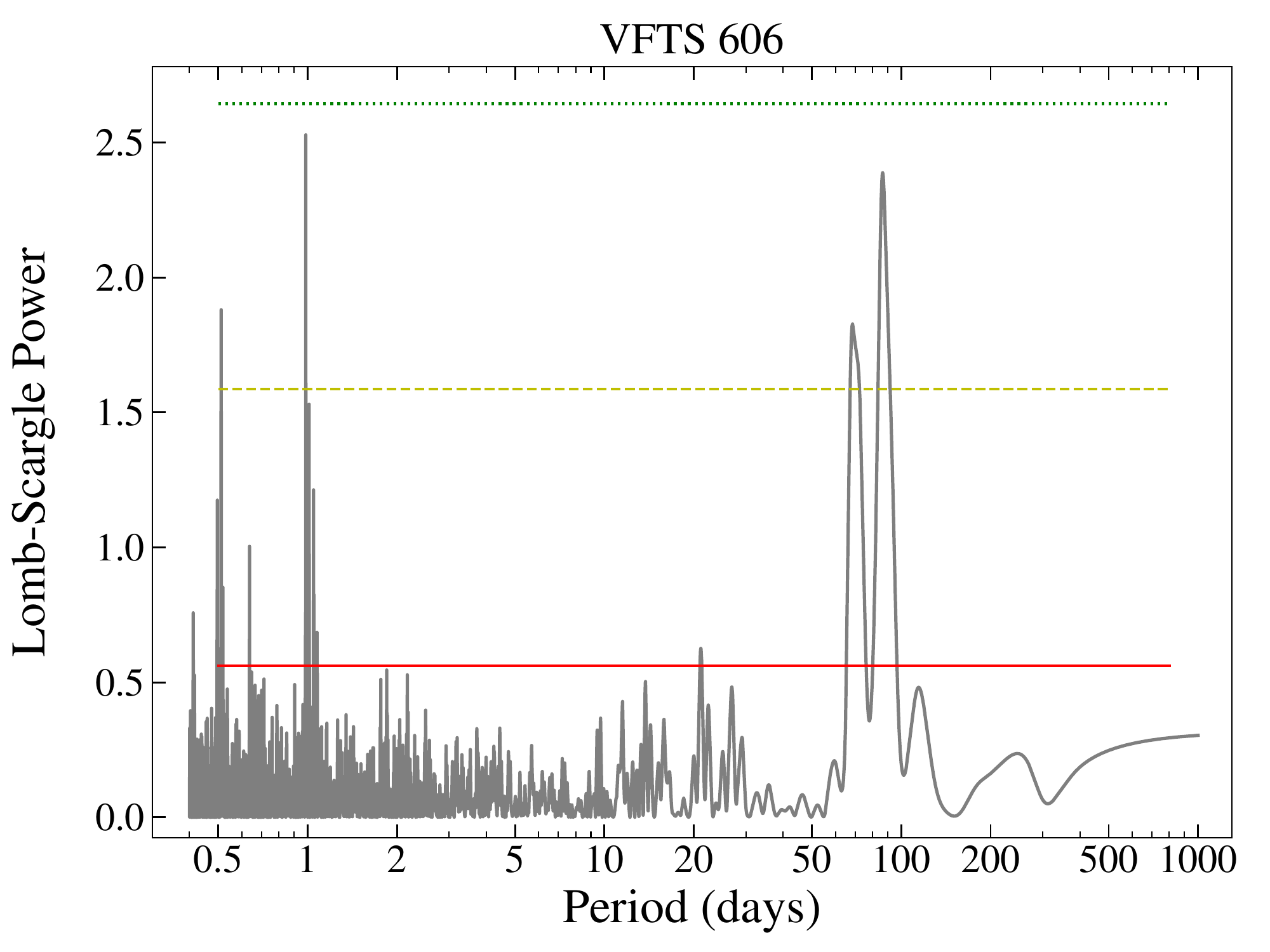}\hfill
    \includegraphics[width=0.31\textwidth]{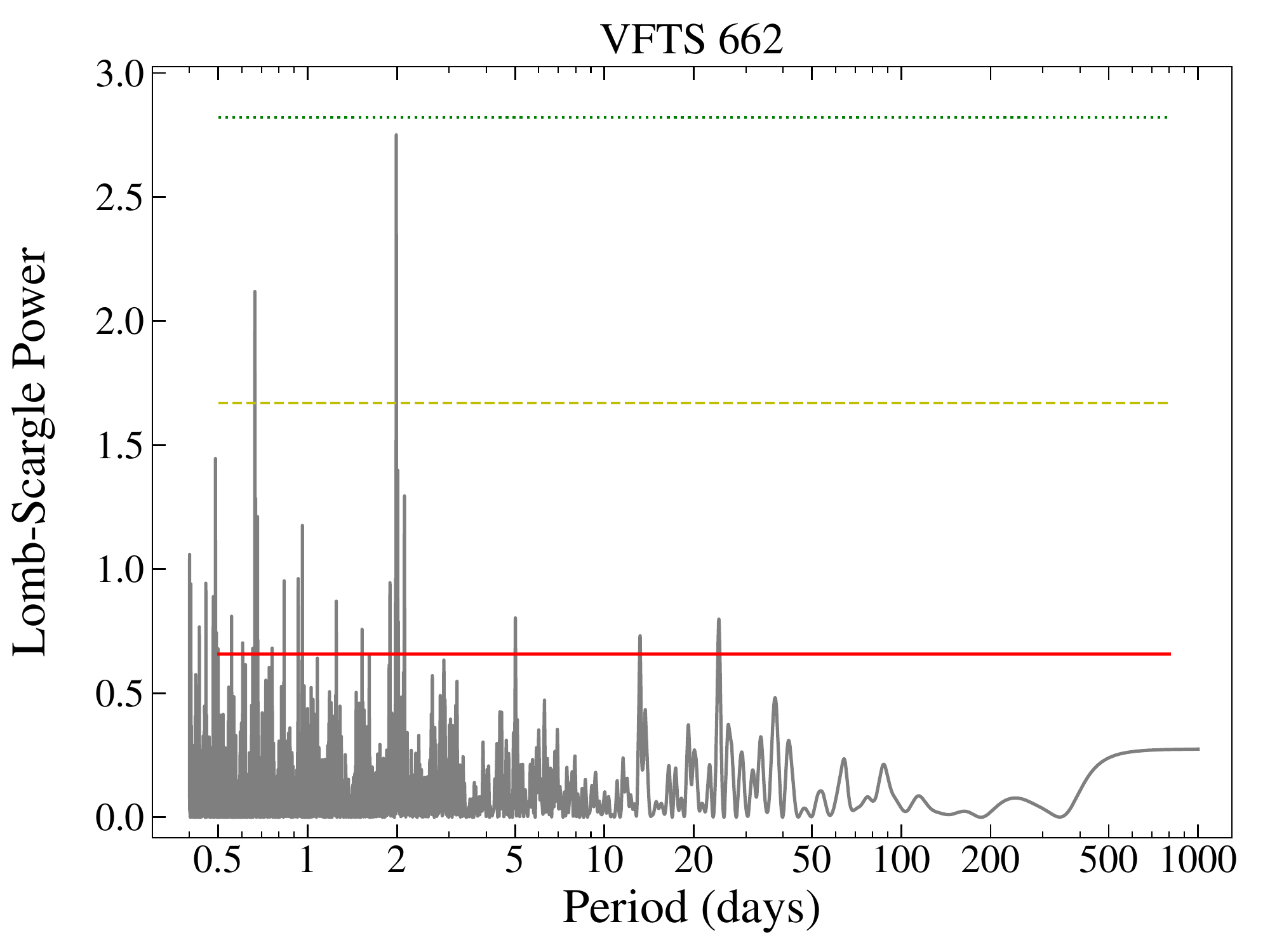}\hfill
    \includegraphics[width=0.31\textwidth]{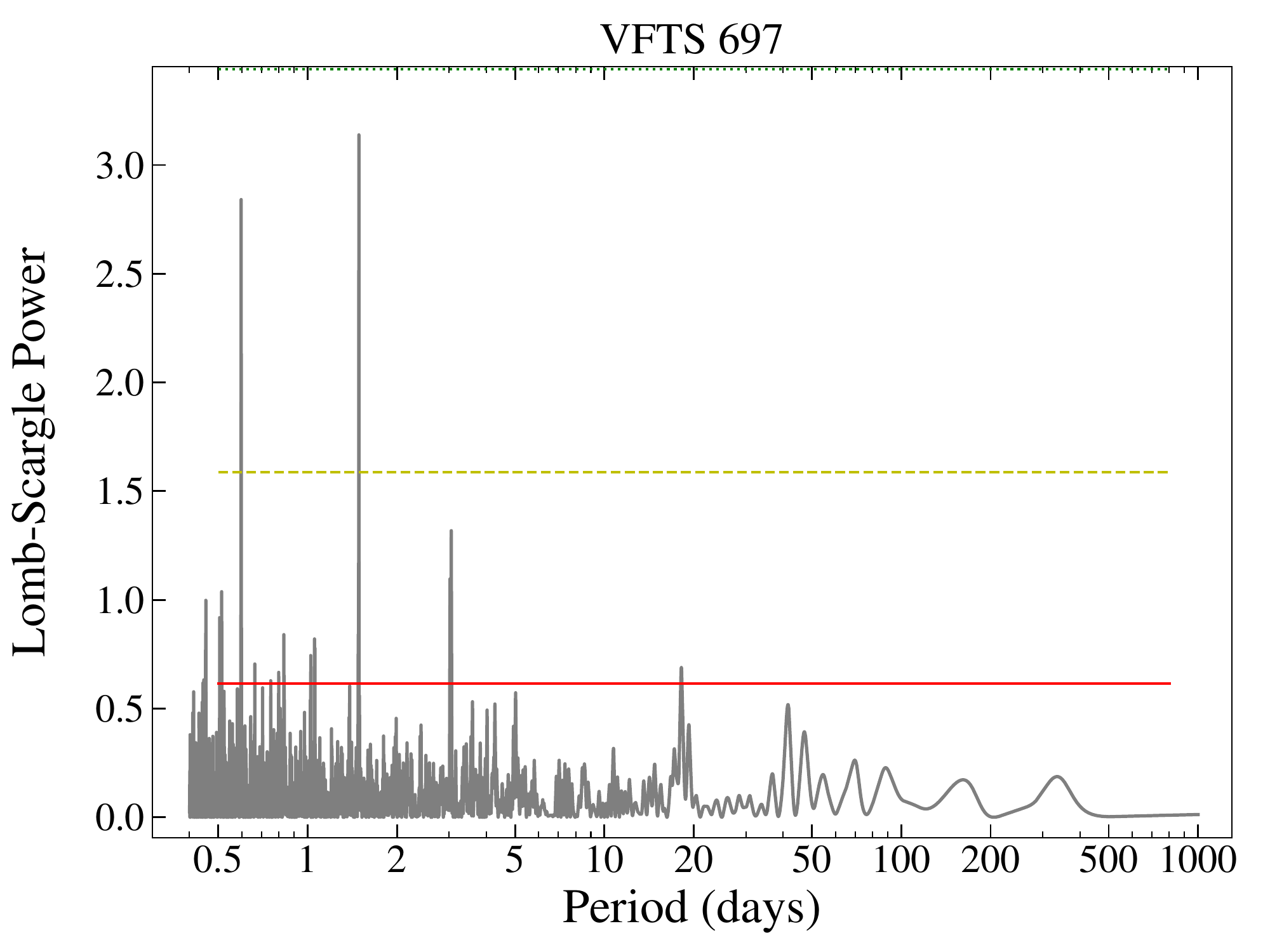}\hfill
    \includegraphics[width=0.31\textwidth]{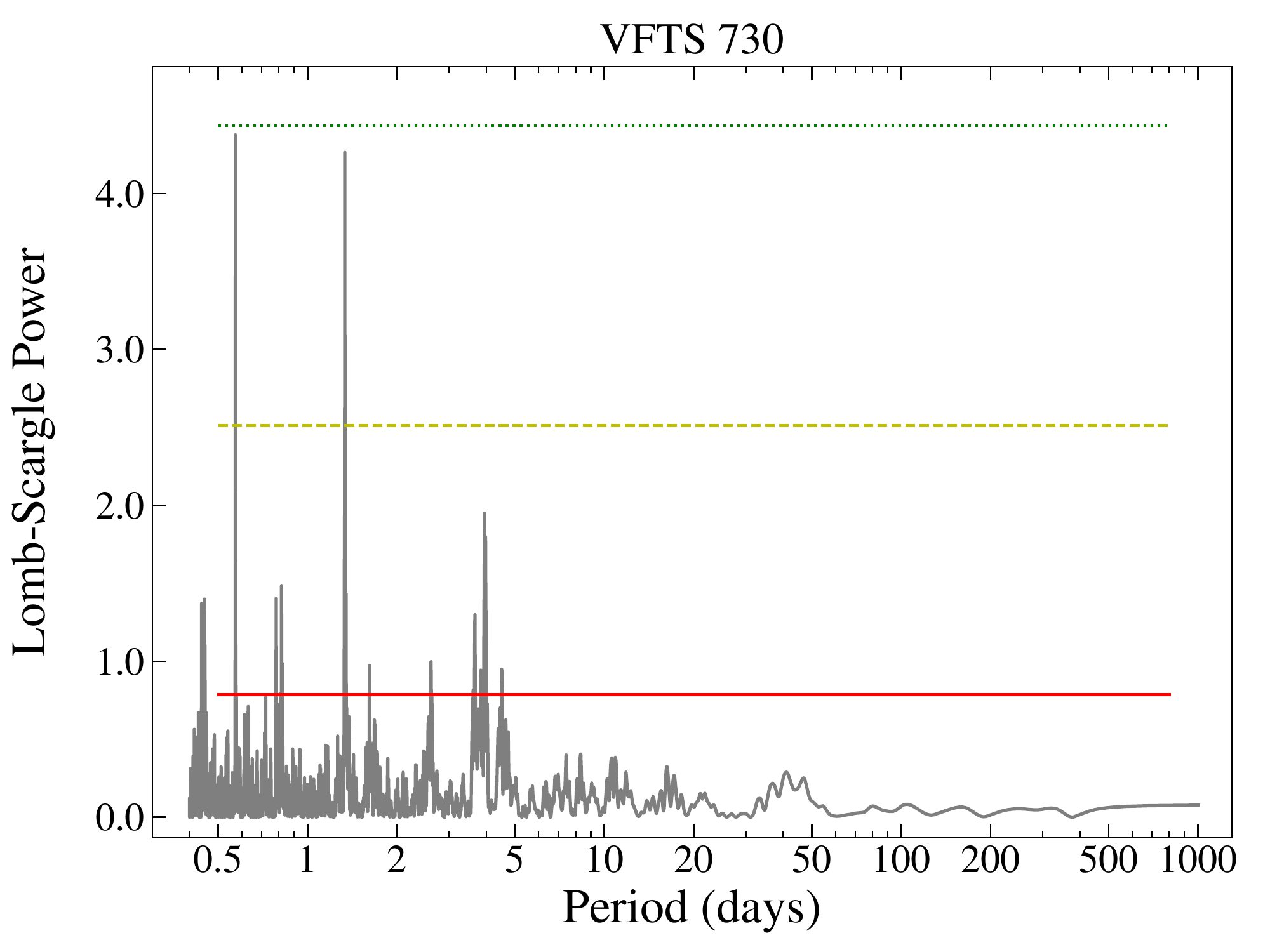}\hfill
\caption{Periodograms of the SB1* systems (possible periods)}
\label{figAp:LSsb1*}
\end{figure*}

\begin{figure*}
\ContinuedFloat
    \centering
    \includegraphics[width=0.31\textwidth]{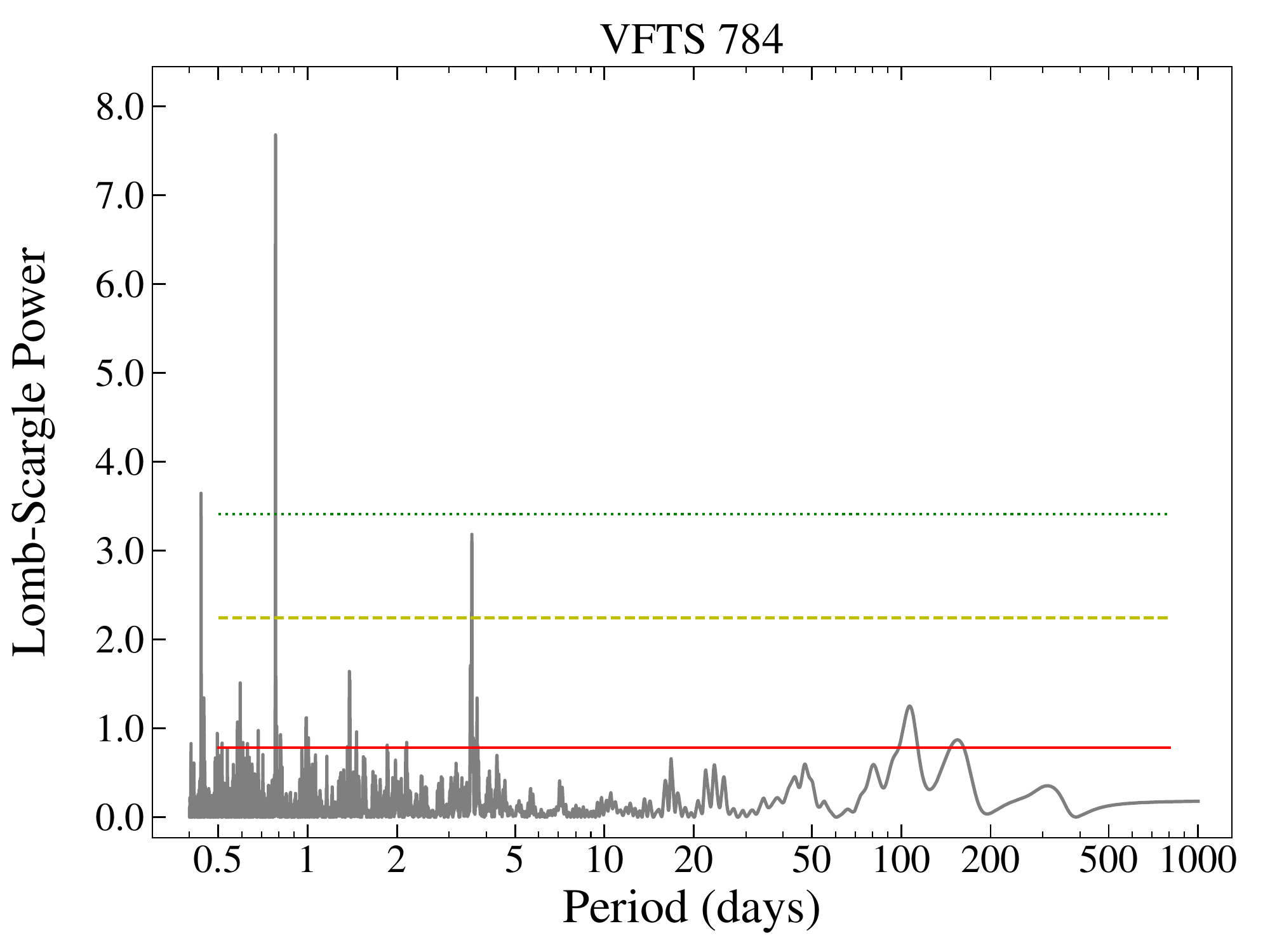}\hfill
    \includegraphics[width=0.31\textwidth]{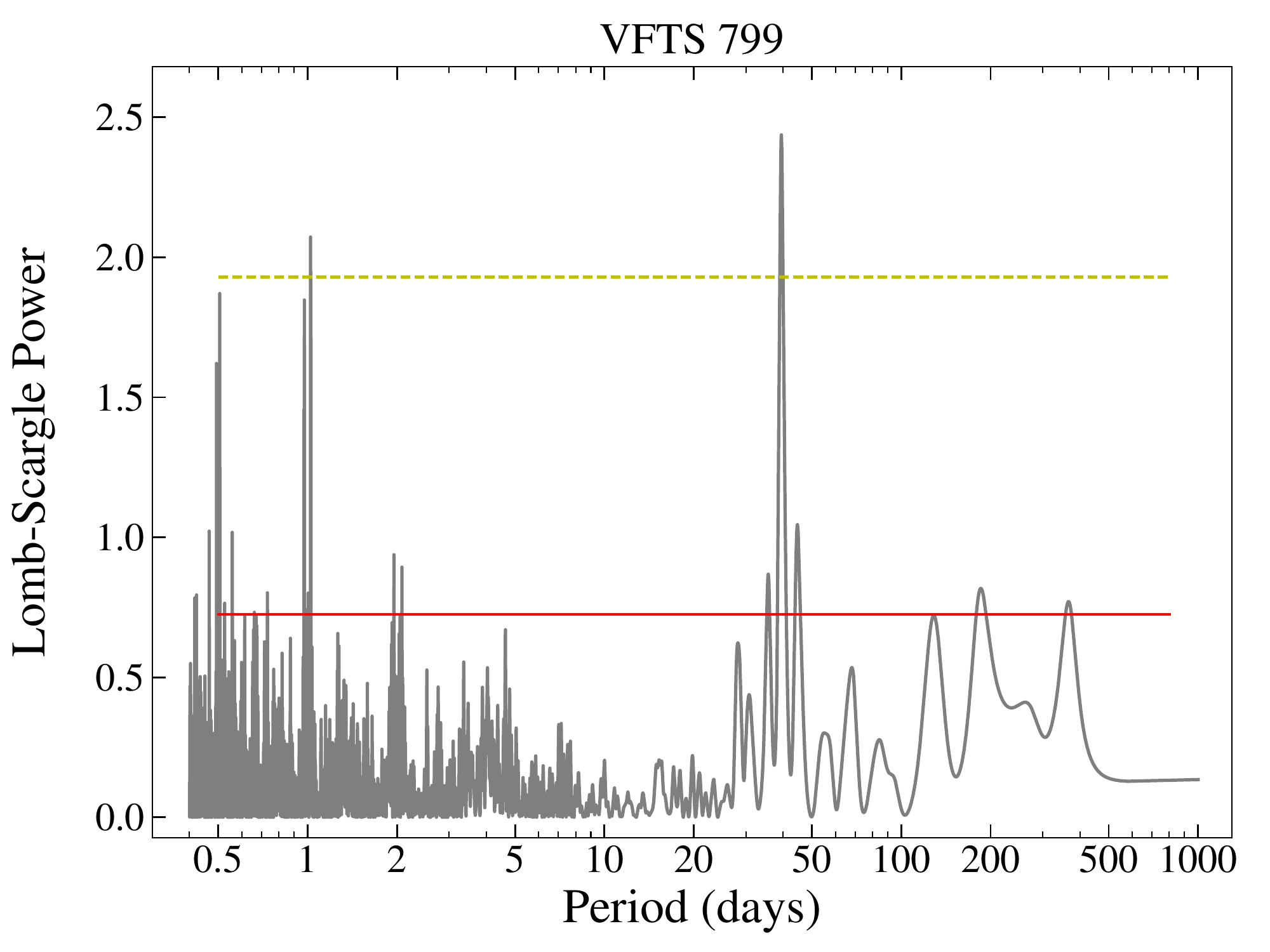}\hfill
    \includegraphics[width=0.31\textwidth]{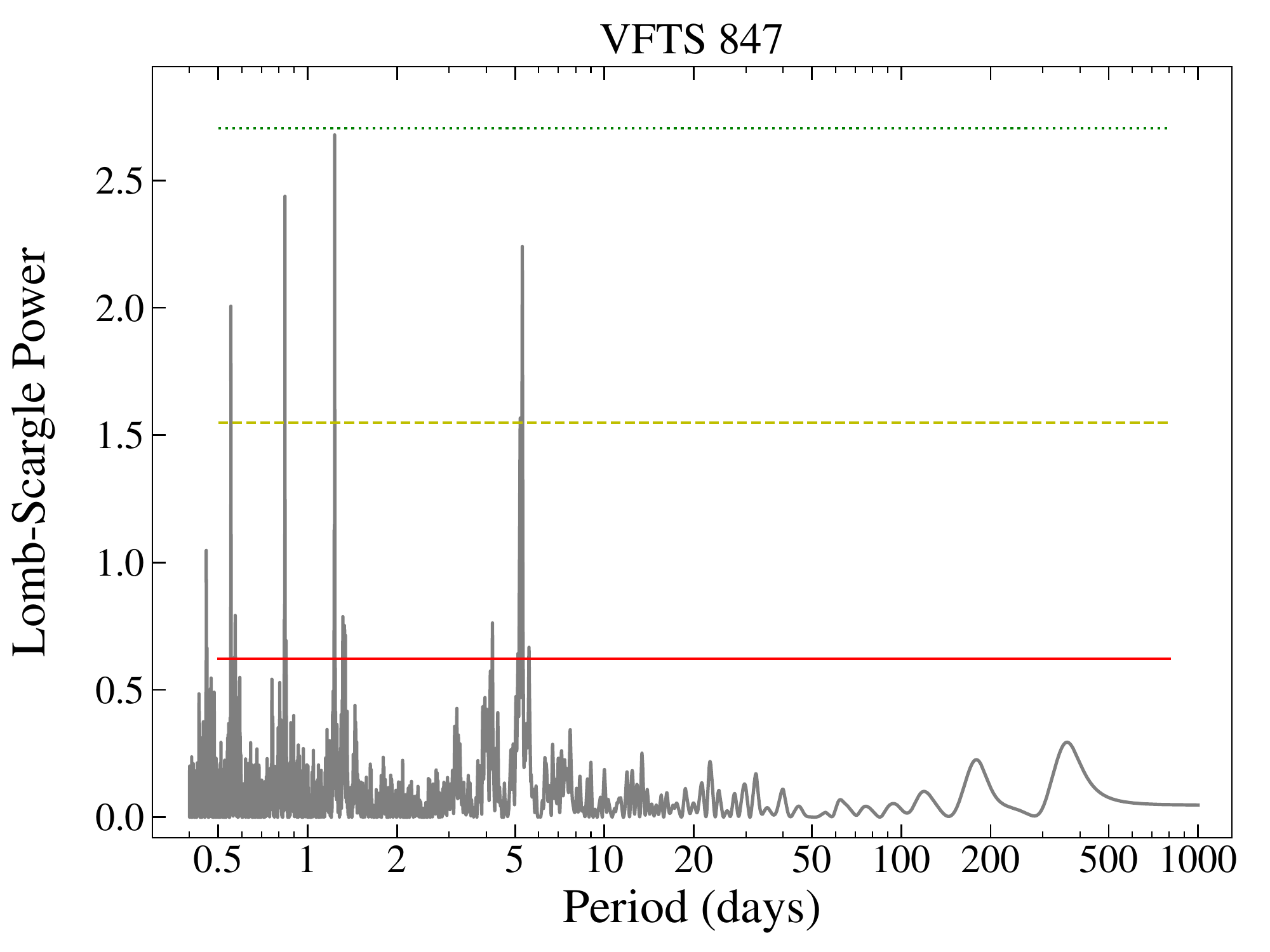}\hfill
    \includegraphics[width=0.31\textwidth]{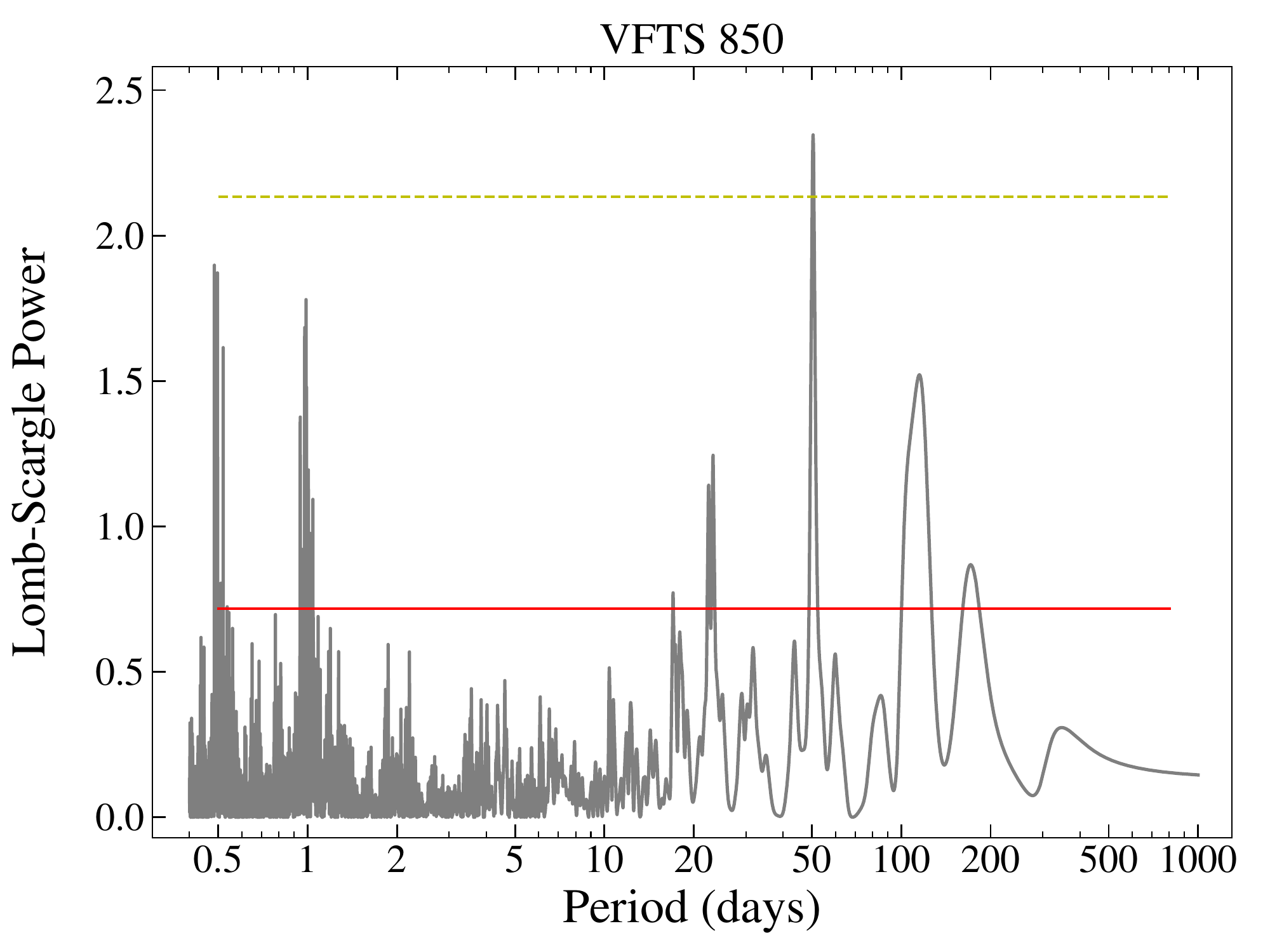}\hspace{0.03\textwidth}
    \includegraphics[width=0.31\textwidth]{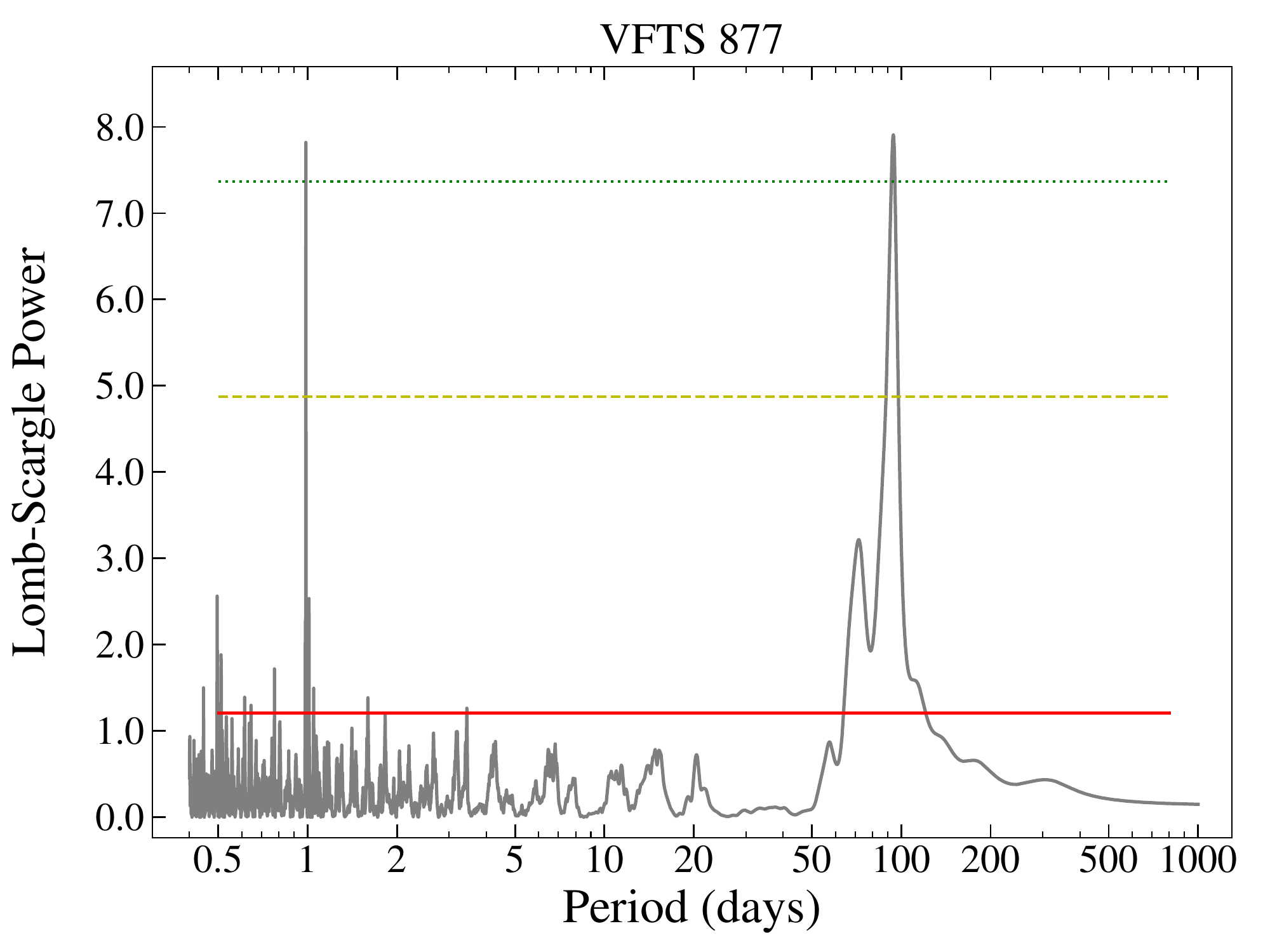}
    \caption{$-$ \it continued}
\end{figure*}

\begin{figure*}
    \centering
    \includegraphics[width=0.31\textwidth]{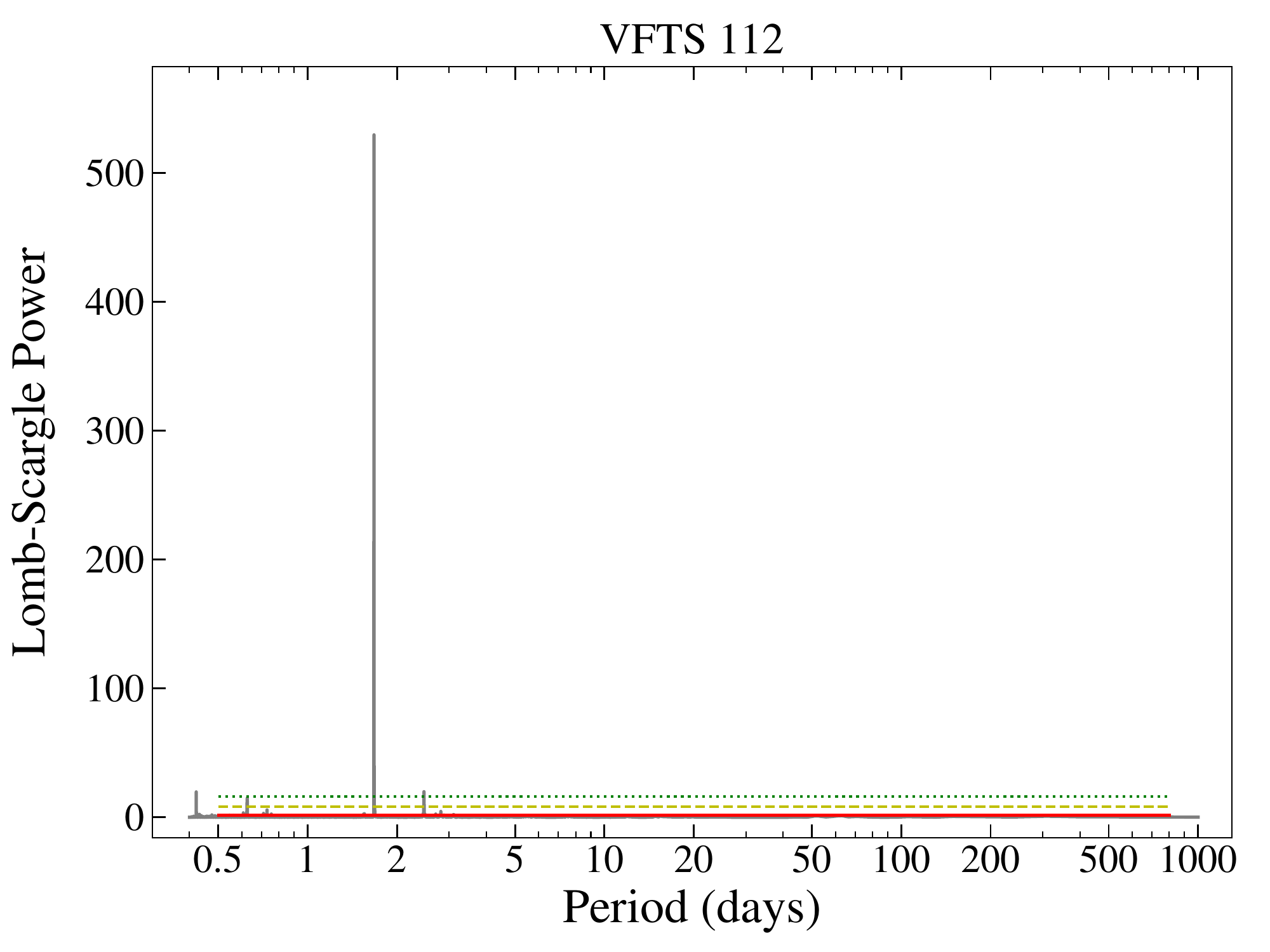}\hfill
    \includegraphics[width=0.31\textwidth]{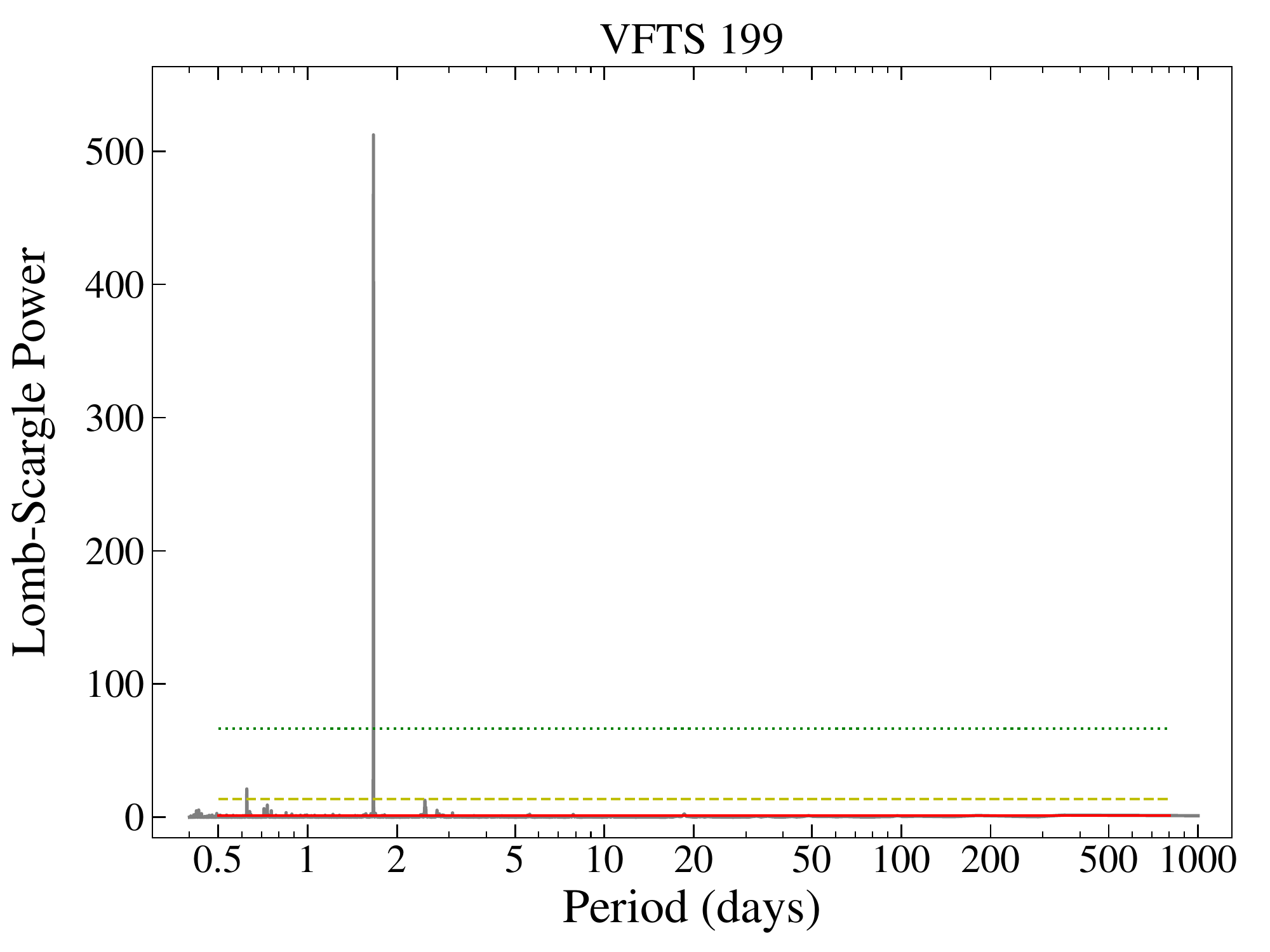}\hfill
    \includegraphics[width=0.31\textwidth]{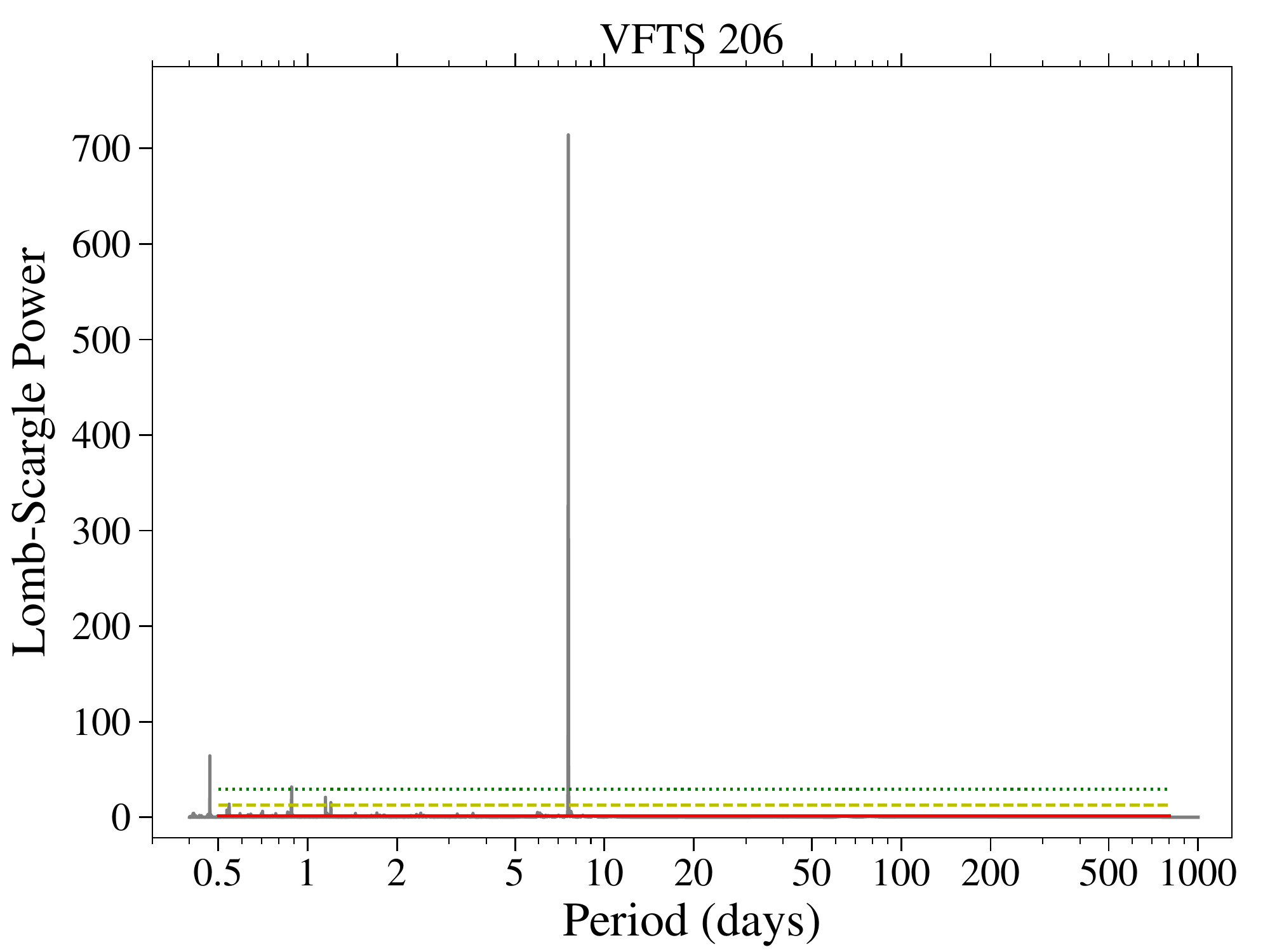}\hfill
    \includegraphics[width=0.31\textwidth]{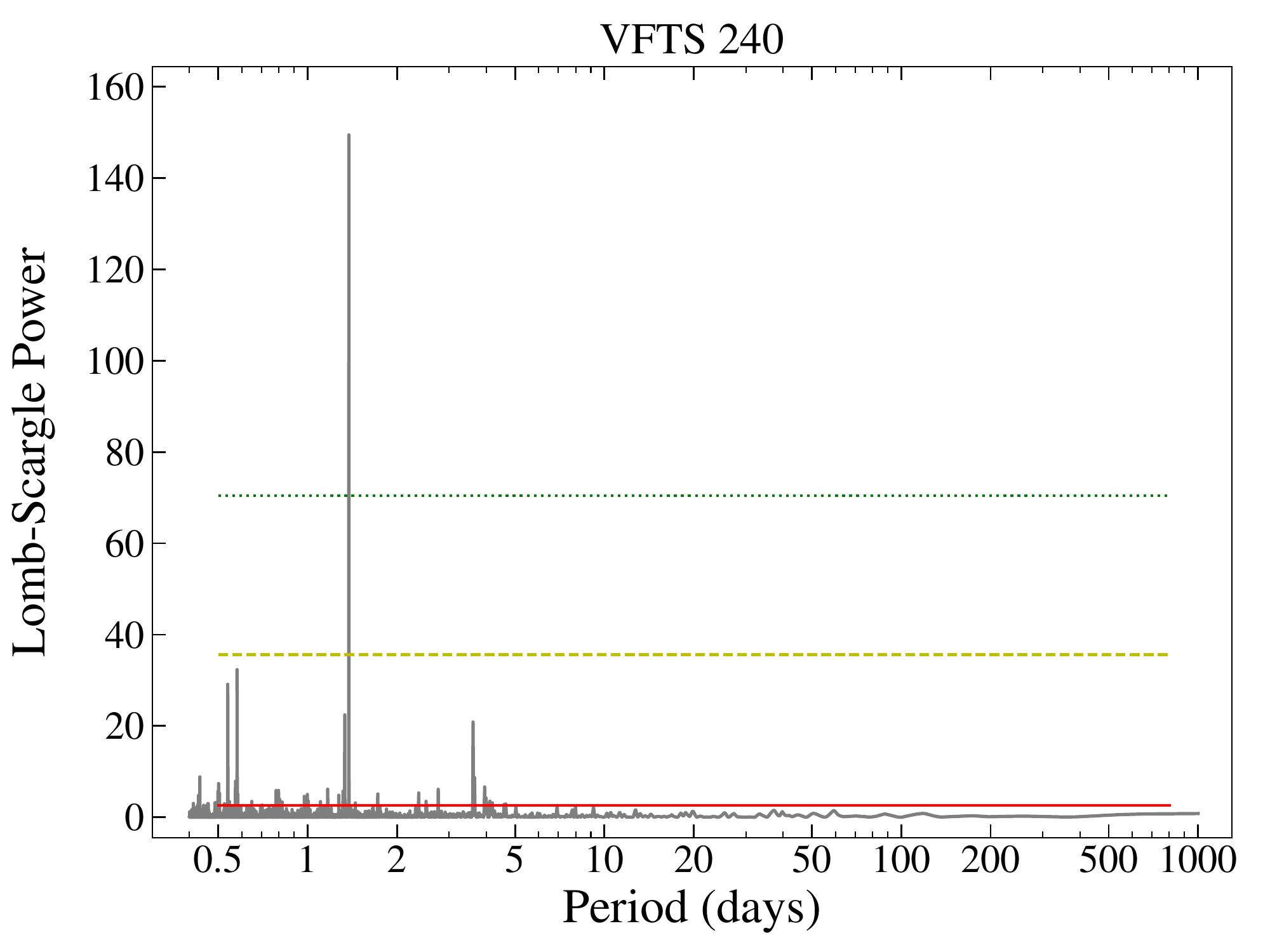}\hfill
    \includegraphics[width=0.31\textwidth]{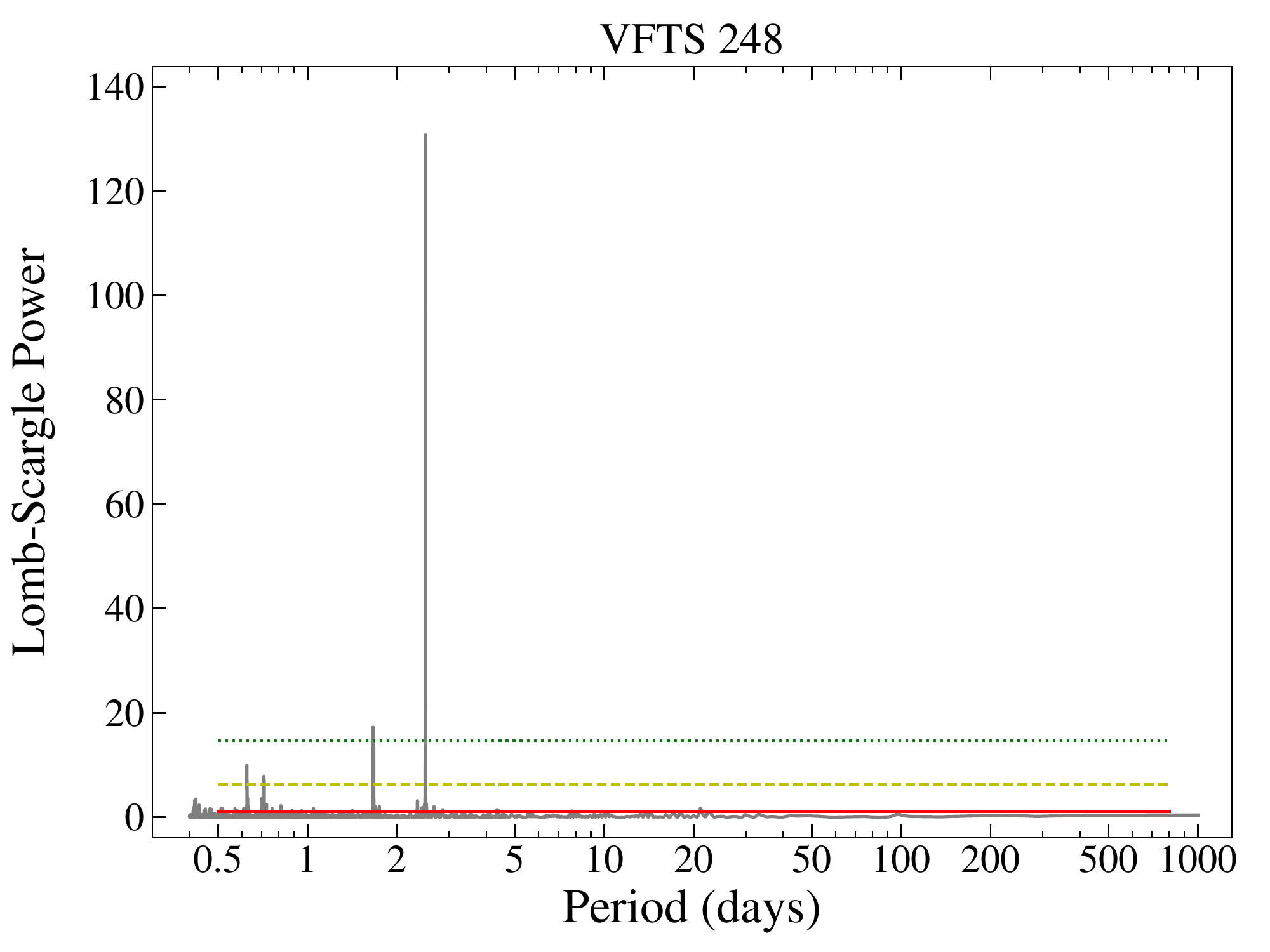}\hfill
    \includegraphics[width=0.31\textwidth]{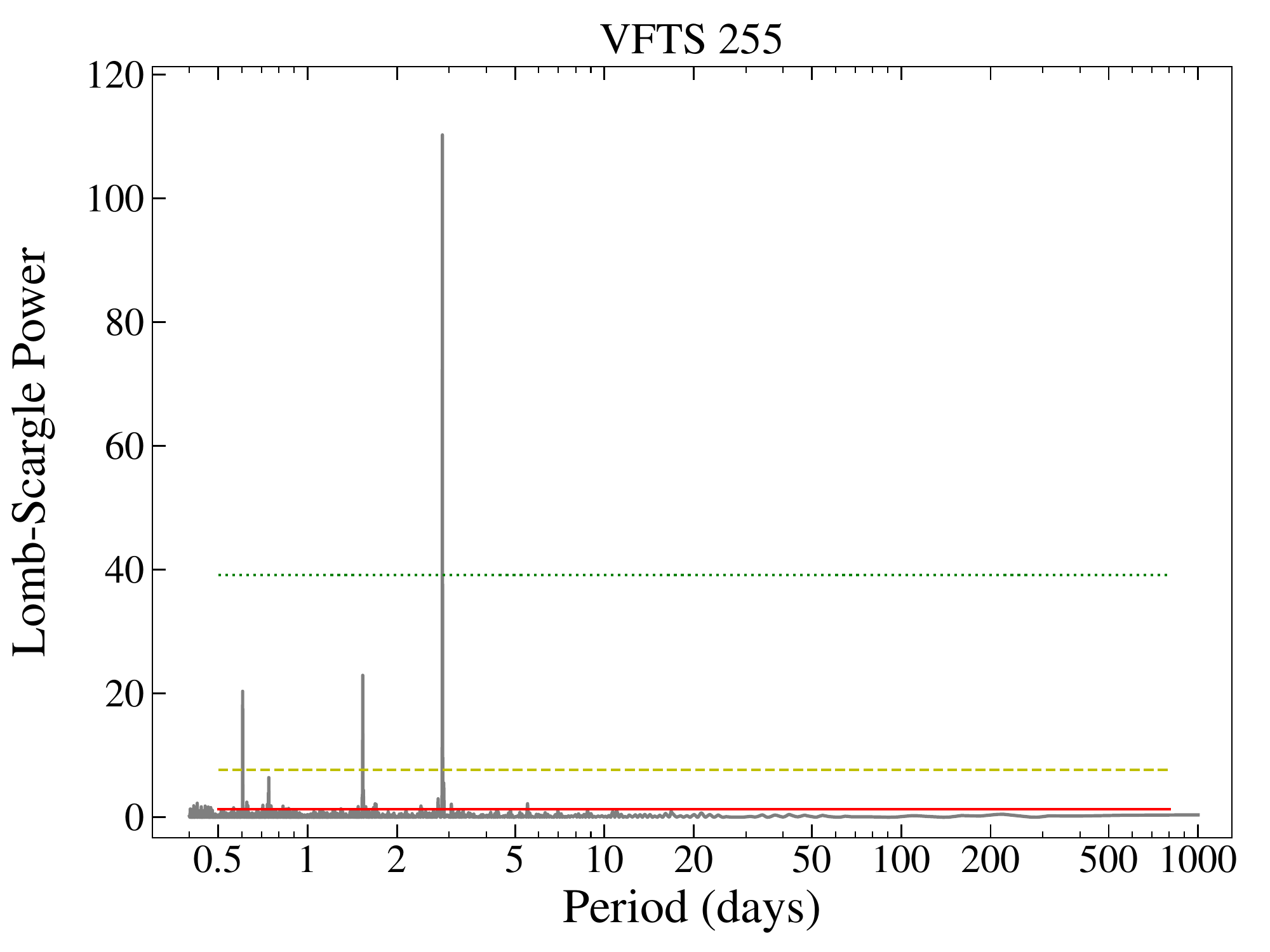}\hfill
    \includegraphics[width=0.31\textwidth]{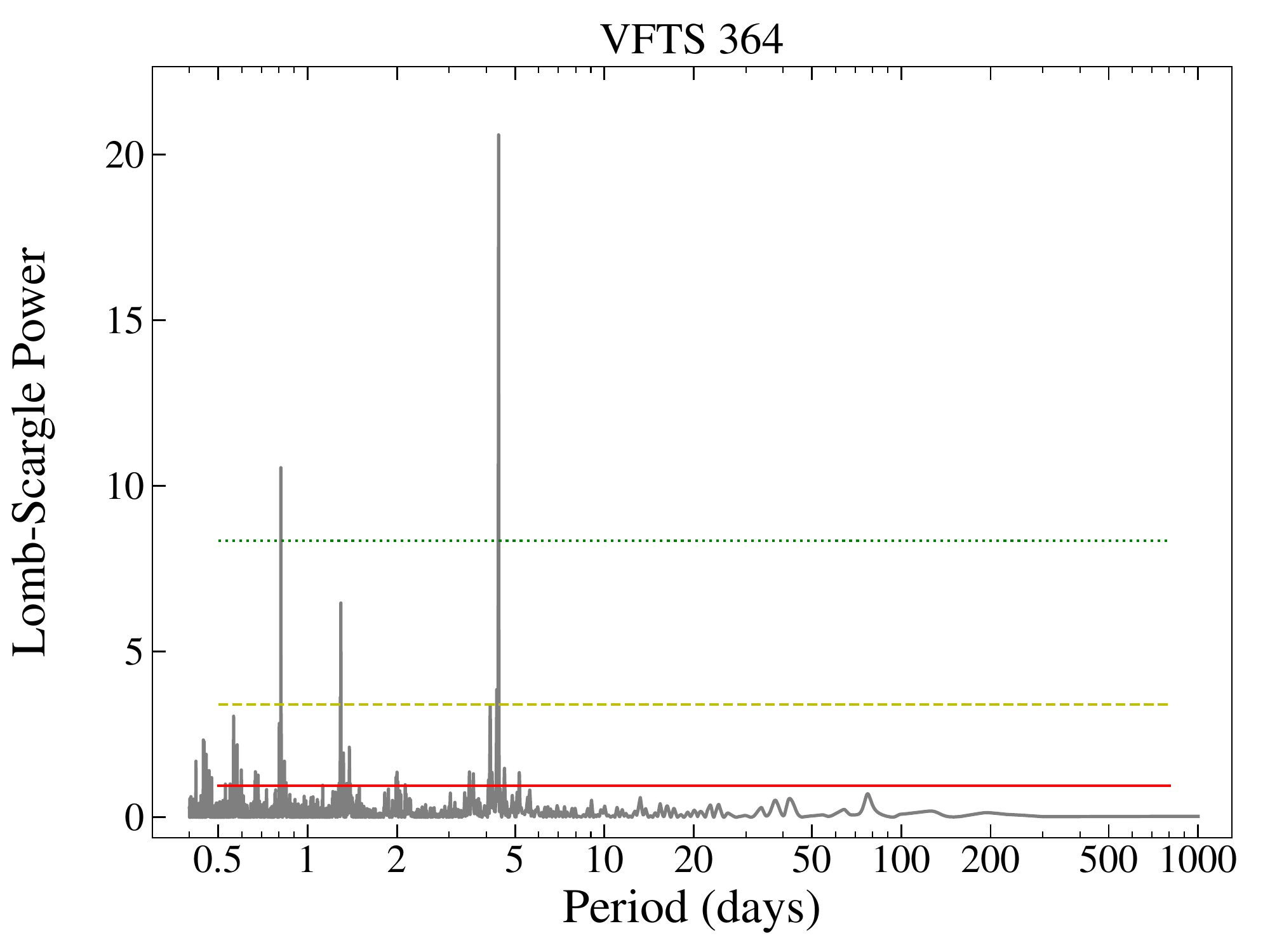}\hfill
    \includegraphics[width=0.31\textwidth]{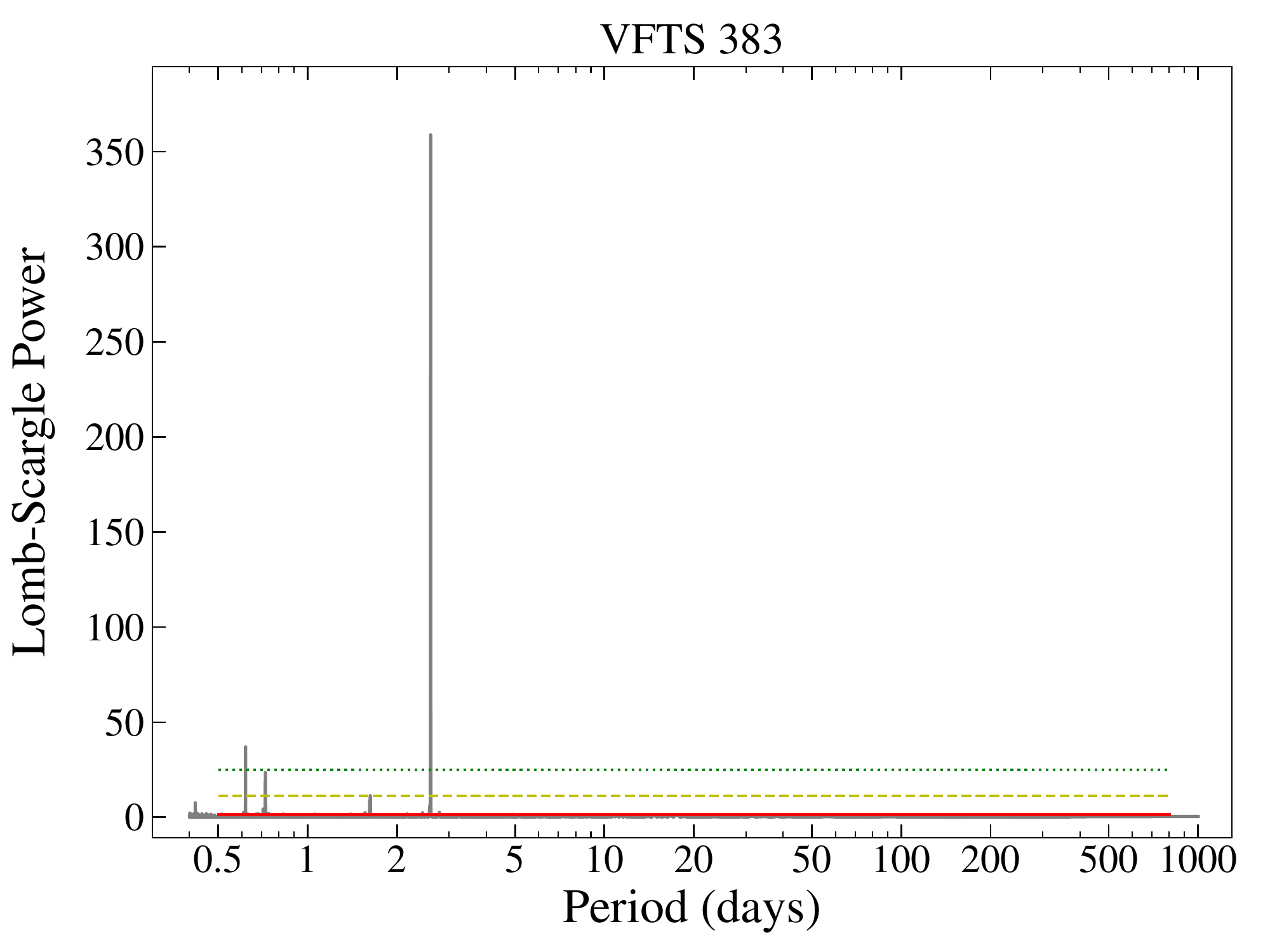}\hfill
    \includegraphics[width=0.31\textwidth]{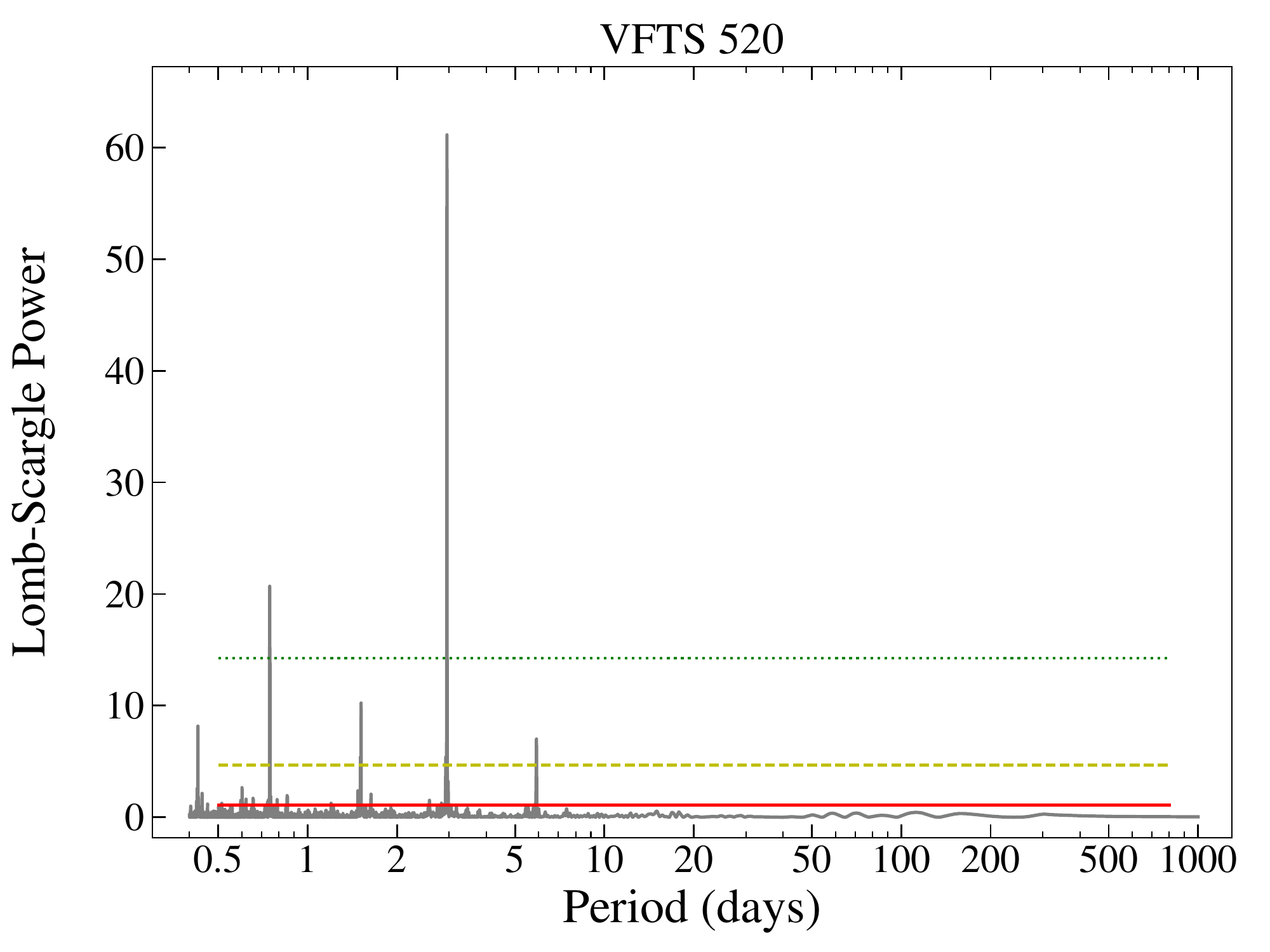}\hfill
    \includegraphics[width=0.31\textwidth]{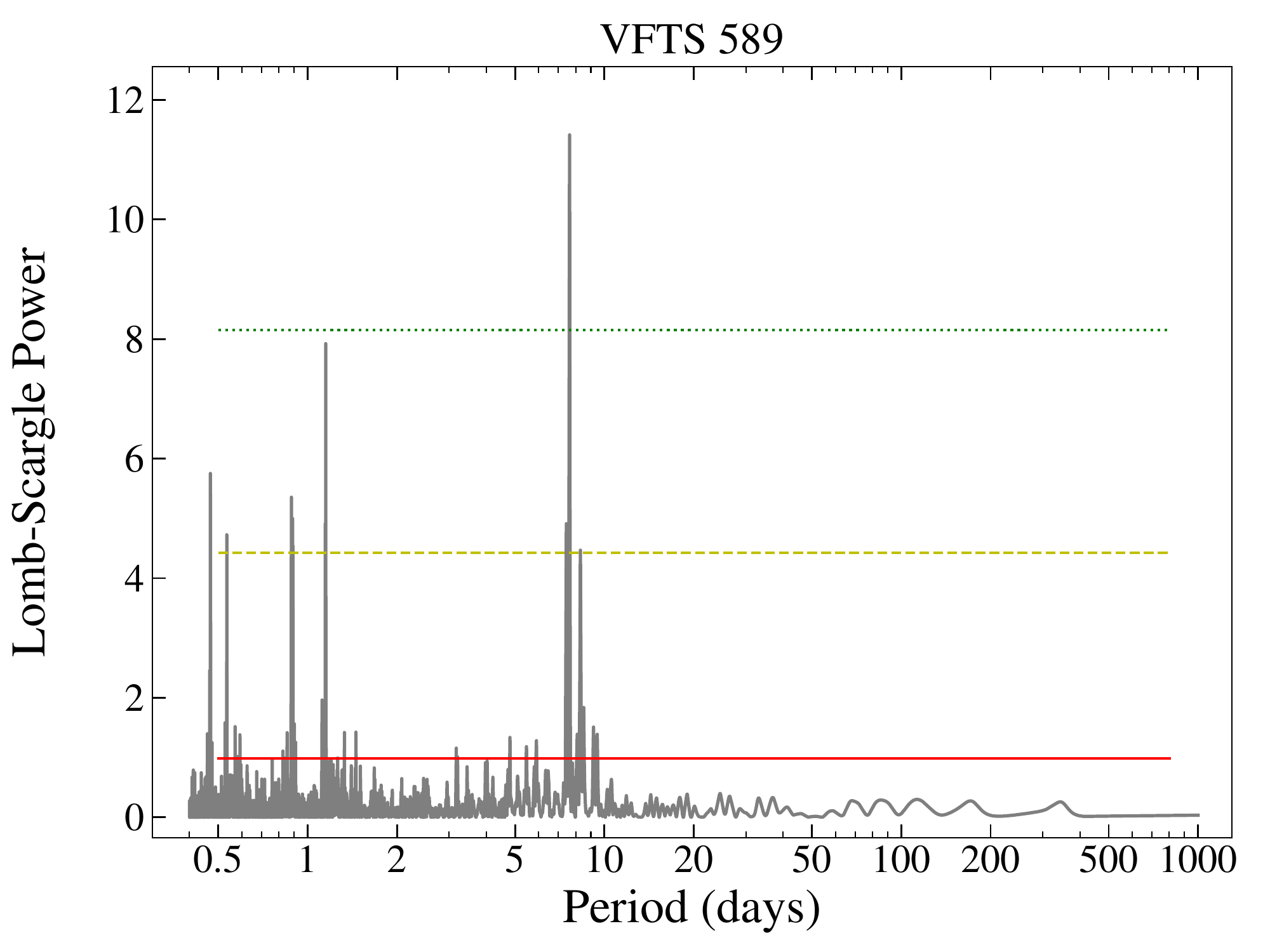}\hfill
    \includegraphics[width=0.31\textwidth]{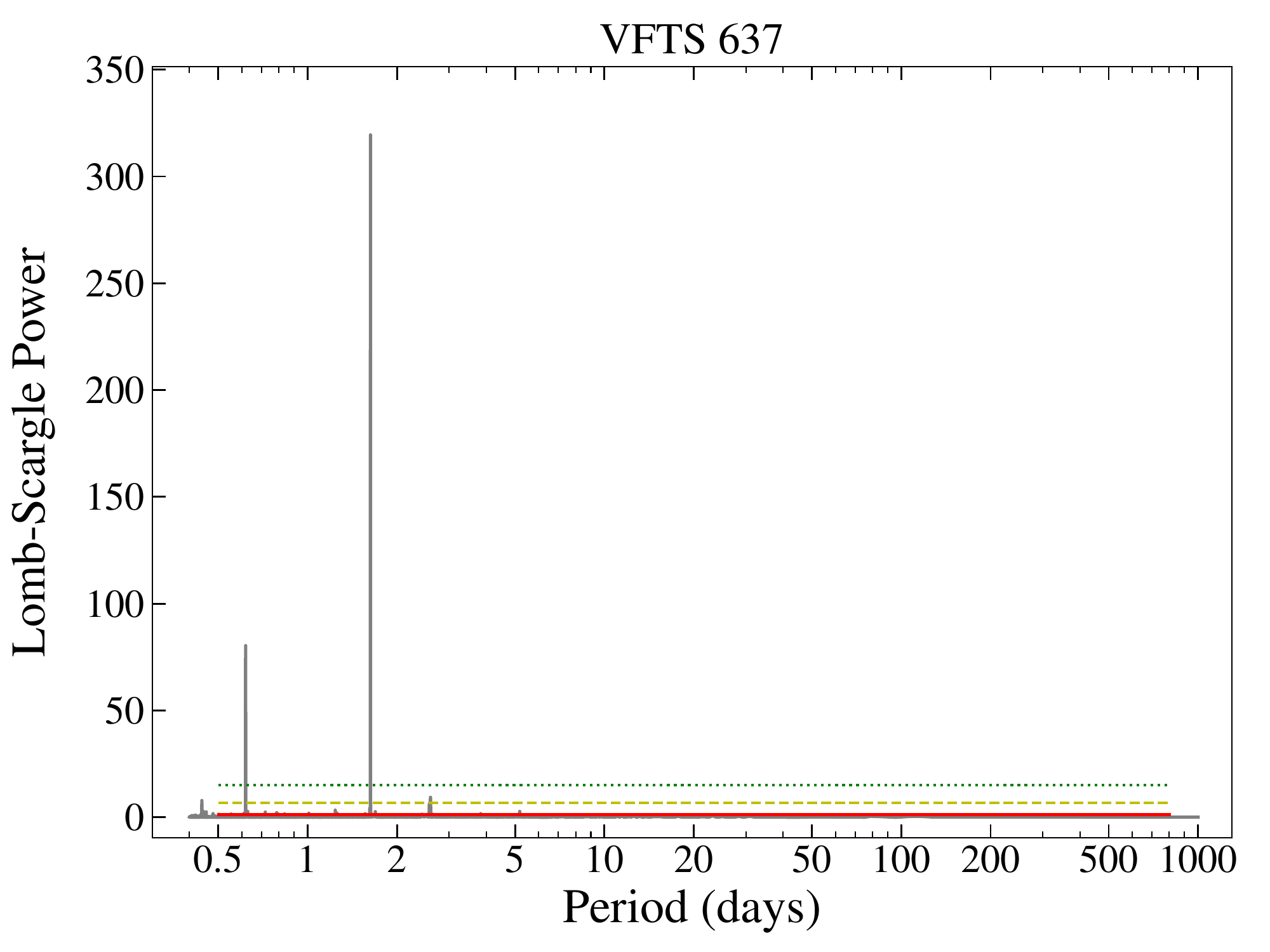}\hfill
    \includegraphics[width=0.31\textwidth]{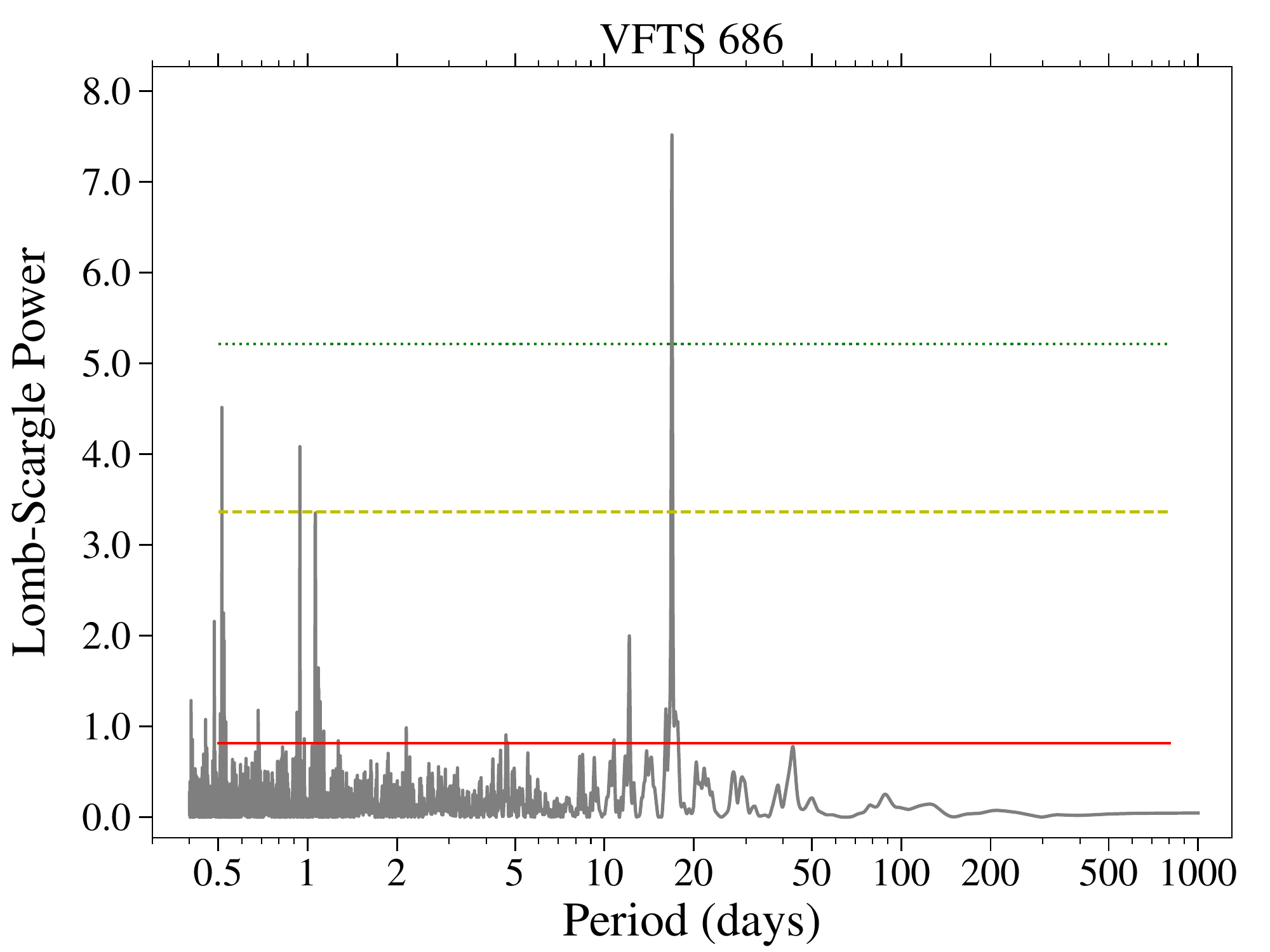}\hfill
    \includegraphics[width=0.31\textwidth]{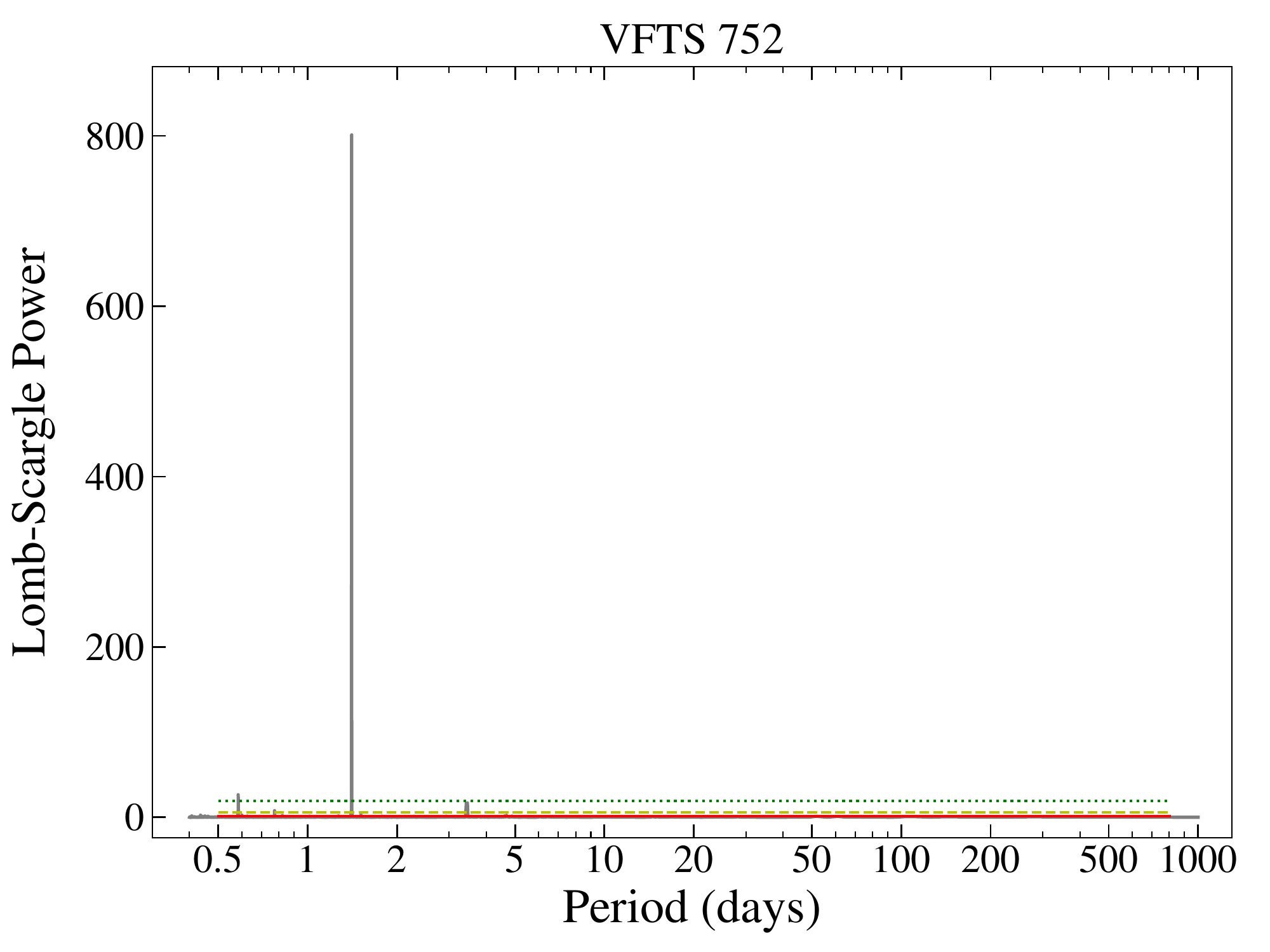}\hspace{0.03\textwidth}
    \includegraphics[width=0.31\textwidth]{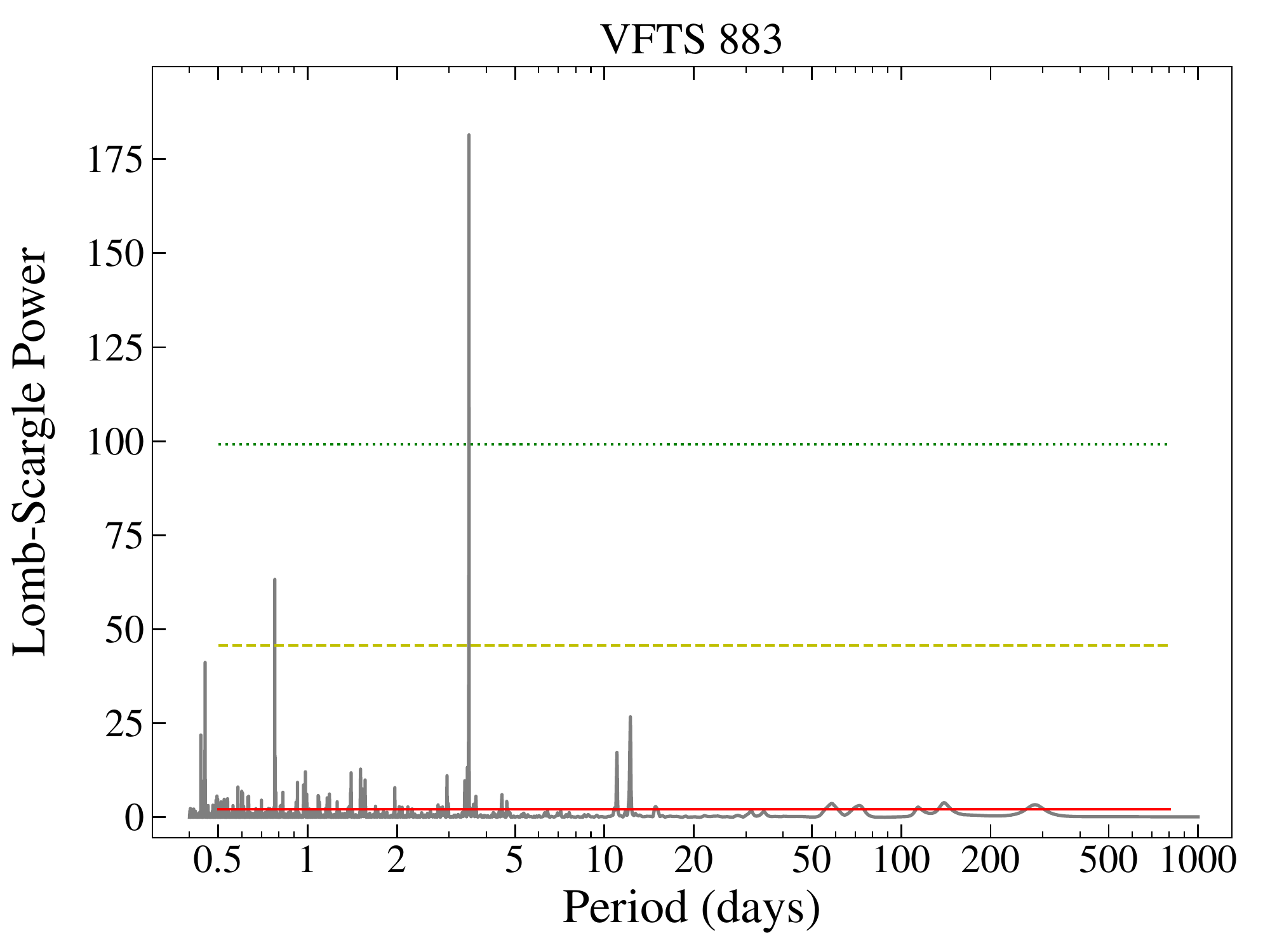}
\caption{Periodograms of the SB2 systems.}
\label{figAp:LSsb2}
\end{figure*}

\clearpage

% Appendix D
\section{RADIAL VELOCITY CURVES}\label{apx:RVcurves}
\setcounter{figure}{1}
\begin{myfloat}
    \centering
    \includegraphics[width=0.31\textwidth]{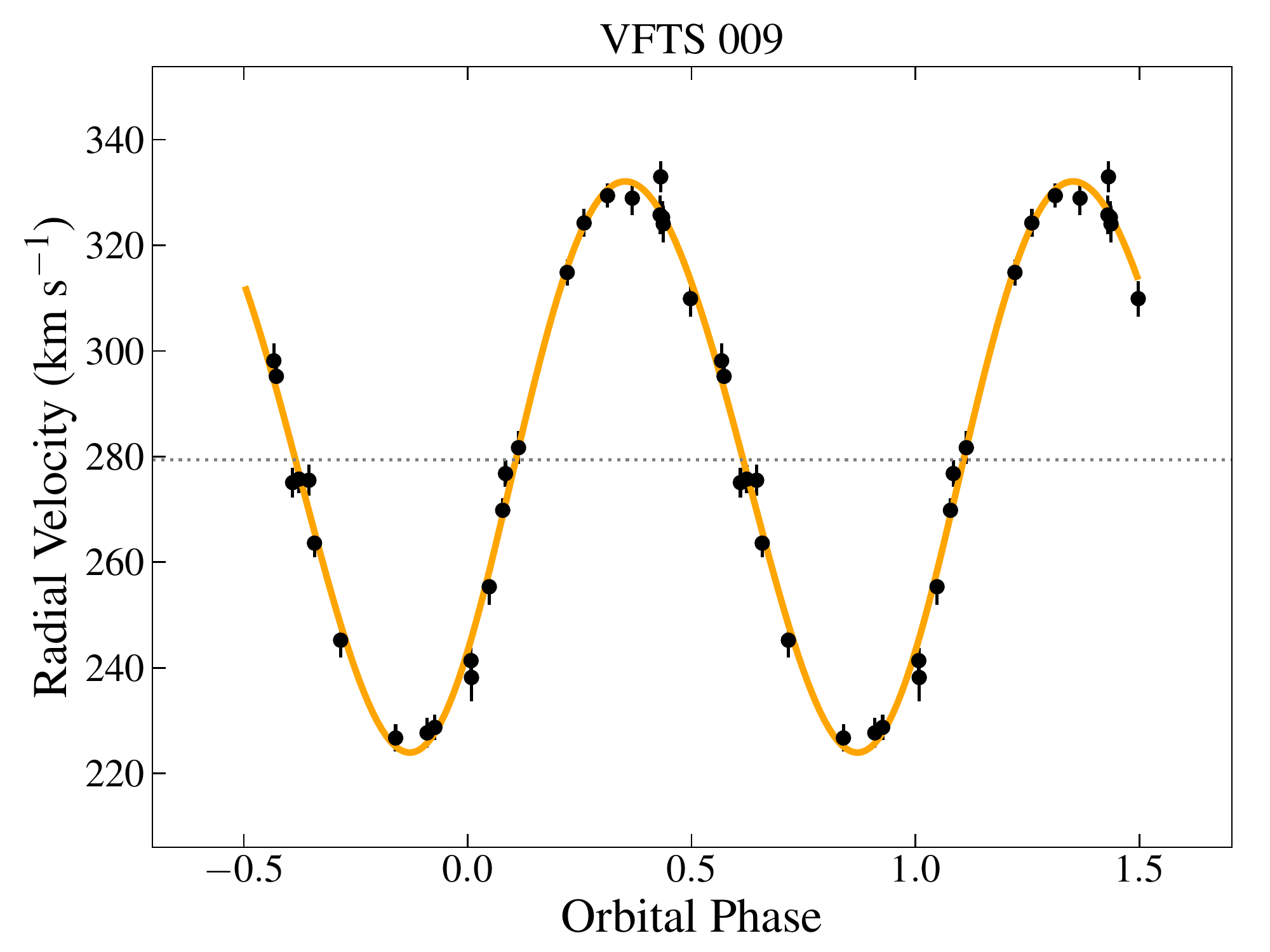}\hfill
    \includegraphics[width=0.31\textwidth]{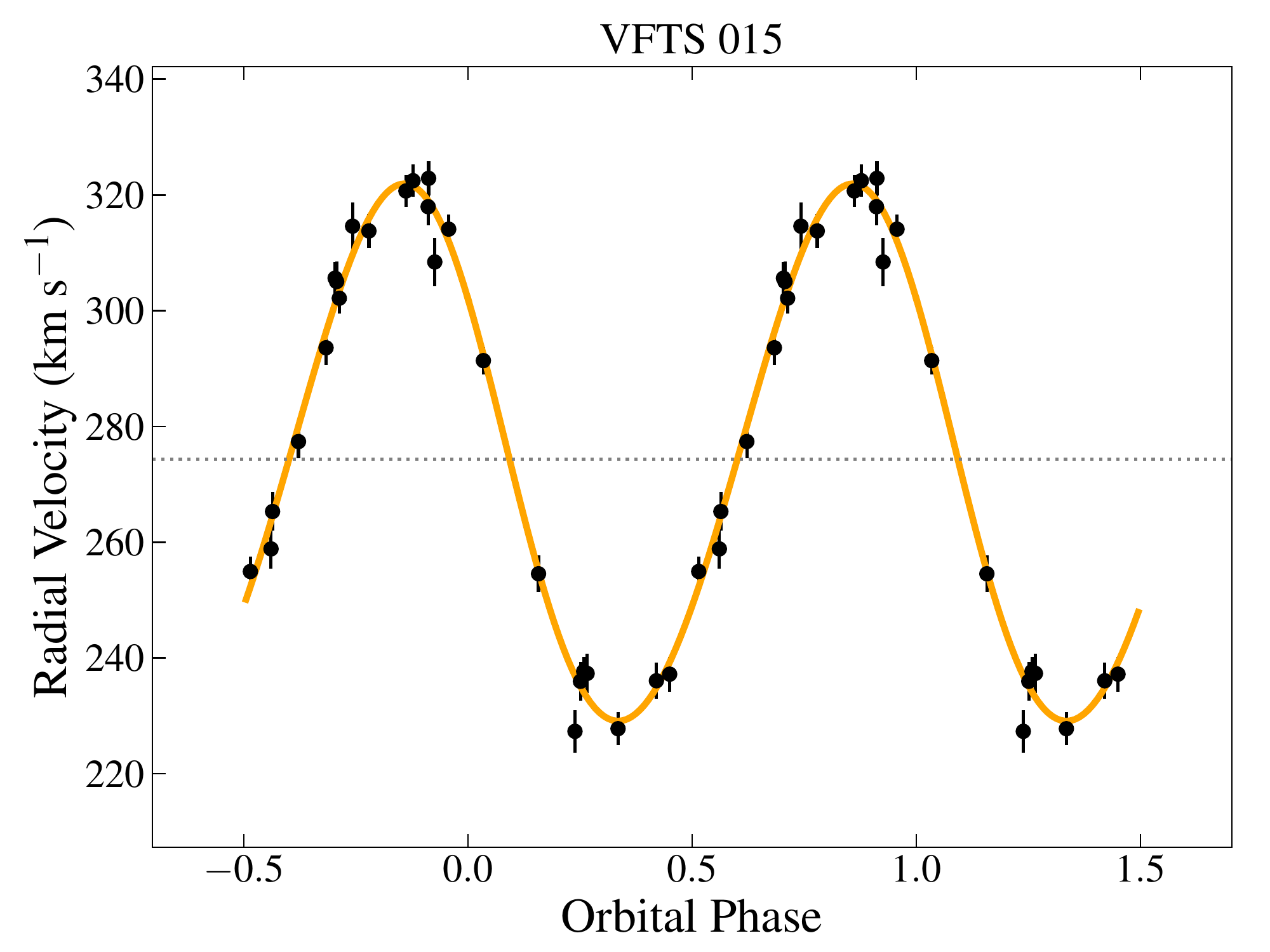}\hfill
    \includegraphics[width=0.31\textwidth]{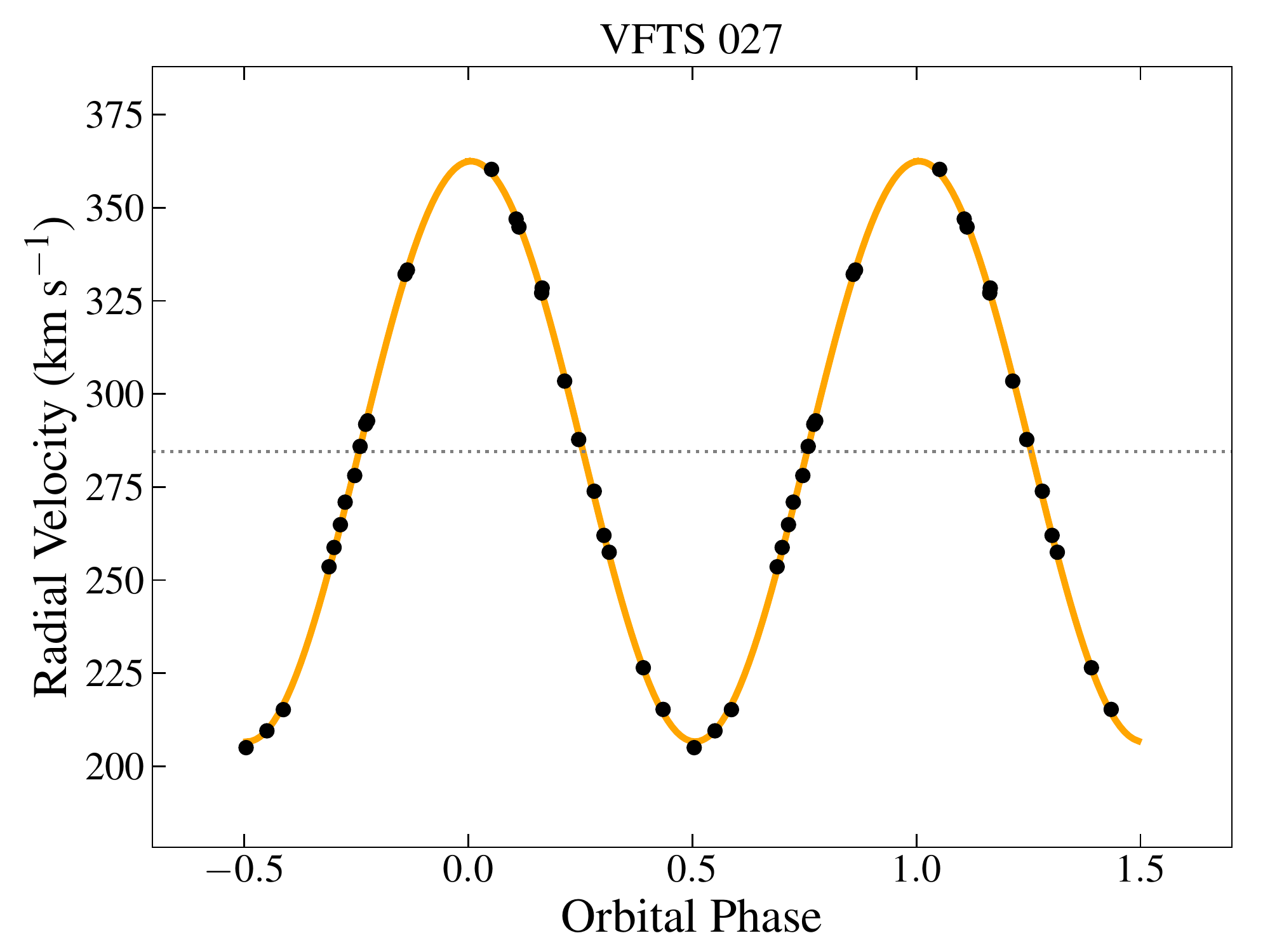}\hfill
    \includegraphics[width=0.31\textwidth]{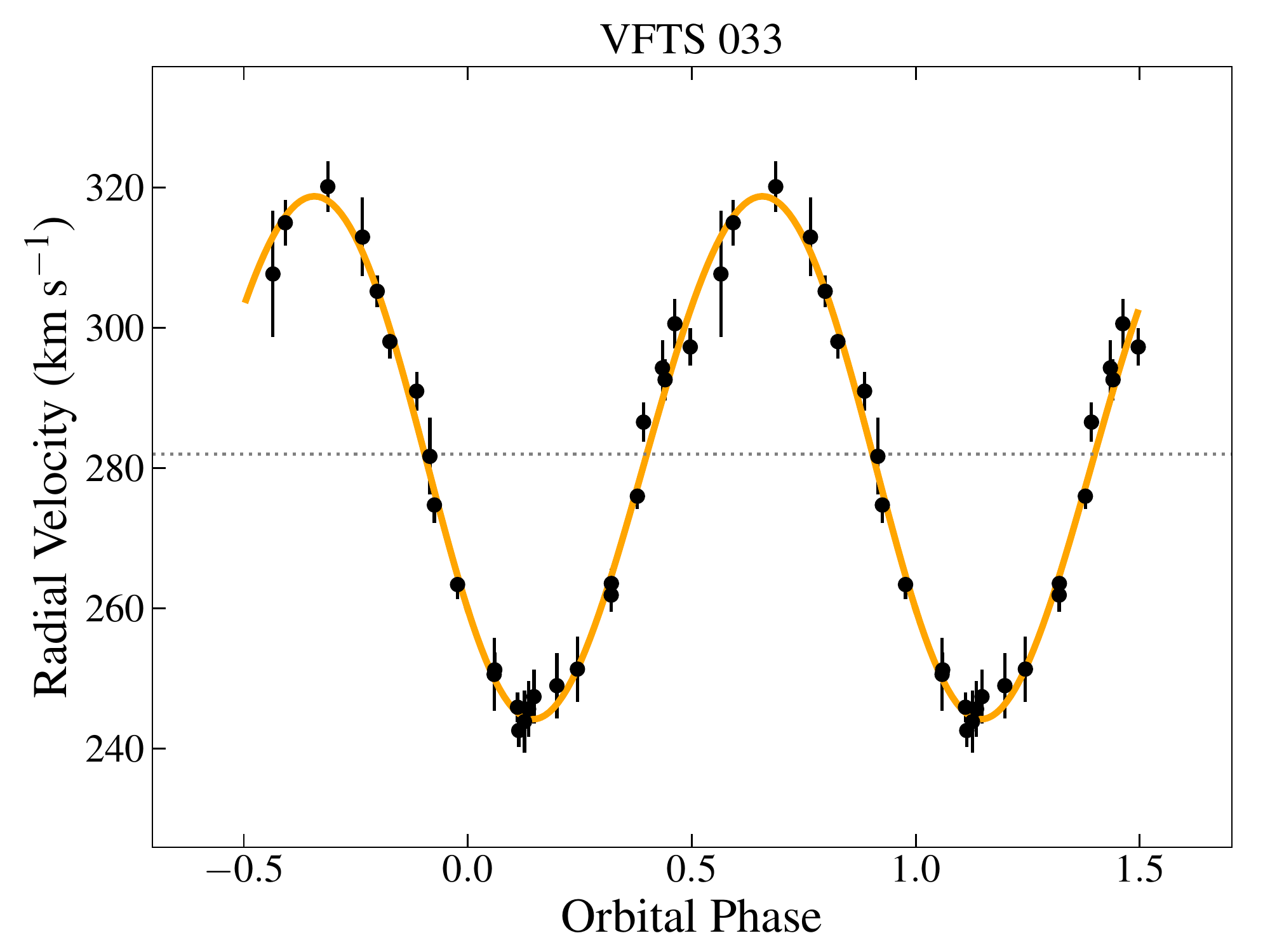}\hfill
    \includegraphics[width=0.31\textwidth]{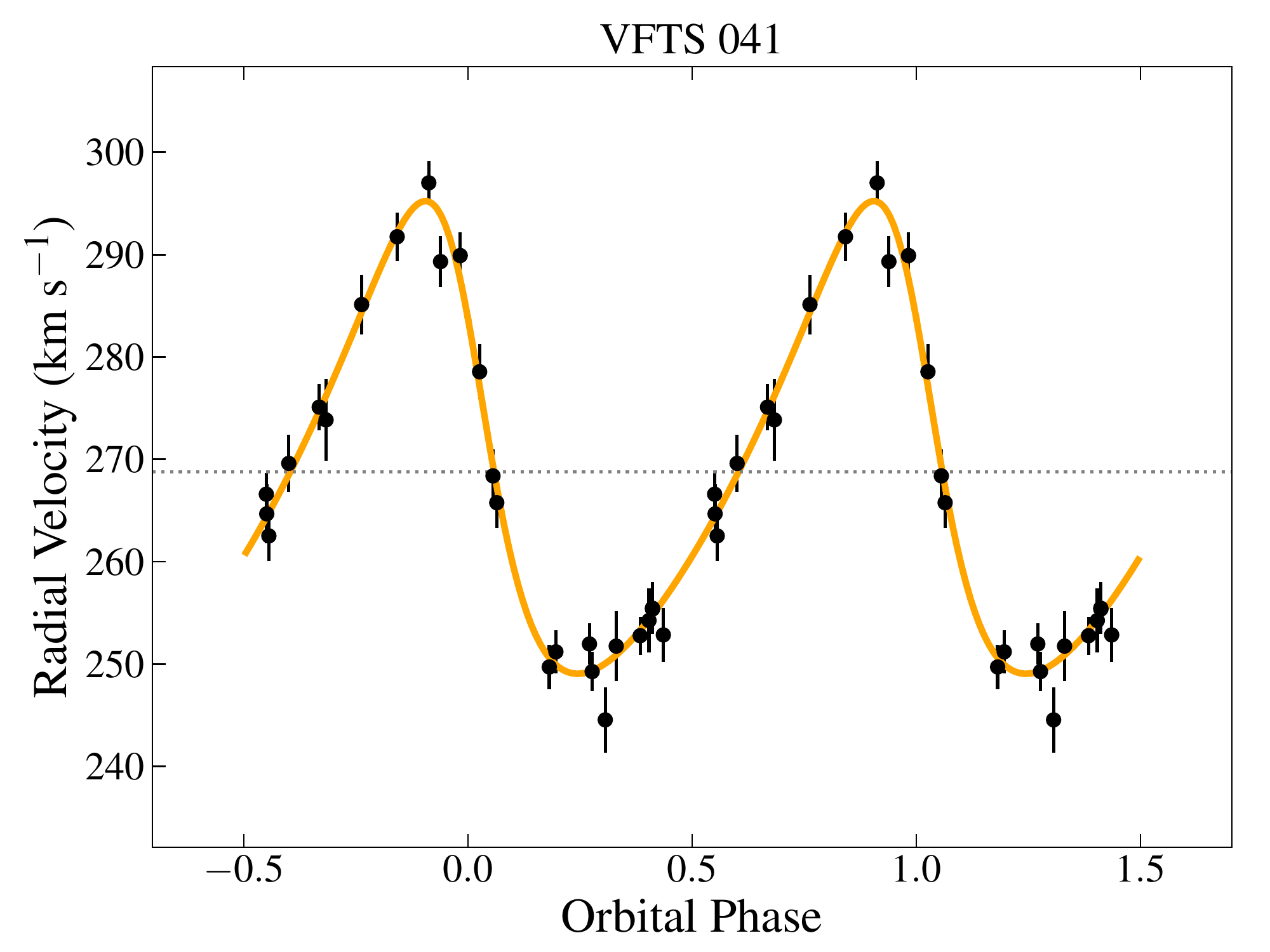}\hfill
    \includegraphics[width=0.31\textwidth]{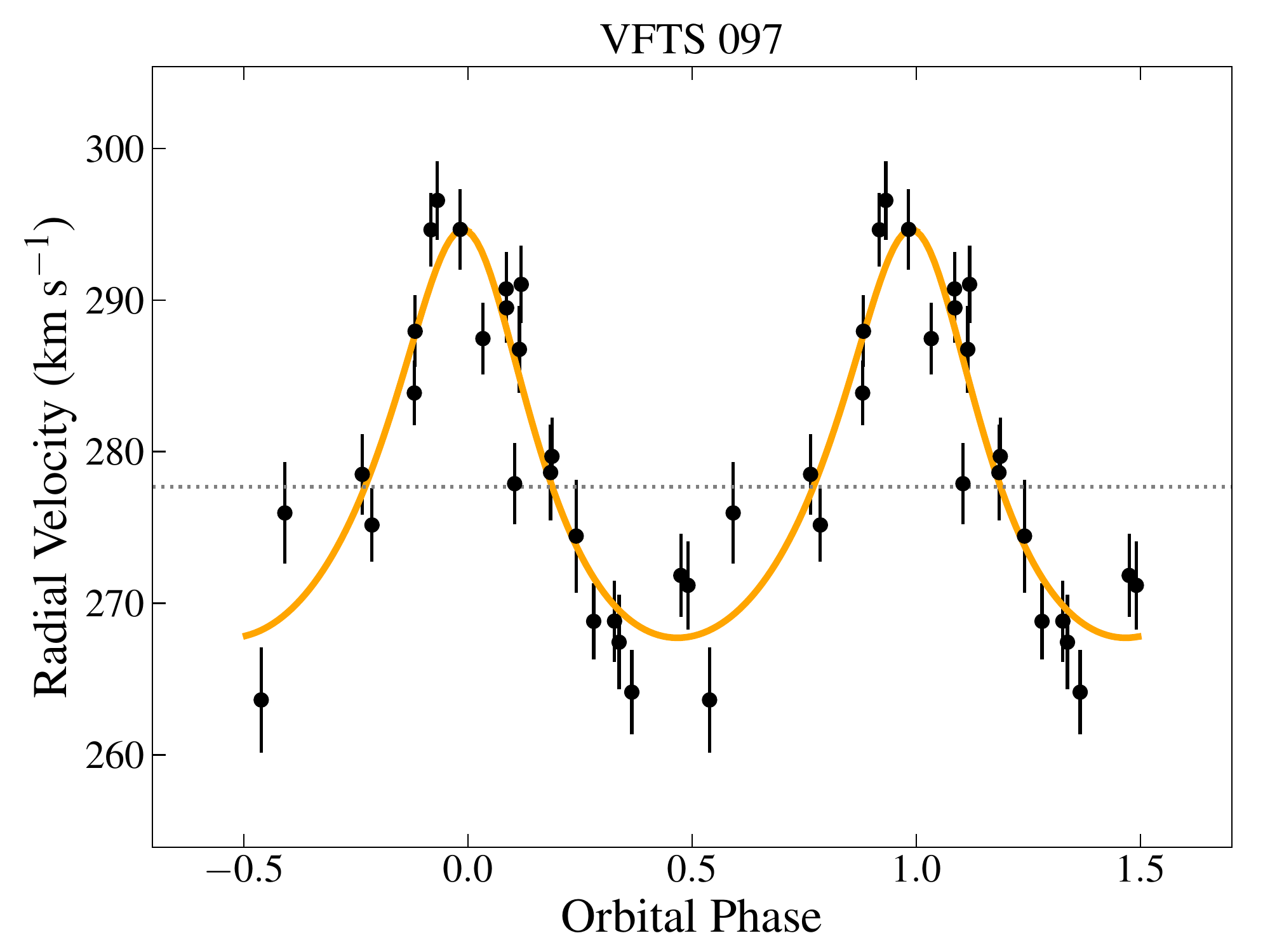}\hfill
    \includegraphics[width=0.31\textwidth]{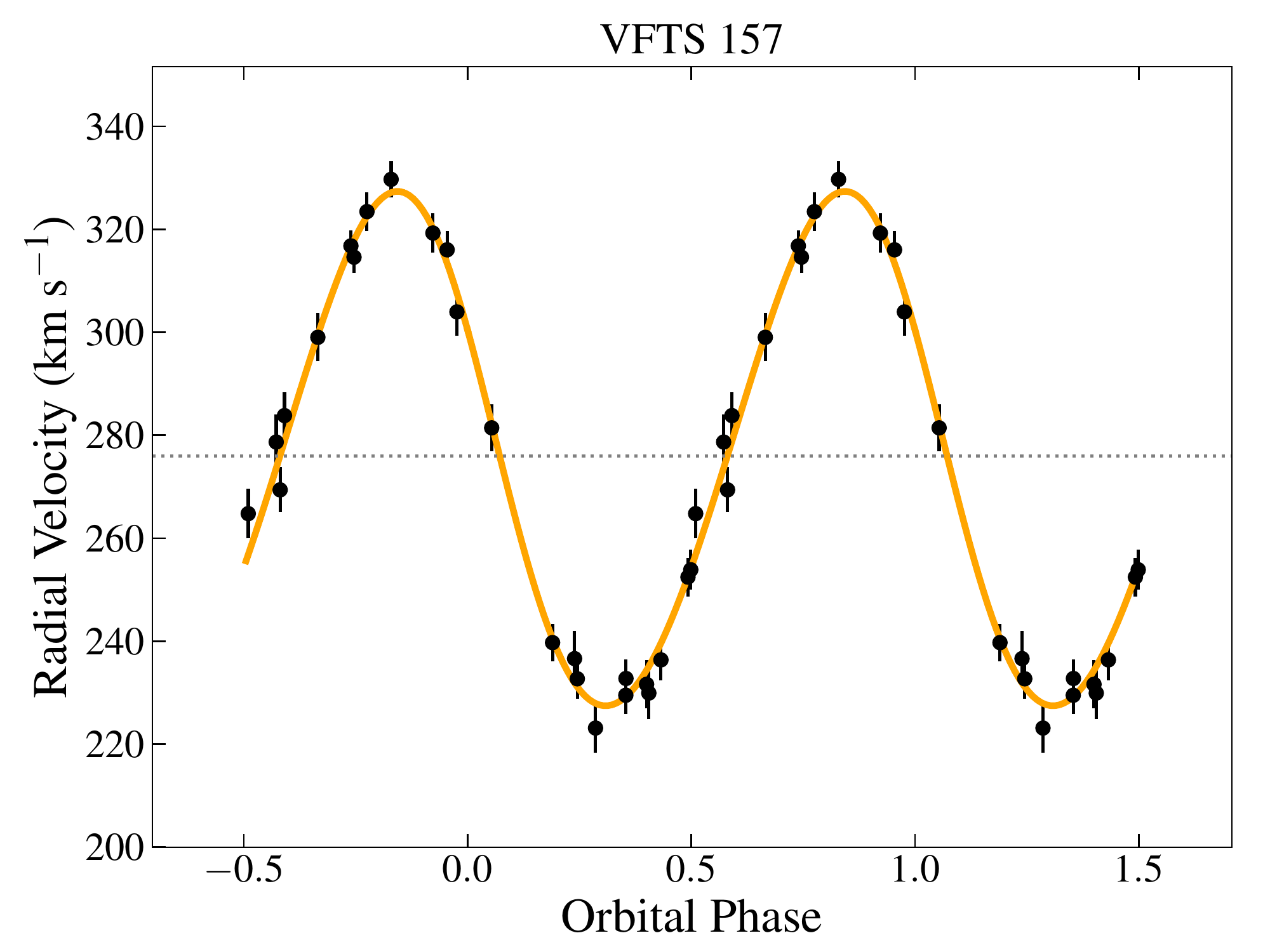}\hfill
    \includegraphics[width=0.31\textwidth]{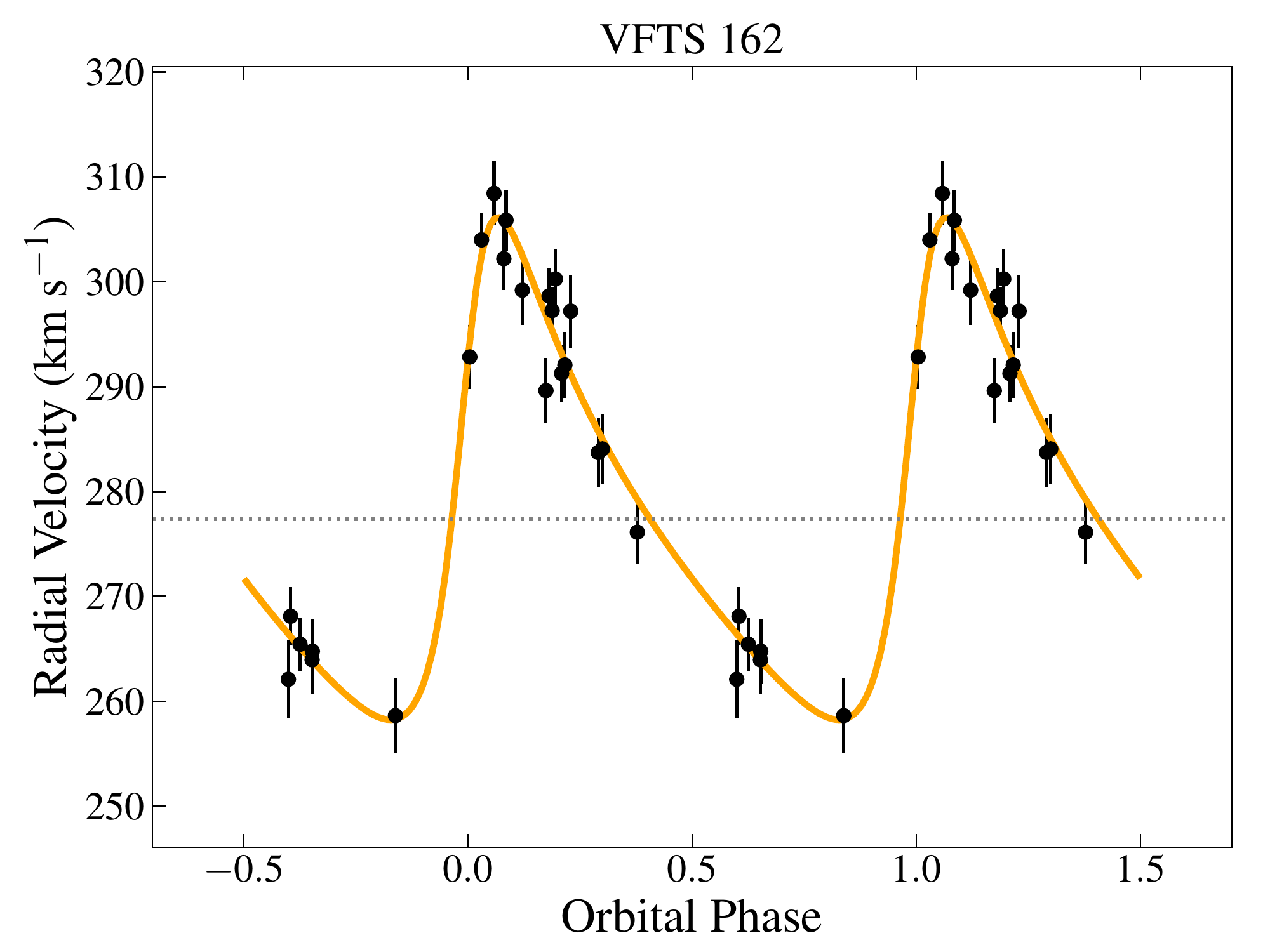}\hfill
    \includegraphics[width=0.31\textwidth]{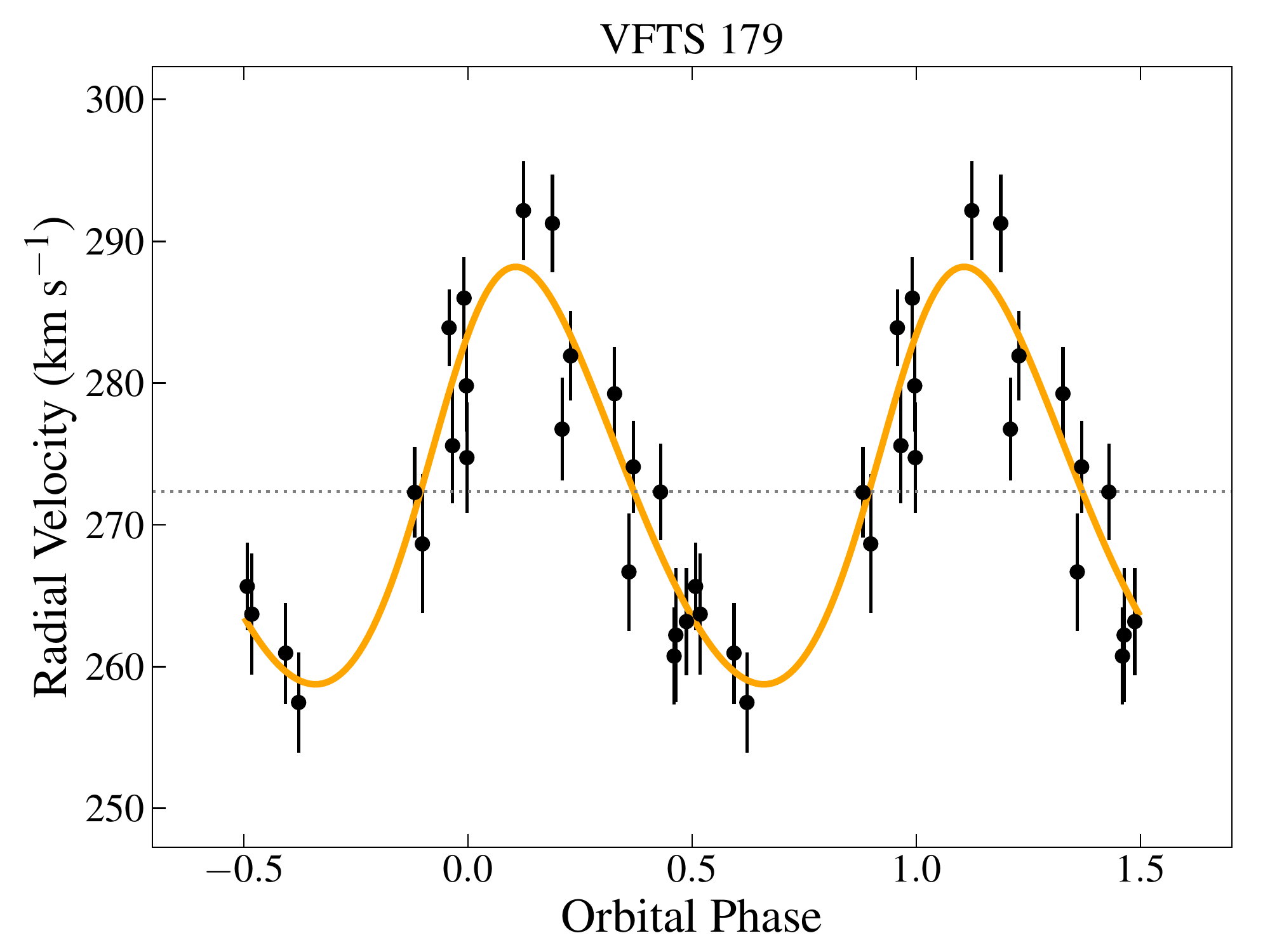}\hfill
    \includegraphics[width=0.31\textwidth]{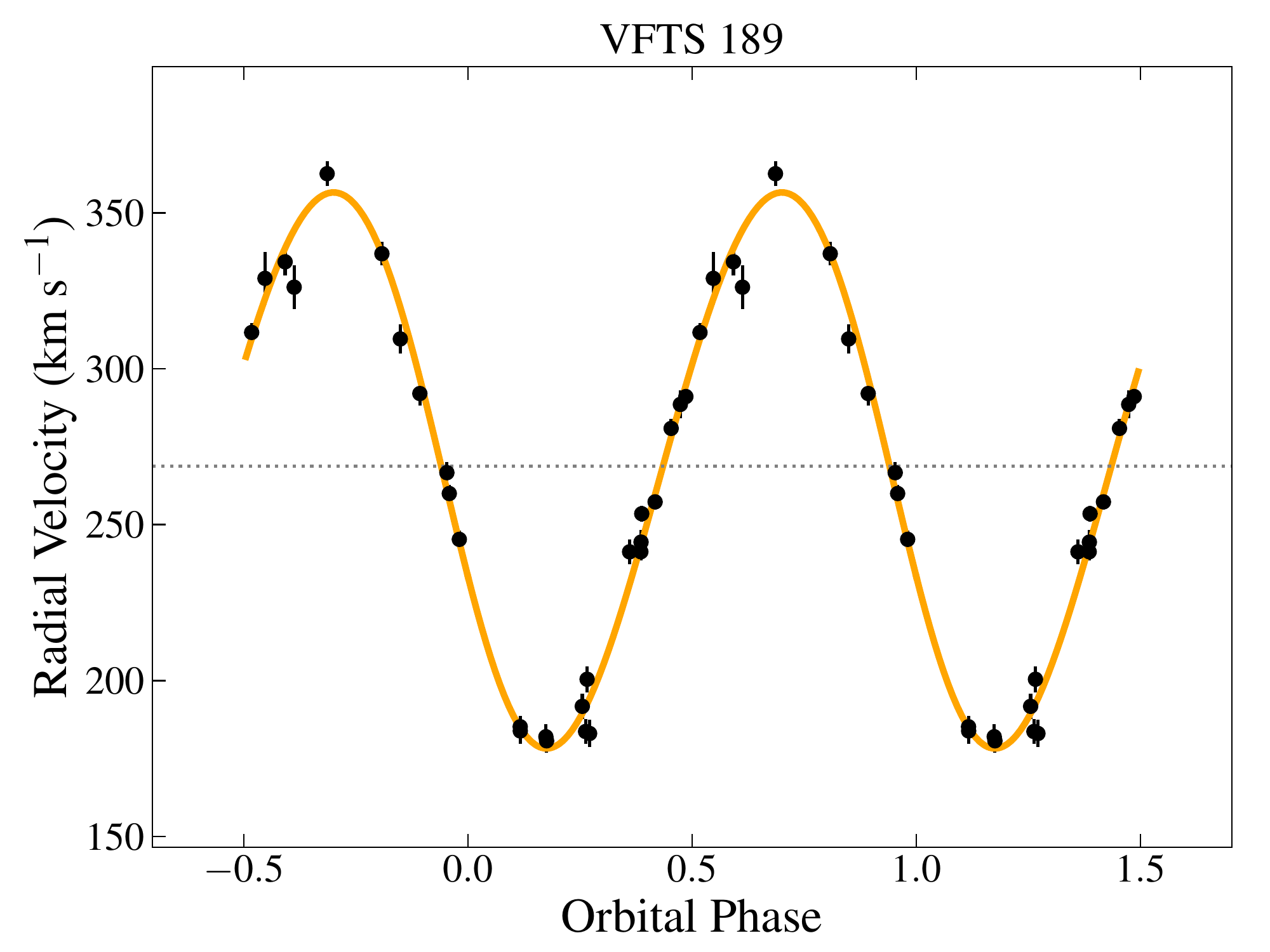}\hfill
    \includegraphics[width=0.31\textwidth]{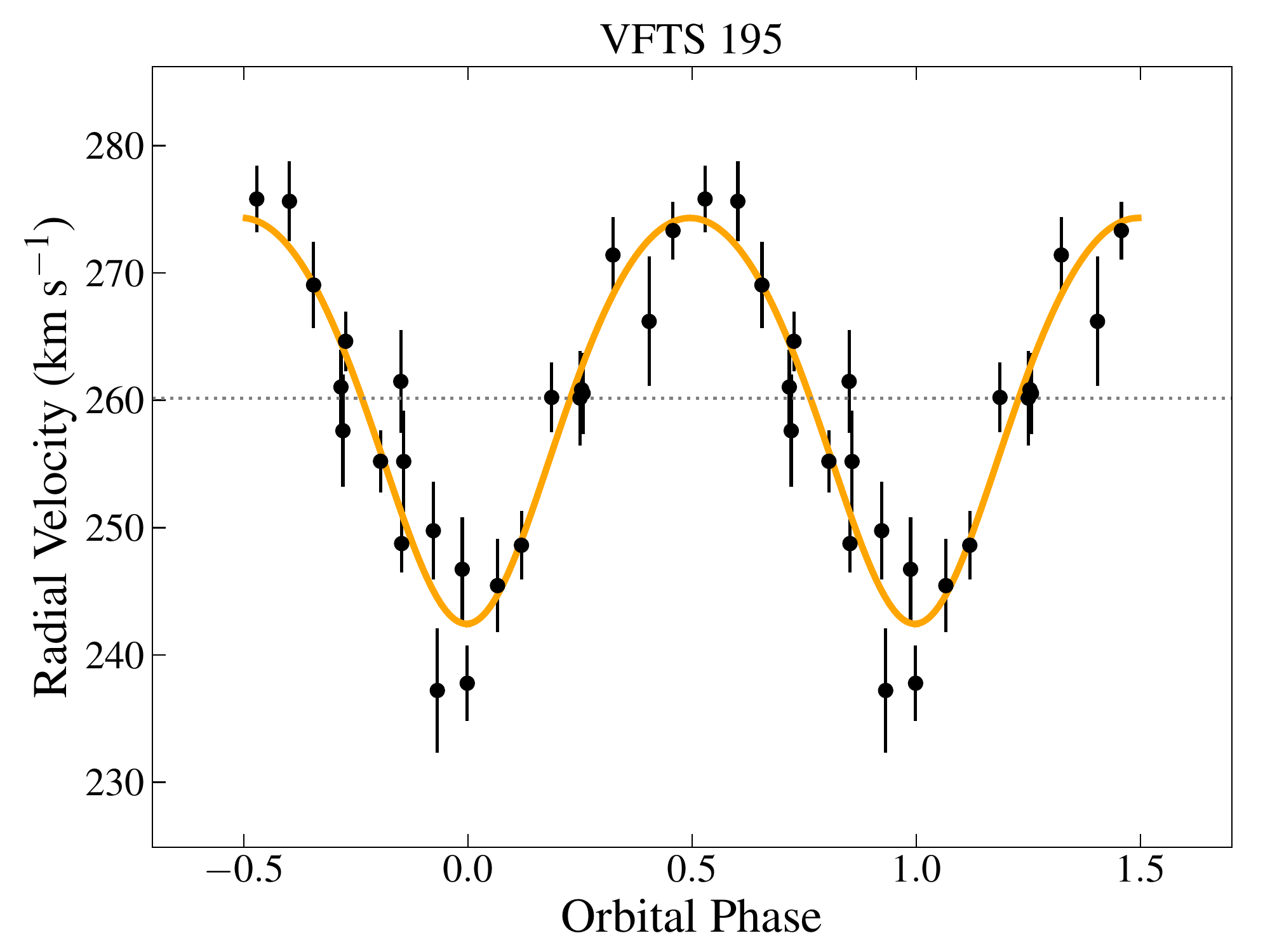}\hfill
    \includegraphics[width=0.31\textwidth]{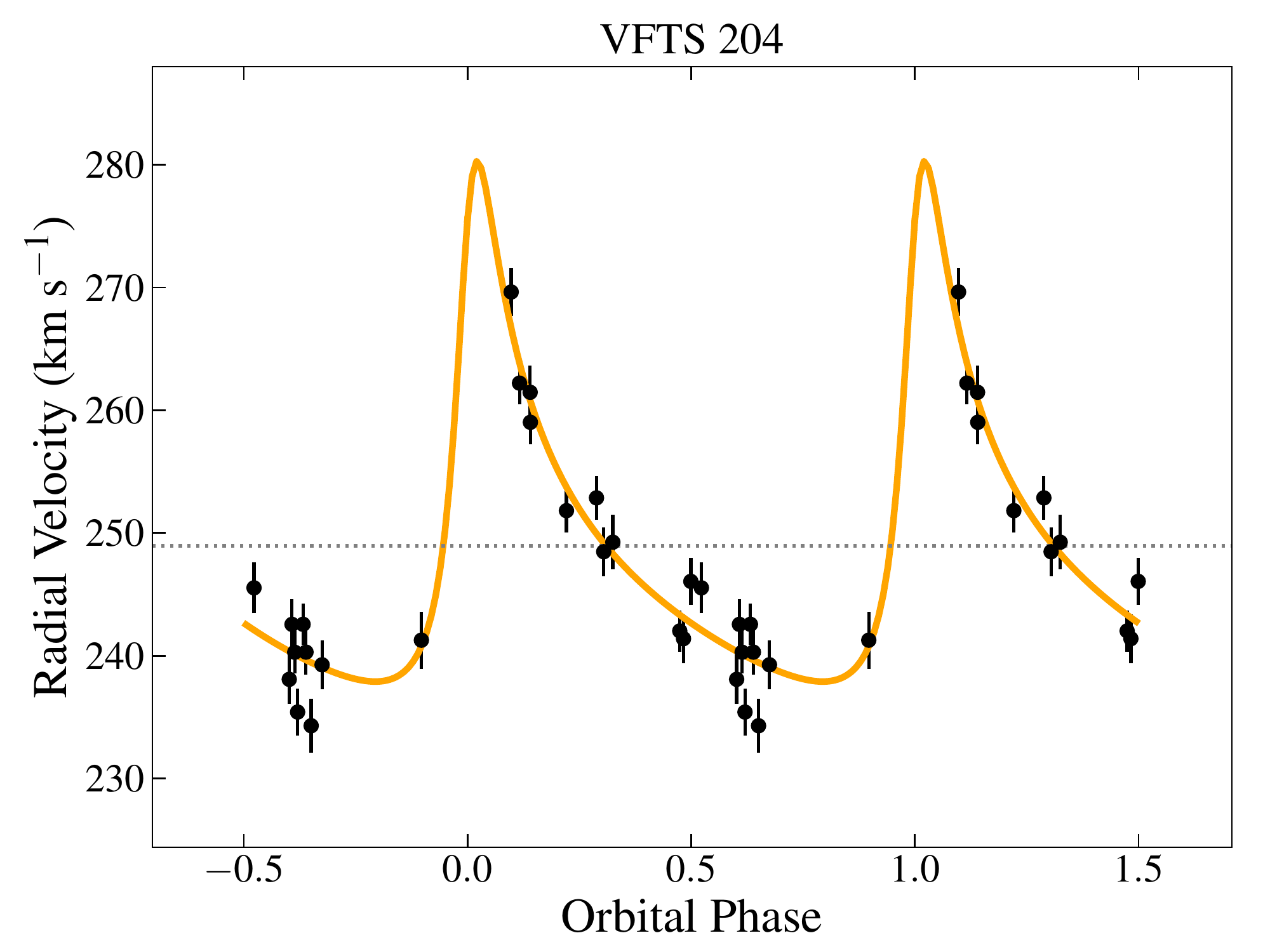}\hfill
    \includegraphics[width=0.31\textwidth]{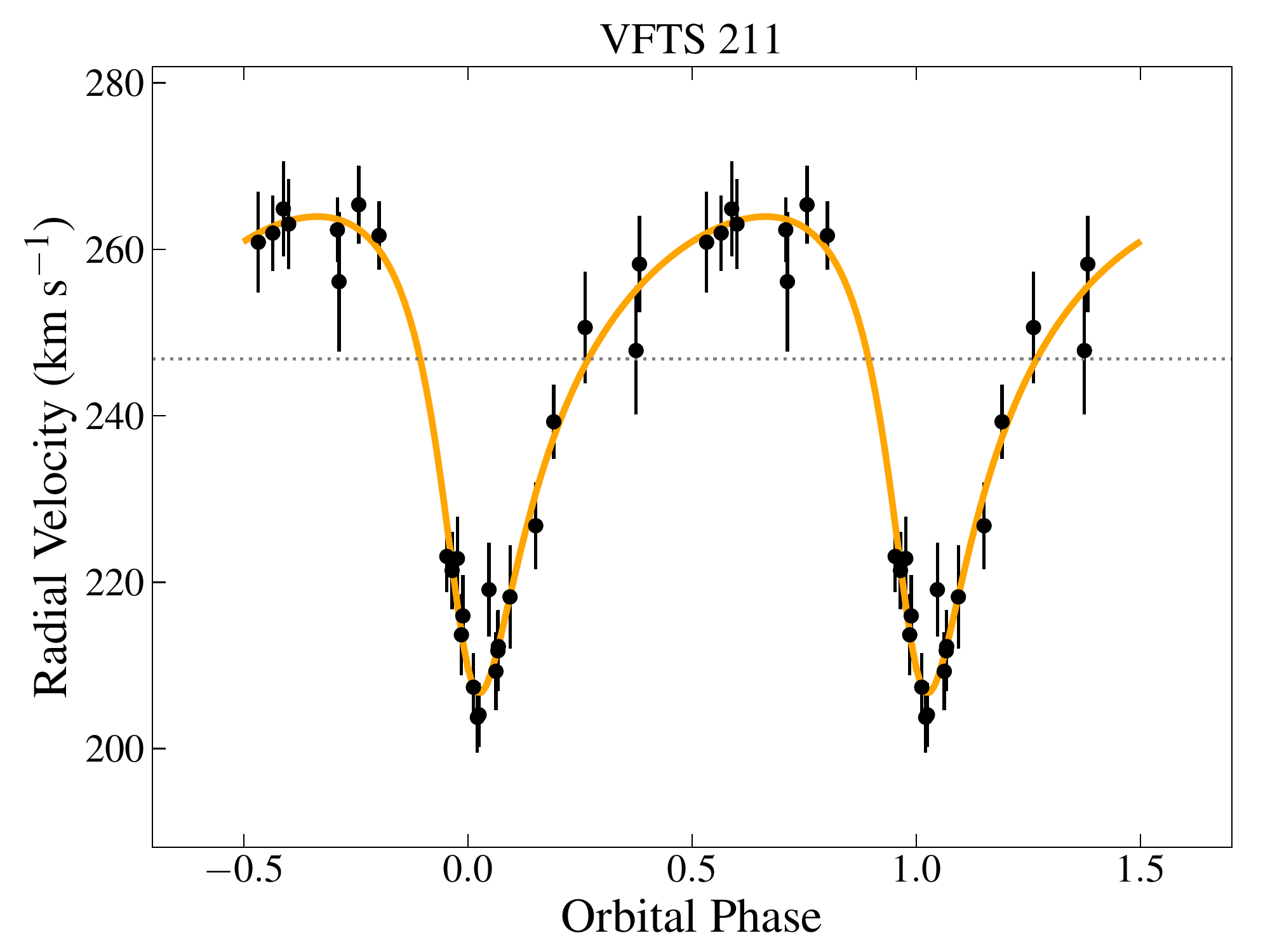}\hfill
    \includegraphics[width=0.31\textwidth]{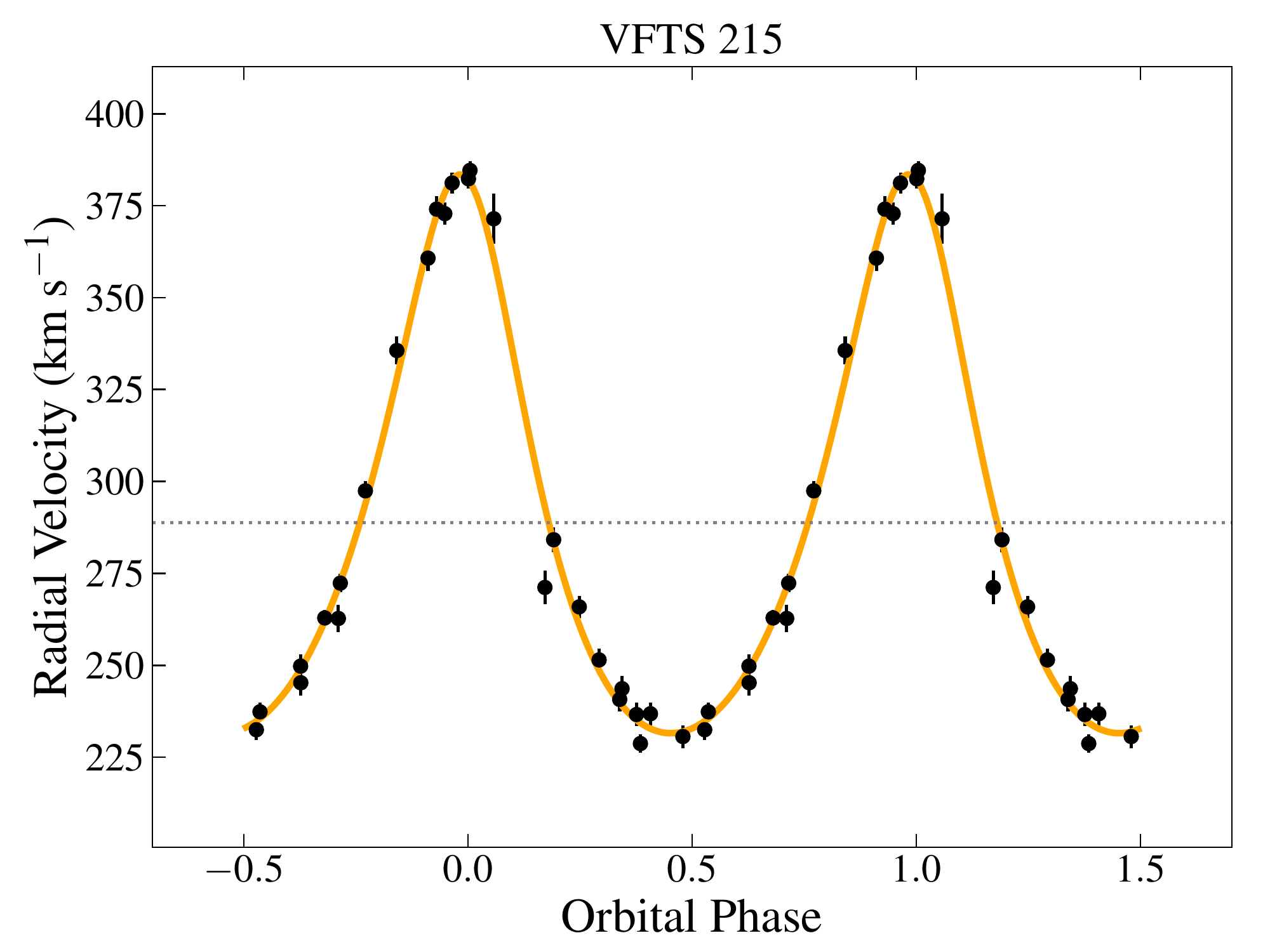}\hfill
    \includegraphics[width=0.31\textwidth]{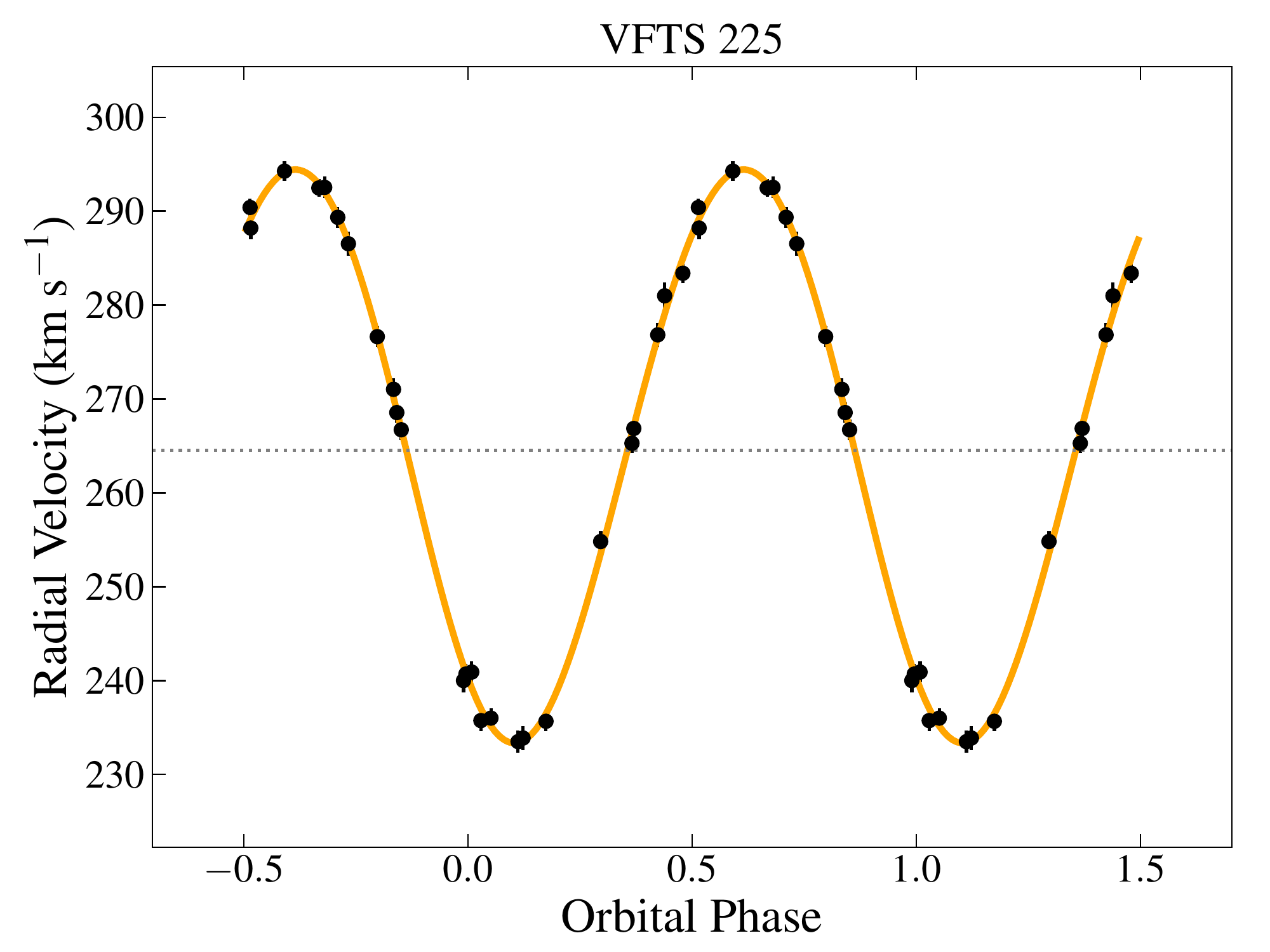}\hfill
\caption{Radial velocity curves of the SB1 systems.}
\end{myfloat}

\begin{myfloat}
\ContinuedFloat
    \centering
    \includegraphics[width=0.31\textwidth]{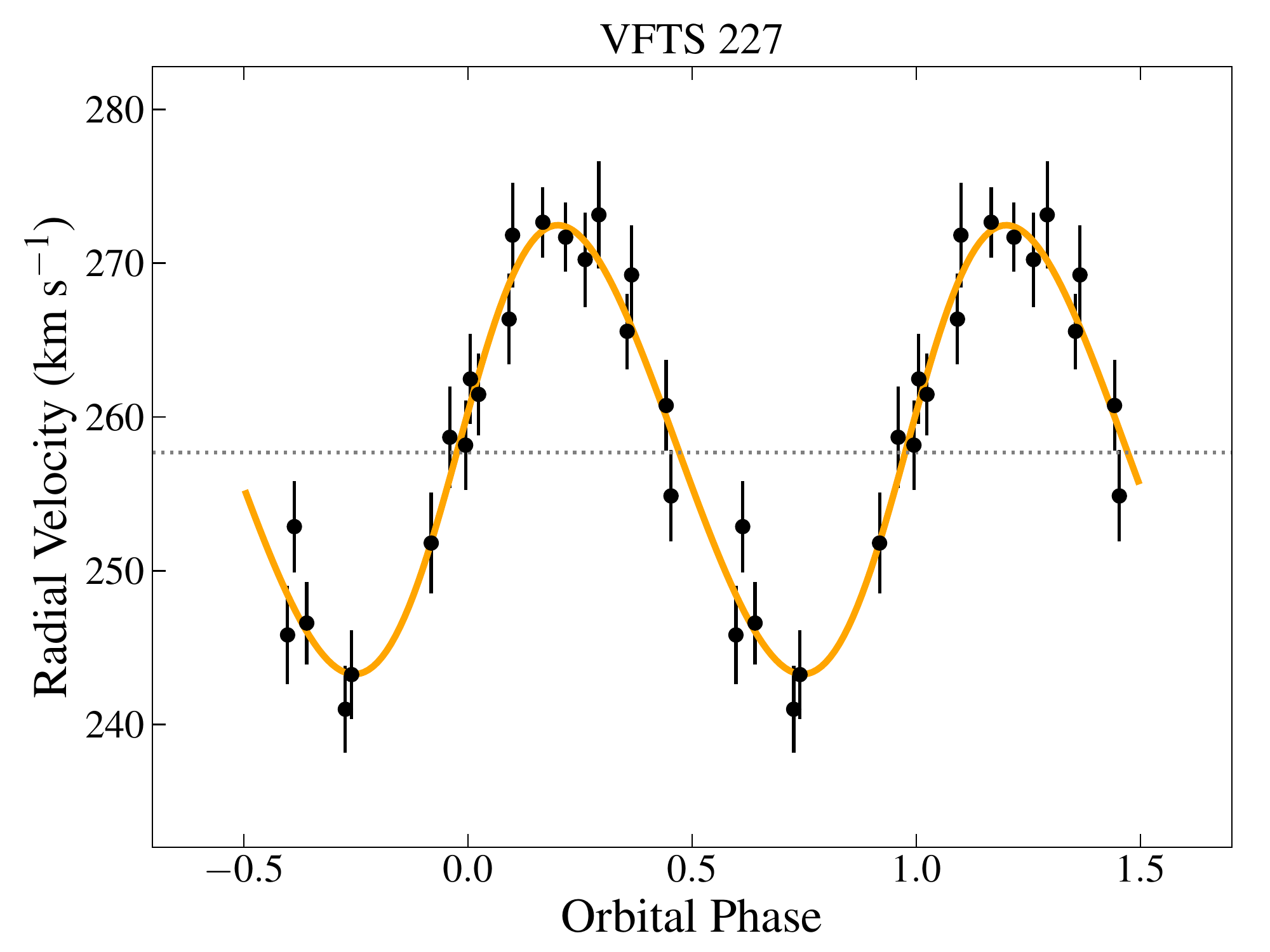}\hfill
    \includegraphics[width=0.31\textwidth]{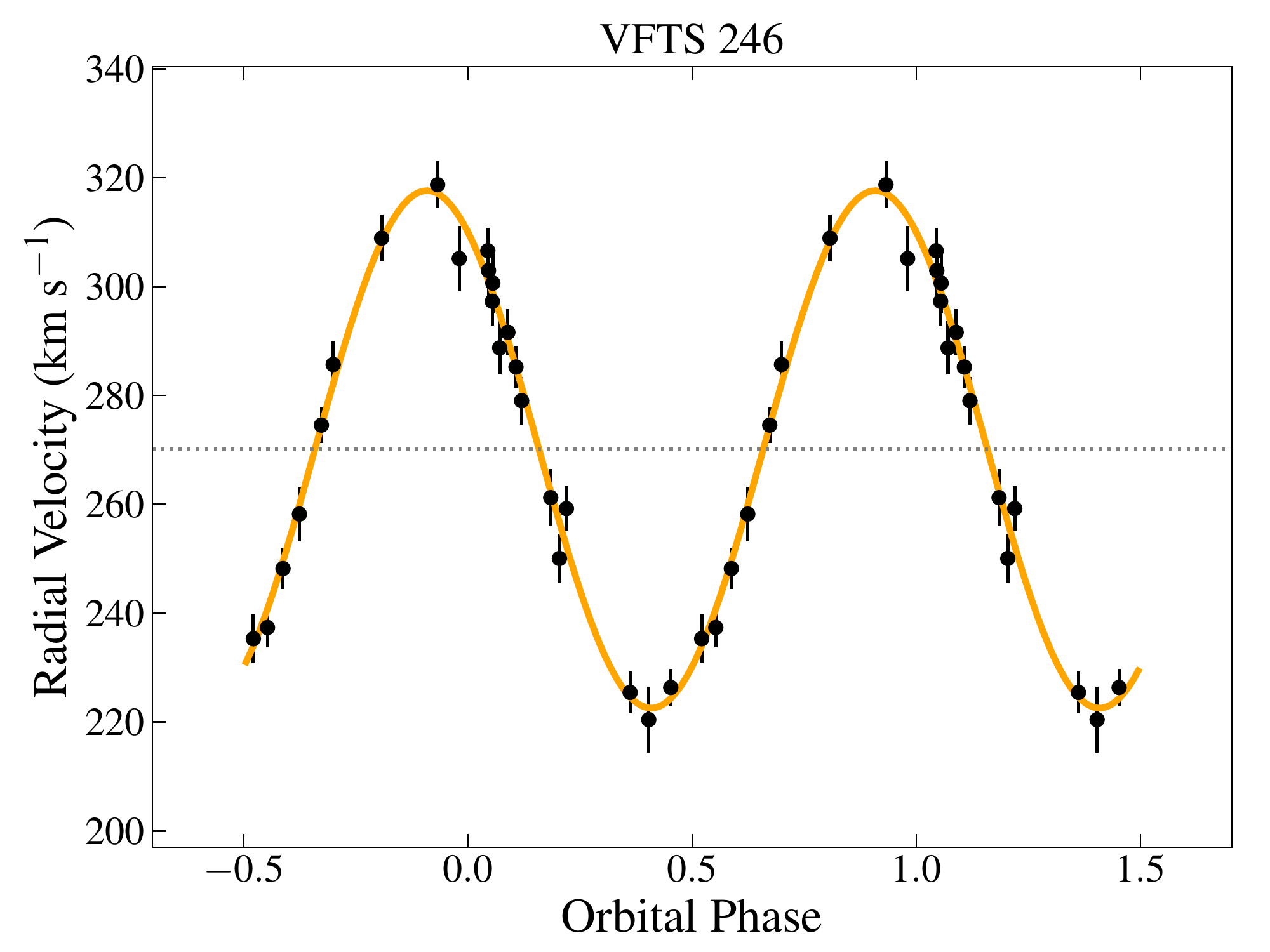}\hfill
    \includegraphics[width=0.31\textwidth]{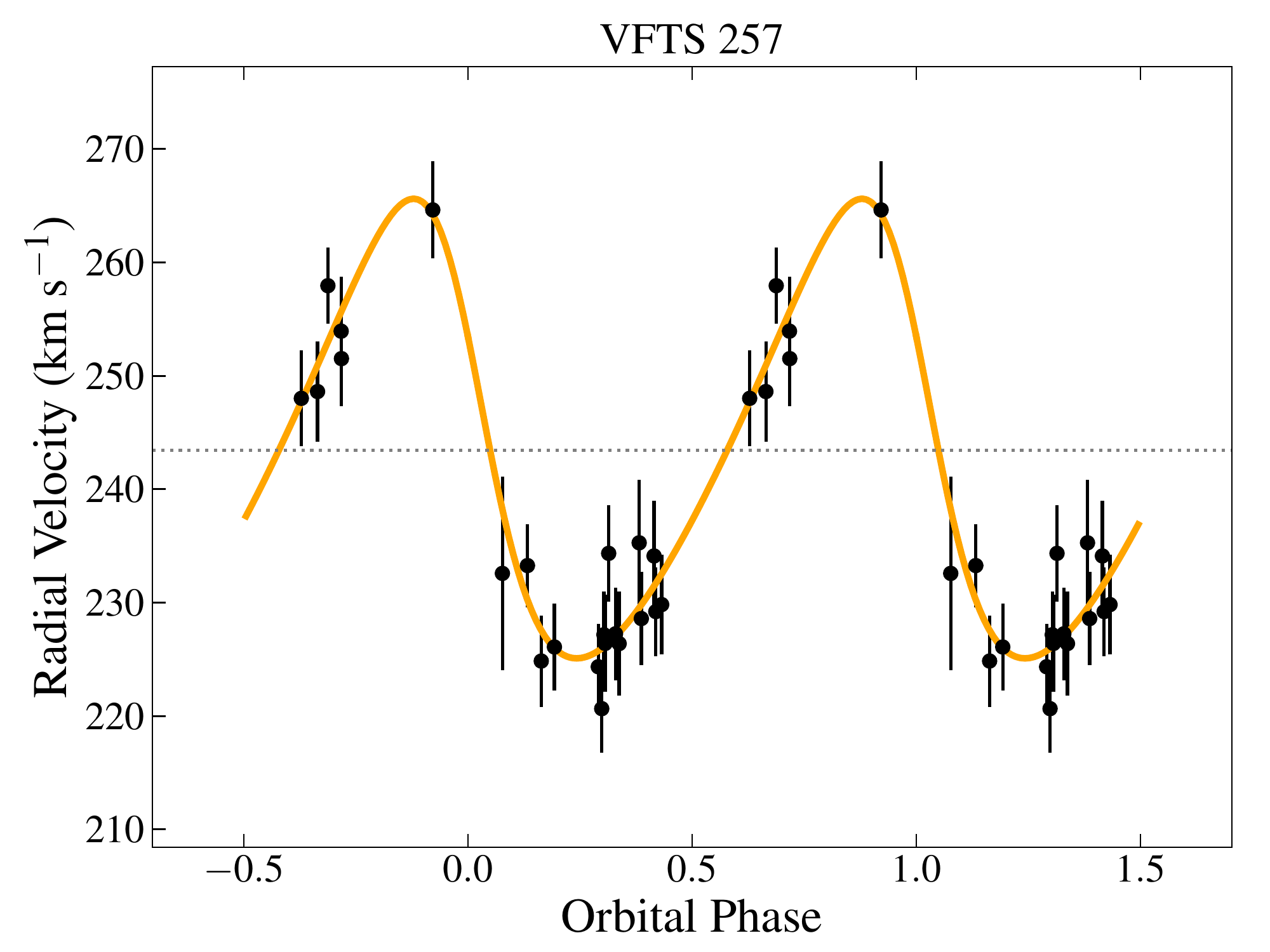}\hfill
    \includegraphics[width=0.31\textwidth]{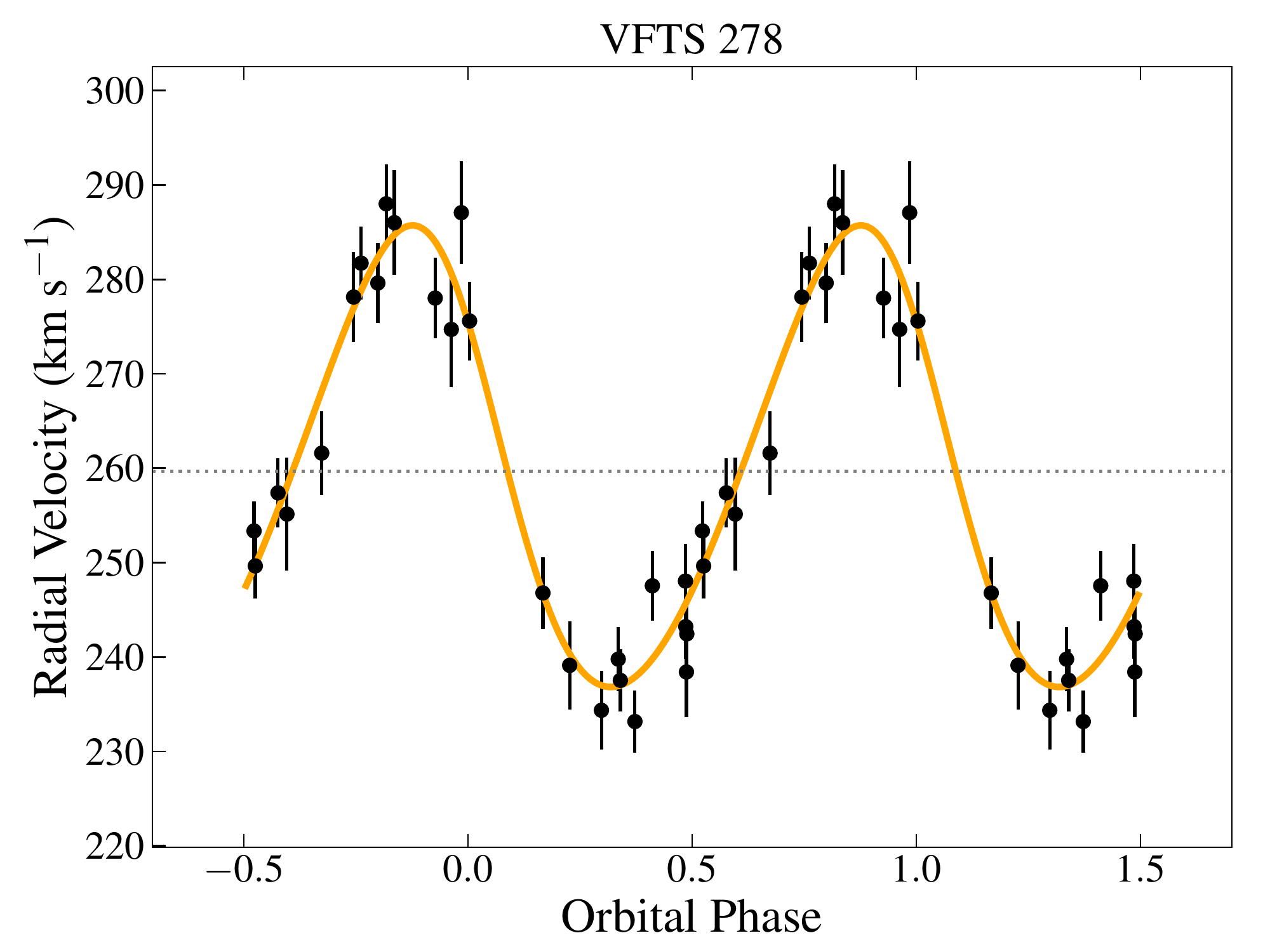}\hfill
    \includegraphics[width=0.31\textwidth]{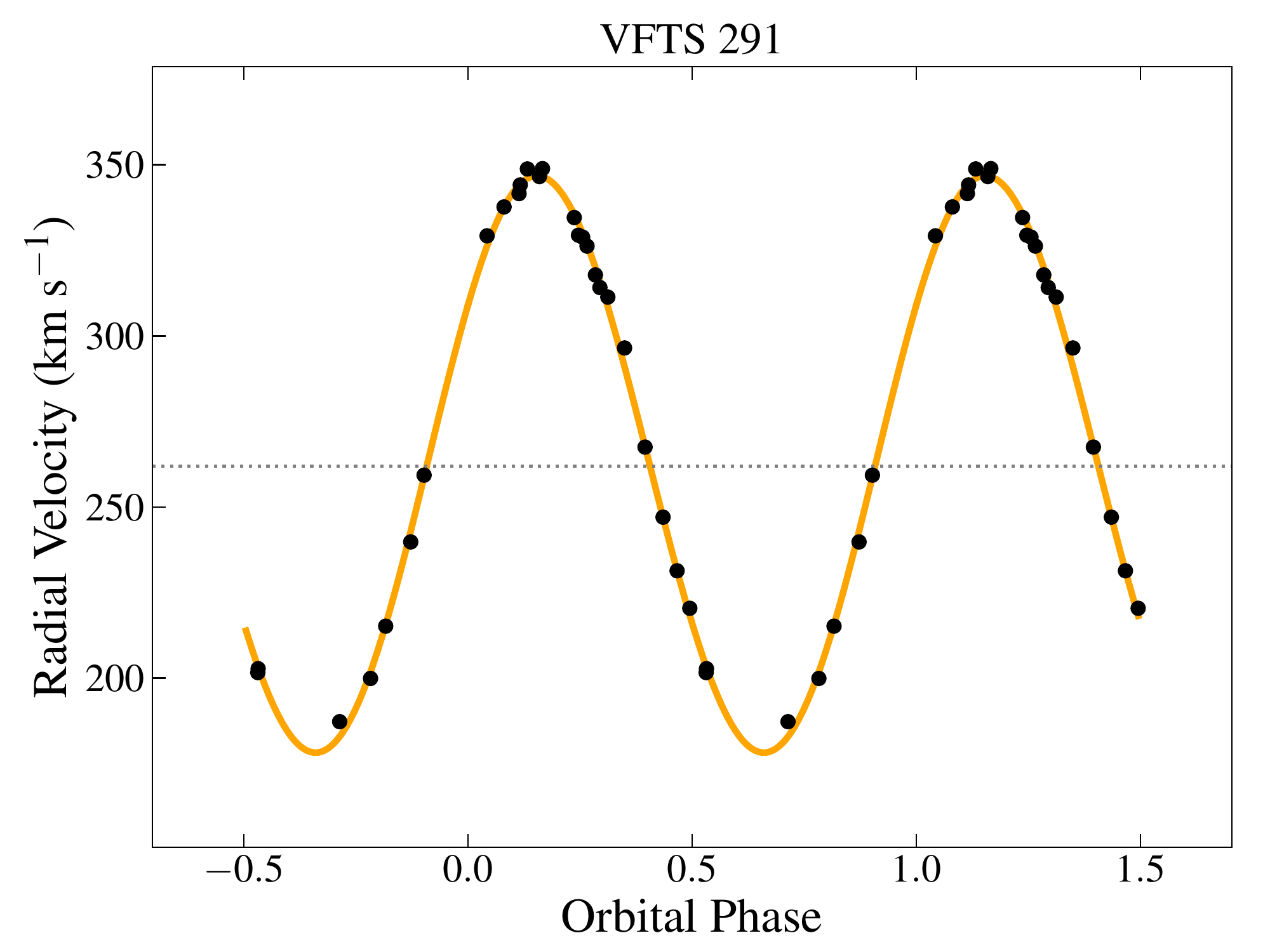}\hfill
    \includegraphics[width=0.31\textwidth]{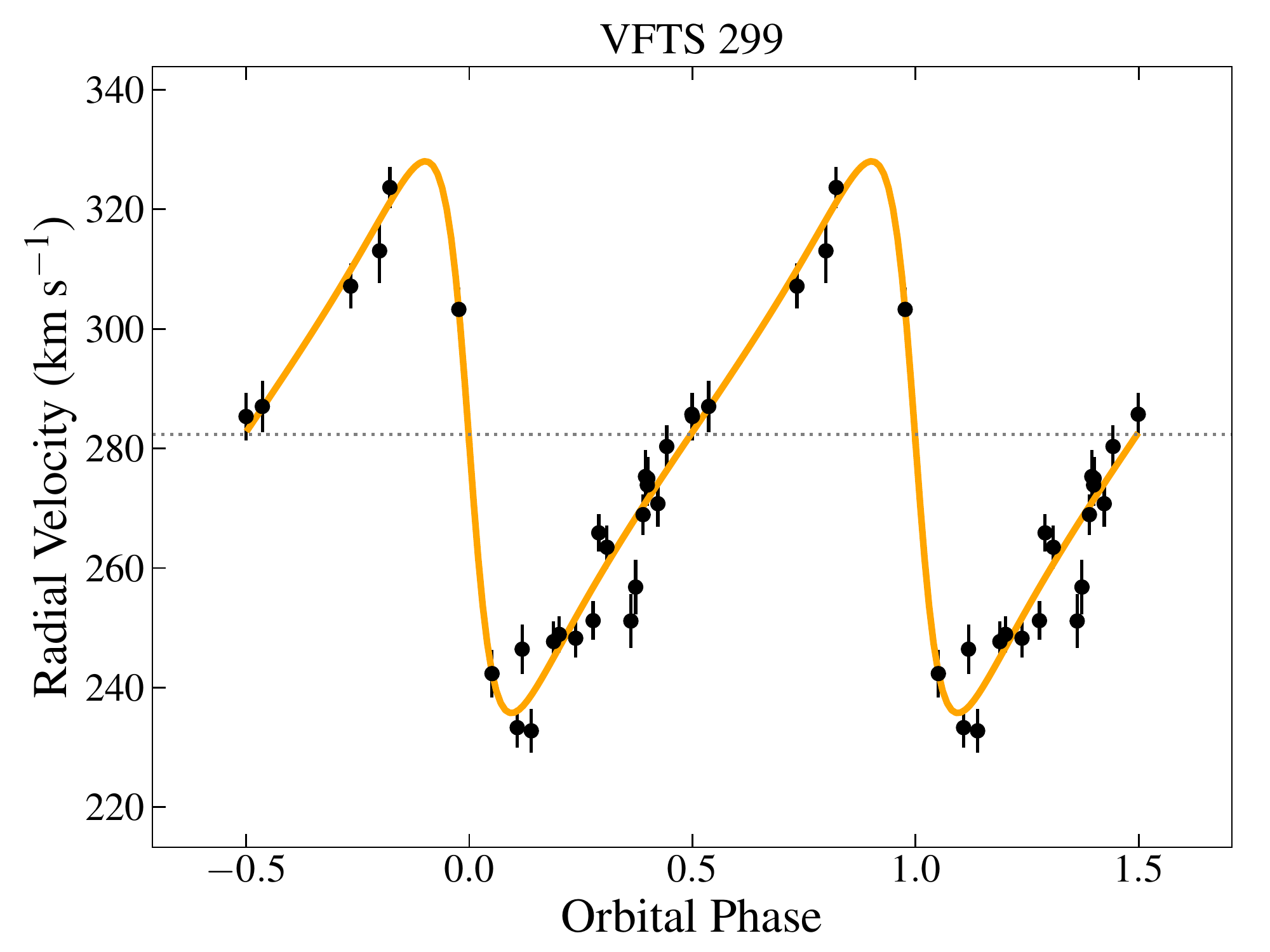}\hfill
    \includegraphics[width=0.31\textwidth]{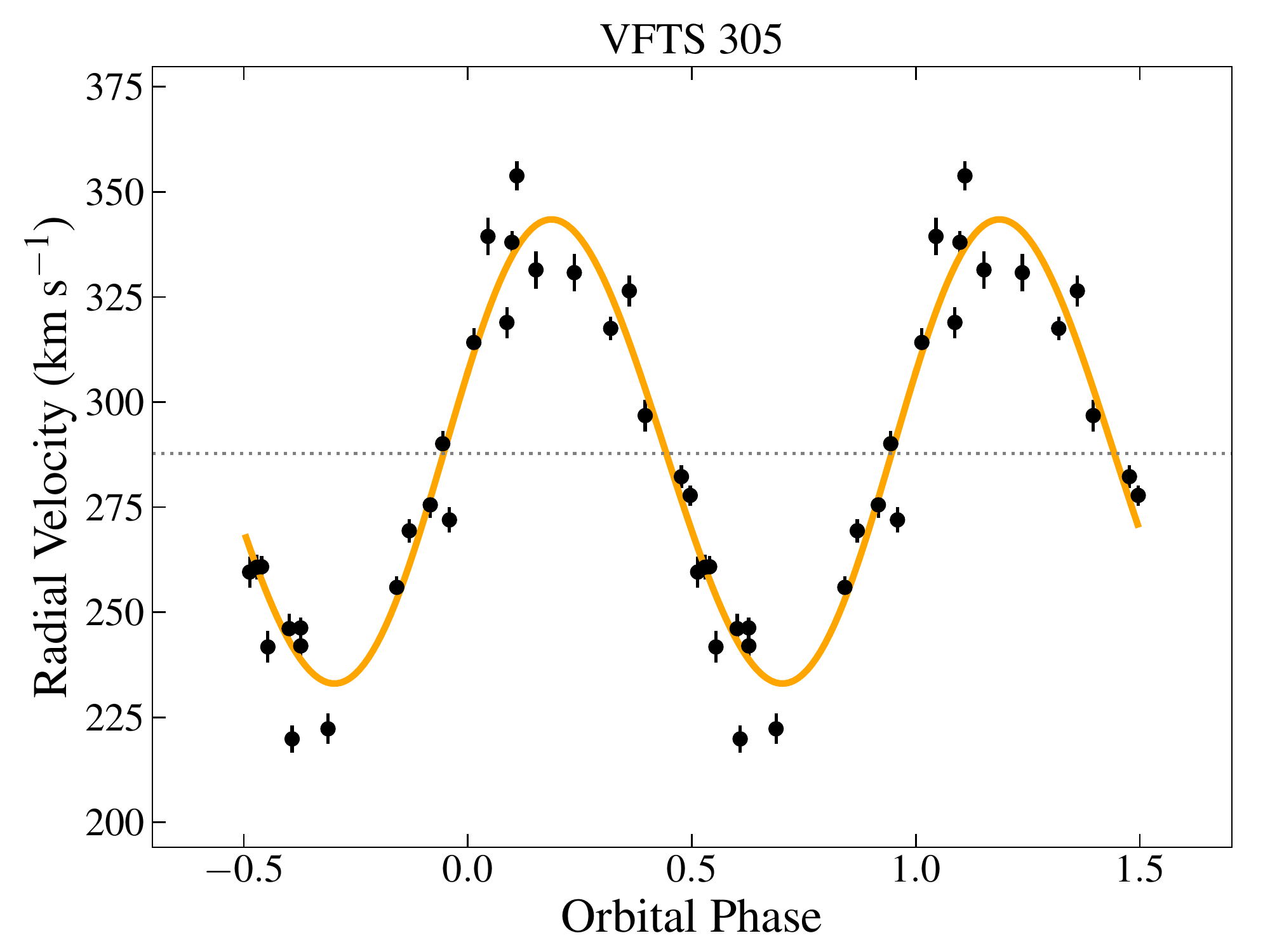}\hfill
    \includegraphics[width=0.31\textwidth]{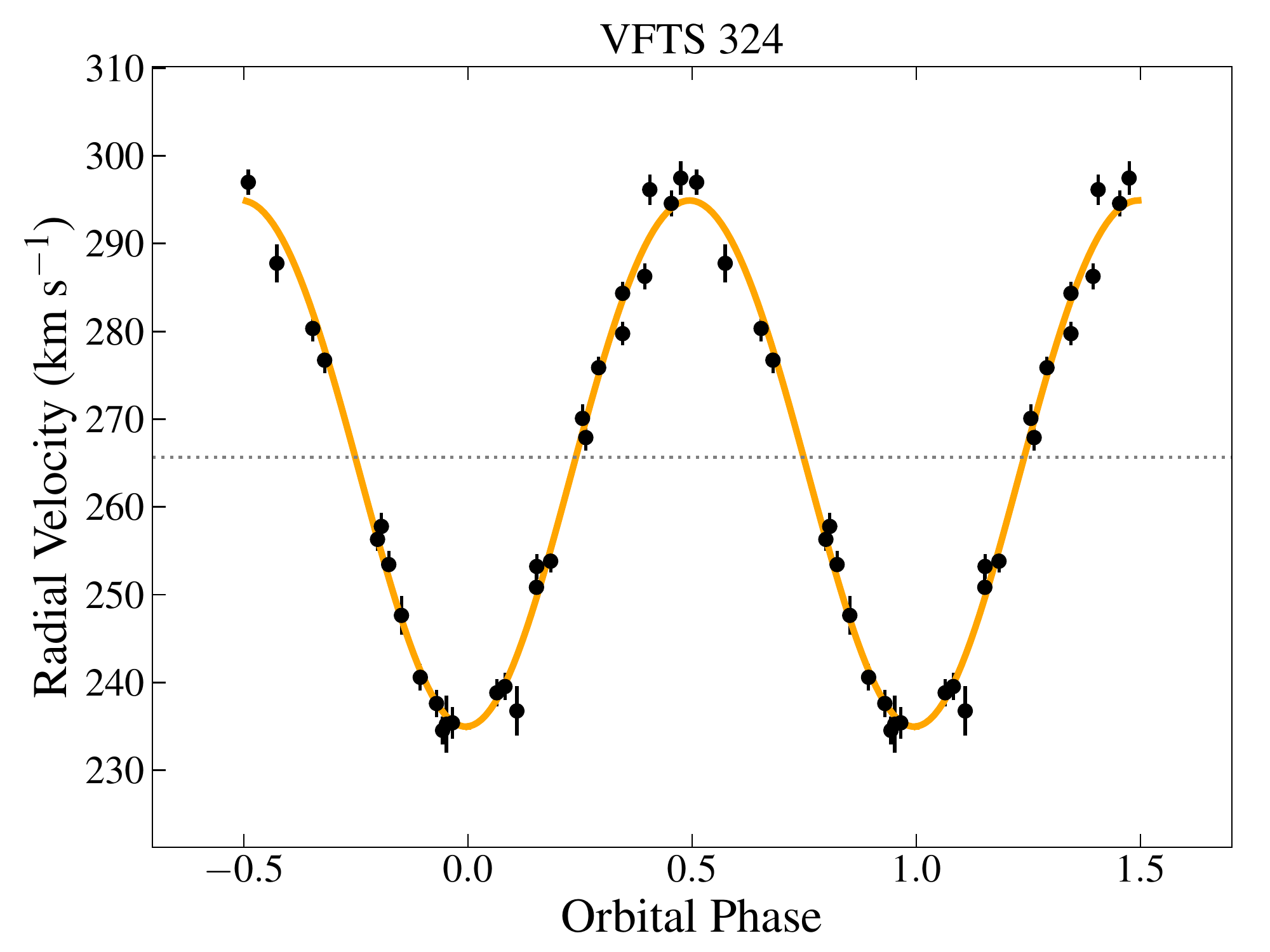}\hfill
    \includegraphics[width=0.31\textwidth]{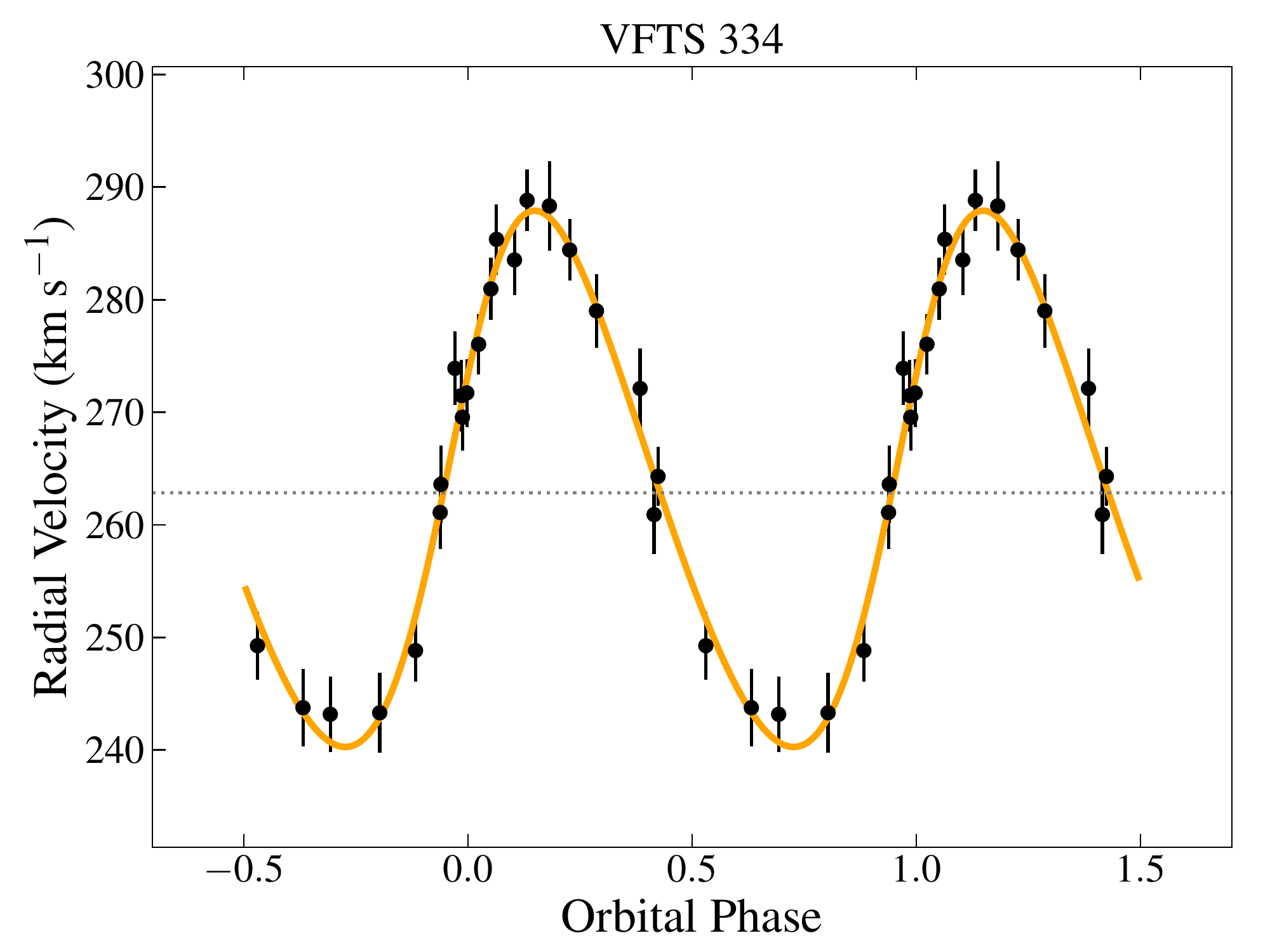}\hfill
    \includegraphics[width=0.31\textwidth]{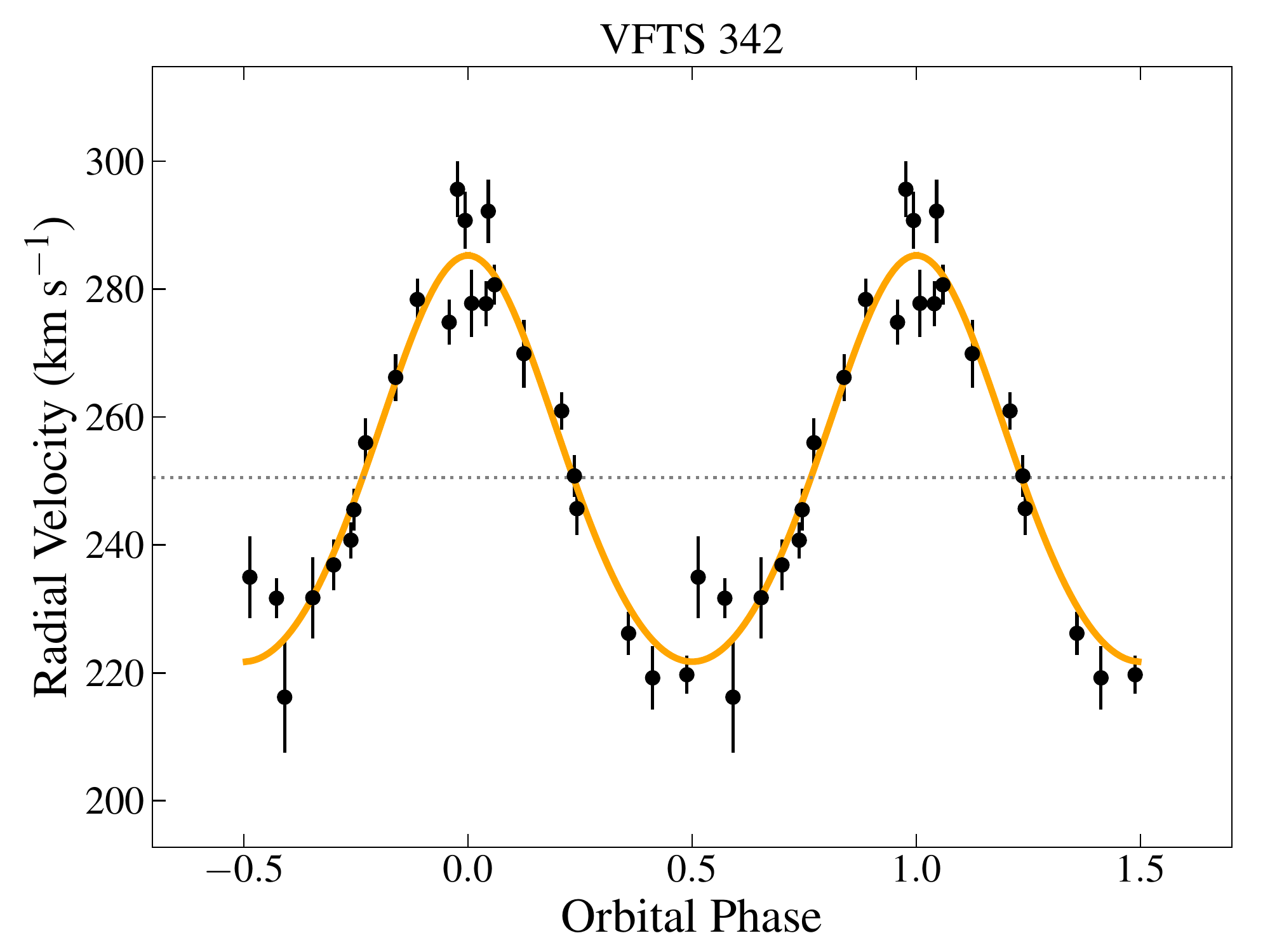}\hfill
    \includegraphics[width=0.31\textwidth]{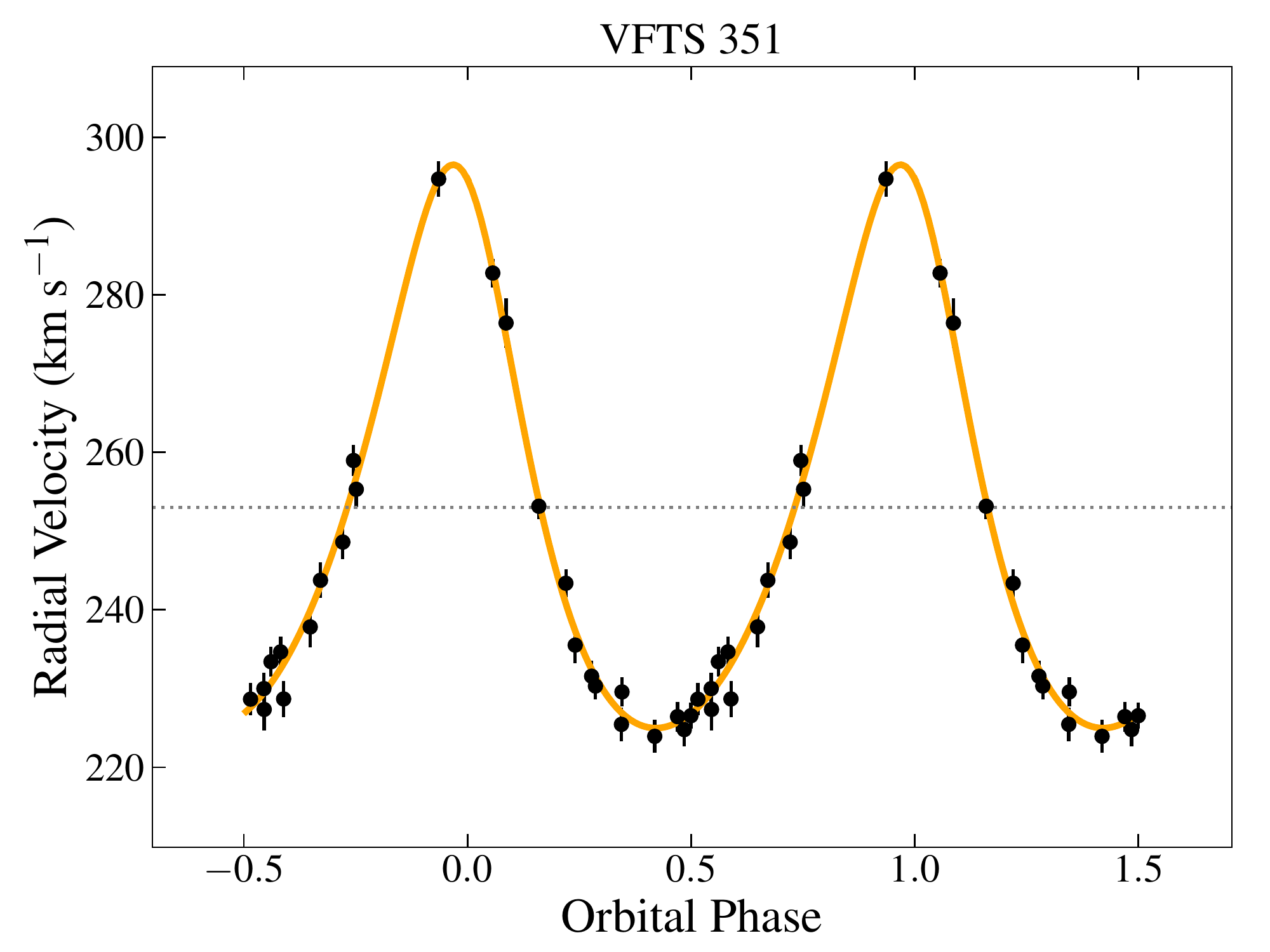}\hfill
    \includegraphics[width=0.31\textwidth]{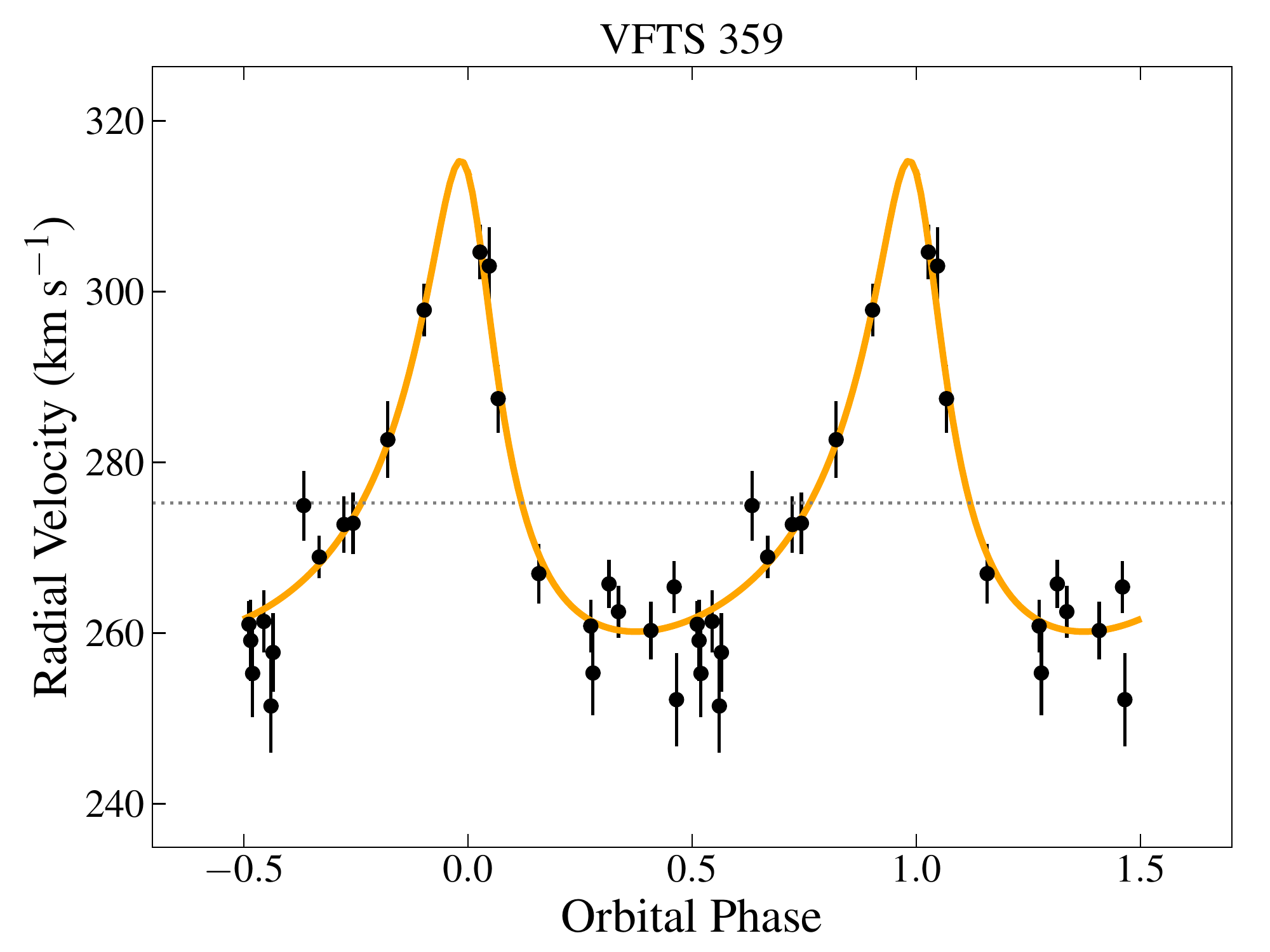}\hfill
    \includegraphics[width=0.31\textwidth]{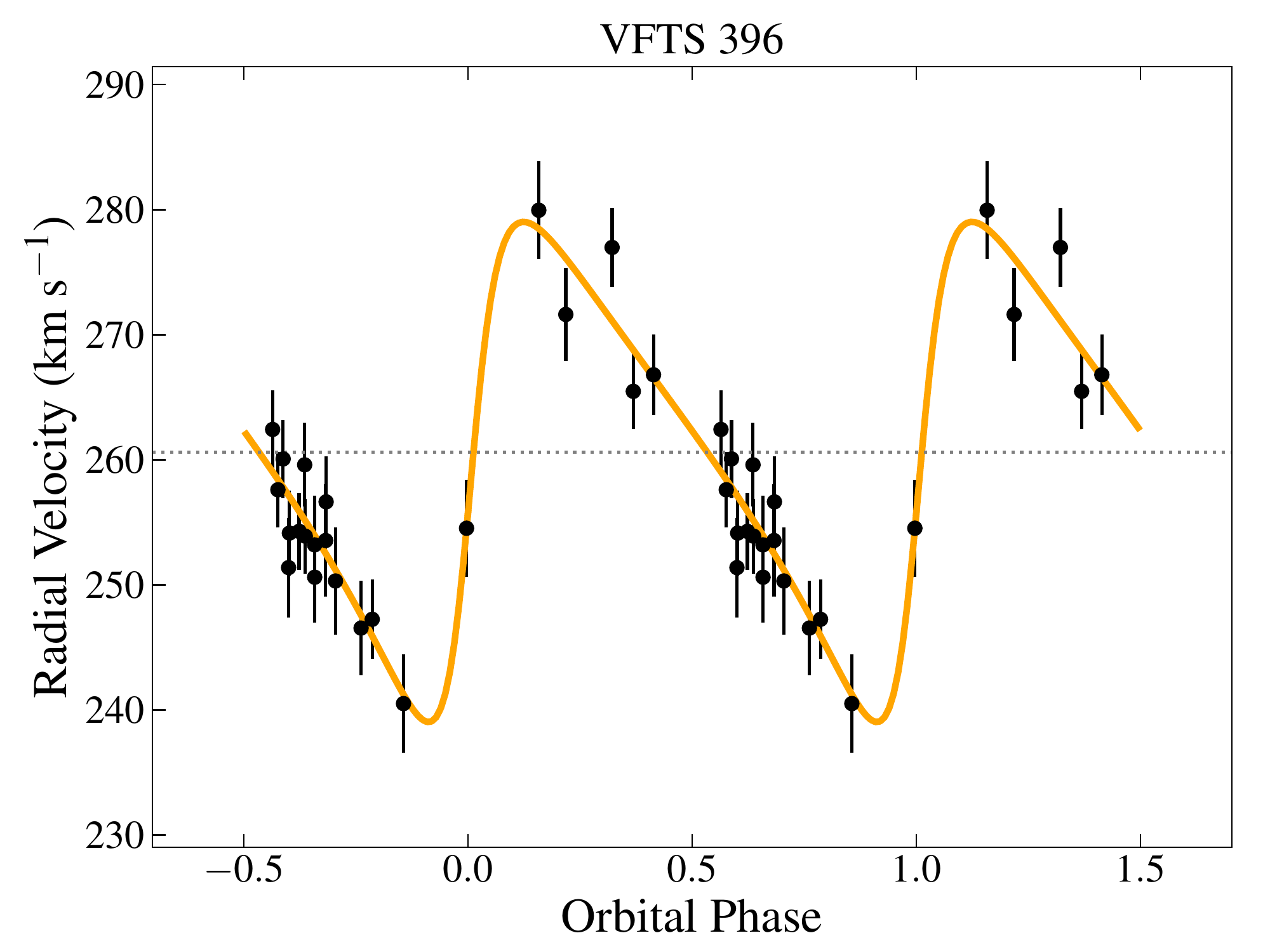}\hfill
    \includegraphics[width=0.31\textwidth]{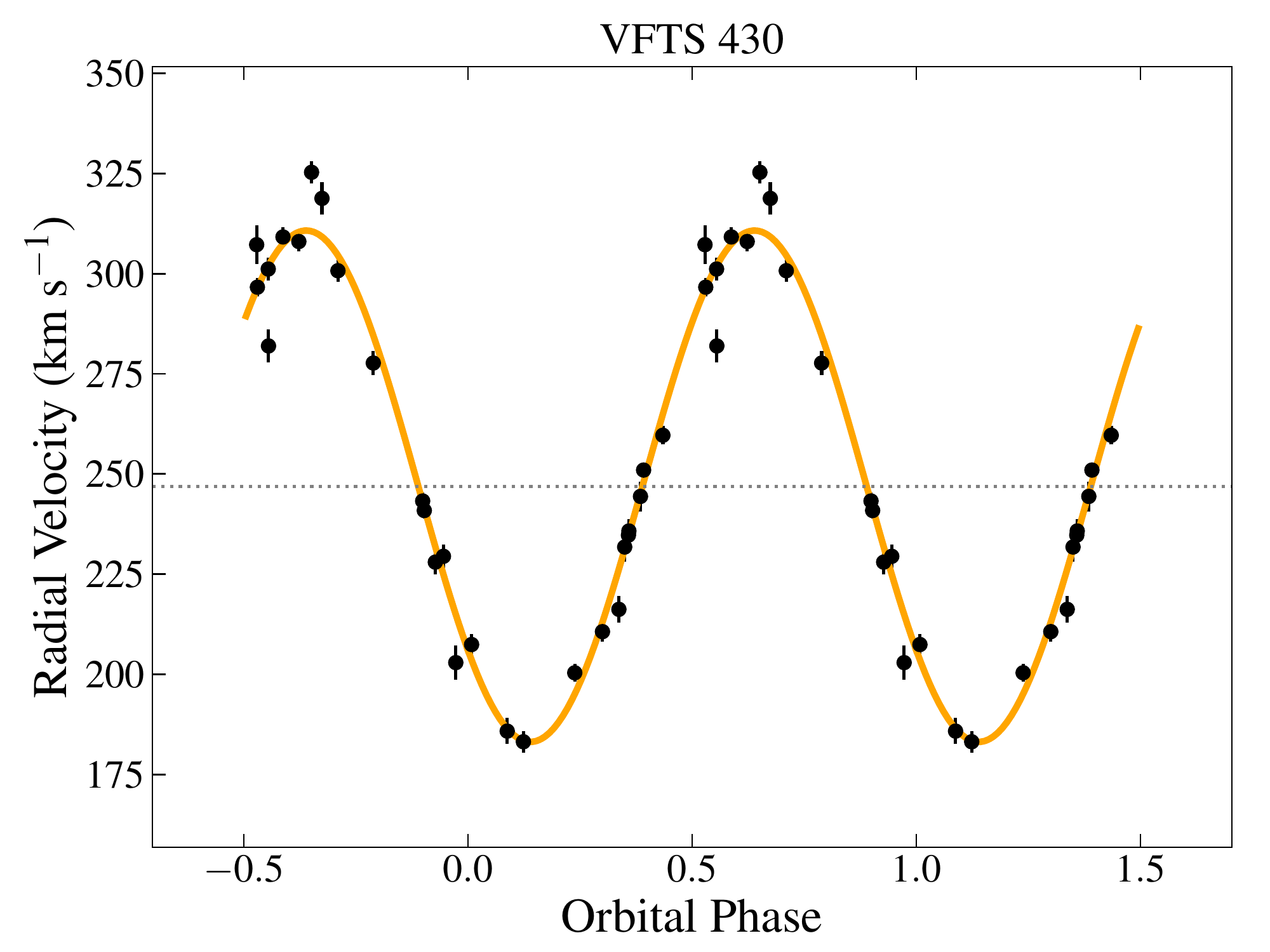}\hfill
    \includegraphics[width=0.31\textwidth]{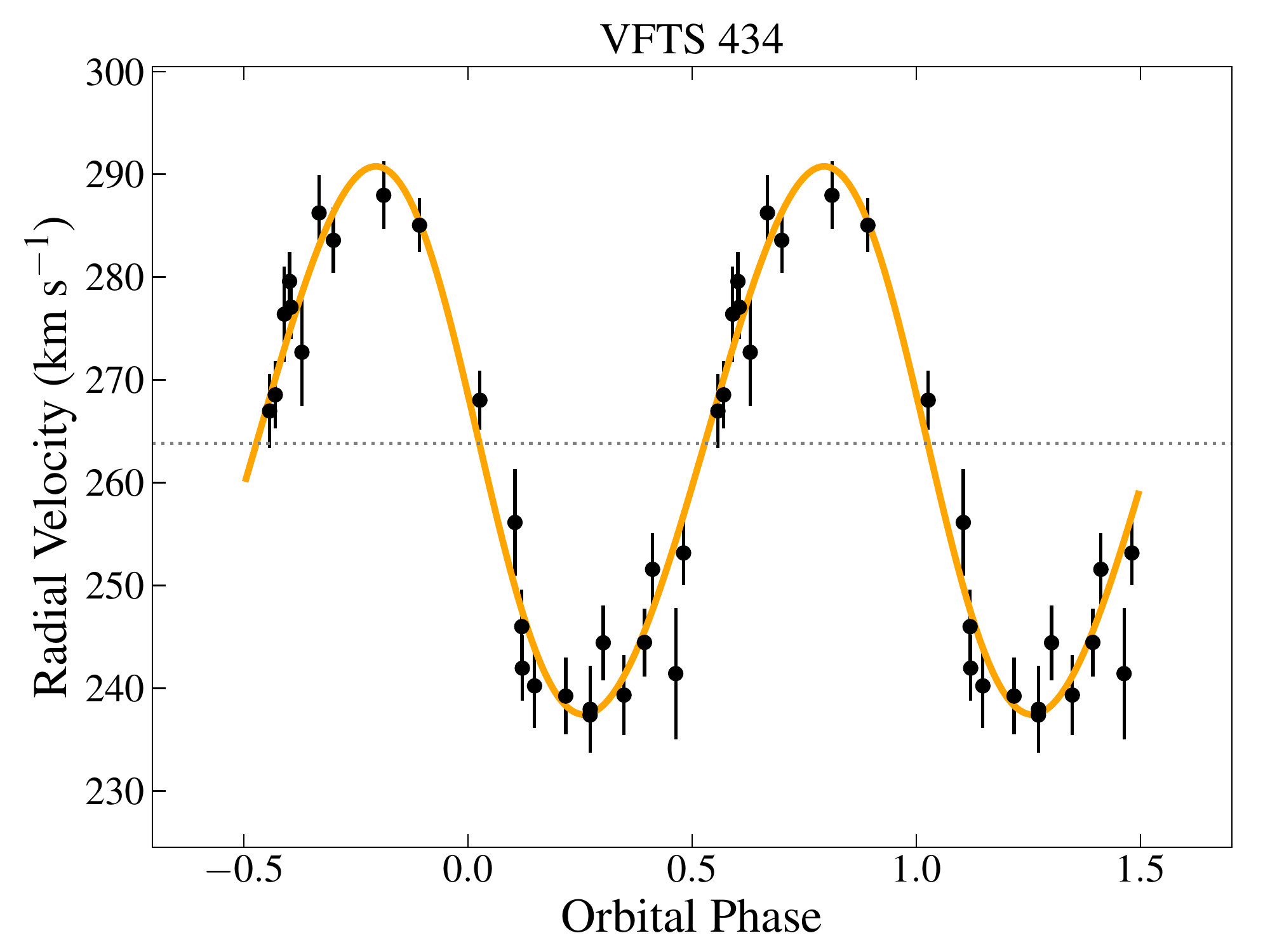}\hfill
    \caption{$-$ \it continued}
\end{myfloat}

\begin{myfloat}
\ContinuedFloat
    \centering
    \includegraphics[width=0.31\textwidth]{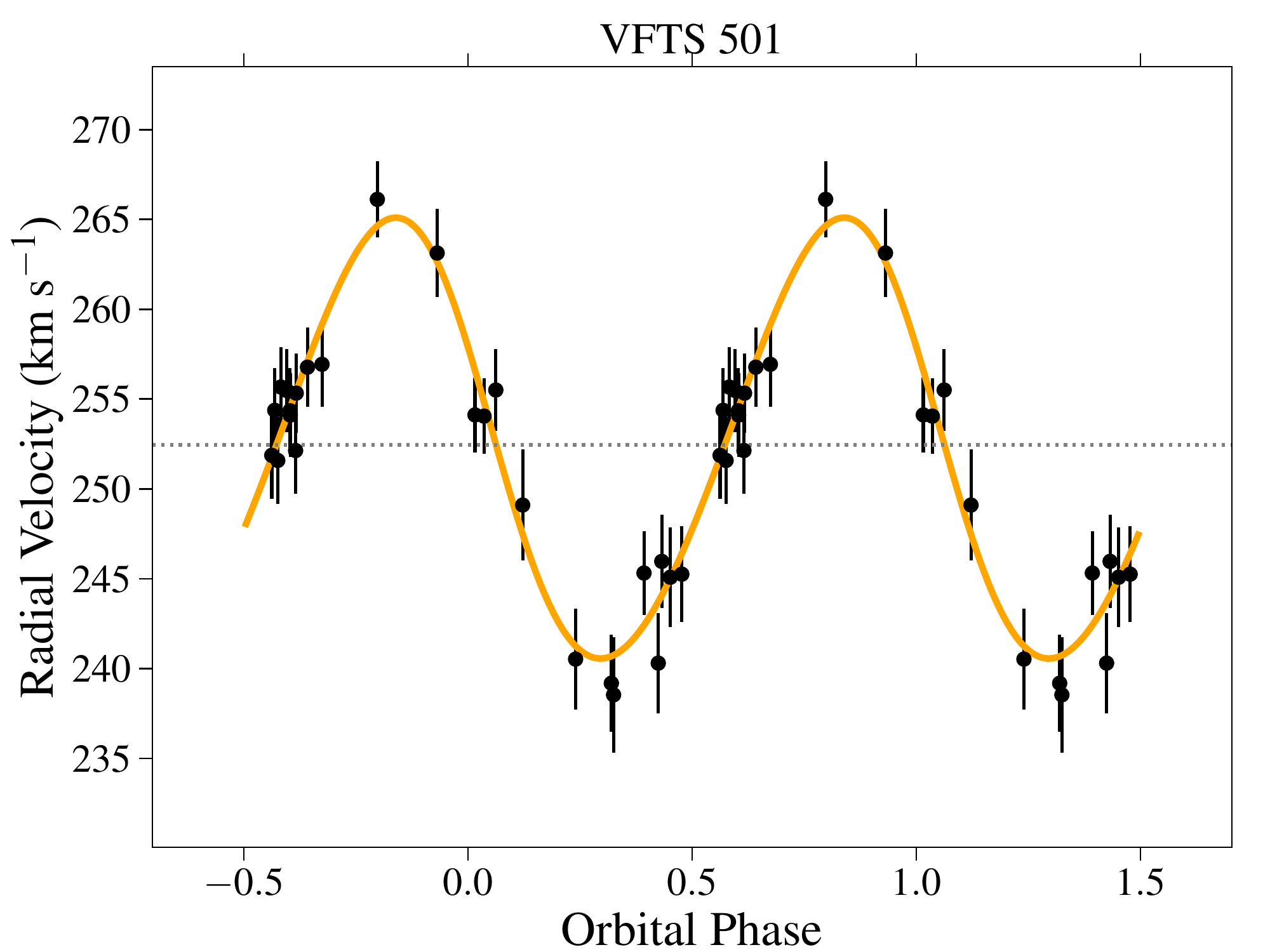}\hfill
    \includegraphics[width=0.31\textwidth]{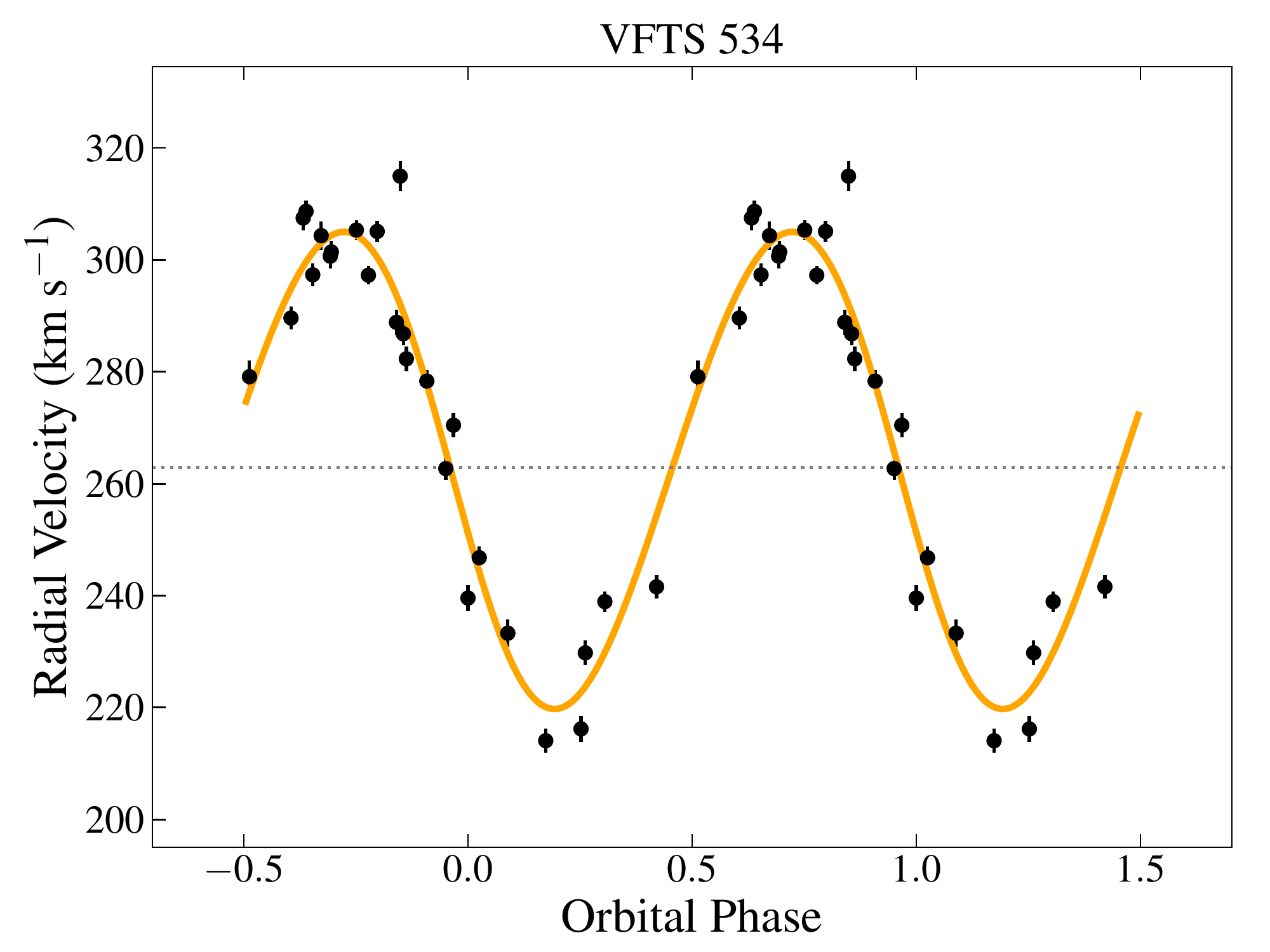}\hfill
    \includegraphics[width=0.31\textwidth]{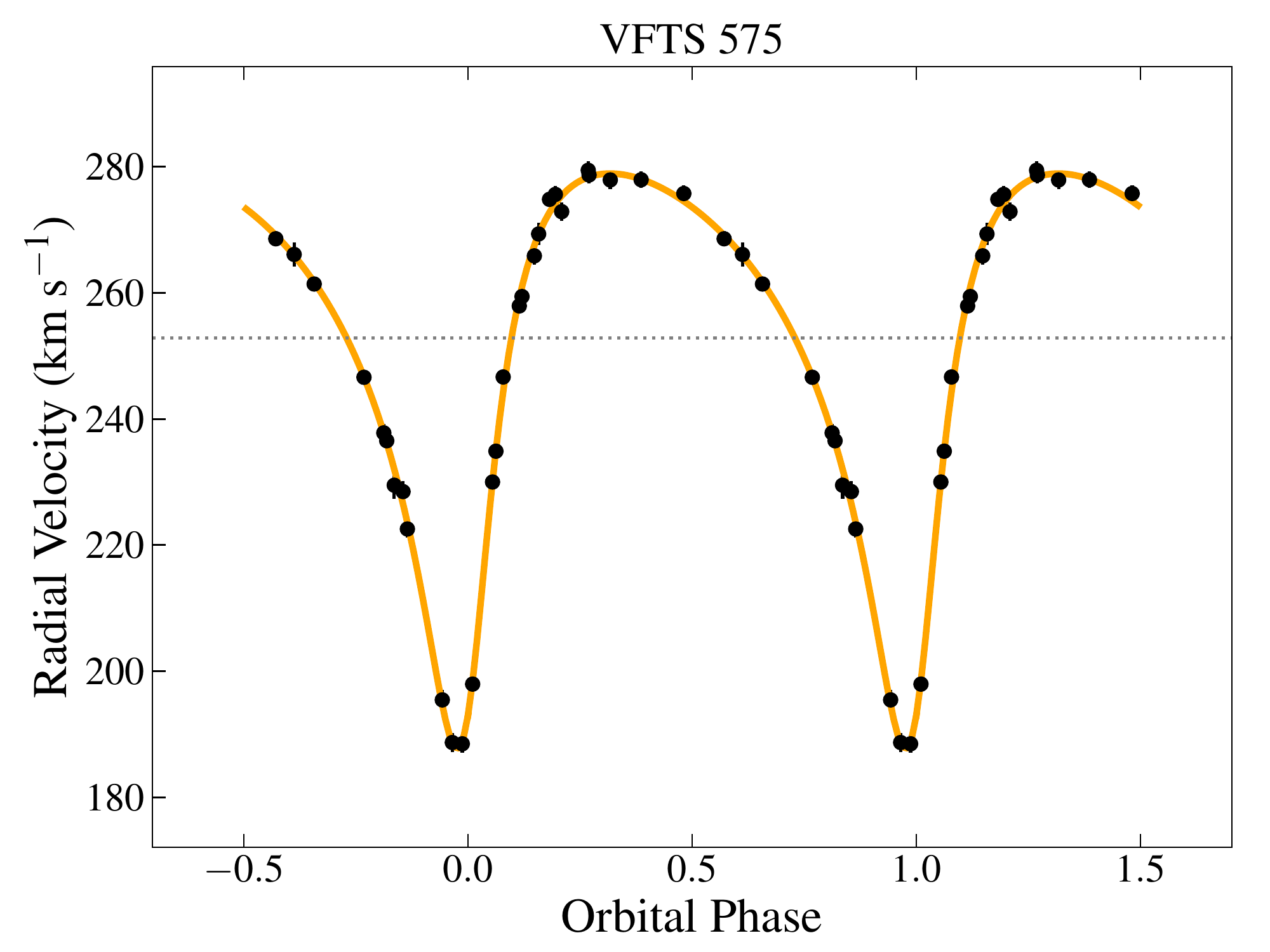}\hfill
    \includegraphics[width=0.31\textwidth]{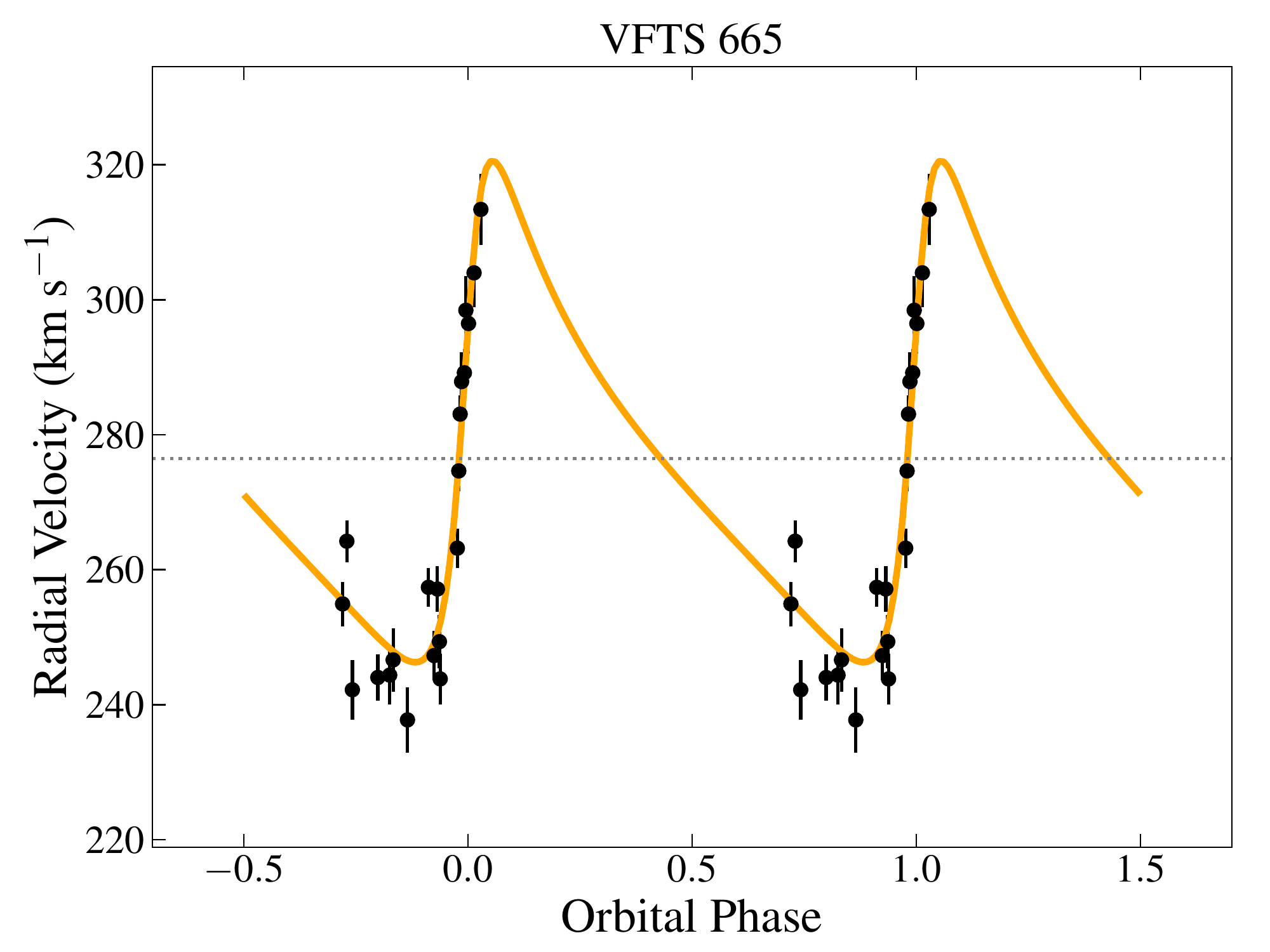}\hfill
    \includegraphics[width=0.31\textwidth]{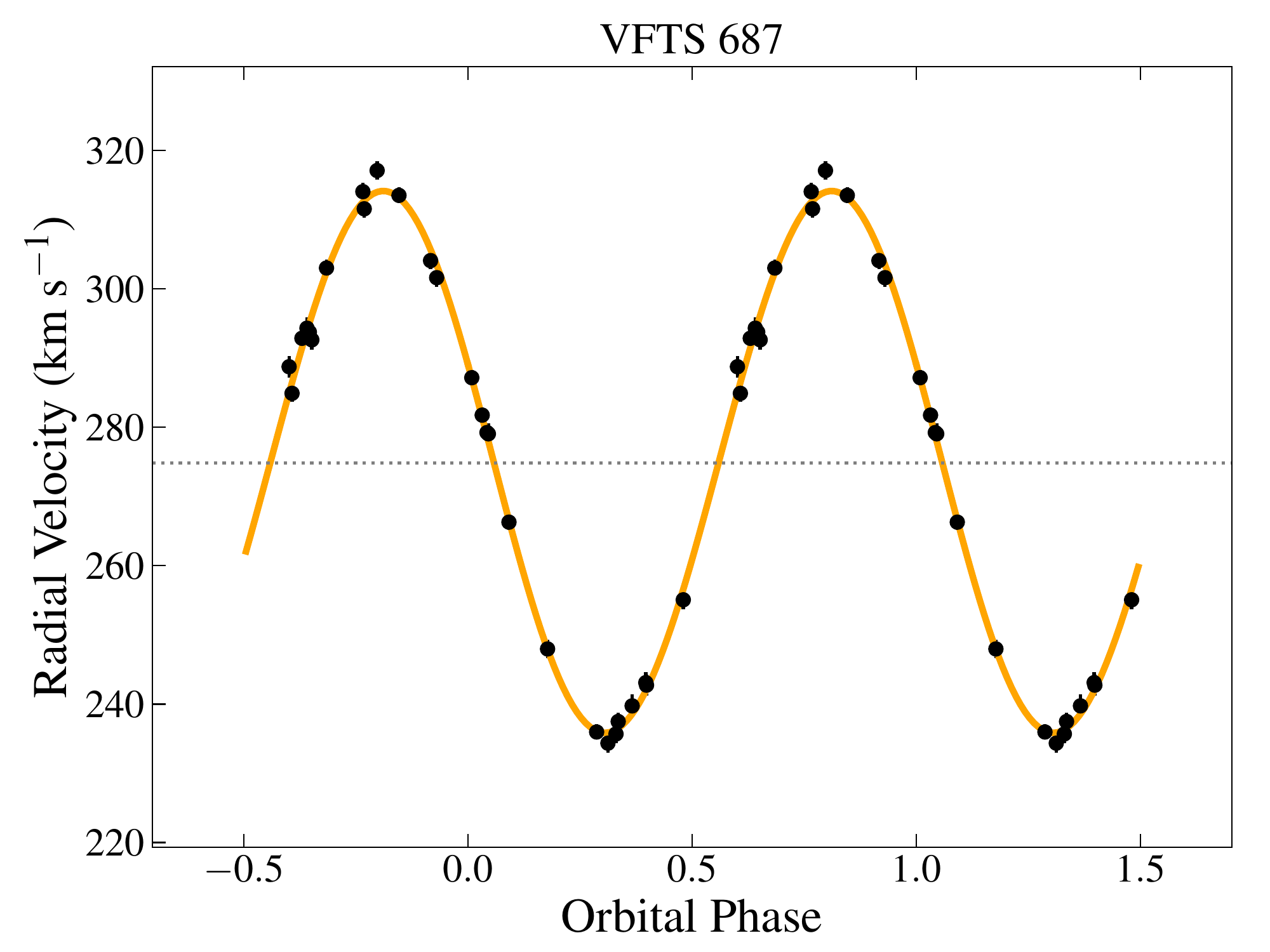}\hfill
    \includegraphics[width=0.31\textwidth]{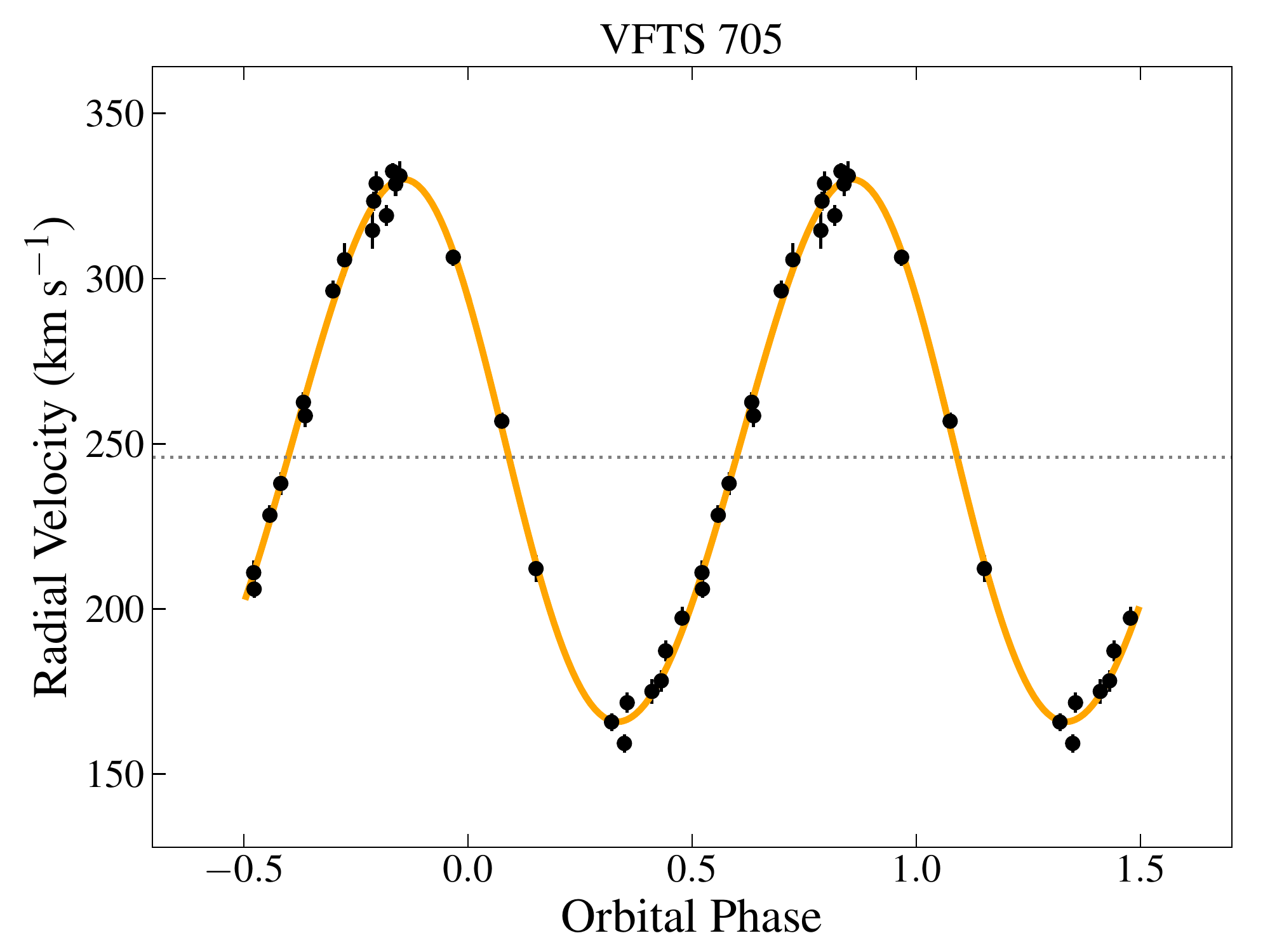}\hfill
    \includegraphics[width=0.31\textwidth]{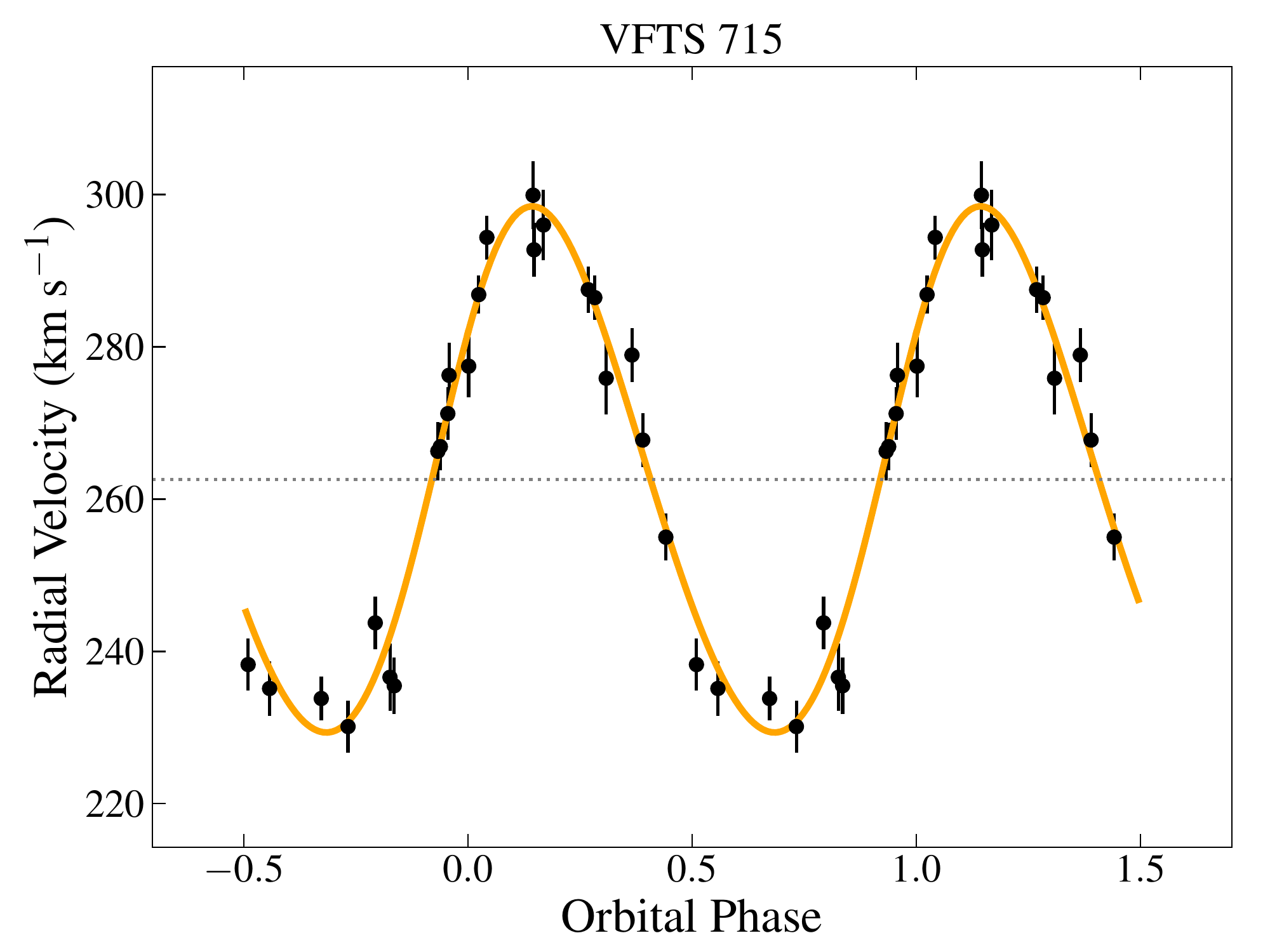}\hfill
    \includegraphics[width=0.31\textwidth]{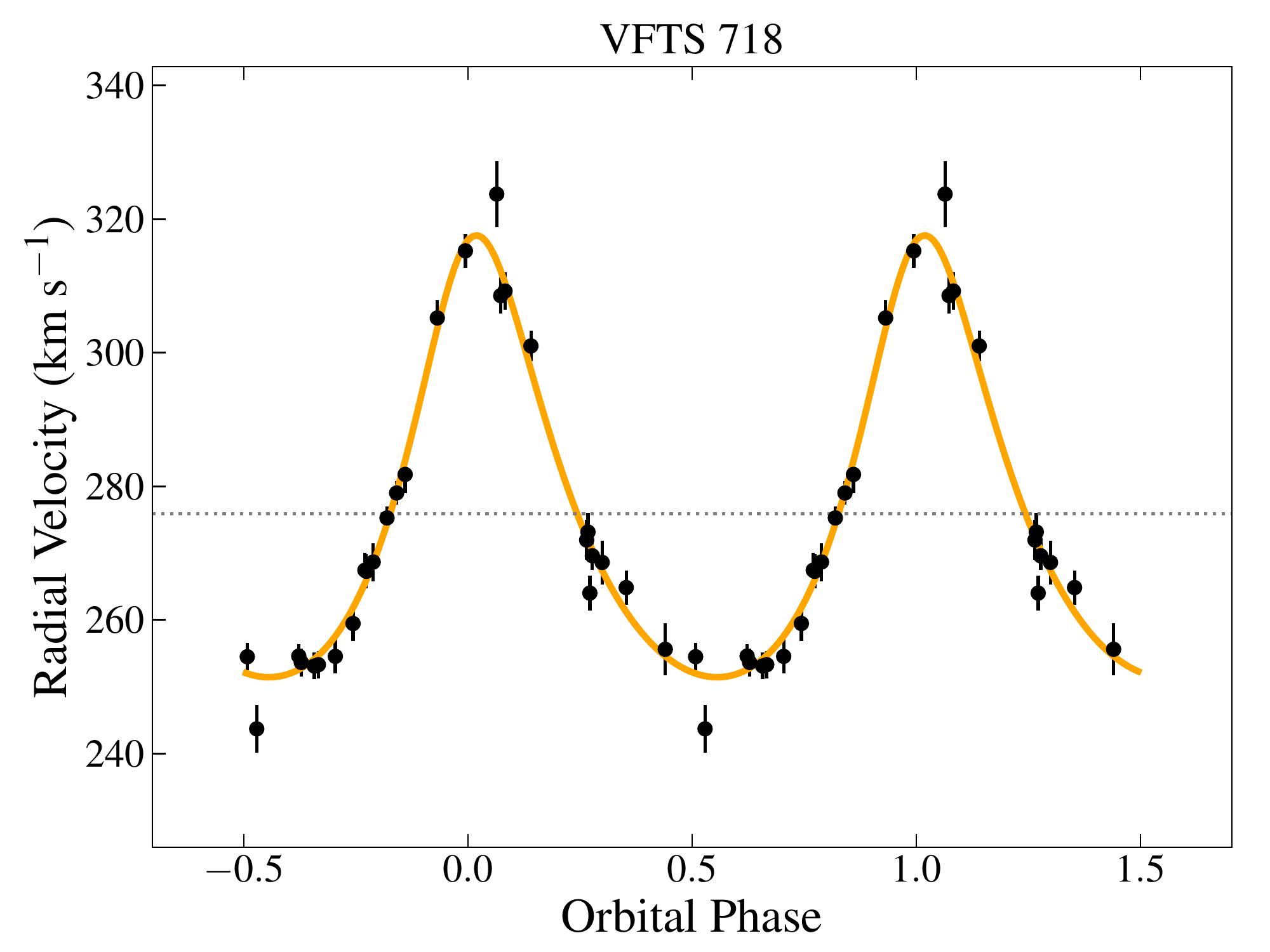}\hfill
    \includegraphics[width=0.31\textwidth]{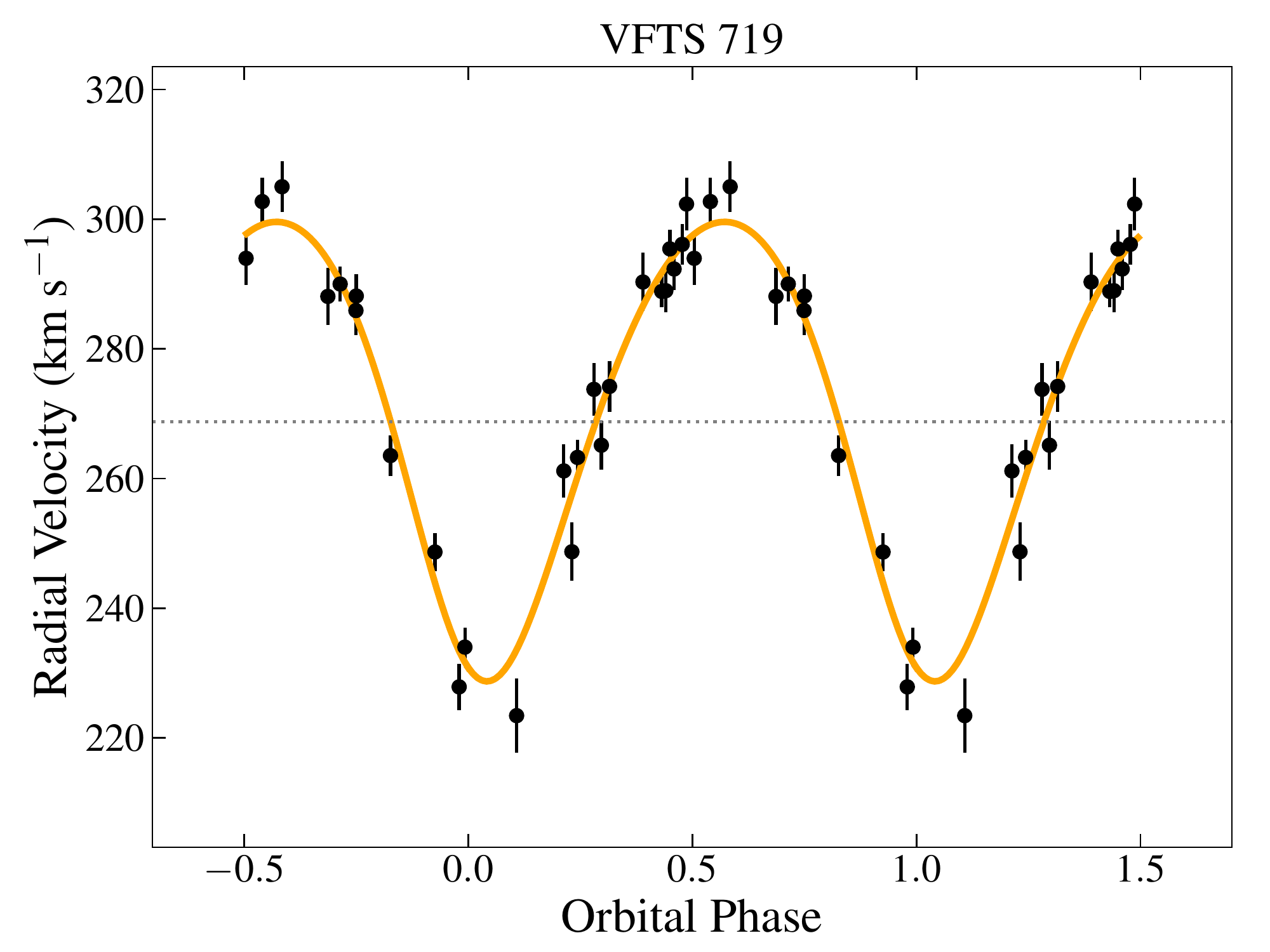}\hfill
    \includegraphics[width=0.31\textwidth]{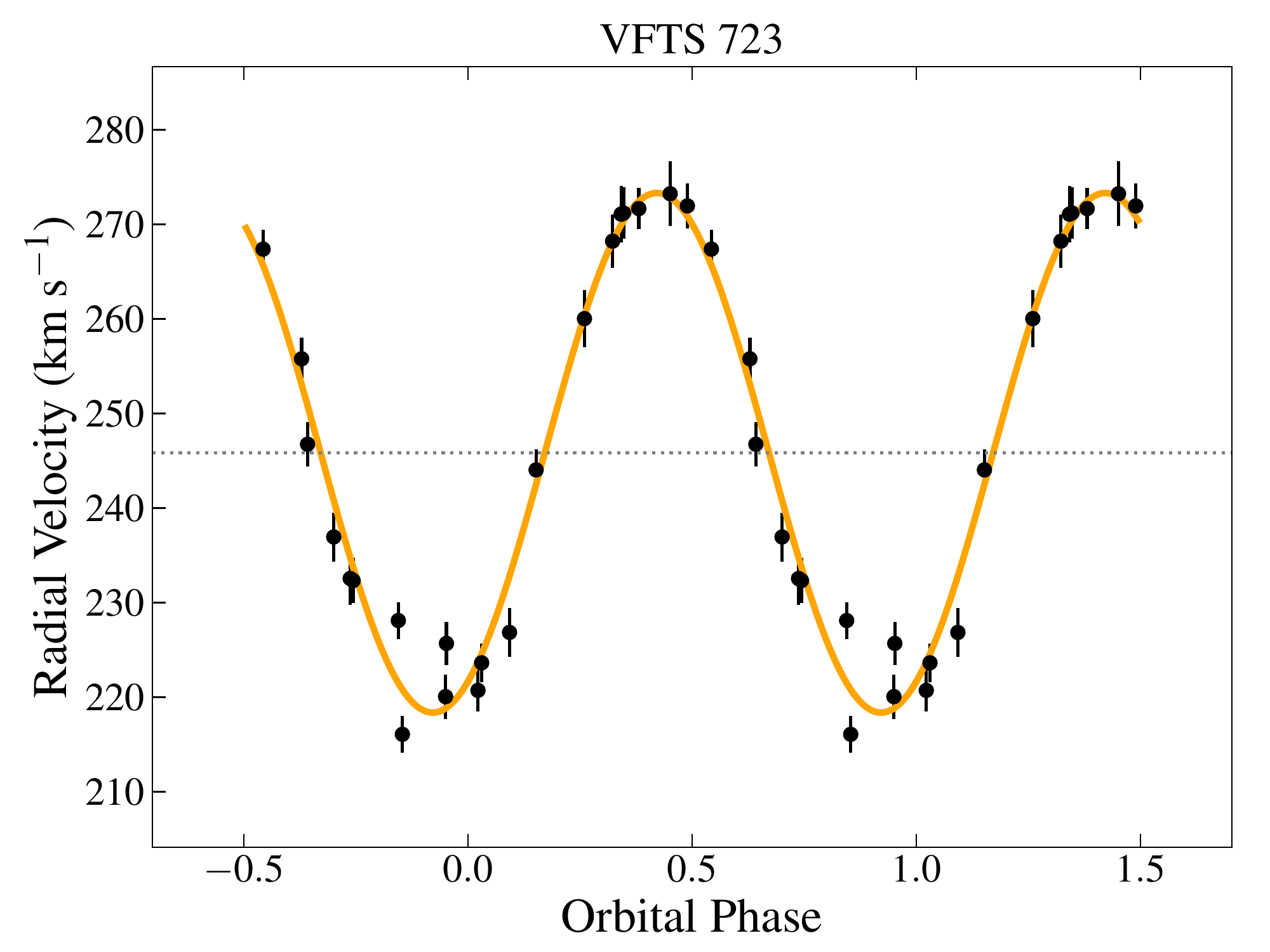}\hfill
    \includegraphics[width=0.31\textwidth]{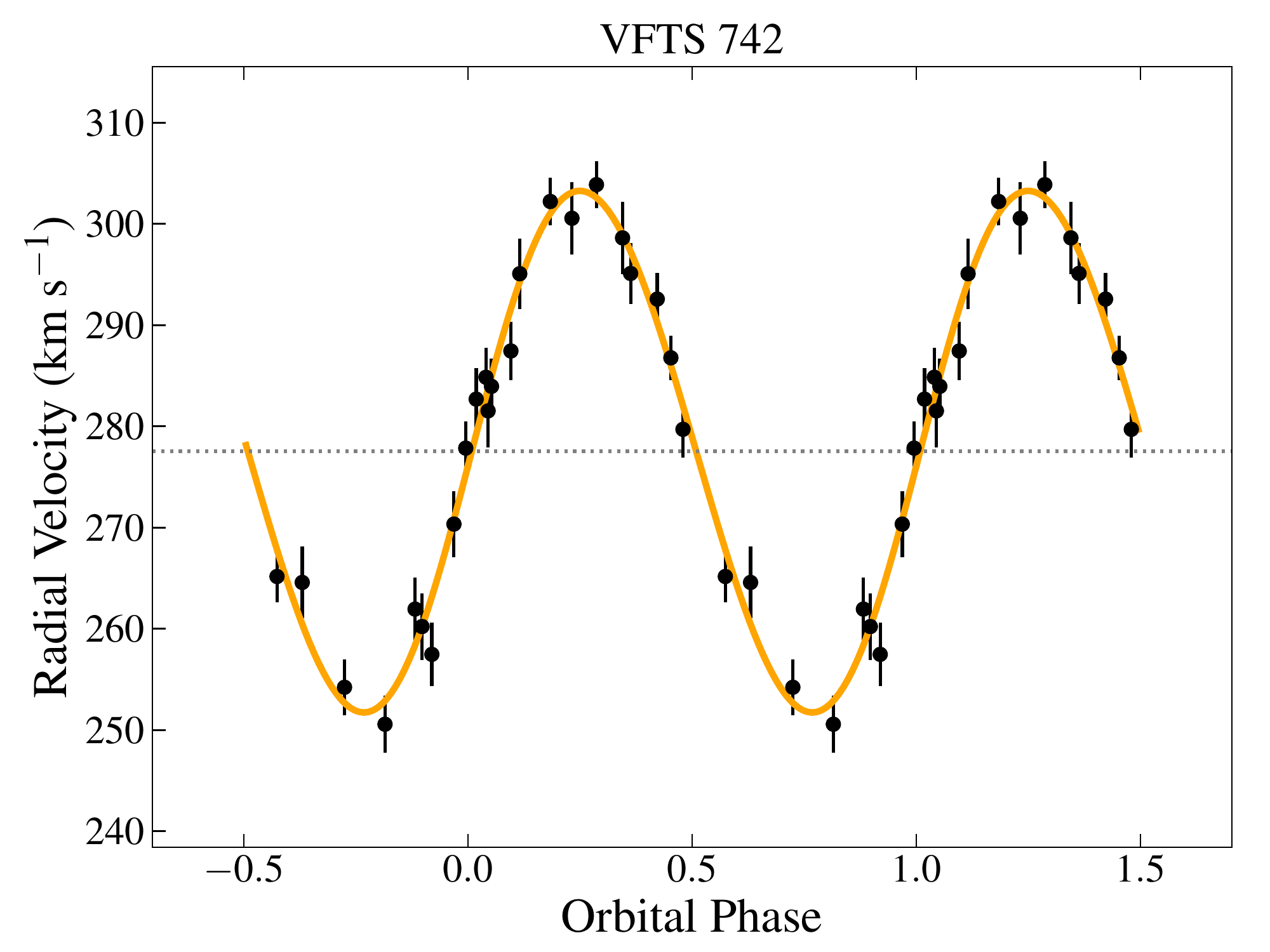}\hfill
    \includegraphics[width=0.31\textwidth]{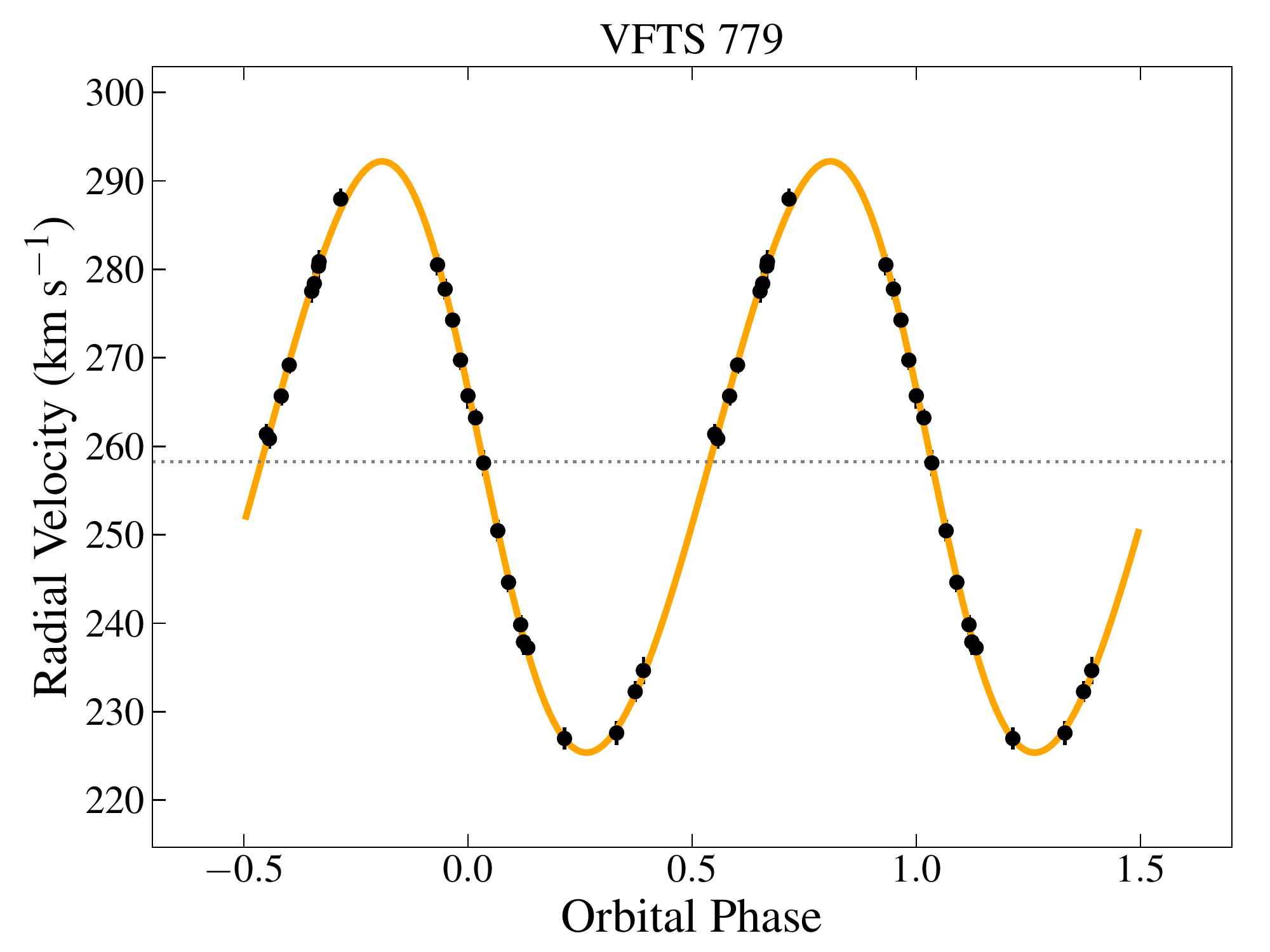}\hfill
    \includegraphics[width=0.31\textwidth]{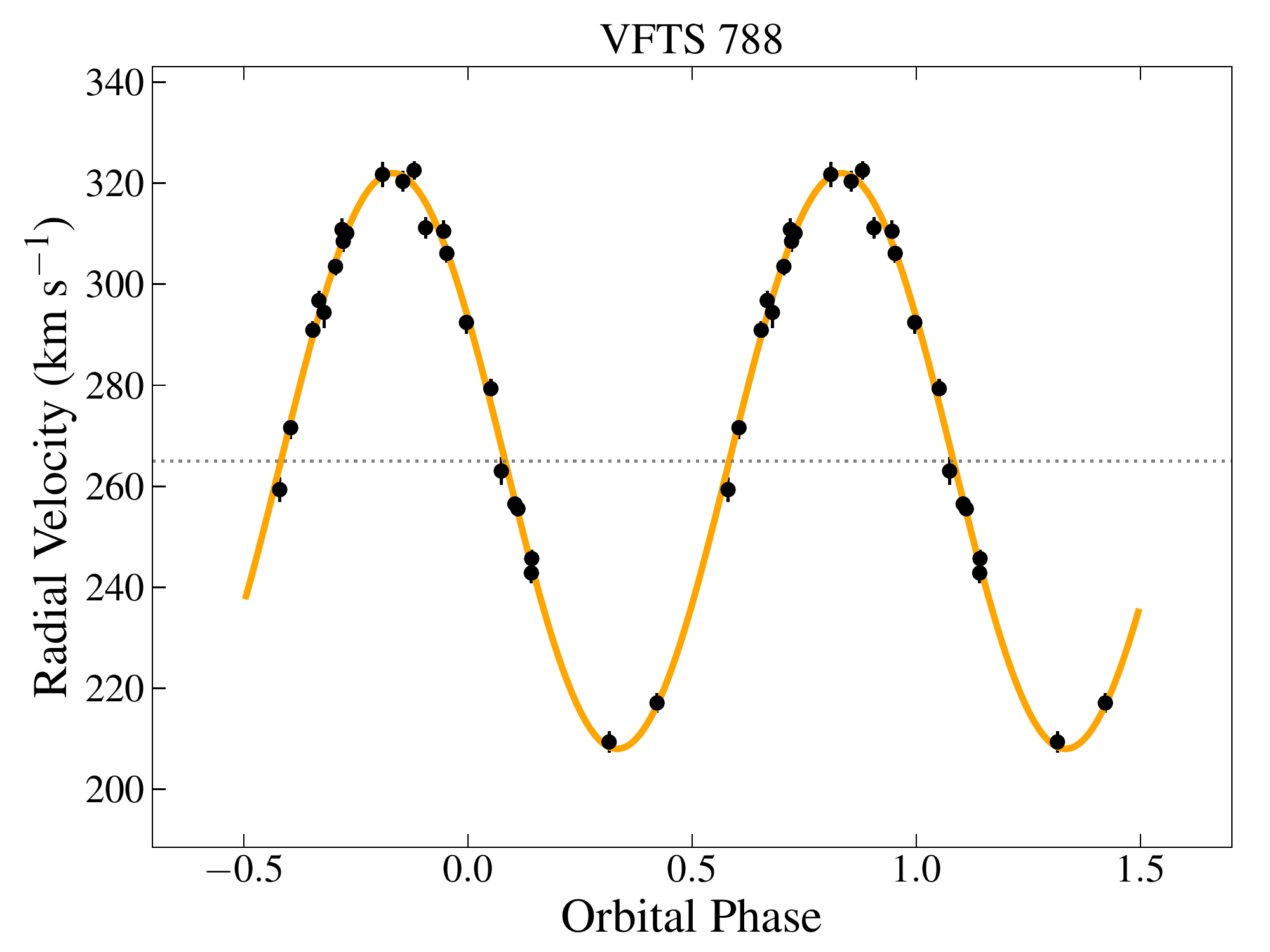}\hfill
    \includegraphics[width=0.31\textwidth]{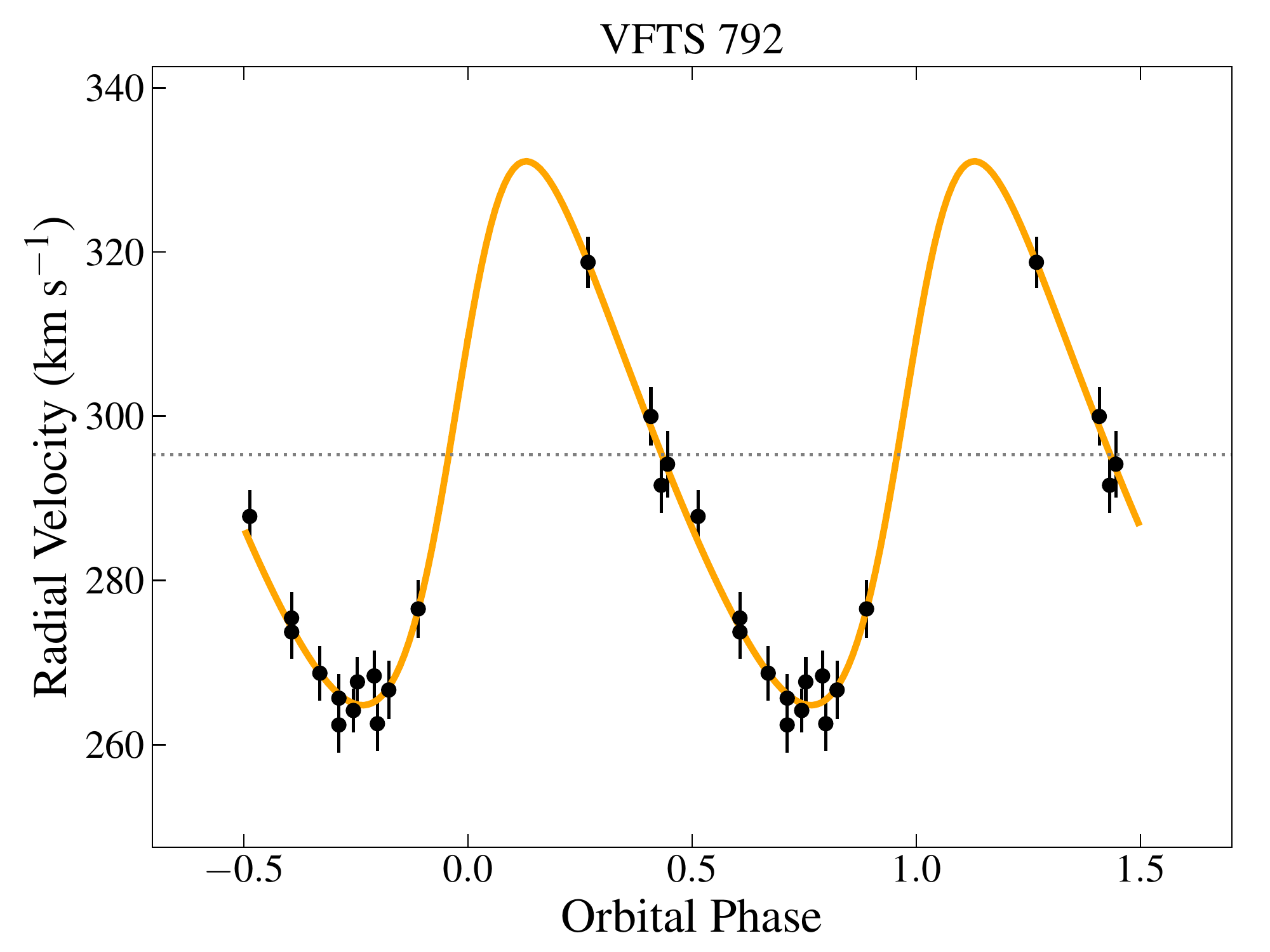}\hfill
    \includegraphics[width=0.31\textwidth]{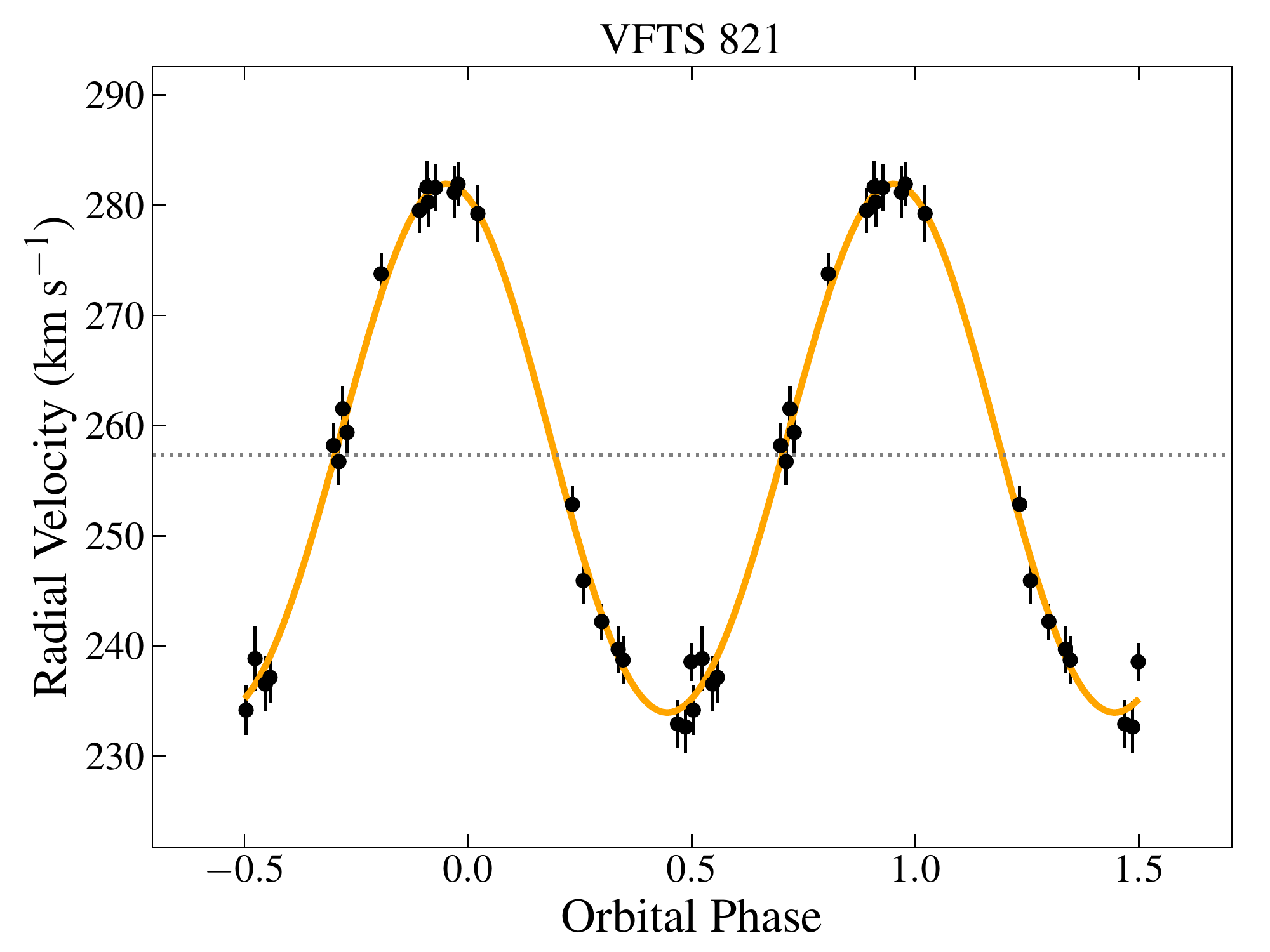}\hfill
    \caption{$-$ \it continued}
\end{myfloat}

\begin{myfloat}
\ContinuedFloat
    \centering
    \includegraphics[width=0.31\textwidth]{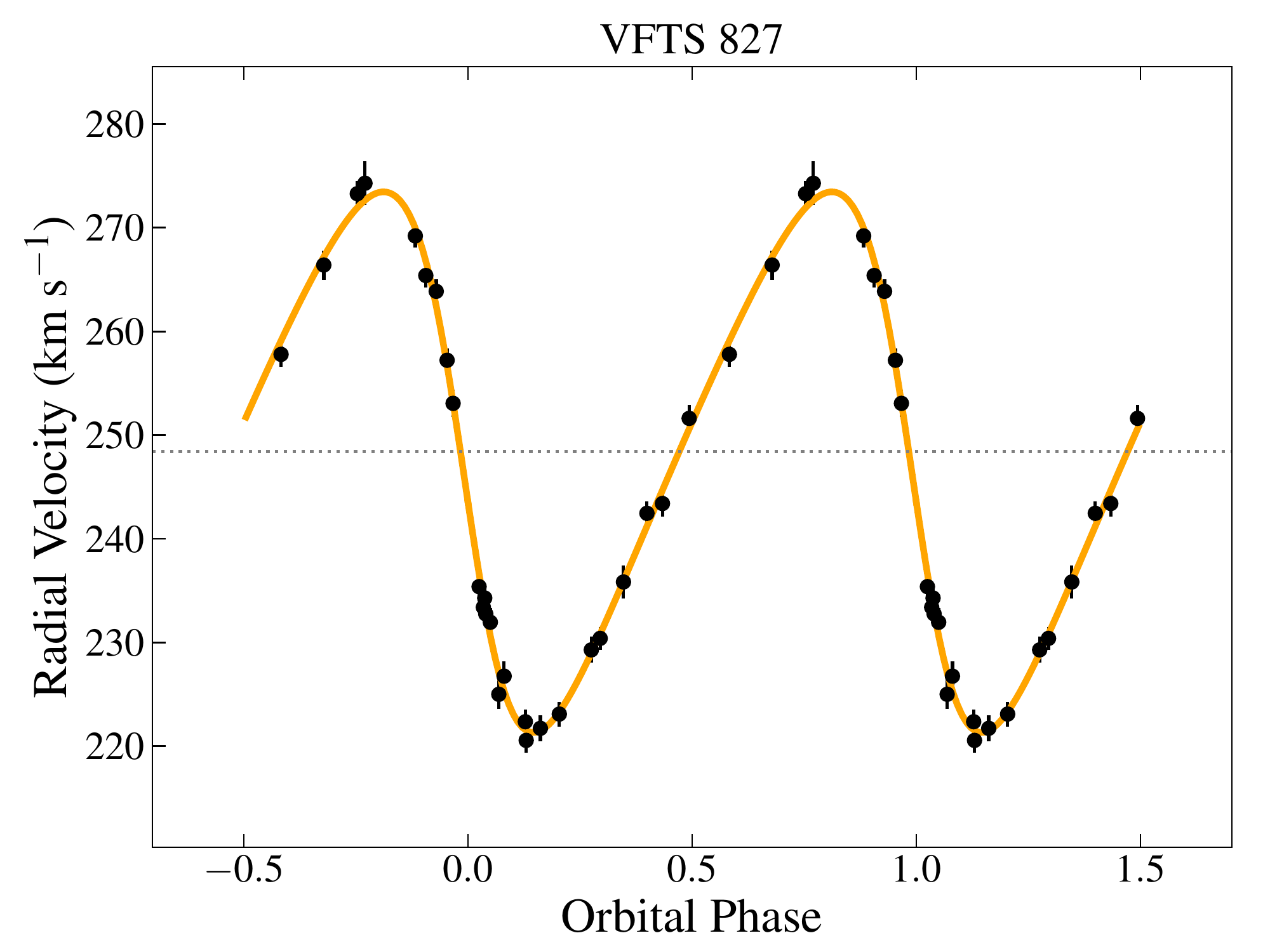}\hfill
    \includegraphics[width=0.31\textwidth]{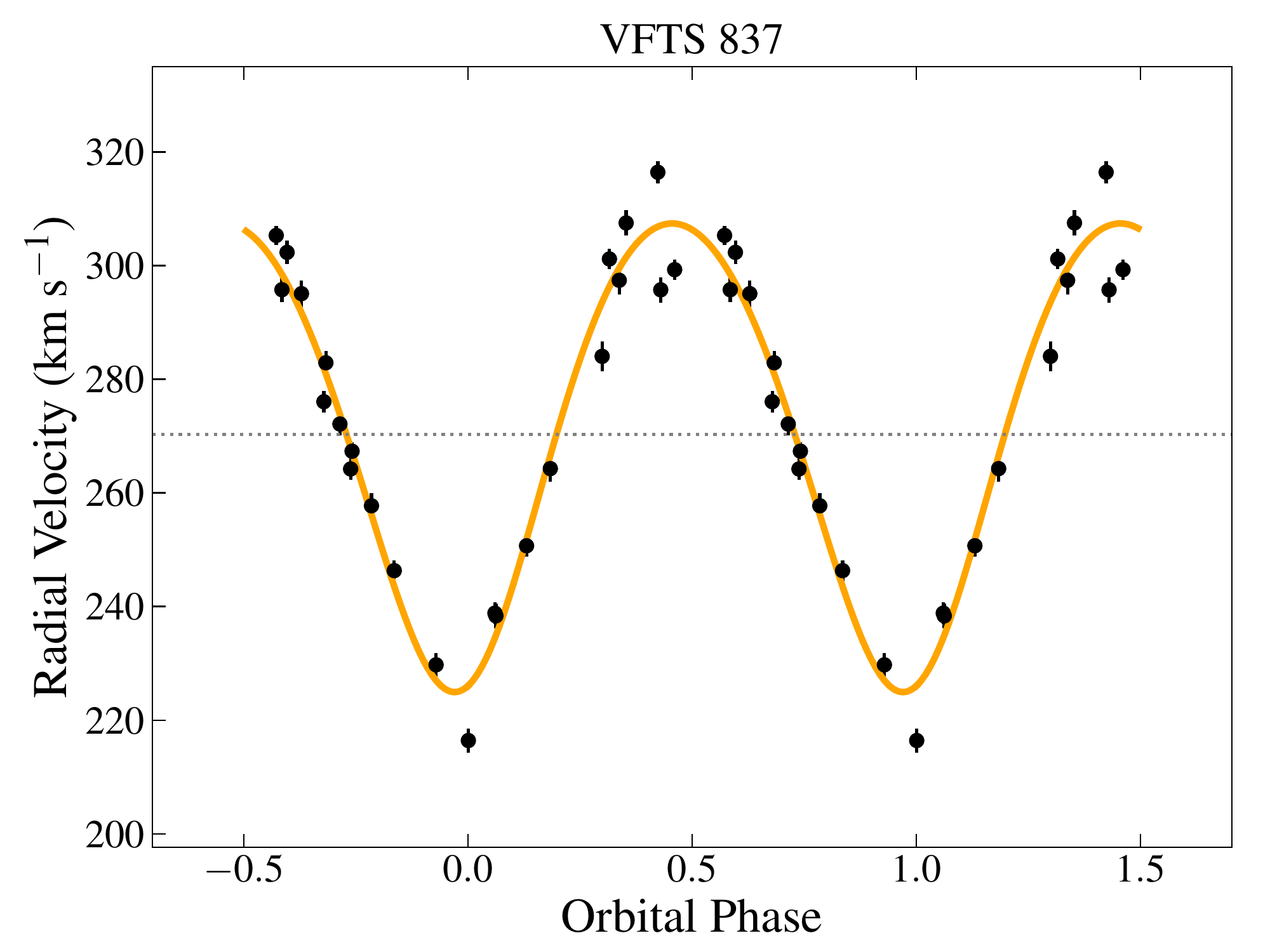}\hfill
    \includegraphics[width=0.31\textwidth]{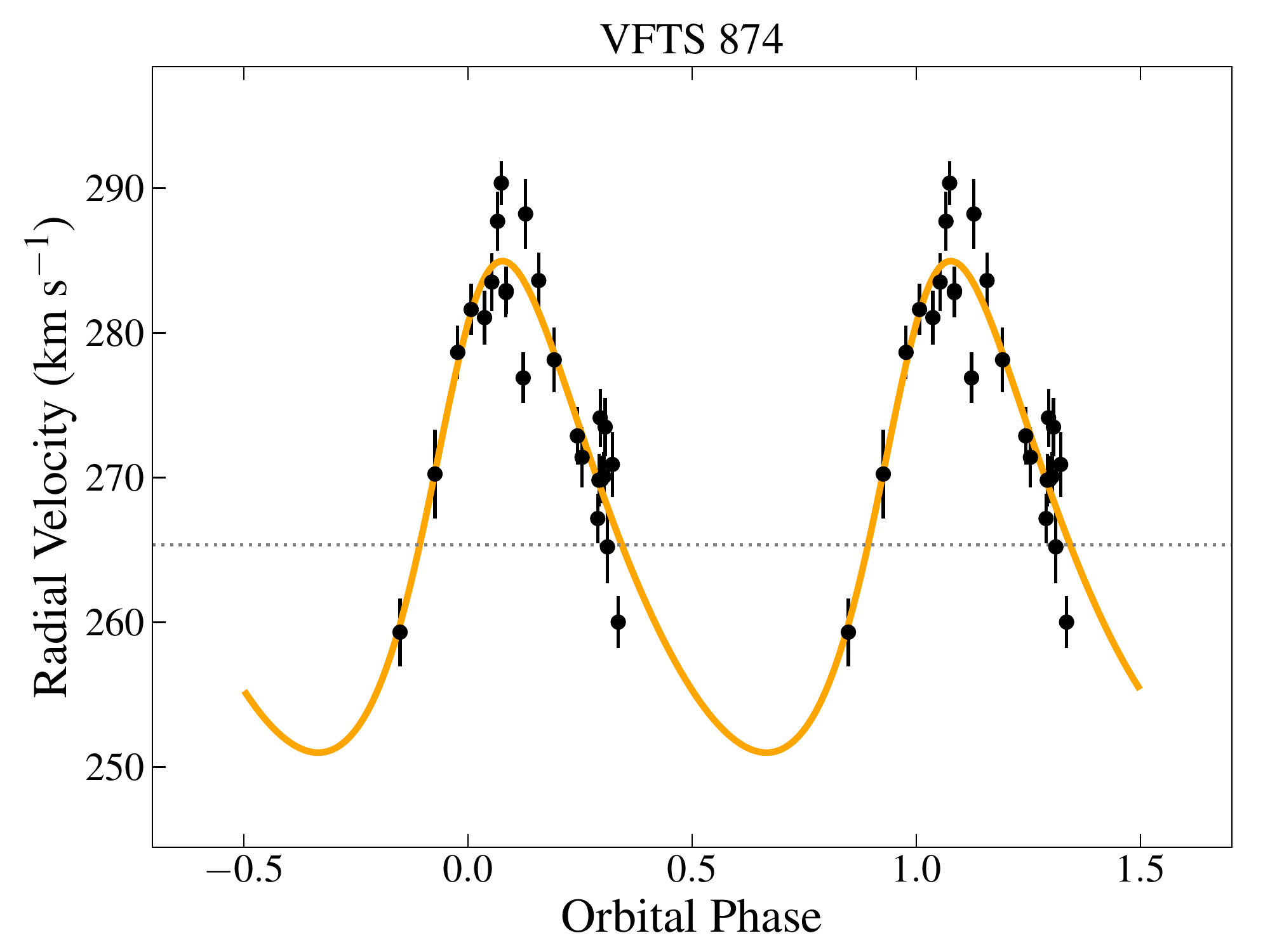}\hfill
    \includegraphics[width=0.31\textwidth]{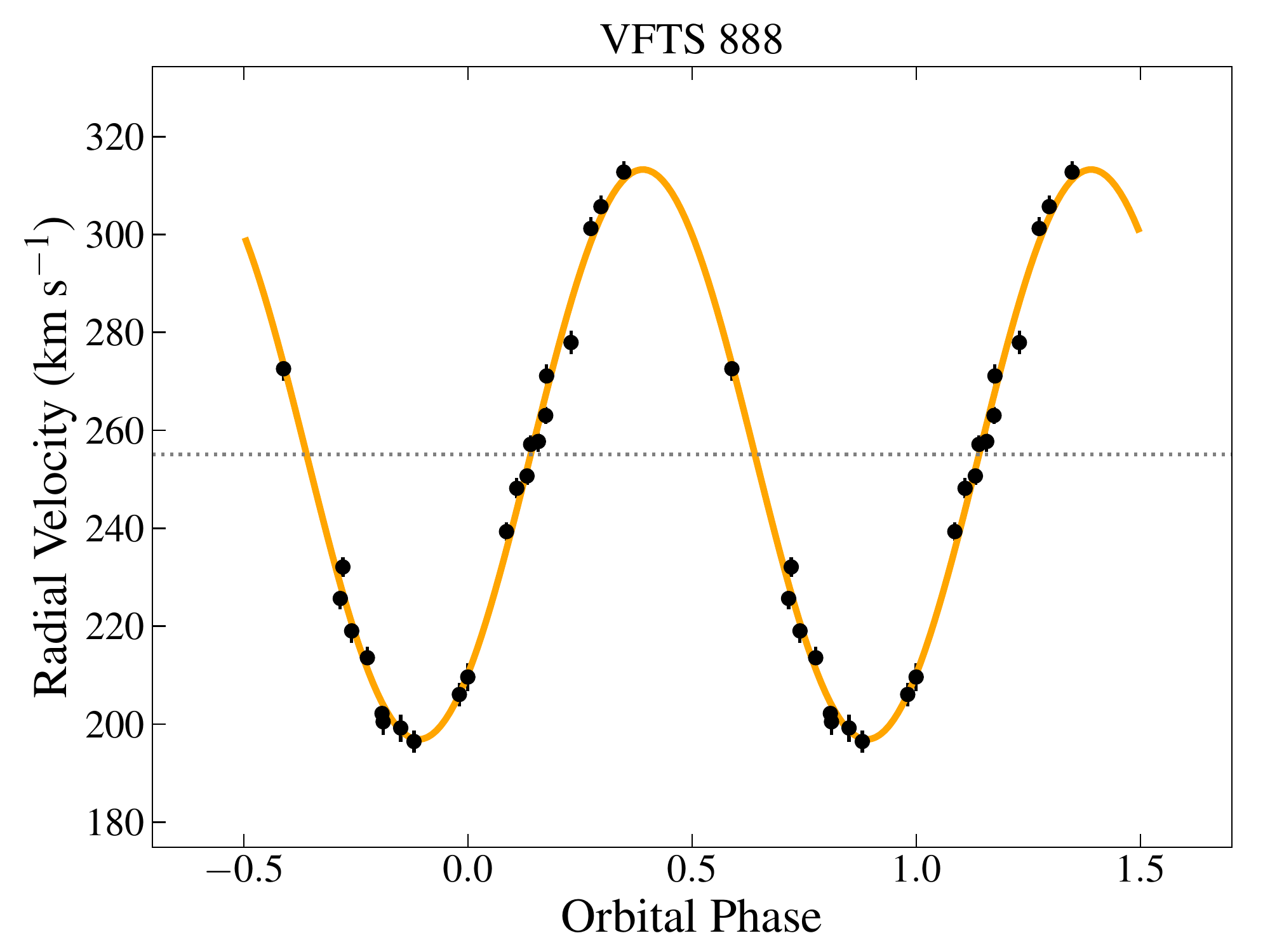}\hspace{0.035\textwidth}
    \includegraphics[width=0.31\textwidth]{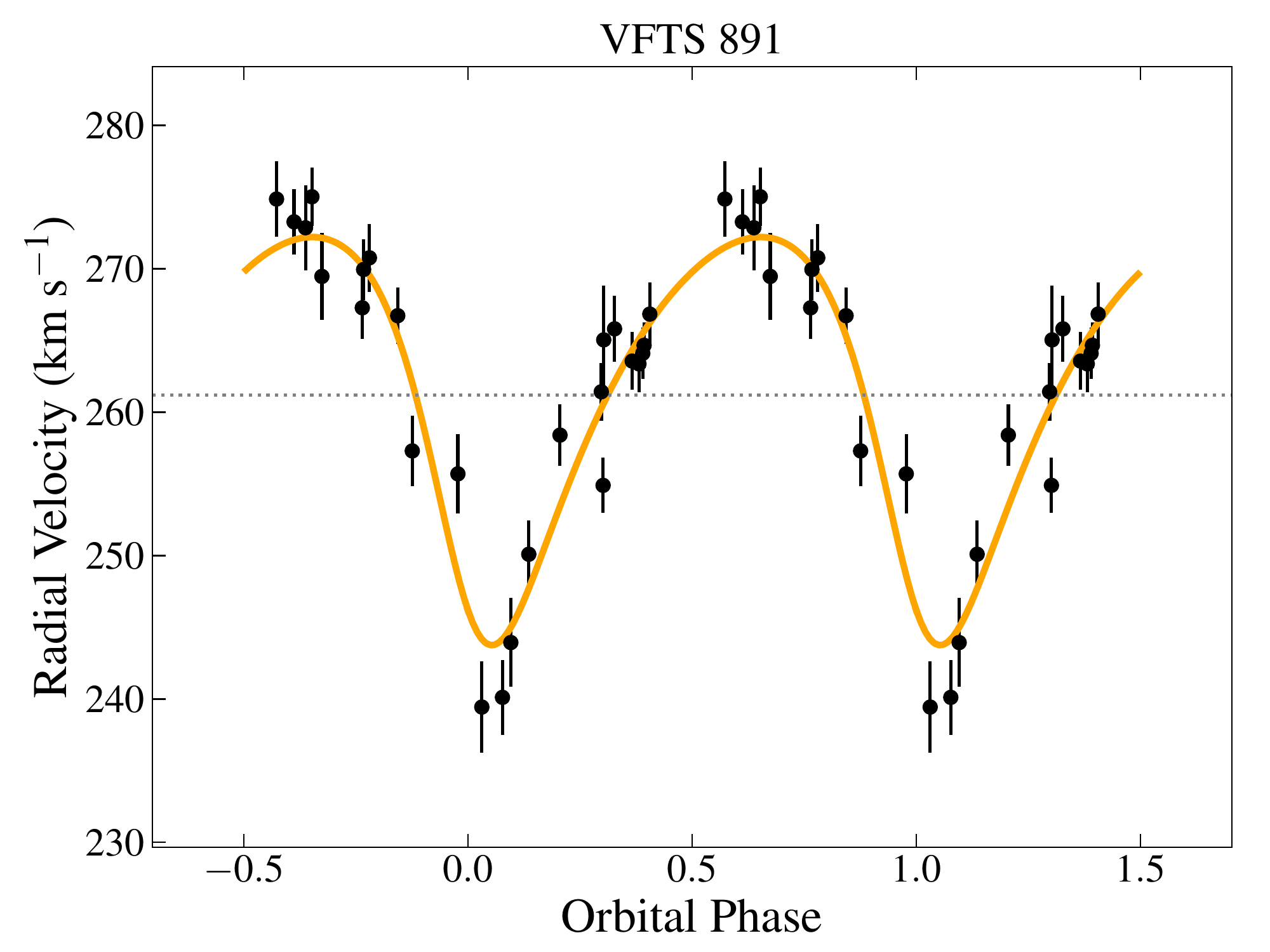}
    \caption{$-$ \it continued}
\end{myfloat}
\clearpage
\begin{figure*}
    \centering
    \includegraphics[width=0.31\textwidth]{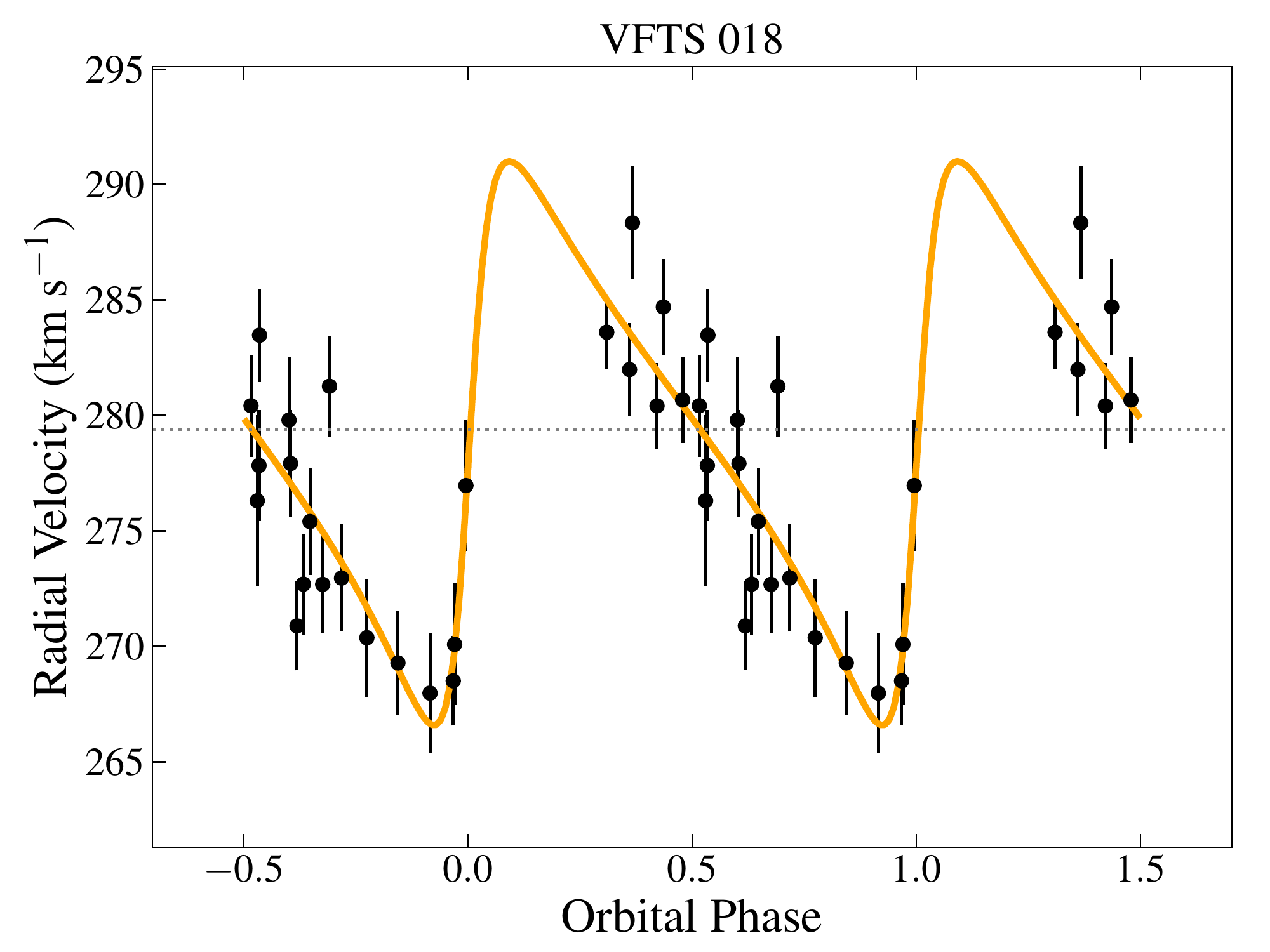}\hfill
    \includegraphics[width=0.31\textwidth]{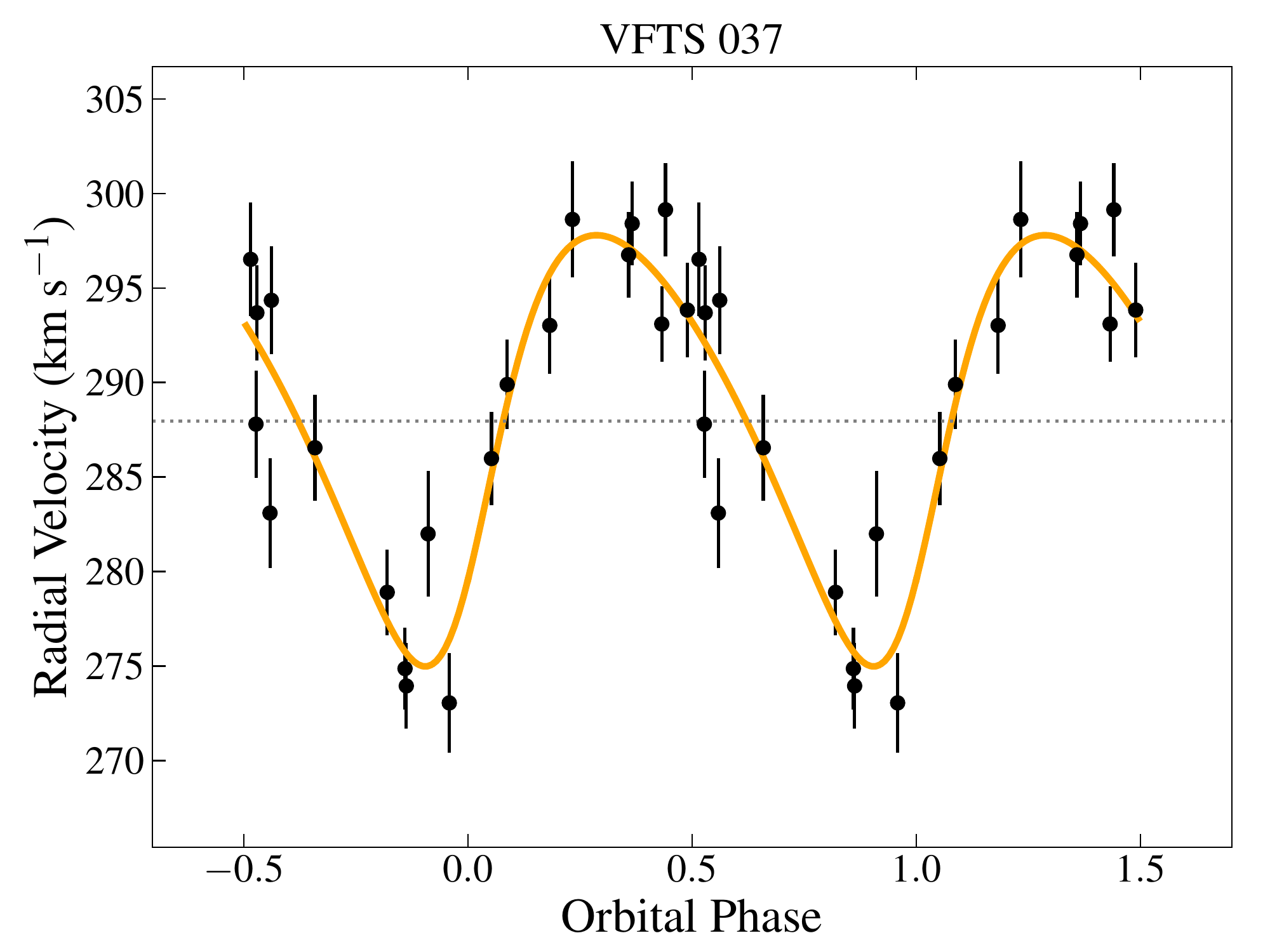}\hfill
    \includegraphics[width=0.31\textwidth]{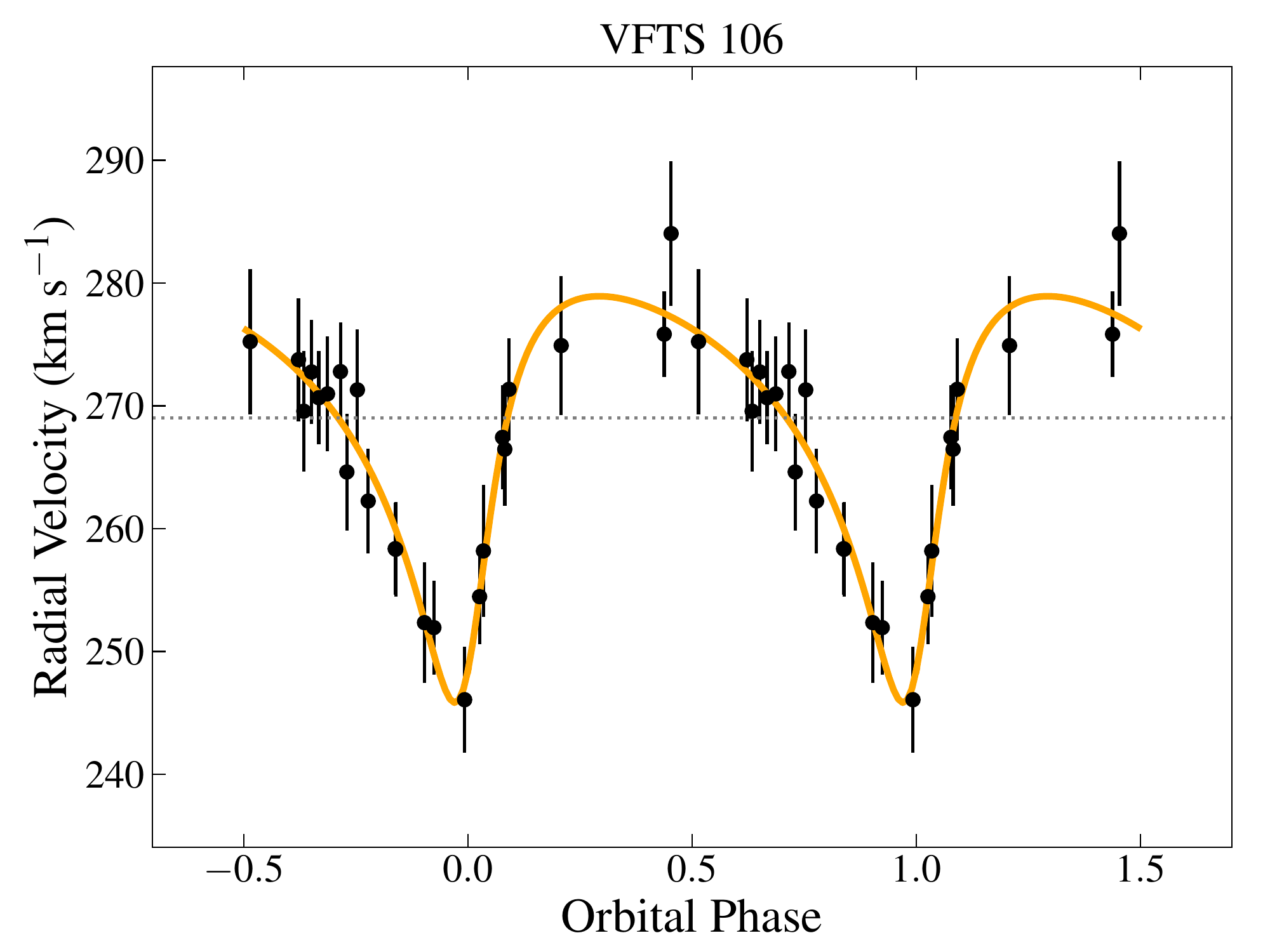}\hfill
    \includegraphics[width=0.31\textwidth]{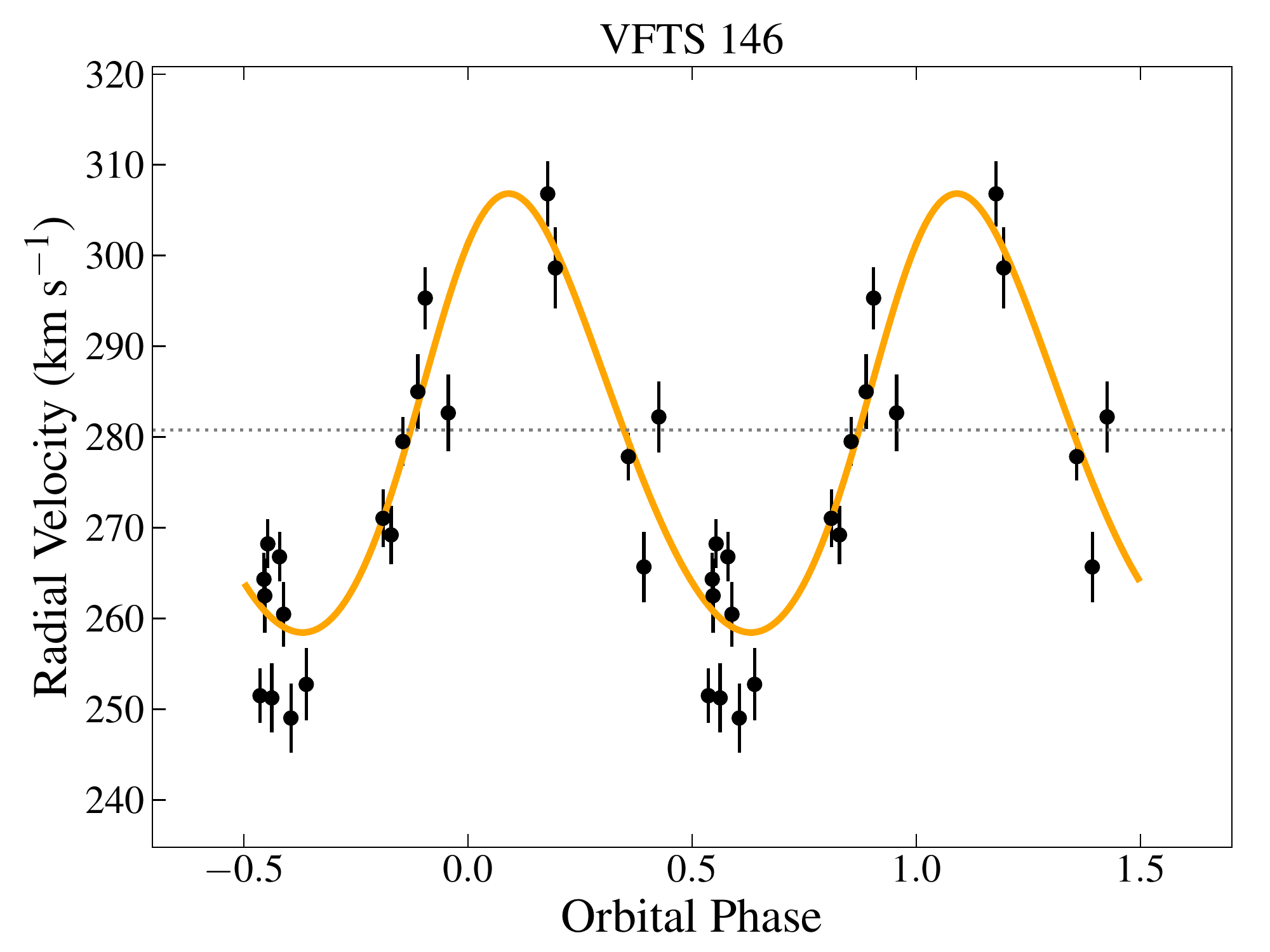}\hfill
    \includegraphics[width=0.31\textwidth]{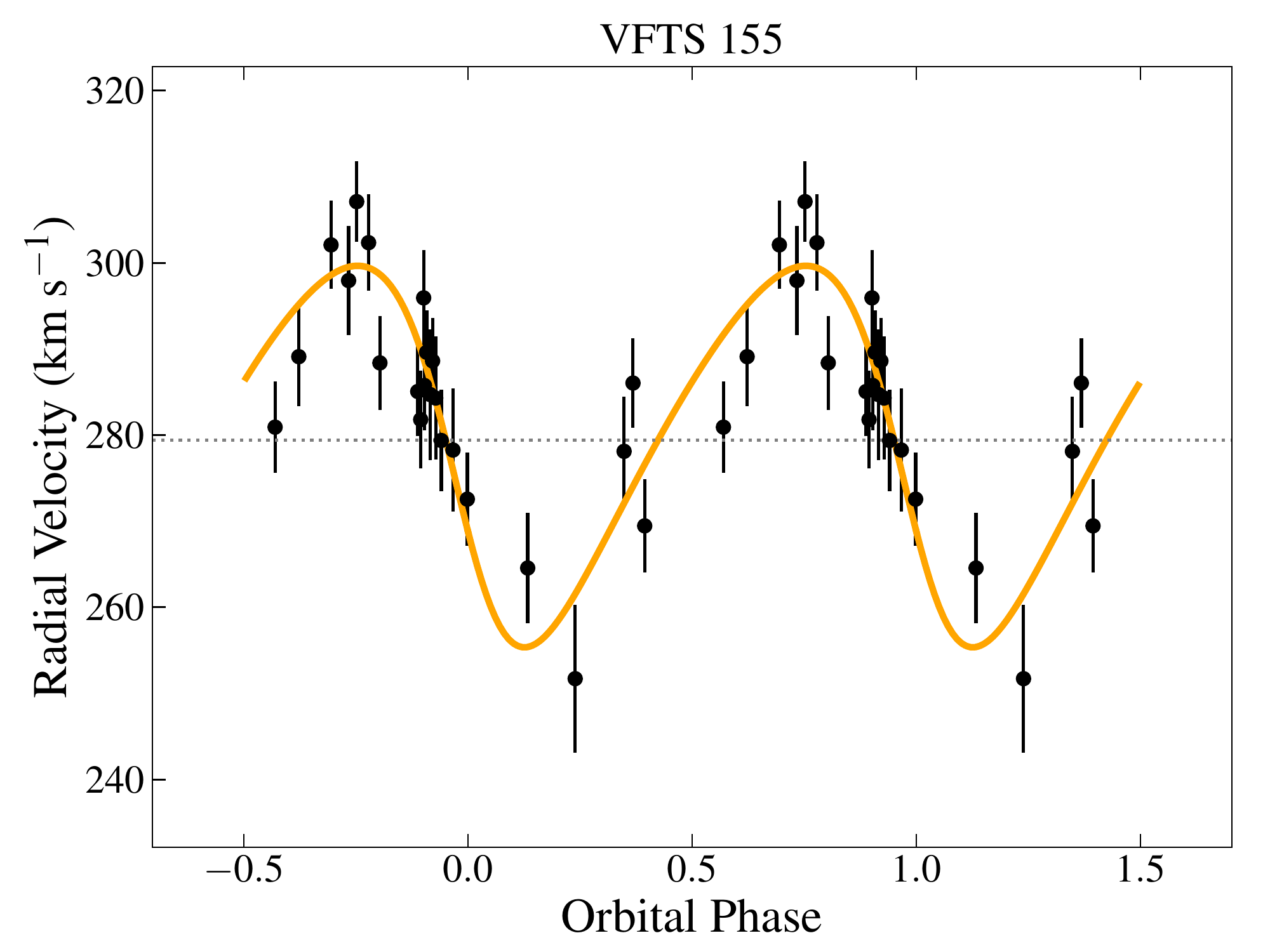}\hfill
    \includegraphics[width=0.31\textwidth]{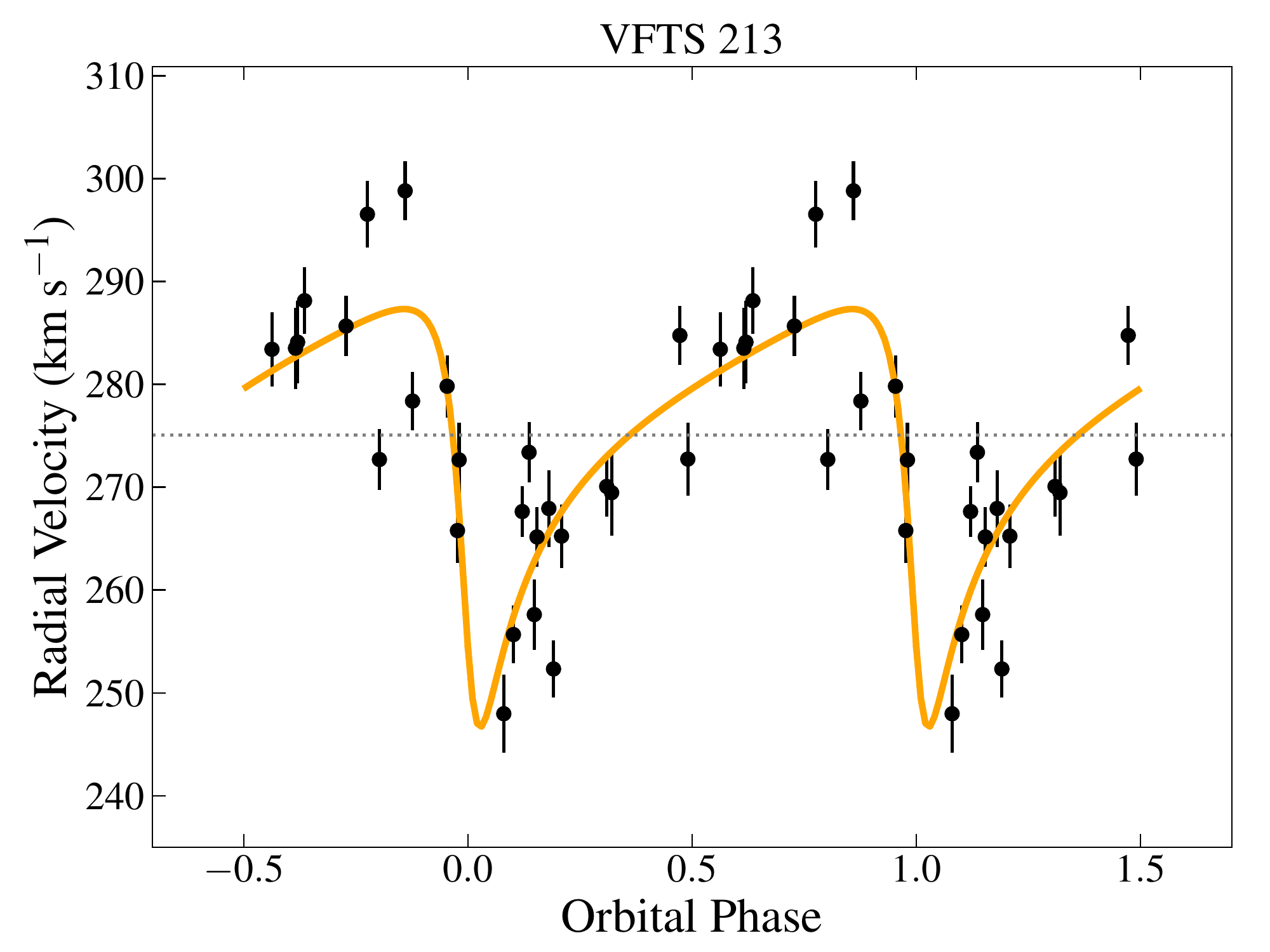}\hfill
    \includegraphics[width=0.31\textwidth]{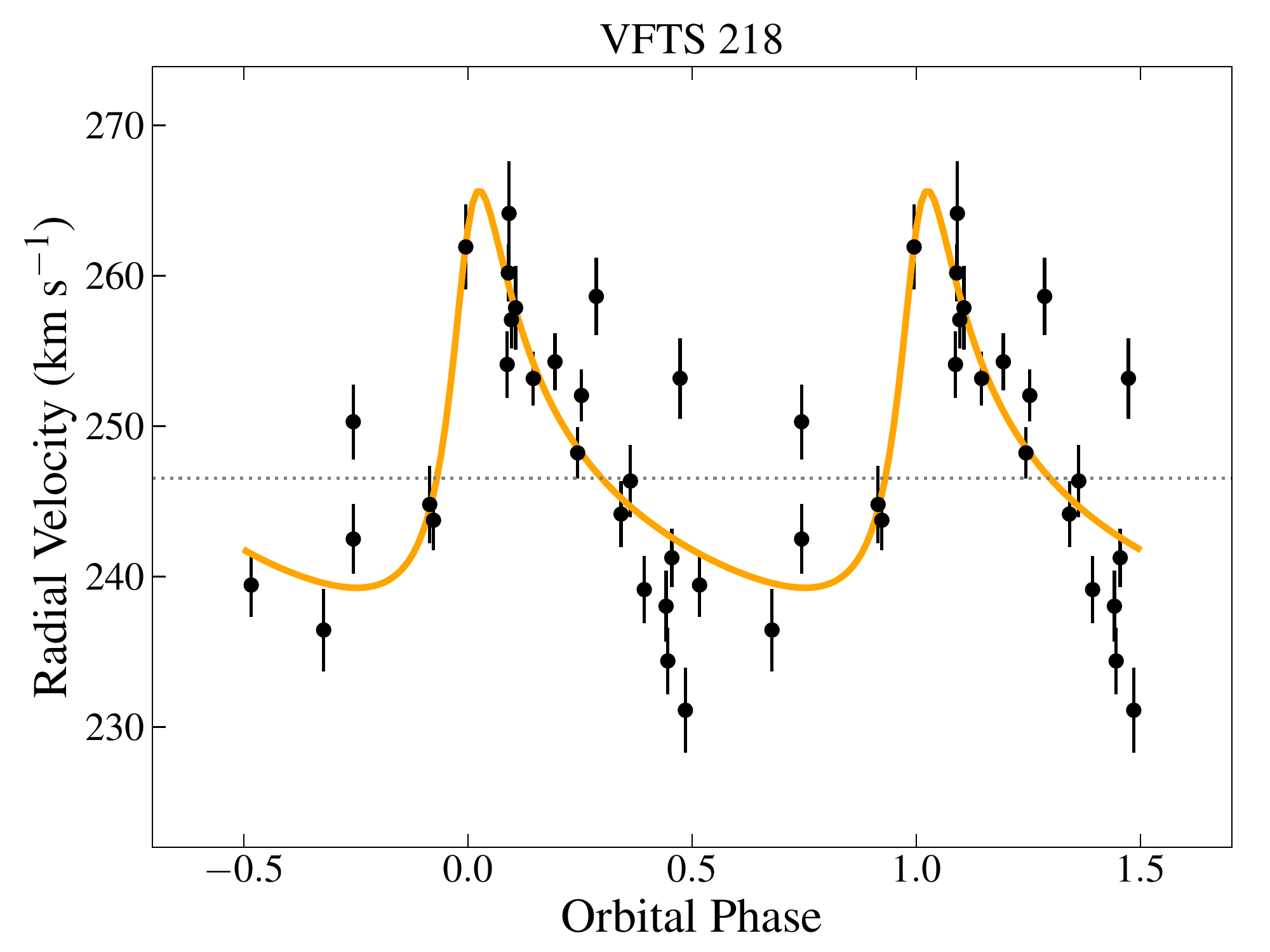}\hfill
    \includegraphics[width=0.31\textwidth]{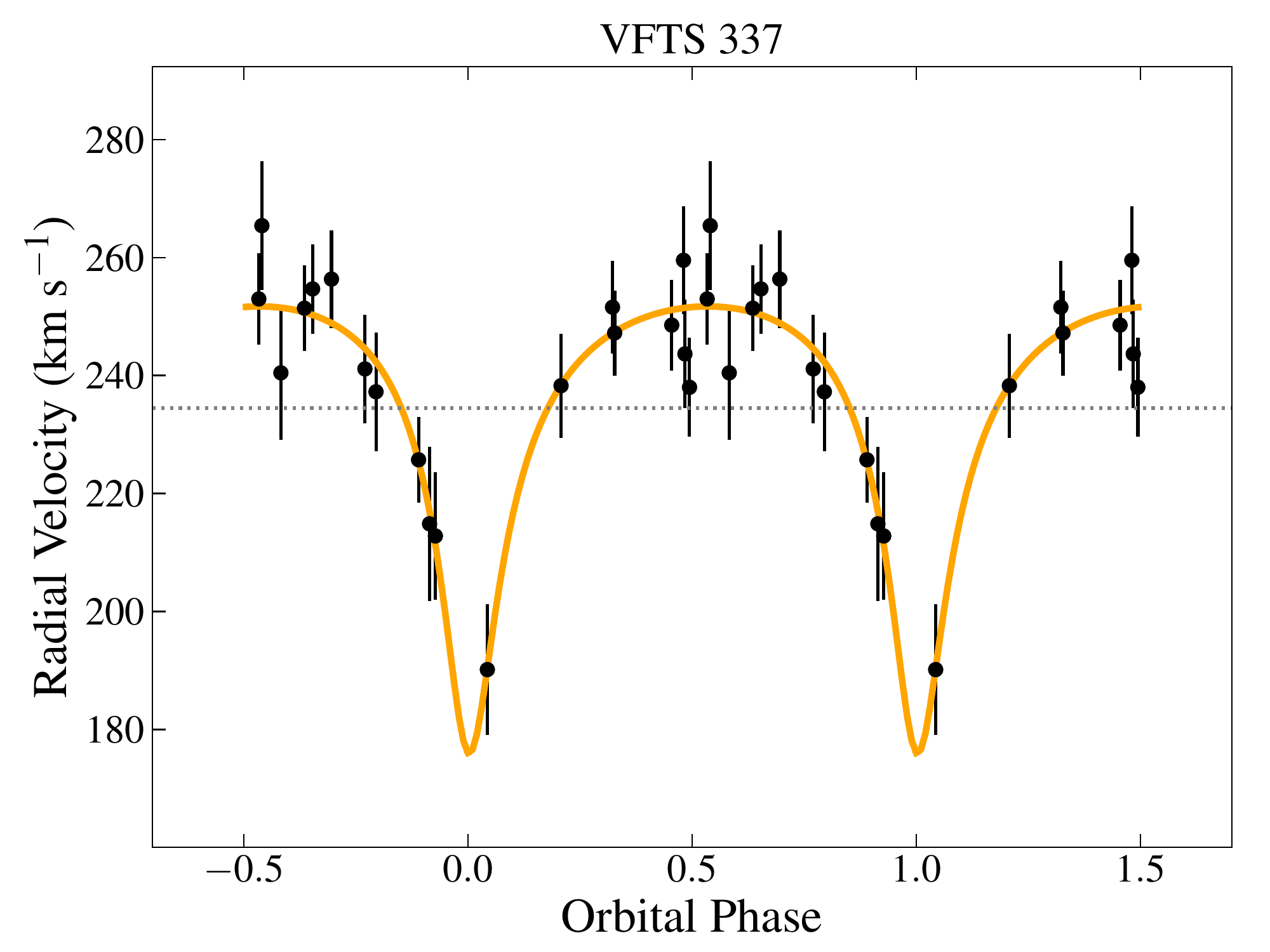}\hfill
    \includegraphics[width=0.31\textwidth]{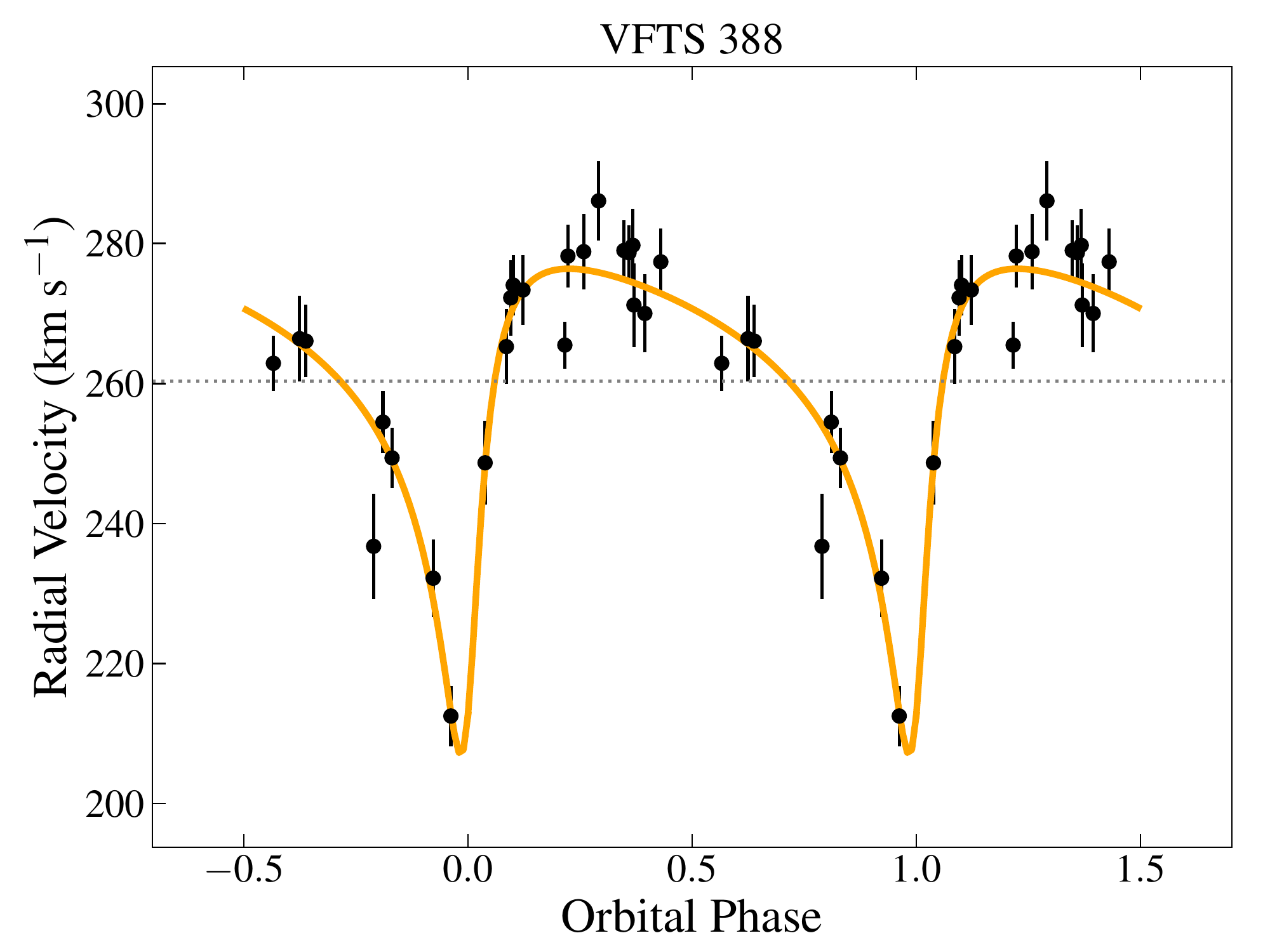}\hfill
    \includegraphics[width=0.31\textwidth]{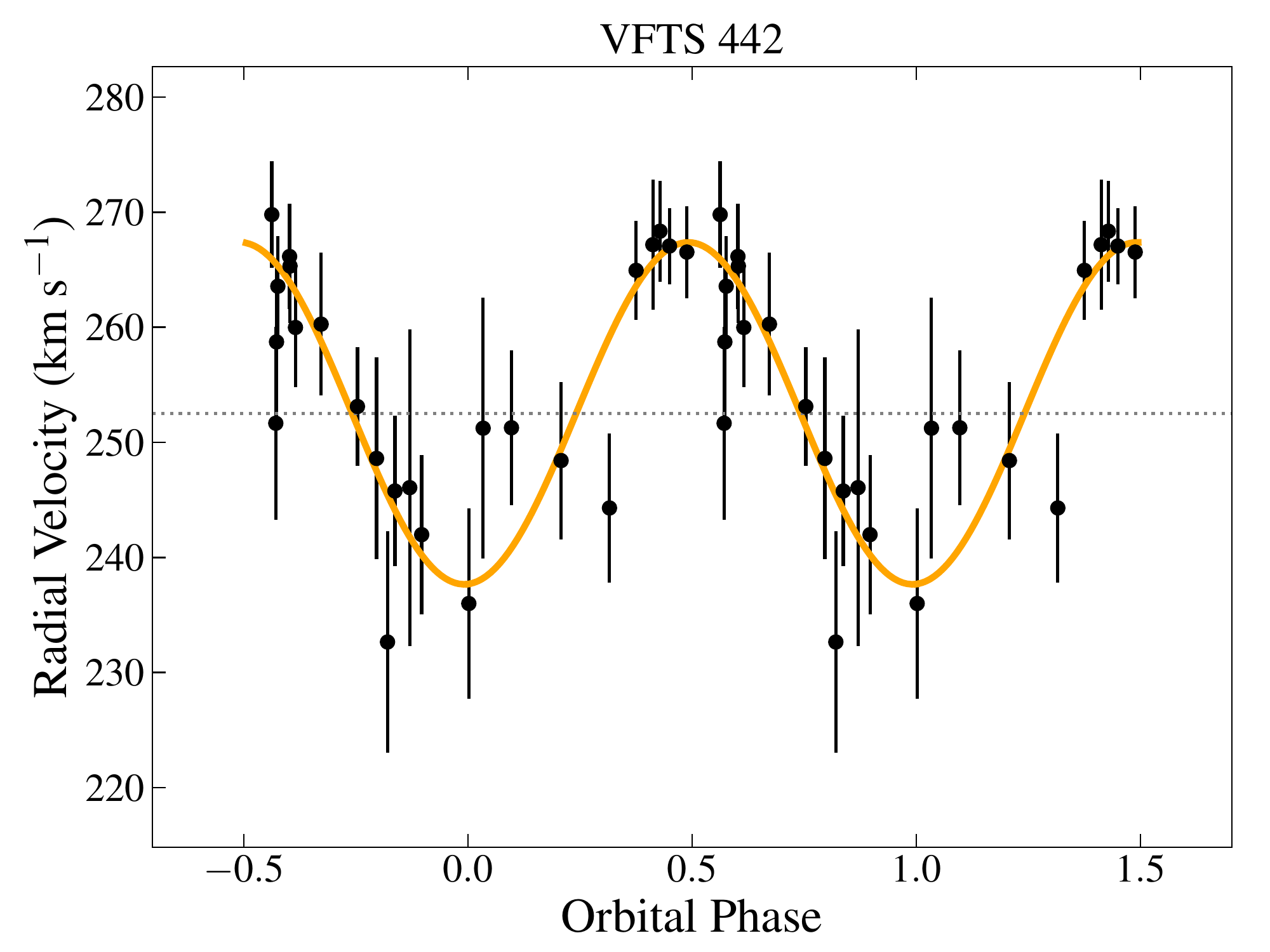}\hfill
    \includegraphics[width=0.31\textwidth]{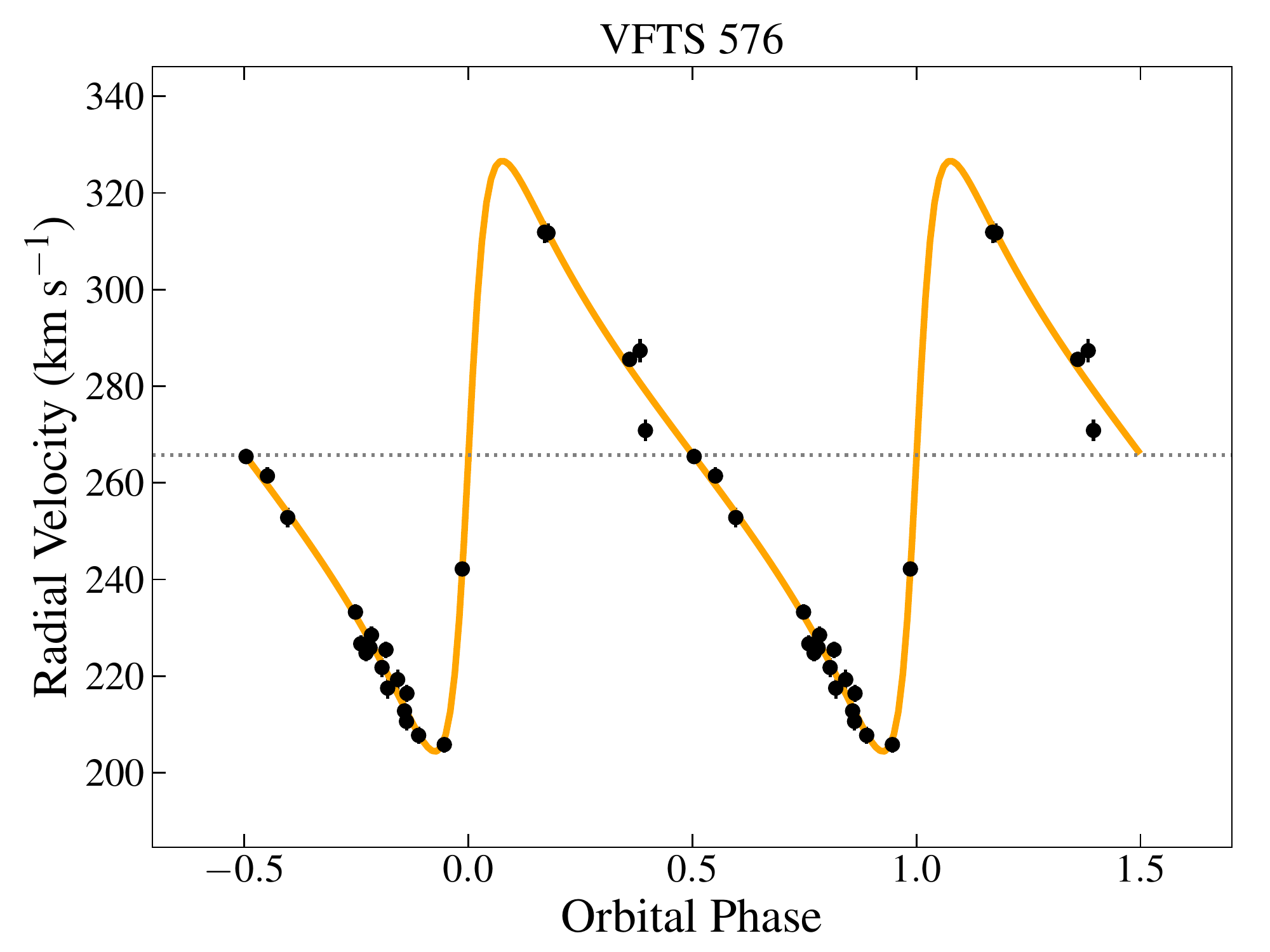}\hfill
    \includegraphics[width=0.31\textwidth]{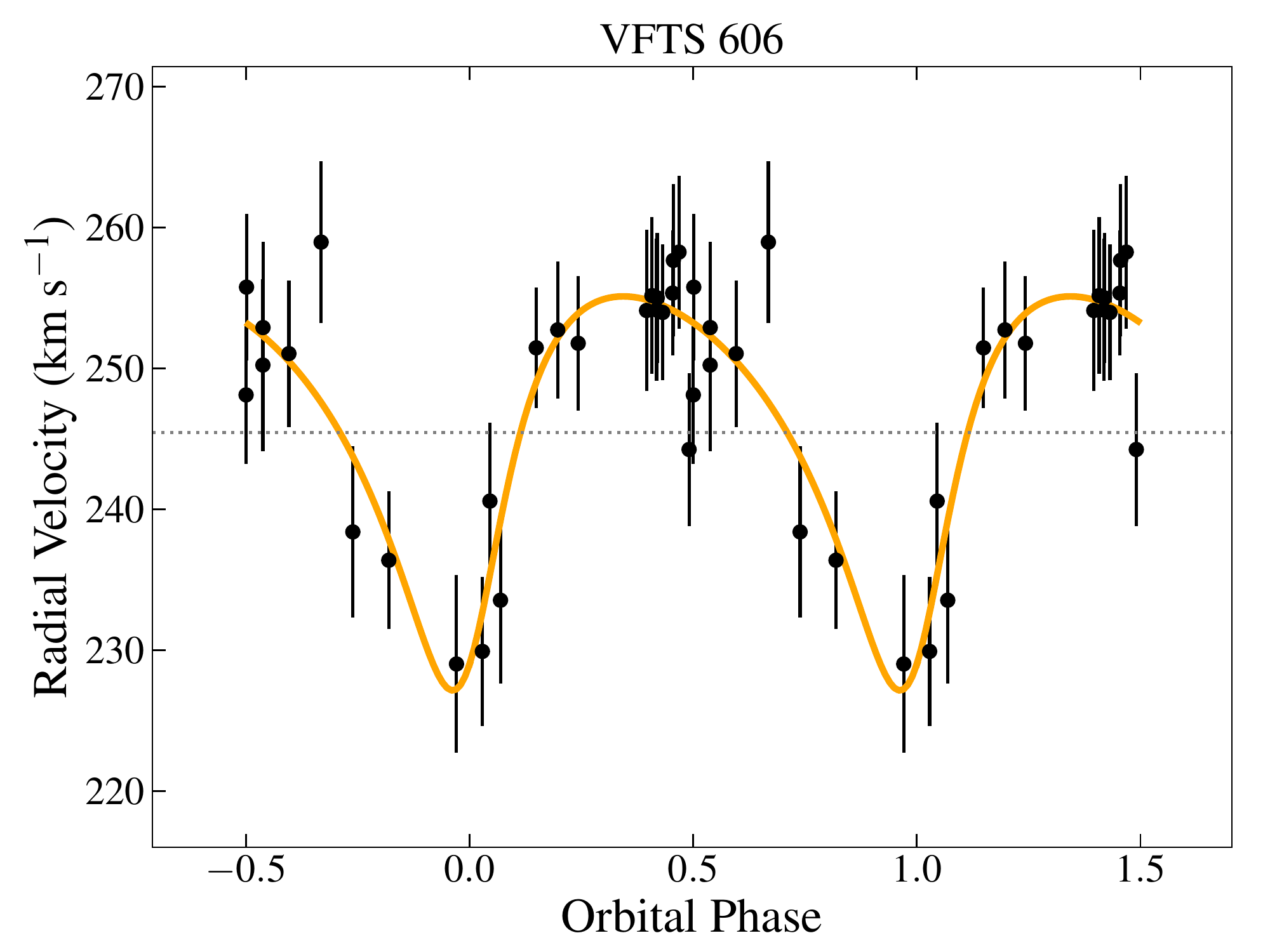}\hfill
    \includegraphics[width=0.31\textwidth]{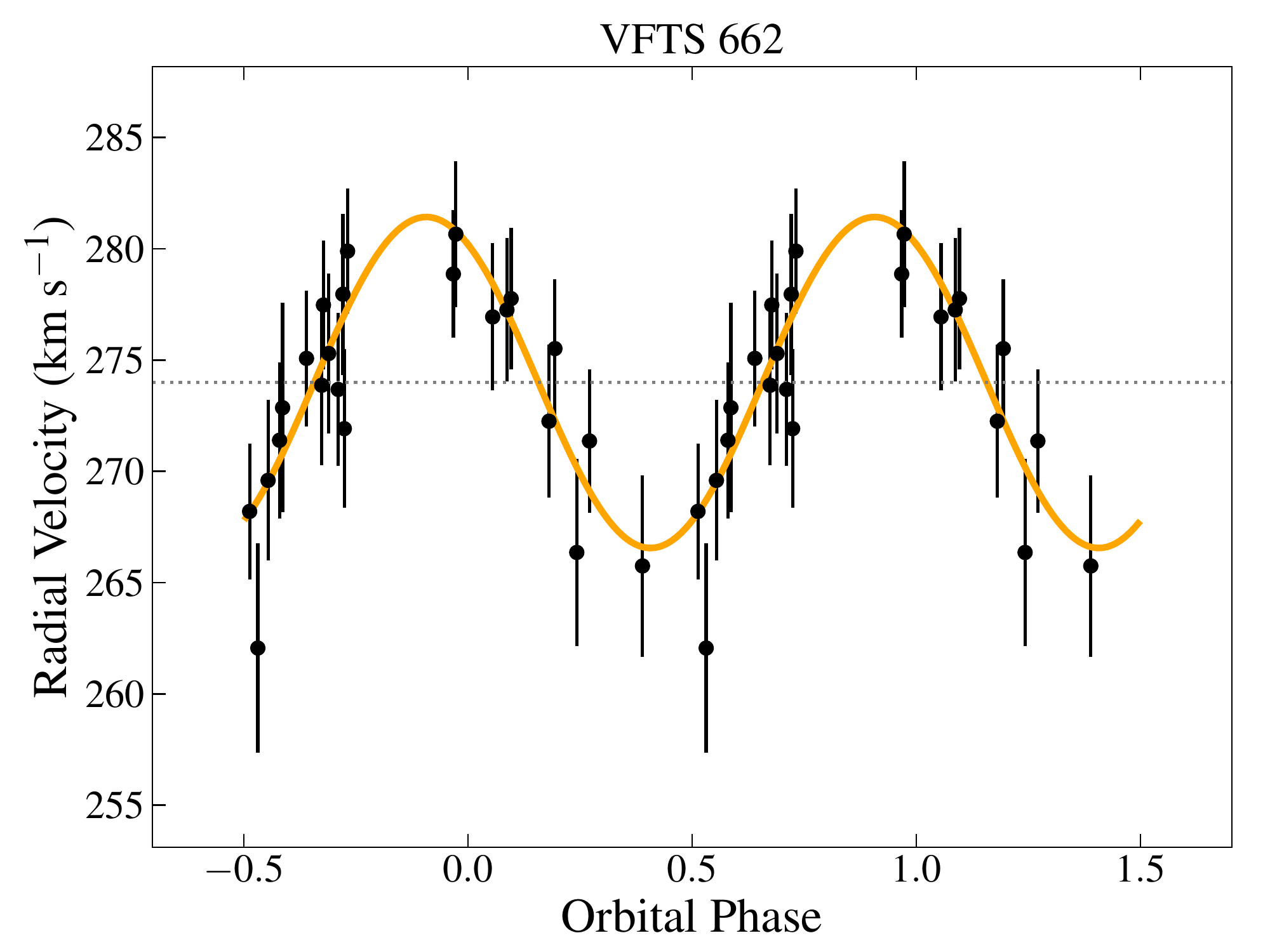}\hfill
    \includegraphics[width=0.31\textwidth]{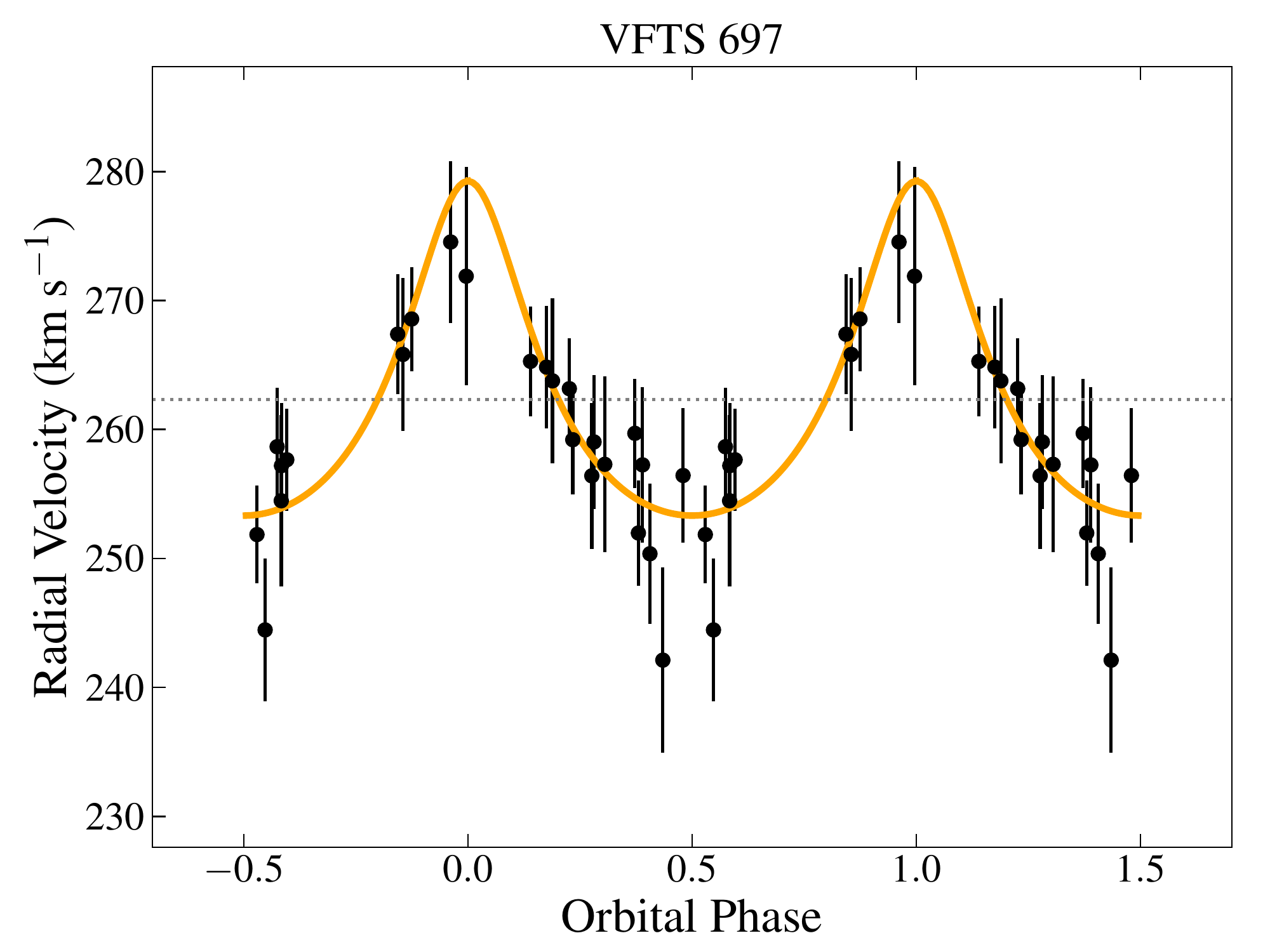}\hfill
    \includegraphics[width=0.31\textwidth]{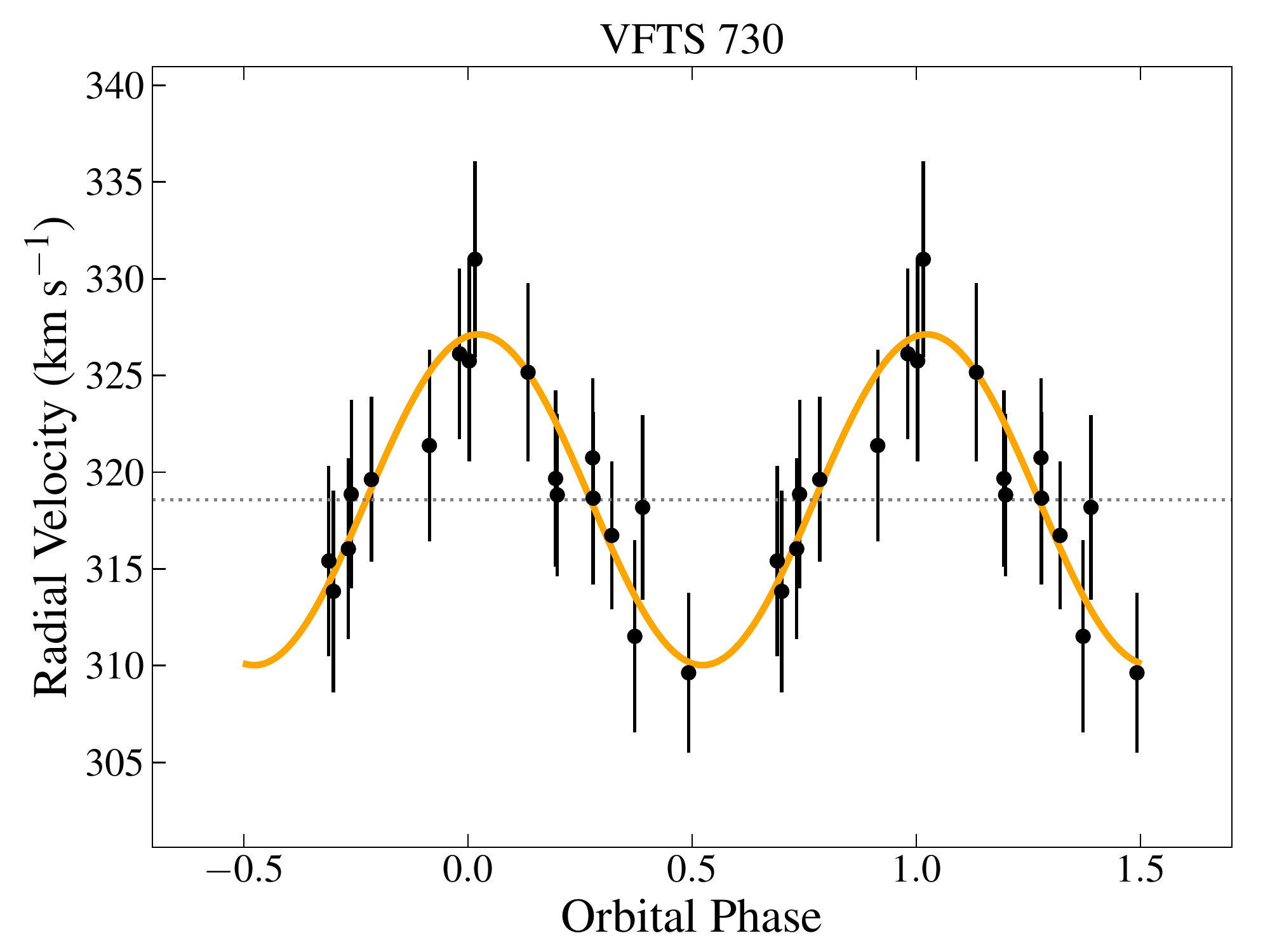}\hfill
\caption{Radial velocity curves of the SB1* systems (possible periods)}
\end{figure*}

\begin{figure*}
\ContinuedFloat
    \centering
    \includegraphics[width=0.31\textwidth]{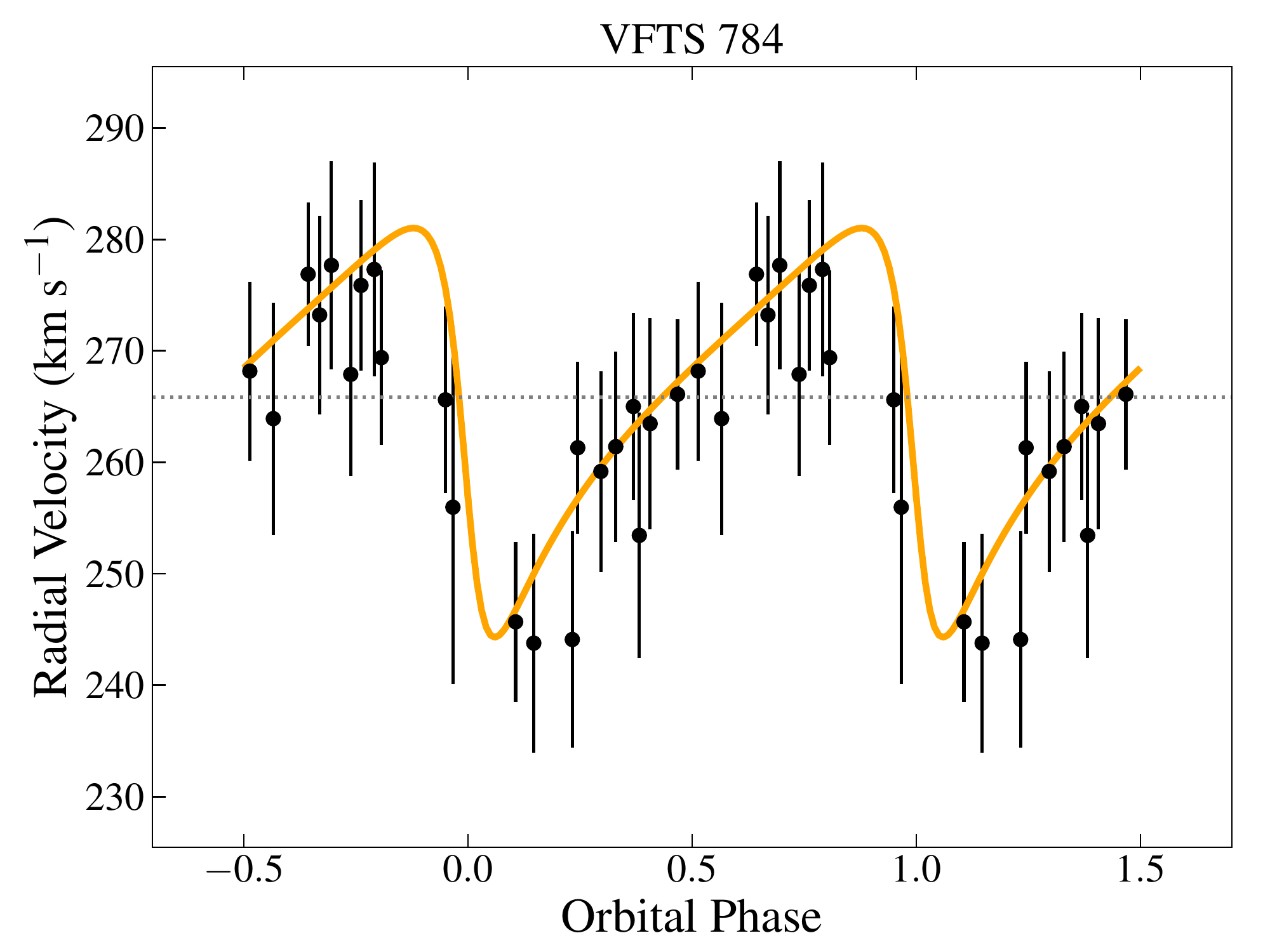}\hfill
    \includegraphics[width=0.31\textwidth]{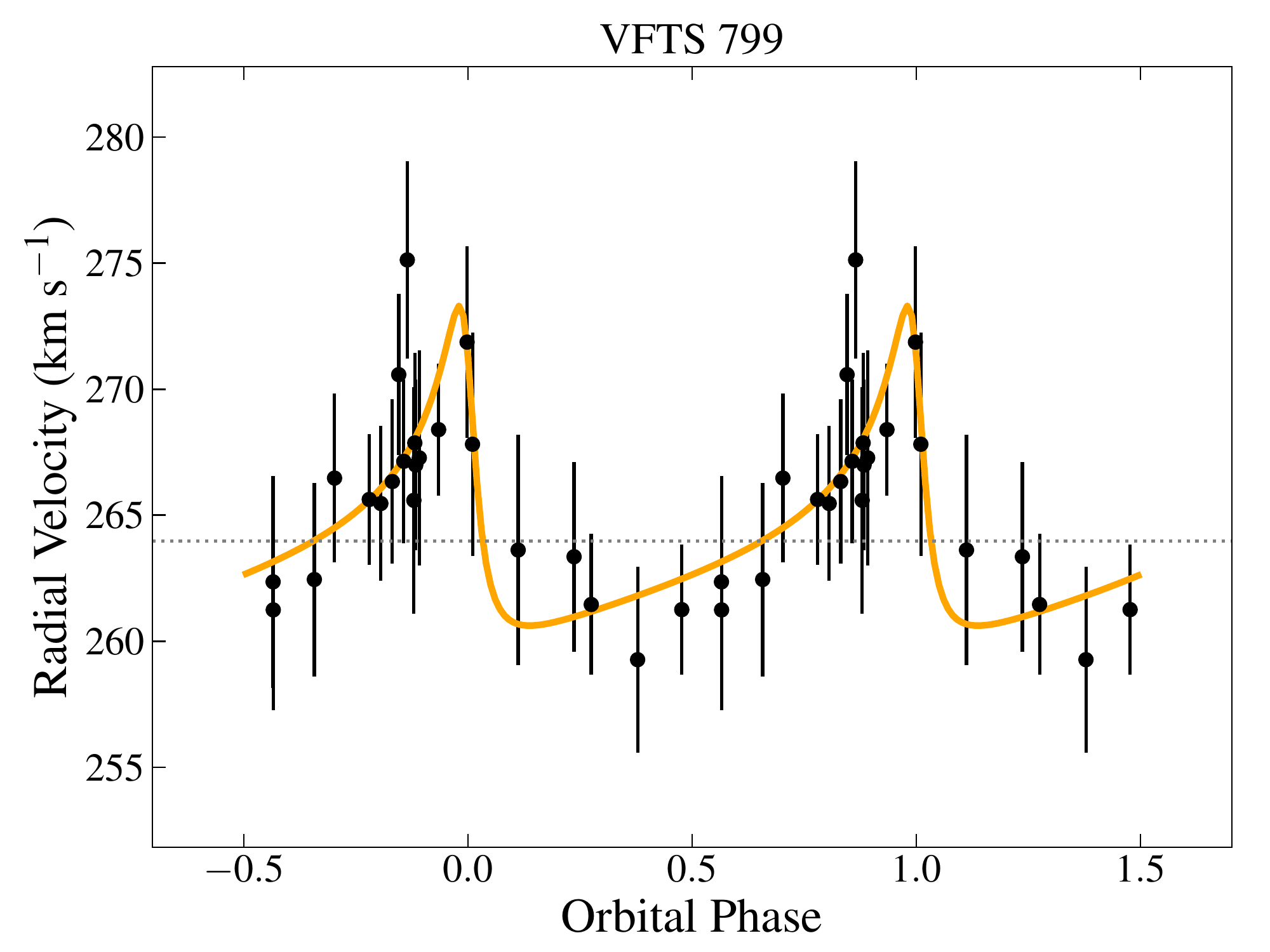}\hfill
    \includegraphics[width=0.31\textwidth]{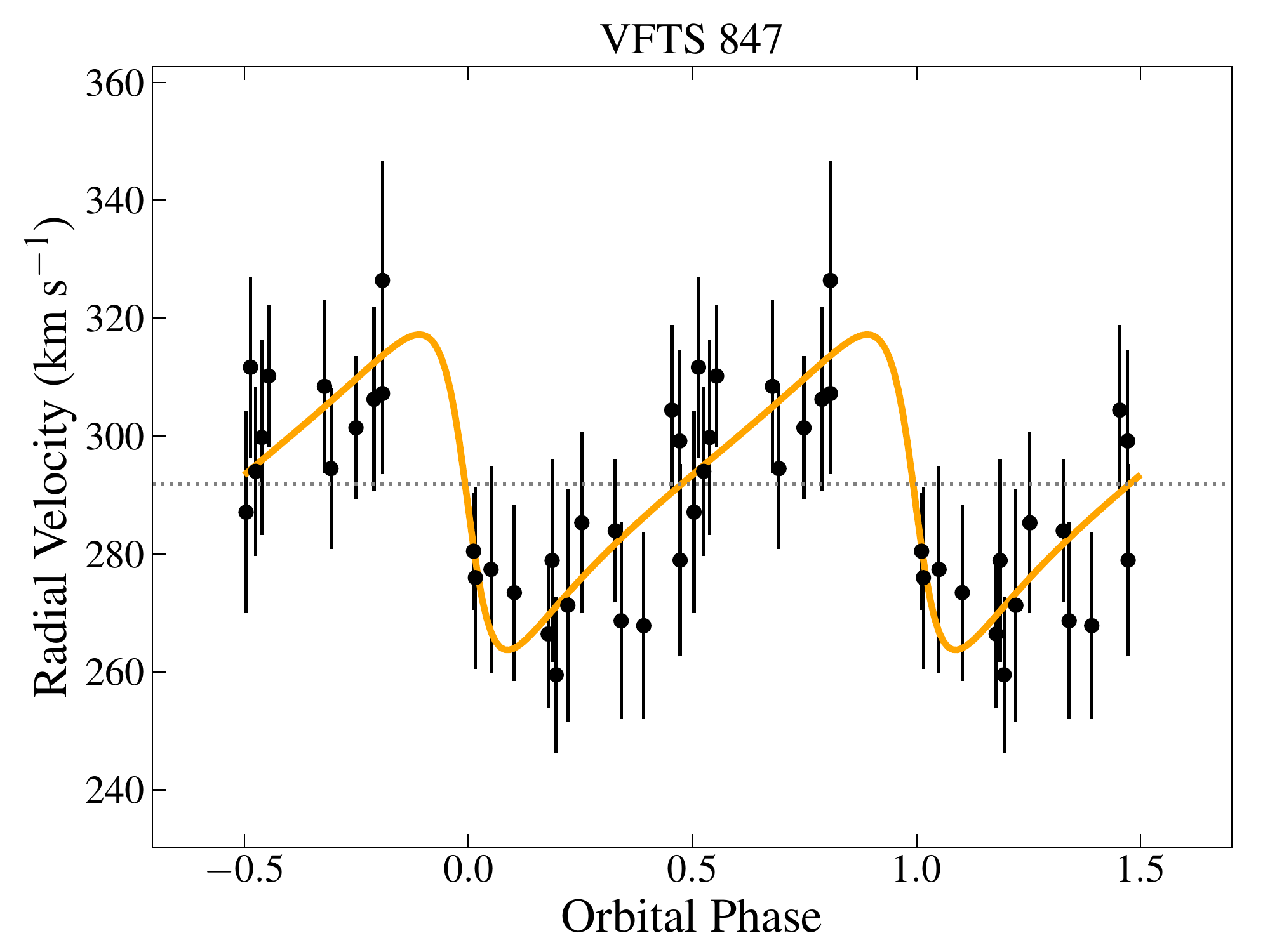}\hfill
    \includegraphics[width=0.31\textwidth]{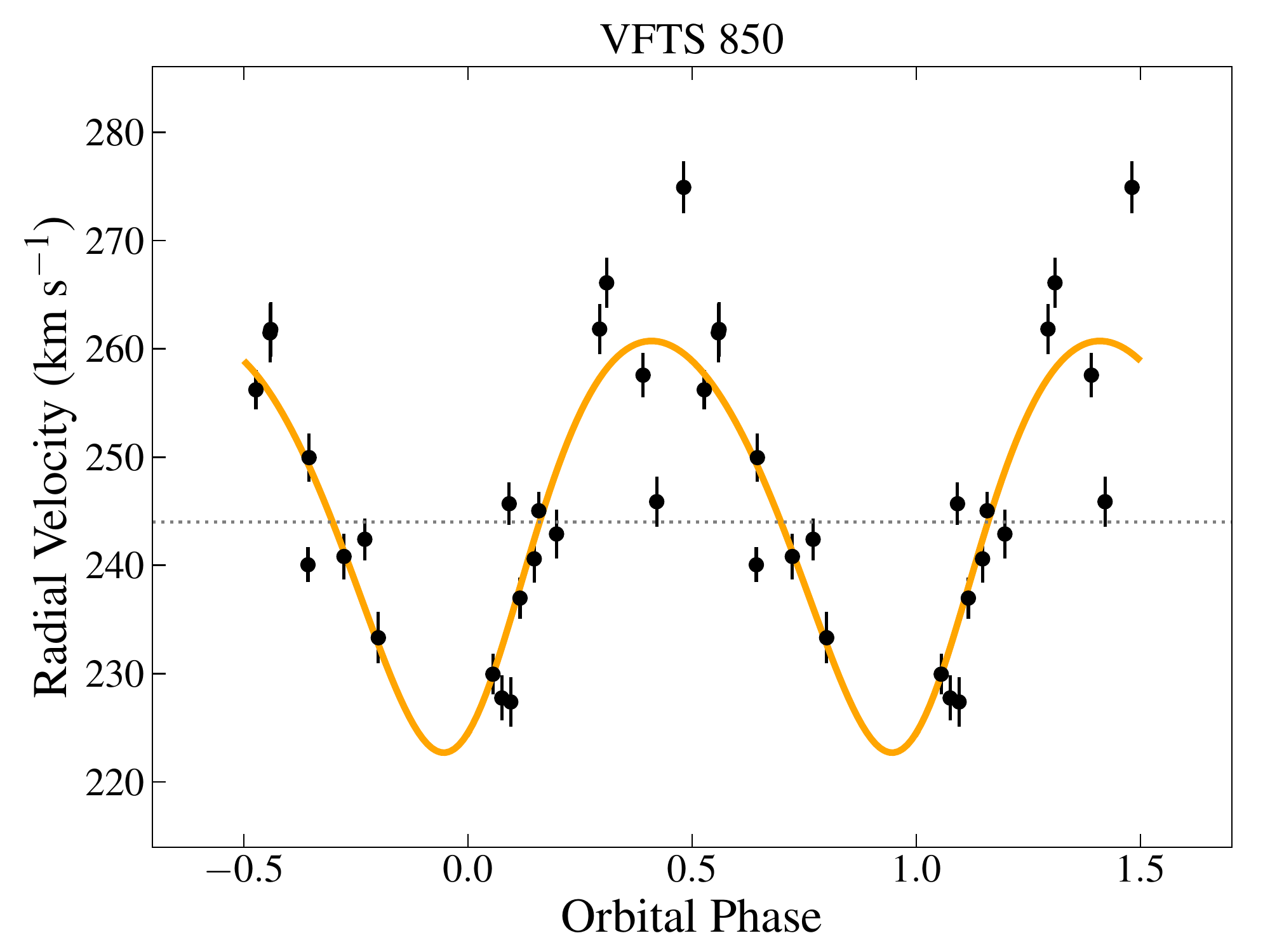}\hspace{0.035\textwidth}
    \includegraphics[width=0.31\textwidth]{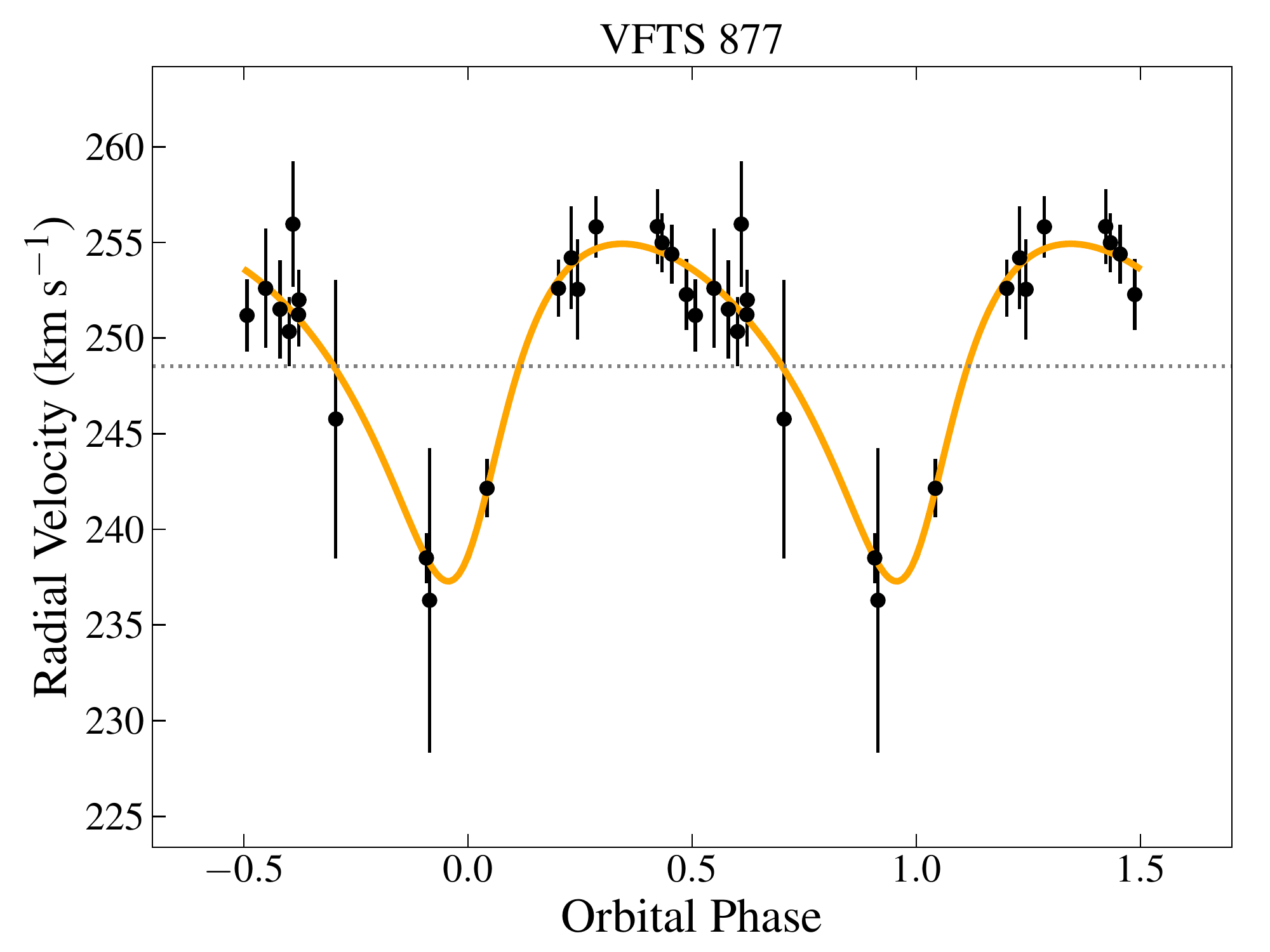}
\caption{$-$ \it continued}
\end{figure*}

\begin{figure*}
    \centering
    \includegraphics[width=0.31\textwidth]{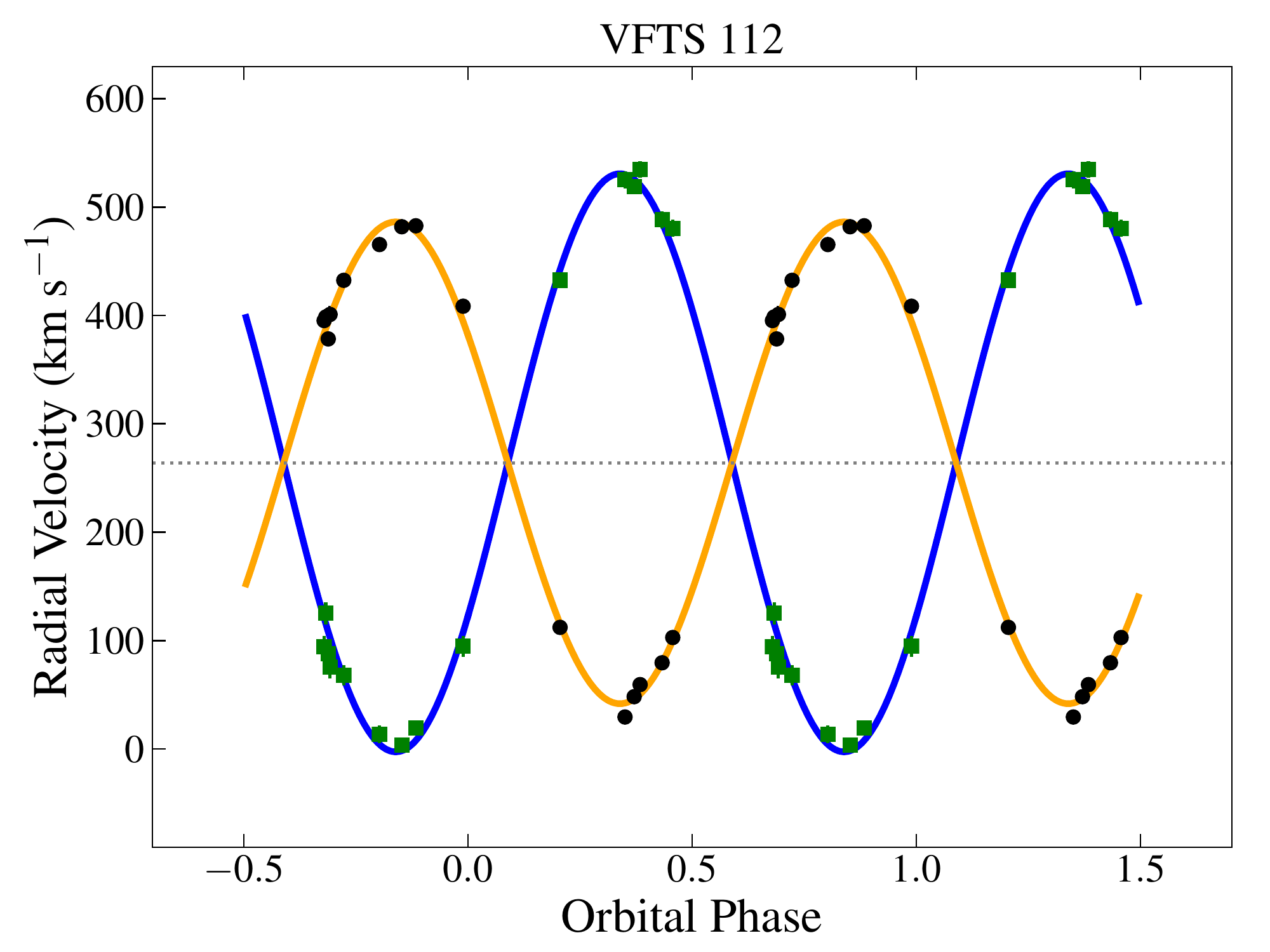}\hfill
    \includegraphics[width=0.31\textwidth]{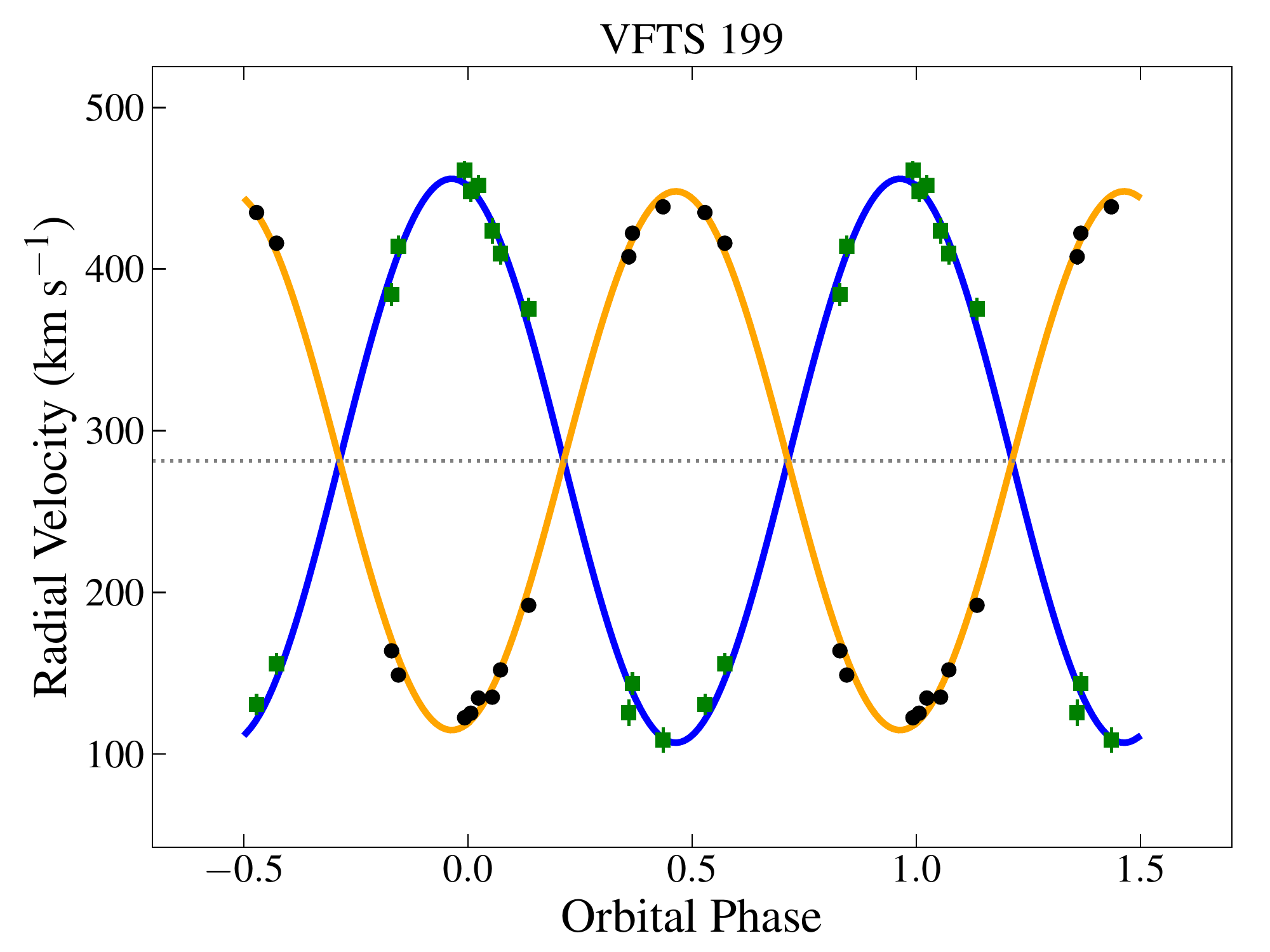}\hfill
    \includegraphics[width=0.31\textwidth]{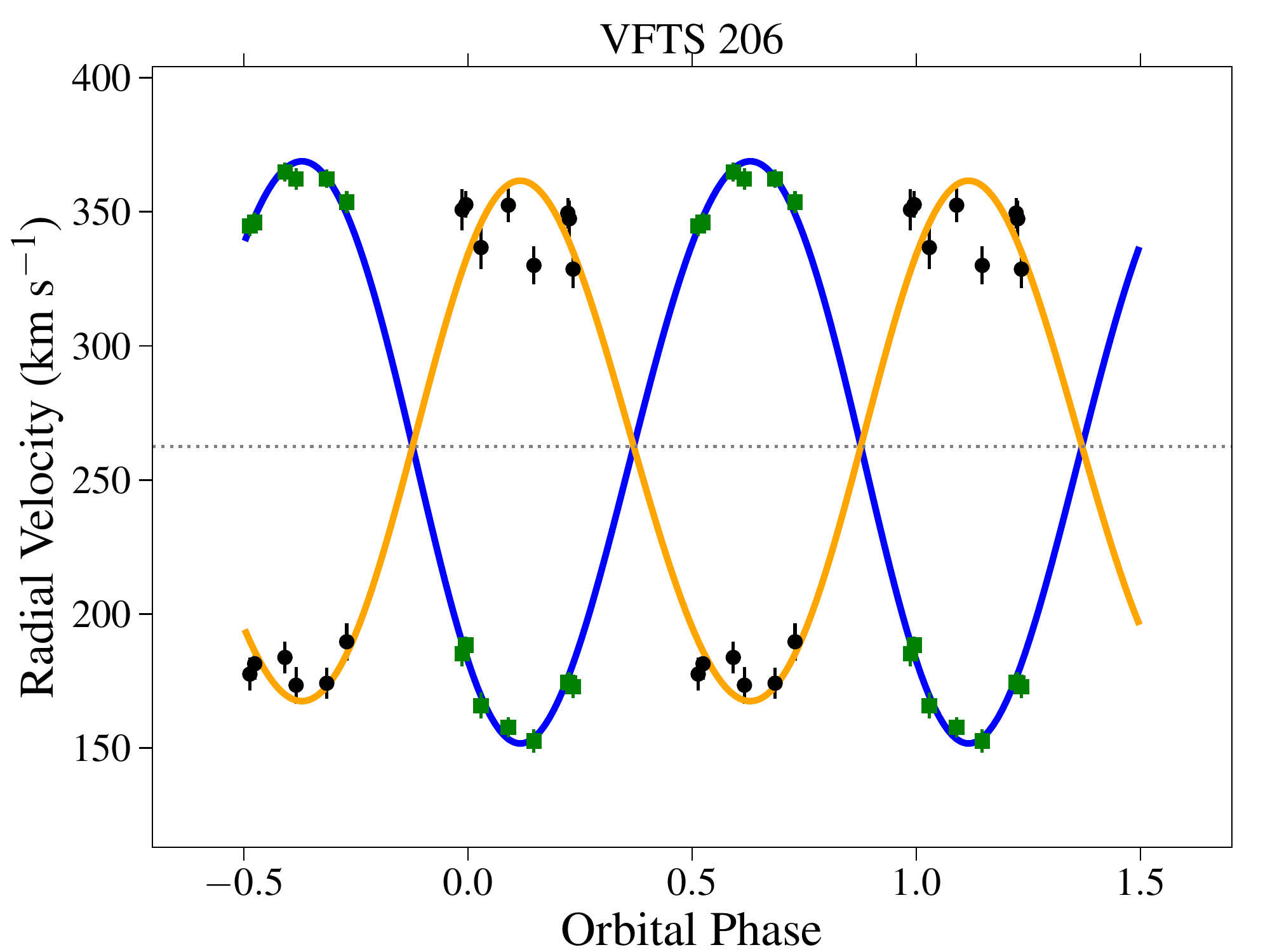}\hfill
    \includegraphics[width=0.31\textwidth]{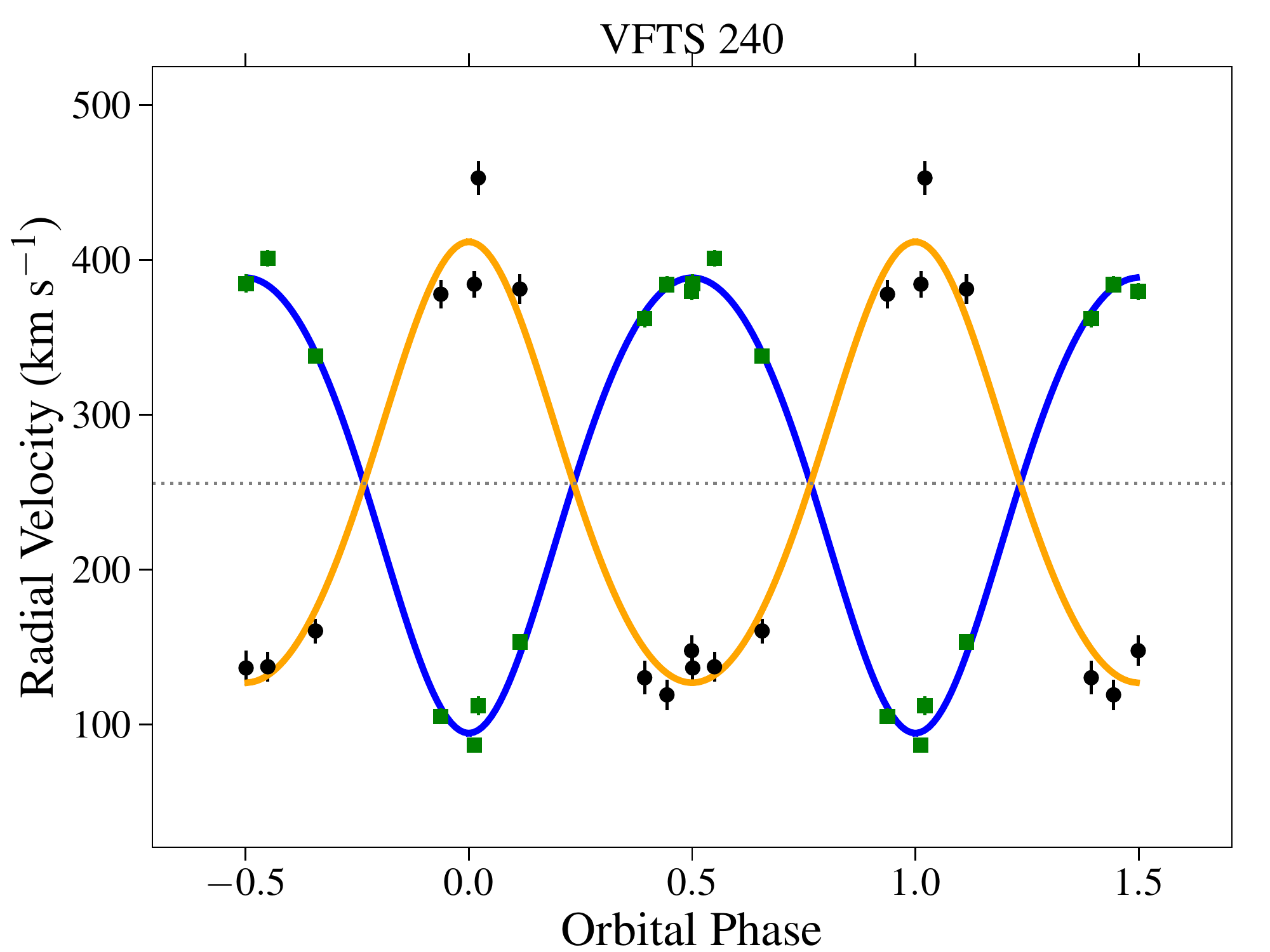}\hfill
    \includegraphics[width=0.31\textwidth]{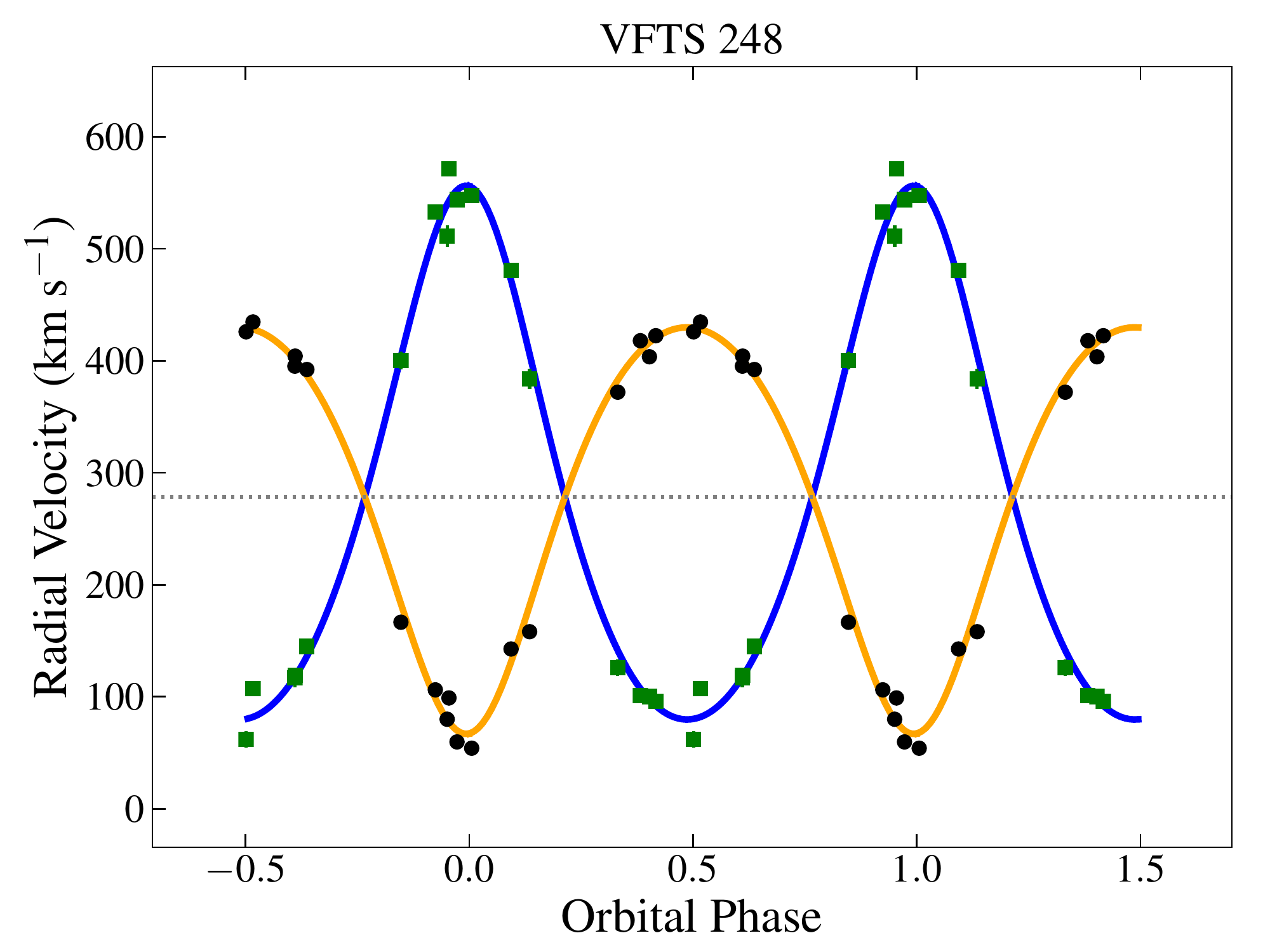}\hfill
    \includegraphics[width=0.31\textwidth]{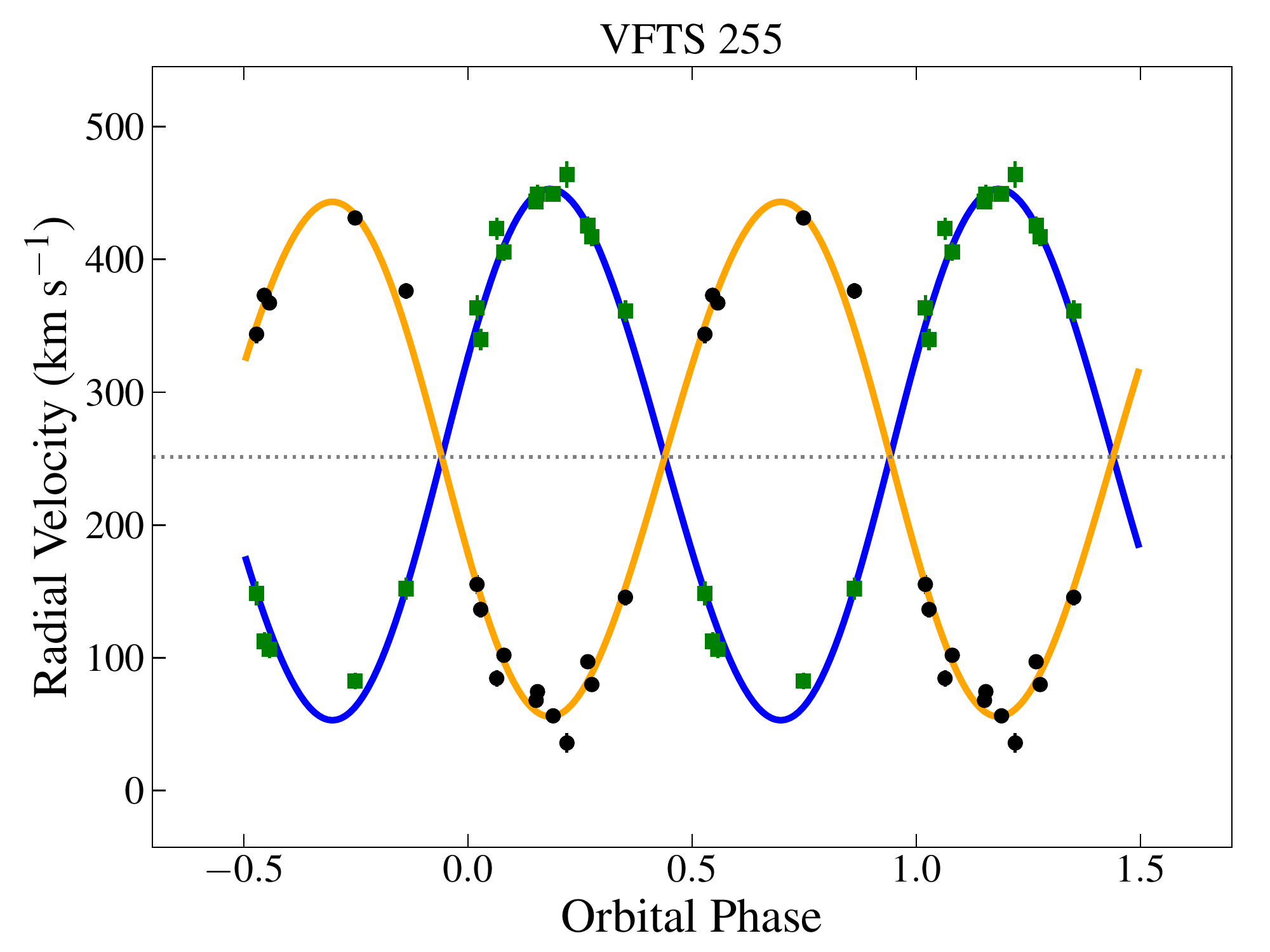}\hfill
    \includegraphics[width=0.31\textwidth]{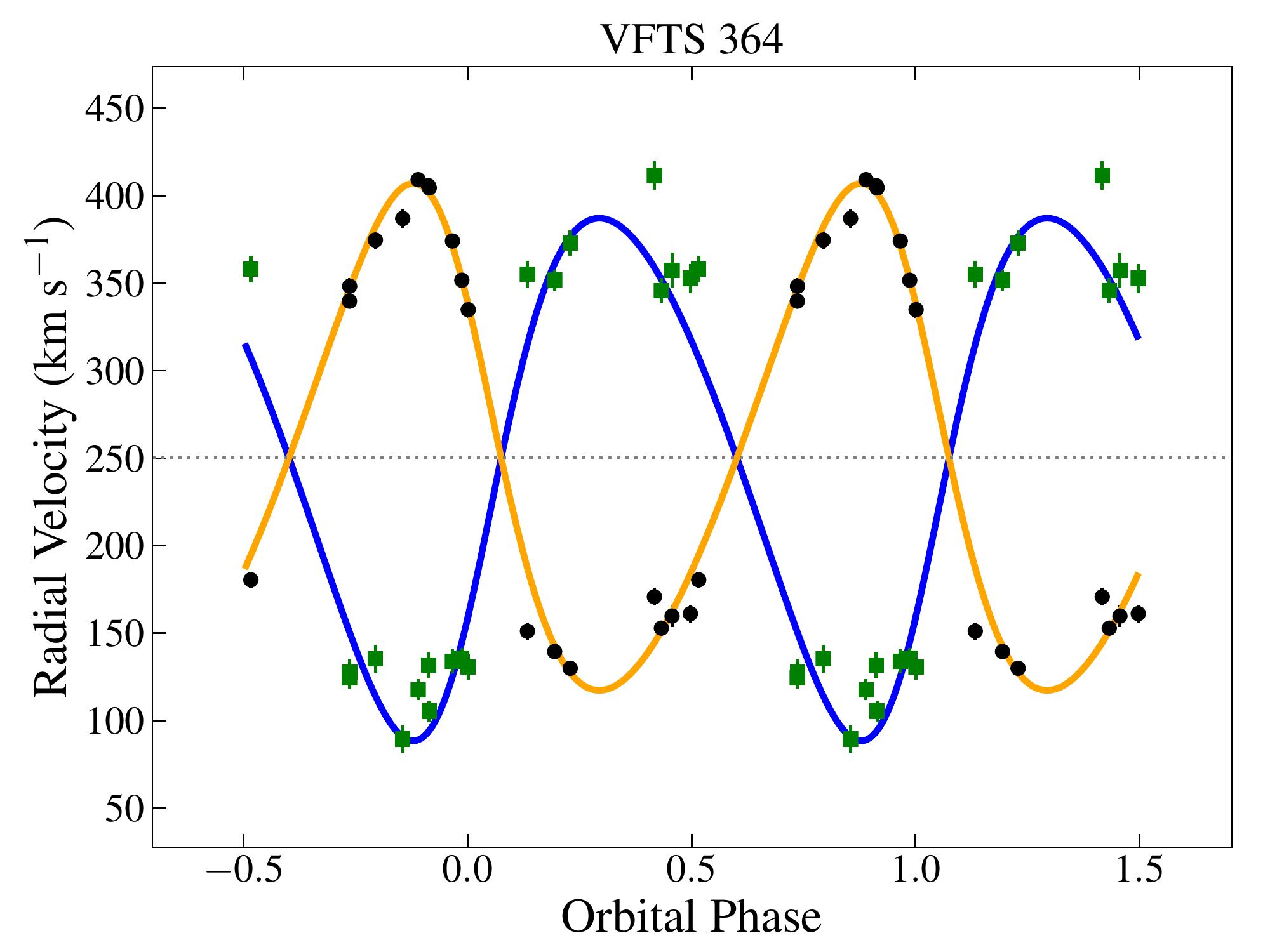}\hfill
    \includegraphics[width=0.31\textwidth]{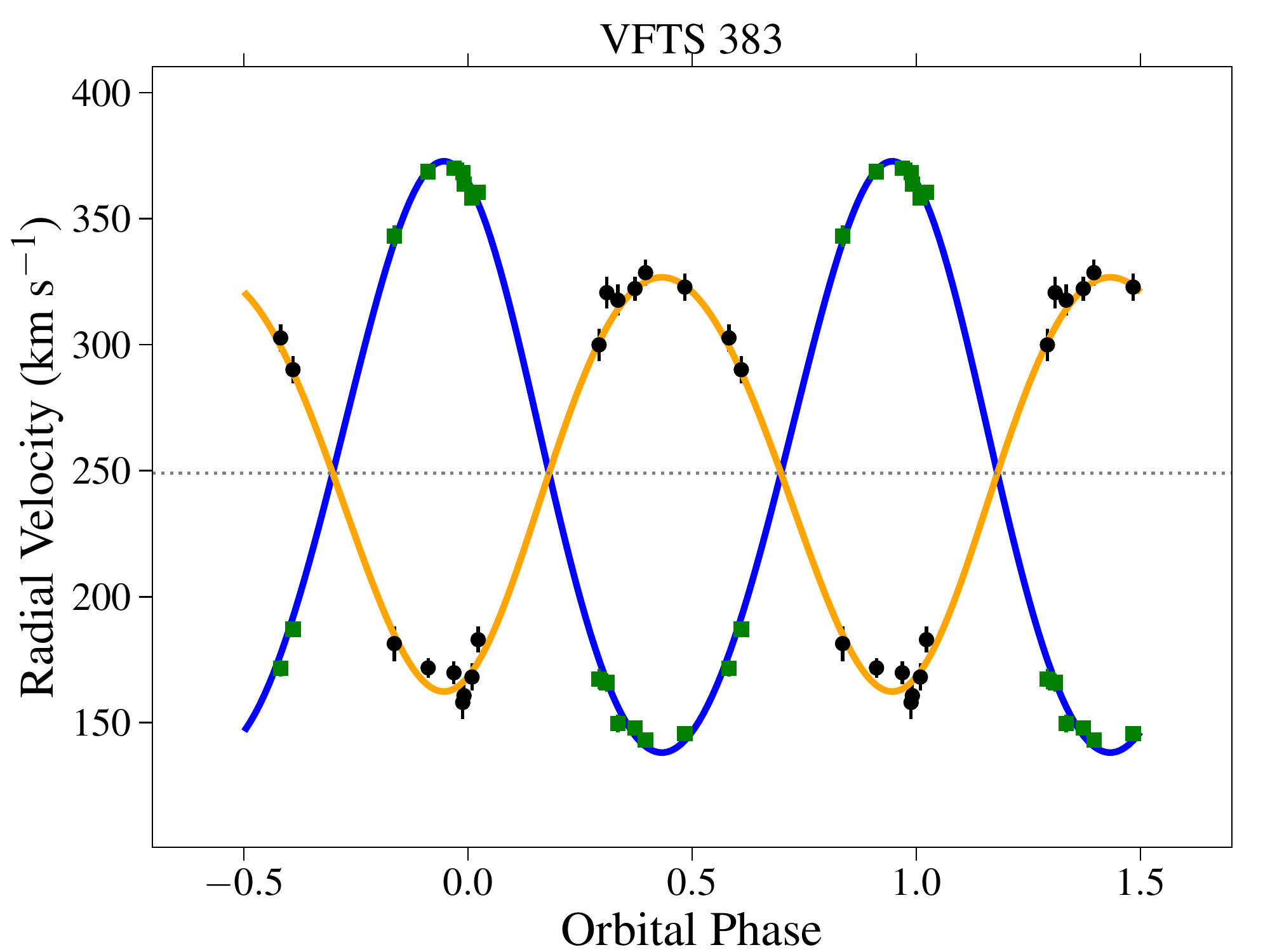}\hfill
    \includegraphics[width=0.31\textwidth]{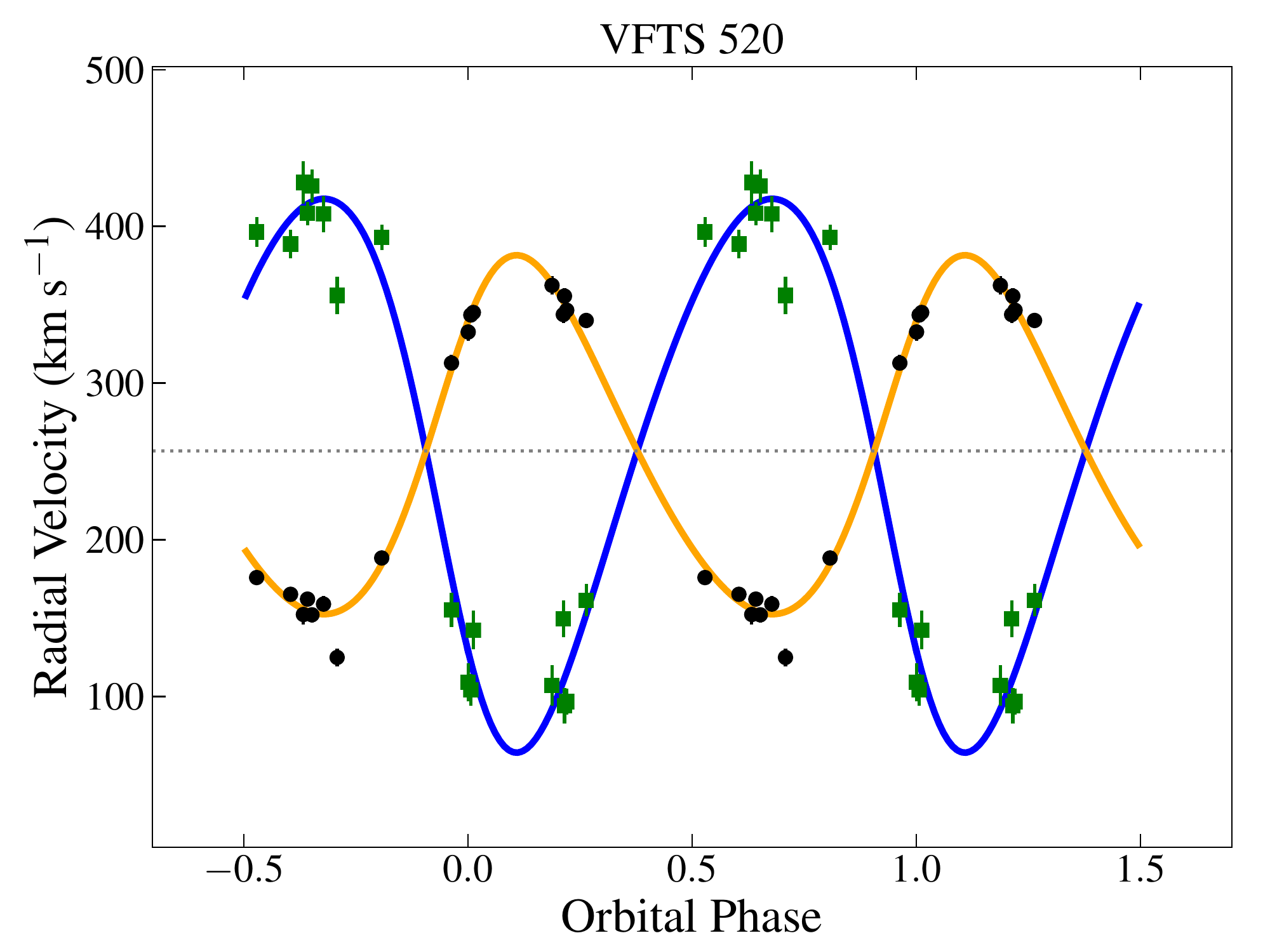}\hfill
    \includegraphics[width=0.31\textwidth]{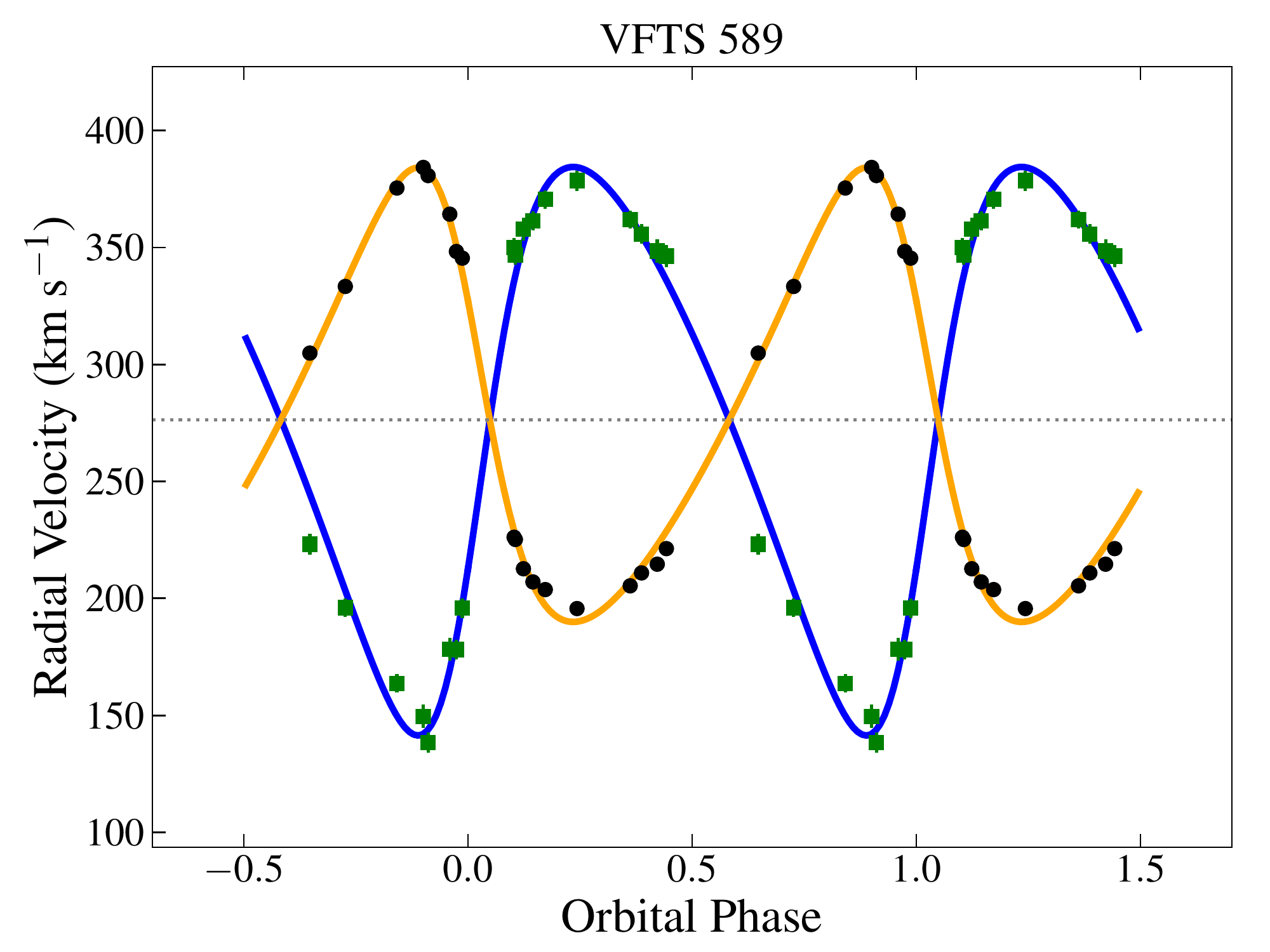}\hfill
    \includegraphics[width=0.31\textwidth]{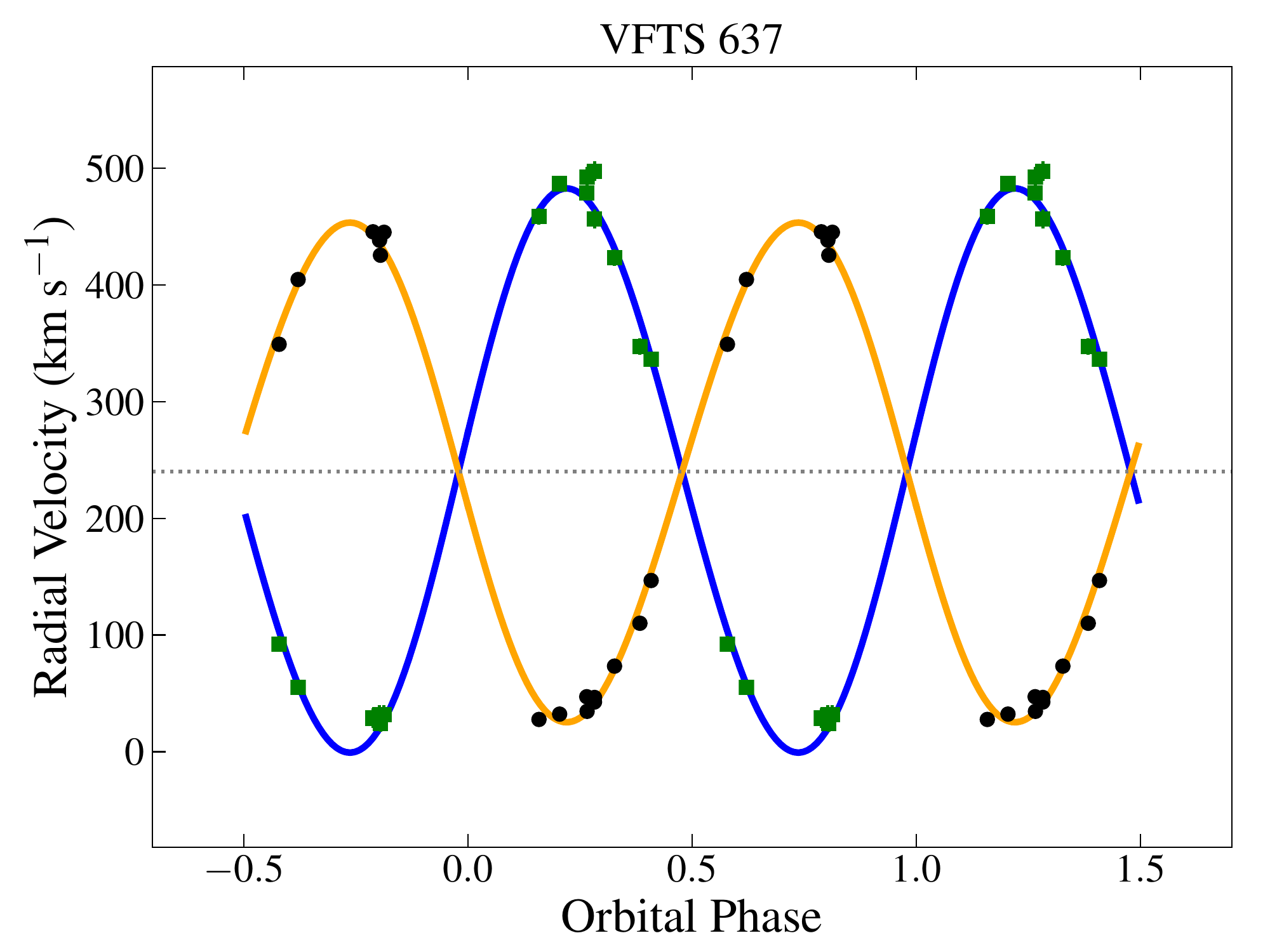}\hfill
    \includegraphics[width=0.31\textwidth]{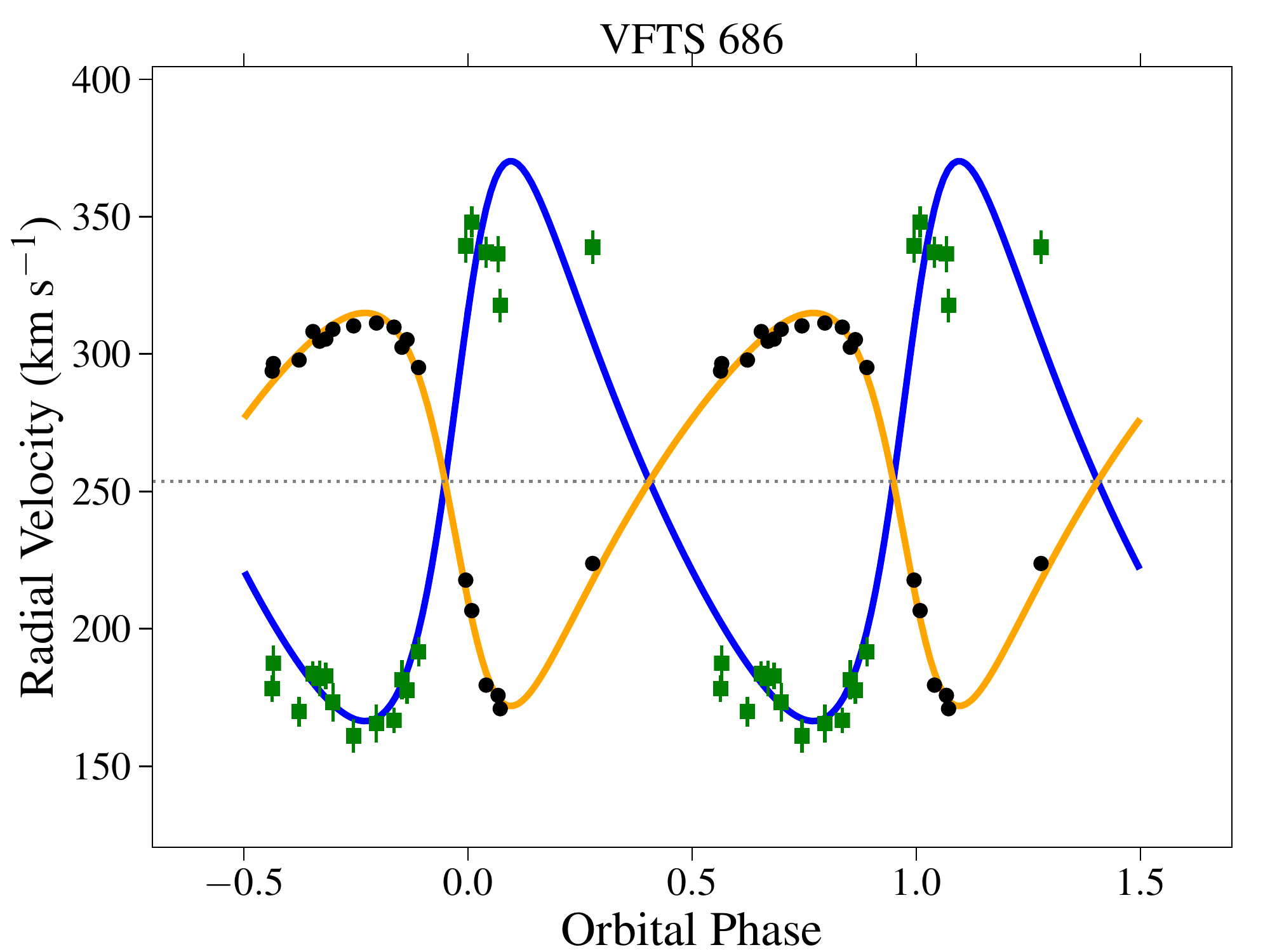}\hfill
    \includegraphics[width=0.31\textwidth]{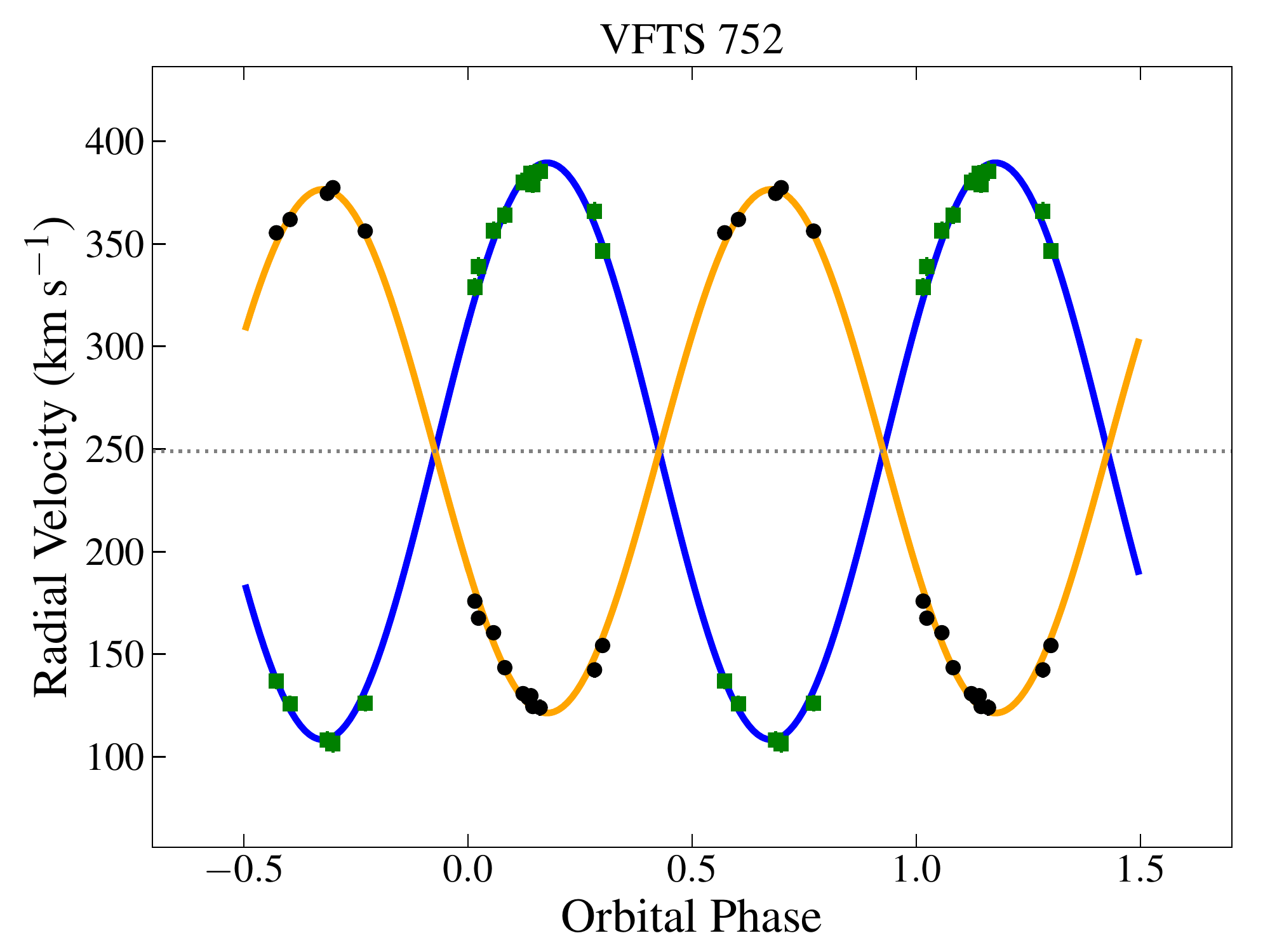}\hspace{0.035\textwidth}
    \includegraphics[width=0.31\textwidth]{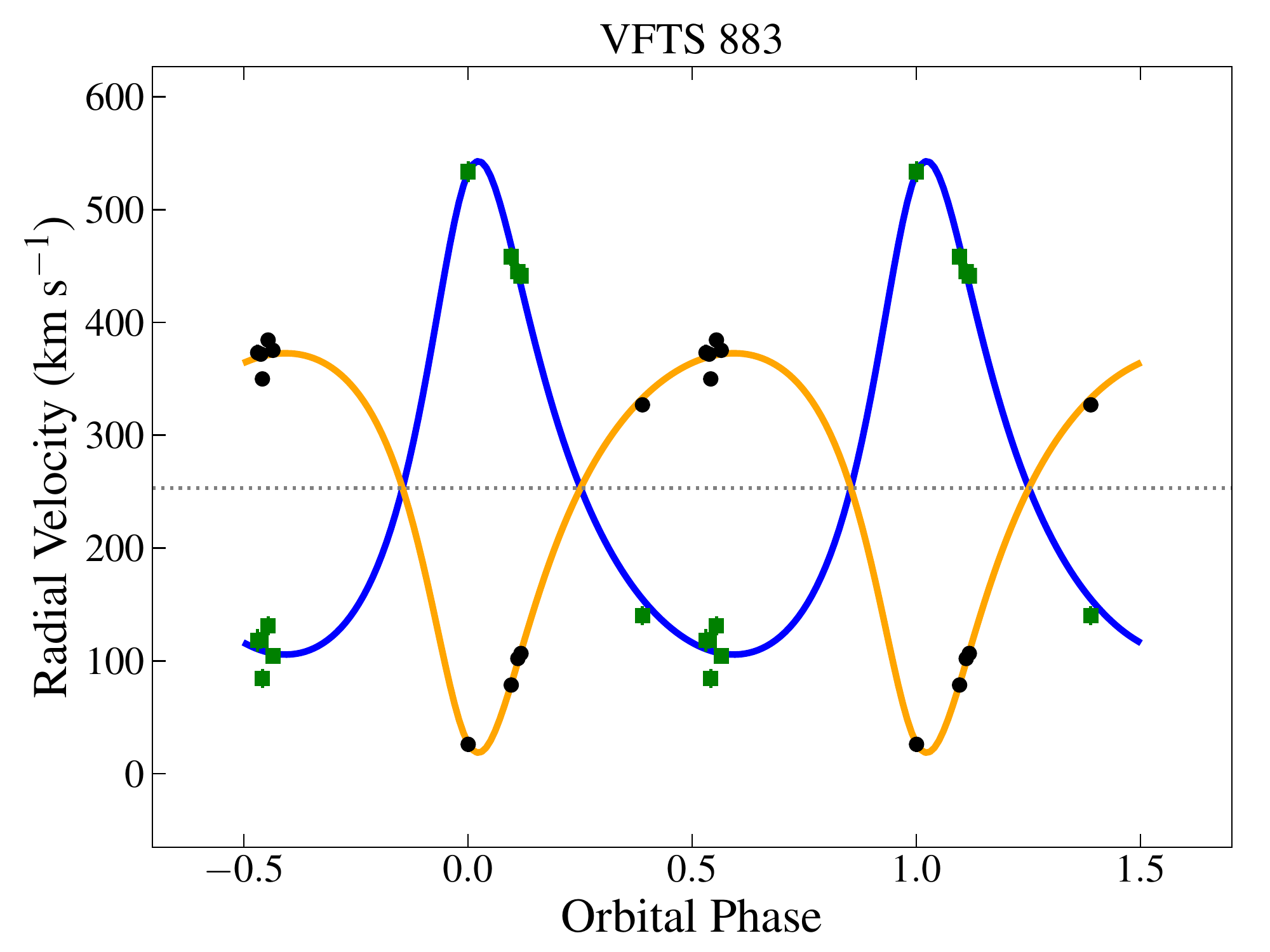}
\caption{Radial velocity curves of the SB2 systems.}
\end{figure*}

\clearpage

%%%%%%%%%%%%%%%%%%%%%%%%%%%%%%%%%%%%%%%%%%%%%%%%%%

% Don't change these lines

\label{lastpage}
\end{document}